\documentclass[a4paper, 10pt]{article}

\usepackage{jheppub}

\usepackage{amsthm}

\usepackage[numbers,sort&compress]{natbib}
\usepackage{dsfont}
\usepackage{slashed}
\usepackage{color}
\usepackage{empheq}

\usepackage{adjustbox}

\usepackage{bibgerm}

\usepackage{soul}

\usepackage[english]{babel}

\usepackage[utf8x]{inputenc}

\usepackage{graphicx,epsfig}
\usepackage{pstricks}

\usepackage{tikz}

\usepackage{bashful}

\usepackage{subfigure}

\usepackage{setspace}

\usepackage{booktabs}

\usepackage{float}

\usepackage{nextpage}

\usepackage{pdfpages}

\usepackage{hypcap}

\usepackage[small,bf]{caption}

\usepackage{longtable}

\usepackage{fancyhdr}

\usepackage[makeroom]{cancel}

\usepackage{microtype}

\usepackage{feynmp-auto-mod}

\usepackage{xparse}
\usepackage{etoolbox}

\usepackage[subfigure]{tocloft}

\usepackage{listings}

\newcommand{\p}{\partial}
\newcommand{\pl}{\overleftarrow{\partial}\hspace{-0.75mm}}

\renewcommand{\Re}{\mathrm{Re}}
\renewcommand{\Im}{\mathrm{Im}}

\newcommand{\<}{\langle}
\renewcommand{\>}{\rangle}

\renewcommand{\O}{\mathcal{O}}

\newcommand{\tr}{\mathrm{Tr}}
\newcommand{\M}{\mathcal{M}}
\newcommand{\W}{\mathcal{W}}
\newcommand{\Z}{\mathcal{Z}}
\newcommand{\B}{\mathcal{B}}
\newcommand{\A}{\mathcal{A}}
\newcommand{\C}{\mathcal{C}}

\newcommand{\D}{\mathcal{D}}
\newcommand{\G}{\mathcal{G}}
\renewcommand{\L}{\mathcal{L}}
\newcommand{\E}{\mathcal{E}}

\newcommand{\gz}{\bar g_\mathcal{Z}}
\newcommand{\mw}{M_\mathcal{W}}
\newcommand{\mz}{M_\mathcal{Z}}
\newcommand{\chkin}{c_{H,\mathrm{kin}}}
\newcommand{\cgkin}{c_{G,\mathrm{kin}}}
\newcommand{\nn}{\nonumber\\ }
\newcommand{\nnw}{\nonumber\\[0.2cm]}

\newcommand{\msbar}{$\overline{\text{MS}}$}

\newcommand{\La}{\Lambda_\epsilon}

\newcommand{\hc}{\mathrm{h.c.}}

\NewDocumentCommand{\g}{ O{} }{
	\ifblank{#1}{\bar g}{\bar g_{#1}}
}

\newcommand{\op}[3]{\O^{#2,#3}_{#1}}

\NewDocumentCommand{\Op}{ m m O{} o }{
	\O^{\ifblank{#3}{}{#3,}#2 }_{\IfNoValueTF{#4}{#1}{\substack{#1\\#4}}}
}
\NewDocumentCommand{\Ev}{ m m O{} o }{
	\E^{\ifblank{#3}{}{#3,}#2 }_{\IfNoValueTF{#4}{#1}{\substack{#1\\#4}}}
}
\NewDocumentCommand{\lwc}{ m m O{} o }{
	L^{\ifblank{#3}{}{#3,}#2 }_{\IfNoValueTF{#4}{#1}{\substack{#1\\#4}}}
}
\NewDocumentCommand{\kwc}{ m m O{} o }{
	K^{\ifblank{#3}{}{#3,}#2 }_{\IfNoValueTF{#4}{#1}{\substack{#1\\#4}}}
}
\NewDocumentCommand{\dlwc}{ m m O{} o }{
	{\dot L}^{\ifblank{#3}{}{#3,}#2 }_{\IfNoValueTF{#4}{#1}{\substack{#1\\#4}}}
}
\NewDocumentCommand{\cwc}{ m m O{} o }{
	C^{\ifblank{#3}{}{#3,}#2 }_{\IfNoValueTF{#4}{#1}{\substack{#1\\#4}}}
}

\newcommand{\XXint}[3]{{\setbox0=\hbox{$#1{#2#3}{\int}$}
\vcenter{\hbox{$#2#3$ }}\kern-.65\wd0}}

\newcommand{\remark}[1]{}

\newcommand{\mytag}{\\[-\baselineskip] \stepcounter{equation}\tag{\theequation}}

\definecolor{darkgreen}{rgb}{0,0.5,0}
\definecolor{darkblue}{rgb}{0,0,0.5}
\definecolor{darkred}{rgb}{0.5,0,0}
\definecolor{beige}{rgb}{0.7,0.4,0.3}

\makeatletter
  \def\my@tag@font{\normalsize}
  \def\maketag@@@#1{\hbox{\m@th\normalfont\my@tag@font#1}}
  \let\amsmath@eqref\eqref
  \renewcommand\eqref[1]{{\let\my@tag@font\relax\amsmath@eqref{#1}}}
\makeatother

\newenvironment{myfmf}[1]
{\begin{fmffile}{#1}
\fmfcmd{%
  style_def wboson expr p =
  cdraw (wiggly p);
  shrink (1);
  cfill (arrow p);
  endshrink;
  enddef;}
\fmfcmd{%
  style_def momins expr p =
  drawarrow p;
  enddef;}
  }
{
\end{fmffile} 
}

\makeatletter
\renewcommand\paragraph{\@startsection{paragraph}{4}{\z@}%
  {-3.25ex\@plus -1ex \@minus -.2ex}%
  {1.5ex \@plus .2ex}%
  {\normalfont\normalsize\bfseries}}
\makeatother

\cftsetindents{section}{0em}{2em}
\cftsetindents{subsection}{2em}{3em}
\cftsetindents{subsubsection}{5em}{4em}

\allowdisplaybreaks[1]

\title{\boldmath Low-energy effective field theory below the electroweak scale: matching at one loop}

\author{Wouter Dekens,}

\author{Peter Stoffer}

\affiliation{Department of Physics, University of California at San Diego, 9500 Gilman Drive,\\ La Jolla, CA 92093-0319, USA}

\abstract{
We compute the one-loop matching between the Standard Model Effective Field Theory and the low-energy effective field theory below the electroweak scale, where the heavy gauge bosons, the Higgs particle, and the top quark are integrated out. The complete set of matching equations is derived including effects up to dimension six in the power counting of both theories. We present the results for general flavor structures and include both the $CP$-even and $CP$-odd sectors. The matching equations express the masses, gauge couplings, as well as the coefficients of dipole, three-gluon, and four-fermion operators in the low-energy theory in terms of the parameters of the Standard Model Effective Field Theory. Using momentum insertion, we also obtain the matching for the $CP$-violating theta angles.
Our results provide an ingredient for a  model-independent analysis of constraints on physics beyond the Standard Model. They can be used for fixed-order calculations at one-loop accuracy and represent a first step towards a systematic next-to-leading-log analysis.
}

\numberwithin{equation}{section}

\begin{document}

	\maketitle


	\begin{myfmf}{diags/diags}


\section{Introduction}

The fact that so far no particles beyond the Standard Model (SM) have been observed at the LHC indicates that New Physics (NP) is either very weakly coupled or consists of heavy particles with masses well above the electroweak scale. In the second scenario, the effects of NP on experiments at energies below the threshold of the new heavy particles can be described by an effective field theory (EFT) that contains the SM particles only. This effective description of heavy NP remains valid even in the presence of additional very weakly coupled light NP, unless one is performing a dedicated experiment at the resonance energy of the new light degree of freedom~\cite{Manohar:HEFT2019}. Under the assumption that the Higgs boson and the Goldstone bosons of spontaneous electroweak symmetry breaking transform as a complex scalar doublet of the weak gauge group, the appropriate EFT is the Standard Model Effective Field Theory (SMEFT), consisting of the SM Lagrangian and all higher-dimension operators that are invariant under the SM gauge group~\cite{Jackson:1957zz,Weinberg:1979sa,Wilczek:1979hc,Weldon:1980gi,Buchmuller:1985jz,Grzadkowski:2010es}, see~\cite{Brivio:2017vri} for a detailed review. For processes at energies below the electroweak scale, another EFT should be used, in which the heavy SM particles are integrated out: these are the top quark, the $W$ and $Z$ gauge bosons, as well as the Higgs scalar. Since these particles have comparable masses, no large logarithms are generated by performing the matching at one common scale. The resulting low-energy EFT (LEFT) corresponds to the Fermi theory of weak interactions~\cite{Fermi:1934sk}. It is given by the QED and QCD Lagrangian and all higher-dimension operators that are invariant under the unbroken subgroup $SU(3)_c \otimes U(1)_\mathrm{em}$. Working with these EFTs has the advantage that scales are separated, which both simplifies loop calculations and, through renormalization-group equations (RGEs), provides the means to systematically resum large logarithms in a leading-log expansion known as RG-improved perturbation theory. The one-loop running in the SMEFT up to dimension six was calculated in~\cite{Jenkins:2013zja,Jenkins:2013wua,Alonso:2013hga}. The low-energy operator basis and RGEs relevant for $B$-meson mixing and decay were derived in~\cite{Aebischer:2017gaw}. In~\cite{Jenkins:2017jig}, the complete LEFT operator basis up to dimension six was constructed and the tree-level matching equations to the SMEFT were provided. The one-loop running in the LEFT up to dimension-six effects was calculated in~\cite{Jenkins:2017dyc}, completing the framework for a consistent leading-logarithm analysis of the effects of, and constraints on, NP. These RGEs allow one to connect energy scales ranging from the scale of NP down to the hadronic scale, where QCD becomes non-perturbative. Observables are then calculated by combining the running of the Wilson coefficients with the matrix elements of the effective operators, thereby explicitly separating the high- and low-energy scales in the problem.
At the hadronic scale, the matrix elements in general require a direct non-perturbative calculation or a non-perturbative matching to chiral perturbation theory~\cite{Weinberg:1968de,Gasser:1983yg,Gasser:1984gg}. Interestingly, the non-perturbative nature of QCD at low energies has an impact even on processes without any external hadrons, which is of importance e.g.\ when constraining lepton-flavor-violating NP in the process $\mu\to e\gamma$~\cite{Dekens:2018pbu}.

EFT analyses of constraints on NP is a rapidly developing field. For example, the leading-log calculations have been implemented into software tools that provide automated calculations of bounds on Wilson coefficients~\cite{Celis:2017hod,Aebischer:2018bkb,Straub:2018kue} or support an automated calculation of processes at the tree and loop level~\cite{Brivio:2017btx,Dedes:2019uzs}.

The absence of evidence for NP on the one hand as well as the increasing precision in experiments on the other hand require an improved accuracy in the theoretical calculations. The constraining power e.g.\ of searches for lepton-flavor-violating processes~\cite{Adam:2013mnn,TheMEG:2016wtm,Bertl:2006up,Cui:2009zz,Kutschke:2011ux,Kuno:2013mha} or electric dipole moments~\cite{Baron:2013eja,Andreev:2018ayy} calls for an analysis beyond leading logarithms~\cite{Pruna:2014asa,Crivellin:2017rmk,Panico:2018hal}. The motivation to go to higher loop orders is not to obtain a small correction to a coefficient that is already constrained to be tiny, but to obtain new constraints due to a richer mixing structure, as well as to assess the uncertainties in the theoretical calculation. Here, we present the calculation of the matching equations at the weak scale to one-loop accuracy, which is the natural next step in this direction. The results enable simplified calculations at fixed one-loop order with EFT methods for cases where no resumming of logarithms is required. They also provide one ingredient in an extension to next-to-leading-logarithmic accuracy, which however will require the two-loop anomalous dimensions as well. Results for the two- and three-loop anomalous dimensions of the three-gluon operator in the SMEFT have recently been presented~\cite{deVries:2019nsu}.
One-loop matching at the weak scale has been previously considered in the context of specific processes, i.e.\ restricted to a limited subset of operators. The most complete partial results so far were given in the context of $B$ physics~\cite{Aebischer:2015fzz,Hurth:2019ula}.

The structure of the paper is as follows. In Sect.~\ref{sec:SMEFT}, we recall the properties of the SMEFT in the broken phase, mainly to set the conventions for the parameters, choice of basis, and gauge fixing. In Sect.~\ref{sec:LEFT}, we describe the basis choice for the LEFT that we use for the one-loop matching. In Sect.~\ref{sec:LoopCalculation}, we discuss the technical details of the loop calculation including regularization and scheme definitions. In Sect.~\ref{sec:Matching}, we describe the matching procedure and derive the generic matching equations. We also comment on the calculation of observables. In Sect.~\ref{sec:Results}, we highlight some aspects of the results of the calculation. Since the explicit expressions for the matching equations are very long, we provide them in electronic form as supplemental material. We offer our conclusions in Sect.~\ref{sec:Conclusions}, while the appendices provide some further comments on conventions, Majorana fermions, as well as the complete list of Feynman diagrams that were calculated.


\section{SMEFT in the broken phase}
\label{sec:SMEFT}

While the calculation of the divergence structure of the SMEFT can be performed in the unbroken phase~\cite{Jenkins:2013zja,Jenkins:2013wua,Alonso:2013hga}, in the matching to the LEFT one has to integrate out the heavy particles, which implies that one is required to work within the broken phase. In the following, we discuss the transition from the unbroken to the broken phase and the corrections that are induced by the presence of higher-dimension operators compared to the SM, in order to specify our conventions. The SMEFT in the broken phase has been previously discussed in~\cite{Alonso:2013hga,Dedes:2017zog,Jenkins:2017jig}.

We start from the SMEFT Lagrangian
\begin{align}
	\L^\mathrm{SMEFT} = \L^\mathrm{SM} + \sum_i C_i Q_i \, ,
\end{align}
where the sum runs over all higher-dimension operators and the SM Lagrangian is
\begin{align}
	\label{eq:SM}
	\L _{\rm SM} &= -\frac14 G_{\mu \nu}^A G^{A\mu \nu}-\frac14 W_{\mu \nu}^I W^{I \mu \nu} -\frac14 B_{\mu \nu} B^{\mu \nu}
		+ (D_\mu H^\dagger)(D^\mu H)
		+\sum_{\mathclap{\psi=q,u,d,l,e}} \overline \psi\, i \slashed{D} \, \psi\nn
		&-\lambda \left(H^\dagger H -\frac12 v^2\right)^2- \biggl[ H^{\dagger i} \overline d\, Y_d\, q_{i} + \widetilde H^{\dagger i} \overline u\, Y_u\, q_{i} + H^{\dagger i} \overline e\, Y_e\,  l_{i} + \hc \biggr]\nn
		&+ \frac{\theta_3 g_3^2}{32 \pi^2} \, G^A_{\mu \nu} \widetilde G^{A\, \mu \nu} + \frac{\theta_2 g_2^2}{32 \pi^2} \, W^I_{\mu \nu} \widetilde W^{I\, \mu \nu} + \frac{\theta_1 g_1^2}{32 \pi^2} \, B_{\mu \nu} \widetilde B^{\mu \nu} \, .
\end{align}
The covariant derivative is given by
\begin{align}
	D_\mu = \p_\mu + i g_3 T^A G_\mu^A + i g_2 t^I W_\mu^I + i g_1 y B_\mu \, ,
\end{align}
with the $SU(3)$ generators $T^A$, the $SU(2)$ generators $t^I = \tau^I / 2$, and the $U(1)$ hypercharge generator $y$. The Yukawa matrices $Y_{u,d,e}$ are $3\times3$ matrices in flavor space. We reproduce the SMEFT operators up to dimension six in App.~\ref{sec:SMEFTBasis}.

\subsection{Scalar sector}

The scalar potential and kinetic energy terms
\begin{align}
	\L_H = (D_\mu H^\dagger)(D^\mu H) - \lambda \left( H^\dagger H - \frac{1}{2} v^2 \right)^2
\end{align}
get modified by dimension-six operators in SMEFT:
\begin{align}
	\begin{split}
		\L^{(6)}_H &= C_{H\Box} Q_{H\Box} + C_{HD} Q_{HD} + C_H Q_H \\
			&= C_{H\Box} (H^\dagger H) \Box (H^\dagger H) + C_{HD} (H^\dagger D_\mu H)^* (H^\dagger D^\mu H) + C_H (H^\dagger H)^3 .
	\end{split}
\end{align}
The minimum of the potential at tree level is given by the modified vacuum-expectation value (vev)~\cite{Alonso:2013hga}
\begin{align}
	\label{eq:Vev}
	\<H^\dagger H\> = \frac{1}{2}v_T^2 , \quad v = \left( 1 - \frac{3 C_H v_T^2}{8\lambda} \right) v_T .
\end{align}
The Higgs doublet is given by
\begin{align}
	H = \frac{1}{\sqrt{2}} \begin{pmatrix} \phi_2 + i \phi_1 \\ \phi_4  -i \phi_3  \end{pmatrix} \, ,
\end{align}
where 
\begin{align}
	\phi_4 = [ 1 + \chkin ] h + v_T \, , \quad \phi_3 = [ 1 + \cgkin ] G^0 \, , \quad \frac{1}{\sqrt{2}} ( \phi_2 \pm i \phi_1 ) = G^\pm \, .
\end{align}
The rescaling factors
\begin{align}
	\chkin &= \left( C_{H\Box} - \frac{1}{4} C_{HD} \right) v_T^2 \, , \quad
	\cgkin = - \frac{1}{4} C_{HD} v_T^2
\end{align}
ensure canonically normalized kinetic terms in the presence of the dimension-six operators~\cite{Alonso:2013hga,Dedes:2017zog}.\footnote{Note that~\cite{Dedes:2017zog} uses a different convention for the scalar self-coupling: $\lambda^\text{\cite{Dedes:2017zog}} = 2 \lambda$.}

The Higgs mass term is given by
\begin{align}
	\label{eq:HiggsMass}
	\L_{h,\mathrm{mass}} = - \frac{1}{2} M_H^2 h^2 \, , \quad M_H^2 = 2 \lambda v_T^2 \left( 1 + 2 \chkin - \frac{3C_H v_T^2}{2\lambda}\right) .
\end{align}

\subsection{Gauge sector}
\label{sec:SMEFTGaugeBosons}

The Higgs and gauge kinetic terms at dimension four are given by
\begin{align}
	\L_{H,\mathrm{kin}} + \L_\mathrm{gauge} = (D_\mu H^\dagger)(D^\mu H)  - \frac{1}{4} G_{\mu\nu}^A G^{A\mu\nu} - \frac{1}{4} W_{\mu\nu}^I W^{I\mu\nu} - \frac{1}{4} B_{\mu\nu} B^{\mu\nu}.
\end{align}
At dimension six, the $CP$-even operators of the class $X^2 H^2$ lead to kinetic terms of the gauge fields in the broken phase that are not canonically normalized:
\begin{align}
	\L_{X^2 H^2}^{(6),\mathrm{even}} = C_{HG} Q_{HG} + C_{HW} Q_{HW} + C_{HB} Q_{HB} + C_{HWB} Q_{HWB}  \, .
\end{align}
We define the following rescaled gauge fields and couplings~\cite{Alonso:2013hga}:
\begin{align}
	\label{eq:GaugeFieldRescaling}
	G_\mu^A &= \G_\mu^A ( 1 + v_T^2 C_{HG}) , \quad & W_\mu^I &= \W_\mu^I ( 1 + v_T^2 C_{HW} ) , \quad & B_\mu &= \B_\mu ( 1 + v_T^2 C_{HB} ) \, , \nn
	\bar g_3 &= g_3 ( 1 + v_T^2 C_{HG} ) , & \g[2] &= g_2 ( 1 + v_T^2 C_{HW} ) , & \g[1] &= g_1 ( 1 + v_T^2 C_{HB} ) \, ,
\end{align}
where the fields $\G_\mu^A$, $\W_\mu^I$, and $\B_\mu$ have canonically normalized kinetic terms and the products of couplings and fields is unchanged, $g_3 G_\mu^A = \bar g_3 \G_\mu^A$, $g_2 W_\mu^I = \g[2] \W_\mu^I$, $g_1 B_\mu = \g[1] \B_\mu$.
Hence, the covariant derivative takes the same form in terms of the rescaled quantities:
\begin{align}
	D_\mu &= \p_\mu + i g_3 T^A G^A_\mu + i g_2 t^I W_\mu^I + i g_1 y B_\mu = \p_\mu + i \bar g_3 T^A \G^A_\mu + i \g[2] t^I \W_\mu^I + i \g[1] y \B_\mu \, .
\end{align}
In the $CP$-odd sector, the theta terms
\begin{align}
	\L_\theta = \frac{\theta_3 g_3^2}{32 \pi^2} \, G^A_{\mu \nu} \widetilde G^{A\, \mu \nu} + \frac{\theta_2 g_2^2}{32 \pi^2} \, W^I_{\mu \nu} \widetilde W^{I\, \mu \nu} + \frac{\theta_1 g_1^2}{32 \pi^2} \, B_{\mu \nu} \widetilde B^{\mu \nu}
\end{align}
get modified in the broken phase by contributions from the operators
\begin{align}
	\L_{X^2 H^2}^{(6),\mathrm{odd}} = C_{H \widetilde G} Q_{H\widetilde G} + C_{H\widetilde W} Q_{H\widetilde W} + C_{H\widetilde B} Q_{H\widetilde B} + C_{H\widetilde WB} Q_{H\widetilde WB}  \, .
\end{align}
Similarly to the gauge couplings, we define rescaled theta angles
\begin{align}
	\label{eq:ThetaParameterRescaling}
	\bar\theta_3 &= \theta_3 + \frac{16\pi^2}{g_3^2} v_T^2 C_{H\widetilde G}  \, , \quad 
	\bar\theta_2 = \theta_2 + \frac{16\pi^2}{g_2^2} v_T^2 C_{H\widetilde W} \, , \quad 
	\bar\theta_1 = \theta_1 + \frac{16\pi^2}{g_1^2} v_T^2 C_{H\widetilde B} \, .
\end{align}

The gauge-boson mass and mixing terms at dimension four are
\begin{align}
	(D_\mu H^\dagger)(D^\mu H) = \frac{v_T^2}{8} \left( \g[2]^2 \W_\mu^I \W^\mu_I +  \g[1]^2 \B_\mu \B^\mu - 2 \g[2] \g[1] \W_\mu^3 \B^\mu \right) + \ldots \, .
\end{align}
The operator $Q_{HWB}$ induces an additional kinetic mixing of the weak gauge bosons:
\begin{align}
	C_{HWB} Q_{HWB} = - \frac{v_T^2}{2} C_{HWB}  \W_{\mu\nu}^3 \B^{\mu\nu} + \ldots \, .
\end{align}
Furthermore, the operator $Q_{HD}$ contributes to the gauge-boson mass terms:
\begin{align}
	C_{HD} Q_{HD} = C_{HD} \frac{v_T^4}{16} ( \g[2]^2 \W^\mu_3 \W_\mu^3 + \g[1]^2 \B^\mu \B_\mu - 2 \g[1] \g[2] \W_\mu^3 \B^\mu ) + \ldots \, .
\end{align}
Defining $\epsilon := v_T^2 C_{HWB}$, we perform the transformation~\cite{Grinstein:1991cd}
\begin{align}
	\left( \begin{matrix} \Z^\mu \\ \A^\mu \end{matrix} \right) = \left( \begin{matrix} \bar c - \frac{\epsilon}{2} \bar s & - \bar s + \frac{\epsilon}{2} \bar c \\ \bar s + \frac{\epsilon}{2} \bar c & \bar c + \frac{\epsilon}{2} \bar s \end{matrix} \right) \left( \begin{matrix} \W_3^\mu \\ \B^\mu \end{matrix} \right) ,
\end{align}
where
\begin{align}
	\bar c := \cos \bar\theta , \quad \bar s := \sin\bar\theta .
\end{align}
This diagonalizes both the kinetic terms and the mass matrix of the neutral gauge bosons, provided that
\begin{align}
	\bar c &= \frac{\g[2]}{\sqrt{\g[1]^2 + \g[2]^2}} \left( 1 - \frac{\epsilon}{2} \frac{\g[1]}{\g[2]} \frac{ \g[2]^2 - \g[1]^2}{\g[1]^2 + \g[2]^2} \right) , \quad
	\bar s = \frac{\g[1]}{\sqrt{\g[1]^2 + \g[2]^2}} \left( 1 + \frac{\epsilon}{2} \frac{\g[2]}{\g[1]} \frac{ \g[2]^2 - \g[1]^2}{\g[1]^2 + \g[2]^2} \right) .
\end{align}
The mass terms become
\begin{align}
	\frac{v_T^2}{8} \g[2]^2 \left( \W_\mu^1 \W^\mu_1 + \W_\mu^2 \W^\mu_2 \right) &+ \frac{v_T^2}{8} \left( 1 + \frac{v_T^2}{2} C_{HD} \right) \left( \g[2]^2 \W_\mu^3 \W^\mu_3 +  \g[1]^2 \B_\mu \B^\mu - 2 \g[2] \g[1] \W_\mu^3 \B^\mu \right) \nn
		&= \mw^2 \W_\mu^+ \W^\mu_- + \frac{1}{2} \mz^2 \Z_\mu \Z^\mu ,
\end{align}
where
\begin{align}
	\label{eq:GaugeBosonMasses}
	\W_\mu^\pm := \frac{1}{\sqrt{2}} ( \W_\mu^1 \mp i \W_\mu^2 ) , \quad \mw^2 = \frac{v_T^2  \g[2]^2}{4} , \quad \mz^2 = \frac{v_T^2}{4} \left( 1 + \frac{v_T^2}{2} C_{HD} \right) \left( \g[2]^2 + \g[1]^2 \right) + \frac{v_T^2}{2} \epsilon \g[1] \g[2] ,
\end{align}
and the photon field $\A_\mu$ remains massless.

The covariant derivative can be rewritten in the mass basis as
\begin{align}
	D_\mu = \p_\mu + i \bar g_3 T^A \G_\mu^A + i \frac{\bar g_2}{\sqrt{2}} [ \W^+_\mu t^+ + \W^-_\mu t^- ] + i \bar g_\Z [t^3 - \bar s^2 Q] \Z_\mu + i \bar e Q \A_\mu \, ,
\end{align}
where $t^\pm = t^1 \pm i t^2$ and the effective couplings are given by
\begin{align}
	\bar e &= \bar g_2 \bar s - \frac{1}{2} \bar c  \bar g_2 v_T^2 C_{HWB} \, , \quad \gz = \frac{\bar e}{\bar s \bar c} \left[ 1 + \frac{\bar g_1^2 + \bar g_2^2}{2 \bar g_1 \bar g_2} v_T^2 C_{HWB} \right] \, .
\end{align}
Including dimension-six effects, the theta terms in the broken phase read
\begin{align}
	\label{eq:ThetaTerms}
	\L_\theta^{(6)} &= \bar \theta_3 \frac{\g[3]^2}{32 \pi^2} \, \G^A_{\mu \nu} \widetilde \G^{A\, \mu \nu} + \bar \theta_2 \frac{\g[2]^2}{32 \pi^2} \, \W^+_{\mu \nu} \widetilde \W^{-\mu \nu} \nn
		&\quad + \bar \theta_\Z \frac{\g_\Z^2}{32 \pi^2} \, \Z_{\mu \nu} \widetilde \Z^{\mu \nu} + \bar \theta_{\Z\A} \frac{\g_\Z \bar e}{32 \pi^2} \, \Z_{\mu \nu} \widetilde \A^{\mu \nu} + \bar \theta_\A \frac{\bar e^2}{32 \pi^2} \, \A_{\mu \nu} \widetilde \A^{\mu \nu} \, ,
\end{align}
where the theta angles are given by
\begin{align}
	\bar\theta_\A &= \bar\theta_1 + \bar\theta_2 - \frac{16\pi^2}{\g[1] \g[2]} v_T^2 C_{H\widetilde W B} \, , \nn
	\bar\theta_\Z &= \frac{2 \left(\g[1]^4 \bar\theta_1 + \g[2]^4 \bar\theta_2 \right)}{\left(\g[1]^2+\g[2]^2\right)^2} - \frac{4 \g[1] \g[2] (\g[1]^2-\g[2]^2) \left(\g[1]^2 \bar\theta_1-\g[2]^2 \bar\theta_2\right)}{\left(\g[1]^2+\g[2]^2\right)^3} v_T^2 C_{HWB} + \frac{32 \pi ^2 \g[1] \g[2]}{\left(\g[1]^2+\g[2]^2\right)^2} v_T^2 C_{H\widetilde WB} \, , \nn
	\bar\theta_{\Z\A} &= \frac{4 (\g[2]^2 \bar\theta_2 - \g[1]^2 \bar\theta_1)}{\g[1]^2+\g[2]^2} + \frac{4 \g[1] \g[2] \left(\g[1]^2-\g[2]^2\right) (\bar\theta_1+\bar\theta_2)}{\left(\g[1]^2+\g[2]^2\right)^2} v_T^2 C_{HWB}  +\frac{32 \pi ^2 \left(\g[1]^2-\g[2]^2\right)}{\g[1] \g[2] \left(\g[1]^2+\g[2]^2\right)} v_T^2 C_{H\widetilde WB} \, .
\end{align}

\subsection{Background-field method, gauge fixing, and ghosts}

For the one-loop matching calculation, we will employ the background-field method~\cite{Abbott:1980hw,Abbott:1983zw,Denner:1994xt,Denner:1996wn,Helset:2018fgq}. All the fields are split into a quantum field $F$ and a classical background field $\hat F$,
\begin{align}
	F \mapsto F + \hat F \, ,
\end{align}
where the quantum fields are the variables of integration in the functional integral. In Feynman diagrams, the external legs as well as internal tree-level propagators correspond to the background fields~\cite{Abbott:1983zw}, while internal loop propagators are quantum fields. Splitting the fermion fields into background and quantum fields does not change the Feynman rules, hence one does not need to distinguish them~\cite{Denner:1994xt,Denner:1996wn}.

Compared to conventional gauge fixing, the background-field method has several advantages: gauge invariance for Green's functions of background fields is retained explicitly, while in the conventional method, gauge invariance is broken to BRST invariance, resulting in gauge-variant (but BRST invariant) counterterms. Note that it is not necessary to renormalize Green's functions involving quantum fields~\cite{Abbott:1980hw}. The background-field method also results in a simplification of Ward identities~\cite{Denner:1994xt,Denner:1996wn}. Finally, the gauge for the background fields can be fixed independently of the gauge fixing for the quantum fields. We choose unitary gauge for the background fields and linear $R_\xi$ gauge for the quantum fields~\cite{Denner:1994xt}. On the one hand, this reduces the number of possible diagrams, on the other hand, the independence of the gauge-fixing parameter $\xi$ provides a strong check on the loop calculation.
As an example, the Higgs field is split into a background Higgs field $\hat h$ and a quantum field $h$, while due to unitary background-field gauge, the Goldstone bosons are pure quantum fields:
\begin{align}
	h \mapsto h + \hat h \, , \quad G^\pm \mapsto G^\pm \, , \quad G^0 \mapsto G^0 \, .
\end{align}

The gauge-fixing term for the SMEFT in the background-field method has been derived in~\cite{Helset:2018fgq}. The generating functional is
\begin{align}
	Z[\hat F,J] = \int \D F \det\left[ \frac{\delta \G^A}{\delta \alpha^B} \right] \exp\left[ i \int d^4x \left( \L(F+\hat F) + \L_\mathrm{GF} + \text{source terms} \right) \right] \, ,
\end{align}
with sources $J$ coupled to the quantum fields only. The gauge-fixing term is
\begin{align}
	\label{eq:GaugeFixingTerm}
	\L_\mathrm{GF} &= - \frac{\hat g_{AB}}{2\xi} \G^A \G^B \, ,
\end{align}
where $\hat g$ accounts for the curved field space due to the higher-dimension SMEFT contributions and only involves the background fields, see~\cite{Helset:2018fgq} for its definition. $\L_\mathrm{GF}$ fixes the gauge of the quantum fields, while $Z[\hat F,J]$ is still invariant with respect to background-fields gauge transformations. The determinant can be replaced by the Faddeev--Popov ghost term given in~\cite{Helset:2018fgq}.

$Z[\hat F,J]$ generates disconnected Green's functions, while the generating functional of connected Green's functions is given by
\begin{align}
	W[\hat F,J] = -i \log Z[\hat F,J] \, .
\end{align}
The Legendre transform thereof is the effective action\footnote{We denote all fields by $F$ and disregard any subtleties with fermion minus signs in functional derivatives for simplicity.}
\begin{align}
	\Gamma[ \hat F, \tilde F] = W[\hat F, J] - \int d^4x \, J \cdot \tilde F \bigg|_{\tilde F = \frac{\delta W}{\delta J} } \, ,
\end{align}
generating the one-particle-irreducible (1PI) Green's functions. The $S$-matrix elements can be constructed from
\begin{align}
	\Gamma^\text{full} = \Gamma[\hat F, 0] + i \int d^4x \, \L_\mathrm{GF}^\mathrm{BG}
\end{align}
and its Legendre transform, the generating functional for connected Green's functions
\begin{align}
	W[\hat J] = \Gamma^\text{full} - \int d^4x \, \hat J \cdot \hat F \bigg|_{\hat J = \frac{\delta \Gamma^\text{full}}{\delta \hat F}} \, ,
\end{align}
by considering all tree-level diagrams, where the 1PI vertices are connected by background-field propagators. The gauge-fixing term for the background fields is independent of the gauge-fixing for the quantum fields. As mentioned above, we choose unitary gauge for the background fields of the weak sector. The gauge for the unbroken background $SU(3)_c \otimes U(1)_\mathrm{em}$ has to be fixed as well---here again, we are using $R_\xi$ gauge.

For the extraction of the vertex Feynman rules, we note that terms linear in the quantum fields do not contribute to the 1PI Green's functions. Furthermore, vertices with more than two quantum fields only contribute beyond one loop, hence we disregard them as well.

The gauge-fixing term~\eqref{eq:GaugeFixingTerm} generates the following masses for the Goldstone bosons:
\begin{align}
	M_{G^\pm}^2 = \xi \mw^2 \, , \quad M_{G^0} = \xi \mz^2 \, .
\end{align}
It also generates a mixing between the Goldstone and gauge bosons that exactly cancels the mixing term from $\L_{H,\mathrm{kin}}$. The kinetic terms from $\L_\mathrm{GF}$ are given by
\begin{align}
	\L_\mathrm{GF}^\mathrm{kin} = -\frac{1}{2\xi} (\p^\mu \A_\mu)^2 -\frac{1}{2\xi} (\p^\mu \Z_\mu)^2 - \frac{1}{\xi} (\p^\mu \W^+_\mu)(\p^\nu \W^-_\nu) \, ,
\end{align}
which lead to standard $R_\xi$ propagators for the quantum gauge bosons. The gauge-fixing term does not contribute to the kinetic terms of the background weak gauge fields, which have propagators as in standard unitary gauge, see App.~\ref{sec:Propagators}.

Beyond the bilinear terms, $\L_\mathrm{GF}$ generates interaction vertices with two quantum fields and up to four background fields.

The Faddeev--Popov ghost term for the SMEFT with background-field method is given in~\cite{Helset:2018fgq}. It inherits the curved field-space metric of the gauge sector. Hence, in analogy to the gauge-field rescaling~\eqref{eq:GaugeFieldRescaling}, we define rescaled ghost fields to account for the dimension-six effects,\footnote{We change the sign of the anti-ghost compared to~\cite{Helset:2018fgq} in order to obtain the more conventional form of the ghost propagator.}
\begin{align}
	\begin{alignedat}{2}
		u_{W_i} &= \eta_i ( 1 + v_T^2 C_{HW} ) \, , \quad & u_B &= \eta_B ( 1 + v_T^2 C_{HB} ) \, , \\
		\bar u_{W_i} &= -\bar \eta_i ( 1 + v_T^2 C_{HW} ) \, , \quad &  \bar u_B &= -\bar \eta_B ( 1 + v_T^2 C_{HB} ) \, ,
	\end{alignedat}
\end{align}
and we rotate the neutral ghost fields
\begin{align}
	\begin{pmatrix} \eta_Z \\ \eta_A \end{pmatrix} = \begin{pmatrix} \bar c - \frac{\epsilon}{2} \bar s & - \bar s + \frac{\epsilon}{2} \bar c \\ \bar s + \frac{\epsilon}{2} \bar c & \bar c + \frac{\epsilon}{2} \bar s \end{pmatrix} \begin{pmatrix} \eta_3 \\ \eta_B \end{pmatrix} ,
\end{align}
and similarly for the anti-ghosts. The charged ghosts are defined as
\begin{align}
	\eta_\pm = \frac{1}{\sqrt{2}} ( \eta_1 \mp i \eta_2 ) \, , \quad \bar\eta_\pm = \frac{1}{\sqrt{2}} ( \bar\eta_1 \pm i \bar\eta_2 ) \, .
\end{align}
This leads to canonically normalized kinetic and mass terms for the ghosts:
\begin{align}
	\L_\mathrm{FP} &= - \bar\eta_+ \Box \eta_+ - \bar\eta_- \Box \eta_- - \bar\eta_Z \Box \eta_Z - \bar\eta_A \Box \eta_A \nn
		&\quad - M_{\eta_W}^2 ( \bar\eta_+ \eta_+  + \bar\eta_- \eta_- ) - M_{\eta_Z}^2 \bar\eta_Z \eta_Z \, ,
\end{align}
where
\begin{align}
	M_{\eta_W}^2 = \xi \mw^2 \, , \quad M_{\eta_Z}^2 = \xi \mz^2 \, .
\end{align}
In addition to the bilinear terms, the Faddeev--Popov term in the background-field method leads to interaction vertices involving up to six fields (in contrast to the ghost term in conventional $R_\xi$ gauge~\cite{Dedes:2017zog}).

The gauge fixing in the QCD sector is independent of the electroweak gauge fixing. In terms of the canonically normalized gluon fields~\eqref{eq:GaugeFieldRescaling} it is identical to the SM case~\cite{Abbott:1980hw}:
\begin{align}
	\L_\mathrm{GF}^\mathrm{QCD} = - \frac{1}{2\xi_g} (G^A)^2 \, , \quad G^A = \p^\mu \G_\mu^A - \bar g_3 f^{ABC} \hat \G_\mu^B \G^{C\mu} \, .
\end{align}
The ghost Lagrangian reads
\begin{align}
	\L_\mathrm{FP}^\mathrm{QCD} = -\bar\eta_G^A & \bigg[ \Box \delta^{AB} + \bar g_3 \pl_\mu f^{ACB} ( \hat \G^{C\mu} + \G^{C\mu} ) \nn
														& - \bar g_3 f^{ACB} \hat\G_\mu^C \p^\mu
														 + \bar g_3^2 f^{ACX} f^{XDB} \hat\G_\mu^C(\hat \G^{D\mu} + \G^{D\mu} ) \bigg] \eta_G^B \, .
\end{align}
The following Hermiticity properties of the ghosts ensure that the Lagrangian is Hermitian~\cite{Kugo:1979gm}:
\begin{align}
	\eta_{Z,A}^\dagger = \eta_{Z,A} \, , \quad \eta_\pm^\dagger = \eta_\mp \, , \quad \eta_G^A{}^\dagger = \eta_G^A \, , \quad \bar\eta_{Z,A}^\dagger = -\bar\eta_{Z,A} \, , \quad \bar\eta_\pm^\dagger = -\bar\eta_\mp \, , \quad \bar\eta_G^A{}^\dagger = -\bar \eta_G^A \, .
\end{align}

Unitary gauge for the background fields does not fix the $SU(3)_c\otimes U(1)_\mathrm{em}$ background-field gauge. We use a background-field gauge-fixing Lagrangian
\begin{align}
	\L_\mathrm{GF}^\mathrm{BG} = - \frac{1}{2\hat\xi_\gamma} (\p_\mu \hat\A^\mu)^2 - \frac{1}{2\hat\xi_g} ( \p^\mu \hat \G^A_\mu)^2 \, .
\end{align}
Since the background fields only enter at tree level, background ghost terms are irrelevant.

\subsection{Fermions}

The fermion mass matrices and Yukawa couplings
\begin{align}
	\L = - \left[ H^{\dagger i} \bar d Y_d q_i + \tilde H^{\dagger i} \bar u Y_u q_i + H^{\dagger i} \bar e Y_e l_i + \hc \right]
\end{align}
are modified in the SMEFT by the dimension-six operators of the class $\psi^2 H^3$:
\begin{align}
	\L_{\psi^2 H^3}^{(6)} = C_{\substack{eH\\rs}} Q_{\substack{eH\\rs}} +  C_{\substack{uH\\rs}} Q_{\substack{uH\\rs}} +  C_{\substack{dH\\rs}} Q_{\substack{dH\\rs}} + \hc
\end{align}
The bilinear terms are
\begin{align}
	Q_{\substack{eH\\rs}} &= (H^\dagger H) (\bar l_r e_s H) = \frac{v_T^3}{2^{3/2}} \bar e_{Lr} e_{Rs} + \ldots \, , \nn
	Q_{\substack{uH\\rs}} &= (H^\dagger H) (\bar q_r u_s \tilde H) = \frac{v_T^3}{2^{3/2}} \bar u_{Lr} u_{Rs} + \ldots \, , \nn
	Q_{\substack{dH\\rs}} &= (H^\dagger H) (\bar q_r d_s H) = \frac{v_T^3}{2^{3/2}} \bar d_{Lr} d_{Rs} + \ldots \, .
\end{align}
Therefore, in the spontaneously broken SMEFT, the fermion mass terms are given by
\begin{align}
	\label{eq:FermionMassTerms}
	\L_M = - [M_\psi]_{rs} \bar \psi_{Rr} \psi_{Ls} + \hc \, , \quad [M_\psi]_{rs} = \frac{v_T}{\sqrt{2}} \left( [Y_\psi]_{rs} - \frac{v_T^2}{2}C_{\substack{\psi H\\sr}}^* \right) \, , \quad \psi = u, d, e \, .
\end{align}
The Yukawa interaction in the broken phase is modified to
\begin{align}
	\L_Y &= -  [\mathcal{Y}_\psi]_{rs} h \bar \psi_{Rr} \psi_{Ls} + \hc \, , \quad \psi = u, d, e \, , \nn
	[\mathcal{Y}_\psi]_{rs} &= \frac{1}{\sqrt{2}} \left( [1 + \chkin] [Y_\psi]_{rs} - \frac{3}{2} v_T^2 C_{\substack{\psi H\\sr}}^* \right) = \frac{1 + \chkin}{v_T} [ M_\psi ]_{rs} - \frac{v_T^2}{\sqrt{2}} C_{\substack{\psi H\\sr}}^* \, .
\end{align}

The dimension-five Weinberg operator generates a neutrino mass in the broken phase:
\begin{align}
	\label{eq:nuMass}
	\L &= - \frac12 \left[ M_\nu \right]_{rs} \left( \nu^T_{Lr} C \nu_{Ls} \right) + \hc \,, \quad
	\left[ M_\nu \right]_{rs} = - C_{\substack{5 \\ rs}} v_T^2 \, .
\end{align}
The Higgs boson couples to the neutrinos via $\L = h \left[ \mathcal{Y}_5 \right]_{rs} (\nu^T_{L r} C \nu_{L s} )+ \hc$, where
\begin{align}
\label{eq:nuYuk}
\left[ \mathcal{Y}_5 \right]_{rs} & := v_T \left[ C_5 \right]_{rs} \left[ 1+ \chkin \right] 
\end{align}
is proportional to the Majorana-neutrino mass matrix when keeping only operators up to dimension six in the SMEFT.

We choose a basis, where the weak eigenstates coincide with mass eigenstates for the up-type quarks, charged leptons, and right-handed down-type quarks. The weak eigenstates and mass eigenstates of the left-handed down-type quarks are related by a unitary transformation $V$, while the neutrino mass matrix is diagonalized by another unitary transformation $U$ (see App.~\ref{app:Neutrinos}). Although in the SM these matrices correspond to the usual CKM and PMNS matrices, respectively, they receive corrections within the SMEFT~\cite{Descotes-Genon:2018foz}. 
The mass matrices in~\eqref{eq:FermionMassTerms}~and~\eqref{eq:nuMass} are
\begin{align}
	\label{eq:TreeLevelSMEFTMasses}
	M_u &= \mathrm{diag}(m_u, m_c, m_t) \, , \quad M_d = \mathrm{diag}(m_d, m_s, m_b) V^\dagger \, , \nn
	M_e &= \mathrm{diag}(m_e, m_\mu, m_\tau ) \, , \quad M_\nu = U^* \mathrm{diag}(m_{\nu_1}, m_{\nu_2}, m_{\nu_3}) U^\dagger \, .
\end{align}
The mass eigenstates are related to the weak eigenstates by
\begin{align}
	\label{eq:BasisChange}
	d_{Lr}' = V^\dagger_{rs} d_{Ls} \, , \quad \nu_{Lr}' = U^\dagger_{rs} \nu_{Ls} \, .
\end{align}
For notational simplicity, we will mostly drop the prime on the mass eigenstates.

The interactions of the fermions with the weak gauge bosons get modified by the dimension-six operators of the class $\psi^2 H^2 D$. These contributions are given by
\begin{align}
	\begin{split}
		C_{\substack{H\psi \\ rs}}^{(1)} (H^\dagger i \overleftrightarrow D_\mu H) (\bar \psi_r \gamma^\mu \psi_s) &= \gz  \frac{v_T^2}{2} C_{\substack{H\psi\\rs}}^{(1)} \mathcal{Z}_\mu ( \bar \psi_r \gamma^\mu \psi_s ) + \ldots , \\
		C_{\substack{H\psi \\ rs}}^{(3)} (H^\dagger i \overleftrightarrow D_\mu^I H) (\bar \psi_r \tau^I \gamma^\mu \psi_s) &= - \bar g_2  \frac{v_T^2}{\sqrt{2}} C_{\substack{H\psi\\rs}}^{(3)} \mathcal{W}_\mu^+ ( \bar \psi_r t^+ \gamma^\mu \psi_s ) \\
			&\quad - \bar g_2  \frac{v_T^2}{\sqrt{2}} C_{\substack{H\psi\\rs}}^{(3)} \mathcal{W}_\mu^- ( \bar \psi_r t^- \gamma^\mu \psi_s ) \\
			&\quad - \gz v_T^2 C_{\substack{H\psi\\rs}}^{(3)} \mathcal{Z}_\mu ( \bar \psi_r t^3 \gamma^\mu \psi_s ) + \ldots , \\
		C_{\substack{Hud \\ rs}} (\tilde H^\dagger i D_\mu H) (\bar u_{Rr} \gamma^\mu d_{Rs}) &= - \bar g_2  \frac{v_T^2}{2\sqrt{2}} C_{\substack{Hud\\rs}} \mathcal{W}_\mu^+ ( \bar u_{Rr} \gamma^\mu d_{Rs} ) + \ldots ,
	\end{split}
\end{align}
where the ellipses denote terms involving scalar fields.

After the rotation to the mass-eigenstate basis, we introduce Majorana spinors for the neutrinos, see App.~\ref{app:Neutrinos}. The extraction of the Feynman rules and the calculation of diagrams involving Majorana neutrinos are performed according to~\cite{Denner:1992vza}. We note that in loop calculations with Majorana fermions, the appearance of evanescent operators leads to a subtlety that we will comment on in Sect.~\ref{sec:EvancescentOperators}.


\section{LEFT}

\label{sec:LEFT}

Below the weak scale, we are working with an EFT that is an $SU(3)_c \otimes U(1)_\mathrm{em}$ gauge theory and contains the SM fermions, apart from the top quark, as matter fields. Therefore, the LEFT Lagrangian consists of QCD and QED at dimension four, plus a tower of additional effective operators:
\begin{align}
	\label{eq:LEFTLagrangian}
	\mathcal{L}_\mathrm{LEFT} = \mathcal{L}_\mathrm{QCD+QED} + \mathcal{L}^{(3)}_{\slashed L} + \sum_{d\ge5} \sum_i \lwc{i}{(d)} \O_i^{(d)} \, ,
\end{align}
where the QCD and QED Lagrangian is given by
\begin{align}
	\label{eq:qcdqed}
	\mathcal{L}_{\rm QCD + QED} &= - \frac14 G_{\mu \nu}^A G^{A \mu \nu} -\frac14 F_{\mu \nu} F^{\mu\nu} + \theta_{\rm QCD} \frac{g^2}{32 \pi^2} G_{\mu \nu}^A \widetilde G^{A \mu \nu} +  \theta_{\rm QED} \frac{e^2}{32 \pi^2} F_{\mu \nu} \widetilde F^{\mu \nu} \nn
		&+ \sum_{\psi=u,d,e,\nu_L}\overline \psi i \slashed{D} \psi   - \left[ \sum_{\psi=u,d,e}  \overline \psi_{Rr} [M_\psi]_{rs} \psi_{Ls} + \text{h.c.} \right] \, .
\end{align}
The additional operators are the Majorana-neutrino mass terms at dimension three, as well as operators at dimension five and above. The complete list of LEFT operators up to dimension six was derived in~\cite{Jenkins:2017jig}. We reproduce the operator basis in App.~\ref{sec:LEFTBasis}.

The mass matrices $M_\psi$ and $M_\psi^\dagger$ in~\eqref{eq:qcdqed} and the Majorana-neutrino mass matrix are general complex matrices in flavor space. When performing the tree-level matching to the SMEFT, they are set equal to the mass matrices in~\eqref{eq:TreeLevelSMEFTMasses} (with the up-type quark mass matrix restricted to the first two generations). Through field redefinitions, one can perform a basis change between weak eigenstates and mass eigenstates. In the SM, the unitary rotation of the left-handed down-type quarks is the CKM matrix, while in the SMEFT the quark-mixing matrix appearing in the interaction with the $W$ boson is non-unitary due to an additional dimension-six contribution proportional to $\cwc{Hq}{(3)}$~\cite{Dedes:2017zog,Jenkins:2017jig}. In the LEFT, the $W$ boson is integrated out, hence in the basis change from weak eigenstates to mass eigenstates the unitary rotations that diagonalize the mass matrices are reabsorbed into the Wilson coefficients of the higher-dimension operators.

At one loop, the corrections to the propagators again generate general complex mass matrices, which have to be rediagonalized in the mass eigenstate basis. In our calculation, we choose a LEFT basis where the mass matrices contain complex off-diagonal contributions from the one-loop matching. An additional rotation that diagonalizes the mass matrices is always possible and reshuffles the off-diagonal contributions into external-leg corrections for three- and four-point functions. The choice of basis at the matching scale simply is a convention---when running to the scale of an experiment, the RGEs generate off-diagonal contributions to the mass matrices~\cite{Jenkins:2017dyc} and one has to perform another diagonalization.


\section{Loop calculation}
\label{sec:LoopCalculation}

\subsection{Dimensional regularization and scheme definition}
\label{sec:DimReg}

In order to treat both UV and IR divergences in the loop diagrams, we use dimensional regularization and work in $D = 4 - 2\epsilon$ dimensions. As usual for chiral gauge theories, one immediately faces the problem how to handle intrinsically four-dimensional objects like $\gamma_5$ or the Levi-Civita tensor $\epsilon^{\mu\nu\lambda\sigma}$. As is well known, the existence of the chiral anomaly implies that there is no symmetry-preserving regulator. Therefore, any scheme faces problems at some point in the calculation. For a detailed review, we refer to~\cite{Jegerlehner:2000dz}. The $\gamma_5$ problem in the context of SMEFT has recently been discussed in~\cite{Boughezal:2019xpp}.

The simplest and most popular scheme is to use an anticommuting $\gamma_5$ in $D$ dimensions, called naive dimensional regularization (NDR). This scheme does not work for certain divergent diagrams involving $\gamma_5$-odd traces, because for $D\neq4$ it implies
\begin{align}
	\tr[ \gamma^\mu\gamma^\nu\gamma^\lambda\gamma^\sigma\gamma_5 ] = 0 \, ,
\end{align}
at odds with the required limit for $D\to4$. However, in many cases it is possible to avoid the calculation of divergent integrals involving these traces. E.g.\ the ABJ triangle anomaly can be obtained by imposing gauge invariance beforehand, requiring then only the calculation of a finite integral~\cite{Adler:1970}.

Alternatively, one can give up the anticommutation properties of $\gamma_5$ with gamma matrices in $\epsilon$ dimensions, as in the 't\,Hooft--Veltman (HV) scheme~\cite{tHooft:1972tcz,Breitenlohner:1977hr}. In this case, chiral invariance is explicitly broken, which leads to the appearance of spurious anomalies that violate the Ward or Slavnov--Taylor identities. These spurious anomalies have to be removed again by symmetry-restoring local counterterms, order by order in the perturbative expansion~\cite{Breitenlohner:1977hr,Ferrari:1994ct}. This can be achieved most elegantly by algebraic renormalization~\cite{Grassi:1999tp}. In some cases, on-shell renormalization conditions automatically take care of the symmetry-restoring counterterms, but the problem is especially severe in connection with minimal subtraction~\cite{Trueman:1995ca}. With the inclusion of symmetry-restoring counterterms, the HV scheme is the only scheme known to be fully consistent to all orders in perturbation theory, but it remains cumbersome to use in practical applications. However, as shown in~\cite{Trueman:1995ca} for many classes of diagrams the HV scheme with symmetry-restoring counterterms leads to identical results as NDR, in particular in the case of open fermion lines and traces with an even number of $\gamma_5$ matrices.

As observed in~\cite{Jegerlehner:2000dz} the anticommutator of $\gamma_5$ with the gamma matrices,
\begin{align}
	\mathrm{AC}(\mu) := \{ \gamma^\mu, \gamma_5 \} \, ,
\end{align}
is a matrix of rank $D-4$, but has matrix elements of $\O(1)$. This implies that products of $\mathrm{AC}(\mu)$ are still matrices of rank $D-4$ and not contributions of higher order in $D-4$, hence in the HV scheme, they cannot be discarded. As shown in~\cite{Jegerlehner:2000dz}, a modified HV scheme leads to the same results as the NDR scheme if one uses chiral vertices~\cite{Korner:1989is} and performs a formal expansion to linear order in $\mathrm{AC}(\mu)$. Therefore, the appearance of spurious anomalies in the HV scheme can be traced back to the presence of higher orders of $\mathrm{AC}(\mu)$. The mentioned chiral vertices are defined as the vertex rules that follow from consistently using chiral fermions in the Lagrangian, i.e.\ the left-handed vertices involve $P_R \gamma^\mu P_L$, the right-handed vertices $P_L \gamma^\mu P_R$. Vector- and axial-vector vertices are given by the sum and difference of left- and right-handed vertices. In the HV scheme, projectors on both sides of a single gamma matrix reduce the vertex to the four-dimensional one.

For our calculation, we will use the \msbar{} scheme (apart from evanescent operators, see Sect.~\ref{sec:EvancescentOperators}), hence the need for symmetry-restoring counterterms would make the HV scheme cumbersome. Therefore, we use the NDR scheme whenever possible, with only a few extra prescriptions and exceptions as explained in the following subsections. The \msbar{} renormalization prescription amounts to the subtraction of divergent terms of the form
\begin{align}
	\La = - \frac{1}{2} \mu^{D-4} \left( \frac{1}{\epsilon} + \log(4\pi) - \gamma_E \right)  \, .
\end{align}
Although the problem with $\gamma_5$ requires special attention in the calculation of the finite pieces of the diagrams,  the divergences can be directly compared to the UV counterterms in the SMEFT and LEFT. The bare parameters are split into renormalized ones and counterterms
\begin{align}
	C^\text{bare} = C^\text{ren}(\mu) + \La \; C^\text{ct} \, .
\end{align}
The divergences of loop diagrams that are expanded in the light scales have to match the difference of the SMEFT and LEFT UV divergences, see Sect.~\ref{sec:ExpandingLoops}. This allows for nontrivial cross-checks with the SMEFT and LEFT renormalization-group equations derived in~\cite{Jenkins:2013wua,Jenkins:2013zja,Alonso:2013hga,Jenkins:2017dyc}, see Sect.~\ref{sec:Results}, which are given in terms of
\begin{align}
	\dot C := 16 \pi^2 \mu \frac{d}{d\mu} C^\text{ren}(\mu) \, ,
\end{align}
hence the counterterms are directly given by
\begin{align}
	C^\text{ct} = - \frac{1}{16\pi^2} \dot C \, .
\end{align}

\subsubsection{Levi-Civita tensor}

The totally antisymmetric tensor $\epsilon^{\mu\nu\lambda\sigma}$ is an intrinsically four-dimensional object, although there are schemes that promote contractions of two Levi-Civita tensors to $D$ dimensions~\cite{Larin:1993tq,Bednyakov:2015ooa,Zoller:2015tha}. In the one-loop matching calculation, the Levi-Civita tensor appears explicitly in vertices due to $CP$-odd dimension-six operators involving the dual field-strength tensors. At the order we are considering, only single insertions of these operators are needed, hence we do not encounter contractions of two Levi-Civita tensors. The $CP$-odd vertices appear in two-point functions with momentum insertion, required to extract the matching for the theta terms. Here, the Levi-Civita symbol can be kept as an external quantity, present on both sides of the matching equation. Vertices with dual field-strength tensors further appear in fermionic three-point functions, where the external boson is attached to a triple-gauge-boson or Higgs-gauge vertex. In the fermionic vertices, only the SM part contributes at the considered order, i.e.\ the fermion line contains at most three gamma matrices (plus $\gamma_5$). In the HV scheme with chiral vertices, only the gamma matrix from the propagator has a component in $D-4$ dimensions. Since after performing the tensor loop integrals, its index can be contracted only with four-dimensional quantities, no $\mathrm{AC}(\mu)$ terms appear. In practice, the same result is obtained by using the NDR scheme: at first we keep the Levi-Civita tensor uncontracted, perform the loop integrals including all necessary tensor reductions in $D$ dimensions until only contractions with external momenta, polarization vectors, and the Levi-Civita tensor remain. Similarly to the HV scheme, we then consider these contractions in four dimensions and perform the remaining algebra in four dimensions. In order to ensure the correctness of our results, we verify the Ward identities for the vertex functions with insertion of the $CP$-odd operator.

We also note that the exact form of the $CP$-odd triple-gauge operators is part of the scheme definition: if the Lagrangian is transformed by four-dimensional relations like the Schouten identity, evanescent operators are introduced that can give finite contributions at one loop, see the discussion at the end of Sect.~\ref{sec:Results}.

\subsubsection{Closed fermion loops}
\label{sec:ClosedFermionLoops}

The second potentially critical case involves diagrams with closed fermion loops. They appear in the matching calculation in several places. In regular bosonic two-point functions, no problem occurs: there are two open Lorentz indices for the polarization of the two gauge bosons, the remaining Lorentz indices are all contracted with the same external momentum, hence they cannot be antisymmetrized and no ill-defined trace appears in the NDR scheme. In bosonic two-point functions with momentum insertion into $CP$-odd dipole operators, Lorentz invariance only allows the structure $\epsilon^{\mu\nu\lambda\sigma} p_\lambda p_\sigma'$, which is transverse to both momenta. No symmetry-breaking terms can appear even though, due to the presence of $\gamma_5$-odd traces, we make use of the HV scheme. In the matching for the three-gluon operators, top loops appear in bulb and triangle diagrams. In principle, problematic traces appear in these diagrams, however the matching equations only depend on terms that are trilinear in the external momenta, which only contain UV finite integrals. Therefore, the HV scheme leads in this case to the same result as working directly in four dimensions. Finally, closed fermion loops also appear in penguin topologies in the fermion three-point functions with a four-fermion operator insertion. Since the loop depends only on one external momentum, problematic traces only show up in the case of a Dirac tensor structure in the four-fermion vertex. The only SMEFT four-fermion operator with a $\sigma^{\mu\nu}$ structure is $Q_{lequ}^{(3)}$. Furthermore, the penguin diagrams with heavy external gauge bosons are not critical: they only contribute to the matching of the four-fermion operators, where the external momentum can be set to zero, as explained in Sect.\ \ref{sec:PowerCounting}. Therefore, the only problematic diagram is the $Q_{lequ}^{(3)}$ part of the $\bar e e \gamma$ penguin. We evaluate it in the HV scheme and verify the Ward identity in order to ensure that no symmetry-breaking terms are generated.

\subsection{Evanescent operators}
\label{sec:EvancescentOperators}

In the case of fermion four-point functions, the loop calculation in $D$ dimensions generates counterterms that correspond to elements of the operator basis only in four dimensions, either due to the appearance of higher products of gamma matrices in both fermion bilinears or due to Fierz relations that only hold in four dimensions. The differences between the $D$-dimensional counterterms and the four-dimensional operators are evanescent operators. We use a scheme where the contribution of evanescent operators is compensated by local counterterms~\cite{Buras:1989xd,Dugan:1990df,Herrlich:1994kh}, so that evanescent operators do not contribute to one-loop matrix elements. In order to define the scheme, the evanescent operators have to be specified. We do not use pure \msbar{} for the renormalization of four-fermion operators, but rather use the following definition of evanescent operators for the NDR scheme:
\begin{align}
	\label{eq:EvanescentOperators}
	P_L \gamma^\mu \gamma^\nu P_L \otimes P_L \gamma_\mu \gamma_\nu P_L &= ( 4- 2 \epsilon ) P_L \otimes P_L - P_L \sigma^{\mu\nu} P_L \otimes P_L \sigma_{\mu\nu} P_L \, , \nn
	P_L \gamma^\mu \gamma^\nu P_L \otimes P_R \gamma_\mu \gamma_\nu P_R &= 4( 1+ a_\mathrm{ev} \epsilon ) P_L \otimes P_R + E_{LR}^{(2)} \, , \nn
	P_R \gamma^\mu \gamma^\nu \gamma^\lambda P_L \otimes P_R \gamma_\mu \gamma_\nu \gamma_\lambda P_L &= 4( 4- b_\mathrm{ev} \epsilon ) P_R \gamma^\mu P_L \otimes P_R \gamma_\mu P_L + E_{LL}^{(3)} \, , \nn
	P_R \gamma^\mu \gamma^\nu \gamma^\lambda P_L \otimes P_L \gamma_\mu \gamma_\nu \gamma_\lambda P_R &= 4( 1+ c_\mathrm{ev} \epsilon ) P_R \gamma^\mu P_L \otimes P_L \gamma_\mu P_R + E_{LR}^{(3)} \, , \nn
	P_L \gamma^\mu \gamma^\nu \gamma^\lambda \gamma^\sigma P_L \otimes P_L \gamma_\mu \gamma_\nu \gamma_\lambda \gamma_\sigma P_L &= 32( 2-3 d_\mathrm{ev} \epsilon ) P_L \otimes P_L \nn
		&\quad - 8( 2 - e_\mathrm{ev} \epsilon ) P_L \sigma^{\mu\nu} P_L \otimes P_L \sigma_{\mu\nu} P_L + E_{LL}^{(4)} \, , \nn
	P_L \gamma^\mu \gamma^\nu \gamma^\lambda \gamma^\sigma P_L \otimes P_R \gamma_\mu \gamma_\nu \gamma_\lambda \gamma_\sigma P_R &= 16( 1+8 f_\mathrm{ev} \epsilon ) P_L \otimes P_R + E_{LR}^{(4)} \, ,
\end{align}
and analogous definitions for opposite chirality. The coefficients $a_\mathrm{ev}$, $\ldots$, $f_\mathrm{ev}$ are kept generic. They can be assigned any value, defining different schemes~\cite{Herrlich:1994kh}. This definition of evanescent operators is motivated by the following derivation~\cite{Tracas:1982gp,Buras:1989xd,Herrlich:1994kh}. We define the set of physical operators in the chiral basis
\begin{align}
	\{ \Gamma_1^i \otimes \Gamma_2^i \}= \{ & P_L \otimes P_L, P_R \otimes P_R, P_L \otimes P_R, P_R \otimes P_L , \nn
			& P_R \gamma^\mu P_L \otimes P_R \gamma_\mu P_L, P_L \gamma^\mu P_R \otimes P_L \gamma_\mu P_R, P_R \gamma^\mu P_L \otimes P_L \gamma_\mu P_R, P_L \gamma^\mu P_R \otimes P_R \gamma_\mu P_L , \nn
			& P_L \sigma^{\mu\nu} P_L \otimes P_L \sigma_{\mu\nu} P_L, P_R \sigma^{\mu\nu} P_R \otimes P_R \sigma_{\mu\nu} P_R \}^i
\end{align}
and write the operators with higher numbers of gamma matrices as linear combinations
\begin{align}
	A_1 \otimes A_2 = \sum_i a_i \Gamma_1^i \otimes \Gamma_2^i \, .
\end{align}
The coefficients $a_i$ are then determined by computing the traces
\begin{align}
	\tr[ \Gamma_1^j \Gamma_1^i \Gamma_2^j \Gamma_2^i ] = c_{ij} , \quad \tr[ \Gamma_1^j A_1 \Gamma_2^j A_2 ] = b_j
\end{align}
and solving the system
\begin{align}
	b_i = \sum_i a_i c_{ij}
\end{align}
for $a_i$ up to $\O(\epsilon)$. The calculation of these traces in NDR is possible since all Lorentz indices are contracted. This derivation leads to evanescent operators as defined in~\eqref{eq:EvanescentOperators} with $a_\mathrm{ev} = b_\mathrm{ev} = c_\mathrm{ev} = d_\mathrm{ev} = e_\mathrm{ev} = f_\mathrm{ev} = 1$. We abstain from adding further terms linear in $\epsilon$ that mix even with odd numbers of gamma matrices. Such terms show up if one uses the above recipe within the parity basis instead of the chiral basis as in~\cite{Herrlich:1994kh}, e.g.\ the operators corresponding to $E_{LR}^{(3)}$ and $E_{RL}^{(3)}$ would involve contributions with an even number of gamma matrices, multiplied by $\epsilon$.
 
Operators with an even higher number of Dirac matrices in both bilinears do not appear in our calculation and operators with a different ordering of the Lorentz indices can be related to the ones given in~\eqref{eq:EvanescentOperators} by only making use of the anticommutation relations for the gamma matrices.

Further evanescent operators appear in our calculation: they are given by the differences of structures that are related by the usual Fierz identities in four dimensions:
\begin{align}
	\label{eq:ChiralFierzIdentities}
	( P_R \gamma^\mu P_L ) \otimes [ P_R \gamma_\mu P_L ] &= - ( P_R \gamma^\mu P_L ] \otimes [ P_R \gamma_\mu P_L ) + E_{LL}^{(F1)} \, , \nn
	( P_R \gamma^\mu P_L ) \otimes [ P_L \gamma_\mu P_R ] &= 2 ( P_R ] \otimes [ P_L ) + E_{LR}^{(F1)} \, , \nn
	( P_L \sigma^{\mu\nu} P_L ) \otimes [ P_L \sigma_{\mu\nu} P_L ] &= 8 ( P_L ] \otimes [ P_L ) -  4 ( P_L ) \otimes [ P_L ]  + E_{LL}^{(F2)} \, ,
\end{align}
and analogously for opposite chirality. The parentheses and brackets abbreviate Dirac indices and we have not included the minus sign from anticommuting the fermion fields. We choose not to introduce any explicit $\epsilon$-dependence into the definition of these Fierz-evanescent operators. Note that the definitions~\eqref{eq:EvanescentOperators} imply
\begin{align}
	P_L \sigma^{\mu\nu} P_L \otimes P_R \sigma_{\mu\nu} P_R &= - (2+4 a_\mathrm{ev}) \epsilon P_L \otimes P_R - E_{LR}^{(2)} \, .
\end{align}

We observe that with the appearance of evanescent operators, a subtlety occurs in connection with Majorana neutrinos. The formalism of~\cite{Denner:1992vza} for Majorana fermions and fermion-number-violating interactions has the advantage to simplify Feynman rules, the determination of relative signs of interfering diagrams, and to reduce the number of diagrams since Majorana neutrinos can be treated similarly to real scalar fields, by introducing a fermion flow direction on each fermion line. The direction of the fermion flow can be chosen arbitrarily. However, it turns out that reversing the fermion flow direction can lead to non-trivial relations between evanescent structures, such that the naive application of crossing relations and dropping evanescent operators at one loop do not lead to correct results. Consider a divergent loop diagram with an insertion of a $\gamma^\mu P_L \otimes \gamma_\mu P_L$ four-fermion operator and a vector-boson correction connecting the two incoming fermion lines. The tensor reduction of the loop integral leads to Dirac structures of the form
\begin{align}
	(\bar u_1 \gamma^\mu \gamma^\nu \gamma^\lambda P_L u_2) (\bar u_3 \gamma_\mu \gamma_\nu \gamma_\lambda P_L u_4) = 4( 4- b_\mathrm{ev} \epsilon ) (\bar u_1 \gamma^\mu P_L u_2) (\bar u_3 \gamma_\mu P_L u_4) + \< E_{LL}^{(3)} \>_{1234} \, .
\end{align}
However, we could also assign a reversed flow direction to the second fermion line, leading to an expression
\begin{align}
	-(\bar u_1 \gamma^\mu \gamma^\nu \gamma^\lambda P_L u_2) &(\bar v_4 \gamma_\lambda \gamma_\nu \gamma_\mu P_R v_3) = -4( 4- (3+c_\mathrm{ev}) \epsilon ) (\bar u_1 \gamma^\mu P_L u_2) (\bar v_4 \gamma_\mu P_R v_3) - \< E_{LR}^{(3)} \>_{1243} \nn
		&= 4( 4- (3+c_\mathrm{ev}) \epsilon ) (\bar u_1 \gamma^\mu P_L u_2) (\bar u_3 \gamma_\mu P_L u_4) - \< E_{LR}^{(3)} \>_{1243} \, .
\end{align}
This implies that the evanescent structures $\< E_{LL}^{(3)} \>_{1234}$ and $- \< E_{LR}^{(3)} \>_{1243}$ are trivially related only for $b_\mathrm{ev} = 3 + c_\mathrm{ev}$, while in general one has to be careful keeping track of crossed evanescent structures. A scheme preserving Fierz relations at the one-loop level requires $b_\mathrm{ev} = c_\mathrm{ev} = 1$ instead~\cite{Herrlich:1996vf}.

In the matching between the SMEFT and LEFT, the following bare evanescent operators arise already at tree level due to the application of Fierz identities in the $(\overline L L)(\overline L L)$ and $(\overline R R)(\overline R R)$ sectors:
\begin{align}
	\label{eq:TreeLevelEvanescents}
	\Ev{\nu e}{LL}[(F1)][prst] &= \Op{\nu e}{LL}[V][prst] - (\bar\nu_{Lp} \gamma^\mu e_{Lt})(\bar e_{Ls} \gamma_\mu \nu_{Lr}) \, , \nn
	\Ev{ee}{LL}[(F1)][prst] &= \Op{ee}{LL}[V][prst] - \Op{ee}{LL}[V][ptsr] \, , \nn
	\Ev{ud}{LL}[(F1)1][prst] &= \Op{ud}{LL}[V1][prst] - \frac{1}{N_c} ( \bar u_{Lp} \gamma^\mu d_{Lt} )( \bar d_{Ls} \gamma_\mu u_{Lr} ) - 2 ( \bar u_{Lp} \gamma^\mu T^A d_{Lt} )( \bar d_{Ls} \gamma_\mu T^A u_{Lr} )  \, , \nn
	\Ev{ud}{LL}[(F1)8][prst] &= \Op{ud}{LL}[V8][prst] - \frac{N_c^2-1}{2N_c^2} ( \bar u_{Lp} \gamma^\mu d_{Lt} )( \bar d_{Ls} \gamma_\mu u_{Lr} ) + \frac{1}{N_c} ( \bar u_{Lp} \gamma^\mu T^A d_{Lt} )( \bar d_{Ls} \gamma_\mu T^A u_{Lr} )  \, , \nn
	\Ev{ee}{RR}[(F1)][prst] &= \Op{ee}{RR}[V][prst] - \Op{ee}{RR}[V][ptsr] \, , \nn
	\Ev{ud}{RR}[(F1)1][prst] &= \Op{ud}{RR}[V1][prst] - \frac{1}{N_c} ( \bar u_{Rp} \gamma^\mu d_{Rt} )( \bar d_{Rs} \gamma_\mu u_{Rr} ) - 2 ( \bar u_{Rp} \gamma^\mu T^A d_{Rt} )( \bar d_{Rs} \gamma_\mu T^A u_{Rr} )  \, , \nn
	\Ev{ud}{RR}[(F1)8][prst] &= \Op{ud}{RR}[V8][prst] - \frac{N_c^2-1}{2N_c^2} ( \bar u_{Rp} \gamma^\mu d_{Rt} )( \bar d_{Rs} \gamma_\mu u_{Rr} ) + \frac{1}{N_c} ( \bar u_{Rp} \gamma^\mu T^A d_{Rt} )( \bar d_{Rs} \gamma_\mu T^A u_{Rr} )  \, .
\end{align}
Fierz relations are also used in the $\Delta B = \Delta L = 1$ sector, generating the evanescent operators
\begin{align}
	\label{eq:TreeLevelEvanescentsBL}
	\Ev{udd}{LL}[(F2)][prst] &= \Op{udd}{LL}[S][prst] - \Op{udd}{LL}[S][psrt] - \epsilon_{\alpha\beta\gamma} (d^{\alpha T}_{Ls} C d^\beta_{Lr})(u^{\gamma T}_{Lp} C \nu_{Lt}) \, , \nn
	\Ev{duu}{LL}[(F2)][prst] &= \Op{duu}{LL}[S][prst] - \Op{duu}{LL}[S][psrt] - \epsilon_{\alpha\beta\gamma} (u^{\alpha T}_{Ls} C u^\beta_{Lr})(d^{\gamma T}_{Lp} C e_{Lt})
\end{align}
and their Hermitian conjugates. When inserted into divergent one-loop matrix elements in the LEFT, the bare evanescent operators lead to a finite shift. The usual convention for renormalized evanescent operators~\cite{Buras:1989xd,Dugan:1990df,Herrlich:1994kh} corresponds to a finite renormalization of the physical operator coefficients that compensates this non-vanishing contribution of bare evanescent operators to one-loop matrix elements in the LEFT, see also the discussions in~\cite{Aebischer:2022aze,Fuentes-Martin:2022vvu}.

In the case of vector operators, the insertions of the Fierz-evanescent operators into 1PI four-point functions vanish for $b_\mathrm{ev} = 1$~\cite{Herrlich:1996vf}, but penguin corrections give a finite contribution independent of $b_\mathrm{ev}$~\cite{Buras:2000if}. 
In the original version of the present paper the finite renormalizations due to insertions of bare evanescent operators were not taken into account, which corresponds to the \msbar{} scheme and would require the inclusion of the evanescent operators~\eqref{eq:TreeLevelEvanescents} and~\eqref{eq:TreeLevelEvanescentsBL} in one-loop matrix elements in the LEFT. The results of the present version are given in the usual scheme where the insertion of the evanescent operators is compensated by finite renormalizations of the physical operator coefficients. In this scheme, the renormalization-group equations of physical parameters are independent of the coefficients of evanescent operators and the contribution of evanescent operators in one-loop matrix elements can be effectively ignored~\cite{Buras:1989xd,Dugan:1990df,Herrlich:1994kh}.

\subsection{Tadpoles}

Different treatments of tadpoles can be found in the literature, some of which are not compatible with the \msbar{} scheme, as they would introduce gauge dependences into the $S$-matrix, see~\cite{Denner:2016etu} for a detailed discussion. We follow the most direct approach and simply calculate all tadpole topologies explicitly, with diagrams shown in App.~\ref{sec:TadpoleDiagrams}. In this scheme, the top-loop tadpole generates relatively large corrections due to an enhancement factor $\propto \frac{N_c m_t^4}{M_H^2 v_T^2}$ in relations between observables and \msbar{} Lagrangian parameters~\cite{Bednyakov:2016onn,Cullen:2019nnr}. However, these enhanced corrections have to cancel again in relations between observables~\cite{Hartmann:2015oia}.

\subsection{Diagrams and required vertices}

For any diagram, the Euler formula relates the number of loops $L$ with the number of internal propagators $I$ and vertices $V$:
\begin{align}
	I = L + V - 1 \, .
\end{align}
Counting propagators, the number of internal propagators $I$ and external legs $E$ are related by
\begin{align}
	E + 2 I = \sum_n n V_n \, ,
\end{align}
where $V_n$ is the number of vertices with $n$ legs. This results in
\begin{align}
	E + 2 L - 2 = \sum_n (n-2) V_n \, .
\end{align}
For the matching of the SMEFT to the LEFT up to dimension six, we are interested in diagrams with up to $E=4$ external legs. Hence, at one loop the one-point function is given by the tadpole topology:
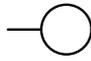
\begin{align*}
	\begin{gathered}
		\begin{fmfgraph*}(50,30)
			\fmfleft{l1} \fmfright{r1}
			\fmf{plain,tension=3}{l1,v1}
			\fmf{phantom,tension=2}{v2,r1}
			\fmf{plain,right}{v1,v2,v1}
		\end{fmfgraph*}
	\end{gathered}
\end{align*}
For the two-point functions we only need vertices with three and four legs, resulting in the following topologies:
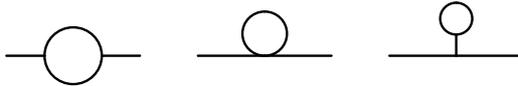
\begin{align*}
	\begin{gathered}
		\begin{fmfgraph*}(50,40)
			\fmfleft{l1} \fmfright{r1}
			\fmf{plain,tension=3}{l1,v1}
			\fmf{plain,tension=3}{v2,r1}
			\fmf{plain,right}{v1,v2,v1}
			\fmffreeze
		\end{fmfgraph*}
	\end{gathered}
\qquad
	\begin{gathered}
		\begin{fmfgraph*}(50,40)
			\fmfleft{l1} \fmfright{r1} \fmftop{t1}
			\fmf{plain,tension=3}{l1,v1}
			\fmf{plain,tension=3}{v1,r1}
			\fmffreeze
			\fmf{phantom,tension=5}{t1,v2}
			\fmf{phantom,tension=0.8}{v1,v1}
			\fmf{plain,tension=0.5,right=1}{v1,v2,v1}
		\end{fmfgraph*}
	\end{gathered}
\qquad
	\begin{gathered}
		\begin{fmfgraph*}(50,40)
			\fmfset{curly_len}{2mm}
			\fmfleft{l1} \fmfright{r1} \fmftop{t1}
			\fmf{plain,tension=3}{l1,v1}
			\fmf{plain,tension=3}{v1,r1}
			\fmffreeze
			\fmf{plain,tension=3}{v1,v2}
			\fmf{plain,right}{v2,t1,v2}
		\end{fmfgraph*}
	\end{gathered}
\end{align*}
For the three-point function vertices with three, four, and five legs are required, leading to four different topologies:
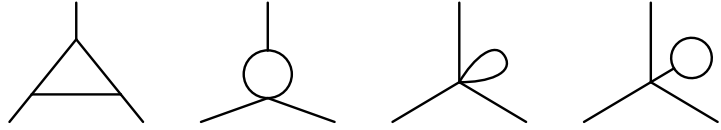
\begin{align*}
	\begin{gathered}
		\begin{fmfgraph*}(50,50)
			\fmfset{curly_len}{2mm}
			\fmftop{t1} \fmfbottom{b1,b2}
			\fmf{plain,tension=2}{b1,v1}
			\fmf{plain}{v3,v2}
			\fmf{plain}{v2,v1}
			\fmf{plain,tension=2}{v3,b2}
			\fmf{plain,tension=3}{t1,v2}
			\fmffreeze
			\fmf{plain}{v1,v3}
		\end{fmfgraph*}
	\end{gathered}
\qquad
	\begin{gathered}
		\begin{fmfgraph*}(50,50)
			\fmfset{curly_len}{2mm}
			\fmftop{t1} \fmfbottom{b1,b2}
			\fmf{plain,tension=2}{b1,v2}
			\fmf{plain,tension=2}{v2,b2}
			\fmf{plain,tension=2}{t1,v1}
			\fmf{plain,right}{v1,v2,v1}
		\end{fmfgraph*}
	\end{gathered}
\qquad
	\begin{gathered}
		\begin{fmfgraph*}(50,50)
			\fmfset{curly_len}{2mm}
			\fmftop{t1} \fmfbottom{b1,b2}
			\fmf{plain,tension=2}{b1,v1}
			\fmf{plain,tension=2}{v1,b2}
			\fmf{plain,tension=2}{t1,v1}
			\fmf{plain}{v1,v1}
		\end{fmfgraph*}
	\end{gathered}
\qquad
	\begin{gathered}
		\begin{fmfgraph*}(50,50)
			\fmfset{curly_len}{2mm}
			\fmftop{t1,t2,t3} \fmfbottom{b1,b2}
			\fmf{plain,tension=2}{b1,v1}
			\fmf{plain,tension=2}{v1,b2}
			\fmf{plain,tension=2}{t2,v1}
			\fmffreeze
			\fmf{plain,tension=3}{v1,v2}
			\fmf{plain,right}{v2,v3,v2}
			\fmf{phantom,tension=6}{t3,v3}
			\fmf{phantom,tension=2}{v3,b2}
		\end{fmfgraph*}
	\end{gathered}
\end{align*}
In the case of the four-point function, we need vertices with up to six legs. Omitting crossed versions, there are 14 topologies:
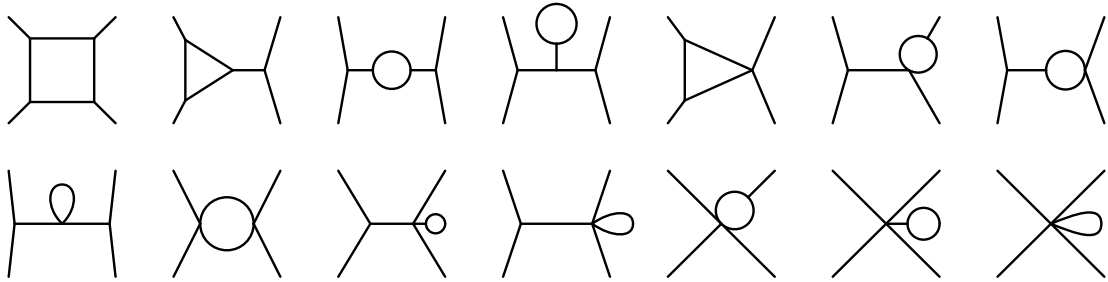
\begin{align*}
	& \begin{gathered}
		\begin{fmfgraph*}(40,50)
			\fmftop{t1,t2} \fmfbottom{b1,b2}
			\fmf{plain,tension=3}{b1,v1}
			\fmf{plain,tension=3}{b2,v2}
			\fmf{plain,tension=3}{t1,v3}
			\fmf{plain,tension=3}{t2,v4}
			\fmf{plain}{v1,v2,v4,v3,v1}
		\end{fmfgraph*}
	\end{gathered}
\qquad
	\begin{gathered}
		\begin{fmfgraph*}(40,50)
			\fmftop{t1,t2} \fmfbottom{b1,b2}
			\fmf{plain,tension=4}{b1,v1}
			\fmf{plain,tension=4}{t1,v2}
			\fmf{plain,tension=3}{b2,v3}
			\fmf{plain,tension=3}{t2,v3}
			\fmf{plain}{v4,v1,v2,v4}
			\fmf{plain,tension=3}{v3,v4}
		\end{fmfgraph*}
	\end{gathered}
\qquad
	\begin{gathered}
		\begin{fmfgraph*}(40,50)
			\fmftop{t1,t2} \fmfbottom{b1,b2}
			\fmf{plain,tension=4}{b1,v1,t1}
			\fmf{plain,tension=4}{b2,v2,t2}
			\fmf{plain,tension=3}{v1,v3}
			\fmf{plain,tension=3}{v4,v2}
			\fmf{plain,right}{v3,v4,v3}
		\end{fmfgraph*}
	\end{gathered}
\qquad
	\begin{gathered}
		\begin{fmfgraph*}(40,50)
			\fmftop{t1,t2,t3} \fmfbottom{b1,b2}
			\fmf{plain,tension=4}{b1,v1,t1}
			\fmf{plain,tension=4}{b2,v2,t3}
			\fmf{plain,tension=3}{v1,v3}
			\fmf{plain,tension=3}{v3,v2}
			\fmffreeze
			\fmf{plain,tension=3}{v3,v4}
			\fmf{plain,right}{v4,t2,v4}
		\end{fmfgraph*}
	\end{gathered}
\qquad
	\begin{gathered}
		\begin{fmfgraph*}(40,50)
			\fmftop{t1,t2} \fmfbottom{b1,b2}
			\fmf{plain,tension=4}{b1,v1}
			\fmf{plain,tension=4}{t1,v2}
			\fmf{plain,tension=3}{b2,v3}
			\fmf{plain,tension=3}{t2,v3}
			\fmf{plain}{v3,v1,v2,v3}
		\end{fmfgraph*}
	\end{gathered}
\qquad
	\begin{gathered}
		\begin{fmfgraph*}(40,50)
			\fmftop{t1,t2} \fmfbottom{b1,b2}
			\fmf{plain,tension=4}{b1,v1}
			\fmf{plain,tension=4}{t1,v1}
			\fmf{plain,tension=2}{b2,v2}
			\fmf{phantom,tension=2}{t2,v2}
			\fmf{plain,tension=2}{v1,v2}
			\fmffreeze
			\fmf{plain,tension=3}{t2,v3}
			\fmf{plain,right}{v2,v3,v2}
		\end{fmfgraph*}
	\end{gathered}
\qquad
	\begin{gathered}
		\begin{fmfgraph*}(40,50)
			\fmftop{t1,t2} \fmfbottom{b1,b2}
			\fmf{plain,tension=4}{b1,v1}
			\fmf{plain,tension=4}{t1,v1}
			\fmf{plain,tension=2}{b2,v2}
			\fmf{plain,tension=2}{t2,v2}
			\fmf{plain,tension=2}{v1,v3}
			\fmf{plain,right}{v2,v3,v2}
		\end{fmfgraph*}
	\end{gathered}
\nn
	& \begin{gathered}
		\begin{fmfgraph*}(40,50)
			\fmftop{t1,t2} \fmfbottom{b1,b2}
			\fmf{plain,tension=4}{b1,v1,t1}
			\fmf{plain,tension=4}{b2,v2,t2}
			\fmf{plain}{v1,v3,v2}
			\fmf{plain,right,tension=0.8}{v3,v3}
		\end{fmfgraph*}
	\end{gathered}
\qquad
	\begin{gathered}
		\begin{fmfgraph*}(40,50)
			\fmftop{t1,t2} \fmfbottom{b1,b2}
			\fmf{plain,tension=2}{b1,v1}
			\fmf{plain,tension=2}{t1,v1}
			\fmf{plain,tension=2}{b2,v2}
			\fmf{plain,tension=2}{t2,v2}
			\fmf{plain,right}{v1,v2,v1}
		\end{fmfgraph*}
	\end{gathered}
\qquad
	\begin{gathered}
		\begin{fmfgraph*}(40,50)
			\fmftop{t1,t2} \fmfbottom{b1,b2} \fmfright{r1}
			\fmf{plain,tension=2}{b1,v1}
			\fmf{plain,tension=2}{t1,v1}
			\fmf{plain,tension=2}{b2,v2}
			\fmf{plain,tension=2}{t2,v2}
			\fmf{plain,tension=3}{v1,v2}
			\fmffreeze
			\fmf{plain,tension=3}{v2,v3}
			\fmf{plain,right}{v3,r1,v3}
		\end{fmfgraph*}
	\end{gathered}
\qquad
	\begin{gathered}
		\begin{fmfgraph*}(40,50)
			\fmftop{t1,t2} \fmfbottom{b1,b2}
			\fmf{plain,tension=2}{b1,v1}
			\fmf{plain,tension=2}{t1,v1}
			\fmf{plain,tension=2}{b2,v2}
			\fmf{plain,tension=2}{t2,v2}
			\fmf{plain}{v1,v2}
			\fmf{plain,right}{v2,v2}
		\end{fmfgraph*}
	\end{gathered}
\qquad
	\begin{gathered}
		\begin{fmfgraph*}(40,50)
			\fmftop{t1,t2} \fmfbottom{b1,b2}
			\fmf{plain,tension=2}{b1,v1}
			\fmf{plain,tension=2}{t1,v1}
			\fmf{plain,tension=2}{b2,v1}
			\fmf{phantom,tension=2}{t2,v1}
			\fmffreeze
			\fmf{plain,right}{v1,v2,v1}
			\fmf{plain,tension=2}{t2,v2}
		\end{fmfgraph*}
	\end{gathered}
\qquad
	\begin{gathered}
		\begin{fmfgraph*}(40,50)
			\fmftop{t1,t2} \fmfbottom{b1,b2} \fmfright{r1}
			\fmf{plain,tension=2}{b1,v1}
			\fmf{plain,tension=2}{t1,v1}
			\fmf{plain,tension=2}{b2,v1}
			\fmf{plain,tension=2}{t2,v1}
			\fmffreeze
			\fmf{plain,tension=3}{v1,v2}
			\fmf{plain,right}{v2,r1,v2}
		\end{fmfgraph*}
	\end{gathered}
\qquad
	\begin{gathered}
		\begin{fmfgraph*}(40,50)
			\fmftop{t1,t2} \fmfbottom{b1,b2}
			\fmf{plain,tension=2}{b1,v1}
			\fmf{plain,tension=2}{t1,v1}
			\fmf{plain,tension=2}{b2,v1}
			\fmf{plain,tension=2}{t2,v1}
			\fmf{plain,right}{v1,v1}
		\end{fmfgraph*}
	\end{gathered}
\end{align*}
The only four-point functions we need to calculate are four-fermion Green's functions. As we only include vertices up to dimension six, the last three topologies do not appear and we do not need any vertices with six legs in the matching calculation.
The diagrams are best organized in terms of 1PI subgraphs. Distinguishing between fermions (solid lines) and bosons (dashed lines), we find the following classification (crossed diagrams and external-leg corrections are not shown):
\begin{align*}
	\begin{gathered}
		\begin{fmfgraph*}(50,50)
			\fmfleft{l1} \fmfright{r1}
			\fmf{dashes,tension=3}{l1,v1}
			\fmf{phantom,tension=2}{v1,r1}
			\fmfblob{5mm}{v1}
		\end{fmfgraph*}
	\end{gathered} \hspace{-0.5cm} &= \quad
	\begin{gathered}
		\begin{fmfgraph*}(50,50)
			\fmfleft{l1} \fmfright{r1}
			\fmf{dashes,tension=3}{l1,v1}
			\fmf{phantom,tension=2}{v2,r1}
			\fmf{plain,right}{v1,v2,v1}
		\end{fmfgraph*}
	\end{gathered} \hspace{-0.5cm} + \quad
	\begin{gathered}
		\begin{fmfgraph*}(50,50)
			\fmfleft{l1} \fmfright{r1}
			\fmf{dashes,tension=3}{l1,v1}
			\fmf{phantom,tension=2}{v2,r1}
			\fmf{dashes,right}{v1,v2,v1}
		\end{fmfgraph*}
	\end{gathered} \hspace{-0.4cm} ,
\nn
	\begin{gathered}
		\begin{fmfgraph*}(50,50)
			\fmfleft{l1} \fmfright{r1}
			\fmf{plain,tension=3}{l1,v1}
			\fmf{plain,tension=3}{v1,r1}
			\fmfblob{5mm}{v1}
		\end{fmfgraph*}
	\end{gathered} \; &= \;
	\begin{gathered}
		\begin{fmfgraph*}(50,50)
			\fmfleft{l1} \fmfright{r1}
			\fmf{plain,tension=3}{l1,v1}
			\fmf{plain,tension=3}{v2,r1}
			\fmf{plain,right}{v1,v2}
			\fmf{dashes,right}{v2,v1}
			\fmffreeze
		\end{fmfgraph*}
	\end{gathered} \; + \;
	\begin{gathered}
		\begin{fmfgraph*}(50,50)
			\fmfleft{l1} \fmfright{r1} \fmftop{t1}
			\fmf{plain,tension=3}{l1,v1}
			\fmf{plain,tension=3}{v1,r1}
			\fmffreeze
			\fmf{phantom,tension=5}{t1,v2}
			\fmf{phantom,tension=0.8}{v1,v1}
			\fmf{plain,tension=0.5,right=1}{v1,v2,v1}
		\end{fmfgraph*}
	\end{gathered} \; + \;
	\begin{gathered}
		\begin{fmfgraph*}(50,50)
			\fmfleft{l1} \fmfright{r1} \fmftop{t1}
			\fmf{plain,tension=3}{l1,v1}
			\fmf{plain,tension=3}{v1,r1}
			\fmffreeze
			\fmf{phantom,tension=5}{t1,v2}
			\fmf{phantom,tension=0.8}{v1,v1}
			\fmf{dashes,tension=0.5,right=1}{v1,v2,v1}
		\end{fmfgraph*}
	\end{gathered} \; + \;
	\begin{gathered}
		\begin{fmfgraph*}(50,50)
			\fmfset{curly_len}{2mm}
			\fmfleft{l1} \fmfright{r1} \fmftop{t1}
			\fmf{plain,tension=3}{l1,v1}
			\fmf{plain,tension=3}{v1,r1}
			\fmffreeze
			\fmf{dashes}{v1,v2}
			\fmf{phantom,tension=2}{v2,t1}
			\fmfblob{5mm}{v2}
		\end{fmfgraph*}
	\end{gathered} \;\; ,
\nn
	\begin{gathered}
		\begin{fmfgraph*}(50,50)
			\fmfleft{l1} \fmfright{r1}
			\fmf{dashes,tension=3}{l1,v1}
			\fmf{dashes,tension=3}{v1,r1}
			\fmfblob{5mm}{v1}
		\end{fmfgraph*}
	\end{gathered} \; &= \;
	\begin{gathered}
		\begin{fmfgraph*}(50,50)
			\fmfleft{l1} \fmfright{r1}
			\fmf{dashes,tension=3}{l1,v1}
			\fmf{dashes,tension=3}{v2,r1}
			\fmf{plain,right}{v1,v2,v1}
			\fmffreeze
		\end{fmfgraph*}
	\end{gathered} \; + \;
	\begin{gathered}
		\begin{fmfgraph*}(50,50)
			\fmfleft{l1} \fmfright{r1}
			\fmf{dashes,tension=3}{l1,v1}
			\fmf{dashes,tension=3}{v2,r1}
			\fmf{dashes,right}{v1,v2}
			\fmf{dashes,right}{v2,v1}
			\fmffreeze
		\end{fmfgraph*}
	\end{gathered} \; + \;
	\begin{gathered}
		\begin{fmfgraph*}(50,50)
			\fmfleft{l1} \fmfright{r1} \fmftop{t1}
			\fmf{dashes,tension=3}{l1,v1}
			\fmf{dashes,tension=3}{v1,r1}
			\fmffreeze
			\fmf{phantom,tension=5}{t1,v2}
			\fmf{phantom,tension=0.8}{v1,v1}
			\fmf{plain,tension=0.5,right=1}{v1,v2,v1}
		\end{fmfgraph*}
	\end{gathered} \; + \;
	\begin{gathered}
		\begin{fmfgraph*}(50,50)
			\fmfleft{l1} \fmfright{r1} \fmftop{t1}
			\fmf{dashes,tension=3}{l1,v1}
			\fmf{dashes,tension=3}{v1,r1}
			\fmffreeze
			\fmf{phantom,tension=5}{t1,v2}
			\fmf{phantom,tension=0.8}{v1,v1}
			\fmf{dashes,tension=0.5,right=1}{v1,v2,v1}
		\end{fmfgraph*}
	\end{gathered} \; + \;
	\begin{gathered}
		\begin{fmfgraph*}(50,50)
			\fmfset{curly_len}{2mm}
			\fmfleft{l1} \fmfright{r1} \fmftop{t1}
			\fmf{dashes,tension=3}{l1,v1}
			\fmf{dashes,tension=3}{v1,r1}
			\fmffreeze
			\fmf{dashes}{v1,v2}
			\fmf{phantom,tension=2}{v2,t1}
			\fmfblob{5mm}{v2}
		\end{fmfgraph*}
	\end{gathered} \;\; ,
\nn
	\begin{gathered}
		\begin{fmfgraph*}(40,40)
			\fmfset{curly_len}{2mm}
			\fmftop{t1,t2,t3} \fmfbottom{b1,b2}
			\fmf{plain,tension=2}{b1,v1}
			\fmf{plain,tension=2}{v1,b2}
			\fmf{dashes,tension=2}{t2,v1}
			\fmfblob{5mm}{v1}
		\end{fmfgraph*}
	\end{gathered} \; &= \;
	\begin{gathered}
		\begin{fmfgraph*}(40,40)
			\fmfset{curly_len}{2mm}
			\fmftop{t1} \fmfbottom{b1,b2}
			\fmf{plain,tension=2}{b1,v1}
			\fmf{plain}{v3,v2}
			\fmf{plain}{v2,v1}
			\fmf{plain,tension=2}{v3,b2}
			\fmf{dashes,tension=3}{t1,v2}
			\fmffreeze
			\fmf{dashes}{v1,v3}
		\end{fmfgraph*}
	\end{gathered} \; + \;
	\begin{gathered}
		\begin{fmfgraph*}(40,40)
			\fmfset{curly_len}{2mm}
			\fmftop{t1} \fmfbottom{b1,b2}
			\fmf{plain,tension=2}{b1,v1}
			\fmf{dashes}{v3,v2}
			\fmf{dashes}{v2,v1}
			\fmf{plain,tension=2}{v3,b2}
			\fmf{dashes,tension=3}{t1,v2}
			\fmffreeze
			\fmf{plain}{v1,v3}
		\end{fmfgraph*}
	\end{gathered} \; + \;
	\begin{gathered}
		\begin{fmfgraph*}(40,40)
			\fmfset{curly_len}{2mm}
			\fmftop{t1} \fmfbottom{b1,b2}
			\fmf{plain,tension=2}{b1,v2}
			\fmf{plain,tension=2}{v2,b2}
			\fmf{dashes,tension=2}{t1,v1}
			\fmf{plain,right}{v1,v2,v1}
		\end{fmfgraph*}
	\end{gathered} \; + \;
	\begin{gathered}
		\begin{fmfgraph*}(40,40)
			\fmfset{curly_len}{2mm}
			\fmftop{t1} \fmfbottom{b1,b2}
			\fmf{plain,tension=2}{b1,v2}
			\fmf{plain,tension=2}{v2,b2}
			\fmf{dashes,tension=2}{t1,v1}
			\fmf{dashes,right}{v1,v2,v1}
		\end{fmfgraph*}
	\end{gathered} \; + \;
	\begin{gathered}
		\begin{fmfgraph*}(40,40)
			\fmfset{curly_len}{2mm}
			\fmftop{t1} \fmfbottom{b1,b2}
			\fmf{plain,tension=2}{b1,v2}
			\fmf{plain,tension=0.67}{v1,b2}
			\fmf{dashes,tension=0.67}{t1,v1}
			\fmf{plain,tension=1}{v2,v1}
			\fmffreeze
			\fmf{dashes,left}{v2,v1}
		\end{fmfgraph*}
	\end{gathered} \; + \;
	\begin{gathered}
		\begin{fmfgraph*}(40,40)
			\fmfset{curly_len}{2mm}
			\fmftop{t1} \fmfbottom{b1,b2}
			\fmf{plain,tension=2}{b1,v1}
			\fmf{plain,tension=2}{v1,b2}
			\fmf{dashes,tension=2}{t1,v1}
			\fmf{dashes}{v1,v1}
		\end{fmfgraph*}
	\end{gathered} \; \nn
	&\quad + \;
	\begin{gathered}
		\begin{fmfgraph*}(40,40)
			\fmfset{curly_len}{2mm}
			\fmftop{t1,t2,t3} \fmfbottom{b1,b2}
			\fmf{plain,tension=2}{b1,v1}
			\fmf{plain,tension=2}{v1,b2}
			\fmf{dashes,tension=2}{t2,v1}
			\fmffreeze
			\fmf{dashes,tension=3}{v1,v2}
			\fmf{phantom,tension=6}{t3,v2}
			\fmf{phantom,tension=2}{v2,b2}
			\fmfblob{5mm}{v2}
		\end{fmfgraph*}
	\end{gathered} \;\; ,
\nn
	\begin{gathered}
		\begin{fmfgraph*}(40,40)
			\fmfset{curly_len}{2mm}
			\fmftop{t1,t2,t3} \fmfbottom{b1,b2}
			\fmf{dashes,tension=2}{b1,v1}
			\fmf{dashes,tension=2}{v1,b2}
			\fmf{dashes,tension=2}{t2,v1}
			\fmfblob{5mm}{v1}
		\end{fmfgraph*}
	\end{gathered} \; &= \;
	\begin{gathered}
		\begin{fmfgraph*}(40,40)
			\fmfset{curly_len}{2mm}
			\fmftop{t1} \fmfbottom{b1,b2}
			\fmf{dashes,tension=2}{b1,v1}
			\fmf{plain}{v3,v2}
			\fmf{plain}{v2,v1}
			\fmf{dashes,tension=2}{v3,b2}
			\fmf{dashes,tension=3}{t1,v2}
			\fmffreeze
			\fmf{plain}{v1,v3}
		\end{fmfgraph*}
	\end{gathered} \; + \;
	\begin{gathered}
		\begin{fmfgraph*}(40,40)
			\fmfset{curly_len}{2mm}
			\fmftop{t1} \fmfbottom{b1,b2}
			\fmf{dashes,tension=2}{b1,v1}
			\fmf{dashes}{v3,v2}
			\fmf{dashes}{v2,v1}
			\fmf{dashes,tension=2}{v3,b2}
			\fmf{dashes,tension=3}{t1,v2}
			\fmffreeze
			\fmf{dashes}{v1,v3}
		\end{fmfgraph*}
	\end{gathered} \; + \;
	\begin{gathered}
		\begin{fmfgraph*}(40,40)
			\fmfset{curly_len}{2mm}
			\fmftop{t1} \fmfbottom{b1,b2}
			\fmf{dashes,tension=2}{b1,v2}
			\fmf{dashes,tension=2}{v2,b2}
			\fmf{dashes,tension=2}{t1,v1}
			\fmf{plain,right}{v1,v2,v1}
		\end{fmfgraph*}
	\end{gathered} \; + \;
	\begin{gathered}
		\begin{fmfgraph*}(40,40)
			\fmfset{curly_len}{2mm}
			\fmftop{t1} \fmfbottom{b1,b2}
			\fmf{dashes,tension=2}{b1,v2}
			\fmf{dashes,tension=2}{v2,b2}
			\fmf{dashes,tension=2}{t1,v1}
			\fmf{dashes,right}{v1,v2,v1}
		\end{fmfgraph*}
	\end{gathered} \; + \;
	\begin{gathered}
		\begin{fmfgraph*}(40,40)
			\fmfset{curly_len}{2mm}
			\fmftop{t1} \fmfbottom{b1,b2}
			\fmf{dashes,tension=2}{b1,v1}
			\fmf{dashes,tension=2}{v1,b2}
			\fmf{dashes,tension=2}{t1,v1}
			\fmf{dashes}{v1,v1}
		\end{fmfgraph*}
	\end{gathered} \; + \;
	\begin{gathered}
		\begin{fmfgraph*}(40,40)
			\fmfset{curly_len}{2mm}
			\fmftop{t1,t2,t3} \fmfbottom{b1,b2}
			\fmf{dashes,tension=2}{b1,v1}
			\fmf{dashes,tension=2}{v1,b2}
			\fmf{dashes,tension=2}{t2,v1}
			\fmffreeze
			\fmf{dashes,tension=3}{v1,v2}
			\fmf{phantom,tension=6}{t3,v2}
			\fmf{phantom,tension=2}{v2,b2}
			\fmfblob{5mm}{v2}
		\end{fmfgraph*}
	\end{gathered} \;\; ,
\nn
	\begin{gathered}
		\begin{fmfgraph*}(40,50)
			\fmftop{t1,t2} \fmfbottom{b1,b2}
			\fmf{plain}{b1,v1}
			\fmf{plain}{t1,v1}
			\fmf{plain}{b2,v1}
			\fmf{plain}{t2,v1}
			\fmfblob{5mm}{v1}
		\end{fmfgraph*}
	\end{gathered} \; &= \;
	\begin{gathered}
		\begin{fmfgraph*}(40,50)
			\fmftop{t1,t2} \fmfbottom{b1,b2}
			\fmf{plain,tension=3}{b1,v1}
			\fmf{plain,tension=3}{b2,v2}
			\fmf{plain,tension=3}{t1,v3}
			\fmf{plain,tension=3}{t2,v4}
			\fmf{dashes}{v1,v2}
			\fmf{plain}{v2,v4}
			\fmf{dashes}{v4,v3}
			\fmf{plain}{v3,v1}
		\end{fmfgraph*}
	\end{gathered} \; + \;
	\begin{gathered}
		\begin{fmfgraph*}(40,50)
			\fmftop{t1,t2} \fmfbottom{b1,b2}
			\fmf{plain,tension=4}{b1,v1}
			\fmf{plain,tension=4}{t1,v2}
			\fmf{plain,tension=3}{b2,v3}
			\fmf{plain,tension=3}{t2,v3}
			\fmf{plain}{v1,v2}
			\fmf{dashes}{v1,v3,v2}
		\end{fmfgraph*}
	\end{gathered} \; + \;
	\begin{gathered}
		\begin{fmfgraph*}(40,50)
			\fmftop{t1,t2} \fmfbottom{b1,b2}
			\fmf{plain,tension=4}{b1,v1}
			\fmf{plain,tension=4}{t1,v2}
			\fmf{plain,tension=3}{b2,v3}
			\fmf{plain,tension=3}{t2,v3}
			\fmf{dashes}{v1,v2}
			\fmf{plain}{v1,v3,v2}
		\end{fmfgraph*}
	\end{gathered} \; + \;
	\begin{gathered}
		\begin{fmfgraph*}(40,50)
			\fmftop{t1,t2} \fmfbottom{b1,b2}
			\fmf{plain,tension=2}{b1,v1,t1}
			\fmf{plain,tension=2}{b2,v2,t2}
			\fmf{dashes,right}{v1,v2,v1}
		\end{fmfgraph*}
	\end{gathered} \; + \;
	\begin{gathered}
		\begin{fmfgraph*}(40,50)
			\fmftop{t1,t2} \fmfbottom{b1,b2}
			\fmf{plain}{b1,v1}
			\fmf{plain}{t1,v1}
			\fmf{plain}{b2,v2}
			\fmf{plain}{t2,v2}
			\fmf{dashes}{v1,v2}
			\fmfblob{5mm}{v1}
		\end{fmfgraph*}
	\end{gathered} \; + \;
	\begin{gathered}
		\begin{fmfgraph*}(40,50)
			\fmftop{t1,t2} \fmfbottom{b1,b2}
			\fmf{plain,tension=4}{b1,v1,t1}
			\fmf{plain,tension=4}{b2,v2,t2}
			\fmf{dashes,tension=3}{v1,v3}
			\fmf{dashes,tension=3}{v3,v2}
			\fmfblob{5mm}{v3}
		\end{fmfgraph*}
	\end{gathered} \;\;.
\end{align*}
The fourth (bulb) diagram in the four-point function involves two dimension-five operators, hence it is proportional to $m_\nu^2$ and neglected in our calculation as we will explain in Sect.~\ref{sec:PowerCounting}.

All diagrams that contribute to the matching equations at the considered order (and some diagrams that turn out to vanish) are listed in App.~\ref{sec:Diagrams}.

\subsection{Computational tools}

The diagrammatic matching of the SMEFT and LEFT involves a large number of one-loop diagrams. In order to handle this complexity, we made use of several computer algebra systems. First, we implemented, using dummy fields, a version of the SMEFT Lagrangian that encodes the type of fields and possible vertices in \textsc{FeynRules}~\cite{Christensen:2008py,Alloul:2013bka}. We then used \textsc{FeynArts}~\cite{Hahn:2000kx} to generate the complete set of diagrams. The extraction of the actual Feynman rules for the SMEFT in background-field gauge was performed using manually coded \textsc{Mathematica} routines and the \texttt{FeynRule[]} command of \textsc{FeynCalc}~\cite{Mertig:1990an,Shtabovenko:2016sxi}. Several commands of the same package were used in different steps of the loop calculation, e.g.\ for the tensor reduction (\texttt{TID[]}) and Dirac algebra (\texttt{DiracReduce[]}). For the analytic evaluation of scalar loop functions, we performed cross-checks with \textsc{Package-X}~\cite{Patel:2015tea}.
We compared the results of two independent implementations of the complete computation.


\section{Matching procedure}
\label{sec:Matching}

\subsection{Expanding loops}
\label{sec:ExpandingLoops}

For the matching procedure, we follow standard EFT techniques, see e.g.~\cite{Manohar:1996cq,Manohar:1997qy,Manohar:2018aog} for detailed discussions.
The matching condition at the electroweak scale requires that the LEFT and SMEFT $S$-matrix elements for the light-particle processes agree. This can be achieved by matching one-light-particle irreducible (1LPI) amplitudes including wave-function renormalization:
\begin{align}
	\label{eq:Matching1LPIWF}
	\M^\text{LEFT} = \M^\text{SMEFT} \, .
\end{align}
The goal of the matching is to express the renormalized LEFT coefficients in term of the renormalized SMEFT coefficients. This relation obviously is scheme dependent, hence it is important that the presented results are used only in connection with loop calculations that employ the same scheme definitions as the matching calculation, defined in Sect.~\ref{sec:LoopCalculation}, or that perform an explicit scheme translation.

Each side of~\eqref{eq:Matching1LPIWF} consists of tree-level and one-loop diagrams, where the bare parameters in the tree-level part are split into renormalized parameters plus counterterms that cancel the UV divergences of the loops:
\begin{align}
	\M^\text{LEFT}_{\text{tree, ren.}} + \M^\text{LEFT}_{\text{ct}} + \M^\text{LEFT}_{\text{loop}} = \M^\text{SMEFT}_{\text{tree, ren.}} + \M^\text{SMEFT}_{\text{ct}} + \M^\text{SMEFT}_{\text{loop}} \, .
\end{align}
The part of the LEFT tree-level amplitude that only contains the renormalized parameters is given by
\begin{align}
	\M^\text{LEFT}_{\text{tree, ren.}} = \M^\text{SMEFT}_{\text{tree, ren.}} + \M^\text{SMEFT}_{\text{ct}} + \M^\text{SMEFT}_{\text{loop}} - ( \M^\text{LEFT}_{\text{ct}} + \M^\text{LEFT}_{\text{loop}}  ) \, .
\end{align}
It is analytic in the low-energy scales, i.e.\ a polynomial. On the right-hand side of the equation, both SMEFT and LEFT loop contributions contain non-analytic pieces in the low scales that cancel in the difference. In principle, the integrands of the loop integrals cannot be expanded in the low-energy scales, because this alters the structure of the non-analytic pieces of the integrals. However, if we expand the integrands, both non-analytic pieces in the SMEFT and the LEFT loops are altered in the same way and the analytic difference is unaffected. Therefore
\begin{align}
	\M^\text{LEFT}_{\text{tree, ren.}} = \M^\text{SMEFT}_{\text{tree, ren.}} + \M^\text{SMEFT}_{\text{ct}} + \M^\text{SMEFT}_{\text{loop, exp.}} - ( \M^\text{LEFT}_{\text{ct}} + \M^\text{LEFT}_{\text{loop, exp.}}  ) \, .
\end{align}
The LEFT integrals with integrands expanded in the low-energy scales are scaleless integrals, which vanish in dimensional regularization. This leads us to the matching prescription
\begin{align}
	\label{eq:Matching}
	\M^\text{LEFT}_{\text{tree, ren.}} = \M^\text{SMEFT}_{\text{tree, ren.}} + \M^\text{SMEFT}_{\text{loop, exp.}} + \M^\text{SMEFT}_{\text{ct}} - \M^\text{LEFT}_{\text{ct}} \, .
\end{align}
Note that the SMEFT and LEFT loop contributions can both contain IR divergences. Since the LEFT reproduces the IR structure of the SMEFT, these divergences are identical and cancel in the difference. After the expansion of the loop integrands in the light scales, the structure of the IR-divergences are altered both in the SMEFT and LEFT loops in the same way. Since the expanded LEFT integrals vanish, the altered IR divergences are equal to minus the LEFT UV divergences. The expansion in the light scales does not affect the UV structure. Therefore, in~\eqref{eq:Matching} the UV divergences of the expanded SMEFT loop integrals are cancelled by the SMEFT counterterms, whereas the IR divergences of the expanded SMEFT loops are cancelled by subtracting the LEFT UV counterterms. The finite part of the expanded loops gives the matching contribution. Note that in this matching procedure, the only diagrams that have to be calculated are the expanded SMEFT loops. The SMEFT counterterms can be extracted from the RGEs in~\cite{Jenkins:2013wua,Jenkins:2013zja,Alonso:2013hga}, the counterterms for the LEFT are obtained from the RGEs in~\cite{Jenkins:2017dyc}.

\subsection{Power counting}
\label{sec:PowerCounting}

The power counting of both the SMEFT and the LEFT is given in terms of canonical mass dimensions. The dimensionless expansion parameter of the SMEFT is $v/\Lambda$ or $p/\Lambda$, where $\Lambda$ is the heavy mass scale of NP, $v$ denotes the vacuum expectation values (vev) or the mass of a SM particle, and $p$ an external momentum. Similarly, in the LEFT the expansion parameter is $m/v$ or $p/v$, where $m$ stands for a light SM particle mass. By matching the SMEFT and the LEFT at the weak scale, the low-energy parameters are expressed in terms of SMEFT parameters and inherit the SMEFT power couting~\cite{Jenkins:2017jig}. Through the matching equations, the scale $\Lambda$ appears in the LEFT in the form
\begin{align}
	\frac{m}{\Lambda} = \frac{m}{v} \frac{v}{\Lambda} \, ,
\end{align}
which separates LEFT and SMEFT power counting. Following the procedure outlined in the previous section, the matching determines the Wilson coefficients of the LEFT in terms of SMEFT parameters. In the SMEFT, the scale $\Lambda$ of NP only appears explicitly through the expansion parameter as a polynomial in $1/\Lambda$. An implicit dependence on NP parameters is absorbed into the values of the Wilson coefficients $C_i$. The SMEFT contains the full non-analytic dependence on the heavy and light SM scales, $M$ and $m$. The LEFT reproduces the same IR physics as the SMEFT, i.e.\ it explicitly contains the same non-analytic structure in terms of the light SM scales. Therefore, the contribution of the LEFT Wilson coefficients $L_i$ is analytic in the low scales $m$. In the LEFT, the vev $v$ explicitly appears only as polynomials in $1/v$ due to the LEFT expansion parameter, while any further dependence on $v$ is absorbed into the values of the Wilson coefficients $L_i$. Hence, the matching equations depend on two expansion parameters
\begin{align}
	\epsilon_\mathrm{LEFT} = \frac{m}{v} \, , \quad \epsilon_\mathrm{SMEFT} = \frac{v}{\Lambda}
\end{align}
in a polynomial way, and in addition contain pieces that are non-analytic in the ratios of the heavy SM scales and the matching scale $\mu_W$. In the following, we will work out these equations at one loop and up to dimension six in both the LEFT and SMEFT expansions, i.e.\ we consider contributions to the LEFT Lagrangian parameters of the order $\epsilon_\mathrm{LEFT}^2$ and $\epsilon_\mathrm{SMEFT}^2$. This implies that we in principle take into account double insertions of the dimension-five SMEFT operator and single insertions of the dimension-six SMEFT operators. At the same time, we will consider corrections to the LEFT masses and gauge couplings containing up to two additional powers of light scales. For corrections to the dimension-five LEFT dipole coefficients, we will work up to linear order in the light scales, while for the dimension-six operator coefficients we will set the light scales to zero. A special case is the Majorana-neutrino mass term, which is induced at tree level by the dimension-five SMEFT operator,
\begin{align}
	\left[ M_\nu \right]_{rs} = - C_{\substack{5 \\ rs}} v_T^2 \, .
\end{align}
Hence, this term can be counted either as $v \times \epsilon_\mathrm{LEFT}$ or as $v \times \epsilon_\mathrm{SMEFT}$. The smallness of the neutrino masses shows that the scale of lepton-number-violating NP is huge and could be much larger that the generic scale $\Lambda$ for lepton- and baryon-number-conserving NP. Therefore, we consider both counting schemes and only take into account single insertions of $Q_5$ in the matching for the neutrino dipole operator, but neglect insertions of $Q_5$ in the matching for the dimension-six operators. This implies that for the four-fermion matching equations, the neutrinos can be treated as massless Weyl fermions. We also neglect corrections to the neutrino masses of order $\delta m_\nu \sim m_\nu^3$: since $m_\nu \sim v \times \epsilon_\mathrm{SMEFT}$, these are dimension-seven contributions in the SMEFT power counting. In order to include double insertions of $Q_5$ in the matching of the four-fermion operators, one should go to dimension eight in the LEFT counting. Vice versa, in order to consistently include $m_\nu^3$ contributions to the neutrino-mass corrections, one should work at dimension seven in the SMEFT counting. In the context of RGEs, double insertions of the dimension-five operator have been considered in~\cite{Broncano:2004tz,Davidson:2018zuo}.

\subsection{Fermion two-point functions}

The fermion two-point functions provide the matching conditions for the mass matrices. Through field-redefinitions, they also contribute to higher-dimension operators as external-leg corrections. In Sect.~\ref{sec:FermionFieldRedefinitions}, we describe the matching equations for the mass matrices and derive the necessary field redefinitions. We are working in a basis with non-diagonal complex mass matrices. In Sect.~\ref{sec:WaveFunctionMixing}, we discuss the diagonalization of the mass matrices and the relation to $S$-matrix elements.

\subsubsection{Field redefinitions and mass matrices}
\label{sec:FermionFieldRedefinitions}

We assume that the SMEFT tree-level mass matrices have been diagonalized as in~\eqref{eq:TreeLevelSMEFTMasses}. The inverse fermion propagator is calculated from the two-point function. For Dirac fermions we find the structure
\begin{align}
	\label{eq:FermionInversePropagator}
	S^{-1}_{pr} = -i ( \slashed p - m + \slashed p P_L \Sigma^L + \slashed p P_R \Sigma^R - P_L \M^L - P_R \M^R  )_{pr} \, ,
\end{align}
where $\Sigma^L$, $\Sigma^R$, $\M^L$, $\M^R$, are loop-generated matrices in flavor space and the tree-level mass term is given by $m_{pr} = m_p \delta_{pr}$, with $p,r$ flavor indices. For Dirac fermions, the matrices fulfill $\M := \M^L = \M^{R\dagger}$, $\Sigma^L = \Sigma^{L\dagger}$, and  $\Sigma^R = \Sigma^{R\dagger}$.
In the case of Majorana neutrinos, the additional Majorana condition
\begin{align}
	\nu_M = C \bar\nu_M^T
\end{align}
implies the relations
\begin{align}
	\Sigma^M := \Sigma^L = (\Sigma^R)^T \, , \quad \M = \M^T \, ,
\end{align}
i.e.\ the inverse Majorana propagator has the form
\begin{align}
	S^{-1}_{pr} = -i ( \slashed p - m + \slashed p P_L \Sigma^M + \slashed p P_R \Sigma^{M*} - P_L \M - P_R \M^*  )_{pr} \, .
\end{align}

If the loops are expanded in all the light scales as discussed in Sect.~\ref{sec:ExpandingLoops}, the inverse propagator~\eqref{eq:FermionInversePropagator} can be directly identified with Lagrangian terms of the LEFT. Terms proportional to $p^2$ in $\Sigma^{L,R}$ and $\M^{L,R}$ correspond to higher-derivative operators that are not in the canonical LEFT operator basis. They can be removed by performing an explicit field redefinition, which is often referred to as using the equations of motion (EOM). As discussed in~\cite{Jenkins:2017dyc}, the naive application of the EOM can lead to apparently ambiguous results. The ambiguity is due to the fact that derivatives can be integrated by parts before the application of the EOM. By properly writing the field redefinitions, one sees that the ambiguities are simply related by chiral field redefinitions~\cite{Jenkins:2017dyc}, which are always possible.

Consider the most general field redefinition that removes terms with up to three derivatives:
\begin{align}
	\label{eq:FermionFieldRedefinition}
	\psi_{L,R} \mapsto \psi_{L,R} + A_{L,R} \psi_{L,R} + B_{L,R} i \slashed D \psi_{R,L} + C_{L,R} (i \slashed D)^2 \psi_{L,R} \, ,
\end{align}
where $A_{L,R}$, $B_{L,R}$, and $C_{L,R}$ are generic matrices in flavor space of $\O(\epsilon)$ and $\epsilon$ formally keeps track of power counting. The induced change in the Lagrangian
\begin{align}
	\L = \bar\psi ( i \slashed D - M P_L - M^\dagger P_R ) \psi + \O(\epsilon)
\end{align}
is given by
\begin{align}
	\label{eq:LagrangianFermionFieldRedefinition}
	\delta \L &= \bar\psi_L ( C_L + C_L^\dagger ) (i \slashed D)^3 \psi_L + \bar\psi_R ( C_R + C_R^\dagger ) (i \slashed D)^3 \psi_R \nn
		&\quad + \bar\psi_L ( B_L + B_R^\dagger - M^\dagger C_R - C_L^\dagger M^\dagger) (i \slashed D)^2 \psi_R + \bar\psi_R ( B_R + B_L^\dagger - M C_L - C_R^\dagger M ) (i \slashed D)^2 \psi_L \nn
		&\quad + \bar\psi_L ( A_L + A_L^\dagger - M^\dagger B_R - B_R^\dagger M ) i \slashed D \psi_L +  \bar\psi_R ( A_R + A_R^\dagger  - M B_L - B_L^\dagger M^\dagger ) i \slashed D \psi_R \nn
		&\quad - \bar\psi_R ( M A_L + A_R^\dagger M ) \psi_L - \bar\psi_L ( M^\dagger A_R + A_L^\dagger M^\dagger ) \psi_R \, .
\end{align}
Up to dimension-six effects, the one-loop fermion two-point function will generate terms that can be written as a Lagrangian
\begin{align}
	\L = \bar\psi &\Big( i \slashed D - M P_L - M^\dagger P_R + i \slashed D P_L \Sigma^L(0) + (i \slashed D)^3 P_L {\Sigma^L}'(0) + i \slashed D P_R \Sigma^R(0) + (i \slashed D)^3 P_R {\Sigma^R}'(0) \nn
		& - P_L \M(0) - P_R \M^\dagger(0) - (i\slashed D)^2 P_L \M'(0) - (i \slashed D)^2 P_R {\M'}^\dagger(0)  \Big) \psi \, .
\end{align}
where the primes denote a derivative with respect to $p^2$.

In order to remove all the terms with multiple derivatives, we choose a field redefinition with
\begin{align}
	C_L + C_L^\dagger &= - {\Sigma^L}'(0) \, , \quad C_R + C_R^\dagger = - {\Sigma^R}'(0) \, , \nn
	B_R + B_L^\dagger &= \M'(0) - \frac{1}{2} M {\Sigma^L}'(0) - \frac{1}{2} {\Sigma^R}'(0) M + \frac{1}{2} M (C_L - C_L^\dagger) - \frac{1}{2} (C_R - C_R^\dagger) M \, , \nn
	A_L + A_L^\dagger &= - \Sigma^L(0) + B_R^\dagger M + M^\dagger B_R \, , \quad A_R + A_R^\dagger = - \Sigma^R(0) + B_L^\dagger M^\dagger + M B_L \, .
\end{align}
The anti-Hermitian parts of $A_{L,R}$, $C_{L,R}$ and the combination $B_R-B_L^\dagger$ are not fixed and represent chiral transformations. The axial part of the chiral transformation induces a shift in the theta parameters, which however is of two-loop order, see Sect.~\ref{sec:ThetaParameters}. The induced phase is proportional to
\begin{align}
	\arg \det &\left( (1+A_R)^\dagger (1+A_L) \right) = \arg\det\left( 1+A_R^\dagger + A_L \right) = \arg\left( 1 + \tr[A_R^\dagger] + \tr[A_L] \right) \nn
		&= \Im\left( \tr[A_R^\dagger] + \tr[A_L] \right) = \frac{1}{2i} \left( \tr[ A_L - A_L^\dagger] - \tr[A_R - A_R^\dagger] \right) \, .
\end{align}
We choose
\begin{align}
	C_L - C_L^\dagger &= C_R - C_R^\dagger = 0 \, , \nn
	B_R - B_L^\dagger &= \frac{3}{2} \Big( {\Sigma^R}'(0) M - M {\Sigma^L}'(0) \Big) \, , \nn
	A_L - A_L^\dagger &= \frac{1}{2} \Big( {\M'}^\dagger(0) M - M^\dagger \M'(0) \Big) \, , \quad A_R - A_R^\dagger = \frac{1}{2} \Big( M {\M'}^\dagger(0) - \M'(0) M^\dagger \Big) \, .
\end{align}
Since $\tr[A_L - A_L^\dagger] = \tr[A_R - A_R^\dagger]$, the chiral rotation does not contain an axial part and does not affect the theta terms.
The transformed Lagrangian then becomes
\begin{align}
	\L + \delta \L &= \bar\psi \Big( i \slashed D - M P_L - M^\dagger P_R - \M(0) P_L - \M^\dagger(0) P_R \Big) \psi \nn
		&\quad + \frac{1}{2} \bar \psi_R \Big( M \Sigma^L(0) + M M^\dagger M {\Sigma^L}'(0) + \Sigma^R(0) M + {\Sigma^R}'(0) M M^\dagger M \nn
			&\qquad\qquad - M {\M'}^\dagger(0) M - \M'(0) M^\dagger M \Big) \psi_L \nn
		&\quad + \frac{1}{2} \bar \psi_L \Big( \Sigma^L(0) M^\dagger + {\Sigma^L}'(0) M^\dagger M M^\dagger  + M^\dagger \Sigma^R(0) + M^\dagger M M^\dagger {\Sigma^R}'(0) \nn
			&\qquad\qquad - M^\dagger \M'(0) M^\dagger - M^\dagger M {\M'}^\dagger(0) \Big) \psi_R \, .
\end{align}
The one-loop mass matrix is given by
\begin{align}
	\widetilde M^\text{1-loop} &= M - M \frac{1}{2} \left( \Sigma^L(0) + M^\dagger M {\Sigma^L}'(0) \right) - \frac{1}{2} \left( \Sigma^R(0) + {\Sigma^R}'(0) M M^\dagger \right) M \nn
		&\quad + \M(0) + \frac{1}{2} \left( M {\M'}^\dagger(0) M + \M'(0) M^\dagger M \right) \, .
\end{align}
The mass matrix defined in this way is not only complex and non-diagonal, but it even contains gauge-dependent parts~\cite{Espriu:2002xv,Kniehl:2006rc,Kniehl:2009kk}. The gauge dependence drops out if the mass matrices are diagonalized. We choose to perform only a partial diagonalization that removes the gauge dependence from the non-diagonal mass matrices. In order to do so, we perform a bi-unitary transformation
\begin{align}
	\psi_L \mapsto L \psi_L \, , \quad \psi_R \mapsto R \psi_R \, .
\end{align}
We specialize again to real diagonal tree-level mass matrices $M = M^\dagger = m$. Writing
\begin{align}
	 \widetilde M^\text{1-loop}(\xi) = m + \delta M + \delta \widetilde M(\xi) \, , \quad L = 1 + \delta L \, , \quad R = 1 + \delta R \, ,
\end{align}
where $\delta M$ is independent of $\xi$, we define the rotations
\begin{align}
	\label{eq:MassMatrixDiagonalizationXi}
	\delta L_{pr} &= \delta_{pr} i \Im \delta L_{pp} - (1-\delta_{pr}) \frac{m_p \delta \widetilde M_{pr} + \delta \widetilde M_{pr}^\dagger m_r}{m_p^2 - m_r^2} \, , \nn
	\delta R_{pr} &= \delta_{pr} i \Im \delta R_{pp} - (1-\delta_{pr}) \frac{m_p \delta \widetilde M_{pr}^\dagger + \delta \widetilde M_{pr} m_r}{m_p^2 - m_r^2} \, .
\end{align}
If only the $\xi$-dependent part of the mass matrix is rotated away, the numerators of the second terms in~\eqref{eq:MassMatrixDiagonalizationXi} are proportional to $m_p^2 - m_r^2$, cancelling the denominator. Of course, the splitting of $\xi$-dependent and $\xi$-independent parts is ambiguous and the explicit form of the non-diagonal mass matrices defines the particular choice of basis. Since we are integrating out the top quark, we bring the mass matrix of the up-type quarks to the block-diagonal form
\begin{align}
	\arraycolsep=0.15cm
	M_u^\text{1-loop} = m + \delta M = \left( \begin{array}{cc|c} * & * & 0 \\ * & * & 0 \\ \hline 0 & 0 & * \end{array} \right) \, ,
\end{align}
i.e.\ we add in~\eqref{eq:MassMatrixDiagonalizationXi} not only $\xi$-dependent pieces but the full off-diagonal part for the top quark. In this case, the denominators in~\eqref{eq:MassMatrixDiagonalizationXi} can again be expanded in the light scales.

We restrict~\eqref{eq:MassMatrixDiagonalizationXi} to non-axial rotations, $\arg\det(R^\dagger L) = 0$, which implies $\tr[\delta L] = \tr[\delta R]$. The diagonalized mass matrix is directly given by the diagonal entries of $\widetilde M^\text{1-loop}$, which is $\xi$-independent but includes a $CP$-violating contribution:
\begin{align}
	\label{eq:MassMatrixCP}
	M^\text{1-loop}_s &= m_s \Big( 1 - \frac{1}{2} \Sigma^L(m_s^2) - \frac{1}{2} \Sigma^R(m_s^2) \Big) + \Re \M_{ss}(m_s^2) + i \Im \M_{ss}(0) \, .
\end{align}
The $CP$-violating parts of the mass matrix can always be rotated away by an axial transformation, reshuffling them into the $\theta$ parameters at the two-loop level, see Sect.~\ref{sec:ThetaParameters}.

\subsubsection{Fermion wave-function renormalization and mixing}
\label{sec:WaveFunctionMixing}

Alternatively to the field redefinitions specified in Sect.~\ref{sec:FermionFieldRedefinitions}, one could perform the matching by equating observables, i.e.\ the pole masses and $S$-matrix elements in the SMEFT and LEFT. We briefly review the wave-function renormalization procedure for a chiral theory in the presence of fermion mixing~\cite{Aoki:1982ed,Denner:1990yz,Kniehl:1996bd,Pilaftsis:2002nc}, which is needed to compute observables in the SMEFT and LEFT. We are working with fields that are renormalized in the \msbar{} scheme. Consider a set of massive chiral fermion fields that annihilate one-particle states as follows:
\begin{align}
	\< 0 | \psi_{Lp}(0) | p_r \> &= \zeta_{Lp}^r u_{Lr}(p) \, , \nn
	\< 0 | \psi_{Rp}(0) | p_r \> &= \zeta_{Rp}^r u_{Rr}(p) \, ,
\end{align}
where $p,r,s,t$ are used as flavor indices. The two-point functions are matrices in spinor and flavor space and they have poles at the values of the physical masses $M_s$. Close to the poles at $p^2 = M_s^2$, they are given by
\begin{align}
	\< 0 | T\{ \psi_{Lp}(x) \bar\psi_{Lr}(0) \} | 0 \> &= i \int \frac{d^4p}{(2\pi)^4} e^{-i p\cdot x} \, \zeta_{Lp}^s \frac{P_L(\slashed p + M_s) P_R}{p^2 - M_s^2} \zeta_{Lr}^{s*} + \text{(regular at $p^2=M_s^2$)} \, , \nn
	\< 0 | T\{ \psi_{Rp}(x) \bar\psi_{Rr}(0) \} | 0 \> &= i \int \frac{d^4p}{(2\pi)^4} e^{-i p\cdot x} \, \zeta_{Rp}^s \frac{P_R(\slashed p + M_s) P_L}{p^2 - M_s^2} \zeta_{Rr}^{s*} + \text{(regular at $p^2=M_s^2$)} \, , \nn
	\< 0 | T\{ \psi_{Lp}(x) \bar\psi_{Rr}(0) \} | 0 \> &= i \int \frac{d^4p}{(2\pi)^4} e^{-i p\cdot x} \, \zeta_{Lp}^s \frac{P_L(\slashed p + M_s) P_L}{p^2 - M_s^2} \zeta_{Rr}^{s*} + \text{(regular at $p^2=M_s^2$)} \, , \nn
	\< 0 | T\{ \psi_{Rp}(x) \bar\psi_{Lr}(0) \} | 0 \> &= i \int \frac{d^4p}{(2\pi)^4} e^{-i p\cdot x} \, \zeta_{Rp}^s \frac{P_R(\slashed p + M_s) P_R}{p^2 - M_s^2} \zeta_{Lr}^{s*} + \text{(regular at $p^2=M_s^2$)} \, .
\end{align}
In matrix form, the propagator is
\begin{align}
	\label{eq:PropagatorPole}
	S_{pr} \stackrel{p^2 \to M_s^2}{\sim} \frac{i}{p^2 - M_s^2} \begin{pmatrix} \zeta_{Lp}^s \zeta_{Rr}^{s*} M_s  & \zeta_{Lp}^s \zeta_{Lr}^{s*} \slashed p \\ \zeta_{Rp}^s \zeta_{Rr}^{s*}  \slashed p & \zeta_{Rp}^s \zeta_{Lr}^{s*} M_s \end{pmatrix} \, .
\end{align}
We allow for the case where the propagator contains off-diagonal, i.e.\ flavor-changing elements.
The LSZ formula~\cite{Lehmann:1954rq} relates as usual $S$-matrix elements to amputated Green's functions:
\begin{align}
	\< p_s \ldots | \mathcal{S} | \ldots \> = \sum_r \zeta_r^{s*} \ldots \int d^4x e^{i p_s \cdot x} \ldots \< 0 | T \{ \psi_{r}(x) \ldots \} | 0 \>_\text{amp.}
\end{align}
We need to sum over all fields $r$, whose two-point function with the interpolating field for $s$ has a pole at $p^2 = M_s^2$. In matrix form, the inverse propagator~\eqref{eq:FermionInversePropagator} reads
\begin{align}
	\label{eq:InverseFermionPropagatorMatrix}
	S^{-1} = -i \begin{pmatrix} -m - \M^L & \slashed p ( 1 + \Sigma^R ) \\ \slashed p ( 1 + \Sigma^L ) & -m - \M^R \end{pmatrix} \, ,
\end{align}
which has zero eigenvalues at the physical masses of the fermions:
\begin{align}
	0 = \det\Big( p^2 (1 + \Sigma^R )(m+\M^R)^{-1}(1+\Sigma^L)(m+\M^R) - (m + \M^L) (m + \M^R) \Big) \, .
\end{align}
To one-loop accuracy, this gives
\begin{align}
	0 &= \det\Big( p^2 - m^2 - m \M^R - \M^L m + p^2(\Sigma^R + m^{-1} \Sigma^L m ) \Big) \nn
		&= \det( p^2 - m^2 ) + \tr\Big( \mathrm{adj}(p^2-m^2) \big[ p^2( \Sigma^R + m^{-1} \Sigma^L m ) - m \M^R - \M^L m \big] \Big)  \, ,
\end{align}
with the solution
\begin{align}
	\label{eq:PoleMass}
	M_s &= m_s \Big( 1 - \frac{1}{2} \Sigma^R_{ss}(m_s^2) - \frac{1}{2} \Sigma^L_{ss}(m_s^2) \Big) + \frac{1}{2} \Big( \M^L_{ss}(m_s^2) + \M^R_{ss}(m_s^2) \Big) \, ,
\end{align}
depending only on the diagonal entries of the matrices $\Sigma^L$, $\Sigma^R$, $\M^L$, $\M^R$. This agrees with the $CP$-even part of the mass matrix~\eqref{eq:MassMatrixCP} that we diagonalized by field redefinitions.
The propagator is given by the inverse of~\eqref{eq:InverseFermionPropagatorMatrix}:
\begin{align}
	S = i \begin{pmatrix} S^{LL} & \slashed p S^{LR} \\ \slashed p S^{RL} & S^{RR} \end{pmatrix} \, ,
\end{align}
where
\begin{align}
	S^{LL} &= \Big( p^2(1+\Sigma^R)(m+\M^R)^{-1}(1+\Sigma^L) - (m+\M^L) \Big)^{-1} \, , \nn
	S^{RR} &= \Big( p^2(1+\Sigma^L)(m+\M^L)^{-1}(1+\Sigma^R) - (m+\M^R) \Big)^{-1} \, , \nn
	S^{RL} &= (m+\M^R)^{-1}(1+\Sigma^L) S^{LL} \, , \nn
	S^{LR} &= (m+\M^L)^{-1}(1+\Sigma^R) S^{RR} \, .
\end{align}
In the vicinity of the poles, the propagator matrix has the form~\eqref{eq:PropagatorPole}. The square-rooted residues can be extracted as
\begin{align}
	\lim_{p^2\to M_s^2} (p^2 - M_s^2) S = \frac{1}{\frac{d}{dp^2} \det(S^{-1}) \big|_{p^2 = M_s^2}} \mathrm{adj}(S^{-1}) \Big|_{p^2 = M_s^2} \, ,
\end{align}
leading to
\begin{align}
	\zeta_{Lp}^s &= \delta_{ps} \Bigg[ 1 + \begin{aligned}[t]
		& m_s \Re \M_{ss}{}'(m_s^2) - i \, (1-\kappa) \frac{\Im \M_{ss}(m_s^2)}{m_s} - \frac{1}{2} \Sigma^L_{ss}(m_s^2) - \frac{m_s^2}{2} \left( \Sigma^L_{ss}{}'(m_s^2) + \Sigma^R_{ss}{}'(m_s^2) \right) \Bigg] \end{aligned} \nn
		&\quad + (1 - \delta_{ps} ) \begin{aligned}[t]
			& \Bigg[ \frac{m_p \M_{ps}(m_s^2) + m_s \M^\dagger_{ps}(m_s^2) - m_p m_s \Sigma^R_{ps}(m_s^2) - m_s^2 \Sigma^L_{ps}(m_s^2)}{m_s^2-m_p^2} \Bigg] \, , \end{aligned} \nn
	\zeta_{Rr}^s &= \delta_{rs} \Bigg[ 1 + \begin{aligned}[t]
		& m_s \Re \M_{ss}{}'(m_s^2) + i \, \kappa \frac{ \Im \M_{ss}(m_s^2)}{m_s} - \frac{1}{2} \Sigma^R_{ss}(m_s^2) - \frac{m_s^2}{2} \left( \Sigma^L_{ss}{}'(m_s^2) + \Sigma^R_{ss}{}'(m_s^2) \right) \Bigg] \end{aligned} \nn
		&\quad + (1 - \delta_{rs} ) \begin{aligned}[t]
			& \Bigg[ \frac{m_r \M^\dagger_{rs}(m_s^2) + m_s \M_{rs}(m_s^2) - m_s m_r \Sigma^L_{rs}(m_s^2) - m_s^2 \Sigma^R_{rs}(m_s^2)}{m_s^2-m_r^2} \Bigg] \, . \end{aligned}
\end{align}
Here, $\kappa$ is a free phase parameter for the physical states. The off-diagonal terms with the denomiators $1/(m_s^2 - m_p^2)$ correspond to the chiral rotations~\eqref{eq:MassMatrixDiagonalizationXi} that diagonalize the mass matrices. The denominators appear as usual in non-degenerate perturbation theory and e.g.\ show up in the renormalization of the mixing matrices in \msbar{} schemes~\cite{Balzereit:1998id,Pilaftsis:2002nc,Denner:2004bm,Sirlin:2012mh,Denner:2018opp}.

In the case of Majorana neutrinos, the physical mass at one-loop order is given by
\begin{align}
	M_s &= m_s \Big( 1 - \Sigma^M_{ss}(m_s^2) \Big) + \frac{1}{2} \Big( \M_{ss}(m_s^2) + \M^*_{ss}(m_s^2) \Big)
\end{align}
and the square-rooted residues are
\begin{align}
	\zeta_{Lp}^s = \zeta_{Rp}^{s*} &=  \delta_{ps} \Bigg[ 1 + \begin{aligned}[t]
		& m_s \Re \M_{ss}{}'(m_s^2) - \frac{i}{2} \frac{\Im \M_{ss}(m_s^2)}{m_s} - \frac{1}{2} \Sigma^M_{ss}(m_s^2) - m_s^2 \Sigma^M_{ss}{}'(m_s^2) \Bigg] \end{aligned} \nn
		&\quad + (1 - \delta_{ps} ) \begin{aligned}[t]
			& \Bigg[ \frac{m_p \M_{ps}(m_s^2) + m_s \M^*_{ps}(m_s^2) - m_p m_s \Sigma^{M*}_{ps}(m_s^2) - m_s^2 \Sigma^M_{ps}(m_s^2)}{m_s^2-m_p^2} \Bigg] \, . \end{aligned}
\end{align}

\subsection{Gauge-boson two-point functions}

We proceed with the gauge-boson two-point functions and we first discuss one-loop mixing between the neutral gauge bosons in Sect.~\ref{sec:GaugeBosonMixing}. The gauge-boson two-point functions are then used to extract the matching expressions for the gauge couplings in Sect.~\ref{sec:GaugeCouplings} and the theta parameters in Sect.~\ref{sec:ThetaParameters}.

\subsubsection{Gauge-boson mixing}
\label{sec:GaugeBosonMixing}

We consider mixing of vector fields, as it is present at one loop between the neutral background fields $\hat\A$ and $\hat\Z$~\cite{Fleischer:1980ub,Baulieu:1981ux,Bohm:2001yx}.
The inverse propagator can be written in terms of transverse and longitudinal contributions:
\begin{align}
	\label{eq:BosonInversePropagator}
	\Sigma^{ab}_{\mu\nu}(p) = i \Big[ \Big( g_{\mu\nu} - \frac{p_\mu p_\nu}{p^2} \Big) \big( t^{ab}(p^2) + \Sigma^{ab}_T(p^2) \big) + \frac{p_\mu p_\nu}{p^2} \big( l^{ab}(p^2) + \Sigma^{ab}_L(p^2) \big) \Big] \, ,
\end{align}
where the tree-level contributions are (in unitary gauge for the $\hat\Z$ field and with general gauge parameter $\hat\xi_\gamma$ for the $\hat\A$ field)
\begin{align}
	t^{AA}(p^2) &= p^2 \, , \quad t^{AZ}(p^2) = t^{ZA}(p^2) = 0 \, , \quad t^{ZZ}(p^2) = p^2 - M_\Z^2 \, , \nn
	l^{AA}(p^2) &= \frac{p^2}{\hat\xi_\gamma} \, , \quad l^{AZ}(p^2) = l^{ZA}(p^2) = 0 \, , \quad l^{ZZ}(p^2) = - M_\Z^2
\end{align}
and $\Sigma_{T,L}^{ab}(p^2)$ denote the one-loop corrections. The Ward identity requires $\Sigma_L^{AA}(p^2) = 0$, i.e.\ higher-order corrections are transverse. The absence of poles then implies $\Sigma_T^{AA}(0) = 0$. The propagator matrix is given by
\begin{align}
	\Pi^{ab}_{\mu\nu}(p) = -i \Big[ \Big( g_{\mu\nu} - \frac{p_\mu p_\nu}{p^2} \Big) \Pi^{ab}_T(p^2) + \frac{p_\mu p_\nu}{p^2} \Pi^{ab}_L(p^2) \Big] \, ,
\end{align}
satisfying
\begin{align}
	\Sigma_{\mu\nu}(p) \Pi^{\nu\lambda}(p) = g_\mu^\lambda \, ,
\end{align}
which implies
\begin{align}
	\Pi^{ab}_T(p^2) = \big( t(p^2) + \Sigma_T(p^2) \big)^{-1}_{ab} \, , \quad \Pi^{ab}_L(p^2) = \big( l(p^2) + \Sigma_L(p^2) \big)^{-1}_{ab} \, .
\end{align}
The physical masses of the gauge bosons are determined by
\begin{align}
	\det\big( t^{ab}(p^2) + \Sigma^{ab}_T(p^2) \big) = 0 \, .
\end{align}
In the vicinity of the poles, the transverse part of the propagator matrix is given by
\begin{align}
	\Pi_T^{ab}(p^2) \stackrel{p^2 \to M_{c,\text{ph}}^2}{\sim} \frac{\zeta^a_c \zeta^b_c}{p^2 - M_{c,\text{ph}}^2} \, .
\end{align}
At one loop, the physical masses are given by
\begin{align}
	M_{\A,\text{ph}}^2 = 0 \, , \quad M_{\Z,\text{ph}}^2 = M_\Z^2 - \Sigma^{ZZ}_T(M_\Z^2) \, ,
\end{align}
and for the square-rooted residues, we find
\begin{align}
	\zeta^A_A &= 1 - \frac{1}{2} {\Sigma_T^{AA}}'(0) \, , \quad \zeta^Z_A = \frac{1}{M_\Z^2} \Sigma_T^{AZ}(0) = \frac{1}{M_\Z^2} \Sigma_T^{ZA}(0) \, , \nn
	\zeta^Z_Z &= 1 - \frac{1}{2} {\Sigma_T^{ZZ}}'(M_\Z^2) \, , \quad \zeta^A_Z = -\frac{1}{M_\Z^2} \Sigma_T^{AZ}(M_\Z^2) = -\frac{1}{M_\Z^2} \Sigma_T^{ZA}(M_\Z^2)  \, .
\end{align}
In the background-field method, gauge invariance requires $\Sigma_L^{AZ}(p^2) = \Sigma_L^{ZA}(p^2) = 0$, which together with analyticity at $p^2=0$ implies that $\Sigma_T^{ZA}(0) = \Sigma_T^{AZ}(0) = 0$, i.e.\ there is no $\A$--$\Z$ one-loop mixing for on-shell photons even in minimal schemes, while in the conventional gauge, this is only enforced by on-shell renormalization~\cite{Denner:1994xt}.

Note that there is also one-loop mixing between the $\Z$ and the Higgs scalar. In background-field gauge, the mixing of the photon with the Higgs scalar vanishes due to gauge invariance.

\subsubsection{Gauge couplings}
\label{sec:GaugeCouplings}

The matching equations for the gauge couplings can be directly obtained from the gauge-boson two-point functions by performing field redefinitions, similarly to the case of the fermion masses. Expanded in all light scales up to dimension-six effects, the one-loop inverse gauge-boson propagator~\eqref{eq:BosonInversePropagator} can be interpreted as terms in the LEFT Lagrangian
\begin{align}
	\label{eq:BosonTwoPointLagrangian}
	\L &= -\frac{1}{4} F_{\mu\nu} F^{\mu\nu} \Big(1 + {\Sigma^{AA}_T}'(0) \Big) - \frac{1}{4} (\p_\mu F^{\mu\nu}) (\p^\lambda F_{\lambda\nu}) {\Sigma^{AA}_T}''(0) \nn
		&\quad -\frac{1}{4} G_{\mu\nu}^A G^{A\mu\nu} \Big(1 + {\Sigma^{GG}_T}'(0) \Big) - \frac{1}{4} (D_\mu G^{\mu\nu})^A (D^\lambda G_{\lambda\nu})^A {\Sigma^{GG}_T}''(0)  + \L_\mathrm{GF} \, .
\end{align}
The kinetic terms can be made canonical by a redefinition of the fields:
\begin{align}
	\label{eq:GaugeBosonFieldRedefinition}
	A_\mu &\mapsto \Big( 1 - \frac{1}{2} {\Sigma^{AA}_T}'(0) \Big) A_\mu \, , \quad
	G_\mu^A \mapsto \Big( 1 - \frac{1}{2} {\Sigma^{GG}_T}'(0) \Big) G_\mu^A \, .
\end{align}
Matching the covariant derivative then directly leads to the matching equations for the LEFT gauge couplings:
\begin{align}
	e &= \bar e \Big( 1 - \frac{1}{2} {\Sigma^{AA}_T}'(0) \Big) \, , \quad 
	g = \bar g_3 \Big( 1 - \frac{1}{2} {\Sigma^{GG}_T}'(0) \Big) \, .
\end{align}
By another field redefinition (or by applying the EOM), the dimension-six terms in~\eqref{eq:BosonTwoPointLagrangian} are transformed into four-fermion operators.

\subsubsection{Theta parameters}
\label{sec:ThetaParameters}

The matching of the theta parameters is complicated by the fact that the theta terms in the Lagrangian are total derivatives and therefore usually do not contribute in Feynman diagrams of ordinary perturbation theory. However, the matching can nevertheless be performed with standard methods by using a trick proposed in~\cite{Georgi:1980cn}: the $CP$-violating parameters in the SMEFT can be multiplied by a non-propagating scalar dummy field that acquires a vev equal to one. By calculating diagrams with a single external dummy field, which allows one to insert momentum into the diagram, the matching can be performed in perturbation theory and the contribution of the $CP$-odd SMEFT terms to the LEFT theta terms can be calculated. The general method of~\cite{Georgi:1980cn} is simplified in the current setting as we are only interested in terms linear in the higher-dimension SMEFT operators. In practice, the method amounts to inserting momentum into the $CP$-odd SMEFT operators or theta terms and matching it with the LEFT theta terms, where the momentum is inserted in the theta parameters. Because on the SMEFT side, momentum is inserted not only into the dimension-six operators but also into the SMEFT theta terms, they appear as new vertices in this modified perturbation theory as they are no longer total derivatives. All relevant diagrams are shown in App.~\ref{sec:TwoPointMomentumInsertion}.

In~\cite{Georgi:1980cn} a potential problem of the method was mentioned: since the theta terms are total derivatives, the generated counterterms could either be of the form
\begin{align}
	F_{\mu\nu} \widetilde F^{\mu\nu} \phi = (\p^\mu K_\mu) \phi
\end{align}
or
\begin{align}
	-K_\mu \p^\mu\phi \, .
\end{align}
While the first variant gives a contribution to the LEFT theta terms, the second variant vanishes when the dummy field $\phi$ is set to its vev. In~\cite{Georgi:1980cn} it was argued that the second variant should not appear because the current $K_\mu$ is not gauge invariant. Using the background-field method~\cite{Abbott:1980hw,Abbott:1983zw} puts this argument on solid ground: it ensures that gauge invariance with respect to the background fields is maintained in the loop calculation.

In addition to the direct contributions to the theta terms, additional effects are generated through the $CP$-odd contributions to the mass matrices. We work in a basis, where these contributions are left in the non-diagonal complex mass matrices. One could choose to rotate the $CP$-violating mass terms away by an axial rotation and would expect an additional contribution to the theta terms through the chiral anomaly. However, the result of the anomalous rotation can be obtained as in~\cite{Georgi:1980cn} by inserting momentum into a $CP$-violating mass term and calculating the fermion-loop contribution to the gauge-boson two- or three-point function. Since the $CP$-violating mass term is generated only at one loop, its insertion into the fermion loop corresponds to a two-loop effect in the matching equations for the theta angles and here it can be disregarded. We remark that in running or matching calculations at the two-loop level, the effect of axial field redefinitions on the theta terms requires special attention.

\subsection{Fermion three-point functions}

The matching equations for the electromagnetic and chromomagnetic dipole operators are extracted from the three-point functions with two fermions and a photon or gluon. All relevant diagrams are shown in App.~\ref{sec:ThreePointDiagrams}. The matching can be performed in several ways. One could calculate the expanded fully off-shell three-point functions, identify the result with a set of operators and perform a field redefinition (or use the EOM) to arrive at the LEFT operator basis. In this case, many contributions would directly correspond to the same operators found in the matching of the two-point functions, due to relations required by gauge invariance. After having performed the field redefinitions as implied by the matching of the two-point functions, these operators are removed from the result for the off-shell three-point functions as well. One can then directly work with on-shell fermions and only keep the gauge boson off-shell, in order to identify the field redefinitions that generate four-fermion operators through the gauge-field EOM.

In addition to the terms~\eqref{eq:LagrangianFermionFieldRedefinition} affecting the two-point functions, the field redefinition~\eqref{eq:FermionFieldRedefinition} generates additional terms that only show up in three-point (and higher-point) functions: the original Lagrangian
\begin{align}
	\label{eq:DipolesLEFT}
	\L &= \bar\psi_L \sigma^{\mu\nu} \lwc{\psi\gamma}{} \psi_R F_{\mu\nu} + \bar\psi_R \sigma^{\mu\nu} \lwc{\psi\gamma}{\dagger} \psi_L F_{\mu\nu} 
\end{align}
gets shifted up to dimension-six terms by
\begin{align}
	\label{eq:DipolesExternalLegShift}
	\delta \L &= \bar\psi_L \sigma^{\mu\nu} ( \lwc{\psi\gamma}{} A_R + A^\dagger_L \lwc{\psi\gamma}{}) \psi_R F_{\mu\nu} +  \bar\psi_R \sigma^{\mu\nu} ( \lwc{\psi\gamma}{\dagger} A_L + A^\dagger_R \lwc{\psi\gamma}{\dagger}) \psi_L F_{\mu\nu} \nn
		&\quad + \bar\psi_L \sigma^{\mu\nu} \lwc{\psi\gamma}{} B_R i \slashed D \psi_L F_{\mu\nu} - \bar\psi_L i \overleftarrow{\slashed D} \sigma^{\mu\nu} B_R^\dagger \lwc{\psi\gamma}{\dagger} \psi_L F_{\mu\nu} \nn
		&\quad + \bar\psi_R \sigma^{\mu\nu} \lwc{\psi\gamma}{\dagger} B_L i \slashed D \psi_R F_{\mu\nu} - \bar\psi_R i \overleftarrow{\slashed D} \sigma^{\mu\nu} B_L^\dagger \lwc{\psi\gamma}{} \psi_R F_{\mu\nu} \, .
\end{align}
In~\eqref{eq:DipolesLEFT}, we have employed the tree-level matching to express the coefficients of the dipole operators already in terms of LEFT coefficients (using matrix notation in flavor space). Analogous shifts are present for the gluonic dipole operators. The derivative operators can be eliminated by the EOM or another field redefinition
\begin{align}
	\psi_L &\mapsto \psi_L - F_{\mu\nu} \sigma^{\mu\nu} B_R^\dagger \lwc{\psi\gamma}{\dagger} \psi_L \, , \nn
	\psi_R &\mapsto \psi_R - F_{\mu\nu} \sigma^{\mu\nu} B_L^\dagger \lwc{\psi\gamma}{} \psi_R \, ,
\end{align}
which does not affect the two-point functions. The Lagrangian shifts~\eqref{eq:LagrangianFermionFieldRedefinition} and~\eqref{eq:DipolesExternalLegShift} together with the shift induced by the redefinition of the gauge bosons~\eqref{eq:GaugeBosonFieldRedefinition} correspond to the external-leg corrections to the three-point function.

The general structure of a photonic or gluonic Dirac fermion three-point function is given by
\begin{align}
	i \M( \psi_r(p) \gamma(k) \to \psi_p(p') ) &= \bar u_p(p') i \Gamma_{pr}^\mu(p,p') u_r(p) \epsilon_\mu(k) \, , \nn
	i \M( \psi_r(p) g^A(k) \to \psi_p(p') ) &= \bar u_p(p') i \Gamma_{pr}^{A\mu}(p,p') u_r(p) \epsilon_\mu(k) \, , \quad k = p'-p \, ,
\end{align}
where the vertex function can be decomposed into the following Lorentz structures:
\begin{align}
	\Gamma_\mu^{(A)}(p,p') &= \gamma_\mu A_1^{(A)} + \gamma_\mu \gamma_5 A_2^{(A)} + k_\mu A_3^{(A)} + k_\mu \gamma_5 A_4^{(A)} + P_\mu A_5^{(A)} + P_\mu \gamma_5 A_6^{(A)} \, ,
\end{align}
with $P=p+p'$. In order to match the vertex functions to the LEFT operators, we first write down an intermediate set of operators up to dimension six, where field redefinitions have already removed fermion EOM operators:
\begin{align}
	\L &= \lwc{\psi\gamma}{}[][pr] \bar\psi_{Lp} \sigma^{\mu\nu} \psi_{Rr} F_{\mu\nu} + \lwc{\psi G}{}[][pr] \bar\psi_{Lp} \sigma^{\mu\nu} T^A \psi_{Rr} G_{\mu\nu}^A \nn
		&\quad + \lwc{\psi\gamma}{\dagger}[][pr] \bar\psi_{Rp} \sigma^{\mu\nu} \psi_{Lr} F_{\mu\nu} + \lwc{\psi G}{\dagger}[][pr] \bar\psi_{Rp} \sigma^{\mu\nu} T^A \psi_{Lr} G_{\mu\nu}^A \nn
		&\quad + E^L_{pr} \bar\psi_{Lp} \gamma^\mu \psi_{Lr} (\p^\nu F_{\mu\nu}) + \tilde E^L_{pr} \bar\psi_{Lp} \gamma^\mu T^A \psi_{Lr} (D^\nu G_{\mu\nu})^A \nn
		&\quad + E^R_{pr} \bar\psi_{Rp} \gamma^\mu \psi_{Rr} (\p^\nu F_{\mu\nu}) + \tilde E^R_{pr} \bar\psi_{Rp} \gamma^\mu T^A \psi_{Rr} (D^\nu G_{\mu\nu})^A \, .
\end{align}
The Feynman rules for this Lagrangian and the Gordon identity provide the immediate identification
\begin{align}
	\begin{alignedat}{2}
		[A_1]_{pr} &= k^2 \frac{1}{2} \big( E^L_{pr} + E^R_{pr} \big) + (m_p + m_r) \big( \lwc{\psi\gamma}{}[][pr] + \lwc{\psi\gamma}{\dagger}[][pr] \big) \, , \hspace{-4cm} & & \nn
		[A_2]_{pr} &= k^2 \frac{1}{2} \big( E^R_{pr} - E^L_{pr} \big) + (m_p - m_r) \big( \lwc{\psi\gamma}{}[][pr] - \lwc{\psi\gamma}{\dagger}[][pr] \big) \, , \hspace{-4cm} & & \nn
		[A_3]_{pr} &= -\frac{1}{2} (m_p - m_r) \big( E^L_{pr} + E^R_{pr} \big) \, , \quad &
		[A_4]_{pr} &= \frac{1}{2} (m_p + m_r) \big( E^L_{pr} - E^R_{pr} \big) \, , \nn
		[A_5]_{pr} &= - \big( \lwc{\psi\gamma}{}[][pr] + \lwc{\psi\gamma}{\dagger}[][pr] \big) \, , \quad &
		[A_6]_{pr} &= - \big( \lwc{\psi\gamma}{}[][pr] - \lwc{\psi\gamma}{\dagger}[][pr] \big) \, ,
	\end{alignedat}\mytag
\end{align}
and similar relations for the coefficients of the gluonic structures. From these relations we obtain the matching equations for the dipole operators as well as gauge-field EOM contributions to the four-fermion operators.

\subsection{Gluon three-point function}

After having applied the field redefinitions that are determined with the gluon two-point function and render the kinetic terms canonical, the expanded off-shell gluon three-point function can be identified with a set of operators
\begin{align}
	\label{eq:ThreeGluonLagrangian}
	\L = \lwc{G}{} \O_G + \lwc{\widetilde G}{} \O_{\widetilde G} + E (D^\mu G_{\mu\nu})^A (D_\lambda G^{\lambda\nu})^A \, .
\end{align}
Due to gauge invariance, the three-gluon part of the EOM operator reappears that we already identified in the two-point function~\eqref{eq:BosonTwoPointLagrangian}, i.e.\ $E = -\frac{1}{4} {\Sigma_T^{GG}}''(0)$. Through field redefinitions or the gauge-field EOM, it is transformed into a set of four-quark operators. The separation of the $CP$-even and $CP$-odd three-gluon operators from the EOM operator is most easily obtained by writing down the Feynman rules for the Lagrangian~\eqref{eq:ThreeGluonLagrangian} and matching at the amplitude level. Some technicalities concerning the calculation of the fermion loops in the three-gluon amplitude have been discussed in Sect.~\ref{sec:ClosedFermionLoops}.

\subsection{Four-fermion interactions}
\label{sec:FourFermionMatching}

In order to extract the matching contribution to the LEFT four-fermion operators, we have to collect several contributions. First of all, a direct contribution arises from 1PI four-point diagrams, shown in App.~\ref{sec:FourPointDiagrams}. Second, there are contributions from 1LPI four-fermion diagrams with tree-level propagators of heavy gauge bosons. We organize them again in terms of 1PI diagrams with heavy gauge bosons: the diagrams for the two-point functions are collected in App.~\ref{sec:TwoPointFunctions}, the three-point diagrams are listed in App.~\ref{sec:ThreePointDiagramsHeavy}. Third, there are loop contributions to 1LPI two- and three-point functions with massless gauge bosons that generate matching contributions to four-fermion operators: either, these 1LPI diagrams can be considered as sub-diagrams in a four-point function with tree-level propagators of massless gauge bosons, or, equivalently and more conveniently, they can be matched directly off shell. In the second method, EOM operators generated by 1LPI two- and three-point functions are turned into four-fermion operators with the help of a field redefinition, as discussed in the previous sections. Finally, the field redefinitions that we determined with the fermion two-point functions to make the kinetic terms canonical modify the tree-level matching of the four-fermion operators in terms of external-leg corrections.

The tree-level matching of the SMEFT to the LEFT~\cite{Jenkins:2017jig} consists only of contact terms as well as exchange diagrams of weak gauge bosons. Since the dimension-four Higgs coupling is proportional to a light scale, no Higgs-exchange diagrams appear, since it would require two insertions of dimension-six operators. This holds true even at the one-loop level, as we discuss in the following.

Two-point functions with a Higgs boson could in principle appear inside of four-fermion diagrams. 
The Higgs two-point function does not give any contribution to the matching up to dimension six: for a four-fermion Higgs-exchange diagram with one-loop corrections to the Higgs propagator, the same argument as in the tree level matching applies~\cite{Jenkins:2017jig}. The Higgs coupling to the light external fermions is either proportional to a light scale or to a dimension-six SMEFT coefficient. Since in the four-fermion operators all light scales are set to zero due to the LEFT power counting, two insertions of dimension-six SMEFT operators would be required, which gives a dimension-eight contribution in the SMEFT counting.

The Higgs-photon two-point function vanishes in the background-field method due to gauge invariance: the amplitude needs to have the form
\begin{align}
	\M^\mu(p) = p^\mu \M(p^2)
\end{align}
and gauge invariance requires $p_\mu \M^\mu = 0$, hence $\M(p^2) = 0$.

This does not apply directly to the Higgs-$\Z$ two-point function, which receives a non-vanishing contribution due to a top-quark loop. However, Lorentz invariance still requires the amplitude to be proportional to an external momentum, which brings in a light scale that is set to zero in the four-fermion matching.

Finally, at one loop the last possibly relevant Higgs-exchange diagrams consist of a dimension-six coupling to the light fermions, the Higgs propagator, and a SM loop-induced coupling to two other light fermions. At dimension six in the SMEFT, only the SM contributions to the Higgs one-loop three-point functions are needed and we do not consider diagrams that require the insertion of higher-dimensional SMEFT operators. We can also neglect diagrams that involve scalar couplings to light fermions, which are proportional to a light scale. As the Higgs three-point function requires either a chirality change or a derivative coupling, a light scale necessarily appears if the internal fermions are light. Hence, only diagrams with internal top quarks need to be considered, which contribute to the $\bar d d h$ three-point function, as shown in App.~\ref{sec:ThreePointHiggs}. However, in all these diagrams the SM vertex structure either contains an external (light) down-type quark mass or the loop produces an external momentum. Since the light scales are set to zero, all diagrams vanish. External-leg corrections do not contribute either as in these diagrams the Higgs boson couples to a light fermion. Therefore, even at one loop no Higgs-exchange diagrams contribute to the matching.


\section{Results}
\label{sec:Results}

The main result of this work is the complete set of one-loop matching equations connecting the SMEFT with the LEFT, including effects up to dimension six in the power counting of both theories. As the explicit expressions are very long, we only provide them in digital form in the supplemental material, with conventions specified in App.~\ref{app:code}. All the results are given in terms of the LEFT coefficients of the operator basis derived in~\cite{Jenkins:2017jig}, which we reproduce in App.~\ref{sec:LEFTBasis}. In order to illustrate the form of the results, we display the coefficient of the LEFT up-type quark gluonic dipole operator at the matching scale $\mu_W$:
{\small
\begin{align}
	\label{eq:DipoleResult}
	\lwc{uG}{}[][pr] &=
			\delta_{pr} \frac{\bar g_3  \left(32 M_\W^4-40 M_\Z^2 M_\W^2+35 M_\Z^4\right) m_{u_p}}{432 M_\Z^4 \pi ^2 v_T^2} \nn
		&\quad + \frac{v_T}{\sqrt{2}} \cwc{uG}{}[][pr]
		 + \frac{\bar g_3  m_t}{16 \pi ^2} \cwc{qu}{(1)}[][pttr] - \frac{\bar g_3  m_t}{32 N_c \pi ^2} \cwc{qu}{(8)}[][pttr] + \delta_{pr} \Bigg\{ \cwc{HWB}{} \frac{ \bar g_1 \bar g_3 v_T \left(8 M_\W^3-5 M_\W M_\Z^2\right) m_{u_p}}{108 M_\Z^4 \pi ^2} \nn
	   	&\quad + \cwc{HD}{} \frac{ \bar g_3 \left(32 M_\W^4-17 M_\Z^4\right) m_{u_p}}{864 M_\Z^4 \pi ^2} + \left( \cwc{HG}{} + i \cwc{H\widetilde G}{} \right) \frac{\bar g_3 m_{u_p}}{8\pi^2} \left[ \frac{3}{2} + \log \left(\frac{\mu_W^2}{M_H^2}\right) \right] \Bigg\} \nn
		&\quad + \cwc{uW}{}[][pr] \frac{\bar g_3 M_\W}{4 \sqrt{2} \pi^2} \left[-\frac{8 M_\W^2+M_\Z^2}{12 M_\Z^2} - \log\left(\frac{\mu_W^2}{M_\W^2}\right) + \frac{5 M_\Z^2-8 M_\W^2}{6 M_\Z^2} \log \left(\frac{\mu_W^2}{M_\Z^2}\right) \right] \nn
	   	&\quad + \cwc{uB}{}[][pr] \frac{\bar g_1 \bar g_3 v_T \left(8 M_\W^2-5 M_\Z^2\right) }{48 \sqrt{2} M_\Z^2 \pi ^2} \left[ \frac{1}{2} + \log \left(\frac{\mu_W^2}{M_\Z^2}\right) \right] \nn
		&\quad + \cwc{Hu}{}[][pr] \frac{\bar g_3  \left(M_\Z^2-M_\W^2\right) m_{u_p}}{18 M_\Z^2 \pi ^2} + \cwc{Hq}{(1)}[][pr] \frac{\bar g_3  \left(M_\Z^2-4 M_\W^2\right) m_{u_r}}{72 M_\Z^2 \pi ^2} + \cwc{Hq}{(3)}[][pr] \frac{\bar g_3  \left(4 M_\W^2+5 M_\Z^2\right) m_{u_r}}{72 M_\Z^2 \pi ^2} \nn
		&\quad + \cwc{uG}{}[][pr] \begin{aligned}[t]
			&\Bigg\{  \frac{54 M_\Z^2 M_H^4+\left(32 M_\W^4+14 M_\Z^2 M_\W^2+35 M_\Z^4\right) M_H^2+36 M_\Z^2 \left(-4 N_c m_t^4+2 M_\W^4+M_\Z^4\right)}{576 \sqrt{2} M_H^2 M_\Z^2 \pi ^2 v_T} \nn
				& + \frac{3 M_H^2 }{32 \sqrt{2} \pi ^2 v_T} \log \left(\frac{\mu_W^2}{M_H^2}\right) + \frac{\left(27 M_\Z^6-4 M_H^2 \left(4 M_\W^4-5 M_\Z^2 M_\W^2+M_\Z^4\right)\right)}{144 \sqrt{2} M_H^2 M_\Z^2 \pi ^2 v_T} \log \left(\frac{\mu_W^2}{M_\Z^2}\right) \nn
				& + \frac{3 M_\W^4}{8 \sqrt{2} M_H^2 \pi ^2 v_T} \log \left(\frac{\mu_W^2}{M_\W^2}\right) - \left(\frac{N_c m_t^4}{4 \sqrt{2} M_H^2 \pi ^2 v_T} + \frac{\bar g_3 ^2 v_T}{48 \sqrt{2} \pi ^2}\right) \log \left(\frac{\mu_W^2}{m_t^2}\right)  \Bigg\} . \end{aligned}
		\mytag
\end{align}}%
Here, $p$ and $r$ are flavor indices, and the subscript $t$ denotes the explicit top-quark flavor index. The first term is the SM one-loop matching contribution, while the second term represents the tree-level SMEFT matching~\cite{Jenkins:2017jig}. All parameters are understood to be the running (i.e.\ renormalized \msbar{}) SMEFT parameters, with implicit dependence on the matching scale $\mu_W$.

The results are given for a SMEFT basis, where the renormalized \msbar{} mass matrices are diagonalized by $\mu_W$-dependent unitary matrices $V$ and $U$, see~\eqref{eq:TreeLevelSMEFTMasses}. We do not reabsorb the mixing matrices $V$ and $U$ into the definition of the SMEFT Wilson coefficients, but we keep them explicitly. It is also possible to use the matching results in connection with a SMEFT basis with off-diagonal mass matrices---only the top quark needs to be diagonalized: in a fixed-order calculation, the off-diagonal mass matrices then only have to be restored in the tree-level matching contribution. In a next-to-leading-log analysis, one should then either perform the full diagonalization at the matching scale or restore the off-diagonal mass matrices everywhere in the one-loop matching equations. Since the light scales explicitly appear only in the matching to coefficients of LEFT operators of dimension less than six, it should be straightforward to recombine the masses and mixing matrices into off-diagonal mass matrices.

We emphasize again that our results are provided for a LEFT basis where the matching contribution itself generates off-diagonal and complex contributions to the mass matrices. Diagonalizing the mass matrices reshuffles these contributions into the LEFT Wilson coefficients in the form of external-leg corrections. In the case of the dipole operators, the tree-level matching starts at dimension six, therefore only the SM contribution to the mass matrices has to be taken into account. The additional external-leg correction for a basis with real diagonal mass matrices can be obtained from a field redefinition of the form~\eqref{eq:MassMatrixDiagonalizationXi}. In the case of up-type quarks, the off-diagonal SM contributions to the mass matrices are cubic in the light scales, hence the diagonalization~\eqref{eq:MassMatrixDiagonalizationXi} would result in a dimension-7 correction to $\lwc{uG}{}$ and can be neglected (we assume the differences of the light masses that appear in \eqref{eq:MassMatrixDiagonalizationXi} to be of the same order as the light masses themselves). We stress that the same does not apply in the case of down-type quarks: here, a contribution appears from the diagonalization of the mass matrix, corresponding to a SM external-leg correction due to a loop with a $W$ boson and top quark, which is not suppressed by a light scale. Similarly, for the coefficients of the four-fermion operators that receive a SM contribution in the tree-level matching, the diagonalization of the mass matrices results in external-leg corrections with dimension-six SMEFT operator insertions that are not suppressed by light scales.

We briefly comment on the checks that we performed to test the correctness of the results. We compared two independent implementations of the complete loop calculation as an internal cross-check. We checked the Ward identities in order to ensure background-field gauge invariance, see also the discussion of the $\gamma_5$ problem in Sect.~\ref{sec:DimReg}. In the whole calculation, we worked with linear $R_\xi$ gauge for the quantum fields, keeping the gauge parameter $\xi$ generic. The gauge-parameter independence of the results provides a very strong check, both of the correctness of the diagrams (again in the context of the regularization problem) and of the combination of all the contributions to the matching equations. Finally, we checked that the counterterms that cancel the UV and IR divergences in the matching calculation agree with the difference of SMEFT and LEFT counterterms extracted from the respective RGEs~\cite{Jenkins:2013zja,Jenkins:2013wua,Alonso:2013hga,Jenkins:2017dyc}. In this comparison, we used the compilation of the SMEFT RGEs from~\cite{Celis:2017hod}. Note that the divergences only have to agree up to a change of basis~\eqref{eq:MassMatrixDiagonalizationXi}, which can be determined by comparing the divergences of the mass matrices. This basis change includes axial field redefinitions (which affects the theta terms only at higher loop order), feeding e.g.\ into the operator coefficient $\lwc{\nu edu}{LL}[V][]$ that receives a tree-level SM matching contribution. Taking into account the basis change, we find agreement of the complete divergence structure, confirming the consistency of the matching calculation with the SMEFT and LEFT RGEs.

In addition to the checks mentioned above, we compared our results to the literature in several cases. Starting with the flavor-changing parts of the dipole operators $\lwc{d\gamma }{}[][] $ and $\lwc{dG}{}[][] $, we find that the calculation reproduces the known  SM matching \cite{Grinstein:1987vj,Inami:1980fz}, as well as the terms proportional to the $\cwc{Hud}{}[][]$, $\cwc{H\widetilde WB}{}[][]$, and  $\cwc{\widetilde W}{}[][]$ couplings \cite{Cho:1993zb,He:1993hx,Grzadkowski:2008mf,Aebischer:2015fzz,Alioli:2017ces}. In addition, the contributions of  $\cwc{H \widetilde G}{}[][]$, $\cwc{H \widetilde B}{}[][]$, $\cwc{H \widetilde W}{}[][]$, and $\cwc{H \widetilde W B}{}[][]$ to
 the flavor-diagonal dipoles, $\lwc{u\gamma, \,uG }{}[][] $, $\lwc{d\gamma,\,dG }{}[][] $, and $\lwc{e\gamma }{}[][] $, are in agreement with the results of~\cite{DeRujula:1990db,Fan:2013qn,Dekens:2013zca,Cirigliano:2019vfc}. A subtlety appears in the case of the $Q_{\widetilde W}{}$ operator for which the definition that appears in the literature \cite{Boudjema:1990dv,Dekens:2013zca,Gripaios:2013lea} differs by an evanescent operator from the one used here, leading to a result that differs by a finite part.\footnote{
Here we use the operator as defined in Table \ref{tab:smeft6ops}, while in parts of the literature an operator $\propto \widetilde F^{\mu\nu} W^{+}_{\,\nu\rho}W_{\mu}^{-\,\rho}$ was used to compute the matching onto the dipoles \cite{Boudjema:1990dv,Dekens:2013zca,Gripaios:2013lea}. Note that the latter definition corresponds to the triple-gauge coupling defined in~\cite{Gounaris:1996uw}. Since they differ by an evanescent term, both definitions should lead to the same predictions for observables if used consistently.} 
For the  $X^3$ class, we find agreement with~\cite{Braaten:1990gq} for the matching induced by $\cwc{uG}{}[][tt]$. Finally, for the $\Delta F=2$ parts of the four-fermion operators $\lwc{dd}{LL}[V][] $, $\lwc{dd}{LR}[V1][] $, and $\lwc{dd}{LR}[V8][] $, we checked that the matching is in agreement with the known SM result~\cite{Inami:1980fz,Buchalla:1995vs} as well as with the contributions due to four-fermion operators derived in~\cite{Aebischer:2015fzz}.


\section{Summary and conclusions}
\label{sec:Conclusions}

In the absence of evidence for NP in direct searches at high energies, model-independent analyses are becoming more relevant. Under the assumption that new particles have masses above the electroweak scale, effective field theories provide an efficient tool to perform all calculations below the scale of NP in a model-independent manner. Using the SMEFT above and the LEFT below the electroweak scale, all constraints on NP from collider searches and low-energy precision experiments can be evaluated in the same framework and expressed as bounds on the Wilson coefficients of effective operators. Any particular model can then be tested simply by integrating out the heavy new particles and comparing the matched Wilson coefficients to the constraints in the SMEFT. Deviations from the SM prediction found in experiments running at an energy below the scale of NP result in non-vanishing coefficients of the higher-dimension SMEFT operators.

The one-loop running within the SMEFT~\cite{Jenkins:2013zja,Jenkins:2013wua,Alonso:2013hga} and LEFT~\cite{Jenkins:2017dyc} and the tree-level matching at the electroweak scale~\cite{Jenkins:2017jig} provide a consistent framework for EFT analyses at leading-log accuracy. The increasing precision of experimental searches e.g.\ in lepton-flavor-violating processes or $CP$-violating dipole moments calls for EFT treatments of increased accuracy.
We have derived the one-loop matching equations up to dimension six in the power counting of both the SMEFT and LEFT. No further approximations are applied, e.g.\ we keep the full flavor structure and include mass corrections to dipole operators that are of dimension six in the power counting. The results are provided in a basis, where off-diagonal matching corrections to the mass matrices appear. Off-diagonal contributions to mass matrices also arise in the RGEs. When computing a process at a given scale, one has to perform a basis change in order to diagonalize the mass matrices, which can result in additional external-leg corrections to the coefficients of effective operators.

Our results provide the complete one-loop matching equations at the electroweak scale. They can be used for fixed-order one-loop calculations within the LEFT/SMEFT in cases where the logarithms are not large. They also provide a first step towards a next-to-leading-log analysis, which is desirable for some high-precision observables at low energies. A systematic phenomenological analysis of the one-loop matching effects is left for future work. To this end, the inclusion of the matching results in existing SMEFT/LEFT software tools would be useful~\cite{Celis:2017hod,Aebischer:2018bkb}. We provide the complete matching results in electronic form as supplemental material, which should facilitate their further use.

	\end{myfmf}

	
	\section*{Acknowledgements}
	\addcontentsline{toc}{section}{Acknowledgements}

	We thank J.~de Vries, B.~Grinstein, F.~Jegerlehner, E.~E.~Jenkins, B.~A.~Kniehl, A.~V.~Manohar, E.~Mereghetti, S.~Pal, and M.~Trott for useful discussions. In particular, we are grateful to B.~Grinstein and A.~V.~Manohar for comments on the manuscript.
	We thank M.~Trott for drawing our attention to the evanescent-scheme dependence in a fixed-order one-loop calculation in~\cite{Corbett:2021cil}, when using the results of the original version of this paper. 
	Financial support by the DOE (Grant No.\ DE-SC0009919) is gratefully acknowledged.

	
	\appendix
	

\section{Conventions}

The commutator of Dirac matrices is denoted by
\begin{align}
	\sigma_{\mu\nu} = \frac{i}{2} [ \gamma_\mu, \gamma_\nu ] \, .
\end{align}
The matrix $\gamma_5$ and the chiral projectors are defined as
\begin{align}
	\gamma_5 := i \gamma^0 \gamma^1 \gamma^2 \gamma^3 \, , \quad P_L = \frac{1}{2} ( 1 - \gamma_5 ) , \quad P_R = \frac{1}{2} ( 1 + \gamma_5 ) \, .
\end{align}
In four space-time dimensions, we have the relation
\begin{align}
	\sigma^{\mu\nu} \gamma_5 = - \frac{i}{2} \epsilon^{\mu\nu\mu^\prime\nu^\prime} \sigma_{\mu^\prime\nu^\prime} \, ,
\end{align}
where the fully antisymmetric Levi-Civita tensor is normalized as
\begin{align}
	\epsilon_{0123} = 1 \, .
\end{align}
The charge-conjugation matrix is given by $C = i \gamma^2 \gamma^0$, fulfilling
\begin{align}
	C \gamma_\mu C^{-1} = - \gamma_\mu^T , \quad C = C^* = - C^{-1} = - C^\dagger = - C^T \, .
\end{align}

The $SU(3)$ generators are normalized to
\begin{align}
	T^A = \frac{\lambda^A}{2} \, , \quad \mathrm{Tr}(T^A T^B) = \frac{1}{2} \delta^{AB} \, ,
\end{align}
where $\lambda^A$ are the Gell-Mann matrices. The antisymmetric and symmetric structure constants are defined by
\begin{align}
	[T^A, T^B] &= i f^{ABC} T^C \, , \quad
	\{ T^A, T^B \} = \frac{1}{N_c} \delta^{AB} + d^{ABC} T^C \, ,
\end{align}
and the color factors are given by
\begin{align}
	\label{eq:ColorFactors}
	T^A_{\alpha\beta} T^A_{\beta\gamma} &= \delta_{\alpha\gamma} C_F = \delta_{\alpha\gamma} \frac{N_c^2-1}{2N_c} \, , \quad
	f^{ABC} f^{ABD} = \delta^{CD} C_A = \delta^{CD} N_c \, .
\end{align}


\section{Neutrinos}

\label{app:Neutrinos}

We consider a purely left-handed neutrino
\begin{align}
	\nu_L = P_L \nu = \nu \, , \quad \nu_R = P_R \nu = 0 \, ,
\end{align}
i.e.\ in the chiral four-component basis
\begin{align}
	\nu = \begin{pmatrix} \nu_L \\ 0 \end{pmatrix} \, .
\end{align}
We define the charge-conjugate field of a four-spinor by
\begin{align}
	\psi^c := \eta_c C \bar\psi^T \, , \quad \overline{\psi^c} = \eta_c^* \psi^T C \, ,
\end{align}
where $\eta_c$ is a complex phase. The charge-conjugate neutrino is then
\begin{align}
	\nu^c = \begin{pmatrix} 0 \\ -\eta_c i \sigma_2 \nu_L^* \end{pmatrix} \, .
\end{align}
The neutrino kinetic term and (lepton-number-violating) mass term in the broken phase of the SMEFT read
\begin{align}
	\L_\nu = \bar\nu_L i \slashed\p \nu_L - \left( \frac{1}{2} [M_\nu]_{pr} ( \nu_{Lp}^T C \nu_{Lr} ) + \hc \right) = \bar\nu_L i \slashed\p \nu_L  - \frac{1}{2} \left( \nu_L^T C M_\nu \nu_L + \bar\nu_L M_\nu^\dagger C \bar\nu_L^T \right) \, .
\end{align}
The mass matrix can be diagonalized by a unitary rotation in flavor space
\begin{align}
	\nu_L = U \nu_L' \, , \quad U^T M_\nu U = U^\dagger M_\nu^\dagger U^* = M_\nu' = \mathrm{diag}(m_{\nu_1}, m_{\nu_2}, m_{\nu_3} ) \, .
\end{align}
After diagonalization, we can define a Majorana fermion
\begin{align}
	\nu_M := \nu' + \eta_c^* \nu^c{}' \, ,
\end{align}
which fulfills the Majorana condition
\begin{align}
	\nu_M = \eta_c^* \nu_M^c = C \bar\nu_M^T \, .
\end{align}
In terms of the Majorana field, the neutrino Lagrangian can be rewritten as
\begin{align}
	\L_\nu = \frac{1}{2} \bar\nu_M i \slashed\p \nu_M - \frac{1}{2} \bar \nu_M M_\nu' \nu_M = \frac{1}{2} \nu_M^T C i \slashed\p \nu_M - \frac{1}{2} \nu_M^T C M_\nu' \nu_M \, .
\end{align}
The SMEFT bilinear interaction terms of the left-handed neutrinos has the form
\begin{align}
	\L_{\nu,\mathrm{int}} &= \bar\nu_{Lp} \Gamma_{pr} \nu_{Lr} + ( \nu_{Lp}^T C \bar\Gamma_{pr} \nu_{Lr} + \hc ) \nn
		&= \bar\nu_{Lp} \Gamma_{pr} \nu_{Lr} + ( \nu_{Lp}^T C \bar\Gamma_{pr} \nu_{Lr} + \bar\nu_{Lp} \gamma^0 \bar\Gamma^\dagger_{pr} \gamma^0 C \bar\nu_{Lr}^T ) \, ,
\end{align}
where $\Gamma$ and $\bar\Gamma$ collect expressions involving Dirac structures and other fields. In terms of the Majorana fields, this can be rewritten as
\begin{align}
	\L_{\nu,\mathrm{int}} &= \frac{1}{2} \bar\nu_{Mp} \left[  \left( P_R \Gamma_{pr} P_L - P_L C \Gamma_{rp}^T C P_R \right) + 2 \left( P_L \bar\Gamma_{pr} P_L + P_R \gamma^0 \bar\Gamma_{rp}^\dagger \gamma^0 P_R \right) \right] \nu_{Mr} \, ,
\end{align}
where transposition acts only in Dirac space and flavor indices are specified explicitly. By writing the Dirac structure as
\begin{align}
	 \Gamma_{pr} &= \gamma_\mu \C_{pr}^\mu \, , \nn
	 \bar\Gamma_{pr} &= \bar\C_{pr} + \sigma_{\mu\nu} \bar\C^{\mu\nu}_{pr}
\end{align}
and using
\begin{align}
	C \gamma_\mu^T C &= \gamma_\mu \, , \quad
	\gamma^0 \gamma^0 = 1 \, , \quad
	\gamma^0 \sigma_{\mu\nu}^\dagger \gamma^0 = \sigma_{\mu\nu} \, ,
\end{align}
we find
\begin{align}
	\L_{\nu,\mathrm{int}} = \frac{1}{2} \bar\nu_{Mp} \bigg[ & \left( P_R \gamma_\mu P_L \C^\mu_{pr}  - P_L \gamma_\mu P_R  \C^\mu_{rp} \right) \nn
		& + 2 \left( P_L \bar\C_{pr} + P_R \bar\C_{rp}^* \right) + 2 \left( P_L \sigma_{\mu\nu} P_L \bar\C_{pr}^{\mu\nu} + P_R \sigma_{\mu\nu} P_R \bar\C_{rp}^{\mu\nu*} \right) \bigg] \nu_{Mr} \, .
\end{align}

	\begin{myfmf}{diags/feynmanRules}

\section{Propagator Feynman rules}
\label{sec:Propagators}

In this appendix, we list the Feynman rules for the propagators of the background and quantum fields, which agree with the expressions in the SM~\cite{Denner:1994xt}. We do not list the vertex rules for the SMEFT in background-field gauge because of the large number of vertices. Differences to the rules for conventional $R_\xi$ gauge~\cite{Dedes:2017zog,Dedes:2019uzs} obviously arise only in vertices that depend on the gauge-fixing Lagrangian.

\subsection{Background fields}

\begin{align}
	\begin{alignedat}{2}
	\begin{gathered}
		\begin{fmfgraph*}(60,30)
			\fmfset{curly_len}{2mm}
			\fmfleft{l1} \fmfright{r1}
			\fmf{dashes,label=$\hat h$,label.side=left}{l1,r1}
		\end{fmfgraph*}
	\end{gathered} &= \frac{i}{p^2-M_H^2} \, , \\
	\begin{gathered}
		\begin{fmfgraph*}(60,30)
			\fmfset{curly_len}{2mm}
			\fmfleft{l1} \fmfright{r1}
			\fmf{wboson,label=$\hat \W^+$,label.side=left}{l1,r1}
		\end{fmfgraph*}
	\end{gathered} &= \frac{-i}{p^2-\mw^2} \left( g^{\mu\nu} - \frac{p^\mu p^\nu}{\mw^2} \right) \, , \qquad &
	\begin{gathered}
		\begin{fmfgraph*}(60,30)
			\fmfset{curly_len}{2mm}
			\fmfleft{l1} \fmfright{r1}
			\fmf{photon,label=$\hat \Z$,label.side=left}{l1,r1}
		\end{fmfgraph*}
	\end{gathered} &= \frac{-i}{p^2-\mz^2} \left( g^{\mu\nu} - \frac{p^\mu p^\nu}{\mz^2} \right) \, , \\
	\begin{gathered}
		\begin{fmfgraph*}(60,30)
			\fmfset{curly_len}{2mm}
			\fmfleft{l1} \fmfright{r1}
			\fmf{photon,label=$\hat\A$,label.side=left}{l1,r1}
		\end{fmfgraph*}
	\end{gathered} &= \frac{-i}{p^2} \left( g^{\mu\nu} - (1-\hat\xi_\gamma) \frac{p^\mu p^\nu}{p^2} \right) \, , \qquad &
	\begin{gathered}
		\begin{fmfgraph*}(60,30)
			\fmfset{curly_len}{2mm}
			\fmfleft{l1} \fmfright{r1}
			\fmf{gluon,label=$\hat \G$,label.side=left}{l1,r1}
		\end{fmfgraph*}
	\end{gathered} &= \frac{-i\delta^{AB}}{p^2} \left( g^{\mu\nu} - (1-\hat\xi_g) \frac{p^\mu p^\nu}{p^2} \right) \, , \\
	\begin{gathered}
		\begin{fmfgraph*}(60,30)
			\fmfset{curly_len}{2mm}
			\fmfleft{l1} \fmfright{r1}
			\fmf{quark,label=$\hat \psi_i$,label.side=left}{l1,r1}
		\end{fmfgraph*}
	\end{gathered} &= \frac{i ( \slashed p + M_{\psi_i} )}{p^2 - M_{\psi_i}^2} \, , \quad \psi = u, d, e \, , \qquad &
	\begin{gathered}
		\begin{fmfgraph*}(60,30)
			\fmfset{curly_len}{2mm}
			\fmfleft{l1} \fmfright{r1}
			\fmf{plain,label=$\hat \nu_i$,label.side=left}{l1,r1}
		\end{fmfgraph*}
	\end{gathered} &= \frac{i ( \slashed p + M_{\nu_i} )}{p^2 - M_{\nu_i}^2} \, .
	\end{alignedat} \nonumber\\[-0.75cm]
\end{align}

\subsection{Quantum fields}

\begin{align}
	\begin{alignedat}{2}
	\begin{gathered}
		\begin{fmfgraph*}(60,30)
			\fmfset{curly_len}{2mm}
			\fmfleft{l1} \fmfright{r1}
			\fmf{scalar,label=$G^+$,label.side=left}{l1,r1}
		\end{fmfgraph*}
	\end{gathered} &= \frac{i}{p^2-\xi \mw^2} \, , \qquad &
	\begin{gathered}
		\begin{fmfgraph*}(60,30)
			\fmfset{curly_len}{2mm}
			\fmfleft{l1} \fmfright{r1}
			\fmf{dashes,label=$G^0$,label.side=left}{l1,r1}
		\end{fmfgraph*}
	\end{gathered} &= \frac{i}{p^2-\xi \mz^2} \, , \\
	\begin{gathered}
		\begin{fmfgraph*}(60,30)
			\fmfset{curly_len}{2mm}
			\fmfleft{l1} \fmfright{r1}
			\fmf{dashes,label=$h$,label.side=left}{l1,r1}
		\end{fmfgraph*}
	\end{gathered} &= \frac{i}{p^2-M_H^2} \, , \\
	\begin{gathered}
		\begin{fmfgraph*}(60,30)
			\fmfset{curly_len}{2mm}
			\fmfleft{l1} \fmfright{r1}
			\fmf{wboson,label=$\W^+$,label.side=left}{l1,r1}
		\end{fmfgraph*}
	\end{gathered} &= \frac{-i}{p^2-\mw^2} \left( g^{\mu\nu} - (1-\xi) \frac{p^\mu p^\nu}{p^2 - \xi \mw^2} \right) \, , \hspace{-2cm} \\
	\begin{gathered}
		\begin{fmfgraph*}(60,30)
			\fmfset{curly_len}{2mm}
			\fmfleft{l1} \fmfright{r1}
			\fmf{photon,label=$\Z$,label.side=left}{l1,r1}
		\end{fmfgraph*}
	\end{gathered} &= \frac{-i}{p^2-\mz^2} \left( g^{\mu\nu} - (1-\xi) \frac{p^\mu p^\nu}{p^2 - \xi \mz^2} \right) \, , \hspace{-2cm} \\
	\begin{gathered}
		\begin{fmfgraph*}(60,30)
			\fmfset{curly_len}{2mm}
			\fmfleft{l1} \fmfright{r1}
			\fmf{photon,label=$\A$,label.side=left}{l1,r1}
		\end{fmfgraph*}
	\end{gathered} &= \frac{-i}{p^2} \left( g^{\mu\nu} - (1-\xi) \frac{p^\mu p^\nu}{p^2} \right) \, , \qquad &
	\begin{gathered}
		\begin{fmfgraph*}(60,30)
			\fmfset{curly_len}{2mm}
			\fmfleft{l1} \fmfright{r1}
			\fmf{gluon,label=$\G$,label.side=left}{l1,r1}
		\end{fmfgraph*}
	\end{gathered} &= \frac{-i\delta^{AB}}{p^2} \left( g^{\mu\nu} - (1 - \xi_g) \frac{p^\mu p^\nu}{p^2} \right) \, , \\
	\begin{gathered}
		\begin{fmfgraph*}(60,30)
			\fmfset{curly_len}{2mm}
			\fmfleft{l1} \fmfright{r1}
			\fmf{ghost,label=$\eta_\pm$,label.side=left}{l1,r1}
		\end{fmfgraph*}
	\end{gathered} &= \frac{i}{p^2-\xi \mw^2} \, , \qquad &
	\begin{gathered}
		\begin{fmfgraph*}(60,30)
			\fmfset{curly_len}{2mm}
			\fmfleft{l1} \fmfright{r1}
			\fmf{ghost,label=$\eta_Z$,label.side=left}{l1,r1}
		\end{fmfgraph*}
	\end{gathered} &= \frac{i}{p^2-\xi \mz^2} \, , \\
	\begin{gathered}
		\begin{fmfgraph*}(60,30)
			\fmfset{curly_len}{2mm}
			\fmfleft{l1} \fmfright{r1}
			\fmf{ghost,label=$\eta_A$,label.side=left}{l1,r1}
		\end{fmfgraph*}
	\end{gathered} &= \frac{i}{p^2} \, , \qquad &
	\begin{gathered}
		\begin{fmfgraph*}(60,30)
			\fmfset{curly_len}{2mm}
			\fmfleft{l1} \fmfright{r1}
			\fmf{ghost,label=$\eta_G$,label.side=left}{l1,r1}
		\end{fmfgraph*}
	\end{gathered} &= \frac{i \delta^{AB}}{p^2} \, , \\
	\begin{gathered}
		\begin{fmfgraph*}(60,30)
			\fmfset{curly_len}{2mm}
			\fmfleft{l1} \fmfright{r1}
			\fmf{quark,label=$\psi_i$,label.side=left}{l1,r1}
		\end{fmfgraph*}
	\end{gathered} &= \frac{i ( \slashed p + M_{\psi_i} )}{p^2 - M_{\psi_i}^2} \, , \quad \psi = u, d, e \, , \qquad &
	\begin{gathered}
		\begin{fmfgraph*}(60,30)
			\fmfset{curly_len}{2mm}
			\fmfleft{l1} \fmfright{r1}
			\fmf{plain,label=$\nu_i$,label.side=left}{l1,r1}
		\end{fmfgraph*}
	\end{gathered} &= \frac{i ( \slashed p + M_{\nu_i} )}{p^2 - M_{\nu_i}^2} \, .
	\end{alignedat} \nonumber\\[-0.75cm]
\end{align}
In the ghost propagators, the arrow denotes the ghost number flow (in the case of $\eta_-$ the charge flow is in the opposite direction).

	\end{myfmf}
	

\section{Diagrams}
\label{sec:Diagrams}

In this appendix, we provide the complete list of relevant Feynman diagrams that we calculated for the one-loop matching of the SMEFT to the LEFT. We organize the diagrams in terms of 1PI functions and do not explicitly show 1LPI diagrams with tree-level propagators for the heavy background-field gauge bosons.

\begin{myfmf}{diags/onePoint}

\subsection{Tadpoles}
\label{sec:TadpoleDiagrams}


\begin{align*}
	&\begin{gathered}
		\begin{fmfgraph*}(60,60)
			\fmfset{curly_len}{2mm}
			\fmfleft{l1} \fmfright{r1}
			\fmf{dashes,label=$\hat h$,label.side=left,tension=3}{l1,v1}
			\fmf{phantom,label=$t$,label.side=left,tension=2}{v2,r1}
			\fmf{quark,right}{v1,v2,v1}
		\end{fmfgraph*}
	\end{gathered}
\qquad
	\begin{gathered}
		\begin{fmfgraph*}(60,60)
			\fmfset{curly_len}{2mm}
			\fmfleft{l1} \fmfright{r1}
			\fmf{dashes,label=$\hat h$,label.side=left,tension=3}{l1,v1}
			\fmf{phantom,label=$\W^+$,label.side=left,tension=2}{v2,r1}
			\fmf{wboson,right}{v1,v2,v1}
		\end{fmfgraph*}
	\end{gathered}
\qquad
	\begin{gathered}
		\begin{fmfgraph*}(60,60)
			\fmfset{curly_len}{2mm}
			\fmfleft{l1} \fmfright{r1}
			\fmf{dashes,label=$\hat h$,label.side=left,tension=3}{l1,v1}
			\fmf{phantom,label=$\Z^+$,label.side=left,tension=2}{v2,r1}
			\fmf{photon,right}{v1,v2,v1}
		\end{fmfgraph*}
	\end{gathered}
\qquad
	\begin{gathered}
		\begin{fmfgraph*}(60,60)
			\fmfset{curly_len}{2mm}
			\fmfleft{l1} \fmfright{r1}
			\fmf{dashes,label=$\hat h$,label.side=left,tension=3}{l1,v1}
			\fmf{phantom,label=$G^+$,label.side=left,tension=2}{v2,r1}
			\fmf{scalar,right}{v1,v2,v1}
		\end{fmfgraph*}
	\end{gathered}
\qquad
	\begin{gathered}
		\begin{fmfgraph*}(60,60)
			\fmfset{curly_len}{2mm}
			\fmfleft{l1} \fmfright{r1}
			\fmf{dashes,label=$\hat h$,label.side=left,tension=3}{l1,v1}
			\fmf{phantom,label=$G^0$,label.side=left,tension=2}{v2,r1}
			\fmf{dashes,right}{v1,v2,v1}
		\end{fmfgraph*}
	\end{gathered}
\nn
	&\begin{gathered}
		\begin{fmfgraph*}(60,60)
			\fmfset{curly_len}{2mm}
			\fmfleft{l1} \fmfright{r1}
			\fmf{dashes,label=$\hat h$,label.side=left,tension=3}{l1,v1}
			\fmf{phantom,label=$h$,label.side=left,tension=2}{v2,r1}
			\fmf{dashes,right}{v1,v2,v1}
		\end{fmfgraph*}
	\end{gathered}
\qquad
	\begin{gathered}
		\begin{fmfgraph*}(60,60)
			\fmfset{curly_len}{2mm}
			\fmfleft{l1} \fmfright{r1}
			\fmf{dashes,label=$\hat h$,label.side=left,tension=3}{l1,v1}
			\fmf{phantom,label=$\eta_\pm$,label.side=left,tension=2}{v2,r1}
			\fmf{ghost,right}{v1,v2,v1}
		\end{fmfgraph*}
	\end{gathered}
\qquad
	\begin{gathered}
		\begin{fmfgraph*}(60,60)
			\fmfset{curly_len}{2mm}
			\fmfleft{l1} \fmfright{r1}
			\fmf{dashes,label=$\hat h$,label.side=left,tension=3}{l1,v1}
			\fmf{phantom,label=$\eta_Z$,label.side=left,tension=2}{v2,r1}
			\fmf{ghost,right}{v1,v2,v1}
		\end{fmfgraph*}
	\end{gathered}
\end{align*}

\end{myfmf}
	
\begin{myfmf}{diags/twoPoint}

\subsection{Two-point functions}
\label{sec:TwoPointFunctions}


\subsubsection[$\gamma^2$]{\boldmath$\gamma^2$}

\begin{align*}
	&\begin{gathered}
		\begin{fmfgraph*}(60,60)
			\fmfset{curly_len}{2mm}
			\fmfleft{l1} \fmfright{r1}
			\fmf{photon,label=$\hat\A$,label.side=left,tension=3}{l1,v1}
			\fmf{photon,label=$\hat\A$,label.side=left,tension=3}{v2,r1}
			\fmf{quark,label=$t$,label.side=right,right}{v1,v2,v1}
			\fmffreeze
		\end{fmfgraph*}
	\end{gathered}
\qquad
	\begin{gathered}
		\begin{fmfgraph*}(60,60)
			\fmfset{curly_len}{2mm}
			\fmfleft{l1} \fmfright{r1}
			\fmf{photon,label=$\hat\A$,label.side=left,tension=3}{l1,v1}
			\fmf{photon,label=$\hat\A$,label.side=left,tension=3}{v2,r1}
			\fmf{wboson,label=$\W^+$,label.side=right,right}{v1,v2,v1}
			\fmffreeze
		\end{fmfgraph*}
	\end{gathered}
\qquad
	\begin{gathered}
		\begin{fmfgraph*}(60,60)
			\fmfset{curly_len}{2mm}
			\fmfleft{l1} \fmfright{r1}
			\fmf{photon,label=$\hat\A$,label.side=left,tension=3}{l1,v1}
			\fmf{photon,label=$\hat\A$,label.side=left,tension=3}{v2,r1}
			\fmf{scalar,label=$G^+$,label.side=right,right}{v1,v2}
			\fmf{scalar,label=$G^+$,label.side=right,right}{v2,v1}
			\fmffreeze
		\end{fmfgraph*}
	\end{gathered}
\qquad
	\begin{gathered}
		\begin{fmfgraph*}(60,60)
			\fmfset{curly_len}{2mm}
			\fmfleft{l1} \fmfright{r1}
			\fmf{photon,label=$\hat\A$,label.side=left,tension=3}{l1,v1}
			\fmf{photon,label=$\hat\A$,label.side=left,tension=3}{v2,r1}
			\fmf{ghost,label=$\eta_\pm$,label.side=right,right}{v1,v2}
			\fmf{ghost,label=$\eta_\pm$,label.side=right,right}{v2,v1}
			\fmffreeze
		\end{fmfgraph*}
	\end{gathered}
\qquad
	\begin{gathered}
		\begin{fmfgraph*}(60,60)
			\fmfset{curly_len}{2mm}
			\fmfleft{l1} \fmfright{r1} \fmftop{t1}
			\fmf{photon,label=$\hat\A$,label.side=right,tension=3}{l1,v1}
			\fmf{photon,label=$\hat\A$,label.side=right,tension=3}{v1,r1}
			\fmffreeze
			\fmf{phantom,tension=5}{t1,v2}
			\fmf{phantom,label=$\W^+$,label.side=right,tension=0.8}{v1,v1}
			\fmf{wboson,tension=0.5,right=1}{v1,v2,v1}
			\fmffreeze
		\end{fmfgraph*}
	\end{gathered}
\nnw
	& \begin{gathered}
		\begin{fmfgraph*}(60,60)
			\fmfset{curly_len}{2mm}
			\fmfleft{l1} \fmfright{r1} \fmftop{t1}
			\fmf{photon,label=$\hat\A$,label.side=right,tension=3}{l1,v1}
			\fmf{photon,label=$\hat\A$,label.side=right,tension=3}{v1,r1}
			\fmffreeze
			\fmf{phantom,tension=5}{t1,v2}
			\fmf{phantom,label=$G^+$,label.side=right,tension=0.8}{v1,v1}
			\fmf{scalar,tension=0.5,right=1}{v1,v2,v1}
			\fmffreeze
		\end{fmfgraph*}
	\end{gathered}
\qquad
	\begin{gathered}
		\begin{fmfgraph*}(60,60)
			\fmfset{curly_len}{2mm}
			\fmfleft{l1} \fmfright{r1} \fmftop{t1}
			\fmf{photon,label=$\hat\A$,label.side=right,tension=3}{l1,v1}
			\fmf{photon,label=$\hat\A$,label.side=right,tension=3}{v1,r1}
			\fmffreeze
			\fmf{phantom,tension=5}{t1,v2}
			\fmf{phantom,label=$\eta_\pm$,label.side=right,tension=0.8}{v1,v1}
			\fmf{ghost,tension=0.5,right=1}{v1,v2,v1}
			\fmffreeze
		\end{fmfgraph*}
	\end{gathered}
\qquad
	\begin{gathered}
		\begin{fmfgraph*}(60,60)
			\fmfset{curly_len}{2mm}
			\fmfleft{l1} \fmfright{r1} \fmftop{t1}
			\fmf{photon,label=$\hat\A$,label.side=right,tension=3}{l1,v1}
			\fmf{photon,label=$\hat\A$,label.side=right,tension=3}{v1,r1}
			\fmffreeze
			\fmf{phantom,tension=5}{t1,v2}
			\fmf{phantom,label=$G^0$,label.side=right,tension=0.8}{v1,v1}
			\fmf{dashes,tension=0.5,right=1}{v1,v2,v1}
			\fmfv{decoration.shape=square,decoration.size=1.5mm}{v1}
		\end{fmfgraph*}
	\end{gathered}
\qquad
	\begin{gathered}
		\begin{fmfgraph*}(60,60)
			\fmfset{curly_len}{2mm}
			\fmfleft{l1} \fmfright{r1} \fmftop{t1}
			\fmf{photon,label=$\hat\A$,label.side=right,tension=3}{l1,v1}
			\fmf{photon,label=$\hat\A$,label.side=right,tension=3}{v1,r1}
			\fmffreeze
			\fmf{phantom,tension=5}{t1,v2}
			\fmf{phantom,label=$h$,label.side=right,tension=0.8}{v1,v1}
			\fmf{dashes,tension=0.5,right=1}{v1,v2,v1}
			\fmfv{decoration.shape=square,decoration.size=1.5mm}{v1}
		\end{fmfgraph*}
	\end{gathered}
\qquad
	\begin{gathered}
		\begin{fmfgraph*}(60,60)
			\fmfset{curly_len}{2mm}
			\fmfleft{l1} \fmfright{r1} \fmftop{t1}
			\fmf{photon,label=$\hat\A$,label.side=right,tension=3}{l1,v1}
			\fmf{photon,label=$\hat\A$,label.side=right,tension=3}{v1,r1}
			\fmffreeze
			\fmf{dashes,label=$\hat h$}{v1,v2}
			\fmf{phantom,tension=5}{t1,v2}
			\fmfv{decoration.shape=square,decoration.size=1.5mm}{v1}
			\fmfblob{5mm}{v2}
		\end{fmfgraph*}
	\end{gathered}
\end{align*}


\subsubsection[$\gamma Z$]{\boldmath$\gamma Z$}

\begin{align*}
	&\begin{gathered}
		\begin{fmfgraph*}(60,60)
			\fmfset{curly_len}{2mm}
			\fmfleft{l1} \fmfright{r1}
			\fmf{photon,label=$\hat\A$,label.side=left,tension=3}{l1,v1}
			\fmf{photon,label=$\hat\Z$,label.side=left,tension=3}{v2,r1}
			\fmf{quark,label=$t$,label.side=right,right}{v1,v2,v1}
			\fmffreeze
		\end{fmfgraph*}
	\end{gathered}
\qquad
	\begin{gathered}
		\begin{fmfgraph*}(60,60)
			\fmfset{curly_len}{2mm}
			\fmfleft{l1} \fmfright{r1}
			\fmf{photon,label=$\hat\A$,label.side=left,tension=3}{l1,v1}
			\fmf{photon,label=$\hat\Z$,label.side=left,tension=3}{v2,r1}
			\fmf{wboson,label=$\W^+$,label.side=right,right}{v1,v2,v1}
			\fmffreeze
		\end{fmfgraph*}
	\end{gathered}
\qquad
	\begin{gathered}
		\begin{fmfgraph*}(60,60)
			\fmfset{curly_len}{2mm}
			\fmfleft{l1} \fmfright{r1}
			\fmf{photon,label=$\hat\A$,label.side=left,tension=3}{l1,v1}
			\fmf{photon,label=$\hat\Z$,label.side=left,tension=3}{v2,r1}
			\fmf{scalar,label=$G^+$,label.side=right,right}{v1,v2}
			\fmf{scalar,label=$G^+$,label.side=right,right}{v2,v1}
			\fmffreeze
		\end{fmfgraph*}
	\end{gathered}
\qquad
	\begin{gathered}
		\begin{fmfgraph*}(60,60)
			\fmfset{curly_len}{2mm}
			\fmfleft{l1} \fmfright{r1}
			\fmf{photon,label=$\hat\A$,label.side=left,tension=3}{l1,v1}
			\fmf{photon,label=$\hat\Z$,label.side=left,tension=3}{v2,r1}
			\fmf{ghost,label=$\eta_\pm$,label.side=right,right}{v1,v2}
			\fmf{ghost,label=$\eta_\pm$,label.side=right,right}{v2,v1}
			\fmffreeze
		\end{fmfgraph*}
	\end{gathered}
\qquad
	\begin{gathered}
		\begin{fmfgraph*}(60,60)
			\fmfset{curly_len}{2mm}
			\fmfleft{l1} \fmfright{r1} \fmftop{t1}
			\fmf{photon,label=$\hat\A$,label.side=right,tension=3}{l1,v1}
			\fmf{photon,label=$\hat\Z$,label.side=right,tension=3}{v1,r1}
			\fmffreeze
			\fmf{phantom,tension=5}{t1,v2}
			\fmf{phantom,label=$\W^+$,label.side=right,tension=0.8}{v1,v1}
			\fmf{wboson,tension=0.5,right=1}{v1,v2,v1}
			\fmffreeze
		\end{fmfgraph*}
	\end{gathered}
\nnw
	& \begin{gathered}
		\begin{fmfgraph*}(60,60)
			\fmfset{curly_len}{2mm}
			\fmfleft{l1} \fmfright{r1} \fmftop{t1}
			\fmf{photon,label=$\hat\A$,label.side=right,tension=3}{l1,v1}
			\fmf{photon,label=$\hat\Z$,label.side=right,tension=3}{v1,r1}
			\fmffreeze
			\fmf{phantom,tension=5}{t1,v2}
			\fmf{phantom,label=$G^+$,label.side=right,tension=0.8}{v1,v1}
			\fmf{scalar,tension=0.5,right=1}{v1,v2,v1}
			\fmffreeze
		\end{fmfgraph*}
	\end{gathered}
\qquad
	\begin{gathered}
		\begin{fmfgraph*}(60,60)
			\fmfset{curly_len}{2mm}
			\fmfleft{l1} \fmfright{r1} \fmftop{t1}
			\fmf{photon,label=$\hat\A$,label.side=right,tension=3}{l1,v1}
			\fmf{photon,label=$\hat\Z$,label.side=right,tension=3}{v1,r1}
			\fmffreeze
			\fmf{phantom,tension=5}{t1,v2}
			\fmf{phantom,label=$\eta_\pm$,label.side=right,tension=0.8}{v1,v1}
			\fmf{ghost,tension=0.5,right=1}{v1,v2,v1}
			\fmffreeze
		\end{fmfgraph*}
	\end{gathered}
\qquad
	\begin{gathered}
		\begin{fmfgraph*}(60,60)
			\fmfset{curly_len}{2mm}
			\fmfleft{l1} \fmfright{r1}
			\fmf{photon,label=$\hat\A$,label.side=left,tension=3}{l1,v1}
			\fmf{photon,label=$\hat\Z$,label.side=left,tension=3}{v2,r1}
			\fmf{wboson,label=$\W^+$,label.side=right,right}{v1,v2}
			\fmf{scalar,label=$G^+$,label.side=right,right}{v2,v1}
			\fmfv{decoration.shape=square,decoration.size=1.5mm}{v1}
		\end{fmfgraph*}
	\end{gathered}
\qquad
	\begin{gathered}
		\begin{fmfgraph*}(60,60)
			\fmfset{curly_len}{2mm}
			\fmfleft{l1} \fmfright{r1}
			\fmf{photon,label=$\hat\A$,label.side=left,tension=3}{l1,v1}
			\fmf{photon,label=$\hat\Z$,label.side=left,tension=3}{v2,r1}
			\fmf{scalar,label=$G^+$,label.side=right,right}{v1,v2}
			\fmf{wboson,label=$\W^+$,label.side=right,right}{v2,v1}
			\fmfv{decoration.shape=square,decoration.size=1.5mm}{v1}
		\end{fmfgraph*}
	\end{gathered}
\qquad
	\begin{gathered}
		\begin{fmfgraph*}(60,60)
			\fmfset{curly_len}{2mm}
			\fmfleft{l1} \fmfright{r1}
			\fmf{photon,label=$\hat\A$,label.side=left,tension=3}{l1,v1}
			\fmf{photon,label=$\hat\Z$,label.side=left,tension=3}{v2,r1}
			\fmf{photon,label=$\Z$,label.side=right,right}{v1,v2}
			\fmf{dashes,label=$h$,label.side=right,right}{v2,v1}
			\fmfv{decoration.shape=square,decoration.size=1.5mm}{v1}
		\end{fmfgraph*}
	\end{gathered}
\nnw
	& \begin{gathered}
		\begin{fmfgraph*}(60,60)
			\fmfset{curly_len}{2mm}
			\fmfleft{l1} \fmfright{r1} \fmftop{t1}
			\fmf{photon,label=$\hat\A$,label.side=right,tension=3}{l1,v1}
			\fmf{photon,label=$\hat\Z$,label.side=right,tension=3}{v1,r1}
			\fmffreeze
			\fmf{phantom,tension=5}{t1,v2}
			\fmf{phantom,label=$G^0$,label.side=right,tension=0.8}{v1,v1}
			\fmf{dashes,tension=0.5,right=1}{v1,v2,v1}
			\fmfv{decoration.shape=square,decoration.size=1.5mm}{v1}
		\end{fmfgraph*}
	\end{gathered}
\qquad
	\begin{gathered}
		\begin{fmfgraph*}(60,60)
			\fmfset{curly_len}{2mm}
			\fmfleft{l1} \fmfright{r1} \fmftop{t1}
			\fmf{photon,label=$\hat\A$,label.side=right,tension=3}{l1,v1}
			\fmf{photon,label=$\hat\Z$,label.side=right,tension=3}{v1,r1}
			\fmffreeze
			\fmf{phantom,tension=5}{t1,v2}
			\fmf{phantom,label=$h$,label.side=right,tension=0.8}{v1,v1}
			\fmf{dashes,tension=0.5,right=1}{v1,v2,v1}
			\fmfv{decoration.shape=square,decoration.size=1.5mm}{v1}
		\end{fmfgraph*}
	\end{gathered}
\qquad
	\begin{gathered}
		\begin{fmfgraph*}(60,60)
			\fmfset{curly_len}{2mm}
			\fmfleft{l1} \fmfright{r1} \fmftop{t1}
			\fmf{photon,label=$\hat\A$,label.side=right,tension=3}{l1,v1}
			\fmf{photon,label=$\hat\Z$,label.side=right,tension=3}{v1,r1}
			\fmffreeze
			\fmf{dashes,label=$\hat h$}{v1,v2}
			\fmf{phantom,tension=5}{t1,v2}
			\fmfv{decoration.shape=square,decoration.size=1.5mm}{v1}
			\fmfblob{5mm}{v2}
		\end{fmfgraph*}
	\end{gathered}
\end{align*}


\subsubsection[$ZZ$]{\boldmath$ZZ$}

\begin{align*}
	&\begin{gathered}
		\begin{fmfgraph*}(60,60)
			\fmfset{curly_len}{2mm}
			\fmfleft{l1} \fmfright{r1}
			\fmf{photon,label=$\hat\Z$,label.side=left,tension=3}{l1,v1}
			\fmf{photon,label=$\hat\Z$,label.side=left,tension=3}{v2,r1}
			\fmf{quark,label=$t$,label.side=right,right}{v1,v2,v1}
			\fmffreeze
		\end{fmfgraph*}
	\end{gathered}
\qquad
	\begin{gathered}
		\begin{fmfgraph*}(60,60)
			\fmfset{curly_len}{2mm}
			\fmfleft{l1} \fmfright{r1}
			\fmf{photon,label=$\hat\Z$,label.side=left,tension=3}{l1,v1}
			\fmf{photon,label=$\hat\Z$,label.side=left,tension=3}{v2,r1}
			\fmf{wboson,label=$\W^+$,label.side=right,right}{v1,v2,v1}
			\fmffreeze
		\end{fmfgraph*}
	\end{gathered}
\qquad
	\begin{gathered}
		\begin{fmfgraph*}(60,60)
			\fmfset{curly_len}{2mm}
			\fmfleft{l1} \fmfright{r1}
			\fmf{photon,label=$\hat\Z$,label.side=left,tension=3}{l1,v1}
			\fmf{photon,label=$\hat\Z$,label.side=left,tension=3}{v2,r1}
			\fmf{scalar,label=$G^+$,label.side=right,right}{v1,v2}
			\fmf{scalar,label=$G^+$,label.side=right,right}{v2,v1}
			\fmffreeze
		\end{fmfgraph*}
	\end{gathered}
\qquad
	\begin{gathered}
		\begin{fmfgraph*}(60,60)
			\fmfset{curly_len}{2mm}
			\fmfleft{l1} \fmfright{r1}
			\fmf{photon,label=$\hat\Z$,label.side=left,tension=3}{l1,v1}
			\fmf{photon,label=$\hat\Z$,label.side=left,tension=3}{v2,r1}
			\fmf{wboson,label=$\W^+$,label.side=right,right}{v1,v2}
			\fmf{scalar,label=$G^+$,label.side=right,right}{v2,v1}
		\end{fmfgraph*}
	\end{gathered}
\qquad
	\begin{gathered}
		\begin{fmfgraph*}(60,60)
			\fmfset{curly_len}{2mm}
			\fmfleft{l1} \fmfright{r1}
			\fmf{photon,label=$\hat\Z$,label.side=left,tension=3}{l1,v1}
			\fmf{photon,label=$\hat\Z$,label.side=left,tension=3}{v2,r1}
			\fmf{scalar,label=$G^+$,label.side=right,right}{v1,v2}
			\fmf{wboson,label=$\W^+$,label.side=right,right}{v2,v1}
		\end{fmfgraph*}
	\end{gathered}
\nnw
	& \begin{gathered}
		\begin{fmfgraph*}(60,60)
			\fmfset{curly_len}{2mm}
			\fmfleft{l1} \fmfright{r1}
			\fmf{photon,label=$\hat\Z$,label.side=left,tension=3}{l1,v1}
			\fmf{photon,label=$\hat\Z$,label.side=left,tension=3}{v2,r1}
			\fmf{ghost,label=$\eta_\pm$,label.side=right,right}{v1,v2}
			\fmf{ghost,label=$\eta_\pm$,label.side=right,right}{v2,v1}
			\fmffreeze
		\end{fmfgraph*}
	\end{gathered}
\qquad
	\begin{gathered}
		\begin{fmfgraph*}(60,60)
			\fmfset{curly_len}{2mm}
			\fmfleft{l1} \fmfright{r1}
			\fmf{photon,label=$\hat\Z$,label.side=left,tension=3}{l1,v1}
			\fmf{photon,label=$\hat\Z$,label.side=left,tension=3}{v2,r1}
			\fmf{photon,label=$\Z$,label.side=right,right}{v1,v2}
			\fmf{dashes,label=$h$,label.side=right,right}{v2,v1}
		\end{fmfgraph*}
	\end{gathered}
\qquad
	\begin{gathered}
		\begin{fmfgraph*}(60,60)
			\fmfset{curly_len}{2mm}
			\fmfleft{l1} \fmfright{r1}
			\fmf{photon,label=$\hat\Z$,label.side=left,tension=3}{l1,v1}
			\fmf{photon,label=$\hat\Z$,label.side=left,tension=3}{v2,r1}
			\fmf{dashes,label=$G^0$,label.side=right,right}{v1,v2}
			\fmf{dashes,label=$h$,label.side=right,right}{v2,v1}
		\end{fmfgraph*}
	\end{gathered}
\qquad
	\begin{gathered}
		\begin{fmfgraph*}(60,60)
			\fmfset{curly_len}{2mm}
			\fmfleft{l1} \fmfright{r1} \fmftop{t1}
			\fmf{photon,label=$\hat\Z$,label.side=right,tension=3}{l1,v1}
			\fmf{photon,label=$\hat\Z$,label.side=right,tension=3}{v1,r1}
			\fmffreeze
			\fmf{phantom,tension=5}{t1,v2}
			\fmf{phantom,label=$\W^+$,label.side=right,tension=0.8}{v1,v1}
			\fmf{wboson,tension=0.5,right=1}{v1,v2,v1}
			\fmffreeze
		\end{fmfgraph*}
	\end{gathered}
\qquad
	\begin{gathered}
		\begin{fmfgraph*}(60,60)
			\fmfset{curly_len}{2mm}
			\fmfleft{l1} \fmfright{r1} \fmftop{t1}
			\fmf{photon,label=$\hat\Z$,label.side=right,tension=3}{l1,v1}
			\fmf{photon,label=$\hat\Z$,label.side=right,tension=3}{v1,r1}
			\fmffreeze
			\fmf{phantom,tension=5}{t1,v2}
			\fmf{phantom,label=$G^+$,label.side=right,tension=0.8}{v1,v1}
			\fmf{scalar,tension=0.5,right=1}{v1,v2,v1}
			\fmffreeze
		\end{fmfgraph*}
	\end{gathered}
\nnw
	& \begin{gathered}
		\begin{fmfgraph*}(60,60)
			\fmfset{curly_len}{2mm}
			\fmfleft{l1} \fmfright{r1} \fmftop{t1}
			\fmf{photon,label=$\hat\Z$,label.side=right,tension=3}{l1,v1}
			\fmf{photon,label=$\hat\Z$,label.side=right,tension=3}{v1,r1}
			\fmffreeze
			\fmf{phantom,tension=5}{t1,v2}
			\fmf{phantom,label=$\eta_\pm$,label.side=right,tension=0.8}{v1,v1}
			\fmf{ghost,tension=0.5,right=1}{v1,v2,v1}
			\fmffreeze
		\end{fmfgraph*}
	\end{gathered}
\qquad
	\begin{gathered}
		\begin{fmfgraph*}(60,60)
			\fmfset{curly_len}{2mm}
			\fmfleft{l1} \fmfright{r1} \fmftop{t1}
			\fmf{photon,label=$\hat\Z$,label.side=right,tension=3}{l1,v1}
			\fmf{photon,label=$\hat\Z$,label.side=right,tension=3}{v1,r1}
			\fmffreeze
			\fmf{phantom,tension=5}{t1,v2}
			\fmf{phantom,label=$G^0$,label.side=right,tension=0.8}{v1,v1}
			\fmf{dashes,tension=0.5,right=1}{v1,v2,v1}
		\end{fmfgraph*}
	\end{gathered}
\qquad
	\begin{gathered}
		\begin{fmfgraph*}(60,60)
			\fmfset{curly_len}{2mm}
			\fmfleft{l1} \fmfright{r1} \fmftop{t1}
			\fmf{photon,label=$\hat\Z$,label.side=right,tension=3}{l1,v1}
			\fmf{photon,label=$\hat\Z$,label.side=right,tension=3}{v1,r1}
			\fmffreeze
			\fmf{phantom,tension=5}{t1,v2}
			\fmf{phantom,label=$h$,label.side=right,tension=0.8}{v1,v1}
			\fmf{dashes,tension=0.5,right=1}{v1,v2,v1}
		\end{fmfgraph*}
	\end{gathered}
\qquad
	\begin{gathered}
		\begin{fmfgraph*}(60,60)
			\fmfset{curly_len}{2mm}
			\fmfleft{l1} \fmfright{r1} \fmftop{t1}
			\fmf{photon,label=$\hat\Z$,label.side=right,tension=3}{l1,v1}
			\fmf{photon,label=$\hat\Z$,label.side=right,tension=3}{v1,r1}
			\fmffreeze
			\fmf{dashes,label=$\hat h$}{v1,v2}
			\fmf{phantom,tension=5}{t1,v2}
			\fmfblob{5mm}{v2}
		\end{fmfgraph*}
	\end{gathered}
\end{align*}


\subsubsection[$hh$, $\gamma h$, and $Zh$]{\boldmath$hh$, $\gamma h$, and $Zh$}

As discussed in Sect.~\ref{sec:FourFermionMatching}, no diagrams with background Higgs fields contribute to the one-loop matching up to dimension six.


\subsubsection[$WW$]{\boldmath$WW$}

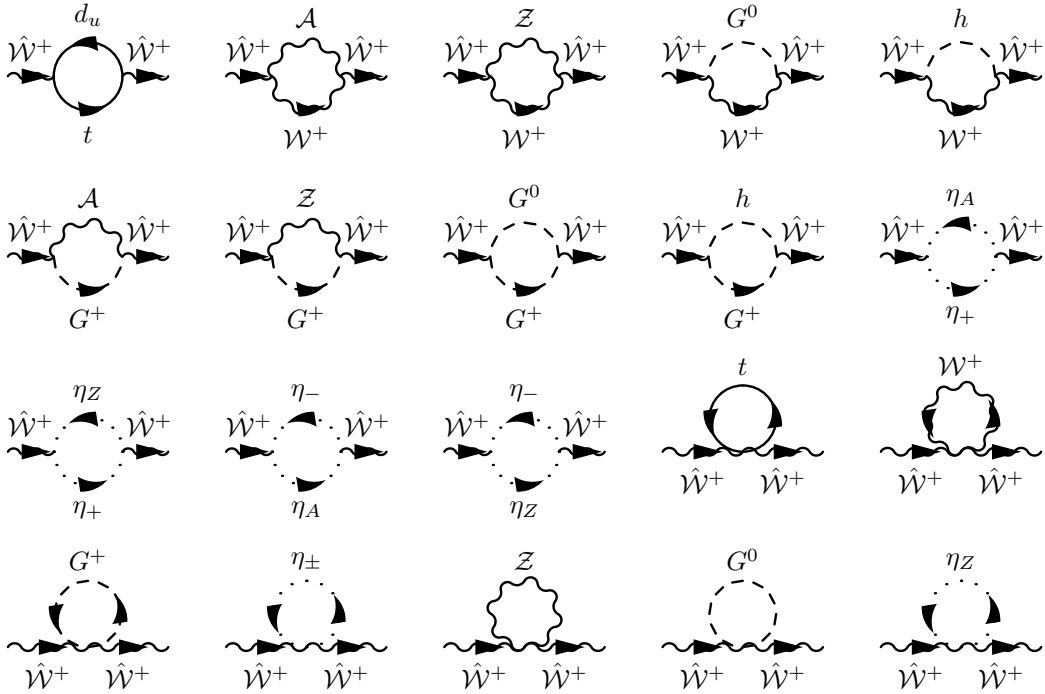
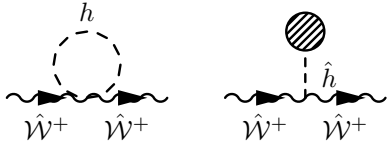
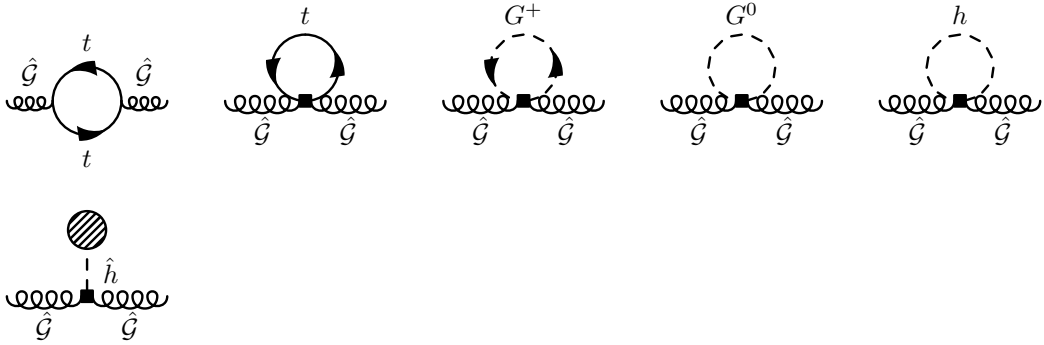
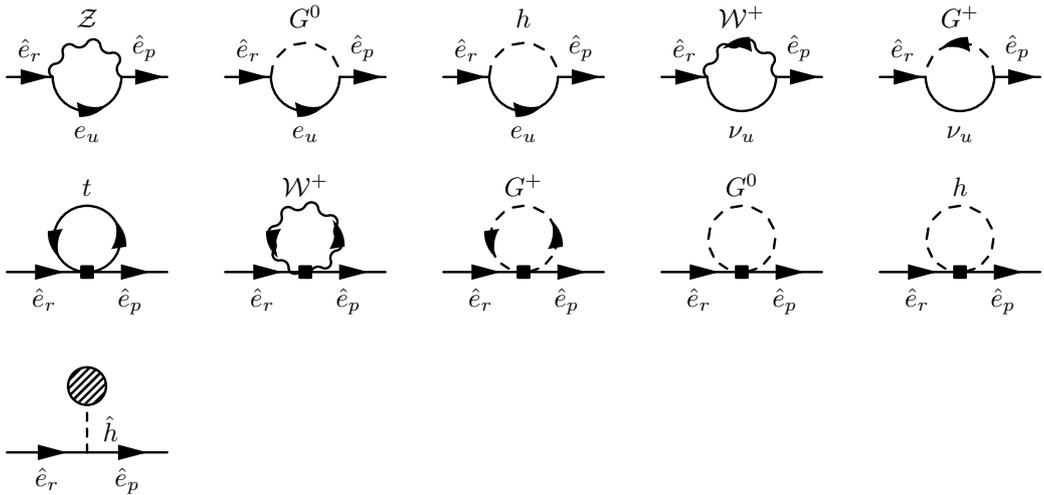
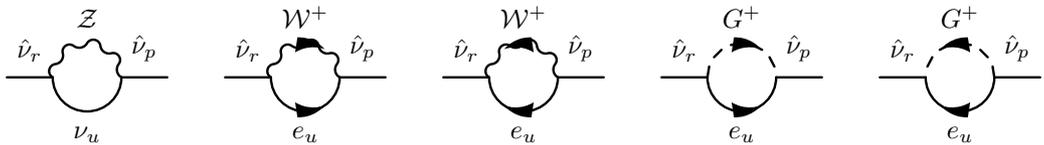
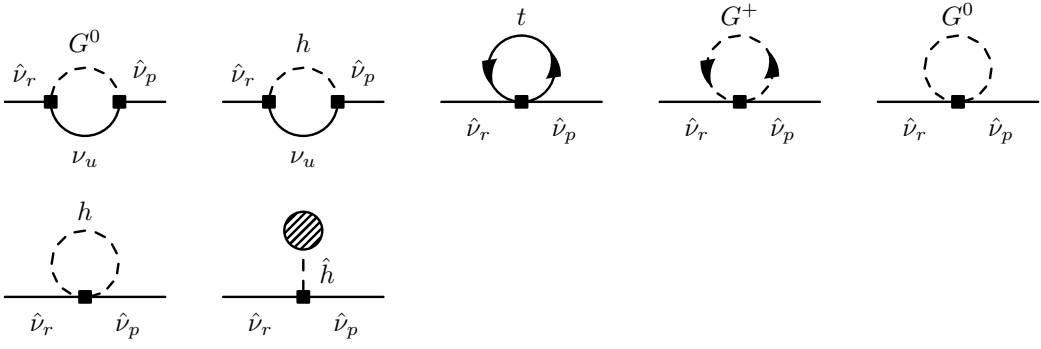
\begin{align*}
	&\begin{gathered}
		\begin{fmfgraph*}(60,60)
			\fmfset{curly_len}{2mm}
			\fmfleft{l1} \fmfright{r1}
			\fmf{wboson,label=$\hat\W^+$,label.side=left,tension=3}{l1,v1}
			\fmf{wboson,label=$\;\hat\W^+$,label.side=left,tension=3}{v2,r1}
			\fmf{quark,label=$t$,label.side=right,right}{v1,v2}
			\fmf{quark,label=$d_u$,label.side=right,right}{v2,v1}
			\fmffreeze
		\end{fmfgraph*}
	\end{gathered}
\qquad
	\begin{gathered}
		\begin{fmfgraph*}(60,60)
			\fmfset{curly_len}{2mm}
			\fmfleft{l1} \fmfright{r1}
			\fmf{wboson,label=$\hat\W^+$,label.side=left,tension=3}{l1,v1}
			\fmf{wboson,label=$\;\hat\W^+$,label.side=left,tension=3}{v2,r1}
			\fmf{wboson,label=$\W^+$,label.side=right,right}{v1,v2}
			\fmf{photon,label=$\A$,label.side=right,right}{v2,v1}
			\fmffreeze
		\end{fmfgraph*}
	\end{gathered}
\qquad
	\begin{gathered}
		\begin{fmfgraph*}(60,60)
			\fmfset{curly_len}{2mm}
			\fmfleft{l1} \fmfright{r1}
			\fmf{wboson,label=$\hat\W^+$,label.side=left,tension=3}{l1,v1}
			\fmf{wboson,label=$\;\hat\W^+$,label.side=left,tension=3}{v2,r1}
			\fmf{wboson,label=$\W^+$,label.side=right,right}{v1,v2}
			\fmf{photon,label=$\Z$,label.side=right,right}{v2,v1}
			\fmffreeze
		\end{fmfgraph*}
	\end{gathered}
\qquad
	\begin{gathered}
		\begin{fmfgraph*}(60,60)
			\fmfset{curly_len}{2mm}
			\fmfleft{l1} \fmfright{r1}
			\fmf{wboson,label=$\hat\W^+$,label.side=left,tension=3}{l1,v1}
			\fmf{wboson,label=$\;\hat\W^+$,label.side=left,tension=3}{v2,r1}
			\fmf{wboson,label=$\W^+$,label.side=right,right}{v1,v2}
			\fmf{dashes,label=$G^0$,label.side=right,right}{v2,v1}
			\fmffreeze
		\end{fmfgraph*}
	\end{gathered}
\qquad
	\begin{gathered}
		\begin{fmfgraph*}(60,60)
			\fmfset{curly_len}{2mm}
			\fmfleft{l1} \fmfright{r1}
			\fmf{wboson,label=$\hat\W^+$,label.side=left,tension=3}{l1,v1}
			\fmf{wboson,label=$\;\hat\W^+$,label.side=left,tension=3}{v2,r1}
			\fmf{wboson,label=$\W^+$,label.side=right,right}{v1,v2}
			\fmf{dashes,label=$h$,label.side=right,right}{v2,v1}
			\fmffreeze
		\end{fmfgraph*}
	\end{gathered}
\nn
	& \begin{gathered}
		\begin{fmfgraph*}(60,60)
			\fmfset{curly_len}{2mm}
			\fmfleft{l1} \fmfright{r1}
			\fmf{wboson,label=$\hat\W^+$,label.side=left,tension=3}{l1,v1}
			\fmf{wboson,label=$\;\hat\W^+$,label.side=left,tension=3}{v2,r1}
			\fmf{scalar,label=$G^+$,label.side=right,right}{v1,v2}
			\fmf{photon,label=$\A$,label.side=right,right}{v2,v1}
			\fmffreeze
		\end{fmfgraph*}
	\end{gathered}
\qquad
	\begin{gathered}
		\begin{fmfgraph*}(60,60)
			\fmfset{curly_len}{2mm}
			\fmfleft{l1} \fmfright{r1}
			\fmf{wboson,label=$\hat\W^+$,label.side=left,tension=3}{l1,v1}
			\fmf{wboson,label=$\;\hat\W^+$,label.side=left,tension=3}{v2,r1}
			\fmf{scalar,label=$G^+$,label.side=right,right}{v1,v2}
			\fmf{photon,label=$\Z$,label.side=right,right}{v2,v1}
			\fmffreeze
		\end{fmfgraph*}
	\end{gathered}
\qquad
	\begin{gathered}
		\begin{fmfgraph*}(60,60)
			\fmfset{curly_len}{2mm}
			\fmfleft{l1} \fmfright{r1}
			\fmf{wboson,label=$\hat\W^+$,label.side=left,tension=3}{l1,v1}
			\fmf{wboson,label=$\;\hat\W^+$,label.side=left,tension=3}{v2,r1}
			\fmf{scalar,label=$G^+$,label.side=right,right}{v1,v2}
			\fmf{dashes,label=$G^0$,label.side=right,right}{v2,v1}
			\fmffreeze
		\end{fmfgraph*}
	\end{gathered}
\qquad
	\begin{gathered}
		\begin{fmfgraph*}(60,60)
			\fmfset{curly_len}{2mm}
			\fmfleft{l1} \fmfright{r1}
			\fmf{wboson,label=$\hat\W^+$,label.side=left,tension=3}{l1,v1}
			\fmf{wboson,label=$\;\hat\W^+$,label.side=left,tension=3}{v2,r1}
			\fmf{scalar,label=$G^+$,label.side=right,right}{v1,v2}
			\fmf{dashes,label=$h$,label.side=right,right}{v2,v1}
			\fmffreeze
		\end{fmfgraph*}
	\end{gathered}
\qquad
	\begin{gathered}
		\begin{fmfgraph*}(60,60)
			\fmfset{curly_len}{2mm}
			\fmfleft{l1} \fmfright{r1}
			\fmf{wboson,label=$\hat\W^+$,label.side=left,tension=3}{l1,v1}
			\fmf{wboson,label=$\;\hat\W^+$,label.side=left,tension=3}{v2,r1}
			\fmf{ghost,label=$\eta_+$,label.side=right,right}{v1,v2}
			\fmf{ghost,label=$\eta_A$,label.side=right,right}{v2,v1}
			\fmffreeze
		\end{fmfgraph*}
	\end{gathered}
\nnw
	& \begin{gathered}
		\begin{fmfgraph*}(60,60)
			\fmfset{curly_len}{2mm}
			\fmfleft{l1} \fmfright{r1}
			\fmf{wboson,label=$\hat\W^+$,label.side=left,tension=3}{l1,v1}
			\fmf{wboson,label=$\;\hat\W^+$,label.side=left,tension=3}{v2,r1}
			\fmf{ghost,label=$\eta_+$,label.side=right,right}{v1,v2}
			\fmf{ghost,label=$\eta_Z$,label.side=right,right}{v2,v1}
			\fmffreeze
		\end{fmfgraph*}
	\end{gathered}
\qquad
	\begin{gathered}
		\begin{fmfgraph*}(60,60)
			\fmfset{curly_len}{2mm}
			\fmfleft{l1} \fmfright{r1}
			\fmf{wboson,label=$\hat\W^+$,label.side=left,tension=3}{l1,v1}
			\fmf{wboson,label=$\;\hat\W^+$,label.side=left,tension=3}{v2,r1}
			\fmf{ghost,label=$\eta_A$,label.side=right,right}{v1,v2}
			\fmf{ghost,label=$\eta_-$,label.side=right,right}{v2,v1}
			\fmffreeze
		\end{fmfgraph*}
	\end{gathered}
\qquad
	\begin{gathered}
		\begin{fmfgraph*}(60,60)
			\fmfset{curly_len}{2mm}
			\fmfleft{l1} \fmfright{r1}
			\fmf{wboson,label=$\hat\W^+$,label.side=left,tension=3}{l1,v1}
			\fmf{wboson,label=$\;\hat\W^+$,label.side=left,tension=3}{v2,r1}
			\fmf{ghost,label=$\eta_Z$,label.side=right,right}{v1,v2}
			\fmf{ghost,label=$\eta_-$,label.side=right,right}{v2,v1}
			\fmffreeze
		\end{fmfgraph*}
	\end{gathered}
\qquad
	\begin{gathered}
		\begin{fmfgraph*}(60,60)
			\fmfset{curly_len}{2mm}
			\fmfleft{l1} \fmfright{r1} \fmftop{t1}
			\fmf{wboson,label=$\hat\W^+$,label.side=right,tension=3}{l1,v1}
			\fmf{wboson,label=$\hat\W^+$,label.side=right,tension=3}{v1,r1}
			\fmffreeze
			\fmf{phantom,tension=5}{t1,v2}
			\fmf{phantom,label=$t$,label.side=right,tension=0.8}{v1,v1}
			\fmf{quark,tension=0.5,right=1}{v1,v2,v1}
			\fmffreeze
		\end{fmfgraph*}
	\end{gathered}
\qquad
	\begin{gathered}
		\begin{fmfgraph*}(60,60)
			\fmfset{curly_len}{2mm}
			\fmfleft{l1} \fmfright{r1} \fmftop{t1}
			\fmf{wboson,label=$\hat\W^+$,label.side=right,tension=3}{l1,v1}
			\fmf{wboson,label=$\hat\W^+$,label.side=right,tension=3}{v1,r1}
			\fmffreeze
			\fmf{phantom,tension=5}{t1,v2}
			\fmf{phantom,label=$\W^+$,label.side=right,tension=0.8}{v1,v1}
			\fmf{wboson,tension=0.5,right=1}{v1,v2,v1}
			\fmffreeze
		\end{fmfgraph*}
	\end{gathered}
\nnw
	& \begin{gathered}
		\begin{fmfgraph*}(60,60)
			\fmfset{curly_len}{2mm}
			\fmfleft{l1} \fmfright{r1} \fmftop{t1}
			\fmf{wboson,label=$\hat\W^+$,label.side=right,tension=3}{l1,v1}
			\fmf{wboson,label=$\hat\W^+$,label.side=right,tension=3}{v1,r1}
			\fmffreeze
			\fmf{phantom,tension=5}{t1,v2}
			\fmf{phantom,label=$G^+$,label.side=right,tension=0.8}{v1,v1}
			\fmf{scalar,tension=0.5,right=1}{v1,v2,v1}
			\fmffreeze
		\end{fmfgraph*}
	\end{gathered}
\qquad
	\begin{gathered}
		\begin{fmfgraph*}(60,60)
			\fmfset{curly_len}{2mm}
			\fmfleft{l1} \fmfright{r1} \fmftop{t1}
			\fmf{wboson,label=$\hat\W^+$,label.side=right,tension=3}{l1,v1}
			\fmf{wboson,label=$\hat\W^+$,label.side=right,tension=3}{v1,r1}
			\fmffreeze
			\fmf{phantom,tension=5}{t1,v2}
			\fmf{phantom,label=$\eta_\pm$,label.side=right,tension=0.8}{v1,v1}
			\fmf{ghost,tension=0.5,right=1}{v1,v2,v1}
			\fmffreeze
		\end{fmfgraph*}
	\end{gathered}
\qquad
	\begin{gathered}
		\begin{fmfgraph*}(60,60)
			\fmfset{curly_len}{2mm}
			\fmfleft{l1} \fmfright{r1} \fmftop{t1}
			\fmf{wboson,label=$\hat\W^+$,label.side=right,tension=3}{l1,v1}
			\fmf{wboson,label=$\hat\W^+$,label.side=right,tension=3}{v1,r1}
			\fmffreeze
			\fmf{phantom,tension=5}{t1,v2}
			\fmf{phantom,label=$\Z$,label.side=right,tension=0.8}{v1,v1}
			\fmf{photon,tension=0.5,right=1}{v1,v2,v1}
			\fmffreeze
		\end{fmfgraph*}
	\end{gathered}
\qquad
	\begin{gathered}
		\begin{fmfgraph*}(60,60)
			\fmfset{curly_len}{2mm}
			\fmfleft{l1} \fmfright{r1} \fmftop{t1}
			\fmf{wboson,label=$\hat\W^+$,label.side=right,tension=3}{l1,v1}
			\fmf{wboson,label=$\hat\W^+$,label.side=right,tension=3}{v1,r1}
			\fmffreeze
			\fmf{phantom,tension=5}{t1,v2}
			\fmf{phantom,label=$G^0$,label.side=right,tension=0.8}{v1,v1}
			\fmf{dashes,tension=0.5,right=1}{v1,v2,v1}
		\end{fmfgraph*}
	\end{gathered}
\qquad
	\begin{gathered}
		\begin{fmfgraph*}(60,60)
			\fmfset{curly_len}{2mm}
			\fmfleft{l1} \fmfright{r1} \fmftop{t1}
			\fmf{wboson,label=$\hat\W^+$,label.side=right,tension=3}{l1,v1}
			\fmf{wboson,label=$\hat\W^+$,label.side=right,tension=3}{v1,r1}
			\fmffreeze
			\fmf{phantom,tension=5}{t1,v2}
			\fmf{phantom,label=$\eta_Z$,label.side=right,tension=0.8}{v1,v1}
			\fmf{ghost,tension=0.5,right=1}{v1,v2,v1}
			\fmffreeze
		\end{fmfgraph*}
	\end{gathered}
\nn
	& \begin{gathered}
		\begin{fmfgraph*}(60,60)
			\fmfset{curly_len}{2mm}
			\fmfleft{l1} \fmfright{r1} \fmftop{t1}
			\fmf{wboson,label=$\hat\W^+$,label.side=right,tension=3}{l1,v1}
			\fmf{wboson,label=$\hat\W^+$,label.side=right,tension=3}{v1,r1}
			\fmffreeze
			\fmf{phantom,tension=5}{t1,v2}
			\fmf{phantom,label=$h$,label.side=right,tension=0.8}{v1,v1}
			\fmf{dashes,tension=0.5,right=1}{v1,v2,v1}
		\end{fmfgraph*}
	\end{gathered}
\qquad
	\begin{gathered}
		\begin{fmfgraph*}(60,60)
			\fmfset{curly_len}{2mm}
			\fmfleft{l1} \fmfright{r1} \fmftop{t1}
			\fmf{wboson,label=$\hat\W^+$,label.side=right,tension=3}{l1,v1}
			\fmf{wboson,label=$\hat\W^+$,label.side=right,tension=3}{v1,r1}
			\fmffreeze
			\fmf{dashes,label=$\hat h$}{v1,v2}
			\fmf{phantom,tension=5}{t1,v2}
			\fmfblob{5mm}{v2}
		\end{fmfgraph*}
	\end{gathered}
\end{align*}

\subsubsection[$g^2$]{\boldmath$g^2$}

\begin{align*}
	&\begin{gathered}
		\begin{fmfgraph*}(60,60)
			\fmfset{curly_len}{2mm}
			\fmfleft{l1} \fmfright{r1}
			\fmf{gluon,label=$\hat\G$,label.side=left,tension=3}{l1,v1}
			\fmf{gluon,label=$\hat\G$,label.side=left,tension=3}{v2,r1}
			\fmf{quark,label=$t$,label.side=right,right}{v1,v2,v1}
			\fmffreeze
		\end{fmfgraph*}
	\end{gathered}
\qquad
	\begin{gathered}
		\begin{fmfgraph*}(60,60)
			\fmfset{curly_len}{2mm}
			\fmfleft{l1} \fmfright{r1} \fmftop{t1}
			\fmf{gluon,label=$\hat\G$,label.side=right,tension=3}{l1,v1}
			\fmf{gluon,label=$\hat\G$,label.side=right,tension=3}{v1,r1}
			\fmffreeze
			\fmf{phantom,tension=5}{t1,v2}
			\fmf{phantom,label=$t$,label.side=right,tension=0.8}{v1,v1}
			\fmf{quark,tension=0.5,right=1}{v1,v2,v1}
			\fmfv{decoration.shape=square,decoration.size=1.5mm}{v1}
		\end{fmfgraph*}
	\end{gathered}
\qquad
	\begin{gathered}
		\begin{fmfgraph*}(60,60)
			\fmfset{curly_len}{2mm}
			\fmfleft{l1} \fmfright{r1} \fmftop{t1}
			\fmf{gluon,label=$\hat\G$,label.side=right,tension=3}{l1,v1}
			\fmf{gluon,label=$\hat\G$,label.side=right,tension=3}{v1,r1}
			\fmffreeze
			\fmf{phantom,tension=5}{t1,v2}
			\fmf{phantom,label=$G^+$,label.side=right,tension=0.8}{v1,v1}
			\fmf{scalar,tension=0.5,right=1}{v1,v2,v1}
			\fmfv{decoration.shape=square,decoration.size=1.5mm}{v1}
		\end{fmfgraph*}
	\end{gathered}
\qquad
	\begin{gathered}
		\begin{fmfgraph*}(60,60)
			\fmfset{curly_len}{2mm}
			\fmfleft{l1} \fmfright{r1} \fmftop{t1}
			\fmf{gluon,label=$\hat\G$,label.side=right,tension=3}{l1,v1}
			\fmf{gluon,label=$\hat\G$,label.side=right,tension=3}{v1,r1}
			\fmffreeze
			\fmf{phantom,tension=5}{t1,v2}
			\fmf{phantom,label=$G^0$,label.side=right,tension=0.8}{v1,v1}
			\fmf{dashes,tension=0.5,right=1}{v1,v2,v1}
			\fmfv{decoration.shape=square,decoration.size=1.5mm}{v1}
		\end{fmfgraph*}
	\end{gathered}
\qquad
	\begin{gathered}
		\begin{fmfgraph*}(60,60)
			\fmfset{curly_len}{2mm}
			\fmfleft{l1} \fmfright{r1} \fmftop{t1}
			\fmf{gluon,label=$\hat\G$,label.side=right,tension=3}{l1,v1}
			\fmf{gluon,label=$\hat\G$,label.side=right,tension=3}{v1,r1}
			\fmffreeze
			\fmf{phantom,tension=5}{t1,v2}
			\fmf{phantom,label=$h$,label.side=right,tension=0.8}{v1,v1}
			\fmf{dashes,tension=0.5,right=1}{v1,v2,v1}
			\fmfv{decoration.shape=square,decoration.size=1.5mm}{v1}
		\end{fmfgraph*}
	\end{gathered}
\nnw
	& \begin{gathered}
		\begin{fmfgraph*}(60,60)
			\fmfset{curly_len}{2mm}
			\fmfleft{l1} \fmfright{r1} \fmftop{t1}
			\fmf{gluon,label=$\hat\G$,label.side=right,tension=3}{l1,v1}
			\fmf{gluon,label=$\hat\G$,label.side=right,tension=3}{v1,r1}
			\fmffreeze
			\fmf{dashes,label=$\hat h$}{v1,v2}
			\fmf{phantom,tension=5}{t1,v2}
			\fmfv{decoration.shape=square,decoration.size=1.5mm}{v1}
			\fmfblob{5mm}{v2}
		\end{fmfgraph*}
	\end{gathered}
\end{align*}


\subsubsection[$\bar e e$]{\boldmath$\bar e e$}

\begin{align*}
	&\begin{gathered}
		\begin{fmfgraph*}(60,60)
			\fmfset{curly_len}{2mm}
			\fmfleft{l1} \fmfright{r1}
			\fmf{quark,label=$\hat e_r$,label.side=left,tension=3}{l1,v1}
			\fmf{quark,label=$\hat e_p$,label.side=left,tension=3}{v2,r1}
			\fmf{quark,label=$e_u$,label.side=right,right}{v1,v2}
			\fmf{photon,label=$\Z$,left}{v1,v2}
			\fmffreeze
		\end{fmfgraph*}
	\end{gathered}
\qquad
	\begin{gathered}
		\begin{fmfgraph*}(60,60)
			\fmfset{curly_len}{2mm}
			\fmfleft{l1} \fmfright{r1}
			\fmf{quark,label=$\hat e_r$,label.side=left,tension=3}{l1,v1}
			\fmf{quark,label=$\hat e_p$,label.side=left,tension=3}{v2,r1}
			\fmf{quark,label=$e_u$,label.side=right,right}{v1,v2}
			\fmf{dashes,label=$G^0$,left}{v1,v2}
			\fmffreeze
		\end{fmfgraph*}
	\end{gathered}
\qquad
	\begin{gathered}
		\begin{fmfgraph*}(60,60)
			\fmfset{curly_len}{2mm}
			\fmfleft{l1} \fmfright{r1}
			\fmf{quark,label=$\hat e_r$,label.side=left,tension=3}{l1,v1}
			\fmf{quark,label=$\hat e_p$,label.side=left,tension=3}{v2,r1}
			\fmf{quark,label=$e_u$,label.side=right,right}{v1,v2}
			\fmf{dashes,label=$h$,left}{v1,v2}
			\fmffreeze
		\end{fmfgraph*}
	\end{gathered}
\qquad
	\begin{gathered}
		\begin{fmfgraph*}(60,60)
			\fmfset{curly_len}{2mm}
			\fmfleft{l1} \fmfright{r1}
			\fmf{quark,label=$\hat e_r$,label.side=left,tension=3}{l1,v1}
			\fmf{quark,label=$\hat e_p$,label.side=left,tension=3}{v2,r1}
			\fmf{plain,label=$\nu_u$,label.side=right,right}{v1,v2}
			\fmf{wboson,label=$\W^+$,right}{v2,v1}
			\fmffreeze
		\end{fmfgraph*}
	\end{gathered}
\qquad
	\begin{gathered}
		\begin{fmfgraph*}(60,60)
			\fmfset{curly_len}{2mm}
			\fmfleft{l1} \fmfright{r1}
			\fmf{quark,label=$\hat e_r$,label.side=left,tension=3}{l1,v1}
			\fmf{quark,label=$\hat e_p$,label.side=left,tension=3}{v2,r1}
			\fmf{plain,label=$\nu_u$,label.side=right,right}{v1,v2}
			\fmf{scalar,label=$G^+$,right}{v2,v1}
			\fmffreeze
		\end{fmfgraph*}
	\end{gathered}
\nnw
	&\begin{gathered}
		\begin{fmfgraph*}(60,60)
			\fmfset{curly_len}{2mm}
			\fmfleft{l1} \fmfright{r1} \fmftop{t1}
			\fmf{quark,label=$\hat e_r$,label.side=right,tension=3}{l1,v1}
			\fmf{quark,label=$\hat e_p$,label.side=right,tension=3}{v1,r1}
			\fmffreeze
			\fmf{phantom,tension=5}{t1,v2}
			\fmf{phantom,label=$t$,label.side=right,tension=0.8}{v1,v1}
			\fmf{quark,tension=0.5,right=1}{v1,v2,v1}
			\fmfv{decoration.shape=square,decoration.size=1.5mm}{v1}
		\end{fmfgraph*}
	\end{gathered}
\qquad
	\begin{gathered}
		\begin{fmfgraph*}(60,60)
			\fmfset{curly_len}{2mm}
			\fmfleft{l1} \fmfright{r1} \fmftop{t1}
			\fmf{quark,label=$\hat e_r$,label.side=right,tension=3}{l1,v1}
			\fmf{quark,label=$\hat e_p$,label.side=right,tension=3}{v1,r1}
			\fmffreeze
			\fmf{phantom,tension=5}{t1,v2}
			\fmf{phantom,label=$\W^+$,label.side=right,tension=0.8}{v1,v1}
			\fmf{wboson,tension=0.5,right=1}{v1,v2,v1}
			\fmfv{decoration.shape=square,decoration.size=1.5mm}{v1}
		\end{fmfgraph*}
	\end{gathered}
\qquad
	\begin{gathered}
		\begin{fmfgraph*}(60,60)
			\fmfset{curly_len}{2mm}
			\fmfleft{l1} \fmfright{r1} \fmftop{t1}
			\fmf{quark,label=$\hat e_r$,label.side=right,tension=3}{l1,v1}
			\fmf{quark,label=$\hat e_p$,label.side=right,tension=3}{v1,r1}
			\fmffreeze
			\fmf{phantom,tension=5}{t1,v2}
			\fmf{phantom,label=$G^+$,label.side=right,tension=0.8}{v1,v1}
			\fmf{scalar,tension=0.5,right=1}{v1,v2,v1}
			\fmfv{decoration.shape=square,decoration.size=1.5mm}{v1}
		\end{fmfgraph*}
	\end{gathered}
\qquad
	\begin{gathered}
		\begin{fmfgraph*}(60,60)
			\fmfset{curly_len}{2mm}
			\fmfleft{l1} \fmfright{r1} \fmftop{t1}
			\fmf{quark,label=$\hat e_r$,label.side=right,tension=3}{l1,v1}
			\fmf{quark,label=$\hat e_p$,label.side=right,tension=3}{v1,r1}
			\fmffreeze
			\fmf{phantom,tension=5}{t1,v2}
			\fmf{phantom,label=$G^0$,label.side=right,tension=0.8}{v1,v1}
			\fmf{dashes,tension=0.5,right=1}{v1,v2,v1}
			\fmfv{decoration.shape=square,decoration.size=1.5mm}{v1}
		\end{fmfgraph*}
	\end{gathered}
\qquad
	\begin{gathered}
		\begin{fmfgraph*}(60,60)
			\fmfset{curly_len}{2mm}
			\fmfleft{l1} \fmfright{r1} \fmftop{t1}
			\fmf{quark,label=$\hat e_r$,label.side=right,tension=3}{l1,v1}
			\fmf{quark,label=$\hat e_p$,label.side=right,tension=3}{v1,r1}
			\fmffreeze
			\fmf{phantom,tension=5}{t1,v2}
			\fmf{phantom,label=$h$,label.side=right,tension=0.8}{v1,v1}
			\fmf{dashes,tension=0.5,right=1}{v1,v2,v1}
			\fmfv{decoration.shape=square,decoration.size=1.5mm}{v1}
		\end{fmfgraph*}
	\end{gathered}
\nn
	& \begin{gathered}
		\begin{fmfgraph*}(60,60)
			\fmfset{curly_len}{2mm}
			\fmfleft{l1} \fmfright{r1} \fmftop{t1}
			\fmf{quark,label=$\hat e_r$,label.side=right,tension=3}{l1,v1}
			\fmf{quark,label=$\hat e_p$,label.side=right,tension=3}{v1,r1}
			\fmffreeze
			\fmf{dashes,label=$\hat h$}{v1,v2}
			\fmf{phantom,tension=5}{t1,v2}
			\fmfblob{5mm}{v2}
		\end{fmfgraph*}
	\end{gathered}
\end{align*}


\subsubsection[$\nu^2$]{\boldmath$\nu^2$}

\begin{align*}
	&\begin{gathered}
		\begin{fmfgraph*}(60,60)
			\fmfset{curly_len}{2mm}
			\fmfleft{l1} \fmfright{r1}
			\fmf{plain,label=$\hat \nu_r$,label.side=left,tension=3}{l1,v1}
			\fmf{plain,label=$\hat \nu_p$,label.side=left,tension=3}{v2,r1}
			\fmf{plain,label=$\nu_u$,label.side=right,right}{v1,v2}
			\fmf{photon,label=$\Z$,left}{v1,v2}
			\fmffreeze
		\end{fmfgraph*}
	\end{gathered}
\qquad
	\begin{gathered}
		\begin{fmfgraph*}(60,60)
			\fmfset{curly_len}{2mm}
			\fmfleft{l1} \fmfright{r1}
			\fmf{plain,label=$\hat \nu_r$,label.side=left,tension=3}{l1,v1}
			\fmf{plain,label=$\hat \nu_p$,label.side=left,tension=3}{v2,r1}
			\fmf{quark,label=$e_u$,label.side=right,right}{v1,v2}
			\fmf{wboson,label=$\W^+$,left}{v1,v2}
			\fmffreeze
		\end{fmfgraph*}
	\end{gathered}
\qquad
	\begin{gathered}
		\begin{fmfgraph*}(60,60)
			\fmfset{curly_len}{2mm}
			\fmfleft{l1} \fmfright{r1}
			\fmf{plain,label=$\hat \nu_r$,label.side=left,tension=3}{l1,v1}
			\fmf{plain,label=$\hat \nu_p$,label.side=left,tension=3}{v2,r1}
			\fmf{quark,label=$e_u$,label.side=left,left}{v2,v1}
			\fmf{wboson,label=$\W^+$,right}{v2,v1}
			\fmffreeze
		\end{fmfgraph*}
	\end{gathered}
\qquad
	\begin{gathered}
		\begin{fmfgraph*}(60,60)
			\fmfset{curly_len}{2mm}
			\fmfleft{l1} \fmfright{r1}
			\fmf{plain,label=$\hat \nu_r$,label.side=left,tension=3}{l1,v1}
			\fmf{plain,label=$\hat \nu_p$,label.side=left,tension=3}{v2,r1}
			\fmf{quark,label=$e_u$,label.side=right,right}{v1,v2}
			\fmf{scalar,label=$G^+$,left}{v1,v2}
			\fmffreeze
		\end{fmfgraph*}
	\end{gathered}
\qquad
	\begin{gathered}
		\begin{fmfgraph*}(60,60)
			\fmfset{curly_len}{2mm}
			\fmfleft{l1} \fmfright{r1}
			\fmf{plain,label=$\hat \nu_r$,label.side=left,tension=3}{l1,v1}
			\fmf{plain,label=$\hat \nu_p$,label.side=left,tension=3}{v2,r1}
			\fmf{quark,label=$e_u$,label.side=left,left}{v2,v1}
			\fmf{scalar,label=$G^+$,right}{v2,v1}
			\fmffreeze
		\end{fmfgraph*}
	\end{gathered}
\nnw
	& \begin{gathered}
		\begin{fmfgraph*}(60,60)
			\fmfset{curly_len}{2mm}
			\fmfleft{l1} \fmfright{r1}
			\fmf{plain,label=$\hat \nu_r$,label.side=left,tension=3}{l1,v1}
			\fmf{plain,label=$\hat \nu_p$,label.side=left,tension=3}{v2,r1}
			\fmf{plain,label=$\nu_u$,label.side=right,right}{v1,v2}
			\fmf{dashes,label=$G^0$,left}{v1,v2}
			\fmfv{decoration.shape=square,decoration.size=1.5mm}{v1,v2}
		\end{fmfgraph*}
	\end{gathered}
\qquad
	\begin{gathered}
		\begin{fmfgraph*}(60,60)
			\fmfset{curly_len}{2mm}
			\fmfleft{l1} \fmfright{r1}
			\fmf{plain,label=$\hat \nu_r$,label.side=left,tension=3}{l1,v1}
			\fmf{plain,label=$\hat \nu_p$,label.side=left,tension=3}{v2,r1}
			\fmf{plain,label=$\nu_u$,label.side=right,right}{v1,v2}
			\fmf{dashes,label=$h$,left}{v1,v2}
			\fmfv{decoration.shape=square,decoration.size=1.5mm}{v1,v2}
		\end{fmfgraph*}
	\end{gathered}
\qquad
	\begin{gathered}
		\begin{fmfgraph*}(60,60)
			\fmfset{curly_len}{2mm}
			\fmfleft{l1} \fmfright{r1} \fmftop{t1}
			\fmf{plain,label=$\hat \nu_r$,label.side=right,tension=3}{l1,v1}
			\fmf{plain,label=$\hat \nu_p$,label.side=right,tension=3}{v1,r1}
			\fmffreeze
			\fmf{phantom,tension=5}{t1,v2}
			\fmf{phantom,label=$t$,label.side=right,tension=0.8}{v1,v1}
			\fmf{quark,tension=0.5,right=1}{v1,v2,v1}
			\fmfv{decoration.shape=square,decoration.size=1.5mm}{v1}
			\fmfv{decoration.shape=square,decoration.size=1.5mm}{v1}
		\end{fmfgraph*}
	\end{gathered}
\qquad
	\begin{gathered}
		\begin{fmfgraph*}(60,60)
			\fmfset{curly_len}{2mm}
			\fmfleft{l1} \fmfright{r1} \fmftop{t1}
			\fmf{plain,label=$\hat \nu_r$,label.side=right,tension=3}{l1,v1}
			\fmf{plain,label=$\hat \nu_p$,label.side=right,tension=3}{v1,r1}
			\fmffreeze
			\fmf{phantom,tension=5}{t1,v2}
			\fmf{phantom,label=$G^+$,label.side=right,tension=0.8}{v1,v1}
			\fmf{scalar,tension=0.5,right=1}{v1,v2,v1}
			\fmfv{decoration.shape=square,decoration.size=1.5mm}{v1}
		\end{fmfgraph*}
	\end{gathered}
\qquad
	\begin{gathered}
		\begin{fmfgraph*}(60,60)
			\fmfset{curly_len}{2mm}
			\fmfleft{l1} \fmfright{r1} \fmftop{t1}
			\fmf{plain,label=$\hat \nu_r$,label.side=right,tension=3}{l1,v1}
			\fmf{plain,label=$\hat \nu_p$,label.side=right,tension=3}{v1,r1}
			\fmffreeze
			\fmf{phantom,tension=5}{t1,v2}
			\fmf{phantom,label=$G^0$,label.side=right,tension=0.8}{v1,v1}
			\fmf{dashes,tension=0.5,right=1}{v1,v2,v1}
			\fmfv{decoration.shape=square,decoration.size=1.5mm}{v1}
		\end{fmfgraph*}
	\end{gathered}
\nnw
	& \begin{gathered}
		\begin{fmfgraph*}(60,60)
			\fmfset{curly_len}{2mm}
			\fmfleft{l1} \fmfright{r1} \fmftop{t1}
			\fmf{plain,label=$\hat \nu_r$,label.side=right,tension=3}{l1,v1}
			\fmf{plain,label=$\hat \nu_p$,label.side=right,tension=3}{v1,r1}
			\fmffreeze
			\fmf{phantom,tension=5}{t1,v2}
			\fmf{phantom,label=$h$,label.side=right,tension=0.8}{v1,v1}
			\fmf{dashes,tension=0.5,right=1}{v1,v2,v1}
			\fmfv{decoration.shape=square,decoration.size=1.5mm}{v1}
		\end{fmfgraph*}
	\end{gathered}
\qquad
	\begin{gathered}
		\begin{fmfgraph*}(60,60)
			\fmfset{curly_len}{2mm}
			\fmfleft{l1} \fmfright{r1} \fmftop{t1}
			\fmf{plain,label=$\hat \nu_r$,label.side=right,tension=3}{l1,v1}
			\fmf{plain,label=$\hat \nu_p$,label.side=right,tension=3}{v1,r1}
			\fmffreeze
			\fmf{dashes,label=$\hat h$}{v1,v2}
			\fmf{phantom,tension=5}{t1,v2}
			\fmfv{decoration.shape=square,decoration.size=1.5mm}{v1}
			\fmfblob{5mm}{v2}
		\end{fmfgraph*}
	\end{gathered}
\end{align*}
Note that the reversed diagrams can be easily obtained from the original diagrams~\cite{Denner:1992vza}:
\begin{align}
	D &= \bar u_p(p) \Gamma_{pr}(p) u_r(p) \, , \nn
	D^\text{rev} &= - \bar v_r(p) \Gamma_{rp}(-p) v_p(p) = \bar u_p(p) C \Gamma^T_{rp}(-p) C^{-1} u_r(p) \, ,
\end{align}
where the transposition acts in Dirac space and transposition in the flavor indices is given explicitly. We use the relations
\begin{align}
	C \mathds{1}^T C^{-1} &= \mathds{1} \, , \quad
	C \gamma_5^T C^{-1} = \gamma_5 \, , \quad
	C \gamma_\mu^T C^{-1} = -\gamma_\mu \, , \quad
	C (\gamma_\mu \gamma_5)^T C^{-1} = \gamma_\mu \gamma_5 \, .
\end{align}


\subsubsection[$\bar u u$]{\boldmath$\bar u u$}

\begin{align*}
	&\begin{gathered}
		\begin{fmfgraph*}(60,60)
			\fmfset{curly_len}{2mm}
			\fmfleft{l1} \fmfright{r1}
			\fmf{quark,label=$\hat u_r$,label.side=left,tension=3}{l1,v1}
			\fmf{quark,label=$\hat u_p$,label.side=left,tension=3}{v2,r1}
			\fmf{quark,label=$t$,label.side=right,right}{v1,v2}
			\fmf{photon,label=$\A$,left}{v1,v2}
			\fmffreeze
		\end{fmfgraph*}
	\end{gathered}
\qquad
	\begin{gathered}
		\begin{fmfgraph*}(60,60)
			\fmfset{curly_len}{2mm}
			\fmfleft{l1} \fmfright{r1}
			\fmf{quark,label=$\hat u_r$,label.side=left,tension=3}{l1,v1}
			\fmf{quark,label=$\hat u_p$,label.side=left,tension=3}{v2,r1}
			\fmf{quark,label=$t$,label.side=right,right}{v1,v2}
			\fmf{gluon,label=$\G$,left}{v1,v2}
			\fmffreeze
		\end{fmfgraph*}
	\end{gathered}
\qquad
	\begin{gathered}
		\begin{fmfgraph*}(60,60)
			\fmfset{curly_len}{2mm}
			\fmfleft{l1} \fmfright{r1}
			\fmf{quark,label=$\hat u_r$,label.side=left,tension=3}{l1,v1}
			\fmf{quark,label=$\hat u_p$,label.side=left,tension=3}{v2,r1}
			\fmf{quark,label=$u_u$,label.side=right,right}{v1,v2}
			\fmf{photon,label=$\Z$,left}{v1,v2}
			\fmffreeze
		\end{fmfgraph*}
	\end{gathered}
\qquad
	\begin{gathered}
		\begin{fmfgraph*}(60,60)
			\fmfset{curly_len}{2mm}
			\fmfleft{l1} \fmfright{r1}
			\fmf{quark,label=$\hat u_r$,label.side=left,tension=3}{l1,v1}
			\fmf{quark,label=$\hat u_p$,label.side=left,tension=3}{v2,r1}
			\fmf{quark,label=$u_u$,label.side=right,right}{v1,v2}
			\fmf{dashes,label=$G^0$,left}{v1,v2}
			\fmffreeze
		\end{fmfgraph*}
	\end{gathered}
\qquad
	\begin{gathered}
		\begin{fmfgraph*}(60,60)
			\fmfset{curly_len}{2mm}
			\fmfleft{l1} \fmfright{r1}
			\fmf{quark,label=$\hat u_r$,label.side=left,tension=3}{l1,v1}
			\fmf{quark,label=$\hat u_p$,label.side=left,tension=3}{v2,r1}
			\fmf{quark,label=$u_u$,label.side=right,right}{v1,v2}
			\fmf{dashes,label=$h$,left}{v1,v2}
			\fmffreeze
		\end{fmfgraph*}
	\end{gathered}
\nnw
	& \begin{gathered}
		\begin{fmfgraph*}(60,60)
			\fmfset{curly_len}{2mm}
			\fmfleft{l1} \fmfright{r1}
			\fmf{quark,label=$\hat u_r$,label.side=left,tension=3}{l1,v1}
			\fmf{quark,label=$\hat u_p$,label.side=left,tension=3}{v2,r1}
			\fmf{quark,label=$d_u$,label.side=right,right}{v1,v2}
			\fmf{wboson,label=$\W^+$,left}{v1,v2}
			\fmffreeze
		\end{fmfgraph*}
	\end{gathered}
\qquad
	\begin{gathered}
		\begin{fmfgraph*}(60,60)
			\fmfset{curly_len}{2mm}
			\fmfleft{l1} \fmfright{r1}
			\fmf{quark,label=$\hat u_r$,label.side=left,tension=3}{l1,v1}
			\fmf{quark,label=$\hat u_p$,label.side=left,tension=3}{v2,r1}
			\fmf{quark,label=$d_u$,label.side=right,right}{v1,v2}
			\fmf{scalar,label=$G^+$,left}{v1,v2}
			\fmffreeze
		\end{fmfgraph*}
	\end{gathered}
\qquad
	\begin{gathered}
		\begin{fmfgraph*}(60,60)
			\fmfset{curly_len}{2mm}
			\fmfleft{l1} \fmfright{r1} \fmftop{t1}
			\fmf{quark,label=$\hat u_r$,label.side=right,tension=3}{l1,v1}
			\fmf{quark,label=$\hat u_p$,label.side=right,tension=3}{v1,r1}
			\fmffreeze
			\fmf{phantom,tension=5}{t1,v2}
			\fmf{phantom,label=$t$,label.side=right,tension=0.8}{v1,v1}
			\fmf{quark,tension=0.5,right=1}{v1,v2,v1}
			\fmfv{decoration.shape=square,decoration.size=1.5mm}{v1}
		\end{fmfgraph*}
	\end{gathered}
\qquad
	\begin{gathered}
		\begin{fmfgraph*}(60,60)
			\fmfset{curly_len}{2mm}
			\fmfleft{l1} \fmfright{r1} \fmftop{t1}
			\fmf{quark,label=$\hat u_r$,label.side=right,tension=3}{l1,v1}
			\fmf{quark,label=$\hat u_p$,label.side=right,tension=3}{v1,r1}
			\fmffreeze
			\fmf{phantom,tension=5}{t1,v2}
			\fmf{phantom,label=$\W^+$,label.side=right,tension=0.8}{v1,v1}
			\fmf{wboson,tension=0.5,right=1}{v1,v2,v1}
			\fmfv{decoration.shape=square,decoration.size=1.5mm}{v1}
		\end{fmfgraph*}
	\end{gathered}
\qquad
	\begin{gathered}
		\begin{fmfgraph*}(60,60)
			\fmfset{curly_len}{2mm}
			\fmfleft{l1} \fmfright{r1} \fmftop{t1}
			\fmf{quark,label=$\hat u_r$,label.side=right,tension=3}{l1,v1}
			\fmf{quark,label=$\hat u_p$,label.side=right,tension=3}{v1,r1}
			\fmffreeze
			\fmf{phantom,tension=5}{t1,v2}
			\fmf{phantom,label=$G^+$,label.side=right,tension=0.8}{v1,v1}
			\fmf{scalar,tension=0.5,right=1}{v1,v2,v1}
			\fmfv{decoration.shape=square,decoration.size=1.5mm}{v1}
		\end{fmfgraph*}
	\end{gathered}
\nnw
	& \begin{gathered}
		\begin{fmfgraph*}(60,60)
			\fmfset{curly_len}{2mm}
			\fmfleft{l1} \fmfright{r1} \fmftop{t1}
			\fmf{quark,label=$\hat u_r$,label.side=right,tension=3}{l1,v1}
			\fmf{quark,label=$\hat u_p$,label.side=right,tension=3}{v1,r1}
			\fmffreeze
			\fmf{phantom,tension=5}{t1,v2}
			\fmf{phantom,label=$G^0$,label.side=right,tension=0.8}{v1,v1}
			\fmf{dashes,tension=0.5,right=1}{v1,v2,v1}
			\fmfv{decoration.shape=square,decoration.size=1.5mm}{v1}
		\end{fmfgraph*}
	\end{gathered}
\qquad
	\begin{gathered}
		\begin{fmfgraph*}(60,60)
			\fmfset{curly_len}{2mm}
			\fmfleft{l1} \fmfright{r1} \fmftop{t1}
			\fmf{quark,label=$\hat u_r$,label.side=right,tension=3}{l1,v1}
			\fmf{quark,label=$\hat u_p$,label.side=right,tension=3}{v1,r1}
			\fmffreeze
			\fmf{phantom,tension=5}{t1,v2}
			\fmf{phantom,label=$h$,label.side=right,tension=0.8}{v1,v1}
			\fmf{dashes,tension=0.5,right=1}{v1,v2,v1}
			\fmfv{decoration.shape=square,decoration.size=1.5mm}{v1}
		\end{fmfgraph*}
	\end{gathered}
\qquad
	\begin{gathered}
		\begin{fmfgraph*}(60,60)
			\fmfset{curly_len}{2mm}
			\fmfleft{l1} \fmfright{r1} \fmftop{t1}
			\fmf{quark,label=$\hat u_r$,label.side=right,tension=3}{l1,v1}
			\fmf{quark,label=$\hat u_p$,label.side=right,tension=3}{v1,r1}
			\fmffreeze
			\fmf{dashes,label=$\hat h$}{v1,v2}
			\fmf{phantom,tension=5}{t1,v2}
			\fmfblob{5mm}{v2}
		\end{fmfgraph*}
	\end{gathered}
\end{align*}


\subsubsection[$\bar d d$]{\boldmath$\bar d d$}

\begin{align*}
	&\begin{gathered}
		\begin{fmfgraph*}(60,60)
			\fmfset{curly_len}{2mm}
			\fmfleft{l1} \fmfright{r1}
			\fmf{quark,label=$\hat d_r$,label.side=left,tension=3}{l1,v1}
			\fmf{quark,label=$\hat d_p$,label.side=left,tension=3}{v2,r1}
			\fmf{quark,label=$d_u$,label.side=right,right}{v1,v2}
			\fmf{photon,label=$\Z$,left}{v1,v2}
			\fmffreeze
		\end{fmfgraph*}
	\end{gathered}
\qquad
	\begin{gathered}
		\begin{fmfgraph*}(60,60)
			\fmfset{curly_len}{2mm}
			\fmfleft{l1} \fmfright{r1}
			\fmf{quark,label=$\hat d_r$,label.side=left,tension=3}{l1,v1}
			\fmf{quark,label=$\hat d_p$,label.side=left,tension=3}{v2,r1}
			\fmf{quark,label=$d_u$,label.side=right,right}{v1,v2}
			\fmf{dashes,label=$G^0$,left}{v1,v2}
			\fmffreeze
		\end{fmfgraph*}
	\end{gathered}
\qquad
	\begin{gathered}
		\begin{fmfgraph*}(60,60)
			\fmfset{curly_len}{2mm}
			\fmfleft{l1} \fmfright{r1}
			\fmf{quark,label=$\hat d_r$,label.side=left,tension=3}{l1,v1}
			\fmf{quark,label=$\hat d_p$,label.side=left,tension=3}{v2,r1}
			\fmf{quark,label=$d_u$,label.side=right,right}{v1,v2}
			\fmf{dashes,label=$h$,left}{v1,v2}
			\fmffreeze
		\end{fmfgraph*}
	\end{gathered}
\qquad
	\begin{gathered}
		\begin{fmfgraph*}(60,60)
			\fmfset{curly_len}{2mm}
			\fmfleft{l1} \fmfright{r1}
			\fmf{quark,label=$\hat d_r$,label.side=left,tension=3}{l1,v1}
			\fmf{quark,label=$\hat d_p$,label.side=left,tension=3}{v2,r1}
			\fmf{quark,label=$u_u$,label.side=right,right}{v1,v2}
			\fmf{wboson,label=$\W^+$,right}{v2,v1}
			\fmffreeze
		\end{fmfgraph*}
	\end{gathered}
\qquad
	\begin{gathered}
		\begin{fmfgraph*}(60,60)
			\fmfset{curly_len}{2mm}
			\fmfleft{l1} \fmfright{r1}
			\fmf{quark,label=$\hat d_r$,label.side=left,tension=3}{l1,v1}
			\fmf{quark,label=$\hat d_p$,label.side=left,tension=3}{v2,r1}
			\fmf{quark,label=$u_u$,label.side=right,right}{v1,v2}
			\fmf{scalar,label=$G^+$,right}{v2,v1}
			\fmffreeze
		\end{fmfgraph*}
	\end{gathered}
\nnw
	& \begin{gathered}
		\begin{fmfgraph*}(60,60)
			\fmfset{curly_len}{2mm}
			\fmfleft{l1} \fmfright{r1} \fmftop{t1}
			\fmf{quark,label=$\hat d_r$,label.side=right,tension=3}{l1,v1}
			\fmf{quark,label=$\hat d_p$,label.side=right,tension=3}{v1,r1}
			\fmffreeze
			\fmf{phantom,tension=5}{t1,v2}
			\fmf{phantom,label=$t$,label.side=right,tension=0.8}{v1,v1}
			\fmf{quark,tension=0.5,right=1}{v1,v2,v1}
			\fmfv{decoration.shape=square,decoration.size=1.5mm}{v1}
		\end{fmfgraph*}
	\end{gathered}
\qquad
	\begin{gathered}
		\begin{fmfgraph*}(60,60)
			\fmfset{curly_len}{2mm}
			\fmfleft{l1} \fmfright{r1} \fmftop{t1}
			\fmf{quark,label=$\hat d_r$,label.side=right,tension=3}{l1,v1}
			\fmf{quark,label=$\hat d_p$,label.side=right,tension=3}{v1,r1}
			\fmffreeze
			\fmf{phantom,tension=5}{t1,v2}
			\fmf{phantom,label=$\W^+$,label.side=right,tension=0.8}{v1,v1}
			\fmf{wboson,tension=0.5,right=1}{v1,v2,v1}
			\fmfv{decoration.shape=square,decoration.size=1.5mm}{v1}
		\end{fmfgraph*}
	\end{gathered}
\qquad
	\begin{gathered}
		\begin{fmfgraph*}(60,60)
			\fmfset{curly_len}{2mm}
			\fmfleft{l1} \fmfright{r1} \fmftop{t1}
			\fmf{quark,label=$\hat d_r$,label.side=right,tension=3}{l1,v1}
			\fmf{quark,label=$\hat d_p$,label.side=right,tension=3}{v1,r1}
			\fmffreeze
			\fmf{phantom,tension=5}{t1,v2}
			\fmf{phantom,label=$G^+$,label.side=right,tension=0.8}{v1,v1}
			\fmf{scalar,tension=0.5,right=1}{v1,v2,v1}
			\fmfv{decoration.shape=square,decoration.size=1.5mm}{v1}
		\end{fmfgraph*}
	\end{gathered}
\qquad
	\begin{gathered}
		\begin{fmfgraph*}(60,60)
			\fmfset{curly_len}{2mm}
			\fmfleft{l1} \fmfright{r1} \fmftop{t1}
			\fmf{quark,label=$\hat d_r$,label.side=right,tension=3}{l1,v1}
			\fmf{quark,label=$\hat d_p$,label.side=right,tension=3}{v1,r1}
			\fmffreeze
			\fmf{phantom,tension=5}{t1,v2}
			\fmf{phantom,label=$G^0$,label.side=right,tension=0.8}{v1,v1}
			\fmf{dashes,tension=0.5,right=1}{v1,v2,v1}
			\fmfv{decoration.shape=square,decoration.size=1.5mm}{v1}
		\end{fmfgraph*}
	\end{gathered}
\qquad
	\begin{gathered}
		\begin{fmfgraph*}(60,60)
			\fmfset{curly_len}{2mm}
			\fmfleft{l1} \fmfright{r1} \fmftop{t1}
			\fmf{quark,label=$\hat d_r$,label.side=right,tension=3}{l1,v1}
			\fmf{quark,label=$\hat d_p$,label.side=right,tension=3}{v1,r1}
			\fmffreeze
			\fmf{phantom,tension=5}{t1,v2}
			\fmf{phantom,label=$h$,label.side=right,tension=0.8}{v1,v1}
			\fmf{dashes,tension=0.5,right=1}{v1,v2,v1}
			\fmfv{decoration.shape=square,decoration.size=1.5mm}{v1}
		\end{fmfgraph*}
	\end{gathered}
\nn
	& \begin{gathered}
		\begin{fmfgraph*}(60,60)
			\fmfset{curly_len}{2mm}
			\fmfleft{l1} \fmfright{r1} \fmftop{t1}
			\fmf{quark,label=$\hat d_r$,label.side=right,tension=3}{l1,v1}
			\fmf{quark,label=$\hat d_p$,label.side=right,tension=3}{v1,r1}
			\fmffreeze
			\fmf{dashes,label=$\hat h$}{v1,v2}
			\fmf{phantom,tension=5}{t1,v2}
			\fmfblob{5mm}{v2}
		\end{fmfgraph*}
	\end{gathered}
\end{align*}


\subsection{Two-point functions with momentum insertion}
\label{sec:TwoPointMomentumInsertion}


\subsubsection[$\gamma^2$]{\boldmath$\gamma^2$}

\begin{align*}
	&\begin{gathered}
		\begin{fmfgraph*}(60,60)
			\fmfset{curly_len}{2mm}
			\fmfleft{l1} \fmfright{r1} \fmftop{t1,t2,t3}
			\fmf{photon,label=$\hat\A$,label.side=right,tension=3}{l1,v1}
			\fmf{photon,label=$\hat\A$,label.side=right,tension=3}{v2,r1}
			\fmf{quark,label=$t$,label.side=right,right}{v1,v2,v1}
			\fmffreeze
			\fmfv{decoration.shape=square,decoration.size=1.5mm}{v1}
			\fmf{momins}{t1,v3}
			\fmf{phantom,tension=2}{v3,v1}
		\end{fmfgraph*}
	\end{gathered}
\qquad
	\begin{gathered}
		\begin{fmfgraph*}(60,60)
			\fmfset{curly_len}{2mm}
			\fmfleft{l1} \fmfright{r1} \fmftop{t1,t2,t3}
			\fmf{photon,label=$\hat\A$,label.side=right,tension=3}{l1,v1}
			\fmf{photon,label=$\hat\A$,label.side=right,tension=3}{v2,r1}
			\fmf{quark,label=$t$,label.side=right,right}{v1,v2,v1}
			\fmffreeze
			\fmfv{decoration.shape=square,decoration.size=1.5mm}{v2}
			\fmf{momins}{t3,v3}
			\fmf{phantom,tension=2}{v3,v2}
		\end{fmfgraph*}
	\end{gathered}
\qquad
	\begin{gathered}
		\begin{fmfgraph*}(60,60)
			\fmfset{curly_len}{2mm}
			\fmfleft{l1} \fmfright{r1} \fmftop{t1,t2,t3}
			\fmf{photon,label=$\hat\A$,label.side=right,tension=3}{l1,v1}
			\fmf{photon,label=$\hat\A$,label.side=right,tension=3}{v2,r1}
			\fmf{wboson,label=$\W^+$,label.side=right,right}{v1,v2,v1}
			\fmffreeze
			\fmfv{decoration.shape=square,decoration.size=1.5mm}{v1}
			\fmf{momins}{t1,v3}
			\fmf{phantom,tension=2}{v3,v1}
		\end{fmfgraph*}
	\end{gathered}
\qquad
	\begin{gathered}
		\begin{fmfgraph*}(60,60)
			\fmfset{curly_len}{2mm}
			\fmfleft{l1} \fmfright{r1} \fmftop{t1,t2,t3}
			\fmf{photon,label=$\hat\A$,label.side=right,tension=3}{l1,v1}
			\fmf{photon,label=$\hat\A$,label.side=right,tension=3}{v2,r1}
			\fmf{wboson,label=$\W^+$,label.side=right,right}{v1,v2,v1}
			\fmffreeze
			\fmfv{decoration.shape=square,decoration.size=1.5mm}{v2}
			\fmf{momins}{t3,v3}
			\fmf{phantom,tension=2}{v3,v2}
		\end{fmfgraph*}
	\end{gathered}
\qquad
	\begin{gathered}
		\begin{fmfgraph*}(60,60)
			\fmfset{curly_len}{2mm}
			\fmfleft{l1} \fmfright{r1} \fmftop{t1,t2,t3}
			\fmf{photon,label=$\hat\A$,label.side=right,tension=3}{l1,v1}
			\fmf{photon,label=$\hat\A$,label.side=right,tension=3}{v2,r1}
			\fmf{phantom,right}{v2,v1}
			\fmf{wboson,label=$\W^+$,label.side=right,right}{v1,v2}
			\fmffreeze
			\fmf{wboson,label=$\W^+$,label.side=right,right=0.5}{v2,v4,v1}
			\fmf{phantom,tension=1.75}{t2,v4}
			\fmfv{decoration.shape=square,decoration.size=1.5mm}{v4}
			\fmffreeze
			\fmf{momins}{t2,v3}
			\fmf{phantom,tension=2}{v3,v4}
		\end{fmfgraph*}
	\end{gathered}
\nn
\nn
	& \begin{gathered}
		\begin{fmfgraph*}(60,60)
			\fmfset{curly_len}{2mm}
			\fmfleft{l1} \fmfright{r1} \fmftop{t1,t2,t3}
			\fmf{photon,label=$\hat\A$,label.side=right,tension=3}{l1,v1}
			\fmf{photon,label=$\hat\A$,label.side=right,tension=3}{v2,r1}
			\fmf{phantom,right}{v2,v1}
			\fmf{wboson,label=$\W^+$,label.side=left,left}{v2,v1}
			\fmffreeze
			\fmf{wboson,label=$\W^+$,label.side=left,left=0.5}{v1,v4,v2}
			\fmf{phantom,tension=1.75}{t2,v4}
			\fmfv{decoration.shape=square,decoration.size=1.5mm}{v4}
			\fmffreeze
			\fmf{momins}{t2,v3}
			\fmf{phantom,tension=2}{v3,v4}
		\end{fmfgraph*}
	\end{gathered}
\qquad
	\begin{gathered}
		\begin{fmfgraph*}(60,60)
			\fmfset{curly_len}{2mm}
			\fmfleft{l1} \fmfright{r1} \fmftop{t1} \fmfbottom{b1}
			\fmf{photon,label=$\hat\A$,label.side=right,tension=3}{l1,v1}
			\fmf{photon,label=$\hat\A$,label.side=right,tension=3}{v1,r1}
			\fmffreeze
			\fmf{phantom,tension=5}{t1,v2}
			\fmf{phantom,label=$\W^+$,label.side=right,tension=0.7}{v1,v1}
			\fmf{wboson,tension=0.5,right=1}{v1,v2,v1}
			\fmffreeze
			\fmfv{decoration.shape=square,decoration.size=1.5mm}{v1}
			\fmf{momins}{b1,v3}
			\fmf{phantom,tension=2}{v3,v1}
		\end{fmfgraph*}
	\end{gathered}
\qquad
	\begin{gathered}
		\\[-2.2cm]
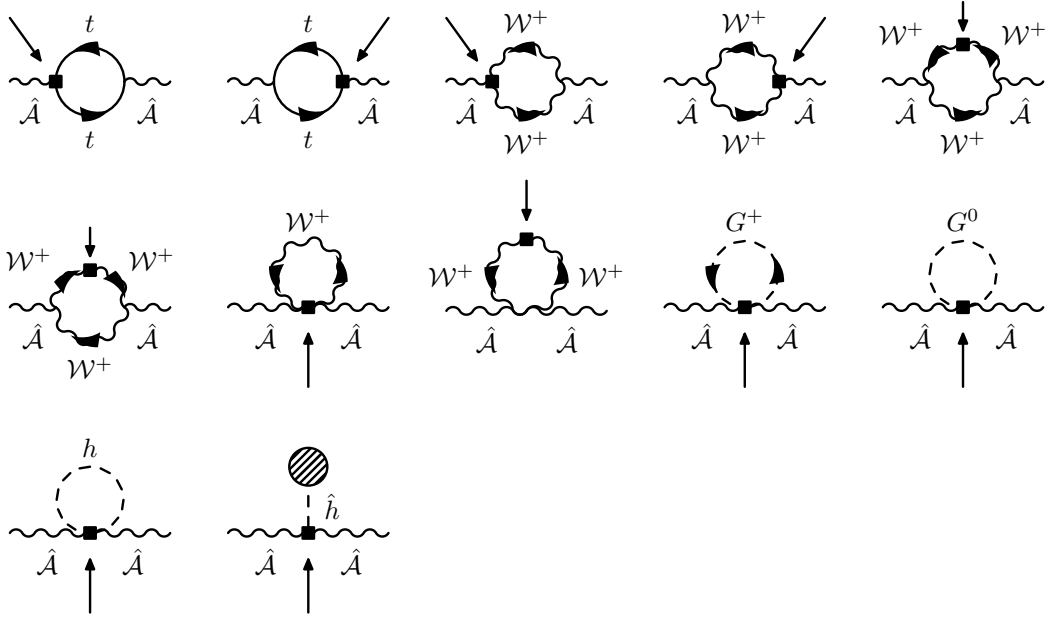
		\begin{fmfgraph*}(60,60)
			\fmfset{curly_len}{2mm}
			\fmfleft{l1} \fmfright{r1} \fmftop{t1} \fmfbottom{b1,b2,b3}
			\fmf{photon,label=$\hat\A$,label.side=right,tension=3}{b1,v1}
			\fmf{photon,label=$\hat\A$,label.side=right,tension=3}{v1,b3}
			\fmffreeze
			\fmf{phantom,tension=1}{t1,v2}
			\fmf{wboson,label=$\W^+$,label.side=right,tension=0.5,right=1}{v1,v2,v1}
			\fmffreeze
			\fmfv{decoration.shape=square,decoration.size=1.5mm}{v2}
			\fmffreeze
			\fmf{phantom,tension=3}{t1,v4}
			\fmf{momins}{v4,v3}
			\fmf{phantom,tension=2}{v3,v2}
		\end{fmfgraph*}
	\end{gathered}
\qquad
	\begin{gathered}
		\begin{fmfgraph*}(60,60)
			\fmfset{curly_len}{2mm}
			\fmfleft{l1} \fmfright{r1} \fmftop{t1} \fmfbottom{b1}
			\fmf{photon,label=$\hat\A$,label.side=right,tension=3}{l1,v1}
			\fmf{photon,label=$\hat\A$,label.side=right,tension=3}{v1,r1}
			\fmffreeze
			\fmf{phantom,tension=5}{t1,v2}
			\fmf{phantom,label=$G^+$,label.side=right,tension=0.7}{v1,v1}
			\fmf{scalar,tension=0.5,right=1}{v1,v2,v1}
			\fmffreeze
			\fmfv{decoration.shape=square,decoration.size=1.5mm}{v1}
			\fmf{momins}{b1,v3}
			\fmf{phantom,tension=2}{v3,v1}
		\end{fmfgraph*}
	\end{gathered}
\qquad
	\begin{gathered}
		\begin{fmfgraph*}(60,60)
			\fmfset{curly_len}{2mm}
			\fmfleft{l1} \fmfright{r1} \fmftop{t1} \fmfbottom{b1}
			\fmf{photon,label=$\hat\A$,label.side=right,tension=3}{l1,v1}
			\fmf{photon,label=$\hat\A$,label.side=right,tension=3}{v1,r1}
			\fmffreeze
			\fmf{phantom,tension=5}{t1,v2}
			\fmf{phantom,label=$G^0$,label.side=right,tension=0.7}{v1,v1}
			\fmf{dashes,tension=0.5,right=1}{v1,v2,v1}
			\fmffreeze
			\fmfv{decoration.shape=square,decoration.size=1.5mm}{v1}
			\fmf{momins}{b1,v3}
			\fmf{phantom,tension=2}{v3,v1}
		\end{fmfgraph*}
	\end{gathered}
\nn
\nn
	& \begin{gathered}
		\begin{fmfgraph*}(60,60)
			\fmfset{curly_len}{2mm}
			\fmfleft{l1} \fmfright{r1} \fmftop{t1} \fmfbottom{b1}
			\fmf{photon,label=$\hat\A$,label.side=right,tension=3}{l1,v1}
			\fmf{photon,label=$\hat\A$,label.side=right,tension=3}{v1,r1}
			\fmffreeze
			\fmf{phantom,tension=5}{t1,v2}
			\fmf{phantom,label=$h$,label.side=right,tension=0.7}{v1,v1}
			\fmf{dashes,tension=0.5,right=1}{v1,v2,v1}
			\fmffreeze
			\fmfv{decoration.shape=square,decoration.size=1.5mm}{v1}
			\fmf{momins}{b1,v3}
			\fmf{phantom,tension=2}{v3,v1}
		\end{fmfgraph*}
	\end{gathered}
\qquad
	\begin{gathered}
		\begin{fmfgraph*}(60,60)
			\fmfset{curly_len}{2mm}
			\fmfleft{l1} \fmfright{r1} \fmftop{t1} \fmfbottom{b1}
			\fmf{photon,label=$\hat\A$,label.side=right,tension=3}{l1,v1}
			\fmf{photon,label=$\hat\A$,label.side=right,tension=3}{v1,r1}
			\fmffreeze
			\fmf{dashes,label=$\hat h$}{v1,v2}
			\fmf{phantom,tension=5}{t1,v2}
			\fmfblob{5mm}{v2}
			\fmffreeze
			\fmfv{decoration.shape=square,decoration.size=1.5mm}{v1}
			\fmf{momins}{b1,v3}
			\fmf{phantom,tension=2}{v3,v1}
		\end{fmfgraph*}
	\end{gathered}
\end{align*}
There are no contributions from external-leg corrections to the $\A\widetilde\A$ and $\Z\widetilde\A$ theta terms~\eqref{eq:ThetaTerms}: the background-field method ensures that there is no mixing between $\A$ and $\Z$, and the external-leg corrections for the photon are compensated by the explicit inclusion of the squared coupling in the operator.


\subsubsection[$g^2$]{\boldmath$g^2$}

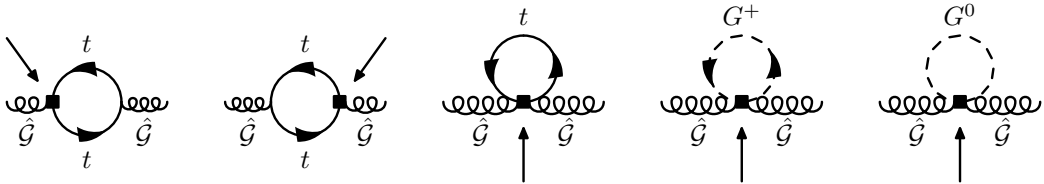
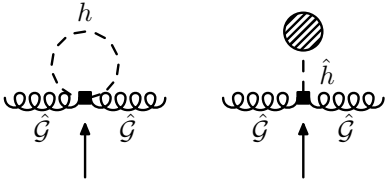
\begin{align*}
	&\begin{gathered}
		\begin{fmfgraph*}(60,60)
			\fmfset{curly_len}{2mm}
			\fmfleft{l1} \fmfright{r1} \fmftop{t1,t2,t3}
			\fmf{gluon,label=$\hat\G$,label.side=right,tension=3}{l1,v1}
			\fmf{gluon,label=$\hat\G$,label.side=right,tension=3}{v2,r1}
			\fmf{quark,label=$t$,label.side=right,right}{v1,v2,v1}
			\fmffreeze
			\fmfv{decoration.shape=square,decoration.size=1.5mm}{v1}
			\fmf{momins}{t1,v3}
			\fmf{phantom,tension=2}{v3,v1}
		\end{fmfgraph*}
	\end{gathered}
\qquad
	\begin{gathered}
		\begin{fmfgraph*}(60,60)
			\fmfset{curly_len}{2mm}
			\fmfleft{l1} \fmfright{r1} \fmftop{t1,t2,t3}
			\fmf{gluon,label=$\hat\G$,label.side=right,tension=3}{l1,v1}
			\fmf{gluon,label=$\hat\G$,label.side=right,tension=3}{v2,r1}
			\fmf{quark,label=$t$,label.side=right,right}{v1,v2,v1}
			\fmffreeze
			\fmfv{decoration.shape=square,decoration.size=1.5mm}{v2}
			\fmf{momins}{t3,v3}
			\fmf{phantom,tension=2}{v3,v2}
		\end{fmfgraph*}
	\end{gathered}
\qquad
	\begin{gathered}
		\begin{fmfgraph*}(60,60)
			\fmfset{curly_len}{2mm}
			\fmfleft{l1} \fmfright{r1} \fmftop{t1} \fmfbottom{b1}
			\fmf{gluon,label=$\hat\G$,label.side=right,tension=3}{l1,v1}
			\fmf{gluon,label=$\hat\G$,label.side=right,tension=3}{v1,r1}
			\fmffreeze
			\fmf{phantom,tension=5}{t1,v2}
			\fmf{phantom,label=$t$,label.side=right,tension=0.7}{v1,v1}
			\fmf{quark,tension=0.5,right=1}{v1,v2,v1}
			\fmfv{decoration.shape=square,decoration.size=1.5mm}{v1}
			\fmffreeze
			\fmfv{decoration.shape=square,decoration.size=1.5mm}{v1}
			\fmf{momins}{b1,v3}
			\fmf{phantom,tension=2}{v3,v1}
		\end{fmfgraph*}
	\end{gathered}
\qquad
	\begin{gathered}
		\begin{fmfgraph*}(60,60)
			\fmfset{curly_len}{2mm}
			\fmfleft{l1} \fmfright{r1} \fmftop{t1} \fmfbottom{b1}
			\fmf{gluon,label=$\hat\G$,label.side=right,tension=3}{l1,v1}
			\fmf{gluon,label=$\hat\G$,label.side=right,tension=3}{v1,r1}
			\fmffreeze
			\fmf{phantom,tension=5}{t1,v2}
			\fmf{phantom,label=$G^+$,label.side=right,tension=0.7}{v1,v1}
			\fmf{scalar,tension=0.5,right=1}{v1,v2,v1}
			\fmfv{decoration.shape=square,decoration.size=1.5mm}{v1}
			\fmffreeze
			\fmfv{decoration.shape=square,decoration.size=1.5mm}{v1}
			\fmf{momins}{b1,v3}
			\fmf{phantom,tension=2}{v3,v1}
		\end{fmfgraph*}
	\end{gathered}
\qquad
	\begin{gathered}
		\begin{fmfgraph*}(60,60)
			\fmfset{curly_len}{2mm}
			\fmfleft{l1} \fmfright{r1} \fmftop{t1} \fmfbottom{b1}
			\fmf{gluon,label=$\hat\G$,label.side=right,tension=3}{l1,v1}
			\fmf{gluon,label=$\hat\G$,label.side=right,tension=3}{v1,r1}
			\fmffreeze
			\fmf{phantom,tension=5}{t1,v2}
			\fmf{phantom,label=$G^0$,label.side=right,tension=0.7}{v1,v1}
			\fmf{dashes,tension=0.5,right=1}{v1,v2,v1}
			\fmfv{decoration.shape=square,decoration.size=1.5mm}{v1}
			\fmffreeze
			\fmfv{decoration.shape=square,decoration.size=1.5mm}{v1}
			\fmf{momins}{b1,v3}
			\fmf{phantom,tension=2}{v3,v1}
		\end{fmfgraph*}
	\end{gathered}
\nn
\nn
	& \begin{gathered}
		\begin{fmfgraph*}(60,60)
			\fmfset{curly_len}{2mm}
			\fmfleft{l1} \fmfright{r1} \fmftop{t1} \fmfbottom{b1}
			\fmf{gluon,label=$\hat\G$,label.side=right,tension=3}{l1,v1}
			\fmf{gluon,label=$\hat\G$,label.side=right,tension=3}{v1,r1}
			\fmffreeze
			\fmf{phantom,tension=5}{t1,v2}
			\fmf{phantom,label=$h$,label.side=right,tension=0.7}{v1,v1}
			\fmf{dashes,tension=0.5,right=1}{v1,v2,v1}
			\fmfv{decoration.shape=square,decoration.size=1.5mm}{v1}
			\fmffreeze
			\fmfv{decoration.shape=square,decoration.size=1.5mm}{v1}
			\fmf{momins}{b1,v3}
			\fmf{phantom,tension=2}{v3,v1}
		\end{fmfgraph*}
	\end{gathered}
\qquad
	\begin{gathered}
		\begin{fmfgraph*}(60,60)
			\fmfset{curly_len}{2mm}
			\fmfleft{l1} \fmfright{r1} \fmftop{t1} \fmfbottom{b1}
			\fmf{gluon,label=$\hat\G$,label.side=right,tension=3}{l1,v1}
			\fmf{gluon,label=$\hat\G$,label.side=right,tension=3}{v1,r1}
			\fmffreeze
			\fmf{dashes,label=$\hat h$}{v1,v2}
			\fmf{phantom,tension=5}{t1,v2}
			\fmfv{decoration.shape=square,decoration.size=1.5mm}{v1}
			\fmfblob{5mm}{v2}
			\fmffreeze
			\fmfv{decoration.shape=square,decoration.size=1.5mm}{v1}
			\fmf{momins}{b1,v3}
			\fmf{phantom,tension=2}{v3,v1}
		\end{fmfgraph*}
	\end{gathered}
\end{align*}
The additional one-loop contributions from external-leg corrections to the gluon theta term~\eqref{eq:ThetaTerms} are compensated by the explicit inclusion of the squared coupling in the operator.


\end{myfmf}
	
\begin{myfmf}{diags/threePoint}

\subsection{Three-point functions with massless bosons}
\label{sec:ThreePointDiagrams}

\subsubsection[$\bar e e \gamma$]{\boldmath$\bar e e \gamma$}

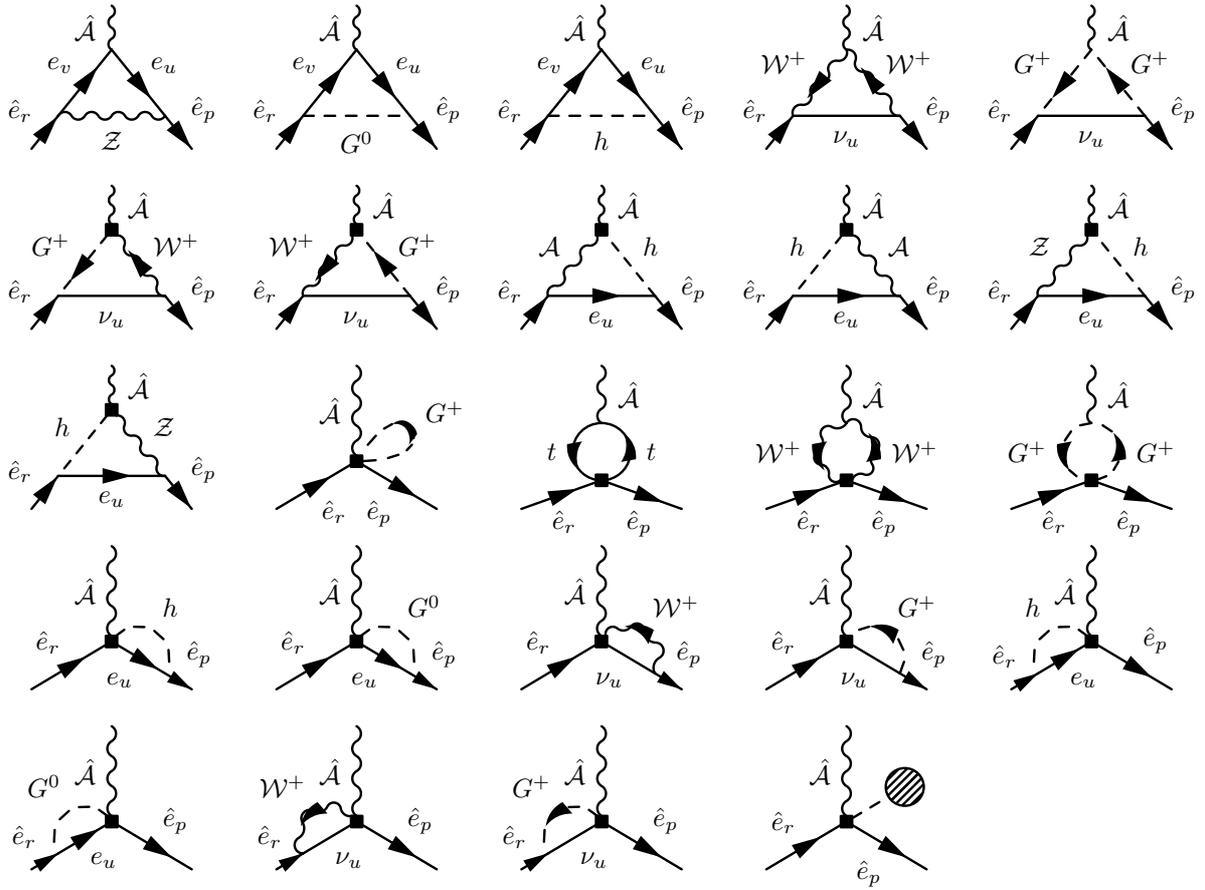
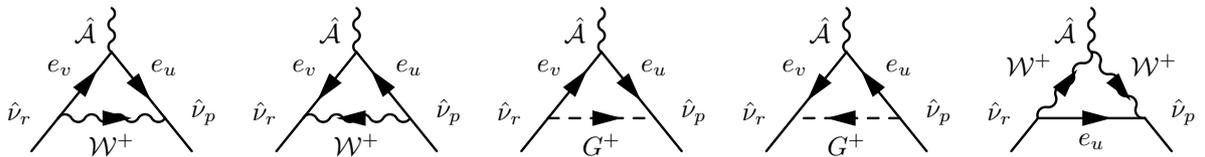
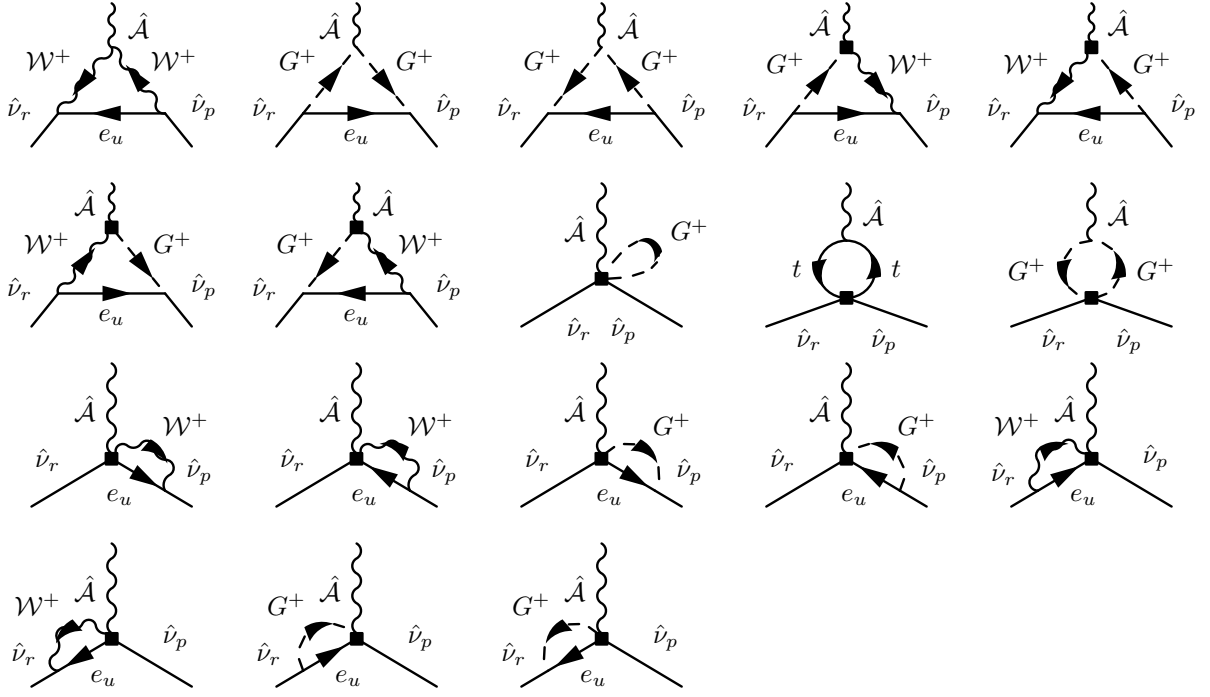
\begin{align*}
	&\begin{gathered}
		\begin{fmfgraph*}(60,60)
			\fmfset{curly_len}{2mm}
			\fmftop{t1} \fmfbottom{b1,b2}
			\fmf{quark,label=$\hat e_r$,label.side=left,tension=2}{b1,v1}
			\fmf{quark,label=$e_v$,label.side=left}{v1,v2}
			\fmf{quark,label=$e_u$,label.side=left}{v2,v3}
			\fmf{quark,label=$\hat e_p$,label.side=left,tension=2}{v3,b2}
			\fmf{photon,label=$\hat \A$,tension=3}{t1,v2}
			\fmffreeze
			\fmf{photon,label=$\Z$}{v1,v3}
		\end{fmfgraph*}
	\end{gathered}
\qquad\quad
	\begin{gathered}
		\begin{fmfgraph*}(60,60)
			\fmfset{curly_len}{2mm}
			\fmftop{t1} \fmfbottom{b1,b2}
			\fmf{quark,label=$\hat e_r$,label.side=left,tension=2}{b1,v1}
			\fmf{quark,label=$e_v$,label.side=left}{v1,v2}
			\fmf{quark,label=$e_u$,label.side=left}{v2,v3}
			\fmf{quark,label=$\hat e_p$,label.side=left,tension=2}{v3,b2}
			\fmf{photon,label=$\hat \A$,tension=3}{t1,v2}
			\fmffreeze
			\fmf{dashes,label=$G^0$}{v1,v3}
		\end{fmfgraph*}
	\end{gathered}
\qquad\quad
	\begin{gathered}
		\begin{fmfgraph*}(60,60)
			\fmfset{curly_len}{2mm}
			\fmftop{t1} \fmfbottom{b1,b2}
			\fmf{quark,label=$\hat e_r$,label.side=left,tension=2}{b1,v1}
			\fmf{quark,label=$e_v$,label.side=left}{v1,v2}
			\fmf{quark,label=$e_u$,label.side=left}{v2,v3}
			\fmf{quark,label=$\hat e_p$,label.side=left,tension=2}{v3,b2}
			\fmf{photon,label=$\hat \A$,tension=3}{t1,v2}
			\fmffreeze
			\fmf{dashes,label=$h$}{v1,v3}
		\end{fmfgraph*}
	\end{gathered}
\qquad\quad
	\begin{gathered}
		\begin{fmfgraph*}(60,60)
			\fmfset{curly_len}{2mm}
			\fmftop{t1} \fmfbottom{b1,b2}
			\fmf{quark,label=$\hat e_r$,label.side=left,tension=2}{b1,v1}
			\fmf{wboson,label=$\W^+$,label.side=right}{v3,v2,v1}
			\fmf{quark,label=$\hat e_p$,label.side=left,tension=2}{v3,b2}
			\fmf{photon,label=$\hat \A$,tension=3}{t1,v2}
			\fmffreeze
			\fmf{plain,label=$\nu_u$}{v1,v3}
		\end{fmfgraph*}
	\end{gathered}
\qquad\quad
	\begin{gathered}
		\begin{fmfgraph*}(60,60)
			\fmfset{curly_len}{2mm}
			\fmftop{t1} \fmfbottom{b1,b2}
			\fmf{quark,label=$\hat e_r$,label.side=left,tension=2}{b1,v1}
			\fmf{scalar,label=$G^+$,label.side=right}{v3,v2}
			\fmf{scalar,label=$G^+$,label.side=right}{v2,v1}
			\fmf{quark,label=$\hat e_p$,label.side=left,tension=2}{v3,b2}
			\fmf{photon,label=$\hat \A$,tension=3}{t1,v2}
			\fmffreeze
			\fmf{plain,label=$\nu_u$}{v1,v3}
		\end{fmfgraph*}
	\end{gathered}
\nn
	& \begin{gathered}
		\begin{fmfgraph*}(60,60)
			\fmfset{curly_len}{2mm}
			\fmftop{t1} \fmfbottom{b1,b2}
			\fmf{quark,label=$\hat e_r$,label.side=left,tension=2}{b1,v1}
			\fmf{wboson,label=$\W^+$,label.side=right}{v3,v2}
			\fmf{scalar,label=$G^+$,label.side=right}{v2,v1}
			\fmf{quark,label=$\hat e_p$,label.side=left,tension=2}{v3,b2}
			\fmf{photon,label=$\hat \A$,tension=3}{t1,v2}
			\fmffreeze
			\fmf{plain,label=$\nu_u$}{v1,v3}
			\fmfv{decoration.shape=square,decoration.size=1.5mm}{v2}
		\end{fmfgraph*}
	\end{gathered}
\qquad\quad
	\begin{gathered}
		\begin{fmfgraph*}(60,60)
			\fmfset{curly_len}{2mm}
			\fmftop{t1} \fmfbottom{b1,b2}
			\fmf{quark,label=$\hat e_r$,label.side=left,tension=2}{b1,v1}
			\fmf{scalar,label=$G^+$,label.side=right}{v3,v2}
			\fmf{wboson,label=$\W^+$,label.side=right}{v2,v1}
			\fmf{quark,label=$\hat e_p$,label.side=left,tension=2}{v3,b2}
			\fmf{photon,label=$\hat \A$,tension=3}{t1,v2}
			\fmffreeze
			\fmf{plain,label=$\nu_u$}{v1,v3}
			\fmfv{decoration.shape=square,decoration.size=1.5mm}{v2}
		\end{fmfgraph*}
	\end{gathered}
\qquad\quad
	\begin{gathered}
		\begin{fmfgraph*}(60,60)
			\fmfset{curly_len}{2mm}
			\fmftop{t1} \fmfbottom{b1,b2}
			\fmf{quark,label=$\hat e_r$,label.side=left,tension=2}{b1,v1}
			\fmf{dashes,label=$h$,label.side=right}{v3,v2}
			\fmf{photon,label=$\A$,label.side=right}{v2,v1}
			\fmf{quark,label=$\hat e_p$,label.side=left,tension=2}{v3,b2}
			\fmf{photon,label=$\hat \A$,tension=3}{t1,v2}
			\fmffreeze
			\fmf{quark,label=$e_u$}{v1,v3}
			\fmfv{decoration.shape=square,decoration.size=1.5mm}{v2}
		\end{fmfgraph*}
	\end{gathered}
\qquad\quad
	\begin{gathered}
		\begin{fmfgraph*}(60,60)
			\fmfset{curly_len}{2mm}
			\fmftop{t1} \fmfbottom{b1,b2}
			\fmf{quark,label=$\hat e_r$,label.side=left,tension=2}{b1,v1}
			\fmf{photon,label=$\A$,label.side=right}{v3,v2}
			\fmf{dashes,label=$h$,label.side=right}{v2,v1}
			\fmf{quark,label=$\hat e_p$,label.side=left,tension=2}{v3,b2}
			\fmf{photon,label=$\hat \A$,tension=3}{t1,v2}
			\fmffreeze
			\fmf{quark,label=$e_u$}{v1,v3}
			\fmfv{decoration.shape=square,decoration.size=1.5mm}{v2}
		\end{fmfgraph*}
	\end{gathered}
\qquad\quad
	\begin{gathered}
		\begin{fmfgraph*}(60,60)
			\fmfset{curly_len}{2mm}
			\fmftop{t1} \fmfbottom{b1,b2}
			\fmf{quark,label=$\hat e_r$,label.side=left,tension=2}{b1,v1}
			\fmf{dashes,label=$h$,label.side=right}{v3,v2}
			\fmf{photon,label=$\Z$,label.side=right}{v2,v1}
			\fmf{quark,label=$\hat e_p$,label.side=left,tension=2}{v3,b2}
			\fmf{photon,label=$\hat \A$,tension=3}{t1,v2}
			\fmffreeze
			\fmf{quark,label=$e_u$}{v1,v3}
			\fmfv{decoration.shape=square,decoration.size=1.5mm}{v2}
		\end{fmfgraph*}
	\end{gathered}
\nn
	& \begin{gathered}
		\begin{fmfgraph*}(60,60)
			\fmfset{curly_len}{2mm}
			\fmftop{t1} \fmfbottom{b1,b2}
			\fmf{quark,label=$\hat e_r$,label.side=left,tension=2}{b1,v1}
			\fmf{photon,label=$\Z$,label.side=right}{v3,v2}
			\fmf{dashes,label=$h$,label.side=right}{v2,v1}
			\fmf{quark,label=$\hat e_p$,label.side=left,tension=2}{v3,b2}
			\fmf{photon,label=$\hat \A$,tension=3}{t1,v2}
			\fmffreeze
			\fmf{quark,label=$e_u$}{v1,v3}
			\fmfv{decoration.shape=square,decoration.size=1.5mm}{v2}
		\end{fmfgraph*}
	\end{gathered}
\qquad\quad
	\begin{gathered}
		\begin{fmfgraph*}(60,60)
			\fmfset{curly_len}{2mm}
			\fmftop{t1} \fmfbottom{b1,b2}
			\fmf{quark,label=$\hat e_r$,label.side=right,tension=2}{b1,v1}
			\fmf{quark,label=$\hat e_p$,label.side=right,tension=2}{v1,b2}
			\fmf{photon,label=$\hat \A$,tension=2}{t1,v1}
			\fmf{scalar,label=$G^+$}{v1,v1}
			\fmffreeze
			\fmfv{decoration.shape=square,decoration.size=1.5mm}{v1}
		\end{fmfgraph*}
	\end{gathered}
\qquad\quad
	\begin{gathered}
		\begin{fmfgraph*}(60,60)
			\fmfset{curly_len}{2mm}
			\fmftop{t1} \fmfbottom{b1,b2}
			\fmf{quark,label=$\hat e_r$,label.side=right,tension=2}{b1,v2}
			\fmf{quark,label=$\hat e_p$,label.side=right,tension=2}{v2,b2}
			\fmf{photon,label=$\hat \A$,tension=2}{t1,v1}
			\fmf{quark,label=$t$,right}{v1,v2,v1}
			\fmffreeze
			\fmfv{decoration.shape=square,decoration.size=1.5mm}{v2}
		\end{fmfgraph*}
	\end{gathered}
\qquad\quad
	\begin{gathered}
		\begin{fmfgraph*}(60,60)
			\fmfset{curly_len}{2mm}
			\fmftop{t1} \fmfbottom{b1,b2}
			\fmf{quark,label=$\hat e_r$,label.side=right,tension=2}{b1,v2}
			\fmf{quark,label=$\hat e_p$,label.side=right,tension=2}{v2,b2}
			\fmf{photon,label=$\hat \A$,tension=2}{t1,v1}
			\fmf{wboson,label=$\W^+$,right}{v1,v2,v1}
			\fmffreeze
			\fmfv{decoration.shape=square,decoration.size=1.5mm}{v2}
		\end{fmfgraph*}
	\end{gathered}
\qquad\quad
	\begin{gathered}
		\begin{fmfgraph*}(60,60)
			\fmfset{curly_len}{2mm}
			\fmftop{t1} \fmfbottom{b1,b2}
			\fmf{quark,label=$\hat e_r$,label.side=right,tension=2}{b1,v2}
			\fmf{quark,label=$\hat e_p$,label.side=right,tension=2}{v2,b2}
			\fmf{photon,label=$\hat \A$,tension=2}{t1,v1}
			\fmf{scalar,label=$G^+$,right}{v1,v2}
			\fmf{scalar,label=$G^+$,right}{v2,v1}
			\fmffreeze
			\fmfv{decoration.shape=square,decoration.size=1.5mm}{v2}
		\end{fmfgraph*}
	\end{gathered}
\nn
	& \begin{gathered}
		\begin{fmfgraph*}(60,60)
			\fmfset{curly_len}{2mm}
			\fmftop{t1} \fmfbottom{b1,b2}
			\fmf{quark,label=$\hat e_r$,label.side=left,tension=0.67}{b1,v1}
			\fmf{quark,label=$\hat e_p$,label.side=left,tension=2}{v2,b2}
			\fmf{photon,label=$\hat \A$,tension=0.67}{t1,v1}
			\fmf{quark,label=$e_u$,label.side=right,tension=1}{v1,v2}
			\fmffreeze
			\fmf{dashes,label=$h$,left}{v1,v2}
			\fmfv{decoration.shape=square,decoration.size=1.5mm}{v1}
		\end{fmfgraph*}
	\end{gathered}
\qquad\quad
	\begin{gathered}
		\begin{fmfgraph*}(60,60)
			\fmfset{curly_len}{2mm}
			\fmftop{t1} \fmfbottom{b1,b2}
			\fmf{quark,label=$\hat e_r$,label.side=left,tension=0.67}{b1,v1}
			\fmf{quark,label=$\hat e_p$,label.side=left,tension=2}{v2,b2}
			\fmf{photon,label=$\hat \A$,tension=0.67}{t1,v1}
			\fmf{quark,label=$e_u$,label.side=right,tension=1}{v1,v2}
			\fmffreeze
			\fmf{dashes,label=$G^0$,left}{v1,v2}
			\fmfv{decoration.shape=square,decoration.size=1.5mm}{v1}
		\end{fmfgraph*}
	\end{gathered}
\qquad\quad
	\begin{gathered}
		\begin{fmfgraph*}(60,60)
			\fmfset{curly_len}{2mm}
			\fmftop{t1} \fmfbottom{b1,b2}
			\fmf{quark,label=$\hat e_r$,label.side=left,tension=0.67}{b1,v1}
			\fmf{quark,label=$\hat e_p$,label.side=left,tension=2}{v2,b2}
			\fmf{photon,label=$\hat \A$,tension=0.67}{t1,v1}
			\fmf{plain,label=$\nu_u$,label.side=right,tension=1}{v1,v2}
			\fmffreeze
			\fmf{wboson,label=$\W^+$,right}{v2,v1}
			\fmfv{decoration.shape=square,decoration.size=1.5mm}{v1}
		\end{fmfgraph*}
	\end{gathered}
\qquad\quad
	\begin{gathered}
		\begin{fmfgraph*}(60,60)
			\fmfset{curly_len}{2mm}
			\fmftop{t1} \fmfbottom{b1,b2}
			\fmf{quark,label=$\hat e_r$,label.side=left,tension=0.67}{b1,v1}
			\fmf{quark,label=$\hat e_p$,label.side=left,tension=2}{v2,b2}
			\fmf{photon,label=$\hat \A$,tension=0.67}{t1,v1}
			\fmf{plain,label=$\nu_u$,label.side=right,tension=1}{v1,v2}
			\fmffreeze
			\fmf{scalar,label=$G^+$,right}{v2,v1}
			\fmfv{decoration.shape=square,decoration.size=1.5mm}{v1}
		\end{fmfgraph*}
	\end{gathered}
\qquad\quad
	\begin{gathered}
		\begin{fmfgraph*}(60,60)
			\fmfset{curly_len}{2mm}
			\fmftop{t1} \fmfbottom{b1,b2}
			\fmf{quark,label=$\hat e_r$,label.side=left,tension=2}{b1,v2}
			\fmf{quark,label=$\hat e_p$,label.side=left,tension=0.67}{v1,b2}
			\fmf{photon,label=$\hat \A$,tension=0.67}{t1,v1}
			\fmf{quark,label=$e_u$,label.side=right,tension=1}{v2,v1}
			\fmffreeze
			\fmf{dashes,label=$h$,right}{v1,v2}
			\fmfv{decoration.shape=square,decoration.size=1.5mm}{v1}
		\end{fmfgraph*}
	\end{gathered}
\nn
	& \begin{gathered}
		\begin{fmfgraph*}(60,60)
			\fmfset{curly_len}{2mm}
			\fmftop{t1} \fmfbottom{b1,b2}
			\fmf{quark,label=$\hat e_r$,label.side=left,tension=2}{b1,v2}
			\fmf{quark,label=$\hat e_p$,label.side=left,tension=0.67}{v1,b2}
			\fmf{photon,label=$\hat \A$,tension=0.67}{t1,v1}
			\fmf{quark,label=$e_u$,label.side=right,tension=1}{v2,v1}
			\fmffreeze
			\fmf{dashes,label=$G^0$,right}{v1,v2}
			\fmfv{decoration.shape=square,decoration.size=1.5mm}{v1}
		\end{fmfgraph*}
	\end{gathered}
\qquad\quad
	\begin{gathered}
		\begin{fmfgraph*}(60,60)
			\fmfset{curly_len}{2mm}
			\fmftop{t1} \fmfbottom{b1,b2}
			\fmf{quark,label=$\hat e_r$,label.side=left,tension=2}{b1,v2}
			\fmf{quark,label=$\hat e_p$,label.side=left,tension=0.67}{v1,b2}
			\fmf{photon,label=$\hat \A$,tension=0.67}{t1,v1}
			\fmf{plain,label=$\nu_u$,label.side=right,tension=1}{v2,v1}
			\fmffreeze
			\fmf{wboson,label=$\W^+$,right}{v1,v2}
			\fmfv{decoration.shape=square,decoration.size=1.5mm}{v1}
		\end{fmfgraph*}
	\end{gathered}
\qquad\quad
	\begin{gathered}
		\begin{fmfgraph*}(60,60)
			\fmfset{curly_len}{2mm}
			\fmftop{t1} \fmfbottom{b1,b2}
			\fmf{quark,label=$\hat e_r$,label.side=left,tension=2}{b1,v2}
			\fmf{quark,label=$\hat e_p$,label.side=left,tension=0.67}{v1,b2}
			\fmf{photon,label=$\hat \A$,tension=0.67}{t1,v1}
			\fmf{plain,label=$\nu_u$,label.side=right,tension=1}{v2,v1}
			\fmffreeze
			\fmf{scalar,label=$G^+$,right}{v1,v2}
			\fmfv{decoration.shape=square,decoration.size=1.5mm}{v1}
		\end{fmfgraph*}
	\end{gathered}
\qquad\quad
	\begin{gathered}
		\begin{fmfgraph*}(60,60)
			\fmfset{curly_len}{2mm}
			\fmftop{t1,t2,t3} \fmfbottom{b1,b2}
			\fmf{quark,label=$\hat e_r$,label.side=left,tension=2}{b1,v1}
			\fmf{quark,label=$\hat e_p$,label.side=right,tension=2}{v1,b2}
			\fmf{photon,label=$\hat \A$,tension=2}{t2,v1}
			\fmffreeze
			\fmf{dashes,tension=3}{v1,v2}
			\fmf{phantom,tension=6}{t3,v2}
			\fmf{phantom,tension=2}{v2,b2}
			\fmfblob{5mm}{v2}
			\fmfv{decoration.shape=square,decoration.size=1.5mm}{v1}
		\end{fmfgraph*}
	\end{gathered}
\end{align*}


\subsubsection[$\nu \nu \gamma$]{\boldmath$\nu \nu \gamma$}

\begin{align*}
	&\begin{gathered}
		\begin{fmfgraph*}(60,60)
			\fmfset{curly_len}{2mm}
			\fmftop{t1} \fmfbottom{b1,b2}
			\fmf{plain,label=$\hat \nu_r$,label.side=left,tension=2}{b1,v1}
			\fmf{quark,label=$e_v$,label.side=left}{v1,v2}
			\fmf{quark,label=$e_u$,label.side=left}{v2,v3}
			\fmf{plain,label=$\hat \nu_p$,label.side=left,tension=2}{v3,b2}
			\fmf{photon,label=$\hat \A$,tension=3}{t1,v2}
			\fmffreeze
			\fmf{wboson,label=$\W^+$}{v1,v3}
		\end{fmfgraph*}
	\end{gathered}
\qquad\quad
	\begin{gathered}
		\begin{fmfgraph*}(60,60)
			\fmfset{curly_len}{2mm}
			\fmftop{t1} \fmfbottom{b1,b2}
			\fmf{plain,label=$\hat \nu_r$,label.side=left,tension=2}{b1,v1}
			\fmf{quark,label=$e_v$,label.side=right}{v2,v1}
			\fmf{quark,label=$e_u$,label.side=right}{v3,v2}
			\fmf{plain,label=$\hat \nu_p$,label.side=left,tension=2}{v3,b2}
			\fmf{photon,label=$\hat \A$,tension=3}{t1,v2}
			\fmffreeze
			\fmf{wboson,label=$\W^+$,label.side=left}{v3,v1}
		\end{fmfgraph*}
	\end{gathered}
\qquad\quad
	\begin{gathered}
		\begin{fmfgraph*}(60,60)
			\fmfset{curly_len}{2mm}
			\fmftop{t1} \fmfbottom{b1,b2}
			\fmf{plain,label=$\hat \nu_r$,label.side=left,tension=2}{b1,v1}
			\fmf{quark,label=$e_v$,label.side=left}{v1,v2}
			\fmf{quark,label=$e_u$,label.side=left}{v2,v3}
			\fmf{plain,label=$\hat \nu_p$,label.side=left,tension=2}{v3,b2}
			\fmf{photon,label=$\hat \A$,tension=3}{t1,v2}
			\fmffreeze
			\fmf{scalar,label=$G^+$}{v1,v3}
		\end{fmfgraph*}
	\end{gathered}
\qquad\quad
	\begin{gathered}
		\begin{fmfgraph*}(60,60)
			\fmfset{curly_len}{2mm}
			\fmftop{t1} \fmfbottom{b1,b2}
			\fmf{plain,label=$\hat \nu_r$,label.side=left,tension=2}{b1,v1}
			\fmf{quark,label=$e_v$,label.side=right}{v2,v1}
			\fmf{quark,label=$e_u$,label.side=right}{v3,v2}
			\fmf{plain,label=$\hat \nu_p$,label.side=left,tension=2}{v3,b2}
			\fmf{photon,label=$\hat \A$,tension=3}{t1,v2}
			\fmffreeze
			\fmf{scalar,label=$G^+$,label.side=left}{v3,v1}
		\end{fmfgraph*}
	\end{gathered}
\qquad\quad
	\begin{gathered}
		\begin{fmfgraph*}(60,60)
			\fmfset{curly_len}{2mm}
			\fmftop{t1} \fmfbottom{b1,b2}
			\fmf{plain,label=$\hat \nu_r$,label.side=left,tension=2}{b1,v1}
			\fmf{wboson,label=$\W^+$,label.side=left}{v1,v2,v3}
			\fmf{plain,label=$\hat \nu_p$,label.side=left,tension=2}{v3,b2}
			\fmf{photon,label=$\hat \A$,tension=3}{t1,v2}
			\fmffreeze
			\fmf{quark,label=$e_u$}{v1,v3}
		\end{fmfgraph*}
	\end{gathered}
\nn
	&\begin{gathered}
		\begin{fmfgraph*}(60,60)
			\fmfset{curly_len}{2mm}
			\fmftop{t1} \fmfbottom{b1,b2}
			\fmf{plain,label=$\hat \nu_r$,label.side=left,tension=2}{b1,v1}
			\fmf{wboson,label=$\W^+$,label.side=right}{v3,v2,v1}
			\fmf{plain,label=$\hat \nu_p$,label.side=left,tension=2}{v3,b2}
			\fmf{photon,label=$\hat \A$,tension=3}{t1,v2}
			\fmffreeze
			\fmf{quark,label=$e_u$,label.side=left}{v3,v1}
		\end{fmfgraph*}
	\end{gathered}
\qquad\quad
	\begin{gathered}
		\begin{fmfgraph*}(60,60)
			\fmfset{curly_len}{2mm}
			\fmftop{t1} \fmfbottom{b1,b2}
			\fmf{plain,label=$\hat \nu_r$,label.side=left,tension=2}{b1,v1}
			\fmf{scalar,label=$G^+$,label.side=left}{v1,v2,v3}
			\fmf{plain,label=$\hat \nu_p$,label.side=left,tension=2}{v3,b2}
			\fmf{photon,label=$\hat \A$,tension=3}{t1,v2}
			\fmffreeze
			\fmf{quark,label=$e_u$}{v1,v3}
		\end{fmfgraph*}
	\end{gathered}
\qquad\quad
	\begin{gathered}
		\begin{fmfgraph*}(60,60)
			\fmfset{curly_len}{2mm}
			\fmftop{t1} \fmfbottom{b1,b2}
			\fmf{plain,label=$\hat \nu_r$,label.side=left,tension=2}{b1,v1}
			\fmf{scalar,label=$G^+$,label.side=right}{v3,v2,v1}
			\fmf{plain,label=$\hat \nu_p$,label.side=left,tension=2}{v3,b2}
			\fmf{photon,label=$\hat \A$,tension=3}{t1,v2}
			\fmffreeze
			\fmf{quark,label=$e_u$,label.side=left}{v3,v1}
		\end{fmfgraph*}
	\end{gathered}
\qquad\quad
	\begin{gathered}
		\begin{fmfgraph*}(60,60)
			\fmfset{curly_len}{2mm}
			\fmftop{t1} \fmfbottom{b1,b2}
			\fmf{plain,label=$\hat \nu_r$,label.side=left,tension=2}{b1,v1}
			\fmf{wboson,label=$\W^+$,label.side=left}{v2,v3}
			\fmf{scalar,label=$G^+$,label.side=left}{v1,v2}
			\fmf{plain,label=$\hat \nu_p$,label.side=left,tension=2}{v3,b2}
			\fmf{photon,label=$\hat \A$,tension=3}{t1,v2}
			\fmffreeze
			\fmf{quark,label=$e_u$}{v1,v3}
			\fmfv{decoration.shape=square,decoration.size=1.5mm}{v2}
		\end{fmfgraph*}
	\end{gathered}
\qquad\quad
	\begin{gathered}
		\begin{fmfgraph*}(60,60)
			\fmfset{curly_len}{2mm}
			\fmftop{t1} \fmfbottom{b1,b2}
			\fmf{plain,label=$\hat \nu_r$,label.side=left,tension=2}{b1,v1}
			\fmf{scalar,label=$G^+$,label.side=right}{v3,v2}
			\fmf{wboson,label=$\W^+$,label.side=right}{v2,v1}
			\fmf{plain,label=$\hat \nu_p$,label.side=left,tension=2}{v3,b2}
			\fmf{photon,label=$\hat \A$,tension=3}{t1,v2}
			\fmffreeze
			\fmf{quark,label=$e_u$,label.side=left}{v3,v1}
			\fmfv{decoration.shape=square,decoration.size=1.5mm}{v2}
		\end{fmfgraph*}
	\end{gathered}
\nn
	& \begin{gathered}
		\begin{fmfgraph*}(60,60)
			\fmfset{curly_len}{2mm}
			\fmftop{t1} \fmfbottom{b1,b2}
			\fmf{plain,label=$\hat \nu_r$,label.side=left,tension=2}{b1,v1}
			\fmf{scalar,label=$G^+$,label.side=left}{v2,v3}
			\fmf{wboson,label=$\W^+$,label.side=left}{v1,v2}
			\fmf{plain,label=$\hat \nu_p$,label.side=left,tension=2}{v3,b2}
			\fmf{photon,label=$\hat \A$,tension=3}{t1,v2}
			\fmffreeze
			\fmf{quark,label=$e_u$}{v1,v3}
			\fmfv{decoration.shape=square,decoration.size=1.5mm}{v2}
		\end{fmfgraph*}
	\end{gathered}
\qquad\quad
	\begin{gathered}
		\begin{fmfgraph*}(60,60)
			\fmfset{curly_len}{2mm}
			\fmftop{t1} \fmfbottom{b1,b2}
			\fmf{plain,label=$\hat \nu_r$,label.side=left,tension=2}{b1,v1}
			\fmf{wboson,label=$\W^+$,label.side=right}{v3,v2}
			\fmf{scalar,label=$G^+$,label.side=right}{v2,v1}
			\fmf{plain,label=$\hat \nu_p$,label.side=left,tension=2}{v3,b2}
			\fmf{photon,label=$\hat \A$,tension=3}{t1,v2}
			\fmffreeze
			\fmf{quark,label=$e_u$,label.side=left}{v3,v1}
			\fmfv{decoration.shape=square,decoration.size=1.5mm}{v2}
		\end{fmfgraph*}
	\end{gathered}
\qquad\quad
	\begin{gathered}
		\begin{fmfgraph*}(60,60)
			\fmfset{curly_len}{2mm}
			\fmftop{t1} \fmfbottom{b1,b2}
			\fmf{plain,label=$\hat \nu_r$,label.side=right,tension=2}{b1,v1}
			\fmf{plain,label=$\hat \nu_p$,label.side=right,tension=2}{v1,b2}
			\fmf{photon,label=$\hat \A$,tension=2}{t1,v1}
			\fmf{scalar,label=$G^+$}{v1,v1}
			\fmffreeze
			\fmfv{decoration.shape=square,decoration.size=1.5mm}{v1}
		\end{fmfgraph*}
	\end{gathered}
\qquad\quad
	\begin{gathered}
		\begin{fmfgraph*}(60,60)
			\fmfset{curly_len}{2mm}
			\fmftop{t1} \fmfbottom{b1,b2}
			\fmf{plain,label=$\hat \nu_r$,label.side=right,tension=2}{b1,v2}
			\fmf{plain,label=$\hat \nu_p$,label.side=right,tension=2}{v2,b2}
			\fmf{photon,label=$\hat \A$,tension=2}{t1,v1}
			\fmf{quark,label=$t$,right}{v1,v2,v1}
			\fmffreeze
			\fmfv{decoration.shape=square,decoration.size=1.5mm}{v2}
		\end{fmfgraph*}
	\end{gathered}
\qquad\quad
	\begin{gathered}
		\begin{fmfgraph*}(60,60)
			\fmfset{curly_len}{2mm}
			\fmftop{t1} \fmfbottom{b1,b2}
			\fmf{plain,label=$\hat \nu_r$,label.side=right,tension=2}{b1,v2}
			\fmf{plain,label=$\hat \nu_p$,label.side=right,tension=2}{v2,b2}
			\fmf{photon,label=$\hat \A$,tension=2}{t1,v1}
			\fmf{scalar,label=$G^+$,right}{v1,v2}
			\fmf{scalar,label=$G^+$,right}{v2,v1}
			\fmffreeze
			\fmfv{decoration.shape=square,decoration.size=1.5mm}{v2}
		\end{fmfgraph*}
	\end{gathered}
\nn
	&\begin{gathered}
		\begin{fmfgraph*}(60,60)
			\fmfset{curly_len}{2mm}
			\fmftop{t1} \fmfbottom{b1,b2}
			\fmf{plain,label=$\hat \nu_r$,label.side=left,tension=0.67}{b1,v1}
			\fmf{plain,label=$\hat \nu_p$,label.side=left,tension=2}{v2,b2}
			\fmf{photon,label=$\hat \A$,tension=0.67}{t1,v1}
			\fmf{quark,label=$e_u$,label.side=right,tension=1}{v1,v2}
			\fmffreeze
			\fmf{wboson,label=$\W^+$,left}{v1,v2}
			\fmfv{decoration.shape=square,decoration.size=1.5mm}{v1}
		\end{fmfgraph*}
	\end{gathered}
\qquad\quad
	\begin{gathered}
		\begin{fmfgraph*}(60,60)
			\fmfset{curly_len}{2mm}
			\fmftop{t1} \fmfbottom{b1,b2}
			\fmf{plain,label=$\hat \nu_r$,label.side=left,tension=0.67}{b1,v1}
			\fmf{plain,label=$\hat \nu_p$,label.side=left,tension=2}{v2,b2}
			\fmf{photon,label=$\hat \A$,tension=0.67}{t1,v1}
			\fmf{quark,label=$e_u$,label.side=left,tension=1}{v2,v1}
			\fmffreeze
			\fmf{wboson,label=$\W^+$,right}{v2,v1}
			\fmfv{decoration.shape=square,decoration.size=1.5mm}{v1}
		\end{fmfgraph*}
	\end{gathered}
\qquad\quad
	\begin{gathered}
		\begin{fmfgraph*}(60,60)
			\fmfset{curly_len}{2mm}
			\fmftop{t1} \fmfbottom{b1,b2}
			\fmf{plain,label=$\hat \nu_r$,label.side=left,tension=0.67}{b1,v1}
			\fmf{plain,label=$\hat \nu_p$,label.side=left,tension=2}{v2,b2}
			\fmf{photon,label=$\hat \A$,tension=0.67}{t1,v1}
			\fmf{quark,label=$e_u$,label.side=right,tension=1}{v1,v2}
			\fmffreeze
			\fmf{scalar,label=$G^+$,left}{v1,v2}
			\fmfv{decoration.shape=square,decoration.size=1.5mm}{v1}
		\end{fmfgraph*}
	\end{gathered}
\qquad\quad
	\begin{gathered}
		\begin{fmfgraph*}(60,60)
			\fmfset{curly_len}{2mm}
			\fmftop{t1} \fmfbottom{b1,b2}
			\fmf{plain,label=$\hat \nu_r$,label.side=left,tension=0.67}{b1,v1}
			\fmf{plain,label=$\hat \nu_p$,label.side=left,tension=2}{v2,b2}
			\fmf{photon,label=$\hat \A$,tension=0.67}{t1,v1}
			\fmf{quark,label=$e_u$,label.side=left,tension=1}{v2,v1}
			\fmffreeze
			\fmf{scalar,label=$G^+$,right}{v2,v1}
			\fmfv{decoration.shape=square,decoration.size=1.5mm}{v1}
		\end{fmfgraph*}
	\end{gathered}
\qquad\quad
	\begin{gathered}
		\begin{fmfgraph*}(60,60)
			\fmfset{curly_len}{2mm}
			\fmftop{t1} \fmfbottom{b1,b2}
			\fmf{plain,label=$\hat \nu_r$,label.side=left,tension=2}{b1,v2}
			\fmf{plain,label=$\hat \nu_p$,label.side=left,tension=0.67}{v1,b2}
			\fmf{photon,label=$\hat \A$,tension=0.67}{t1,v1}
			\fmf{quark,label=$e_u$,label.side=right,tension=1}{v2,v1}
			\fmffreeze
			\fmf{wboson,label=$\W^+$,left}{v2,v1}
			\fmfv{decoration.shape=square,decoration.size=1.5mm}{v1}
		\end{fmfgraph*}
	\end{gathered}
\nn
	& \begin{gathered}
		\begin{fmfgraph*}(60,60)
			\fmfset{curly_len}{2mm}
			\fmftop{t1} \fmfbottom{b1,b2}
			\fmf{plain,label=$\hat \nu_r$,label.side=left,tension=2}{b1,v2}
			\fmf{plain,label=$\hat \nu_p$,label.side=left,tension=0.67}{v1,b2}
			\fmf{photon,label=$\hat \A$,tension=0.67}{t1,v1}
			\fmf{quark,label=$e_u$,label.side=left,tension=1}{v1,v2}
			\fmffreeze
			\fmf{wboson,label=$\W^+$,right}{v1,v2}
			\fmfv{decoration.shape=square,decoration.size=1.5mm}{v1}
		\end{fmfgraph*}
	\end{gathered}
\qquad\quad
	\begin{gathered}
		\begin{fmfgraph*}(60,60)
			\fmfset{curly_len}{2mm}
			\fmftop{t1} \fmfbottom{b1,b2}
			\fmf{plain,label=$\hat \nu_r$,label.side=left,tension=2}{b1,v2}
			\fmf{plain,label=$\hat \nu_p$,label.side=left,tension=0.67}{v1,b2}
			\fmf{photon,label=$\hat \A$,tension=0.67}{t1,v1}
			\fmf{quark,label=$e_u$,label.side=right,tension=1}{v2,v1}
			\fmffreeze
			\fmf{scalar,label=$G^+$,left}{v2,v1}
			\fmfv{decoration.shape=square,decoration.size=1.5mm}{v1}
		\end{fmfgraph*}
	\end{gathered}
\qquad\quad
	\begin{gathered}
		\begin{fmfgraph*}(60,60)
			\fmfset{curly_len}{2mm}
			\fmftop{t1} \fmfbottom{b1,b2}
			\fmf{plain,label=$\hat \nu_r$,label.side=left,tension=2}{b1,v2}
			\fmf{plain,label=$\hat \nu_p$,label.side=left,tension=0.67}{v1,b2}
			\fmf{photon,label=$\hat \A$,tension=0.67}{t1,v1}
			\fmf{quark,label=$e_u$,label.side=left,tension=1}{v1,v2}
			\fmffreeze
			\fmf{scalar,label=$G^+$,right}{v1,v2}
			\fmfv{decoration.shape=square,decoration.size=1.5mm}{v1}
		\end{fmfgraph*}
	\end{gathered}
\end{align*}
Similarly to the two-point functions, the reversed diagrams can be easily obtained from the original diagrams~\cite{Denner:1992vza}:
\begin{align}
	D &= \bar u_p(p') \Gamma_{pr}(p',p) u_r(p) \, , \nn
	D^\text{rev} &= - \bar v_r(p) \Gamma_{rp}(-p,-p') v_p(p') = \bar u_p(p') C \Gamma^T_{rp}(-p,-p') C^{-1} u_r(p) \, ,
\end{align}
where the transposition acts in Dirac space and transposition in the flavor indices is given explicitly.


\subsubsection[$\bar u u \gamma$]{\boldmath$\bar u u \gamma$}

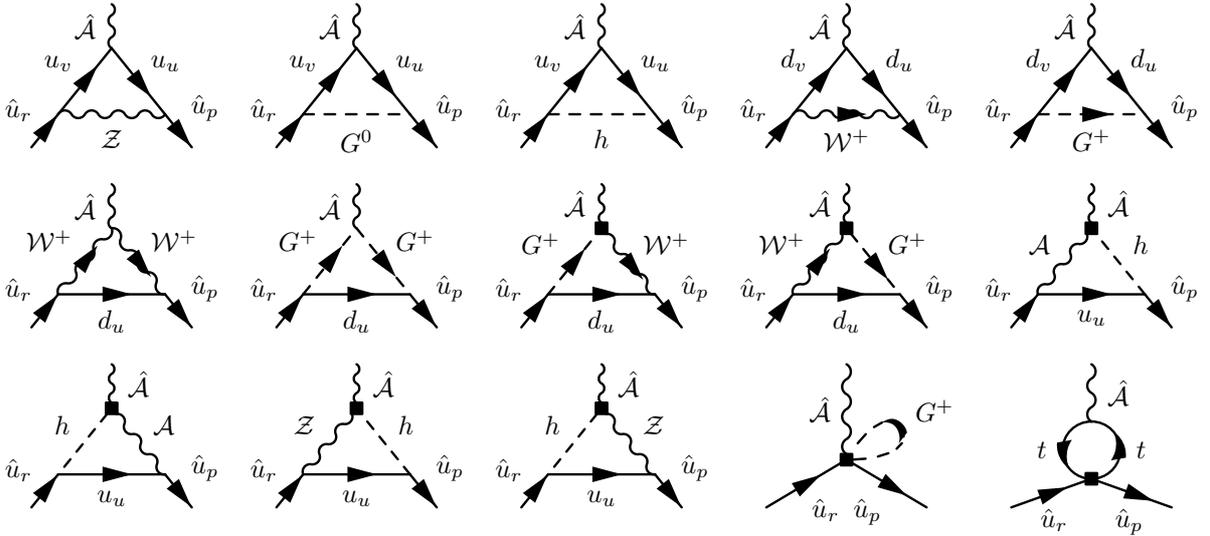
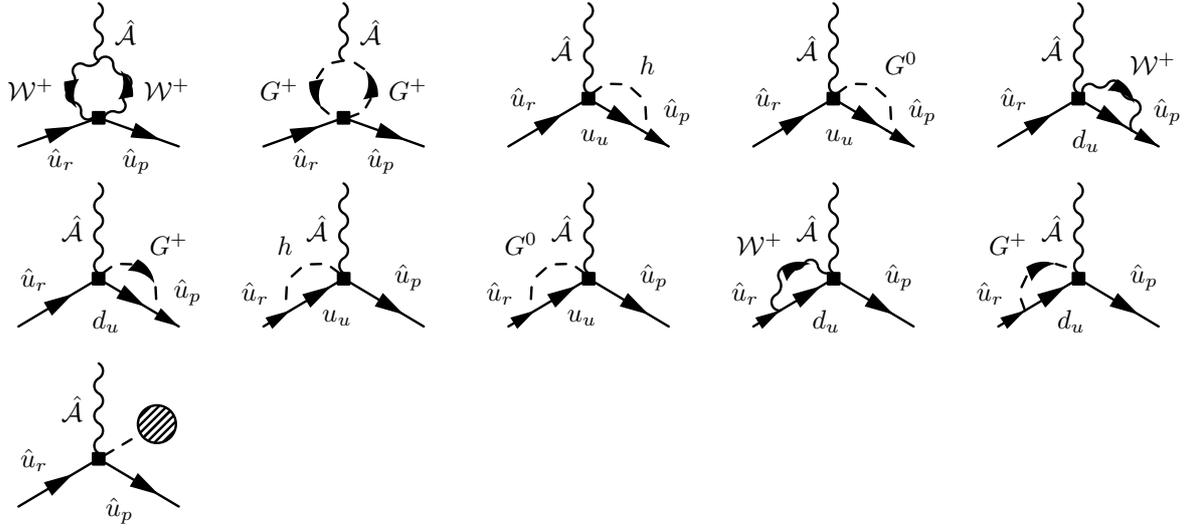
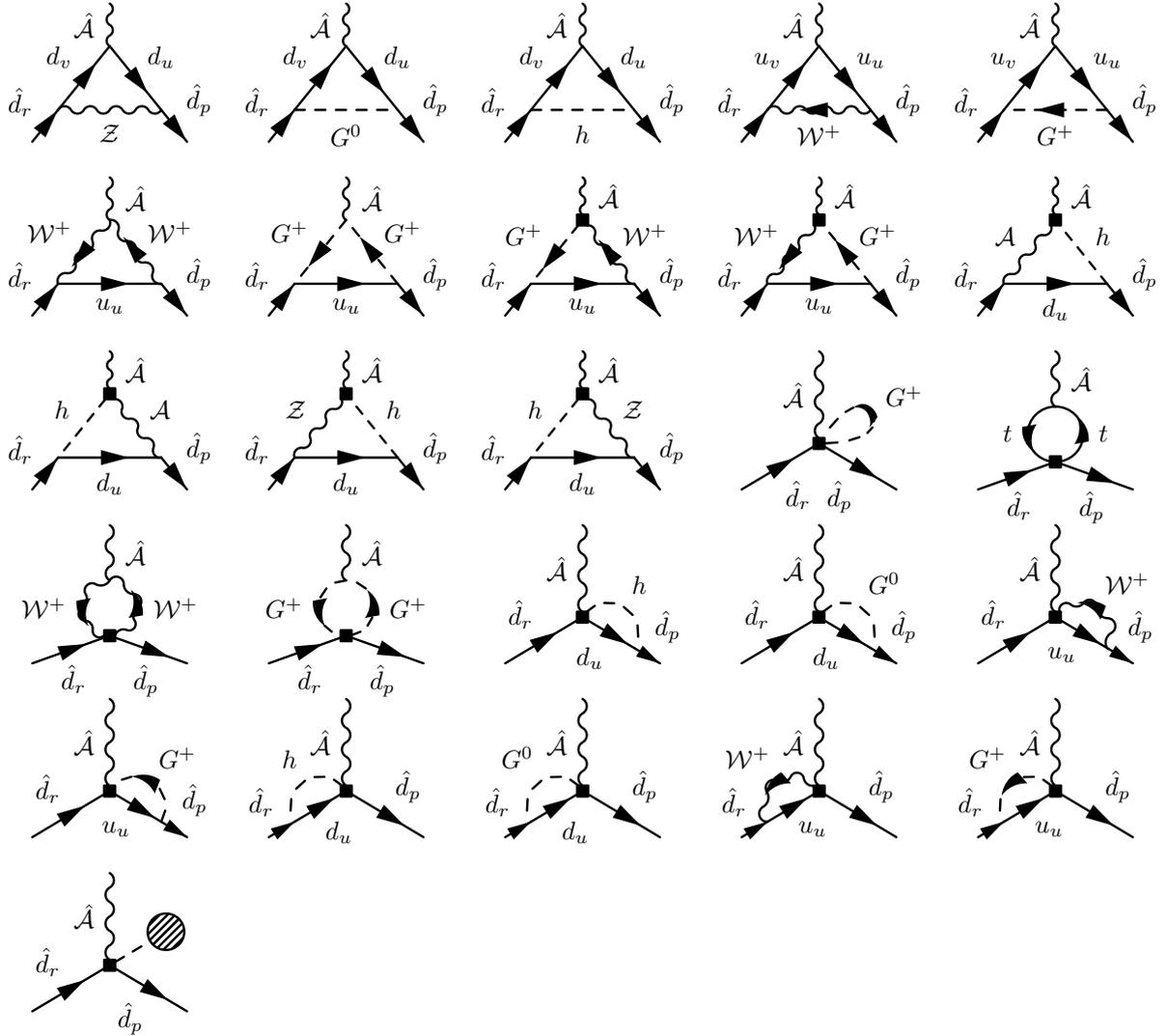
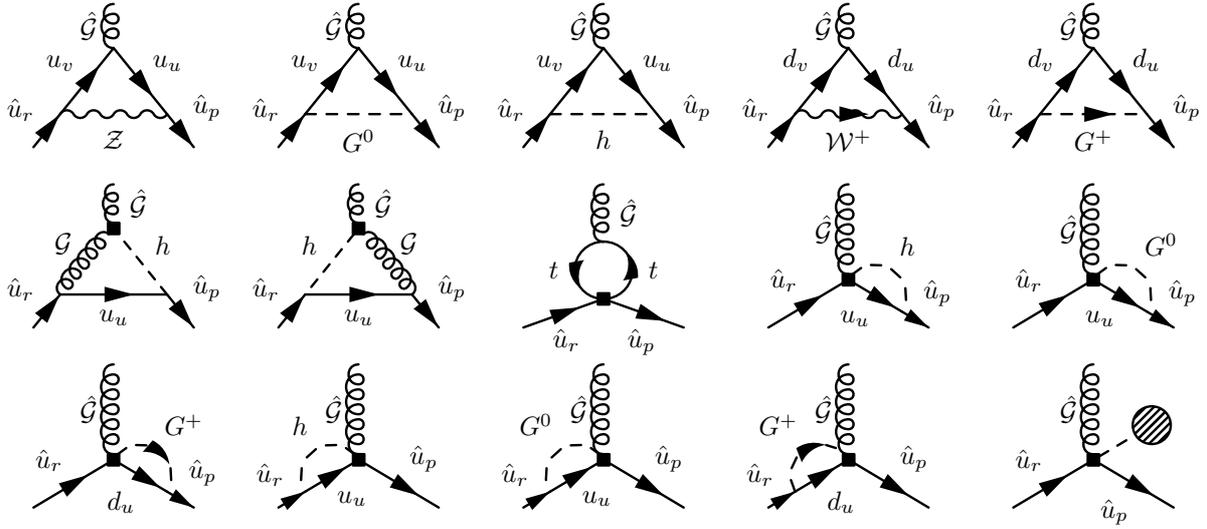
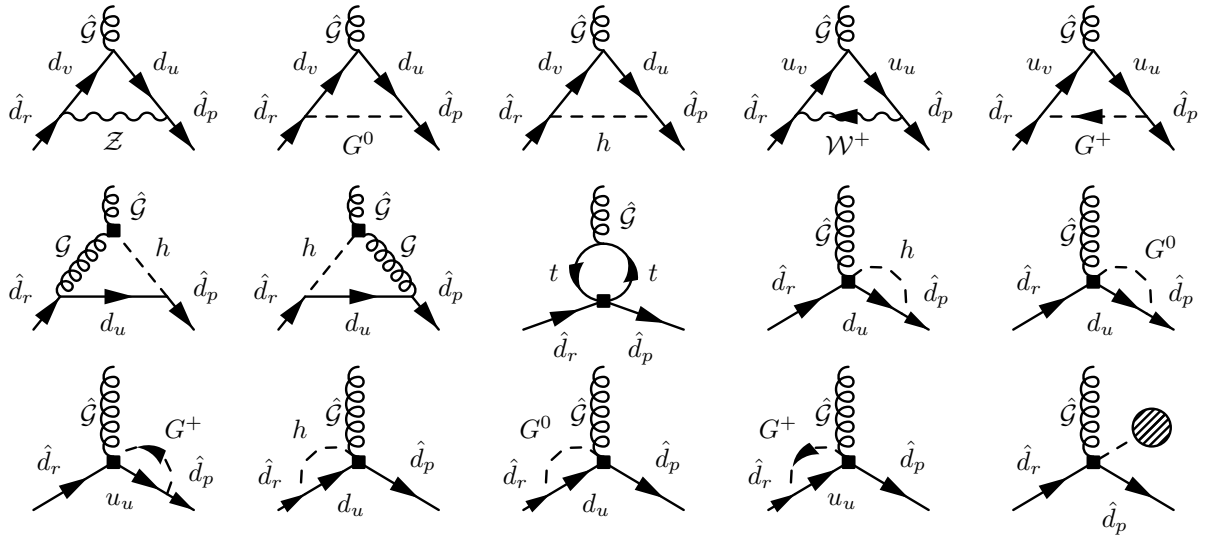
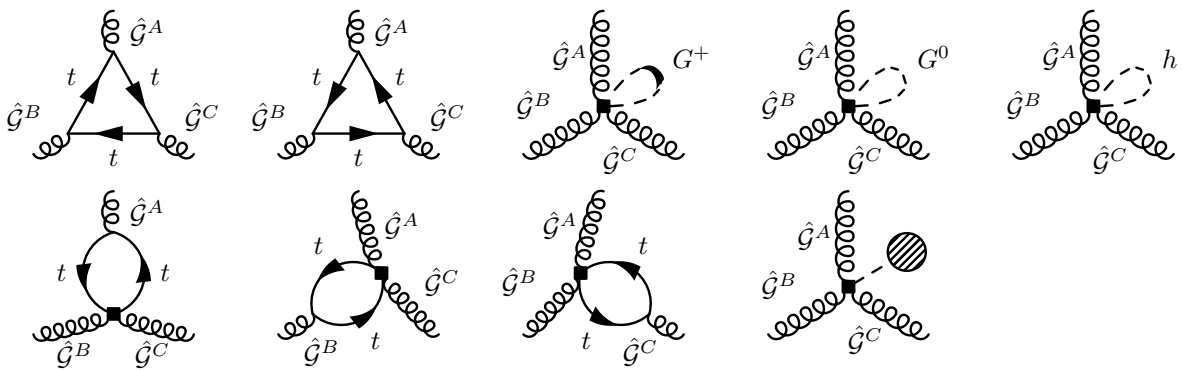
\begin{align*}
	&\begin{gathered}
		\begin{fmfgraph*}(60,60)
			\fmfset{curly_len}{2mm}
			\fmftop{t1} \fmfbottom{b1,b2}
			\fmf{quark,label=$\hat u_r$,label.side=left,tension=2}{b1,v1}
			\fmf{quark,label=$u_v$,label.side=left}{v1,v2}
			\fmf{quark,label=$u_u$,label.side=left}{v2,v3}
			\fmf{quark,label=$\hat u_p$,label.side=left,tension=2}{v3,b2}
			\fmf{photon,label=$\hat \A$,tension=3}{t1,v2}
			\fmffreeze
			\fmf{photon,label=$\Z$}{v1,v3}
		\end{fmfgraph*}
	\end{gathered}
\qquad\quad
	\begin{gathered}
		\begin{fmfgraph*}(60,60)
			\fmfset{curly_len}{2mm}
			\fmftop{t1} \fmfbottom{b1,b2}
			\fmf{quark,label=$\hat u_r$,label.side=left,tension=2}{b1,v1}
			\fmf{quark,label=$u_v$,label.side=left}{v1,v2}
			\fmf{quark,label=$u_u$,label.side=left}{v2,v3}
			\fmf{quark,label=$\hat u_p$,label.side=left,tension=2}{v3,b2}
			\fmf{photon,label=$\hat \A$,tension=3}{t1,v2}
			\fmffreeze
			\fmf{dashes,label=$G^0$}{v1,v3}
		\end{fmfgraph*}
	\end{gathered}
\qquad\quad
	\begin{gathered}
		\begin{fmfgraph*}(60,60)
			\fmfset{curly_len}{2mm}
			\fmftop{t1} \fmfbottom{b1,b2}
			\fmf{quark,label=$\hat u_r$,label.side=left,tension=2}{b1,v1}
			\fmf{quark,label=$u_v$,label.side=left}{v1,v2}
			\fmf{quark,label=$u_u$,label.side=left}{v2,v3}
			\fmf{quark,label=$\hat u_p$,label.side=left,tension=2}{v3,b2}
			\fmf{photon,label=$\hat \A$,tension=3}{t1,v2}
			\fmffreeze
			\fmf{dashes,label=$h$}{v1,v3}
		\end{fmfgraph*}
	\end{gathered}
\qquad\quad
	\begin{gathered}
		\begin{fmfgraph*}(60,60)
			\fmfset{curly_len}{2mm}
			\fmftop{t1} \fmfbottom{b1,b2}
			\fmf{quark,label=$\hat u_r$,label.side=left,tension=2}{b1,v1}
			\fmf{quark,label=$d_v$,label.side=left}{v1,v2}
			\fmf{quark,label=$d_u$,label.side=left}{v2,v3}
			\fmf{quark,label=$\hat u_p$,label.side=left,tension=2}{v3,b2}
			\fmf{photon,label=$\hat \A$,tension=3}{t1,v2}
			\fmffreeze
			\fmf{wboson,label=$\W^+$}{v1,v3}
		\end{fmfgraph*}
	\end{gathered}
\qquad\quad
	\begin{gathered}
		\begin{fmfgraph*}(60,60)
			\fmfset{curly_len}{2mm}
			\fmftop{t1} \fmfbottom{b1,b2}
			\fmf{quark,label=$\hat u_r$,label.side=left,tension=2}{b1,v1}
			\fmf{quark,label=$d_v$,label.side=left}{v1,v2}
			\fmf{quark,label=$d_u$,label.side=left}{v2,v3}
			\fmf{quark,label=$\hat u_p$,label.side=left,tension=2}{v3,b2}
			\fmf{photon,label=$\hat \A$,tension=3}{t1,v2}
			\fmffreeze
			\fmf{scalar,label=$G^+$}{v1,v3}
		\end{fmfgraph*}
	\end{gathered}
\nn
	&\begin{gathered}
		\begin{fmfgraph*}(60,60)
			\fmfset{curly_len}{2mm}
			\fmftop{t1} \fmfbottom{b1,b2}
			\fmf{quark,label=$\hat u_r$,label.side=left,tension=2}{b1,v1}
			\fmf{wboson,label=$\W^+$,label.side=left}{v1,v2,v3}
			\fmf{quark,label=$\hat u_p$,label.side=left,tension=2}{v3,b2}
			\fmf{photon,label=$\hat \A$,tension=3}{t1,v2}
			\fmffreeze
			\fmf{quark,label=$d_u$}{v1,v3}
		\end{fmfgraph*}
	\end{gathered}
\qquad\quad
	\begin{gathered}
		\begin{fmfgraph*}(60,60)
			\fmfset{curly_len}{2mm}
			\fmftop{t1} \fmfbottom{b1,b2}
			\fmf{quark,label=$\hat u_r$,label.side=left,tension=2}{b1,v1}
			\fmf{scalar,label=$G^+$,label.side=left}{v1,v2,v3}
			\fmf{quark,label=$\hat u_p$,label.side=left,tension=2}{v3,b2}
			\fmf{photon,label=$\hat \A$,tension=3}{t1,v2}
			\fmffreeze
			\fmf{quark,label=$d_u$}{v1,v3}
		\end{fmfgraph*}
	\end{gathered}
\qquad\quad
	\begin{gathered}
		\begin{fmfgraph*}(60,60)
			\fmfset{curly_len}{2mm}
			\fmftop{t1} \fmfbottom{b1,b2}
			\fmf{quark,label=$\hat u_r$,label.side=left,tension=2}{b1,v1}
			\fmf{wboson,label=$\W^+$,label.side=left}{v2,v3}
			\fmf{scalar,label=$G^+$,label.side=left}{v1,v2}
			\fmf{quark,label=$\hat u_p$,label.side=left,tension=2}{v3,b2}
			\fmf{photon,label=$\hat \A$,tension=3}{t1,v2}
			\fmffreeze
			\fmf{quark,label=$d_u$}{v1,v3}
			\fmfv{decoration.shape=square,decoration.size=1.5mm}{v2}
		\end{fmfgraph*}
	\end{gathered}
\qquad\quad
	\begin{gathered}
		\begin{fmfgraph*}(60,60)
			\fmfset{curly_len}{2mm}
			\fmftop{t1} \fmfbottom{b1,b2}
			\fmf{quark,label=$\hat u_r$,label.side=left,tension=2}{b1,v1}
			\fmf{scalar,label=$G^+$,label.side=left}{v2,v3}
			\fmf{wboson,label=$\W^+$,label.side=left}{v1,v2}
			\fmf{quark,label=$\hat u_p$,label.side=left,tension=2}{v3,b2}
			\fmf{photon,label=$\hat \A$,tension=3}{t1,v2}
			\fmffreeze
			\fmf{quark,label=$d_u$}{v1,v3}
			\fmfv{decoration.shape=square,decoration.size=1.5mm}{v2}
		\end{fmfgraph*}
	\end{gathered}
\qquad\quad
	\begin{gathered}
		\begin{fmfgraph*}(60,60)
			\fmfset{curly_len}{2mm}
			\fmftop{t1} \fmfbottom{b1,b2}
			\fmf{quark,label=$\hat u_r$,label.side=left,tension=2}{b1,v1}
			\fmf{dashes,label=$h$,label.side=right}{v3,v2}
			\fmf{photon,label=$\A$,label.side=right}{v2,v1}
			\fmf{quark,label=$\hat u_p$,label.side=left,tension=2}{v3,b2}
			\fmf{photon,label=$\hat \A$,tension=3}{t1,v2}
			\fmffreeze
			\fmf{quark,label=$u_u$}{v1,v3}
			\fmfv{decoration.shape=square,decoration.size=1.5mm}{v2}
		\end{fmfgraph*}
	\end{gathered}
\nn
	& \begin{gathered}
		\begin{fmfgraph*}(60,60)
			\fmfset{curly_len}{2mm}
			\fmftop{t1} \fmfbottom{b1,b2}
			\fmf{quark,label=$\hat u_r$,label.side=left,tension=2}{b1,v1}
			\fmf{photon,label=$\A$,label.side=right}{v3,v2}
			\fmf{dashes,label=$h$,label.side=right}{v2,v1}
			\fmf{quark,label=$\hat u_p$,label.side=left,tension=2}{v3,b2}
			\fmf{photon,label=$\hat \A$,tension=3}{t1,v2}
			\fmffreeze
			\fmf{quark,label=$u_u$}{v1,v3}
			\fmfv{decoration.shape=square,decoration.size=1.5mm}{v2}
		\end{fmfgraph*}
	\end{gathered}
\qquad\quad
	\begin{gathered}
		\begin{fmfgraph*}(60,60)
			\fmfset{curly_len}{2mm}
			\fmftop{t1} \fmfbottom{b1,b2}
			\fmf{quark,label=$\hat u_r$,label.side=left,tension=2}{b1,v1}
			\fmf{dashes,label=$h$,label.side=right}{v3,v2}
			\fmf{photon,label=$\Z$,label.side=right}{v2,v1}
			\fmf{quark,label=$\hat u_p$,label.side=left,tension=2}{v3,b2}
			\fmf{photon,label=$\hat \A$,tension=3}{t1,v2}
			\fmffreeze
			\fmf{quark,label=$u_u$}{v1,v3}
			\fmfv{decoration.shape=square,decoration.size=1.5mm}{v2}
		\end{fmfgraph*}
	\end{gathered}
\qquad\quad
	\begin{gathered}
		\begin{fmfgraph*}(60,60)
			\fmfset{curly_len}{2mm}
			\fmftop{t1} \fmfbottom{b1,b2}
			\fmf{quark,label=$\hat u_r$,label.side=left,tension=2}{b1,v1}
			\fmf{photon,label=$\Z$,label.side=right}{v3,v2}
			\fmf{dashes,label=$h$,label.side=right}{v2,v1}
			\fmf{quark,label=$\hat u_p$,label.side=left,tension=2}{v3,b2}
			\fmf{photon,label=$\hat \A$,tension=3}{t1,v2}
			\fmffreeze
			\fmf{quark,label=$u_u$}{v1,v3}
			\fmfv{decoration.shape=square,decoration.size=1.5mm}{v2}
		\end{fmfgraph*}
	\end{gathered}
\qquad\quad
	\begin{gathered}
		\begin{fmfgraph*}(60,60)
			\fmfset{curly_len}{2mm}
			\fmftop{t1} \fmfbottom{b1,b2}
			\fmf{quark,label=$\hat u_r$,label.side=right,tension=2}{b1,v1}
			\fmf{quark,label=$\hat u_p$,label.side=right,tension=2}{v1,b2}
			\fmf{photon,label=$\hat \A$,tension=2}{t1,v1}
			\fmf{scalar,label=$G^+$}{v1,v1}
			\fmffreeze
			\fmfv{decoration.shape=square,decoration.size=1.5mm}{v1}
		\end{fmfgraph*}
	\end{gathered}
\qquad\quad
	\begin{gathered}
		\begin{fmfgraph*}(60,60)
			\fmfset{curly_len}{2mm}
			\fmftop{t1} \fmfbottom{b1,b2}
			\fmf{quark,label=$\hat u_r$,label.side=right,tension=2}{b1,v2}
			\fmf{quark,label=$\hat u_p$,label.side=right,tension=2}{v2,b2}
			\fmf{photon,label=$\hat \A$,tension=2}{t1,v1}
			\fmf{quark,label=$t$,right}{v1,v2,v1}
			\fmffreeze
			\fmfv{decoration.shape=square,decoration.size=1.5mm}{v2}
		\end{fmfgraph*}
	\end{gathered}
\nn
	& \begin{gathered}
		\begin{fmfgraph*}(60,60)
			\fmfset{curly_len}{2mm}
			\fmftop{t1} \fmfbottom{b1,b2}
			\fmf{quark,label=$\hat u_r$,label.side=right,tension=2}{b1,v2}
			\fmf{quark,label=$\hat u_p$,label.side=right,tension=2}{v2,b2}
			\fmf{photon,label=$\hat \A$,tension=2}{t1,v1}
			\fmf{wboson,label=$\W^+$,right}{v1,v2,v1}
			\fmffreeze
			\fmfv{decoration.shape=square,decoration.size=1.5mm}{v2}
		\end{fmfgraph*}
	\end{gathered}
\qquad\quad
	\begin{gathered}
		\begin{fmfgraph*}(60,60)
			\fmfset{curly_len}{2mm}
			\fmftop{t1} \fmfbottom{b1,b2}
			\fmf{quark,label=$\hat u_r$,label.side=right,tension=2}{b1,v2}
			\fmf{quark,label=$\hat u_p$,label.side=right,tension=2}{v2,b2}
			\fmf{photon,label=$\hat \A$,tension=2}{t1,v1}
			\fmf{scalar,label=$G^+$,right}{v1,v2}
			\fmf{scalar,label=$G^+$,right}{v2,v1}
			\fmffreeze
			\fmfv{decoration.shape=square,decoration.size=1.5mm}{v2}
		\end{fmfgraph*}
	\end{gathered}
\qquad\quad
	\begin{gathered}
		\begin{fmfgraph*}(60,60)
			\fmfset{curly_len}{2mm}
			\fmftop{t1} \fmfbottom{b1,b2}
			\fmf{quark,label=$\hat u_r$,label.side=left,tension=0.67}{b1,v1}
			\fmf{quark,label=$\hat u_p$,label.side=left,tension=2}{v2,b2}
			\fmf{photon,label=$\hat \A$,tension=0.67}{t1,v1}
			\fmf{quark,label=$u_u$,label.side=right,tension=1}{v1,v2}
			\fmffreeze
			\fmf{dashes,label=$h$,left}{v1,v2}
			\fmfv{decoration.shape=square,decoration.size=1.5mm}{v1}
		\end{fmfgraph*}
	\end{gathered}
\qquad\quad
	\begin{gathered}
		\begin{fmfgraph*}(60,60)
			\fmfset{curly_len}{2mm}
			\fmftop{t1} \fmfbottom{b1,b2}
			\fmf{quark,label=$\hat u_r$,label.side=left,tension=0.67}{b1,v1}
			\fmf{quark,label=$\hat u_p$,label.side=left,tension=2}{v2,b2}
			\fmf{photon,label=$\hat \A$,tension=0.67}{t1,v1}
			\fmf{quark,label=$u_u$,label.side=right,tension=1}{v1,v2}
			\fmffreeze
			\fmf{dashes,label=$G^0$,left}{v1,v2}
			\fmfv{decoration.shape=square,decoration.size=1.5mm}{v1}
		\end{fmfgraph*}
	\end{gathered}
\qquad\quad
	\begin{gathered}
		\begin{fmfgraph*}(60,60)
			\fmfset{curly_len}{2mm}
			\fmftop{t1} \fmfbottom{b1,b2}
			\fmf{quark,label=$\hat u_r$,label.side=left,tension=0.67}{b1,v1}
			\fmf{quark,label=$\hat u_p$,label.side=left,tension=2}{v2,b2}
			\fmf{photon,label=$\hat \A$,tension=0.67}{t1,v1}
			\fmf{quark,label=$d_u$,label.side=right,tension=1}{v1,v2}
			\fmffreeze
			\fmf{wboson,label=$\W^+$,left}{v1,v2}
			\fmfv{decoration.shape=square,decoration.size=1.5mm}{v1}
		\end{fmfgraph*}
	\end{gathered}
\nn
	& \begin{gathered}
		\begin{fmfgraph*}(60,60)
			\fmfset{curly_len}{2mm}
			\fmftop{t1} \fmfbottom{b1,b2}
			\fmf{quark,label=$\hat u_r$,label.side=left,tension=0.67}{b1,v1}
			\fmf{quark,label=$\hat u_p$,label.side=left,tension=2}{v2,b2}
			\fmf{photon,label=$\hat \A$,tension=0.67}{t1,v1}
			\fmf{quark,label=$d_u$,label.side=right,tension=1}{v1,v2}
			\fmffreeze
			\fmf{scalar,label=$G^+$,left}{v1,v2}
			\fmfv{decoration.shape=square,decoration.size=1.5mm}{v1}
		\end{fmfgraph*}
	\end{gathered}
\qquad\quad
	\begin{gathered}
		\begin{fmfgraph*}(60,60)
			\fmfset{curly_len}{2mm}
			\fmftop{t1} \fmfbottom{b1,b2}
			\fmf{quark,label=$\hat u_r$,label.side=left,tension=2}{b1,v2}
			\fmf{quark,label=$\hat u_p$,label.side=left,tension=0.67}{v1,b2}
			\fmf{photon,label=$\hat \A$,tension=0.67}{t1,v1}
			\fmf{quark,label=$u_u$,label.side=right,tension=1}{v2,v1}
			\fmffreeze
			\fmf{dashes,label=$h$,right}{v1,v2}
			\fmfv{decoration.shape=square,decoration.size=1.5mm}{v1}
		\end{fmfgraph*}
	\end{gathered}
\qquad\quad
	\begin{gathered}
		\begin{fmfgraph*}(60,60)
			\fmfset{curly_len}{2mm}
			\fmftop{t1} \fmfbottom{b1,b2}
			\fmf{quark,label=$\hat u_r$,label.side=left,tension=2}{b1,v2}
			\fmf{quark,label=$\hat u_p$,label.side=left,tension=0.67}{v1,b2}
			\fmf{photon,label=$\hat \A$,tension=0.67}{t1,v1}
			\fmf{quark,label=$u_u$,label.side=right,tension=1}{v2,v1}
			\fmffreeze
			\fmf{dashes,label=$G^0$,right}{v1,v2}
			\fmfv{decoration.shape=square,decoration.size=1.5mm}{v1}
		\end{fmfgraph*}
	\end{gathered}
\qquad\quad
	\begin{gathered}
		\begin{fmfgraph*}(60,60)
			\fmfset{curly_len}{2mm}
			\fmftop{t1} \fmfbottom{b1,b2}
			\fmf{quark,label=$\hat u_r$,label.side=left,tension=2}{b1,v2}
			\fmf{quark,label=$\hat u_p$,label.side=left,tension=0.67}{v1,b2}
			\fmf{photon,label=$\hat \A$,tension=0.67}{t1,v1}
			\fmf{quark,label=$d_u$,label.side=right,tension=1}{v2,v1}
			\fmffreeze
			\fmf{wboson,label=$\W^+$,left}{v2,v1}
			\fmfv{decoration.shape=square,decoration.size=1.5mm}{v1}
		\end{fmfgraph*}
	\end{gathered}
\qquad\quad
	\begin{gathered}
		\begin{fmfgraph*}(60,60)
			\fmfset{curly_len}{2mm}
			\fmftop{t1} \fmfbottom{b1,b2}
			\fmf{quark,label=$\hat u_r$,label.side=left,tension=2}{b1,v2}
			\fmf{quark,label=$\hat u_p$,label.side=left,tension=0.67}{v1,b2}
			\fmf{photon,label=$\hat \A$,tension=0.67}{t1,v1}
			\fmf{quark,label=$d_u$,label.side=right,tension=1}{v2,v1}
			\fmffreeze
			\fmf{scalar,label=$G^+$,left}{v2,v1}
			\fmfv{decoration.shape=square,decoration.size=1.5mm}{v1}
		\end{fmfgraph*}
	\end{gathered}
\nn
	& \begin{gathered}
		\begin{fmfgraph*}(60,60)
			\fmfset{curly_len}{2mm}
			\fmftop{t1,t2,t3} \fmfbottom{b1,b2}
			\fmf{quark,label=$\hat u_r$,label.side=left,tension=2}{b1,v1}
			\fmf{quark,label=$\hat u_p$,label.side=right,tension=2}{v1,b2}
			\fmf{photon,label=$\hat \A$,tension=2}{t2,v1}
			\fmffreeze
			\fmf{dashes,tension=3}{v1,v2}
			\fmf{phantom,tension=6}{t3,v2}
			\fmf{phantom,tension=2}{v2,b2}
			\fmfblob{5mm}{v2}
			\fmfv{decoration.shape=square,decoration.size=1.5mm}{v1}
		\end{fmfgraph*}
	\end{gathered}
\end{align*}


\subsubsection[$\bar d d \gamma$]{\boldmath$\bar d d \gamma$}

\begin{align*}
	&\begin{gathered}
		\begin{fmfgraph*}(60,60)
			\fmfset{curly_len}{2mm}
			\fmftop{t1} \fmfbottom{b1,b2}
			\fmf{quark,label=$\hat d_r$,label.side=left,tension=2}{b1,v1}
			\fmf{quark,label=$d_v$,label.side=left}{v1,v2}
			\fmf{quark,label=$d_u$,label.side=left}{v2,v3}
			\fmf{quark,label=$\hat d_p$,label.side=left,tension=2}{v3,b2}
			\fmf{photon,label=$\hat \A$,tension=3}{t1,v2}
			\fmffreeze
			\fmf{photon,label=$\Z$}{v1,v3}
		\end{fmfgraph*}
	\end{gathered}
\qquad\quad
	\begin{gathered}
		\begin{fmfgraph*}(60,60)
			\fmfset{curly_len}{2mm}
			\fmftop{t1} \fmfbottom{b1,b2}
			\fmf{quark,label=$\hat d_r$,label.side=left,tension=2}{b1,v1}
			\fmf{quark,label=$d_v$,label.side=left}{v1,v2}
			\fmf{quark,label=$d_u$,label.side=left}{v2,v3}
			\fmf{quark,label=$\hat d_p$,label.side=left,tension=2}{v3,b2}
			\fmf{photon,label=$\hat \A$,tension=3}{t1,v2}
			\fmffreeze
			\fmf{dashes,label=$G^0$}{v1,v3}
		\end{fmfgraph*}
	\end{gathered}
\qquad\quad
	\begin{gathered}
		\begin{fmfgraph*}(60,60)
			\fmfset{curly_len}{2mm}
			\fmftop{t1} \fmfbottom{b1,b2}
			\fmf{quark,label=$\hat d_r$,label.side=left,tension=2}{b1,v1}
			\fmf{quark,label=$d_v$,label.side=left}{v1,v2}
			\fmf{quark,label=$d_u$,label.side=left}{v2,v3}
			\fmf{quark,label=$\hat d_p$,label.side=left,tension=2}{v3,b2}
			\fmf{photon,label=$\hat \A$,tension=3}{t1,v2}
			\fmffreeze
			\fmf{dashes,label=$h$}{v1,v3}
		\end{fmfgraph*}
	\end{gathered}
\qquad\quad
	\begin{gathered}
		\begin{fmfgraph*}(60,60)
			\fmfset{curly_len}{2mm}
			\fmftop{t1} \fmfbottom{b1,b2}
			\fmf{quark,label=$\hat d_r$,label.side=left,tension=2}{b1,v1}
			\fmf{quark,label=$u_v$,label.side=left}{v1,v2}
			\fmf{quark,label=$u_u$,label.side=left}{v2,v3}
			\fmf{quark,label=$\hat d_p$,label.side=left,tension=2}{v3,b2}
			\fmf{photon,label=$\hat \A$,tension=3}{t1,v2}
			\fmffreeze
			\fmf{wboson,label=$\W^+$,label.side=left}{v3,v1}
		\end{fmfgraph*}
	\end{gathered}
\qquad\quad
	\begin{gathered}
		\begin{fmfgraph*}(60,60)
			\fmfset{curly_len}{2mm}
			\fmftop{t1} \fmfbottom{b1,b2}
			\fmf{quark,label=$\hat d_r$,label.side=left,tension=2}{b1,v1}
			\fmf{quark,label=$u_v$,label.side=left}{v1,v2}
			\fmf{quark,label=$u_u$,label.side=left}{v2,v3}
			\fmf{quark,label=$\hat d_p$,label.side=left,tension=2}{v3,b2}
			\fmf{photon,label=$\hat \A$,tension=3}{t1,v2}
			\fmffreeze
			\fmf{scalar,label=$G^+$,label.side=left}{v3,v1}
		\end{fmfgraph*}
	\end{gathered}
\nn
	&\begin{gathered}
		\begin{fmfgraph*}(60,60)
			\fmfset{curly_len}{2mm}
			\fmftop{t1} \fmfbottom{b1,b2}
			\fmf{quark,label=$\hat d_r$,label.side=left,tension=2}{b1,v1}
			\fmf{wboson,label=$\W^+$,label.side=right}{v3,v2,v1}
			\fmf{quark,label=$\hat d_p$,label.side=left,tension=2}{v3,b2}
			\fmf{photon,label=$\hat \A$,tension=3}{t1,v2}
			\fmffreeze
			\fmf{quark,label=$u_u$}{v1,v3}
		\end{fmfgraph*}
	\end{gathered}
\qquad\quad
	\begin{gathered}
		\begin{fmfgraph*}(60,60)
			\fmfset{curly_len}{2mm}
			\fmftop{t1} \fmfbottom{b1,b2}
			\fmf{quark,label=$\hat d_r$,label.side=left,tension=2}{b1,v1}
			\fmf{scalar,label=$G^+$,label.side=right}{v3,v2,v1}
			\fmf{quark,label=$\hat d_p$,label.side=left,tension=2}{v3,b2}
			\fmf{photon,label=$\hat \A$,tension=3}{t1,v2}
			\fmffreeze
			\fmf{quark,label=$u_u$}{v1,v3}
		\end{fmfgraph*}
	\end{gathered}
\qquad\quad
	\begin{gathered}
		\begin{fmfgraph*}(60,60)
			\fmfset{curly_len}{2mm}
			\fmftop{t1} \fmfbottom{b1,b2}
			\fmf{quark,label=$\hat d_r$,label.side=left,tension=2}{b1,v1}
			\fmf{wboson,label=$\W^+$,label.side=right}{v3,v2}
			\fmf{scalar,label=$G^+$,label.side=right}{v2,v1}
			\fmf{quark,label=$\hat d_p$,label.side=left,tension=2}{v3,b2}
			\fmf{photon,label=$\hat \A$,tension=3}{t1,v2}
			\fmffreeze
			\fmf{quark,label=$u_u$}{v1,v3}
			\fmfv{decoration.shape=square,decoration.size=1.5mm}{v2}
		\end{fmfgraph*}
	\end{gathered}
\qquad\quad
	\begin{gathered}
		\begin{fmfgraph*}(60,60)
			\fmfset{curly_len}{2mm}
			\fmftop{t1} \fmfbottom{b1,b2}
			\fmf{quark,label=$\hat d_r$,label.side=left,tension=2}{b1,v1}
			\fmf{scalar,label=$G^+$,label.side=right}{v3,v2}
			\fmf{wboson,label=$\W^+$,label.side=right}{v2,v1}
			\fmf{quark,label=$\hat d_p$,label.side=left,tension=2}{v3,b2}
			\fmf{photon,label=$\hat \A$,tension=3}{t1,v2}
			\fmffreeze
			\fmf{quark,label=$u_u$}{v1,v3}
			\fmfv{decoration.shape=square,decoration.size=1.5mm}{v2}
		\end{fmfgraph*}
	\end{gathered}
\qquad\quad
	\begin{gathered}
		\begin{fmfgraph*}(60,60)
			\fmfset{curly_len}{2mm}
			\fmftop{t1} \fmfbottom{b1,b2}
			\fmf{quark,label=$\hat d_r$,label.side=left,tension=2}{b1,v1}
			\fmf{dashes,label=$h$,label.side=right}{v3,v2}
			\fmf{photon,label=$\A$,label.side=right}{v2,v1}
			\fmf{quark,label=$\hat d_p$,label.side=left,tension=2}{v3,b2}
			\fmf{photon,label=$\hat \A$,tension=3}{t1,v2}
			\fmffreeze
			\fmf{quark,label=$d_u$}{v1,v3}
			\fmfv{decoration.shape=square,decoration.size=1.5mm}{v2}
		\end{fmfgraph*}
	\end{gathered}
\nn
	& \begin{gathered}
		\begin{fmfgraph*}(60,60)
			\fmfset{curly_len}{2mm}
			\fmftop{t1} \fmfbottom{b1,b2}
			\fmf{quark,label=$\hat d_r$,label.side=left,tension=2}{b1,v1}
			\fmf{photon,label=$\A$,label.side=right}{v3,v2}
			\fmf{dashes,label=$h$,label.side=right}{v2,v1}
			\fmf{quark,label=$\hat d_p$,label.side=left,tension=2}{v3,b2}
			\fmf{photon,label=$\hat \A$,tension=3}{t1,v2}
			\fmffreeze
			\fmf{quark,label=$d_u$}{v1,v3}
			\fmfv{decoration.shape=square,decoration.size=1.5mm}{v2}
		\end{fmfgraph*}
	\end{gathered}
\qquad\quad
	\begin{gathered}
		\begin{fmfgraph*}(60,60)
			\fmfset{curly_len}{2mm}
			\fmftop{t1} \fmfbottom{b1,b2}
			\fmf{quark,label=$\hat d_r$,label.side=left,tension=2}{b1,v1}
			\fmf{dashes,label=$h$,label.side=right}{v3,v2}
			\fmf{photon,label=$\Z$,label.side=right}{v2,v1}
			\fmf{quark,label=$\hat d_p$,label.side=left,tension=2}{v3,b2}
			\fmf{photon,label=$\hat \A$,tension=3}{t1,v2}
			\fmffreeze
			\fmf{quark,label=$d_u$}{v1,v3}
			\fmfv{decoration.shape=square,decoration.size=1.5mm}{v2}
		\end{fmfgraph*}
	\end{gathered}
\qquad\quad
	\begin{gathered}
		\begin{fmfgraph*}(60,60)
			\fmfset{curly_len}{2mm}
			\fmftop{t1} \fmfbottom{b1,b2}
			\fmf{quark,label=$\hat d_r$,label.side=left,tension=2}{b1,v1}
			\fmf{photon,label=$\Z$,label.side=right}{v3,v2}
			\fmf{dashes,label=$h$,label.side=right}{v2,v1}
			\fmf{quark,label=$\hat d_p$,label.side=left,tension=2}{v3,b2}
			\fmf{photon,label=$\hat \A$,tension=3}{t1,v2}
			\fmffreeze
			\fmf{quark,label=$d_u$}{v1,v3}
			\fmfv{decoration.shape=square,decoration.size=1.5mm}{v2}
		\end{fmfgraph*}
	\end{gathered}
\qquad\quad
	\begin{gathered}
		\begin{fmfgraph*}(60,60)
			\fmfset{curly_len}{2mm}
			\fmftop{t1} \fmfbottom{b1,b2}
			\fmf{quark,label=$\hat d_r$,label.side=right,tension=2}{b1,v1}
			\fmf{quark,label=$\hat d_p$,label.side=right,tension=2}{v1,b2}
			\fmf{photon,label=$\hat \A$,tension=2}{t1,v1}
			\fmf{scalar,label=$G^+$}{v1,v1}
			\fmffreeze
			\fmfv{decoration.shape=square,decoration.size=1.5mm}{v1}
		\end{fmfgraph*}
	\end{gathered}
\qquad\quad
	\begin{gathered}
		\begin{fmfgraph*}(60,60)
			\fmfset{curly_len}{2mm}
			\fmftop{t1} \fmfbottom{b1,b2}
			\fmf{quark,label=$\hat d_r$,label.side=right,tension=2}{b1,v2}
			\fmf{quark,label=$\hat d_p$,label.side=right,tension=2}{v2,b2}
			\fmf{photon,label=$\hat \A$,tension=2}{t1,v1}
			\fmf{quark,label=$t$,right}{v1,v2,v1}
			\fmffreeze
			\fmfv{decoration.shape=square,decoration.size=1.5mm}{v2}
		\end{fmfgraph*}
	\end{gathered}
\nn
	& \begin{gathered}
		\begin{fmfgraph*}(60,60)
			\fmfset{curly_len}{2mm}
			\fmftop{t1} \fmfbottom{b1,b2}
			\fmf{quark,label=$\hat d_r$,label.side=right,tension=2}{b1,v2}
			\fmf{quark,label=$\hat d_p$,label.side=right,tension=2}{v2,b2}
			\fmf{photon,label=$\hat \A$,tension=2}{t1,v1}
			\fmf{wboson,label=$\W^+$,right}{v1,v2,v1}
			\fmffreeze
			\fmfv{decoration.shape=square,decoration.size=1.5mm}{v2}
		\end{fmfgraph*}
	\end{gathered}
\qquad\quad
	\begin{gathered}
		\begin{fmfgraph*}(60,60)
			\fmfset{curly_len}{2mm}
			\fmftop{t1} \fmfbottom{b1,b2}
			\fmf{quark,label=$\hat d_r$,label.side=right,tension=2}{b1,v2}
			\fmf{quark,label=$\hat d_p$,label.side=right,tension=2}{v2,b2}
			\fmf{photon,label=$\hat \A$,tension=2}{t1,v1}
			\fmf{scalar,label=$G^+$,right}{v1,v2}
			\fmf{scalar,label=$G^+$,right}{v2,v1}
			\fmffreeze
			\fmfv{decoration.shape=square,decoration.size=1.5mm}{v2}
		\end{fmfgraph*}
	\end{gathered}
\qquad\quad
	\begin{gathered}
		\begin{fmfgraph*}(60,60)
			\fmfset{curly_len}{2mm}
			\fmftop{t1} \fmfbottom{b1,b2}
			\fmf{quark,label=$\hat d_r$,label.side=left,tension=0.67}{b1,v1}
			\fmf{quark,label=$\hat d_p$,label.side=left,tension=2}{v2,b2}
			\fmf{photon,label=$\hat \A$,tension=0.67}{t1,v1}
			\fmf{quark,label=$d_u$,label.side=right,tension=1}{v1,v2}
			\fmffreeze
			\fmf{dashes,label=$h$,left}{v1,v2}
			\fmfv{decoration.shape=square,decoration.size=1.5mm}{v1}
		\end{fmfgraph*}
	\end{gathered}
\qquad\quad
	\begin{gathered}
		\begin{fmfgraph*}(60,60)
			\fmfset{curly_len}{2mm}
			\fmftop{t1} \fmfbottom{b1,b2}
			\fmf{quark,label=$\hat d_r$,label.side=left,tension=0.67}{b1,v1}
			\fmf{quark,label=$\hat d_p$,label.side=left,tension=2}{v2,b2}
			\fmf{photon,label=$\hat \A$,tension=0.67}{t1,v1}
			\fmf{quark,label=$d_u$,label.side=right,tension=1}{v1,v2}
			\fmffreeze
			\fmf{dashes,label=$G^0$,left}{v1,v2}
			\fmfv{decoration.shape=square,decoration.size=1.5mm}{v1}
		\end{fmfgraph*}
	\end{gathered}
\qquad\quad
	\begin{gathered}
		\begin{fmfgraph*}(60,60)
			\fmfset{curly_len}{2mm}
			\fmftop{t1} \fmfbottom{b1,b2}
			\fmf{quark,label=$\hat d_r$,label.side=left,tension=0.67}{b1,v1}
			\fmf{quark,label=$\hat d_p$,label.side=left,tension=2}{v2,b2}
			\fmf{photon,label=$\hat \A$,tension=0.67}{t1,v1}
			\fmf{quark,label=$u_u$,label.side=right,tension=1}{v1,v2}
			\fmffreeze
			\fmf{wboson,label=$\W^+$,right}{v2,v1}
			\fmfv{decoration.shape=square,decoration.size=1.5mm}{v1}
		\end{fmfgraph*}
	\end{gathered}
\nn
	& \begin{gathered}
		\begin{fmfgraph*}(60,60)
			\fmfset{curly_len}{2mm}
			\fmftop{t1} \fmfbottom{b1,b2}
			\fmf{quark,label=$\hat d_r$,label.side=left,tension=0.67}{b1,v1}
			\fmf{quark,label=$\hat d_p$,label.side=left,tension=2}{v2,b2}
			\fmf{photon,label=$\hat \A$,tension=0.67}{t1,v1}
			\fmf{quark,label=$u_u$,label.side=right,tension=1}{v1,v2}
			\fmffreeze
			\fmf{scalar,label=$G^+$,right}{v2,v1}
			\fmfv{decoration.shape=square,decoration.size=1.5mm}{v1}
		\end{fmfgraph*}
	\end{gathered}
\qquad\quad
	\begin{gathered}
		\begin{fmfgraph*}(60,60)
			\fmfset{curly_len}{2mm}
			\fmftop{t1} \fmfbottom{b1,b2}
			\fmf{quark,label=$\hat d_r$,label.side=left,tension=2}{b1,v2}
			\fmf{quark,label=$\hat d_p$,label.side=left,tension=0.67}{v1,b2}
			\fmf{photon,label=$\hat \A$,tension=0.67}{t1,v1}
			\fmf{quark,label=$d_u$,label.side=right,tension=1}{v2,v1}
			\fmffreeze
			\fmf{dashes,label=$h$,right}{v1,v2}
			\fmfv{decoration.shape=square,decoration.size=1.5mm}{v1}
		\end{fmfgraph*}
	\end{gathered}
\qquad\quad
	\begin{gathered}
		\begin{fmfgraph*}(60,60)
			\fmfset{curly_len}{2mm}
			\fmftop{t1} \fmfbottom{b1,b2}
			\fmf{quark,label=$\hat d_r$,label.side=left,tension=2}{b1,v2}
			\fmf{quark,label=$\hat d_p$,label.side=left,tension=0.67}{v1,b2}
			\fmf{photon,label=$\hat \A$,tension=0.67}{t1,v1}
			\fmf{quark,label=$d_u$,label.side=right,tension=1}{v2,v1}
			\fmffreeze
			\fmf{dashes,label=$G^0$,right}{v1,v2}
			\fmfv{decoration.shape=square,decoration.size=1.5mm}{v1}
		\end{fmfgraph*}
	\end{gathered}
\qquad\quad
	\begin{gathered}
		\begin{fmfgraph*}(60,60)
			\fmfset{curly_len}{2mm}
			\fmftop{t1} \fmfbottom{b1,b2}
			\fmf{quark,label=$\hat d_r$,label.side=left,tension=2}{b1,v2}
			\fmf{quark,label=$\hat d_p$,label.side=left,tension=0.67}{v1,b2}
			\fmf{photon,label=$\hat \A$,tension=0.67}{t1,v1}
			\fmf{quark,label=$u_u$,label.side=right,tension=1}{v2,v1}
			\fmffreeze
			\fmf{wboson,label=$\W^+$,right}{v1,v2}
			\fmfv{decoration.shape=square,decoration.size=1.5mm}{v1}
		\end{fmfgraph*}
	\end{gathered}
\qquad\quad
	\begin{gathered}
		\begin{fmfgraph*}(60,60)
			\fmfset{curly_len}{2mm}
			\fmftop{t1} \fmfbottom{b1,b2}
			\fmf{quark,label=$\hat d_r$,label.side=left,tension=2}{b1,v2}
			\fmf{quark,label=$\hat d_p$,label.side=left,tension=0.67}{v1,b2}
			\fmf{photon,label=$\hat \A$,tension=0.67}{t1,v1}
			\fmf{quark,label=$u_u$,label.side=right,tension=1}{v2,v1}
			\fmffreeze
			\fmf{scalar,label=$G^+$,right}{v1,v2}
			\fmfv{decoration.shape=square,decoration.size=1.5mm}{v1}
		\end{fmfgraph*}
	\end{gathered}
\nn
	& \begin{gathered}
		\begin{fmfgraph*}(60,60)
			\fmfset{curly_len}{2mm}
			\fmftop{t1,t2,t3} \fmfbottom{b1,b2}
			\fmf{quark,label=$\hat d_r$,label.side=left,tension=2}{b1,v1}
			\fmf{quark,label=$\hat d_p$,label.side=right,tension=2}{v1,b2}
			\fmf{photon,label=$\hat \A$,tension=2}{t2,v1}
			\fmffreeze
			\fmf{dashes,tension=3}{v1,v2}
			\fmf{phantom,tension=6}{t3,v2}
			\fmf{phantom,tension=2}{v2,b2}
			\fmfblob{5mm}{v2}
			\fmfv{decoration.shape=square,decoration.size=1.5mm}{v1}
		\end{fmfgraph*}
	\end{gathered}
\end{align*}


\subsubsection[$\bar u u g$]{\boldmath$\bar u u g$}

\begin{align*}
	&\begin{gathered}
		\begin{fmfgraph*}(60,60)
			\fmfset{curly_len}{2mm}
			\fmftop{t1} \fmfbottom{b1,b2}
			\fmf{quark,label=$\hat u_r$,label.side=left,tension=2}{b1,v1}
			\fmf{quark,label=$u_v$,label.side=left}{v1,v2}
			\fmf{quark,label=$u_u$,label.side=left}{v2,v3}
			\fmf{quark,label=$\hat u_p$,label.side=left,tension=2}{v3,b2}
			\fmf{gluon,label=$\hat \G$,tension=3}{t1,v2}
			\fmffreeze
			\fmf{photon,label=$\Z$}{v1,v3}
		\end{fmfgraph*}
	\end{gathered}
\qquad\quad
	\begin{gathered}
		\begin{fmfgraph*}(60,60)
			\fmfset{curly_len}{2mm}
			\fmftop{t1} \fmfbottom{b1,b2}
			\fmf{quark,label=$\hat u_r$,label.side=left,tension=2}{b1,v1}
			\fmf{quark,label=$u_v$,label.side=left}{v1,v2}
			\fmf{quark,label=$u_u$,label.side=left}{v2,v3}
			\fmf{quark,label=$\hat u_p$,label.side=left,tension=2}{v3,b2}
			\fmf{gluon,label=$\hat \G$,tension=3}{t1,v2}
			\fmffreeze
			\fmf{dashes,label=$G^0$}{v1,v3}
		\end{fmfgraph*}
	\end{gathered}
\qquad\quad
	\begin{gathered}
		\begin{fmfgraph*}(60,60)
			\fmfset{curly_len}{2mm}
			\fmftop{t1} \fmfbottom{b1,b2}
			\fmf{quark,label=$\hat u_r$,label.side=left,tension=2}{b1,v1}
			\fmf{quark,label=$u_v$,label.side=left}{v1,v2}
			\fmf{quark,label=$u_u$,label.side=left}{v2,v3}
			\fmf{quark,label=$\hat u_p$,label.side=left,tension=2}{v3,b2}
			\fmf{gluon,label=$\hat \G$,tension=3}{t1,v2}
			\fmffreeze
			\fmf{dashes,label=$h$}{v1,v3}
		\end{fmfgraph*}
	\end{gathered}
\qquad\quad
	\begin{gathered}
		\begin{fmfgraph*}(60,60)
			\fmfset{curly_len}{2mm}
			\fmftop{t1} \fmfbottom{b1,b2}
			\fmf{quark,label=$\hat u_r$,label.side=left,tension=2}{b1,v1}
			\fmf{quark,label=$d_v$,label.side=left}{v1,v2}
			\fmf{quark,label=$d_u$,label.side=left}{v2,v3}
			\fmf{quark,label=$\hat u_p$,label.side=left,tension=2}{v3,b2}
			\fmf{gluon,label=$\hat \G$,tension=3}{t1,v2}
			\fmffreeze
			\fmf{wboson,label=$\W^+$}{v1,v3}
		\end{fmfgraph*}
	\end{gathered}
\qquad\quad
	\begin{gathered}
		\begin{fmfgraph*}(60,60)
			\fmfset{curly_len}{2mm}
			\fmftop{t1} \fmfbottom{b1,b2}
			\fmf{quark,label=$\hat u_r$,label.side=left,tension=2}{b1,v1}
			\fmf{quark,label=$d_v$,label.side=left}{v1,v2}
			\fmf{quark,label=$d_u$,label.side=left}{v2,v3}
			\fmf{quark,label=$\hat u_p$,label.side=left,tension=2}{v3,b2}
			\fmf{gluon,label=$\hat \G$,tension=3}{t1,v2}
			\fmffreeze
			\fmf{scalar,label=$G^+$}{v1,v3}
		\end{fmfgraph*}
	\end{gathered}
\nn
	&\begin{gathered}
		\begin{fmfgraph*}(60,60)
			\fmfset{curly_len}{2mm}
			\fmftop{t1} \fmfbottom{b1,b2}
			\fmf{quark,label=$\hat u_r$,label.side=left,tension=2}{b1,v1}
			\fmf{dashes,label=$h$,label.side=right}{v3,v2}
			\fmf{gluon,label=$\G$,label.side=right}{v2,v1}
			\fmf{quark,label=$\hat u_p$,label.side=left,tension=2}{v3,b2}
			\fmf{gluon,label=$\hat \G$,tension=3}{t1,v2}
			\fmffreeze
			\fmf{quark,label=$u_u$}{v1,v3}
			\fmfv{decoration.shape=square,decoration.size=1.5mm}{v2}
		\end{fmfgraph*}
	\end{gathered}
\qquad\quad
	\begin{gathered}
		\begin{fmfgraph*}(60,60)
			\fmfset{curly_len}{2mm}
			\fmftop{t1} \fmfbottom{b1,b2}
			\fmf{quark,label=$\hat u_r$,label.side=left,tension=2}{b1,v1}
			\fmf{gluon,label=$\G$,label.side=right}{v3,v2}
			\fmf{dashes,label=$h$,label.side=right}{v2,v1}
			\fmf{quark,label=$\hat u_p$,label.side=left,tension=2}{v3,b2}
			\fmf{gluon,label=$\hat \G$,tension=3}{t1,v2}
			\fmffreeze
			\fmf{quark,label=$u_u$}{v1,v3}
			\fmfv{decoration.shape=square,decoration.size=1.5mm}{v2}
		\end{fmfgraph*}
	\end{gathered}
\qquad\quad
	\begin{gathered}
		\begin{fmfgraph*}(60,60)
			\fmfset{curly_len}{2mm}
			\fmftop{t1} \fmfbottom{b1,b2}
			\fmf{quark,label=$\hat u_r$,label.side=right,tension=2}{b1,v2}
			\fmf{quark,label=$\hat u_p$,label.side=right,tension=2}{v2,b2}
			\fmf{gluon,label=$\hat \G$,tension=2}{t1,v1}
			\fmf{quark,label=$t$,right}{v1,v2,v1}
			\fmffreeze
			\fmfv{decoration.shape=square,decoration.size=1.5mm}{v2}
		\end{fmfgraph*}
	\end{gathered}
\qquad\quad
	\begin{gathered}
		\begin{fmfgraph*}(60,60)
			\fmfset{curly_len}{2mm}
			\fmftop{t1} \fmfbottom{b1,b2}
			\fmf{quark,label=$\hat u_r$,label.side=left,tension=0.67}{b1,v1}
			\fmf{quark,label=$\hat u_p$,label.side=left,tension=2}{v2,b2}
			\fmf{gluon,label=$\hat \G$,tension=0.67}{t1,v1}
			\fmf{quark,label=$u_u$,label.side=right,tension=1}{v1,v2}
			\fmffreeze
			\fmf{dashes,label=$h$,left}{v1,v2}
			\fmfv{decoration.shape=square,decoration.size=1.5mm}{v1}
		\end{fmfgraph*}
	\end{gathered}
\qquad\quad
	\begin{gathered}
		\begin{fmfgraph*}(60,60)
			\fmfset{curly_len}{2mm}
			\fmftop{t1} \fmfbottom{b1,b2}
			\fmf{quark,label=$\hat u_r$,label.side=left,tension=0.67}{b1,v1}
			\fmf{quark,label=$\hat u_p$,label.side=left,tension=2}{v2,b2}
			\fmf{gluon,label=$\hat \G$,tension=0.67}{t1,v1}
			\fmf{quark,label=$u_u$,label.side=right,tension=1}{v1,v2}
			\fmffreeze
			\fmf{dashes,label=$G^0$,left}{v1,v2}
			\fmfv{decoration.shape=square,decoration.size=1.5mm}{v1}
		\end{fmfgraph*}
	\end{gathered}
\nn
	& \begin{gathered}
		\begin{fmfgraph*}(60,60)
			\fmfset{curly_len}{2mm}
			\fmftop{t1} \fmfbottom{b1,b2}
			\fmf{quark,label=$\hat u_r$,label.side=left,tension=0.67}{b1,v1}
			\fmf{quark,label=$\hat u_p$,label.side=left,tension=2}{v2,b2}
			\fmf{gluon,label=$\hat \G$,tension=0.67}{t1,v1}
			\fmf{quark,label=$d_u$,label.side=right,tension=1}{v1,v2}
			\fmffreeze
			\fmf{scalar,label=$G^+$,left}{v1,v2}
			\fmfv{decoration.shape=square,decoration.size=1.5mm}{v1}
		\end{fmfgraph*}
	\end{gathered}
\qquad\quad
	\begin{gathered}
		\begin{fmfgraph*}(60,60)
			\fmfset{curly_len}{2mm}
			\fmftop{t1} \fmfbottom{b1,b2}
			\fmf{quark,label=$\hat u_r$,label.side=left,tension=2}{b1,v2}
			\fmf{quark,label=$\hat u_p$,label.side=left,tension=0.67}{v1,b2}
			\fmf{gluon,label=$\hat \G$,tension=0.67}{t1,v1}
			\fmf{quark,label=$u_u$,label.side=right,tension=1}{v2,v1}
			\fmffreeze
			\fmf{dashes,label=$h$,right}{v1,v2}
			\fmfv{decoration.shape=square,decoration.size=1.5mm}{v1}
		\end{fmfgraph*}
	\end{gathered}
\qquad\quad
	\begin{gathered}
		\begin{fmfgraph*}(60,60)
			\fmfset{curly_len}{2mm}
			\fmftop{t1} \fmfbottom{b1,b2}
			\fmf{quark,label=$\hat u_r$,label.side=left,tension=2}{b1,v2}
			\fmf{quark,label=$\hat u_p$,label.side=left,tension=0.67}{v1,b2}
			\fmf{gluon,label=$\hat \G$,tension=0.67}{t1,v1}
			\fmf{quark,label=$u_u$,label.side=right,tension=1}{v2,v1}
			\fmffreeze
			\fmf{dashes,label=$G^0$,right}{v1,v2}
			\fmfv{decoration.shape=square,decoration.size=1.5mm}{v1}
		\end{fmfgraph*}
	\end{gathered}
\qquad\quad
	\begin{gathered}
		\begin{fmfgraph*}(60,60)
			\fmfset{curly_len}{2mm}
			\fmftop{t1} \fmfbottom{b1,b2}
			\fmf{quark,label=$\hat u_r$,label.side=left,tension=2}{b1,v2}
			\fmf{quark,label=$\hat u_p$,label.side=left,tension=0.67}{v1,b2}
			\fmf{gluon,label=$\hat \G$,tension=0.67}{t1,v1}
			\fmf{quark,label=$d_u$,label.side=right,tension=1}{v2,v1}
			\fmffreeze
			\fmf{scalar,label=$G^+$,left}{v2,v1}
			\fmfv{decoration.shape=square,decoration.size=1.5mm}{v1}
		\end{fmfgraph*}
	\end{gathered}
\qquad\quad
	\begin{gathered}
		\begin{fmfgraph*}(60,60)
			\fmfset{curly_len}{2mm}
			\fmftop{t1,t2,t3} \fmfbottom{b1,b2}
			\fmf{quark,label=$\hat u_r$,label.side=left,tension=2}{b1,v1}
			\fmf{quark,label=$\hat u_p$,label.side=right,tension=2}{v1,b2}
			\fmf{gluon,label=$\hat \G$,tension=2}{t2,v1}
			\fmffreeze
			\fmf{dashes,tension=3}{v1,v2}
			\fmf{phantom,tension=6}{t3,v2}
			\fmf{phantom,tension=2}{v2,b2}
			\fmfblob{5mm}{v2}
			\fmfv{decoration.shape=square,decoration.size=1.5mm}{v1}
		\end{fmfgraph*}
	\end{gathered}
\end{align*}


\subsubsection[$\bar d d g$]{\boldmath$\bar d d g$}

\begin{align*}
	&\begin{gathered}
		\begin{fmfgraph*}(60,60)
			\fmfset{curly_len}{2mm}
			\fmftop{t1} \fmfbottom{b1,b2}
			\fmf{quark,label=$\hat d_r$,label.side=left,tension=2}{b1,v1}
			\fmf{quark,label=$d_v$,label.side=left}{v1,v2}
			\fmf{quark,label=$d_u$,label.side=left}{v2,v3}
			\fmf{quark,label=$\hat d_p$,label.side=left,tension=2}{v3,b2}
			\fmf{gluon,label=$\hat \G$,tension=3}{t1,v2}
			\fmffreeze
			\fmf{photon,label=$\Z$}{v1,v3}
		\end{fmfgraph*}
	\end{gathered}
\qquad\quad
	\begin{gathered}
		\begin{fmfgraph*}(60,60)
			\fmfset{curly_len}{2mm}
			\fmftop{t1} \fmfbottom{b1,b2}
			\fmf{quark,label=$\hat d_r$,label.side=left,tension=2}{b1,v1}
			\fmf{quark,label=$d_v$,label.side=left}{v1,v2}
			\fmf{quark,label=$d_u$,label.side=left}{v2,v3}
			\fmf{quark,label=$\hat d_p$,label.side=left,tension=2}{v3,b2}
			\fmf{gluon,label=$\hat \G$,tension=3}{t1,v2}
			\fmffreeze
			\fmf{dashes,label=$G^0$}{v1,v3}
		\end{fmfgraph*}
	\end{gathered}
\qquad\quad
	\begin{gathered}
		\begin{fmfgraph*}(60,60)
			\fmfset{curly_len}{2mm}
			\fmftop{t1} \fmfbottom{b1,b2}
			\fmf{quark,label=$\hat d_r$,label.side=left,tension=2}{b1,v1}
			\fmf{quark,label=$d_v$,label.side=left}{v1,v2}
			\fmf{quark,label=$d_u$,label.side=left}{v2,v3}
			\fmf{quark,label=$\hat d_p$,label.side=left,tension=2}{v3,b2}
			\fmf{gluon,label=$\hat \G$,tension=3}{t1,v2}
			\fmffreeze
			\fmf{dashes,label=$h$}{v1,v3}
		\end{fmfgraph*}
	\end{gathered}
\qquad\quad
	\begin{gathered}
		\begin{fmfgraph*}(60,60)
			\fmfset{curly_len}{2mm}
			\fmftop{t1} \fmfbottom{b1,b2}
			\fmf{quark,label=$\hat d_r$,label.side=left,tension=2}{b1,v1}
			\fmf{quark,label=$u_v$,label.side=left}{v1,v2}
			\fmf{quark,label=$u_u$,label.side=left}{v2,v3}
			\fmf{quark,label=$\hat d_p$,label.side=left,tension=2}{v3,b2}
			\fmf{gluon,label=$\hat \G$,tension=3}{t1,v2}
			\fmffreeze
			\fmf{wboson,label=$\W^+$,label.side=left}{v3,v1}
		\end{fmfgraph*}
	\end{gathered}
\qquad\quad
	\begin{gathered}
		\begin{fmfgraph*}(60,60)
			\fmfset{curly_len}{2mm}
			\fmftop{t1} \fmfbottom{b1,b2}
			\fmf{quark,label=$\hat d_r$,label.side=left,tension=2}{b1,v1}
			\fmf{quark,label=$u_v$,label.side=left}{v1,v2}
			\fmf{quark,label=$u_u$,label.side=left}{v2,v3}
			\fmf{quark,label=$\hat d_p$,label.side=left,tension=2}{v3,b2}
			\fmf{gluon,label=$\hat \G$,tension=3}{t1,v2}
			\fmffreeze
			\fmf{scalar,label=$G^+$,label.side=left}{v3,v1}
		\end{fmfgraph*}
	\end{gathered}
\nn
	&\begin{gathered}
		\begin{fmfgraph*}(60,60)
			\fmfset{curly_len}{2mm}
			\fmftop{t1} \fmfbottom{b1,b2}
			\fmf{quark,label=$\hat d_r$,label.side=left,tension=2}{b1,v1}
			\fmf{dashes,label=$h$,label.side=right}{v3,v2}
			\fmf{gluon,label=$\G$,label.side=right}{v2,v1}
			\fmf{quark,label=$\hat d_p$,label.side=left,tension=2}{v3,b2}
			\fmf{gluon,label=$\hat \G$,tension=3}{t1,v2}
			\fmffreeze
			\fmf{quark,label=$d_u$}{v1,v3}
			\fmfv{decoration.shape=square,decoration.size=1.5mm}{v2}
		\end{fmfgraph*}
	\end{gathered}
\qquad\quad
	\begin{gathered}
		\begin{fmfgraph*}(60,60)
			\fmfset{curly_len}{2mm}
			\fmftop{t1} \fmfbottom{b1,b2}
			\fmf{quark,label=$\hat d_r$,label.side=left,tension=2}{b1,v1}
			\fmf{gluon,label=$\G$,label.side=right}{v3,v2}
			\fmf{dashes,label=$h$,label.side=right}{v2,v1}
			\fmf{quark,label=$\hat d_p$,label.side=left,tension=2}{v3,b2}
			\fmf{gluon,label=$\hat \G$,tension=3}{t1,v2}
			\fmffreeze
			\fmf{quark,label=$d_u$}{v1,v3}
			\fmfv{decoration.shape=square,decoration.size=1.5mm}{v2}
		\end{fmfgraph*}
	\end{gathered}
\qquad\quad
	\begin{gathered}
		\begin{fmfgraph*}(60,60)
			\fmfset{curly_len}{2mm}
			\fmftop{t1} \fmfbottom{b1,b2}
			\fmf{quark,label=$\hat d_r$,label.side=right,tension=2}{b1,v2}
			\fmf{quark,label=$\hat d_p$,label.side=right,tension=2}{v2,b2}
			\fmf{gluon,label=$\hat \G$,tension=2}{t1,v1}
			\fmf{quark,label=$t$,right}{v1,v2,v1}
			\fmffreeze
			\fmfv{decoration.shape=square,decoration.size=1.5mm}{v2}
		\end{fmfgraph*}
	\end{gathered}
\qquad\quad
	\begin{gathered}
		\begin{fmfgraph*}(60,60)
			\fmfset{curly_len}{2mm}
			\fmftop{t1} \fmfbottom{b1,b2}
			\fmf{quark,label=$\hat d_r$,label.side=left,tension=0.67}{b1,v1}
			\fmf{quark,label=$\hat d_p$,label.side=left,tension=2}{v2,b2}
			\fmf{gluon,label=$\hat \G$,tension=0.67}{t1,v1}
			\fmf{quark,label=$d_u$,label.side=right,tension=1}{v1,v2}
			\fmffreeze
			\fmf{dashes,label=$h$,left}{v1,v2}
			\fmfv{decoration.shape=square,decoration.size=1.5mm}{v1}
		\end{fmfgraph*}
	\end{gathered}
\qquad\quad
	\begin{gathered}
		\begin{fmfgraph*}(60,60)
			\fmfset{curly_len}{2mm}
			\fmftop{t1} \fmfbottom{b1,b2}
			\fmf{quark,label=$\hat d_r$,label.side=left,tension=0.67}{b1,v1}
			\fmf{quark,label=$\hat d_p$,label.side=left,tension=2}{v2,b2}
			\fmf{gluon,label=$\hat \G$,tension=0.67}{t1,v1}
			\fmf{quark,label=$d_u$,label.side=right,tension=1}{v1,v2}
			\fmffreeze
			\fmf{dashes,label=$G^0$,left}{v1,v2}
			\fmfv{decoration.shape=square,decoration.size=1.5mm}{v1}
		\end{fmfgraph*}
	\end{gathered}
\nn
	& \begin{gathered}
		\begin{fmfgraph*}(60,60)
			\fmfset{curly_len}{2mm}
			\fmftop{t1} \fmfbottom{b1,b2}
			\fmf{quark,label=$\hat d_r$,label.side=left,tension=0.67}{b1,v1}
			\fmf{quark,label=$\hat d_p$,label.side=left,tension=2}{v2,b2}
			\fmf{gluon,label=$\hat \G$,tension=0.67}{t1,v1}
			\fmf{quark,label=$u_u$,label.side=right,tension=1}{v1,v2}
			\fmffreeze
			\fmf{scalar,label=$G^+$,right}{v2,v1}
			\fmfv{decoration.shape=square,decoration.size=1.5mm}{v1}
		\end{fmfgraph*}
	\end{gathered}
\qquad\quad
	\begin{gathered}
		\begin{fmfgraph*}(60,60)
			\fmfset{curly_len}{2mm}
			\fmftop{t1} \fmfbottom{b1,b2}
			\fmf{quark,label=$\hat d_r$,label.side=left,tension=2}{b1,v2}
			\fmf{quark,label=$\hat d_p$,label.side=left,tension=0.67}{v1,b2}
			\fmf{gluon,label=$\hat \G$,tension=0.67}{t1,v1}
			\fmf{quark,label=$d_u$,label.side=right,tension=1}{v2,v1}
			\fmffreeze
			\fmf{dashes,label=$h$,right}{v1,v2}
			\fmfv{decoration.shape=square,decoration.size=1.5mm}{v1}
		\end{fmfgraph*}
	\end{gathered}
\qquad\quad
	\begin{gathered}
		\begin{fmfgraph*}(60,60)
			\fmfset{curly_len}{2mm}
			\fmftop{t1} \fmfbottom{b1,b2}
			\fmf{quark,label=$\hat d_r$,label.side=left,tension=2}{b1,v2}
			\fmf{quark,label=$\hat d_p$,label.side=left,tension=0.67}{v1,b2}
			\fmf{gluon,label=$\hat \G$,tension=0.67}{t1,v1}
			\fmf{quark,label=$d_u$,label.side=right,tension=1}{v2,v1}
			\fmffreeze
			\fmf{dashes,label=$G^0$,right}{v1,v2}
			\fmfv{decoration.shape=square,decoration.size=1.5mm}{v1}
		\end{fmfgraph*}
	\end{gathered}
\qquad\quad
	\begin{gathered}
		\begin{fmfgraph*}(60,60)
			\fmfset{curly_len}{2mm}
			\fmftop{t1} \fmfbottom{b1,b2}
			\fmf{quark,label=$\hat d_r$,label.side=left,tension=2}{b1,v2}
			\fmf{quark,label=$\hat d_p$,label.side=left,tension=0.67}{v1,b2}
			\fmf{gluon,label=$\hat \G$,tension=0.67}{t1,v1}
			\fmf{quark,label=$u_u$,label.side=right,tension=1}{v2,v1}
			\fmffreeze
			\fmf{scalar,label=$G^+$,right}{v1,v2}
			\fmfv{decoration.shape=square,decoration.size=1.5mm}{v1}
		\end{fmfgraph*}
	\end{gathered}
\qquad\quad
	\begin{gathered}
		\begin{fmfgraph*}(60,60)
			\fmfset{curly_len}{2mm}
			\fmftop{t1,t2,t3} \fmfbottom{b1,b2}
			\fmf{quark,label=$\hat d_r$,label.side=left,tension=2}{b1,v1}
			\fmf{quark,label=$\hat d_p$,label.side=right,tension=2}{v1,b2}
			\fmf{gluon,label=$\hat \G$,tension=2}{t2,v1}
			\fmffreeze
			\fmf{dashes,tension=3}{v1,v2}
			\fmf{phantom,tension=6}{t3,v2}
			\fmf{phantom,tension=2}{v2,b2}
			\fmfblob{5mm}{v2}
			\fmfv{decoration.shape=square,decoration.size=1.5mm}{v1}
		\end{fmfgraph*}
	\end{gathered}
\end{align*}


\subsubsection[$g^3$]{\boldmath$g^3$}

\begin{align*}
	&\begin{gathered}
		\begin{fmfgraph*}(60,60)
			\fmfset{curly_len}{2mm}
			\fmftop{t1} \fmfbottom{b1,b2}
			\fmf{gluon,label=$\hat \G^B$,label.side=left,tension=4}{b1,v1}
			\fmf{gluon,label=$\hat \G^C$,label.side=left,tension=4}{v3,b2}
			\fmf{gluon,label=$\hat \G^A$,label.side=left,tension=4}{t1,v2}
			\fmf{quark,label=$t$,label.side=left}{v1,v2}
			\fmf{quark,label=$t$,label.side=left}{v2,v3}
			\fmf{quark,label=$t$,label.side=left}{v3,v1}
		\end{fmfgraph*}
	\end{gathered}
\qquad\quad
	\begin{gathered}
		\begin{fmfgraph*}(60,60)
			\fmfset{curly_len}{2mm}
			\fmftop{t1} \fmfbottom{b1,b2}
			\fmf{gluon,label=$\hat \G^B$,label.side=left,tension=4}{b1,v1}
			\fmf{gluon,label=$\hat \G^C$,label.side=left,tension=4}{v3,b2}
			\fmf{gluon,label=$\hat \G^A$,label.side=left,tension=4}{t1,v2}
			\fmf{quark,label=$t$,label.side=right}{v1,v3}
			\fmf{quark,label=$t$,label.side=right}{v3,v2}
			\fmf{quark,label=$t$,label.side=right}{v2,v1}
		\end{fmfgraph*}
	\end{gathered}
\qquad\quad
	\begin{gathered}
		\begin{fmfgraph*}(60,60)
			\fmfset{curly_len}{2mm}
			\fmftop{t1} \fmfbottom{b1,b2}
			\fmf{gluon,label=$\hat \G^B$,label.side=left,tension=4}{b1,v1}
			\fmf{gluon,label=$\hat \G^C$,label.side=right,tension=4}{v1,b2}
			\fmf{gluon,label=$\hat \G^A$,label.side=right,tension=4}{t1,v1}
			\fmf{scalar,label=$G^+$}{v1,v1}
			\fmffreeze
			\fmfv{decoration.shape=square,decoration.size=1.5mm}{v1}
		\end{fmfgraph*}
	\end{gathered}
\qquad\quad
	\begin{gathered}
		\begin{fmfgraph*}(60,60)
			\fmfset{curly_len}{2mm}
			\fmftop{t1} \fmfbottom{b1,b2}
			\fmf{gluon,label=$\hat \G^B$,label.side=left,tension=4}{b1,v1}
			\fmf{gluon,label=$\hat \G^C$,label.side=right,tension=4}{v1,b2}
			\fmf{gluon,label=$\hat \G^A$,label.side=right,tension=4}{t1,v1}
			\fmf{dashes,label=$G^0$}{v1,v1}
			\fmffreeze
			\fmfv{decoration.shape=square,decoration.size=1.5mm}{v1}
		\end{fmfgraph*}
	\end{gathered}
\qquad\quad
	\begin{gathered}
		\begin{fmfgraph*}(60,60)
			\fmfset{curly_len}{2mm}
			\fmftop{t1} \fmfbottom{b1,b2}
			\fmf{gluon,label=$\hat \G^B$,label.side=left,tension=4}{b1,v1}
			\fmf{gluon,label=$\hat \G^C$,label.side=right,tension=4}{v1,b2}
			\fmf{gluon,label=$\hat \G^A$,label.side=right,tension=4}{t1,v1}
			\fmf{dashes,label=$h$}{v1,v1}
			\fmffreeze
			\fmfv{decoration.shape=square,decoration.size=1.5mm}{v1}
		\end{fmfgraph*}
	\end{gathered}
\nn
	& \begin{gathered}
		\begin{fmfgraph*}(60,60)
			\fmfset{curly_len}{2mm}
			\fmftop{t1} \fmfbottom{b1,b2}
			\fmf{gluon,label=$\hat \G^B$,label.side=right,tension=4}{b1,v2}
			\fmf{gluon,label=$\hat \G^C$,label.side=right,tension=4}{v2,b2}
			\fmf{gluon,label=$\hat \G^A$,label.side=left,tension=4}{t1,v1}
			\fmf{quark,label=$t$,right=0.75}{v1,v2}
			\fmf{quark,label=$t$,right=0.75}{v2,v1}
			\fmffreeze
			\fmfv{decoration.shape=square,decoration.size=1.5mm}{v2}
		\end{fmfgraph*}
	\end{gathered}
\qquad\quad
	\begin{gathered}
		\begin{fmfgraph*}(60,60)
			\fmfset{curly_len}{2mm}
			\fmftop{t1} \fmfbottom{b1,b2}
			\fmf{gluon,label=$\hat \G^B$,label.side=right,tension=4}{b1,v1}
			\fmf{gluon,label=$\hat \G^C$,label.side=left,tension=4}{v2,b2}
			\fmf{gluon,label=$\hat \G^A$,label.side=left,tension=4}{t1,v2}
			\fmf{quark,label=$t$,right=0.75}{v1,v2}
			\fmf{quark,label=$t$,right=0.75}{v2,v1}
			\fmffreeze
			\fmfv{decoration.shape=square,decoration.size=1.5mm}{v2}
		\end{fmfgraph*}
	\end{gathered}
\qquad\quad
	\begin{gathered}
		\begin{fmfgraph*}(60,60)
			\fmfset{curly_len}{2mm}
			\fmftop{t1} \fmfbottom{b1,b2}
			\fmf{gluon,label=$\hat \G^B$,label.side=left,tension=4}{b1,v2}
			\fmf{gluon,label=$\hat \G^C$,label.side=right,tension=4}{v1,b2}
			\fmf{gluon,label=$\hat \G^A$,label.side=right,tension=4}{t1,v2}
			\fmf{quark,label=$t$,right=0.75}{v1,v2}
			\fmf{quark,label=$t$,right=0.75}{v2,v1}
			\fmffreeze
			\fmfv{decoration.shape=square,decoration.size=1.5mm}{v2}
		\end{fmfgraph*}
	\end{gathered}
\qquad\quad
	\begin{gathered}
		\begin{fmfgraph*}(60,60)
			\fmfset{curly_len}{2mm}
			\fmftop{t1,t2,t3} \fmfbottom{b1,b2}
			\fmf{gluon,label=$\hat \G^B$,label.side=left,tension=4}{b1,v1}
			\fmf{gluon,label=$\hat \G^C$,label.side=right,tension=4}{v1,b2}
			\fmf{gluon,label=$\hat \G^A$,label.side=right,tension=4}{t2,v1}
			\fmffreeze
			\fmf{dashes,tension=3}{v1,v2}
			\fmf{phantom,tension=6}{t3,v2}
			\fmf{phantom,tension=2}{v2,b2}
			\fmfblob{5mm}{v2}
			\fmfv{decoration.shape=square,decoration.size=1.5mm}{v1}
		\end{fmfgraph*}
	\end{gathered} \nn
\end{align*}


\subsection{Three-point functions with heavy bosons}
\label{sec:ThreePointDiagramsHeavy}

The remaining three-point functions only appear as sub-diagrams in the matching onto the four-fermion operators. Up to dimension six, in these diagrams all light scales can be set to zero.


\subsubsection[$\bar e e h$, $\nu^2 h$, $\bar u u h$, and $\bar d d h$]{\boldmath$\bar e e h$, $\nu^2 h$, $\bar u u h$, and $\bar d d h$}
\label{sec:ThreePointHiggs}

As discussed in Sect.~\ref{sec:FourFermionMatching}, the only possibly relevant diagrams with background Higgs fields are SM loop corrections to the $\bar d d h$ three-point function with internal top quarks:
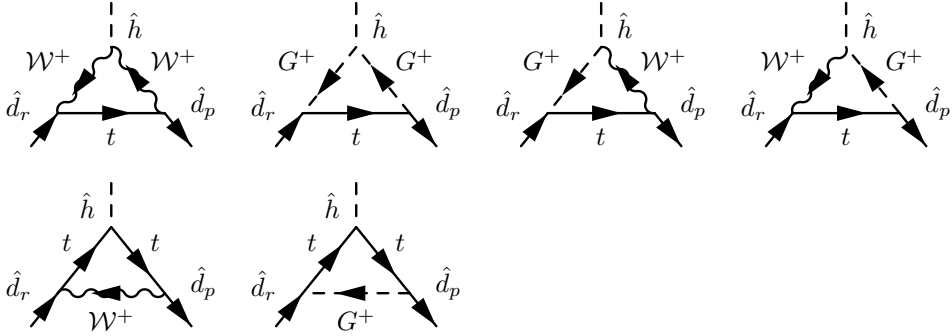
\begin{align*}
	&\begin{gathered}
		\begin{fmfgraph*}(60,60)
			\fmfset{curly_len}{2mm}
			\fmftop{t1} \fmfbottom{b1,b2}
			\fmf{quark,label=$\hat d_r$,label.side=left,tension=2}{b1,v1}
			\fmf{wboson,label=$\W^+$,label.side=right}{v3,v2,v1}
			\fmf{quark,label=$\hat d_p$,label.side=left,tension=2}{v3,b2}
			\fmf{dashes,label=$\hat h$,tension=3}{t1,v2}
			\fmffreeze
			\fmf{quark,label=$t$}{v1,v3}
		\end{fmfgraph*}
	\end{gathered}
\qquad\quad
	\begin{gathered}
		\begin{fmfgraph*}(60,60)
			\fmfset{curly_len}{2mm}
			\fmftop{t1} \fmfbottom{b1,b2}
			\fmf{quark,label=$\hat d_r$,label.side=left,tension=2}{b1,v1}
			\fmf{scalar,label=$G^+$,label.side=right}{v3,v2,v1}
			\fmf{quark,label=$\hat d_p$,label.side=left,tension=2}{v3,b2}
			\fmf{dashes,label=$\hat h$,tension=3}{t1,v2}
			\fmffreeze
			\fmf{quark,label=$t$}{v1,v3}
		\end{fmfgraph*}
	\end{gathered}
\qquad\quad
	\begin{gathered}
		\begin{fmfgraph*}(60,60)
			\fmfset{curly_len}{2mm}
			\fmftop{t1} \fmfbottom{b1,b2}
			\fmf{quark,label=$\hat d_r$,label.side=left,tension=2}{b1,v1}
			\fmf{wboson,label=$\W^+$,label.side=right}{v3,v2}
			\fmf{scalar,label=$G^+$,label.side=right}{v2,v1}
			\fmf{quark,label=$\hat d_p$,label.side=left,tension=2}{v3,b2}
			\fmf{dashes,label=$\hat h$,tension=3}{t1,v2}
			\fmffreeze
			\fmf{quark,label=$t$}{v1,v3}
		\end{fmfgraph*}
	\end{gathered}
\qquad\quad
	\begin{gathered}
		\begin{fmfgraph*}(60,60)
			\fmfset{curly_len}{2mm}
			\fmftop{t1} \fmfbottom{b1,b2}
			\fmf{quark,label=$\hat d_r$,label.side=left,tension=2}{b1,v1}
			\fmf{scalar,label=$G^+$,label.side=right}{v3,v2}
			\fmf{wboson,label=$\W^+$,label.side=right}{v2,v1}
			\fmf{quark,label=$\hat d_p$,label.side=left,tension=2}{v3,b2}
			\fmf{dashes,label=$\hat h$,tension=3}{t1,v2}
			\fmffreeze
			\fmf{quark,label=$t$}{v1,v3}
		\end{fmfgraph*}
	\end{gathered}
\nn
	& \begin{gathered}
		\begin{fmfgraph*}(60,60)
			\fmfset{curly_len}{2mm}
			\fmftop{t1} \fmfbottom{b1,b2}
			\fmf{quark,label=$\hat d_r$,label.side=left,tension=2}{b1,v1}
			\fmf{quark,label=$t$,label.side=left}{v1,v2,v3}
			\fmf{quark,label=$\hat d_p$,label.side=left,tension=2}{v3,b2}
			\fmf{dashes,label=$\hat h$,tension=3}{t1,v2}
			\fmffreeze
			\fmf{wboson,label=$\W^+$,label.side=left}{v3,v1}
		\end{fmfgraph*}
	\end{gathered}
\qquad\quad
	\begin{gathered}
		\begin{fmfgraph*}(60,60)
			\fmfset{curly_len}{2mm}
			\fmftop{t1} \fmfbottom{b1,b2}
			\fmf{quark,label=$\hat d_r$,label.side=left,tension=2}{b1,v1}
			\fmf{quark,label=$t$,label.side=left}{v1,v2,v3}
			\fmf{quark,label=$\hat d_p$,label.side=left,tension=2}{v3,b2}
			\fmf{dashes,label=$\hat h$,tension=3}{t1,v2}
			\fmffreeze
			\fmf{scalar,label=$G^+$,label.side=left}{v3,v1}
		\end{fmfgraph*}
	\end{gathered}
\end{align*}
However, with the light scales set to zero all these diagrams vanish.


\subsubsection[$\bar e e Z$]{\boldmath$\bar e e Z$}

\begin{align*}
	&\begin{gathered}
		\begin{fmfgraph*}(60,60)
			\fmfset{curly_len}{2mm}
			\fmftop{t1} \fmfbottom{b1,b2}
			\fmf{quark,label=$\hat e_r$,label.side=left,tension=2}{b1,v1}
			\fmf{quark,label=$e_v$,label.side=left}{v1,v2}
			\fmf{quark,label=$e_u$,label.side=left}{v2,v3}
			\fmf{quark,label=$\hat e_p$,label.side=left,tension=2}{v3,b2}
			\fmf{photon,label=$\hat \Z$,tension=3}{t1,v2}
			\fmffreeze
			\fmf{photon,label=$\Z$}{v1,v3}
		\end{fmfgraph*}
	\end{gathered}
\qquad\quad
	\begin{gathered}
		\begin{fmfgraph*}(60,60)
			\fmfset{curly_len}{2mm}
			\fmftop{t1} \fmfbottom{b1,b2}
			\fmf{quark,label=$\hat e_r$,label.side=left,tension=2}{b1,v1}
			\fmf{wboson,label=$\W^+$,label.side=right}{v3,v2,v1}
			\fmf{quark,label=$\hat e_p$,label.side=left,tension=2}{v3,b2}
			\fmf{photon,label=$\hat \Z$,tension=3}{t1,v2}
			\fmffreeze
			\fmf{plain,label=$\nu_u$}{v1,v3}
		\end{fmfgraph*}
	\end{gathered}
\qquad\quad
	\begin{gathered}
		\begin{fmfgraph*}(60,60)
			\fmfset{curly_len}{2mm}
			\fmftop{t1} \fmfbottom{b1,b2}
			\fmf{quark,label=$\hat e_r$,label.side=left,tension=2}{b1,v1}
			\fmf{plain,label=$\nu_v$,label.side=left}{v1,v2}
			\fmf{plain,label=$\nu_u$,label.side=left}{v2,v3}
			\fmf{quark,label=$\hat e_p$,label.side=left,tension=2}{v3,b2}
			\fmf{photon,label=$\hat \Z$,tension=3}{t1,v2}
			\fmffreeze
			\fmf{wboson,label=$\W^+$,label.side=left}{v3,v1}
		\end{fmfgraph*}
	\end{gathered}
\qquad\quad
	\begin{gathered}
		\begin{fmfgraph*}(60,60)
			\fmfset{curly_len}{2mm}
			\fmftop{t1} \fmfbottom{b1,b2}
			\fmf{quark,label=$\hat e_r$,label.side=left,tension=2}{b1,v1}
			\fmf{wboson,label=$\W^+$,label.side=right}{v3,v2}
			\fmf{scalar,label=$G^+$,label.side=right}{v2,v1}
			\fmf{quark,label=$\hat e_p$,label.side=left,tension=2}{v3,b2}
			\fmf{photon,label=$\hat \Z$,tension=3}{t1,v2}
			\fmffreeze
			\fmf{plain,label=$\nu_u$}{v1,v3}
			\fmfv{decoration.shape=square,decoration.size=1.5mm}{v1}
		\end{fmfgraph*}
	\end{gathered}
\qquad\quad
	\begin{gathered}
		\begin{fmfgraph*}(60,60)
			\fmfset{curly_len}{2mm}
			\fmftop{t1} \fmfbottom{b1,b2}
			\fmf{quark,label=$\hat e_r$,label.side=left,tension=2}{b1,v1}
			\fmf{scalar,label=$G^+$,label.side=right}{v3,v2}
			\fmf{wboson,label=$\W^+$,label.side=right}{v2,v1}
			\fmf{quark,label=$\hat e_p$,label.side=left,tension=2}{v3,b2}
			\fmf{photon,label=$\hat \Z$,tension=3}{t1,v2}
			\fmffreeze
			\fmf{plain,label=$\nu_u$}{v1,v3}
			\fmfv{decoration.shape=square,decoration.size=1.5mm}{v3}
		\end{fmfgraph*}
	\end{gathered}
\nn
	&\begin{gathered}
		\begin{fmfgraph*}(60,60)
			\fmfset{curly_len}{2mm}
			\fmftop{t1} \fmfbottom{b1,b2}
			\fmf{quark,label=$\hat e_r$,label.side=left,tension=2}{b1,v1}
			\fmf{dashes,label=$h$,label.side=right}{v3,v2}
			\fmf{photon,label=$\Z$,label.side=right}{v2,v1}
			\fmf{quark,label=$\hat e_p$,label.side=left,tension=2}{v3,b2}
			\fmf{photon,label=$\hat \Z$,tension=3}{t1,v2}
			\fmffreeze
			\fmf{quark,label=$e_u$}{v1,v3}
			\fmfv{decoration.shape=square,decoration.size=1.5mm}{v3}
		\end{fmfgraph*}
	\end{gathered}
\qquad\quad
	\begin{gathered}
		\begin{fmfgraph*}(60,60)
			\fmfset{curly_len}{2mm}
			\fmftop{t1} \fmfbottom{b1,b2}
			\fmf{quark,label=$\hat e_r$,label.side=left,tension=2}{b1,v1}
			\fmf{photon,label=$\Z$,label.side=right}{v3,v2}
			\fmf{dashes,label=$h$,label.side=right}{v2,v1}
			\fmf{quark,label=$\hat e_p$,label.side=left,tension=2}{v3,b2}
			\fmf{photon,label=$\hat \Z$,tension=3}{t1,v2}
			\fmffreeze
			\fmf{quark,label=$e_u$}{v1,v3}
			\fmfv{decoration.shape=square,decoration.size=1.5mm}{v1}
		\end{fmfgraph*}
	\end{gathered}
\qquad\quad
	\begin{gathered}
		\begin{fmfgraph*}(60,60)
			\fmfset{curly_len}{2mm}
			\fmftop{t1} \fmfbottom{b1,b2}
			\fmf{quark,label=$\hat e_r$,label.side=right,tension=2}{b1,v1}
			\fmf{quark,label=$\hat e_p$,label.side=right,tension=2}{v1,b2}
			\fmf{photon,label=$\hat \Z$,tension=2}{t1,v1}
			\fmf{scalar,label=$G^+$}{v1,v1}
			\fmffreeze
			\fmfv{decoration.shape=square,decoration.size=1.5mm}{v1}
		\end{fmfgraph*}
	\end{gathered}
\qquad\quad
	\begin{gathered}
		\begin{fmfgraph*}(60,60)
			\fmfset{curly_len}{2mm}
			\fmftop{t1} \fmfbottom{b1,b2}
			\fmf{quark,label=$\hat e_r$,label.side=right,tension=2}{b1,v1}
			\fmf{quark,label=$\hat e_p$,label.side=right,tension=2}{v1,b2}
			\fmf{photon,label=$\hat \Z$,tension=2}{t1,v1}
			\fmf{dashes,label=$G^0$}{v1,v1}
			\fmffreeze
			\fmfv{decoration.shape=square,decoration.size=1.5mm}{v1}
		\end{fmfgraph*}
	\end{gathered}
\qquad\quad
	\begin{gathered}
		\begin{fmfgraph*}(60,60)
			\fmfset{curly_len}{2mm}
			\fmftop{t1} \fmfbottom{b1,b2}
			\fmf{quark,label=$\hat e_r$,label.side=right,tension=2}{b1,v1}
			\fmf{quark,label=$\hat e_p$,label.side=right,tension=2}{v1,b2}
			\fmf{photon,label=$\hat \Z$,tension=2}{t1,v1}
			\fmf{dashes,label=$h$}{v1,v1}
			\fmffreeze
			\fmfv{decoration.shape=square,decoration.size=1.5mm}{v1}
		\end{fmfgraph*}
	\end{gathered}
\nn
	& \begin{gathered}
		\begin{fmfgraph*}(60,60)
			\fmfset{curly_len}{2mm}
			\fmftop{t1} \fmfbottom{b1,b2}
			\fmf{quark,label=$\hat e_r$,label.side=right,tension=2}{b1,v2}
			\fmf{quark,label=$\hat e_p$,label.side=right,tension=2}{v2,b2}
			\fmf{photon,label=$\hat \Z$,tension=2}{t1,v1}
			\fmf{quark,label=$t$,right}{v1,v2,v1}
			\fmffreeze
			\fmfv{decoration.shape=square,decoration.size=1.5mm}{v2}
		\end{fmfgraph*}
	\end{gathered}
\qquad\quad
	\begin{gathered}
		\begin{fmfgraph*}(60,60)
			\fmfset{curly_len}{2mm}
			\fmftop{t1} \fmfbottom{b1,b2}
			\fmf{quark,label=$\hat e_r$,label.side=right,tension=2}{b1,v2}
			\fmf{quark,label=$\hat e_p$,label.side=right,tension=2}{v2,b2}
			\fmf{photon,label=$\hat \Z$,tension=2}{t1,v1}
			\fmf{wboson,label=$\W^+$,right}{v1,v2,v1}
			\fmffreeze
			\fmfv{decoration.shape=square,decoration.size=1.5mm}{v2}
		\end{fmfgraph*}
	\end{gathered}
\qquad\quad
	\begin{gathered}
		\begin{fmfgraph*}(60,60)
			\fmfset{curly_len}{2mm}
			\fmftop{t1} \fmfbottom{b1,b2}
			\fmf{quark,label=$\hat e_r$,label.side=right,tension=2}{b1,v2}
			\fmf{quark,label=$\hat e_p$,label.side=right,tension=2}{v2,b2}
			\fmf{photon,label=$\hat \Z$,tension=2}{t1,v1}
			\fmf{wboson,label=$\W^+$,right}{v1,v2}
			\fmf{scalar,label=$G^+$,right}{v2,v1}
			\fmffreeze
			\fmfv{decoration.shape=square,decoration.size=1.5mm}{v2}
		\end{fmfgraph*}
	\end{gathered}
\qquad\quad
	\begin{gathered}
		\begin{fmfgraph*}(60,60)
			\fmfset{curly_len}{2mm}
			\fmftop{t1} \fmfbottom{b1,b2}
			\fmf{quark,label=$\hat e_r$,label.side=right,tension=2}{b1,v2}
			\fmf{quark,label=$\hat e_p$,label.side=right,tension=2}{v2,b2}
			\fmf{photon,label=$\hat \Z$,tension=2}{t1,v1}
			\fmf{scalar,label=$G^+$,right}{v1,v2}
			\fmf{wboson,label=$\W^+$,right}{v2,v1}
			\fmffreeze
			\fmfv{decoration.shape=square,decoration.size=1.5mm}{v2}
		\end{fmfgraph*}
	\end{gathered}
\qquad\quad
	\begin{gathered}
		\begin{fmfgraph*}(60,60)
			\fmfset{curly_len}{2mm}
			\fmftop{t1} \fmfbottom{b1,b2}
			\fmf{quark,label=$\hat e_r$,label.side=right,tension=2}{b1,v2}
			\fmf{quark,label=$\hat e_p$,label.side=right,tension=2}{v2,b2}
			\fmf{photon,label=$\hat \Z$,tension=2}{t1,v1}
			\fmf{scalar,label=$G^+$,right}{v1,v2}
			\fmf{scalar,label=$G^+$,right}{v2,v1}
			\fmffreeze
			\fmfv{decoration.shape=square,decoration.size=1.5mm}{v2}
		\end{fmfgraph*}
	\end{gathered}
\nn
	& \begin{gathered}
		\begin{fmfgraph*}(60,60)
			\fmfset{curly_len}{2mm}
			\fmftop{t1} \fmfbottom{b1,b2}
			\fmf{quark,label=$\hat e_r$,label.side=right,tension=2}{b1,v2}
			\fmf{quark,label=$\hat e_p$,label.side=right,tension=2}{v2,b2}
			\fmf{photon,label=$\hat \Z$,tension=2}{t1,v1}
			\fmf{dashes,label=$h$,right}{v1,v2}
			\fmf{photon,label=$\Z$,right}{v2,v1}
			\fmffreeze
			\fmfv{decoration.shape=square,decoration.size=1.5mm}{v2}
		\end{fmfgraph*}
	\end{gathered}
\qquad\quad
	\begin{gathered}
		\begin{fmfgraph*}(60,60)
			\fmfset{curly_len}{2mm}
			\fmftop{t1} \fmfbottom{b1,b2}
			\fmf{quark,label=$\hat e_r$,label.side=right,tension=2}{b1,v2}
			\fmf{quark,label=$\hat e_p$,label.side=right,tension=2}{v2,b2}
			\fmf{photon,label=$\hat \Z$,tension=2}{t1,v1}
			\fmf{dashes,label=$h$,right}{v1,v2}
			\fmf{dashes,label=$G^0$,right}{v2,v1}
			\fmffreeze
			\fmfv{decoration.shape=square,decoration.size=1.5mm}{v2}
		\end{fmfgraph*}
	\end{gathered}
\qquad\quad
	\begin{gathered}
		\begin{fmfgraph*}(60,60)
			\fmfset{curly_len}{2mm}
			\fmftop{t1} \fmfbottom{b1,b2}
			\fmf{quark,label=$\hat e_r$,label.side=left,tension=0.67}{b1,v1}
			\fmf{quark,label=$\hat e_p$,label.side=left,tension=2}{v2,b2}
			\fmf{photon,label=$\hat \Z$,tension=0.67}{t1,v1}
			\fmf{plain,label=$\nu_u$,label.side=right,tension=1}{v1,v2}
			\fmffreeze
			\fmf{wboson,label=$\W^+$,right}{v2,v1}
			\fmfv{decoration.shape=square,decoration.size=1.5mm}{v1}
		\end{fmfgraph*}
	\end{gathered}
\qquad\quad
	\begin{gathered}
		\begin{fmfgraph*}(60,60)
			\fmfset{curly_len}{2mm}
			\fmftop{t1} \fmfbottom{b1,b2}
			\fmf{quark,label=$\hat e_r$,label.side=left,tension=2}{b1,v2}
			\fmf{quark,label=$\hat e_p$,label.side=left,tension=0.67}{v1,b2}
			\fmf{photon,label=$\hat \Z$,tension=0.67}{t1,v1}
			\fmf{plain,label=$\nu_u$,label.side=right,tension=1}{v2,v1}
			\fmffreeze
			\fmf{wboson,label=$\W^+$,right}{v1,v2}
			\fmfv{decoration.shape=square,decoration.size=1.5mm}{v1}
		\end{fmfgraph*}
	\end{gathered}
\qquad\quad
	\begin{gathered}
		\begin{fmfgraph*}(60,60)
			\fmfset{curly_len}{2mm}
			\fmftop{t1,t2,t3} \fmfbottom{b1,b2}
			\fmf{quark,label=$\hat e_r$,label.side=left,tension=2}{b1,v1}
			\fmf{quark,label=$\hat e_p$,label.side=right,tension=2}{v1,b2}
			\fmf{photon,label=$\hat \Z$,tension=2}{t2,v1}
			\fmffreeze
			\fmf{dashes,tension=3}{v1,v2}
			\fmf{phantom,tension=6}{t3,v2}
			\fmf{phantom,tension=2}{v2,b2}
			\fmfblob{5mm}{v2}
			\fmfv{decoration.shape=square,decoration.size=1.5mm}{v1}
		\end{fmfgraph*}
	\end{gathered}
\end{align*}


\subsubsection[$\nu^2 Z$]{\boldmath$\nu^2 Z$}

\begin{align*}
	&\begin{gathered}
		\begin{fmfgraph*}(60,60)
			\fmfset{curly_len}{2mm}
			\fmftop{t1} \fmfbottom{b1,b2}
			\fmf{plain,label=$\hat \nu_r$,label.side=left,tension=2}{b1,v1}
			\fmf{plain,label=$\nu_v$,label.side=left}{v1,v2}
			\fmf{plain,label=$\nu_u$,label.side=left}{v2,v3}
			\fmf{plain,label=$\hat \nu_p$,label.side=left,tension=2}{v3,b2}
			\fmf{photon,label=$\hat \Z$,tension=3}{t1,v2}
			\fmffreeze
			\fmf{photon,label=$\Z$}{v1,v3}
		\end{fmfgraph*}
	\end{gathered}
\qquad\quad
	\begin{gathered}
		\begin{fmfgraph*}(60,60)
			\fmfset{curly_len}{2mm}
			\fmftop{t1} \fmfbottom{b1,b2}
			\fmf{plain,label=$\hat \nu_r$,label.side=left,tension=2}{b1,v1}
			\fmf{quark,label=$e_v$,label.side=left}{v1,v2}
			\fmf{quark,label=$e_u$,label.side=left}{v2,v3}
			\fmf{plain,label=$\hat \nu_p$,label.side=left,tension=2}{v3,b2}
			\fmf{photon,label=$\hat \Z$,tension=3}{t1,v2}
			\fmffreeze
			\fmf{wboson,label=$\W^+$}{v1,v3}
		\end{fmfgraph*}
	\end{gathered}
\qquad\quad
	\begin{gathered}
		\begin{fmfgraph*}(60,60)
			\fmfset{curly_len}{2mm}
			\fmftop{t1} \fmfbottom{b1,b2}
			\fmf{plain,label=$\hat \nu_r$,label.side=left,tension=2}{b1,v1}
			\fmf{quark,label=$e_v$,label.side=right}{v2,v1}
			\fmf{quark,label=$e_u$,label.side=right}{v3,v2}
			\fmf{plain,label=$\hat \nu_p$,label.side=left,tension=2}{v3,b2}
			\fmf{photon,label=$\hat \Z$,tension=3}{t1,v2}
			\fmffreeze
			\fmf{wboson,label=$\W^+$,label.side=left}{v3,v1}
		\end{fmfgraph*}
	\end{gathered}
\qquad\quad
	\begin{gathered}
		\begin{fmfgraph*}(60,60)
			\fmfset{curly_len}{2mm}
			\fmftop{t1} \fmfbottom{b1,b2}
			\fmf{plain,label=$\hat \nu_r$,label.side=left,tension=2}{b1,v1}
			\fmf{wboson,label=$\W^+$,label.side=left}{v1,v2,v3}
			\fmf{plain,label=$\hat \nu_p$,label.side=left,tension=2}{v3,b2}
			\fmf{photon,label=$\hat \Z$,tension=3}{t1,v2}
			\fmffreeze
			\fmf{quark,label=$e_u$}{v1,v3}
		\end{fmfgraph*}
	\end{gathered}
\qquad\quad
	\begin{gathered}
		\begin{fmfgraph*}(60,60)
			\fmfset{curly_len}{2mm}
			\fmftop{t1} \fmfbottom{b1,b2}
			\fmf{plain,label=$\hat \nu_r$,label.side=left,tension=2}{b1,v1}
			\fmf{wboson,label=$\W^+$,label.side=right}{v3,v2,v1}
			\fmf{plain,label=$\hat \nu_p$,label.side=left,tension=2}{v3,b2}
			\fmf{photon,label=$\hat \Z$,tension=3}{t1,v2}
			\fmffreeze
			\fmf{quark,label=$e_u$,label.side=left}{v3,v1}
		\end{fmfgraph*}
	\end{gathered}
\nn
	& \begin{gathered}
		\begin{fmfgraph*}(60,60)
			\fmfset{curly_len}{2mm}
			\fmftop{t1} \fmfbottom{b1,b2}
			\fmf{plain,label=$\hat \nu_r$,label.side=left,tension=2}{b1,v1}
			\fmf{wboson,label=$\W^+$,label.side=left}{v2,v3}
			\fmf{scalar,label=$G^+$,label.side=left}{v1,v2}
			\fmf{plain,label=$\hat \nu_p$,label.side=left,tension=2}{v3,b2}
			\fmf{photon,label=$\hat \Z$,tension=3}{t1,v2}
			\fmffreeze
			\fmf{quark,label=$e_u$}{v1,v3}
			\fmfv{decoration.shape=square,decoration.size=1.5mm}{v1}
		\end{fmfgraph*}
	\end{gathered}
\qquad\quad
	\begin{gathered}
		\begin{fmfgraph*}(60,60)
			\fmfset{curly_len}{2mm}
			\fmftop{t1} \fmfbottom{b1,b2}
			\fmf{plain,label=$\hat \nu_r$,label.side=left,tension=2}{b1,v1}
			\fmf{scalar,label=$G^+$,label.side=right}{v3,v2}
			\fmf{wboson,label=$\W^+$,label.side=right}{v2,v1}
			\fmf{plain,label=$\hat \nu_p$,label.side=left,tension=2}{v3,b2}
			\fmf{photon,label=$\hat \Z$,tension=3}{t1,v2}
			\fmffreeze
			\fmf{quark,label=$e_u$,label.side=left}{v3,v1}
			\fmfv{decoration.shape=square,decoration.size=1.5mm}{v3}
		\end{fmfgraph*}
	\end{gathered}
\qquad\quad
	\begin{gathered}
		\begin{fmfgraph*}(60,60)
			\fmfset{curly_len}{2mm}
			\fmftop{t1} \fmfbottom{b1,b2}
			\fmf{plain,label=$\hat \nu_r$,label.side=left,tension=2}{b1,v1}
			\fmf{scalar,label=$G^+$,label.side=left}{v2,v3}
			\fmf{wboson,label=$\W^+$,label.side=left}{v1,v2}
			\fmf{plain,label=$\hat \nu_p$,label.side=left,tension=2}{v3,b2}
			\fmf{photon,label=$\hat \Z$,tension=3}{t1,v2}
			\fmffreeze
			\fmf{quark,label=$e_u$}{v1,v3}
			\fmfv{decoration.shape=square,decoration.size=1.5mm}{v3}
		\end{fmfgraph*}
	\end{gathered}
\qquad\quad
	\begin{gathered}
		\begin{fmfgraph*}(60,60)
			\fmfset{curly_len}{2mm}
			\fmftop{t1} \fmfbottom{b1,b2}
			\fmf{plain,label=$\hat \nu_r$,label.side=left,tension=2}{b1,v1}
			\fmf{wboson,label=$\W^+$,label.side=right}{v3,v2}
			\fmf{scalar,label=$G^+$,label.side=right}{v2,v1}
			\fmf{plain,label=$\hat \nu_p$,label.side=left,tension=2}{v3,b2}
			\fmf{photon,label=$\hat \Z$,tension=3}{t1,v2}
			\fmffreeze
			\fmf{quark,label=$e_u$,label.side=left}{v3,v1}
			\fmfv{decoration.shape=square,decoration.size=1.5mm}{v1}
		\end{fmfgraph*}
	\end{gathered}
\qquad\quad
	\begin{gathered}
		\begin{fmfgraph*}(60,60)
			\fmfset{curly_len}{2mm}
			\fmftop{t1} \fmfbottom{b1,b2}
			\fmf{plain,label=$\hat \nu_r$,label.side=left,tension=2}{b1,v1}
			\fmf{dashes,label=$h$,label.side=left}{v2,v3}
			\fmf{photon,label=$\Z$,label.side=left}{v1,v2}
			\fmf{plain,label=$\hat \nu_p$,label.side=left,tension=2}{v3,b2}
			\fmf{photon,label=$\hat \Z$,tension=3}{t1,v2}
			\fmffreeze
			\fmf{plain,label=$\nu_u$}{v1,v3}
			\fmfv{decoration.shape=square,decoration.size=1.5mm}{v3}
		\end{fmfgraph*}
	\end{gathered}
\nn
	& \begin{gathered}
		\begin{fmfgraph*}(60,60)
			\fmfset{curly_len}{2mm}
			\fmftop{t1} \fmfbottom{b1,b2}
			\fmf{plain,label=$\hat \nu_r$,label.side=left,tension=2}{b1,v1}
			\fmf{photon,label=$\Z$,label.side=right}{v3,v2}
			\fmf{dashes,label=$h$,label.side=right}{v2,v1}
			\fmf{plain,label=$\hat \nu_p$,label.side=left,tension=2}{v3,b2}
			\fmf{photon,label=$\hat \Z$,tension=3}{t1,v2}
			\fmffreeze
			\fmf{plain,label=$\nu_u$,label.side=left}{v3,v1}
			\fmfv{decoration.shape=square,decoration.size=1.5mm}{v1}
		\end{fmfgraph*}
	\end{gathered}
\qquad\quad
	\begin{gathered}
		\begin{fmfgraph*}(60,60)
			\fmfset{curly_len}{2mm}
			\fmftop{t1} \fmfbottom{b1,b2}
			\fmf{plain,label=$\hat \nu_r$,label.side=right,tension=2}{b1,v1}
			\fmf{plain,label=$\hat \nu_p$,label.side=right,tension=2}{v1,b2}
			\fmf{photon,label=$\hat \Z$,tension=2}{t1,v1}
			\fmf{scalar,label=$G^+$}{v1,v1}
			\fmffreeze
			\fmfv{decoration.shape=square,decoration.size=1.5mm}{v1}
		\end{fmfgraph*}
	\end{gathered}
\qquad\quad
	\begin{gathered}
		\begin{fmfgraph*}(60,60)
			\fmfset{curly_len}{2mm}
			\fmftop{t1} \fmfbottom{b1,b2}
			\fmf{plain,label=$\hat \nu_r$,label.side=right,tension=2}{b1,v1}
			\fmf{plain,label=$\hat \nu_p$,label.side=right,tension=2}{v1,b2}
			\fmf{photon,label=$\hat \Z$,tension=2}{t1,v1}
			\fmf{dashes,label=$G^0$}{v1,v1}
			\fmffreeze
			\fmfv{decoration.shape=square,decoration.size=1.5mm}{v1}
		\end{fmfgraph*}
	\end{gathered}
\qquad\quad
	\begin{gathered}
		\begin{fmfgraph*}(60,60)
			\fmfset{curly_len}{2mm}
			\fmftop{t1} \fmfbottom{b1,b2}
			\fmf{plain,label=$\hat \nu_r$,label.side=right,tension=2}{b1,v1}
			\fmf{plain,label=$\hat \nu_p$,label.side=right,tension=2}{v1,b2}
			\fmf{photon,label=$\hat \Z$,tension=2}{t1,v1}
			\fmf{dashes,label=$h$}{v1,v1}
			\fmffreeze
			\fmfv{decoration.shape=square,decoration.size=1.5mm}{v1}
		\end{fmfgraph*}
	\end{gathered}
\qquad\quad
	\begin{gathered}
		\begin{fmfgraph*}(60,60)
			\fmfset{curly_len}{2mm}
			\fmftop{t1} \fmfbottom{b1,b2}
			\fmf{plain,label=$\hat \nu_r$,label.side=right,tension=2}{b1,v2}
			\fmf{plain,label=$\hat \nu_p$,label.side=right,tension=2}{v2,b2}
			\fmf{photon,label=$\hat \Z$,tension=2}{t1,v1}
			\fmf{quark,label=$t$,right}{v1,v2,v1}
			\fmffreeze
			\fmfv{decoration.shape=square,decoration.size=1.5mm}{v2}
		\end{fmfgraph*}
	\end{gathered}
\nn
	& \begin{gathered}
		\begin{fmfgraph*}(60,60)
			\fmfset{curly_len}{2mm}
			\fmftop{t1} \fmfbottom{b1,b2}
			\fmf{plain,label=$\hat \nu_r$,label.side=right,tension=2}{b1,v2}
			\fmf{plain,label=$\hat \nu_p$,label.side=right,tension=2}{v2,b2}
			\fmf{photon,label=$\hat \Z$,tension=2}{t1,v1}
			\fmf{wboson,label=$\W^+$,right}{v1,v2}
			\fmf{scalar,label=$G^+$,right}{v2,v1}
			\fmffreeze
			\fmfv{decoration.shape=square,decoration.size=1.5mm}{v2}
		\end{fmfgraph*}
	\end{gathered}
\qquad\quad
	\begin{gathered}
		\begin{fmfgraph*}(60,60)
			\fmfset{curly_len}{2mm}
			\fmftop{t1} \fmfbottom{b1,b2}
			\fmf{plain,label=$\hat \nu_r$,label.side=right,tension=2}{b1,v2}
			\fmf{plain,label=$\hat \nu_p$,label.side=right,tension=2}{v2,b2}
			\fmf{photon,label=$\hat \Z$,tension=2}{t1,v1}
			\fmf{scalar,label=$G^+$,right}{v1,v2}
			\fmf{wboson,label=$\W^+$,right}{v2,v1}
			\fmffreeze
			\fmfv{decoration.shape=square,decoration.size=1.5mm}{v2}
		\end{fmfgraph*}
	\end{gathered}
\qquad\quad
	\begin{gathered}
		\begin{fmfgraph*}(60,60)
			\fmfset{curly_len}{2mm}
			\fmftop{t1} \fmfbottom{b1,b2}
			\fmf{plain,label=$\hat \nu_r$,label.side=right,tension=2}{b1,v2}
			\fmf{plain,label=$\hat \nu_p$,label.side=right,tension=2}{v2,b2}
			\fmf{photon,label=$\hat \Z$,tension=2}{t1,v1}
			\fmf{scalar,label=$G^+$,right}{v1,v2}
			\fmf{scalar,label=$G^+$,right}{v2,v1}
			\fmffreeze
			\fmfv{decoration.shape=square,decoration.size=1.5mm}{v2}
		\end{fmfgraph*}
	\end{gathered}
\qquad\quad
	\begin{gathered}
		\begin{fmfgraph*}(60,60)
			\fmfset{curly_len}{2mm}
			\fmftop{t1} \fmfbottom{b1,b2}
			\fmf{plain,label=$\hat \nu_r$,label.side=right,tension=2}{b1,v2}
			\fmf{plain,label=$\hat \nu_p$,label.side=right,tension=2}{v2,b2}
			\fmf{photon,label=$\hat \Z$,tension=2}{t1,v1}
			\fmf{dashes,label=$h$,right}{v1,v2}
			\fmf{photon,label=$\Z$,right}{v2,v1}
			\fmffreeze
			\fmfv{decoration.shape=square,decoration.size=1.5mm}{v2}
		\end{fmfgraph*}
	\end{gathered}
\qquad\quad
	\begin{gathered}
		\begin{fmfgraph*}(60,60)
			\fmfset{curly_len}{2mm}
			\fmftop{t1} \fmfbottom{b1,b2}
			\fmf{plain,label=$\hat \nu_r$,label.side=right,tension=2}{b1,v2}
			\fmf{plain,label=$\hat \nu_p$,label.side=right,tension=2}{v2,b2}
			\fmf{photon,label=$\hat \Z$,tension=2}{t1,v1}
			\fmf{dashes,label=$h$,right}{v1,v2}
			\fmf{dashes,label=$G^0$,right}{v2,v1}
			\fmffreeze
			\fmfv{decoration.shape=square,decoration.size=1.5mm}{v2}
		\end{fmfgraph*}
	\end{gathered}
\nn
	& \begin{gathered}
		\begin{fmfgraph*}(60,60)
			\fmfset{curly_len}{2mm}
			\fmftop{t1} \fmfbottom{b1,b2}
			\fmf{plain,label=$\hat \nu_r$,label.side=left,tension=0.67}{b1,v1}
			\fmf{plain,label=$\hat \nu_p$,label.side=left,tension=2}{v2,b2}
			\fmf{photon,label=$\hat \Z$,tension=0.67}{t1,v1}
			\fmf{quark,label=$e_u$,label.side=right,tension=1}{v1,v2}
			\fmffreeze
			\fmf{wboson,label=$\W^+$,left}{v1,v2}
			\fmfv{decoration.shape=square,decoration.size=1.5mm}{v1}
		\end{fmfgraph*}
	\end{gathered}
\qquad\quad
	\begin{gathered}
		\begin{fmfgraph*}(60,60)
			\fmfset{curly_len}{2mm}
			\fmftop{t1} \fmfbottom{b1,b2}
			\fmf{plain,label=$\hat \nu_r$,label.side=left,tension=0.67}{b1,v1}
			\fmf{plain,label=$\hat \nu_p$,label.side=left,tension=2}{v2,b2}
			\fmf{photon,label=$\hat \Z$,tension=0.67}{t1,v1}
			\fmf{quark,label=$e_u$,label.side=left,tension=1}{v2,v1}
			\fmffreeze
			\fmf{wboson,label=$\W^+$,right}{v2,v1}
			\fmfv{decoration.shape=square,decoration.size=1.5mm}{v1}
		\end{fmfgraph*}
	\end{gathered}
\qquad\quad
	\begin{gathered}
		\begin{fmfgraph*}(60,60)
			\fmfset{curly_len}{2mm}
			\fmftop{t1} \fmfbottom{b1,b2}
			\fmf{plain,label=$\hat \nu_r$,label.side=left,tension=2}{b1,v2}
			\fmf{plain,label=$\hat \nu_p$,label.side=left,tension=0.67}{v1,b2}
			\fmf{photon,label=$\hat \Z$,tension=0.67}{t1,v1}
			\fmf{quark,label=$e_u$,label.side=right,tension=1}{v2,v1}
			\fmffreeze
			\fmf{wboson,label=$\W^+$,left}{v2,v1}
			\fmfv{decoration.shape=square,decoration.size=1.5mm}{v1}
		\end{fmfgraph*}
	\end{gathered}
\qquad\quad
	\begin{gathered}
		\begin{fmfgraph*}(60,60)
			\fmfset{curly_len}{2mm}
			\fmftop{t1} \fmfbottom{b1,b2}
			\fmf{plain,label=$\hat \nu_r$,label.side=left,tension=2}{b1,v2}
			\fmf{plain,label=$\hat \nu_p$,label.side=left,tension=0.67}{v1,b2}
			\fmf{photon,label=$\hat \Z$,tension=0.67}{t1,v1}
			\fmf{quark,label=$e_u$,label.side=left,tension=1}{v1,v2}
			\fmffreeze
			\fmf{wboson,label=$\W^+$,right}{v1,v2}
			\fmfv{decoration.shape=square,decoration.size=1.5mm}{v1}
		\end{fmfgraph*}
	\end{gathered}
\qquad\quad
	\begin{gathered}
		\begin{fmfgraph*}(60,60)
			\fmfset{curly_len}{2mm}
			\fmftop{t1,t2,t3} \fmfbottom{b1,b2}
			\fmf{plain,label=$\hat \nu_r$,label.side=left,tension=2}{b1,v1}
			\fmf{plain,label=$\hat \nu_p$,label.side=right,tension=2}{v1,b2}
			\fmf{photon,label=$\hat \Z$,tension=2}{t2,v1}
			\fmffreeze
			\fmf{dashes,tension=3}{v1,v2}
			\fmf{phantom,tension=6}{t3,v2}
			\fmf{phantom,tension=2}{v2,b2}
			\fmfblob{5mm}{v2}
			\fmfv{decoration.shape=square,decoration.size=1.5mm}{v1}
		\end{fmfgraph*}
	\end{gathered}
\end{align*}


\subsubsection[$\bar u u Z$]{\boldmath$\bar u u Z$}

\begin{align*}
	&\begin{gathered}
		\begin{fmfgraph*}(60,60)
			\fmfset{curly_len}{2mm}
			\fmftop{t1} \fmfbottom{b1,b2}
			\fmf{quark,label=$\hat u_r$,label.side=left,tension=2}{b1,v1}
			\fmf{quark,label=$u_v$,label.side=left}{v1,v2}
			\fmf{quark,label=$u_u$,label.side=left}{v2,v3}
			\fmf{quark,label=$\hat u_p$,label.side=left,tension=2}{v3,b2}
			\fmf{photon,label=$\hat \Z$,tension=3}{t1,v2}
			\fmffreeze
			\fmf{photon,label=$\Z$}{v1,v3}
		\end{fmfgraph*}
	\end{gathered}
\qquad\quad
	\begin{gathered}
		\begin{fmfgraph*}(60,60)
			\fmfset{curly_len}{2mm}
			\fmftop{t1} \fmfbottom{b1,b2}
			\fmf{quark,label=$\hat u_r$,label.side=left,tension=2}{b1,v1}
			\fmf{quark,label=$d_v$,label.side=left}{v1,v2}
			\fmf{quark,label=$d_u$,label.side=left}{v2,v3}
			\fmf{quark,label=$\hat u_p$,label.side=left,tension=2}{v3,b2}
			\fmf{photon,label=$\hat \Z$,tension=3}{t1,v2}
			\fmffreeze
			\fmf{wboson,label=$\W^+$}{v1,v3}
		\end{fmfgraph*}
	\end{gathered}
\qquad\quad
	\begin{gathered}
		\begin{fmfgraph*}(60,60)
			\fmfset{curly_len}{2mm}
			\fmftop{t1} \fmfbottom{b1,b2}
			\fmf{quark,label=$\hat u_r$,label.side=left,tension=2}{b1,v1}
			\fmf{wboson,label=$\W^+$,label.side=left}{v1,v2,v3}
			\fmf{quark,label=$\hat u_p$,label.side=left,tension=2}{v3,b2}
			\fmf{photon,label=$\hat \Z$,tension=3}{t1,v2}
			\fmffreeze
			\fmf{quark,label=$d_u$}{v1,v3}
		\end{fmfgraph*}
	\end{gathered}
\qquad\quad
	\begin{gathered}
		\begin{fmfgraph*}(60,60)
			\fmfset{curly_len}{2mm}
			\fmftop{t1} \fmfbottom{b1,b2}
			\fmf{quark,label=$\hat u_r$,label.side=left,tension=2}{b1,v1}
			\fmf{wboson,label=$\W^+$,label.side=left}{v2,v3}
			\fmf{scalar,label=$G^+$,label.side=left}{v1,v2}
			\fmf{quark,label=$\hat u_p$,label.side=left,tension=2}{v3,b2}
			\fmf{photon,label=$\hat \Z$,tension=3}{t1,v2}
			\fmffreeze
			\fmf{quark,label=$d_u$}{v1,v3}
			\fmfv{decoration.shape=square,decoration.size=1.5mm}{v1}
		\end{fmfgraph*}
	\end{gathered}
\qquad\quad
	\begin{gathered}
		\begin{fmfgraph*}(60,60)
			\fmfset{curly_len}{2mm}
			\fmftop{t1} \fmfbottom{b1,b2}
			\fmf{quark,label=$\hat u_r$,label.side=left,tension=2}{b1,v1}
			\fmf{scalar,label=$G^+$,label.side=left}{v2,v3}
			\fmf{wboson,label=$\W^+$,label.side=left}{v1,v2}
			\fmf{quark,label=$\hat u_p$,label.side=left,tension=2}{v3,b2}
			\fmf{photon,label=$\hat \Z$,tension=3}{t1,v2}
			\fmffreeze
			\fmf{quark,label=$d_u$}{v1,v3}
			\fmfv{decoration.shape=square,decoration.size=1.5mm}{v3}
		\end{fmfgraph*}
	\end{gathered}
\nn
	& \begin{gathered}
		\begin{fmfgraph*}(60,60)
			\fmfset{curly_len}{2mm}
			\fmftop{t1} \fmfbottom{b1,b2}
			\fmf{quark,label=$\hat u_r$,label.side=left,tension=2}{b1,v1}
			\fmf{dashes,label=$h$,label.side=right}{v3,v2}
			\fmf{photon,label=$\Z$,label.side=right}{v2,v1}
			\fmf{quark,label=$\hat u_p$,label.side=left,tension=2}{v3,b2}
			\fmf{photon,label=$\hat \Z$,tension=3}{t1,v2}
			\fmffreeze
			\fmf{quark,label=$u_u$}{v1,v3}
			\fmfv{decoration.shape=square,decoration.size=1.5mm}{v3}
		\end{fmfgraph*}
	\end{gathered}
\qquad\quad
	\begin{gathered}
		\begin{fmfgraph*}(60,60)
			\fmfset{curly_len}{2mm}
			\fmftop{t1} \fmfbottom{b1,b2}
			\fmf{quark,label=$\hat u_r$,label.side=left,tension=2}{b1,v1}
			\fmf{photon,label=$\Z$,label.side=right}{v3,v2}
			\fmf{dashes,label=$h$,label.side=right}{v2,v1}
			\fmf{quark,label=$\hat u_p$,label.side=left,tension=2}{v3,b2}
			\fmf{photon,label=$\hat \Z$,tension=3}{t1,v2}
			\fmffreeze
			\fmf{quark,label=$u_u$}{v1,v3}
			\fmfv{decoration.shape=square,decoration.size=1.5mm}{v1}
		\end{fmfgraph*}
	\end{gathered}
\qquad\quad
	\begin{gathered}
		\begin{fmfgraph*}(60,60)
			\fmfset{curly_len}{2mm}
			\fmftop{t1} \fmfbottom{b1,b2}
			\fmf{quark,label=$\hat u_r$,label.side=right,tension=2}{b1,v1}
			\fmf{quark,label=$\hat u_p$,label.side=right,tension=2}{v1,b2}
			\fmf{photon,label=$\hat \Z$,tension=2}{t1,v1}
			\fmf{scalar,label=$G^+$}{v1,v1}
			\fmffreeze
			\fmfv{decoration.shape=square,decoration.size=1.5mm}{v1}
		\end{fmfgraph*}
	\end{gathered}
\qquad\quad
	\begin{gathered}
		\begin{fmfgraph*}(60,60)
			\fmfset{curly_len}{2mm}
			\fmftop{t1} \fmfbottom{b1,b2}
			\fmf{quark,label=$\hat u_r$,label.side=right,tension=2}{b1,v1}
			\fmf{quark,label=$\hat u_p$,label.side=right,tension=2}{v1,b2}
			\fmf{photon,label=$\hat \Z$,tension=2}{t1,v1}
			\fmf{dashes,label=$G^0$}{v1,v1}
			\fmffreeze
			\fmfv{decoration.shape=square,decoration.size=1.5mm}{v1}
		\end{fmfgraph*}
	\end{gathered}
\qquad\quad
	\begin{gathered}
		\begin{fmfgraph*}(60,60)
			\fmfset{curly_len}{2mm}
			\fmftop{t1} \fmfbottom{b1,b2}
			\fmf{quark,label=$\hat u_r$,label.side=right,tension=2}{b1,v1}
			\fmf{quark,label=$\hat u_p$,label.side=right,tension=2}{v1,b2}
			\fmf{photon,label=$\hat \Z$,tension=2}{t1,v1}
			\fmf{dashes,label=$h$}{v1,v1}
			\fmffreeze
			\fmfv{decoration.shape=square,decoration.size=1.5mm}{v1}
		\end{fmfgraph*}
	\end{gathered}
\nn
	& \begin{gathered}
		\begin{fmfgraph*}(60,60)
			\fmfset{curly_len}{2mm}
			\fmftop{t1} \fmfbottom{b1,b2}
			\fmf{quark,label=$\hat u_r$,label.side=right,tension=2}{b1,v2}
			\fmf{quark,label=$\hat u_p$,label.side=right,tension=2}{v2,b2}
			\fmf{photon,label=$\hat \Z$,tension=2}{t1,v1}
			\fmf{quark,label=$t$,right}{v1,v2,v1}
			\fmffreeze
			\fmfv{decoration.shape=square,decoration.size=1.5mm}{v2}
		\end{fmfgraph*}
	\end{gathered}
\qquad\quad
	\begin{gathered}
		\begin{fmfgraph*}(60,60)
			\fmfset{curly_len}{2mm}
			\fmftop{t1} \fmfbottom{b1,b2}
			\fmf{quark,label=$\hat u_r$,label.side=right,tension=2}{b1,v2}
			\fmf{quark,label=$\hat u_p$,label.side=right,tension=2}{v2,b2}
			\fmf{photon,label=$\hat \Z$,tension=2}{t1,v1}
			\fmf{wboson,label=$\W^+$,right}{v1,v2,v1}
			\fmffreeze
			\fmfv{decoration.shape=square,decoration.size=1.5mm}{v2}
		\end{fmfgraph*}
	\end{gathered}
\qquad\quad
	\begin{gathered}
		\begin{fmfgraph*}(60,60)
			\fmfset{curly_len}{2mm}
			\fmftop{t1} \fmfbottom{b1,b2}
			\fmf{quark,label=$\hat u_r$,label.side=right,tension=2}{b1,v2}
			\fmf{quark,label=$\hat u_p$,label.side=right,tension=2}{v2,b2}
			\fmf{photon,label=$\hat \Z$,tension=2}{t1,v1}
			\fmf{wboson,label=$\W^+$,right}{v1,v2}
			\fmf{scalar,label=$G^+$,right}{v2,v1}
			\fmffreeze
			\fmfv{decoration.shape=square,decoration.size=1.5mm}{v2}
		\end{fmfgraph*}
	\end{gathered}
\qquad\quad
	\begin{gathered}
		\begin{fmfgraph*}(60,60)
			\fmfset{curly_len}{2mm}
			\fmftop{t1} \fmfbottom{b1,b2}
			\fmf{quark,label=$\hat u_r$,label.side=right,tension=2}{b1,v2}
			\fmf{quark,label=$\hat u_p$,label.side=right,tension=2}{v2,b2}
			\fmf{photon,label=$\hat \Z$,tension=2}{t1,v1}
			\fmf{scalar,label=$G^+$,right}{v1,v2}
			\fmf{wboson,label=$\W^+$,right}{v2,v1}
			\fmffreeze
			\fmfv{decoration.shape=square,decoration.size=1.5mm}{v2}
		\end{fmfgraph*}
	\end{gathered}
\qquad\quad
	\begin{gathered}
		\begin{fmfgraph*}(60,60)
			\fmfset{curly_len}{2mm}
			\fmftop{t1} \fmfbottom{b1,b2}
			\fmf{quark,label=$\hat u_r$,label.side=right,tension=2}{b1,v2}
			\fmf{quark,label=$\hat u_p$,label.side=right,tension=2}{v2,b2}
			\fmf{photon,label=$\hat \Z$,tension=2}{t1,v1}
			\fmf{scalar,label=$G^+$,right}{v1,v2}
			\fmf{scalar,label=$G^+$,right}{v2,v1}
			\fmffreeze
			\fmfv{decoration.shape=square,decoration.size=1.5mm}{v2}
		\end{fmfgraph*}
	\end{gathered}
\nn
	& \begin{gathered}
		\begin{fmfgraph*}(60,60)
			\fmfset{curly_len}{2mm}
			\fmftop{t1} \fmfbottom{b1,b2}
			\fmf{quark,label=$\hat u_r$,label.side=right,tension=2}{b1,v2}
			\fmf{quark,label=$\hat u_p$,label.side=right,tension=2}{v2,b2}
			\fmf{photon,label=$\hat \Z$,tension=2}{t1,v1}
			\fmf{dashes,label=$h$,right}{v1,v2}
			\fmf{photon,label=$\Z$,right}{v2,v1}
			\fmffreeze
			\fmfv{decoration.shape=square,decoration.size=1.5mm}{v2}
		\end{fmfgraph*}
	\end{gathered}
\qquad\quad
	\begin{gathered}
		\begin{fmfgraph*}(60,60)
			\fmfset{curly_len}{2mm}
			\fmftop{t1} \fmfbottom{b1,b2}
			\fmf{quark,label=$\hat u_r$,label.side=right,tension=2}{b1,v2}
			\fmf{quark,label=$\hat u_p$,label.side=right,tension=2}{v2,b2}
			\fmf{photon,label=$\hat \Z$,tension=2}{t1,v1}
			\fmf{dashes,label=$h$,right}{v1,v2}
			\fmf{dashes,label=$G^0$,right}{v2,v1}
			\fmffreeze
			\fmfv{decoration.shape=square,decoration.size=1.5mm}{v2}
		\end{fmfgraph*}
	\end{gathered}
\qquad\quad
	\begin{gathered}
		\begin{fmfgraph*}(60,60)
			\fmfset{curly_len}{2mm}
			\fmftop{t1} \fmfbottom{b1,b2}
			\fmf{quark,label=$\hat u_r$,label.side=left,tension=0.67}{b1,v1}
			\fmf{quark,label=$\hat u_p$,label.side=left,tension=2}{v2,b2}
			\fmf{photon,label=$\hat \Z$,tension=0.67}{t1,v1}
			\fmf{quark,label=$d_u$,label.side=right,tension=1}{v1,v2}
			\fmffreeze
			\fmf{wboson,label=$\W^+$,left}{v1,v2}
			\fmfv{decoration.shape=square,decoration.size=1.5mm}{v1}
		\end{fmfgraph*}
	\end{gathered}
\qquad\quad
	\begin{gathered}
		\begin{fmfgraph*}(60,60)
			\fmfset{curly_len}{2mm}
			\fmftop{t1} \fmfbottom{b1,b2}
			\fmf{quark,label=$\hat u_r$,label.side=left,tension=2}{b1,v2}
			\fmf{quark,label=$\hat u_p$,label.side=left,tension=0.67}{v1,b2}
			\fmf{photon,label=$\hat \Z$,tension=0.67}{t1,v1}
			\fmf{quark,label=$d_u$,label.side=right,tension=1}{v2,v1}
			\fmffreeze
			\fmf{wboson,label=$\W^+$,left}{v2,v1}
			\fmfv{decoration.shape=square,decoration.size=1.5mm}{v1}
		\end{fmfgraph*}
	\end{gathered}
\qquad\quad
	\begin{gathered}
		\begin{fmfgraph*}(60,60)
			\fmfset{curly_len}{2mm}
			\fmftop{t1,t2,t3} \fmfbottom{b1,b2}
			\fmf{quark,label=$\hat u_r$,label.side=left,tension=2}{b1,v1}
			\fmf{quark,label=$\hat u_p$,label.side=right,tension=2}{v1,b2}
			\fmf{photon,label=$\hat \Z$,tension=2}{t2,v1}
			\fmffreeze
			\fmf{dashes,tension=3}{v1,v2}
			\fmf{phantom,tension=6}{t3,v2}
			\fmf{phantom,tension=2}{v2,b2}
			\fmfblob{5mm}{v2}
			\fmfv{decoration.shape=square,decoration.size=1.5mm}{v1}
		\end{fmfgraph*}
	\end{gathered}
\end{align*}


\subsubsection[$\bar d d Z$]{\boldmath$\bar d d Z$}

\begin{align*}
	&\begin{gathered}
		\begin{fmfgraph*}(60,60)
			\fmfset{curly_len}{2mm}
			\fmftop{t1} \fmfbottom{b1,b2}
			\fmf{quark,label=$\hat d_r$,label.side=left,tension=2}{b1,v1}
			\fmf{quark,label=$d_v$,label.side=left}{v1,v2}
			\fmf{quark,label=$d_u$,label.side=left}{v2,v3}
			\fmf{quark,label=$\hat d_p$,label.side=left,tension=2}{v3,b2}
			\fmf{photon,label=$\hat \Z$,tension=3}{t1,v2}
			\fmffreeze
			\fmf{photon,label=$\Z$}{v1,v3}
		\end{fmfgraph*}
	\end{gathered}
\qquad\quad
	\begin{gathered}
		\begin{fmfgraph*}(60,60)
			\fmfset{curly_len}{2mm}
			\fmftop{t1} \fmfbottom{b1,b2}
			\fmf{quark,label=$\hat d_r$,label.side=left,tension=2}{b1,v1}
			\fmf{quark,label=$u_v$,label.side=left}{v1,v2}
			\fmf{quark,label=$u_u$,label.side=left}{v2,v3}
			\fmf{quark,label=$\hat d_p$,label.side=left,tension=2}{v3,b2}
			\fmf{photon,label=$\hat \Z$,tension=3}{t1,v2}
			\fmffreeze
			\fmf{wboson,label=$\W^+$,label.side=left}{v3,v1}
		\end{fmfgraph*}
	\end{gathered}
\qquad\quad
	\begin{gathered}
		\begin{fmfgraph*}(60,60)
			\fmfset{curly_len}{2mm}
			\fmftop{t1} \fmfbottom{b1,b2}
			\fmf{quark,label=$\hat d_r$,label.side=left,tension=2}{b1,v1}
			\fmf{quark,label=$t$,label.side=left}{v1,v2}
			\fmf{quark,label=$t$,label.side=left}{v2,v3}
			\fmf{quark,label=$\hat d_p$,label.side=left,tension=2}{v3,b2}
			\fmf{photon,label=$\hat \Z$,tension=3}{t1,v2}
			\fmffreeze
			\fmf{scalar,label=$G^+$,label.side=left}{v3,v1}
		\end{fmfgraph*}
	\end{gathered}
\qquad\quad
	\begin{gathered}
		\begin{fmfgraph*}(60,60)
			\fmfset{curly_len}{2mm}
			\fmftop{t1} \fmfbottom{b1,b2}
			\fmf{quark,label=$\hat d_r$,label.side=left,tension=2}{b1,v1}
			\fmf{wboson,label=$\W^+$,label.side=right}{v3,v2,v1}
			\fmf{quark,label=$\hat d_p$,label.side=left,tension=2}{v3,b2}
			\fmf{photon,label=$\hat \Z$,tension=3}{t1,v2}
			\fmffreeze
			\fmf{quark,label=$u_u$}{v1,v3}
		\end{fmfgraph*}
	\end{gathered}
\qquad\quad
	\begin{gathered}
		\begin{fmfgraph*}(60,60)
			\fmfset{curly_len}{2mm}
			\fmftop{t1} \fmfbottom{b1,b2}
			\fmf{quark,label=$\hat d_r$,label.side=left,tension=2}{b1,v1}
			\fmf{wboson,label=$\W^+$,label.side=right}{v3,v2}
			\fmf{scalar,label=$G^+$,label.side=right}{v2,v1}
			\fmf{quark,label=$\hat d_p$,label.side=left,tension=2}{v3,b2}
			\fmf{photon,label=$\hat \Z$,tension=3}{t1,v2}
			\fmffreeze
			\fmf{quark,label=$u_u$}{v1,v3}
		\end{fmfgraph*}
	\end{gathered}
\nn
	& \begin{gathered}
		\begin{fmfgraph*}(60,60)
			\fmfset{curly_len}{2mm}
			\fmftop{t1} \fmfbottom{b1,b2}
			\fmf{quark,label=$\hat d_r$,label.side=left,tension=2}{b1,v1}
			\fmf{scalar,label=$G^+$,label.side=right}{v3,v2}
			\fmf{wboson,label=$\W^+$,label.side=right}{v2,v1}
			\fmf{quark,label=$\hat d_p$,label.side=left,tension=2}{v3,b2}
			\fmf{photon,label=$\hat \Z$,tension=3}{t1,v2}
			\fmffreeze
			\fmf{quark,label=$u_u$}{v1,v3}
		\end{fmfgraph*}
	\end{gathered}
\qquad\quad
	\begin{gathered}
		\begin{fmfgraph*}(60,60)
			\fmfset{curly_len}{2mm}
			\fmftop{t1} \fmfbottom{b1,b2}
			\fmf{quark,label=$\hat d_r$,label.side=left,tension=2}{b1,v1}
			\fmf{scalar,label=$G^+$,label.side=right}{v3,v2,v1}
			\fmf{quark,label=$\hat d_p$,label.side=left,tension=2}{v3,b2}
			\fmf{photon,label=$\hat \Z$,tension=3}{t1,v2}
			\fmffreeze
			\fmf{quark,label=$t$}{v1,v3}
		\end{fmfgraph*}
	\end{gathered}
\qquad\quad
	\begin{gathered}
		\begin{fmfgraph*}(60,60)
			\fmfset{curly_len}{2mm}
			\fmftop{t1} \fmfbottom{b1,b2}
			\fmf{quark,label=$\hat d_r$,label.side=left,tension=2}{b1,v1}
			\fmf{dashes,label=$h$,label.side=right}{v3,v2}
			\fmf{photon,label=$\Z$,label.side=right}{v2,v1}
			\fmf{quark,label=$\hat d_p$,label.side=left,tension=2}{v3,b2}
			\fmf{photon,label=$\hat \Z$,tension=3}{t1,v2}
			\fmffreeze
			\fmf{quark,label=$d_u$}{v1,v3}
			\fmfv{decoration.shape=square,decoration.size=1.5mm}{v3}
		\end{fmfgraph*}
	\end{gathered}
\qquad\quad
	\begin{gathered}
		\begin{fmfgraph*}(60,60)
			\fmfset{curly_len}{2mm}
			\fmftop{t1} \fmfbottom{b1,b2}
			\fmf{quark,label=$\hat d_r$,label.side=left,tension=2}{b1,v1}
			\fmf{photon,label=$\Z$,label.side=right}{v3,v2}
			\fmf{dashes,label=$h$,label.side=right}{v2,v1}
			\fmf{quark,label=$\hat d_p$,label.side=left,tension=2}{v3,b2}
			\fmf{photon,label=$\hat \Z$,tension=3}{t1,v2}
			\fmffreeze
			\fmf{quark,label=$d_u$}{v1,v3}
			\fmfv{decoration.shape=square,decoration.size=1.5mm}{v1}
		\end{fmfgraph*}
	\end{gathered}
\qquad\quad
	\begin{gathered}
		\begin{fmfgraph*}(60,60)
			\fmfset{curly_len}{2mm}
			\fmftop{t1} \fmfbottom{b1,b2}
			\fmf{quark,label=$\hat d_r$,label.side=right,tension=2}{b1,v1}
			\fmf{quark,label=$\hat d_p$,label.side=right,tension=2}{v1,b2}
			\fmf{photon,label=$\hat \Z$,tension=2}{t1,v1}
			\fmf{scalar,label=$G^+$}{v1,v1}
			\fmffreeze
			\fmfv{decoration.shape=square,decoration.size=1.5mm}{v1}
		\end{fmfgraph*}
	\end{gathered}
\nn
	& \begin{gathered}
		\begin{fmfgraph*}(60,60)
			\fmfset{curly_len}{2mm}
			\fmftop{t1} \fmfbottom{b1,b2}
			\fmf{quark,label=$\hat d_r$,label.side=right,tension=2}{b1,v1}
			\fmf{quark,label=$\hat d_p$,label.side=right,tension=2}{v1,b2}
			\fmf{photon,label=$\hat \Z$,tension=2}{t1,v1}
			\fmf{dashes,label=$G^0$}{v1,v1}
			\fmffreeze
			\fmfv{decoration.shape=square,decoration.size=1.5mm}{v1}
		\end{fmfgraph*}
	\end{gathered}
\qquad\quad
	\begin{gathered}
		\begin{fmfgraph*}(60,60)
			\fmfset{curly_len}{2mm}
			\fmftop{t1} \fmfbottom{b1,b2}
			\fmf{quark,label=$\hat d_r$,label.side=right,tension=2}{b1,v1}
			\fmf{quark,label=$\hat d_p$,label.side=right,tension=2}{v1,b2}
			\fmf{photon,label=$\hat \Z$,tension=2}{t1,v1}
			\fmf{dashes,label=$h$}{v1,v1}
			\fmffreeze
			\fmfv{decoration.shape=square,decoration.size=1.5mm}{v1}
		\end{fmfgraph*}
	\end{gathered}
\qquad\quad
	\begin{gathered}
		\begin{fmfgraph*}(60,60)
			\fmfset{curly_len}{2mm}
			\fmftop{t1} \fmfbottom{b1,b2}
			\fmf{quark,label=$\hat d_r$,label.side=right,tension=2}{b1,v2}
			\fmf{quark,label=$\hat d_p$,label.side=right,tension=2}{v2,b2}
			\fmf{photon,label=$\hat \Z$,tension=2}{t1,v1}
			\fmf{quark,label=$t$,right}{v1,v2,v1}
			\fmffreeze
			\fmfv{decoration.shape=square,decoration.size=1.5mm}{v2}
		\end{fmfgraph*}
	\end{gathered}
\qquad\quad
	\begin{gathered}
		\begin{fmfgraph*}(60,60)
			\fmfset{curly_len}{2mm}
			\fmftop{t1} \fmfbottom{b1,b2}
			\fmf{quark,label=$\hat d_r$,label.side=right,tension=2}{b1,v2}
			\fmf{quark,label=$\hat d_p$,label.side=right,tension=2}{v2,b2}
			\fmf{photon,label=$\hat \Z$,tension=2}{t1,v1}
			\fmf{wboson,label=$\W^+$,right}{v1,v2,v1}
			\fmffreeze
			\fmfv{decoration.shape=square,decoration.size=1.5mm}{v2}
		\end{fmfgraph*}
	\end{gathered}
\qquad\quad
	\begin{gathered}
		\begin{fmfgraph*}(60,60)
			\fmfset{curly_len}{2mm}
			\fmftop{t1} \fmfbottom{b1,b2}
			\fmf{quark,label=$\hat d_r$,label.side=right,tension=2}{b1,v2}
			\fmf{quark,label=$\hat d_p$,label.side=right,tension=2}{v2,b2}
			\fmf{photon,label=$\hat \Z$,tension=2}{t1,v1}
			\fmf{wboson,label=$\W^+$,right}{v1,v2}
			\fmf{scalar,label=$G^+$,right}{v2,v1}
			\fmffreeze
			\fmfv{decoration.shape=square,decoration.size=1.5mm}{v2}
		\end{fmfgraph*}
	\end{gathered}
\nn
	& \begin{gathered}
		\begin{fmfgraph*}(60,60)
			\fmfset{curly_len}{2mm}
			\fmftop{t1} \fmfbottom{b1,b2}
			\fmf{quark,label=$\hat d_r$,label.side=right,tension=2}{b1,v2}
			\fmf{quark,label=$\hat d_p$,label.side=right,tension=2}{v2,b2}
			\fmf{photon,label=$\hat \Z$,tension=2}{t1,v1}
			\fmf{scalar,label=$G^+$,right}{v1,v2}
			\fmf{wboson,label=$\W^+$,right}{v2,v1}
			\fmffreeze
			\fmfv{decoration.shape=square,decoration.size=1.5mm}{v2}
		\end{fmfgraph*}
	\end{gathered}
\qquad\quad
	\begin{gathered}
		\begin{fmfgraph*}(60,60)
			\fmfset{curly_len}{2mm}
			\fmftop{t1} \fmfbottom{b1,b2}
			\fmf{quark,label=$\hat d_r$,label.side=right,tension=2}{b1,v2}
			\fmf{quark,label=$\hat d_p$,label.side=right,tension=2}{v2,b2}
			\fmf{photon,label=$\hat \Z$,tension=2}{t1,v1}
			\fmf{scalar,label=$G^+$,right}{v1,v2}
			\fmf{scalar,label=$G^+$,right}{v2,v1}
			\fmffreeze
			\fmfv{decoration.shape=square,decoration.size=1.5mm}{v2}
		\end{fmfgraph*}
	\end{gathered}
\qquad\quad
	\begin{gathered}
		\begin{fmfgraph*}(60,60)
			\fmfset{curly_len}{2mm}
			\fmftop{t1} \fmfbottom{b1,b2}
			\fmf{quark,label=$\hat d_r$,label.side=right,tension=2}{b1,v2}
			\fmf{quark,label=$\hat d_p$,label.side=right,tension=2}{v2,b2}
			\fmf{photon,label=$\hat \Z$,tension=2}{t1,v1}
			\fmf{dashes,label=$h$,right}{v1,v2}
			\fmf{photon,label=$\Z$,right}{v2,v1}
			\fmffreeze
			\fmfv{decoration.shape=square,decoration.size=1.5mm}{v2}
		\end{fmfgraph*}
	\end{gathered}
\qquad\quad
	\begin{gathered}
		\begin{fmfgraph*}(60,60)
			\fmfset{curly_len}{2mm}
			\fmftop{t1} \fmfbottom{b1,b2}
			\fmf{quark,label=$\hat d_r$,label.side=right,tension=2}{b1,v2}
			\fmf{quark,label=$\hat d_p$,label.side=right,tension=2}{v2,b2}
			\fmf{photon,label=$\hat \Z$,tension=2}{t1,v1}
			\fmf{dashes,label=$h$,right}{v1,v2}
			\fmf{dashes,label=$G^0$,right}{v2,v1}
			\fmffreeze
			\fmfv{decoration.shape=square,decoration.size=1.5mm}{v2}
		\end{fmfgraph*}
	\end{gathered}
\qquad\quad
	\begin{gathered}
		\begin{fmfgraph*}(60,60)
			\fmfset{curly_len}{2mm}
			\fmftop{t1} \fmfbottom{b1,b2}
			\fmf{quark,label=$\hat d_r$,label.side=left,tension=0.67}{b1,v1}
			\fmf{quark,label=$\hat d_p$,label.side=left,tension=2}{v2,b2}
			\fmf{photon,label=$\hat \Z$,tension=0.67}{t1,v1}
			\fmf{quark,label=$u_u$,label.side=right,tension=1}{v1,v2}
			\fmffreeze
			\fmf{wboson,label=$\W^+$,right}{v2,v1}
			\fmfv{decoration.shape=square,decoration.size=1.5mm}{v1}
		\end{fmfgraph*}
	\end{gathered}
\nn
	& \begin{gathered}
		\begin{fmfgraph*}(60,60)
			\fmfset{curly_len}{2mm}
			\fmftop{t1} \fmfbottom{b1,b2}
			\fmf{quark,label=$\hat d_r$,label.side=left,tension=0.67}{b1,v1}
			\fmf{quark,label=$\hat d_p$,label.side=left,tension=2}{v2,b2}
			\fmf{photon,label=$\hat \Z$,tension=0.67}{t1,v1}
			\fmf{quark,label=$t$,label.side=right,tension=1}{v1,v2}
			\fmffreeze
			\fmf{scalar,label=$G^+$,right}{v2,v1}
			\fmfv{decoration.shape=square,decoration.size=1.5mm}{v1}
		\end{fmfgraph*}
	\end{gathered}
\qquad\quad
	\begin{gathered}
		\begin{fmfgraph*}(60,60)
			\fmfset{curly_len}{2mm}
			\fmftop{t1} \fmfbottom{b1,b2}
			\fmf{quark,label=$\hat d_r$,label.side=left,tension=2}{b1,v2}
			\fmf{quark,label=$\hat d_p$,label.side=left,tension=0.67}{v1,b2}
			\fmf{photon,label=$\hat \Z$,tension=0.67}{t1,v1}
			\fmf{quark,label=$u_u$,label.side=right,tension=1}{v2,v1}
			\fmffreeze
			\fmf{wboson,label=$\W^+$,right}{v1,v2}
			\fmfv{decoration.shape=square,decoration.size=1.5mm}{v1}
		\end{fmfgraph*}
	\end{gathered}
\qquad\quad
	\begin{gathered}
		\begin{fmfgraph*}(60,60)
			\fmfset{curly_len}{2mm}
			\fmftop{t1} \fmfbottom{b1,b2}
			\fmf{quark,label=$\hat d_r$,label.side=left,tension=2}{b1,v2}
			\fmf{quark,label=$\hat d_p$,label.side=left,tension=0.67}{v1,b2}
			\fmf{photon,label=$\hat \Z$,tension=0.67}{t1,v1}
			\fmf{quark,label=$t$,label.side=right,tension=1}{v2,v1}
			\fmffreeze
			\fmf{scalar,label=$G^+$,right}{v1,v2}
			\fmfv{decoration.shape=square,decoration.size=1.5mm}{v1}
		\end{fmfgraph*}
	\end{gathered}
\qquad\quad
	\begin{gathered}
		\begin{fmfgraph*}(60,60)
			\fmfset{curly_len}{2mm}
			\fmftop{t1,t2,t3} \fmfbottom{b1,b2}
			\fmf{quark,label=$\hat d_r$,label.side=left,tension=2}{b1,v1}
			\fmf{quark,label=$\hat d_p$,label.side=right,tension=2}{v1,b2}
			\fmf{photon,label=$\hat \Z$,tension=2}{t2,v1}
			\fmffreeze
			\fmf{dashes,tension=3}{v1,v2}
			\fmf{phantom,tension=6}{t3,v2}
			\fmf{phantom,tension=2}{v2,b2}
			\fmfblob{5mm}{v2}
			\fmfv{decoration.shape=square,decoration.size=1.5mm}{v1}
		\end{fmfgraph*}
	\end{gathered}
\end{align*}


\subsubsection[$\nu e W^+$]{\boldmath$\nu e W^+$}

\begin{align*}
	&\begin{gathered}
		\begin{fmfgraph*}(60,60)
			\fmfset{curly_len}{2mm}
			\fmftop{t1} \fmfbottom{b1,b2}
			\fmf{quark,label=$\hat e_r$,label.side=left,tension=2}{b1,v1}
			\fmf{quark,label=$e_v$,label.side=left}{v1,v2}
			\fmf{plain,label=$\nu_u$,label.side=left}{v2,v3}
			\fmf{plain,label=$\hat \nu_p$,label.side=left,tension=2}{v3,b2}
			\fmf{wboson,label=$\hat \W^+$,tension=3}{t1,v2}
			\fmffreeze
			\fmf{photon,label=$\Z$}{v1,v3}
		\end{fmfgraph*}
	\end{gathered}
\qquad\quad
	\begin{gathered}
		\begin{fmfgraph*}(60,60)
			\fmfset{curly_len}{2mm}
			\fmftop{t1} \fmfbottom{b1,b2}
			\fmf{quark,label=$\hat e_r$,label.side=left,tension=2}{b1,v1}
			\fmf{plain,label=$\nu_v$,label.side=left}{v1,v2}
			\fmf{quark,label=$e_u$,label.side=right}{v3,v2}
			\fmf{plain,label=$\hat \nu_p$,label.side=left,tension=2}{v3,b2}
			\fmf{wboson,label=$\hat \W^+$,tension=3}{t1,v2}
			\fmffreeze
			\fmf{wboson,label=$\W^+$,label.side=left}{v3,v1}
		\end{fmfgraph*}
	\end{gathered}
\qquad\quad
	\begin{gathered}
		\begin{fmfgraph*}(60,60)
			\fmfset{curly_len}{2mm}
			\fmftop{t1} \fmfbottom{b1,b2}
			\fmf{quark,label=$\hat e_r$,label.side=left,tension=2}{b1,v1}
			\fmf{photon,label=$\Z$,label.side=right}{v2,v1}
			\fmf{wboson,label=$\W^+$,label.side=left}{v2,v3}
			\fmf{plain,label=$\hat \nu_p$,label.side=left,tension=2}{v3,b2}
			\fmf{wboson,label=$\hat \W^+$,tension=3}{t1,v2}
			\fmffreeze
			\fmf{quark,label=$e_u$}{v1,v3}
		\end{fmfgraph*}
	\end{gathered}
\qquad\quad
	\begin{gathered}
		\begin{fmfgraph*}(60,60)
			\fmfset{curly_len}{2mm}
			\fmftop{t1} \fmfbottom{b1,b2}
			\fmf{quark,label=$\hat e_r$,label.side=left,tension=2}{b1,v1}
			\fmf{photon,label=$\A$,label.side=right}{v2,v1}
			\fmf{wboson,label=$\W^+$,label.side=left}{v2,v3}
			\fmf{plain,label=$\hat \nu_p$,label.side=left,tension=2}{v3,b2}
			\fmf{wboson,label=$\hat \W^+$,tension=3}{t1,v2}
			\fmffreeze
			\fmf{quark,label=$e_u$}{v1,v3}
		\end{fmfgraph*}
	\end{gathered}
\qquad\quad
	\begin{gathered}
		\begin{fmfgraph*}(60,60)
			\fmfset{curly_len}{2mm}
			\fmftop{t1} \fmfbottom{b1,b2}
			\fmf{quark,label=$\hat e_r$,label.side=left,tension=2}{b1,v1}
			\fmf{dashes,label=$G^0$,label.side=right}{v2,v1}
			\fmf{wboson,label=$\W^+$,label.side=left}{v2,v3}
			\fmf{plain,label=$\hat \nu_p$,label.side=left,tension=2}{v3,b2}
			\fmf{wboson,label=$\hat \W^+$,tension=3}{t1,v2}
			\fmffreeze
			\fmf{quark,label=$e_u$}{v1,v3}
			\fmfv{decoration.shape=square,decoration.size=1.5mm}{v1}
		\end{fmfgraph*}
	\end{gathered}
\nn
	& \begin{gathered}
		\begin{fmfgraph*}(60,60)
			\fmfset{curly_len}{2mm}
			\fmftop{t1} \fmfbottom{b1,b2}
			\fmf{quark,label=$\hat e_r$,label.side=left,tension=2}{b1,v1}
			\fmf{dashes,label=$h$,label.side=right}{v2,v1}
			\fmf{wboson,label=$\W^+$,label.side=left}{v2,v3}
			\fmf{plain,label=$\hat \nu_p$,label.side=left,tension=2}{v3,b2}
			\fmf{wboson,label=$\hat \W^+$,tension=3}{t1,v2}
			\fmffreeze
			\fmf{quark,label=$e_u$}{v1,v3}
			\fmfv{decoration.shape=square,decoration.size=1.5mm}{v1}
		\end{fmfgraph*}
	\end{gathered}
\qquad\quad
	\begin{gathered}
		\begin{fmfgraph*}(60,60)
			\fmfset{curly_len}{2mm}
			\fmftop{t1} \fmfbottom{b1,b2}
			\fmf{quark,label=$\hat e_r$,label.side=left,tension=2}{b1,v1}
			\fmf{photon,label=$\Z$,label.side=right}{v2,v1}
			\fmf{scalar,label=$G^+$,label.side=left}{v2,v3}
			\fmf{plain,label=$\hat \nu_p$,label.side=left,tension=2}{v3,b2}
			\fmf{wboson,label=$\hat \W^+$,tension=3}{t1,v2}
			\fmffreeze
			\fmf{quark,label=$e_u$}{v1,v3}
			\fmfv{decoration.shape=square,decoration.size=1.5mm}{v3}
		\end{fmfgraph*}
	\end{gathered}
\qquad\quad
	\begin{gathered}
		\begin{fmfgraph*}(60,60)
			\fmfset{curly_len}{2mm}
			\fmftop{t1} \fmfbottom{b1,b2}
			\fmf{quark,label=$\hat e_r$,label.side=left,tension=2}{b1,v1}
			\fmf{photon,label=$\A$,label.side=right}{v2,v1}
			\fmf{scalar,label=$G^+$,label.side=left}{v2,v3}
			\fmf{plain,label=$\hat \nu_p$,label.side=left,tension=2}{v3,b2}
			\fmf{wboson,label=$\hat \W^+$,tension=3}{t1,v2}
			\fmffreeze
			\fmf{quark,label=$e_u$}{v1,v3}
			\fmfv{decoration.shape=square,decoration.size=1.5mm}{v3}
		\end{fmfgraph*}
	\end{gathered}
\qquad\quad
	\begin{gathered}
		\begin{fmfgraph*}(60,60)
			\fmfset{curly_len}{2mm}
			\fmftop{t1} \fmfbottom{b1,b2}
			\fmf{quark,label=$\hat e_r$,label.side=left,tension=2}{b1,v1}
			\fmf{wboson,label=$\W^+$,label.side=right}{v2,v1}
			\fmf{photon,label=$\Z$,label.side=left}{v2,v3}
			\fmf{plain,label=$\hat \nu_p$,label.side=left,tension=2}{v3,b2}
			\fmf{wboson,label=$\hat \W^+$,tension=3}{t1,v2}
			\fmffreeze
			\fmf{plain,label=$\nu_u$}{v1,v3}
		\end{fmfgraph*}
	\end{gathered}
\qquad\quad
	\begin{gathered}
		\begin{fmfgraph*}(60,60)
			\fmfset{curly_len}{2mm}
			\fmftop{t1} \fmfbottom{b1,b2}
			\fmf{quark,label=$\hat e_r$,label.side=left,tension=2}{b1,v1}
			\fmf{wboson,label=$\W^+$,label.side=right}{v2,v1}
			\fmf{dashes,label=$G^0$,label.side=left}{v2,v3}
			\fmf{plain,label=$\hat \nu_p$,label.side=left,tension=2}{v3,b2}
			\fmf{wboson,label=$\hat \W^+$,tension=3}{t1,v2}
			\fmffreeze
			\fmf{plain,label=$\nu_u$}{v1,v3}
			\fmfv{decoration.shape=square,decoration.size=1.5mm}{v3}
		\end{fmfgraph*}
	\end{gathered}
\nn
	& \begin{gathered}
		\begin{fmfgraph*}(60,60)
			\fmfset{curly_len}{2mm}
			\fmftop{t1} \fmfbottom{b1,b2}
			\fmf{quark,label=$\hat e_r$,label.side=left,tension=2}{b1,v1}
			\fmf{wboson,label=$\W^+$,label.side=right}{v2,v1}
			\fmf{dashes,label=$h$,label.side=left}{v2,v3}
			\fmf{plain,label=$\hat \nu_p$,label.side=left,tension=2}{v3,b2}
			\fmf{wboson,label=$\hat \W^+$,tension=3}{t1,v2}
			\fmffreeze
			\fmf{plain,label=$\nu_u$}{v1,v3}
			\fmfv{decoration.shape=square,decoration.size=1.5mm}{v3}
		\end{fmfgraph*}
	\end{gathered}
\qquad\quad
	\begin{gathered}
		\begin{fmfgraph*}(60,60)
			\fmfset{curly_len}{2mm}
			\fmftop{t1} \fmfbottom{b1,b2}
			\fmf{quark,label=$\hat e_r$,label.side=left,tension=2}{b1,v1}
			\fmf{scalar,label=$G^+$,label.side=right}{v2,v1}
			\fmf{photon,label=$\Z$,label.side=left}{v2,v3}
			\fmf{plain,label=$\hat \nu_p$,label.side=left,tension=2}{v3,b2}
			\fmf{wboson,label=$\hat \W^+$,tension=3}{t1,v2}
			\fmffreeze
			\fmf{plain,label=$\nu_u$}{v1,v3}
			\fmfv{decoration.shape=square,decoration.size=1.5mm}{v1}
		\end{fmfgraph*}
	\end{gathered}
\qquad\quad
	\begin{gathered}
		\begin{fmfgraph*}(60,60)
			\fmfset{curly_len}{2mm}
			\fmftop{t1} \fmfbottom{b1,b2}
			\fmf{quark,label=$\hat e_r$,label.side=right,tension=2}{b1,v1}
			\fmf{plain,label=$\hat \nu_p$,label.side=right,tension=2}{v1,b2}
			\fmf{wboson,label=$\hat \W^+$,tension=2}{t1,v1}
			\fmf{scalar,label=$G^+$}{v1,v1}
			\fmffreeze
			\fmfv{decoration.shape=square,decoration.size=1.5mm}{v1}
		\end{fmfgraph*}
	\end{gathered}
\qquad\quad
	\begin{gathered}
		\begin{fmfgraph*}(60,60)
			\fmfset{curly_len}{2mm}
			\fmftop{t1} \fmfbottom{b1,b2}
			\fmf{quark,label=$\hat e_r$,label.side=right,tension=2}{b1,v1}
			\fmf{plain,label=$\hat \nu_p$,label.side=right,tension=2}{v1,b2}
			\fmf{wboson,label=$\hat \W^+$,tension=2}{t1,v1}
			\fmf{dashes,label=$G^0$}{v1,v1}
			\fmffreeze
			\fmfv{decoration.shape=square,decoration.size=1.5mm}{v1}
		\end{fmfgraph*}
	\end{gathered}
\qquad\quad
	\begin{gathered}
		\begin{fmfgraph*}(60,60)
			\fmfset{curly_len}{2mm}
			\fmftop{t1} \fmfbottom{b1,b2}
			\fmf{quark,label=$\hat e_r$,label.side=right,tension=2}{b1,v1}
			\fmf{plain,label=$\hat \nu_p$,label.side=right,tension=2}{v1,b2}
			\fmf{wboson,label=$\hat \W^+$,tension=2}{t1,v1}
			\fmf{dashes,label=$h$}{v1,v1}
			\fmffreeze
			\fmfv{decoration.shape=square,decoration.size=1.5mm}{v1}
		\end{fmfgraph*}
	\end{gathered}
\nn
	& \begin{gathered}
		\begin{fmfgraph*}(60,60)
			\fmfset{curly_len}{2mm}
			\fmftop{t1} \fmfbottom{b1,b2}
			\fmf{quark,label=$\hat e_r$,label.side=right,tension=2}{b1,v2}
			\fmf{plain,label=$\hat \nu_p$,label.side=right,tension=2}{v2,b2}
			\fmf{wboson,label=$\hat \W^+$,tension=2}{t1,v1}
			\fmf{quark,label=$t$,right}{v1,v2}
			\fmf{quark,label=$d_u$,right}{v2,v1}
			\fmffreeze
			\fmfv{decoration.shape=square,decoration.size=1.5mm}{v2}
		\end{fmfgraph*}
	\end{gathered}
\qquad\quad
	\begin{gathered}
		\begin{fmfgraph*}(60,60)
			\fmfset{curly_len}{2mm}
			\fmftop{t1} \fmfbottom{b1,b2}
			\fmf{quark,label=$\hat e_r$,label.side=right,tension=2}{b1,v2}
			\fmf{plain,label=$\hat \nu_p$,label.side=right,tension=2}{v2,b2}
			\fmf{wboson,label=$\hat \W^+$,tension=2}{t1,v1}
			\fmf{wboson,label=$\W^+$,right}{v1,v2}
			\fmf{photon,label=$\Z$,right}{v2,v1}
			\fmffreeze
			\fmfv{decoration.shape=square,decoration.size=1.5mm}{v2}
		\end{fmfgraph*}
	\end{gathered}
\qquad\quad
	\begin{gathered}
		\begin{fmfgraph*}(60,60)
			\fmfset{curly_len}{2mm}
			\fmftop{t1} \fmfbottom{b1,b2}
			\fmf{quark,label=$\hat e_r$,label.side=right,tension=2}{b1,v2}
			\fmf{plain,label=$\hat \nu_p$,label.side=right,tension=2}{v2,b2}
			\fmf{wboson,label=$\hat \W^+$,tension=2}{t1,v1}
			\fmf{wboson,label=$\W^+$,right}{v1,v2}
			\fmf{photon,label=$\A$,right}{v2,v1}
			\fmffreeze
			\fmfv{decoration.shape=square,decoration.size=1.5mm}{v2}
		\end{fmfgraph*}
	\end{gathered}
\qquad\quad
	\begin{gathered}
		\begin{fmfgraph*}(60,60)
			\fmfset{curly_len}{2mm}
			\fmftop{t1} \fmfbottom{b1,b2}
			\fmf{quark,label=$\hat e_r$,label.side=right,tension=2}{b1,v2}
			\fmf{plain,label=$\hat \nu_p$,label.side=right,tension=2}{v2,b2}
			\fmf{wboson,label=$\hat \W^+$,tension=2}{t1,v1}
			\fmf{wboson,label=$\W^+$,right}{v1,v2}
			\fmf{dashes,label=$G^0$,right}{v2,v1}
			\fmffreeze
			\fmfv{decoration.shape=square,decoration.size=1.5mm}{v2}
		\end{fmfgraph*}
	\end{gathered}
\qquad\quad
	\begin{gathered}
		\begin{fmfgraph*}(60,60)
			\fmfset{curly_len}{2mm}
			\fmftop{t1} \fmfbottom{b1,b2}
			\fmf{quark,label=$\hat e_r$,label.side=right,tension=2}{b1,v2}
			\fmf{plain,label=$\hat \nu_p$,label.side=right,tension=2}{v2,b2}
			\fmf{wboson,label=$\hat \W^+$,tension=2}{t1,v1}
			\fmf{wboson,label=$\W^+$,right}{v1,v2}
			\fmf{dashes,label=$h$,right}{v2,v1}
			\fmffreeze
			\fmfv{decoration.shape=square,decoration.size=1.5mm}{v2}
		\end{fmfgraph*}
	\end{gathered}
\nn
	& \begin{gathered}
		\begin{fmfgraph*}(60,60)
			\fmfset{curly_len}{2mm}
			\fmftop{t1} \fmfbottom{b1,b2}
			\fmf{quark,label=$\hat e_r$,label.side=right,tension=2}{b1,v2}
			\fmf{plain,label=$\hat \nu_p$,label.side=right,tension=2}{v2,b2}
			\fmf{wboson,label=$\hat \W^+$,tension=2}{t1,v1}
			\fmf{scalar,label=$G^+$,right}{v1,v2}
			\fmf{photon,label=$\Z$,right}{v2,v1}
			\fmffreeze
			\fmfv{decoration.shape=square,decoration.size=1.5mm}{v2}
		\end{fmfgraph*}
	\end{gathered}
\qquad\quad
	\begin{gathered}
		\begin{fmfgraph*}(60,60)
			\fmfset{curly_len}{2mm}
			\fmftop{t1} \fmfbottom{b1,b2}
			\fmf{quark,label=$\hat e_r$,label.side=right,tension=2}{b1,v2}
			\fmf{plain,label=$\hat \nu_p$,label.side=right,tension=2}{v2,b2}
			\fmf{wboson,label=$\hat \W^+$,tension=2}{t1,v1}
			\fmf{scalar,label=$G^+$,right}{v1,v2}
			\fmf{photon,label=$\A$,right}{v2,v1}
			\fmffreeze
			\fmfv{decoration.shape=square,decoration.size=1.5mm}{v2}
		\end{fmfgraph*}
	\end{gathered}
\qquad\quad
	\begin{gathered}
		\begin{fmfgraph*}(60,60)
			\fmfset{curly_len}{2mm}
			\fmftop{t1} \fmfbottom{b1,b2}
			\fmf{quark,label=$\hat e_r$,label.side=right,tension=2}{b1,v2}
			\fmf{plain,label=$\hat \nu_p$,label.side=right,tension=2}{v2,b2}
			\fmf{wboson,label=$\hat \W^+$,tension=2}{t1,v1}
			\fmf{scalar,label=$G^+$,right}{v1,v2}
			\fmf{dashes,label=$G^0$,right}{v2,v1}
			\fmffreeze
			\fmfv{decoration.shape=square,decoration.size=1.5mm}{v2}
		\end{fmfgraph*}
	\end{gathered}
\qquad\quad
	\begin{gathered}
		\begin{fmfgraph*}(60,60)
			\fmfset{curly_len}{2mm}
			\fmftop{t1} \fmfbottom{b1,b2}
			\fmf{quark,label=$\hat e_r$,label.side=right,tension=2}{b1,v2}
			\fmf{plain,label=$\hat \nu_p$,label.side=right,tension=2}{v2,b2}
			\fmf{wboson,label=$\hat \W^+$,tension=2}{t1,v1}
			\fmf{scalar,label=$G^+$,right}{v1,v2}
			\fmf{dashes,label=$h$,right}{v2,v1}
			\fmffreeze
			\fmfv{decoration.shape=square,decoration.size=1.5mm}{v2}
		\end{fmfgraph*}
	\end{gathered}
\qquad\quad
	\begin{gathered}
		\begin{fmfgraph*}(60,60)
			\fmfset{curly_len}{2mm}
			\fmftop{t1} \fmfbottom{b1,b2}
			\fmf{quark,label=$\hat e_r$,label.side=left,tension=0.67}{b1,v1}
			\fmf{plain,label=$\hat \nu_p$,label.side=left,tension=2}{v2,b2}
			\fmf{wboson,label=$\hat \W^+$,tension=0.67}{t1,v1}
			\fmf{plain,label=$\nu_u$,label.side=right,tension=1}{v1,v2}
			\fmffreeze
			\fmf{photon,label=$\Z$,right}{v2,v1}
			\fmfv{decoration.shape=square,decoration.size=1.5mm}{v1}
		\end{fmfgraph*}
	\end{gathered}
\nn
	& \begin{gathered}
		\begin{fmfgraph*}(60,60)
			\fmfset{curly_len}{2mm}
			\fmftop{t1} \fmfbottom{b1,b2}
			\fmf{quark,label=$\hat e_r$,label.side=left,tension=0.67}{b1,v1}
			\fmf{plain,label=$\hat \nu_p$,label.side=left,tension=2}{v2,b2}
			\fmf{wboson,label=$\hat \W^+$,tension=0.67}{t1,v1}
			\fmf{quark,label=$e_u$,label.side=right,tension=1}{v1,v2}
			\fmffreeze
			\fmf{wboson,label=$\W^+$,left}{v1,v2}
			\fmfv{decoration.shape=square,decoration.size=1.5mm}{v1}
		\end{fmfgraph*}
	\end{gathered}
\qquad\quad
	\begin{gathered}
		\begin{fmfgraph*}(60,60)
			\fmfset{curly_len}{2mm}
			\fmftop{t1} \fmfbottom{b1,b2}
			\fmf{quark,label=$\hat e_r$,label.side=left,tension=2}{b1,v2}
			\fmf{plain,label=$\hat \nu_p$,label.side=left,tension=0.67}{v1,b2}
			\fmf{wboson,label=$\hat \W^+$,label.side=left,tension=0.67}{t1,v1}
			\fmf{quark,label=$e_u$,label.side=right,tension=1}{v2,v1}
			\fmffreeze
			\fmf{photon,label=$\Z$,right}{v1,v2}
			\fmfv{decoration.shape=square,decoration.size=1.5mm}{v1}
		\end{fmfgraph*}
	\end{gathered}
\qquad\quad
	\begin{gathered}
		\begin{fmfgraph*}(60,60)
			\fmfset{curly_len}{2mm}
			\fmftop{t1,t2,t3} \fmfbottom{b1,b2}
			\fmf{quark,label=$\hat e_r$,label.side=left,tension=2}{b1,v1}
			\fmf{plain,label=$\hat \nu_p$,label.side=right,tension=2}{v1,b2}
			\fmf{wboson,label=$\hat \W^+$,tension=2}{t2,v1}
			\fmffreeze
			\fmf{dashes,tension=3}{v1,v2}
			\fmf{phantom,tension=6}{t3,v2}
			\fmf{phantom,tension=2}{v2,b2}
			\fmfblob{5mm}{v2}
			\fmfv{decoration.shape=square,decoration.size=1.5mm}{v1}
		\end{fmfgraph*}
	\end{gathered}
\end{align*}


\subsubsection[$\bar e \nu W^-$]{\boldmath$\bar e \nu W^-$}

These diagrams are trivially related to the $\nu e W^+$ class by Hermitian conjugation.


\subsubsection[$\bar u d W^+$]{\boldmath$\bar u d W^+$}

\begin{align*}
	&\begin{gathered}
		\begin{fmfgraph*}(60,60)
			\fmfset{curly_len}{2mm}
			\fmftop{t1} \fmfbottom{b1,b2}
			\fmf{quark,label=$\hat d_r$,label.side=left,tension=2}{b1,v1}
			\fmf{quark,label=$d_v$,label.side=left}{v1,v2}
			\fmf{quark,label=$t$,label.side=left}{v2,v3}
			\fmf{quark,label=$\hat u_p$,label.side=left,tension=2}{v3,b2}
			\fmf{wboson,label=$\hat \W^+$,tension=3}{t1,v2}
			\fmffreeze
			\fmf{photon,label=$\A$}{v1,v3}
		\end{fmfgraph*}
	\end{gathered}
\qquad\quad
	\begin{gathered}
		\begin{fmfgraph*}(60,60)
			\fmfset{curly_len}{2mm}
			\fmftop{t1} \fmfbottom{b1,b2}
			\fmf{quark,label=$\hat d_r$,label.side=left,tension=2}{b1,v1}
			\fmf{quark,label=$d_v$,label.side=left}{v1,v2}
			\fmf{quark,label=$t$,label.side=left}{v2,v3}
			\fmf{quark,label=$\hat u_p$,label.side=left,tension=2}{v3,b2}
			\fmf{wboson,label=$\hat \W^+$,tension=3}{t1,v2}
			\fmffreeze
			\fmf{gluon,label=$\G$}{v1,v3}
		\end{fmfgraph*}
	\end{gathered}
\qquad\quad
	\begin{gathered}
		\begin{fmfgraph*}(60,60)
			\fmfset{curly_len}{2mm}
			\fmftop{t1} \fmfbottom{b1,b2}
			\fmf{quark,label=$\hat d_r$,label.side=left,tension=2}{b1,v1}
			\fmf{quark,label=$d_v$,label.side=left}{v1,v2}
			\fmf{quark,label=$u_u$,label.side=left}{v2,v3}
			\fmf{quark,label=$\hat u_p$,label.side=left,tension=2}{v3,b2}
			\fmf{wboson,label=$\hat \W^+$,tension=3}{t1,v2}
			\fmffreeze
			\fmf{photon,label=$\Z$}{v1,v3}
		\end{fmfgraph*}
	\end{gathered}
\qquad\quad
	\begin{gathered}
		\begin{fmfgraph*}(60,60)
			\fmfset{curly_len}{2mm}
			\fmftop{t1} \fmfbottom{b1,b2}
			\fmf{quark,label=$\hat d_r$,label.side=left,tension=2}{b1,v1}
			\fmf{photon,label=$\Z$,label.side=right}{v2,v1}
			\fmf{wboson,label=$\W^+$,label.side=left}{v2,v3}
			\fmf{quark,label=$\hat u_p$,label.side=left,tension=2}{v3,b2}
			\fmf{wboson,label=$\hat \W^+$,tension=3}{t1,v2}
			\fmffreeze
			\fmf{quark,label=$d_u$}{v1,v3}
		\end{fmfgraph*}
	\end{gathered}
\qquad\quad
	\begin{gathered}
		\begin{fmfgraph*}(60,60)
			\fmfset{curly_len}{2mm}
			\fmftop{t1} \fmfbottom{b1,b2}
			\fmf{quark,label=$\hat d_r$,label.side=left,tension=2}{b1,v1}
			\fmf{photon,label=$\A$,label.side=right}{v2,v1}
			\fmf{wboson,label=$\W^+$,label.side=left}{v2,v3}
			\fmf{quark,label=$\hat u_p$,label.side=left,tension=2}{v3,b2}
			\fmf{wboson,label=$\hat \W^+$,tension=3}{t1,v2}
			\fmffreeze
			\fmf{quark,label=$d_u$}{v1,v3}
		\end{fmfgraph*}
	\end{gathered}
\nn
	& \begin{gathered}
		\begin{fmfgraph*}(60,60)
			\fmfset{curly_len}{2mm}
			\fmftop{t1} \fmfbottom{b1,b2}
			\fmf{quark,label=$\hat d_r$,label.side=left,tension=2}{b1,v1}
			\fmf{dashes,label=$G^0$,label.side=right}{v2,v1}
			\fmf{wboson,label=$\W^+$,label.side=left}{v2,v3}
			\fmf{quark,label=$\hat u_p$,label.side=left,tension=2}{v3,b2}
			\fmf{wboson,label=$\hat \W^+$,tension=3}{t1,v2}
			\fmffreeze
			\fmf{quark,label=$d_u$}{v1,v3}
			\fmfv{decoration.shape=square,decoration.size=1.5mm}{v1}
		\end{fmfgraph*}
	\end{gathered}
\qquad\quad
	\begin{gathered}
		\begin{fmfgraph*}(60,60)
			\fmfset{curly_len}{2mm}
			\fmftop{t1} \fmfbottom{b1,b2}
			\fmf{quark,label=$\hat d_r$,label.side=left,tension=2}{b1,v1}
			\fmf{dashes,label=$h$,label.side=right}{v2,v1}
			\fmf{wboson,label=$\W^+$,label.side=left}{v2,v3}
			\fmf{quark,label=$\hat u_p$,label.side=left,tension=2}{v3,b2}
			\fmf{wboson,label=$\hat \W^+$,tension=3}{t1,v2}
			\fmffreeze
			\fmf{quark,label=$d_u$}{v1,v3}
			\fmfv{decoration.shape=square,decoration.size=1.5mm}{v1}
		\end{fmfgraph*}
	\end{gathered}
\qquad\quad
	\begin{gathered}
		\begin{fmfgraph*}(60,60)
			\fmfset{curly_len}{2mm}
			\fmftop{t1} \fmfbottom{b1,b2}
			\fmf{quark,label=$\hat d_r$,label.side=left,tension=2}{b1,v1}
			\fmf{photon,label=$\Z$,label.side=right}{v2,v1}
			\fmf{scalar,label=$G^+$,label.side=left}{v2,v3}
			\fmf{quark,label=$\hat u_p$,label.side=left,tension=2}{v3,b2}
			\fmf{wboson,label=$\hat \W^+$,tension=3}{t1,v2}
			\fmffreeze
			\fmf{quark,label=$d_u$}{v1,v3}
			\fmfv{decoration.shape=square,decoration.size=1.5mm}{v3}
		\end{fmfgraph*}
	\end{gathered}
\qquad\quad
	\begin{gathered}
		\begin{fmfgraph*}(60,60)
			\fmfset{curly_len}{2mm}
			\fmftop{t1} \fmfbottom{b1,b2}
			\fmf{quark,label=$\hat d_r$,label.side=left,tension=2}{b1,v1}
			\fmf{photon,label=$\A$,label.side=right}{v2,v1}
			\fmf{scalar,label=$G^+$,label.side=left}{v2,v3}
			\fmf{quark,label=$\hat u_p$,label.side=left,tension=2}{v3,b2}
			\fmf{wboson,label=$\hat \W^+$,tension=3}{t1,v2}
			\fmffreeze
			\fmf{quark,label=$d_u$}{v1,v3}
			\fmfv{decoration.shape=square,decoration.size=1.5mm}{v3}
		\end{fmfgraph*}
	\end{gathered}
\qquad\quad
	\begin{gathered}
		\begin{fmfgraph*}(60,60)
			\fmfset{curly_len}{2mm}
			\fmftop{t1} \fmfbottom{b1,b2}
			\fmf{quark,label=$\hat d_r$,label.side=left,tension=2}{b1,v1}
			\fmf{wboson,label=$\W^+$,label.side=right}{v2,v1}
			\fmf{photon,label=$\Z$,label.side=left}{v2,v3}
			\fmf{quark,label=$\hat u_p$,label.side=left,tension=2}{v3,b2}
			\fmf{wboson,label=$\hat \W^+$,tension=3}{t1,v2}
			\fmffreeze
			\fmf{quark,label=$u_u$}{v1,v3}
		\end{fmfgraph*}
	\end{gathered}
\nn
	& \begin{gathered}
		\begin{fmfgraph*}(60,60)
			\fmfset{curly_len}{2mm}
			\fmftop{t1} \fmfbottom{b1,b2}
			\fmf{quark,label=$\hat d_r$,label.side=left,tension=2}{b1,v1}
			\fmf{wboson,label=$\W^+$,label.side=right}{v2,v1}
			\fmf{photon,label=$\A$,label.side=left}{v2,v3}
			\fmf{quark,label=$\hat u_p$,label.side=left,tension=2}{v3,b2}
			\fmf{wboson,label=$\hat \W^+$,tension=3}{t1,v2}
			\fmffreeze
			\fmf{quark,label=$u_u$}{v1,v3}
		\end{fmfgraph*}
	\end{gathered}
\qquad\quad
	\begin{gathered}
		\begin{fmfgraph*}(60,60)
			\fmfset{curly_len}{2mm}
			\fmftop{t1} \fmfbottom{b1,b2}
			\fmf{quark,label=$\hat d_r$,label.side=left,tension=2}{b1,v1}
			\fmf{wboson,label=$\W^+$,label.side=right}{v2,v1}
			\fmf{dashes,label=$G^0$,label.side=left}{v2,v3}
			\fmf{quark,label=$\hat u_p$,label.side=left,tension=2}{v3,b2}
			\fmf{wboson,label=$\hat \W^+$,tension=3}{t1,v2}
			\fmffreeze
			\fmf{quark,label=$u_u$}{v1,v3}
			\fmfv{decoration.shape=square,decoration.size=1.5mm}{v3}
		\end{fmfgraph*}
	\end{gathered}
\qquad\quad
	\begin{gathered}
		\begin{fmfgraph*}(60,60)
			\fmfset{curly_len}{2mm}
			\fmftop{t1} \fmfbottom{b1,b2}
			\fmf{quark,label=$\hat d_r$,label.side=left,tension=2}{b1,v1}
			\fmf{wboson,label=$\W^+$,label.side=right}{v2,v1}
			\fmf{dashes,label=$h$,label.side=left}{v2,v3}
			\fmf{quark,label=$\hat u_p$,label.side=left,tension=2}{v3,b2}
			\fmf{wboson,label=$\hat \W^+$,tension=3}{t1,v2}
			\fmffreeze
			\fmf{quark,label=$u_u$}{v1,v3}
			\fmfv{decoration.shape=square,decoration.size=1.5mm}{v3}
		\end{fmfgraph*}
	\end{gathered}
\qquad\quad
	\begin{gathered}
		\begin{fmfgraph*}(60,60)
			\fmfset{curly_len}{2mm}
			\fmftop{t1} \fmfbottom{b1,b2}
			\fmf{quark,label=$\hat d_r$,label.side=left,tension=2}{b1,v1}
			\fmf{scalar,label=$G^+$,label.side=right}{v2,v1}
			\fmf{photon,label=$\Z$,label.side=left}{v2,v3}
			\fmf{quark,label=$\hat u_p$,label.side=left,tension=2}{v3,b2}
			\fmf{wboson,label=$\hat \W^+$,tension=3}{t1,v2}
			\fmffreeze
			\fmf{quark,label=$u_u$}{v1,v3}
		\end{fmfgraph*}
	\end{gathered}
\qquad\quad
	\begin{gathered}
		\begin{fmfgraph*}(60,60)
			\fmfset{curly_len}{2mm}
			\fmftop{t1} \fmfbottom{b1,b2}
			\fmf{quark,label=$\hat d_r$,label.side=left,tension=2}{b1,v1}
			\fmf{scalar,label=$G^+$,label.side=right}{v2,v1}
			\fmf{photon,label=$\A$,label.side=left}{v2,v3}
			\fmf{quark,label=$\hat u_p$,label.side=left,tension=2}{v3,b2}
			\fmf{wboson,label=$\hat \W^+$,tension=3}{t1,v2}
			\fmffreeze
			\fmf{quark,label=$u_u$}{v1,v3}
		\end{fmfgraph*}
	\end{gathered}
\nn
	& \begin{gathered}
		\begin{fmfgraph*}(60,60)
			\fmfset{curly_len}{2mm}
			\fmftop{t1} \fmfbottom{b1,b2}
			\fmf{quark,label=$\hat d_r$,label.side=left,tension=2}{b1,v1}
			\fmf{scalar,label=$G^+$,label.side=right}{v2,v1}
			\fmf{dashes,label=$G^0$,label.side=left}{v2,v3}
			\fmf{quark,label=$\hat u_p$,label.side=left,tension=2}{v3,b2}
			\fmf{wboson,label=$\hat \W^+$,tension=3}{t1,v2}
			\fmffreeze
			\fmf{quark,label=$t$}{v1,v3}
			\fmfv{decoration.shape=square,decoration.size=1.5mm}{v3}
		\end{fmfgraph*}
	\end{gathered}
\qquad\quad
	\begin{gathered}
		\begin{fmfgraph*}(60,60)
			\fmfset{curly_len}{2mm}
			\fmftop{t1} \fmfbottom{b1,b2}
			\fmf{quark,label=$\hat d_r$,label.side=left,tension=2}{b1,v1}
			\fmf{scalar,label=$G^+$,label.side=right}{v2,v1}
			\fmf{dashes,label=$h$,label.side=left}{v2,v3}
			\fmf{quark,label=$\hat u_p$,label.side=left,tension=2}{v3,b2}
			\fmf{wboson,label=$\hat \W^+$,tension=3}{t1,v2}
			\fmffreeze
			\fmf{quark,label=$t$}{v1,v3}
			\fmfv{decoration.shape=square,decoration.size=1.5mm}{v3}
		\end{fmfgraph*}
	\end{gathered}
\qquad\quad
	\begin{gathered}
		\begin{fmfgraph*}(60,60)
			\fmfset{curly_len}{2mm}
			\fmftop{t1} \fmfbottom{b1,b2}
			\fmf{quark,label=$\hat d_r$,label.side=right,tension=2}{b1,v1}
			\fmf{quark,label=$\hat u_p$,label.side=right,tension=2}{v1,b2}
			\fmf{wboson,label=$\hat \W^+$,tension=2}{t1,v1}
			\fmf{scalar,label=$G^+$}{v1,v1}
			\fmffreeze
			\fmfv{decoration.shape=square,decoration.size=1.5mm}{v1}
		\end{fmfgraph*}
	\end{gathered}
\qquad\quad
	\begin{gathered}
		\begin{fmfgraph*}(60,60)
			\fmfset{curly_len}{2mm}
			\fmftop{t1} \fmfbottom{b1,b2}
			\fmf{quark,label=$\hat d_r$,label.side=right,tension=2}{b1,v1}
			\fmf{quark,label=$\hat u_p$,label.side=right,tension=2}{v1,b2}
			\fmf{wboson,label=$\hat \W^+$,tension=2}{t1,v1}
			\fmf{dashes,label=$G^0$}{v1,v1}
			\fmffreeze
			\fmfv{decoration.shape=square,decoration.size=1.5mm}{v1}
		\end{fmfgraph*}
	\end{gathered}
\qquad\quad
	\begin{gathered}
		\begin{fmfgraph*}(60,60)
			\fmfset{curly_len}{2mm}
			\fmftop{t1} \fmfbottom{b1,b2}
			\fmf{quark,label=$\hat d_r$,label.side=right,tension=2}{b1,v1}
			\fmf{quark,label=$\hat u_p$,label.side=right,tension=2}{v1,b2}
			\fmf{wboson,label=$\hat \W^+$,tension=2}{t1,v1}
			\fmf{dashes,label=$h$}{v1,v1}
			\fmffreeze
			\fmfv{decoration.shape=square,decoration.size=1.5mm}{v1}
		\end{fmfgraph*}
	\end{gathered}
\nn
	& \begin{gathered}
		\begin{fmfgraph*}(60,60)
			\fmfset{curly_len}{2mm}
			\fmftop{t1} \fmfbottom{b1,b2}
			\fmf{quark,label=$\hat d_r$,label.side=right,tension=2}{b1,v2}
			\fmf{quark,label=$\hat u_p$,label.side=right,tension=2}{v2,b2}
			\fmf{wboson,label=$\hat \W^+$,tension=2}{t1,v1}
			\fmf{quark,label=$t$,right}{v1,v2}
			\fmf{quark,label=$d_u$,right}{v2,v1}
			\fmffreeze
			\fmfv{decoration.shape=square,decoration.size=1.5mm}{v2}
		\end{fmfgraph*}
	\end{gathered}
\qquad\quad
	\begin{gathered}
		\begin{fmfgraph*}(60,60)
			\fmfset{curly_len}{2mm}
			\fmftop{t1} \fmfbottom{b1,b2}
			\fmf{quark,label=$\hat d_r$,label.side=right,tension=2}{b1,v2}
			\fmf{quark,label=$\hat u_p$,label.side=right,tension=2}{v2,b2}
			\fmf{wboson,label=$\hat \W^+$,tension=2}{t1,v1}
			\fmf{wboson,label=$\W^+$,right}{v1,v2}
			\fmf{photon,label=$\Z$,right}{v2,v1}
			\fmffreeze
			\fmfv{decoration.shape=square,decoration.size=1.5mm}{v2}
		\end{fmfgraph*}
	\end{gathered}
\qquad\quad
	\begin{gathered}
		\begin{fmfgraph*}(60,60)
			\fmfset{curly_len}{2mm}
			\fmftop{t1} \fmfbottom{b1,b2}
			\fmf{quark,label=$\hat d_r$,label.side=right,tension=2}{b1,v2}
			\fmf{quark,label=$\hat u_p$,label.side=right,tension=2}{v2,b2}
			\fmf{wboson,label=$\hat \W^+$,tension=2}{t1,v1}
			\fmf{wboson,label=$\W^+$,right}{v1,v2}
			\fmf{photon,label=$\A$,right}{v2,v1}
			\fmffreeze
			\fmfv{decoration.shape=square,decoration.size=1.5mm}{v2}
		\end{fmfgraph*}
	\end{gathered}
\qquad\quad
	\begin{gathered}
		\begin{fmfgraph*}(60,60)
			\fmfset{curly_len}{2mm}
			\fmftop{t1} \fmfbottom{b1,b2}
			\fmf{quark,label=$\hat d_r$,label.side=right,tension=2}{b1,v2}
			\fmf{quark,label=$\hat u_p$,label.side=right,tension=2}{v2,b2}
			\fmf{wboson,label=$\hat \W^+$,tension=2}{t1,v1}
			\fmf{wboson,label=$\W^+$,right}{v1,v2}
			\fmf{dashes,label=$G^0$,right}{v2,v1}
			\fmffreeze
			\fmfv{decoration.shape=square,decoration.size=1.5mm}{v2}
		\end{fmfgraph*}
	\end{gathered}
\qquad\quad
	\begin{gathered}
		\begin{fmfgraph*}(60,60)
			\fmfset{curly_len}{2mm}
			\fmftop{t1} \fmfbottom{b1,b2}
			\fmf{quark,label=$\hat d_r$,label.side=right,tension=2}{b1,v2}
			\fmf{quark,label=$\hat u_p$,label.side=right,tension=2}{v2,b2}
			\fmf{wboson,label=$\hat \W^+$,tension=2}{t1,v1}
			\fmf{wboson,label=$\W^+$,right}{v1,v2}
			\fmf{dashes,label=$h$,right}{v2,v1}
			\fmffreeze
			\fmfv{decoration.shape=square,decoration.size=1.5mm}{v2}
		\end{fmfgraph*}
	\end{gathered}
\nn
	& \begin{gathered}
		\begin{fmfgraph*}(60,60)
			\fmfset{curly_len}{2mm}
			\fmftop{t1} \fmfbottom{b1,b2}
			\fmf{quark,label=$\hat d_r$,label.side=right,tension=2}{b1,v2}
			\fmf{quark,label=$\hat u_p$,label.side=right,tension=2}{v2,b2}
			\fmf{wboson,label=$\hat \W^+$,tension=2}{t1,v1}
			\fmf{scalar,label=$G^+$,right}{v1,v2}
			\fmf{photon,label=$\Z$,right}{v2,v1}
			\fmffreeze
			\fmfv{decoration.shape=square,decoration.size=1.5mm}{v2}
		\end{fmfgraph*}
	\end{gathered}
\qquad\quad
	\begin{gathered}
		\begin{fmfgraph*}(60,60)
			\fmfset{curly_len}{2mm}
			\fmftop{t1} \fmfbottom{b1,b2}
			\fmf{quark,label=$\hat d_r$,label.side=right,tension=2}{b1,v2}
			\fmf{quark,label=$\hat u_p$,label.side=right,tension=2}{v2,b2}
			\fmf{wboson,label=$\hat \W^+$,tension=2}{t1,v1}
			\fmf{scalar,label=$G^+$,right}{v1,v2}
			\fmf{photon,label=$\A$,right}{v2,v1}
			\fmffreeze
			\fmfv{decoration.shape=square,decoration.size=1.5mm}{v2}
		\end{fmfgraph*}
	\end{gathered}
\qquad\quad
	\begin{gathered}
		\begin{fmfgraph*}(60,60)
			\fmfset{curly_len}{2mm}
			\fmftop{t1} \fmfbottom{b1,b2}
			\fmf{quark,label=$\hat d_r$,label.side=right,tension=2}{b1,v2}
			\fmf{quark,label=$\hat u_p$,label.side=right,tension=2}{v2,b2}
			\fmf{wboson,label=$\hat \W^+$,tension=2}{t1,v1}
			\fmf{scalar,label=$G^+$,right}{v1,v2}
			\fmf{dashes,label=$G^0$,right}{v2,v1}
			\fmffreeze
			\fmfv{decoration.shape=square,decoration.size=1.5mm}{v2}
		\end{fmfgraph*}
	\end{gathered}
\qquad\quad
	\begin{gathered}
		\begin{fmfgraph*}(60,60)
			\fmfset{curly_len}{2mm}
			\fmftop{t1} \fmfbottom{b1,b2}
			\fmf{quark,label=$\hat d_r$,label.side=right,tension=2}{b1,v2}
			\fmf{quark,label=$\hat u_p$,label.side=right,tension=2}{v2,b2}
			\fmf{wboson,label=$\hat \W^+$,tension=2}{t1,v1}
			\fmf{scalar,label=$G^+$,right}{v1,v2}
			\fmf{dashes,label=$h$,right}{v2,v1}
			\fmffreeze
			\fmfv{decoration.shape=square,decoration.size=1.5mm}{v2}
		\end{fmfgraph*}
	\end{gathered}
\qquad\quad
	\begin{gathered}
		\begin{fmfgraph*}(60,60)
			\fmfset{curly_len}{2mm}
			\fmftop{t1} \fmfbottom{b1,b2}
			\fmf{quark,label=$\hat d_r$,label.side=left,tension=0.67}{b1,v1}
			\fmf{quark,label=$\hat u_p$,label.side=left,tension=2}{v2,b2}
			\fmf{wboson,label=$\hat \W^+$,tension=0.67}{t1,v1}
			\fmf{quark,label=$u_u$,label.side=right,tension=1}{v1,v2}
			\fmffreeze
			\fmf{photon,label=$\Z$,right}{v2,v1}
			\fmfv{decoration.shape=square,decoration.size=1.5mm}{v1}
		\end{fmfgraph*}
	\end{gathered}
\nn
	& \begin{gathered}
		\begin{fmfgraph*}(60,60)
			\fmfset{curly_len}{2mm}
			\fmftop{t1} \fmfbottom{b1,b2}
			\fmf{quark,label=$\hat d_r$,label.side=left,tension=0.67}{b1,v1}
			\fmf{quark,label=$\hat u_p$,label.side=left,tension=2}{v2,b2}
			\fmf{wboson,label=$\hat \W^+$,tension=0.67}{t1,v1}
			\fmf{quark,label=$d_u$,label.side=right,tension=1}{v1,v2}
			\fmffreeze
			\fmf{wboson,label=$\W^+$,left}{v1,v2}
			\fmfv{decoration.shape=square,decoration.size=1.5mm}{v1}
		\end{fmfgraph*}
	\end{gathered}
\qquad\quad
	\begin{gathered}
		\begin{fmfgraph*}(60,60)
			\fmfset{curly_len}{2mm}
			\fmftop{t1} \fmfbottom{b1,b2}
			\fmf{quark,label=$\hat d_r$,label.side=left,tension=2}{b1,v2}
			\fmf{quark,label=$\hat u_p$,label.side=left,tension=0.67}{v1,b2}
			\fmf{wboson,label=$\hat \W^+$,label.side=left,tension=0.67}{t1,v1}
			\fmf{quark,label=$d_u$,label.side=right,tension=1}{v2,v1}
			\fmffreeze
			\fmf{photon,label=$\Z$,right}{v1,v2}
			\fmfv{decoration.shape=square,decoration.size=1.5mm}{v1}
		\end{fmfgraph*}
	\end{gathered}
\qquad\quad
	\begin{gathered}
		\begin{fmfgraph*}(60,60)
			\fmfset{curly_len}{2mm}
			\fmftop{t1} \fmfbottom{b1,b2}
			\fmf{quark,label=$\hat d_r$,label.side=left,tension=2}{b1,v2}
			\fmf{quark,label=$\hat u_p$,label.side=left,tension=0.67}{v1,b2}
			\fmf{wboson,label=$\hat \W^+$,label.side=left,tension=0.67}{t1,v1}
			\fmf{quark,label=$u_u$,label.side=right,tension=1}{v2,v1}
			\fmffreeze
			\fmf{wboson,label=$\W^+$,right}{v1,v2}
			\fmfv{decoration.shape=square,decoration.size=1.5mm}{v1}
		\end{fmfgraph*}
	\end{gathered}
\qquad\quad
	\begin{gathered}
		\begin{fmfgraph*}(60,60)
			\fmfset{curly_len}{2mm}
			\fmftop{t1} \fmfbottom{b1,b2}
			\fmf{quark,label=$\hat d_r$,label.side=left,tension=2}{b1,v2}
			\fmf{quark,label=$\hat u_p$,label.side=left,tension=0.67}{v1,b2}
			\fmf{wboson,label=$\hat \W^+$,label.side=left,tension=0.67}{t1,v1}
			\fmf{quark,label=$t$,label.side=right,tension=1}{v2,v1}
			\fmffreeze
			\fmf{scalar,label=$G^+$,right}{v1,v2}
			\fmfv{decoration.shape=square,decoration.size=1.5mm}{v1}
		\end{fmfgraph*}
	\end{gathered}
\qquad\quad
	\begin{gathered}
		\begin{fmfgraph*}(60,60)
			\fmfset{curly_len}{2mm}
			\fmftop{t1,t2,t3} \fmfbottom{b1,b2}
			\fmf{quark,label=$\hat d_r$,label.side=left,tension=2}{b1,v1}
			\fmf{quark,label=$\hat u_p$,label.side=right,tension=2}{v1,b2}
			\fmf{wboson,label=$\hat \W^+$,tension=2}{t2,v1}
			\fmffreeze
			\fmf{dashes,tension=3}{v1,v2}
			\fmf{phantom,tension=6}{t3,v2}
			\fmf{phantom,tension=2}{v2,b2}
			\fmfblob{5mm}{v2}
			\fmfv{decoration.shape=square,decoration.size=1.5mm}{v1}
		\end{fmfgraph*}
	\end{gathered}
\end{align*}


\subsubsection[$\bar d u W^-$]{\boldmath$\bar d u W^-$}

These diagrams are trivially related to the $\bar u d W^+$ class by Hermitian conjugation.

\end{myfmf}
	
\begin{myfmf}{diags/fourPoint}

\subsection{Four-point functions}
\label{sec:FourPointDiagrams}

\subsubsection[$\nu\nu\nu\nu$]{\boldmath$\nu\nu\nu\nu$}

Crossed versions of the diagrams are not shown explicitly but denoted by ${}\times x$.
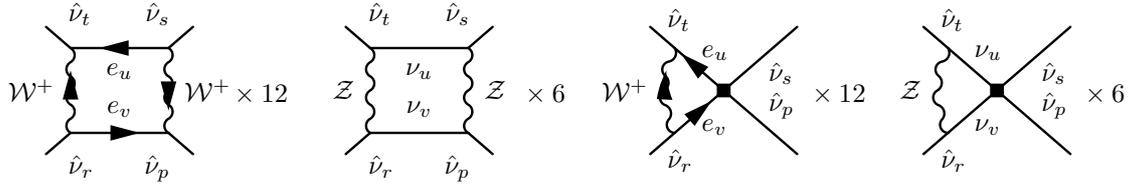
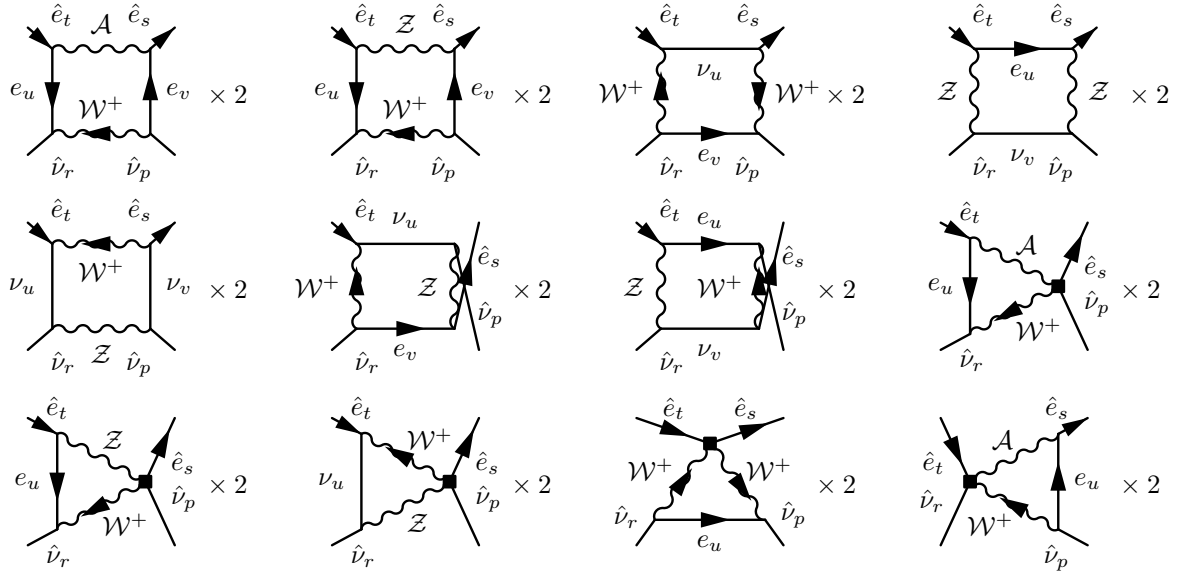
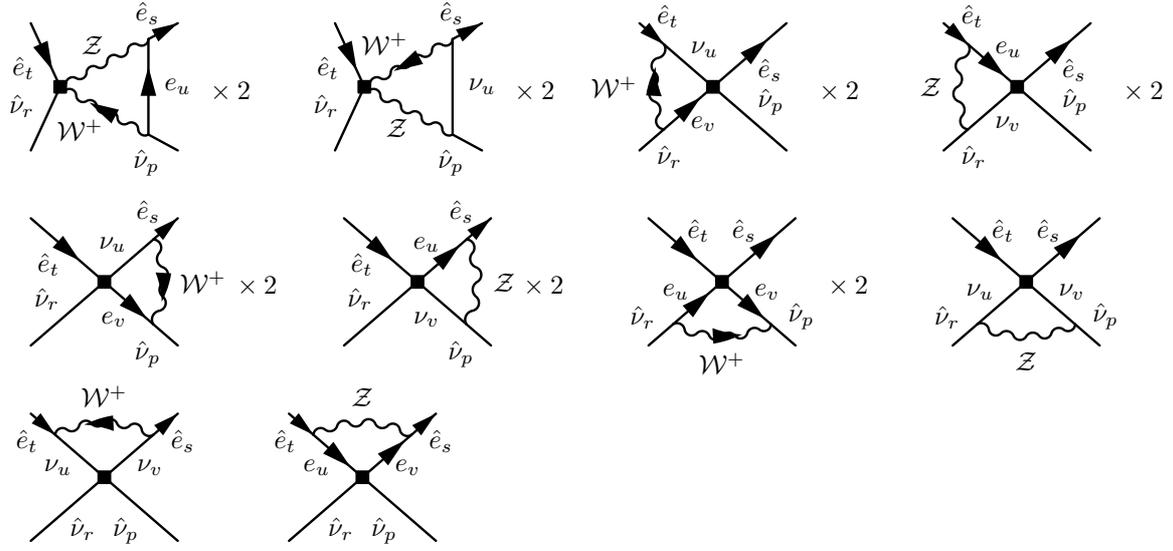
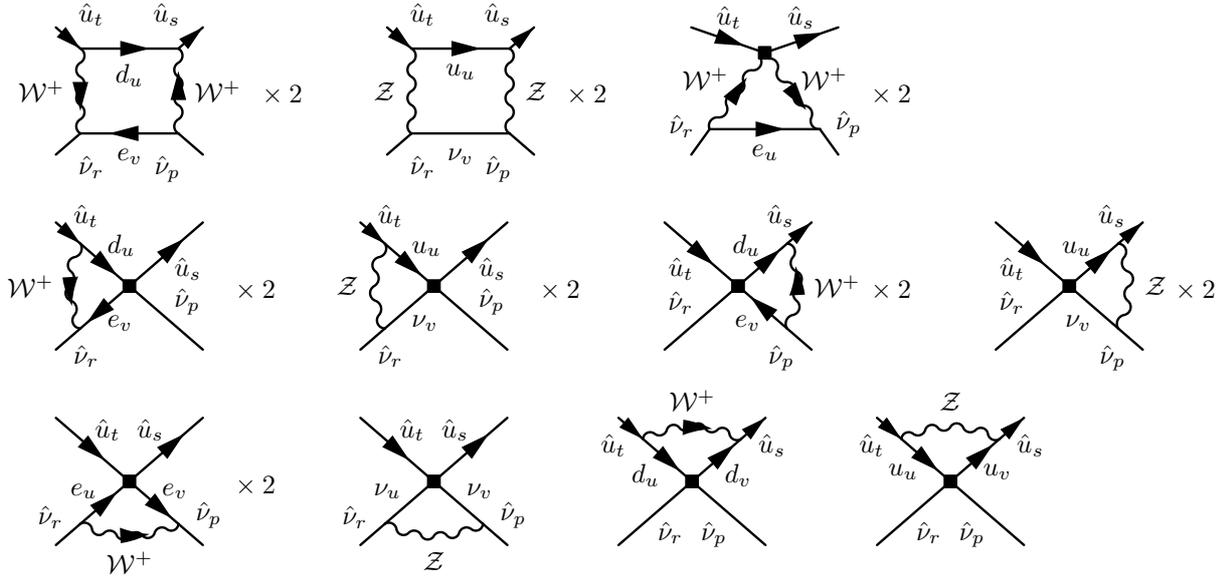
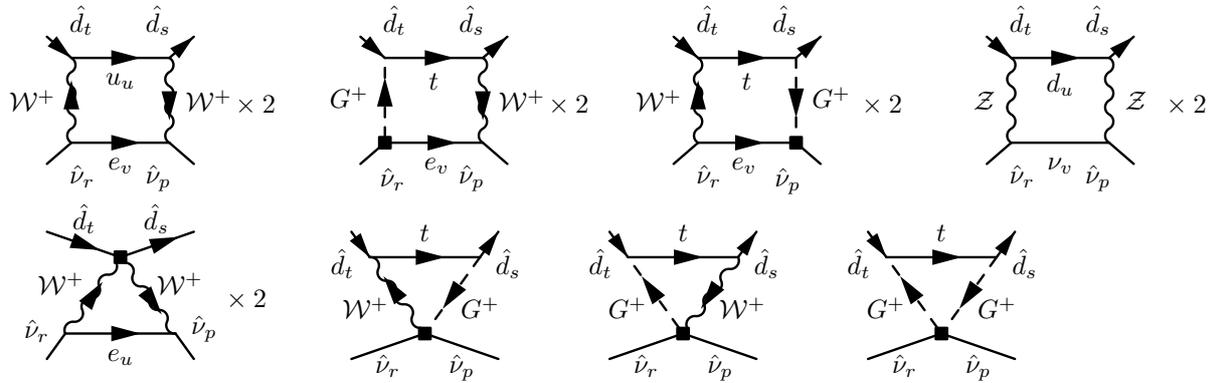
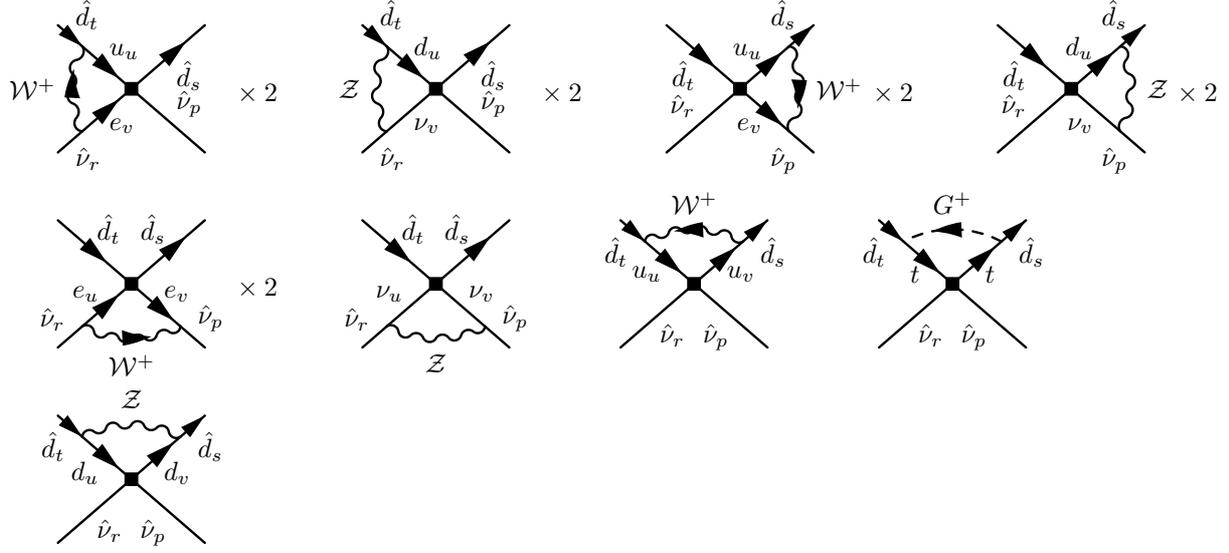
\begin{align*}
	& \begin{gathered}
		\begin{fmfgraph*}(55,60)
			\fmfset{curly_len}{2mm}
			\fmftop{t1,t2} \fmfbottom{b1,b2}
			\fmf{plain,label=$\hat \nu_t$,label.side=left,tension=4}{t1,v1}
			\fmf{plain,label=$\hat \nu_s$,label.side=right,tension=4}{t2,v2}
			\fmf{plain,label=$\hat \nu_r$,label.side=right,tension=4}{b1,v3}
			\fmf{plain,label=$\hat \nu_p$,label.side=left,tension=4}{b2,v4}
			\fmf{quark,label=$e_u$,label.side=left}{v2,v1}
			\fmf{quark,label=$e_v$,label.side=left}{v3,v4}
			\fmf{wboson,label=$\W^+$,label.side=left}{v3,v1}
			\fmf{wboson,label=$\W^+$,label.side=left}{v2,v4}
		\end{fmfgraph*}
	\end{gathered} \quad\,\, \times 12
\quad\quad
	\begin{gathered}
		\begin{fmfgraph*}(55,60)
			\fmfset{curly_len}{2mm}
			\fmftop{t1,t2} \fmfbottom{b1,b2}
			\fmf{plain,label=$\hat \nu_t$,label.side=left,tension=4}{t1,v1}
			\fmf{plain,label=$\hat \nu_s$,label.side=right,tension=4}{t2,v2}
			\fmf{plain,label=$\hat \nu_r$,label.side=right,tension=4}{b1,v3}
			\fmf{plain,label=$\hat \nu_p$,label.side=left,tension=4}{b2,v4}
			\fmf{plain,label=$\nu_u$,label.side=left}{v2,v1}
			\fmf{plain,label=$\nu_v$,label.side=left}{v3,v4}
			\fmf{photon,label=$\Z$,label.side=left}{v3,v1}
			\fmf{photon,label=$\Z$,label.side=left}{v2,v4}
		\end{fmfgraph*}
	\end{gathered} \quad \times 6
\qquad\quad
	\begin{gathered}
		\begin{fmfgraph*}(55,60)
			\fmfset{curly_len}{2mm}
			\fmftop{t1,t2} \fmfbottom{b1,b2}
			\fmf{plain,label=$\hat \nu_t$,label.side=left,label.dist=3,tension=6}{t1,v2}
			\fmf{plain,label=$\hat \nu_s$,label.side=left,label.dist=3,tension=2}{t2,v1}
			\fmf{plain,label=$\hat \nu_r$,label.side=right,label.dist=3,tension=6}{b1,v3}
			\fmf{plain,label=$\hat \nu_p$,label.side=right,label.dist=3,tension=2}{b2,v1}
			\fmf{quark,label=$e_u$,label.side=right,label.dist=3,tension=3}{v1,v2}
			\fmf{quark,label=$e_v$,label.side=right,label.dist=3,tension=3}{v3,v1}
			\fmffreeze
			\fmf{wboson,label=$\W^+$,label.side=left,left=0.25}{v3,v2}
			\fmfv{decoration.shape=square,decoration.size=1.5mm}{v1}
		\end{fmfgraph*}
	\end{gathered} \,\, \times 12
\quad\quad
	\begin{gathered}
		\begin{fmfgraph*}(55,60)
			\fmfset{curly_len}{2mm}
			\fmftop{t1,t2} \fmfbottom{b1,b2}
			\fmf{plain,label=$\hat \nu_t$,label.side=left,label.dist=3,tension=6}{t1,v2}
			\fmf{plain,label=$\hat \nu_s$,label.side=left,label.dist=3,tension=2}{t2,v1}
			\fmf{plain,label=$\hat \nu_r$,label.side=right,label.dist=3,tension=6}{b1,v3}
			\fmf{plain,label=$\hat \nu_p$,label.side=right,label.dist=3,tension=2}{b2,v1}
			\fmf{plain,label=$\nu_u$,label.side=left,label.dist=3,tension=3}{v2,v1}
			\fmf{plain,label=$\nu_v$,label.side=right,label.dist=3,tension=3}{v3,v1}
			\fmffreeze
			\fmf{photon,label=$\Z$,label.side=left,left=0.25}{v3,v2}
			\fmfv{decoration.shape=square,decoration.size=1.5mm}{v1}
		\end{fmfgraph*}
	\end{gathered} \,\, \times 6
\end{align*}


\subsubsection[$\nu \nu \bar ee$]{\boldmath$\nu \nu \bar ee$}

\begin{align*}
	&\begin{gathered}
		\begin{fmfgraph*}(55,60)
			\fmfset{curly_len}{2mm}
			\fmftop{t1,t2} \fmfbottom{b1,b2}
			\fmf{quark,label=$\hat e_t$,label.side=left,tension=4}{t1,v1}
			\fmf{quark,label=$\hat e_s$,label.side=left,tension=4}{v2,t2}
			\fmf{plain,label=$\hat \nu_r$,label.side=right,tension=4}{b1,v3}
			\fmf{plain,label=$\hat \nu_p$,label.side=left,tension=4}{b2,v4}
			\fmf{quark,label=$e_u$,label.side=right}{v1,v3}
			\fmf{quark,label=$e_v$,label.side=right}{v4,v2}
			\fmf{wboson,label=$\W^+$,label.side=right}{v4,v3}
			\fmf{photon,label=$\A$,label.side=left}{v1,v2}
		\end{fmfgraph*}
	\end{gathered} \quad \times 2
\qquad\quad
	\begin{gathered}
		\begin{fmfgraph*}(55,60)
			\fmfset{curly_len}{2mm}
			\fmftop{t1,t2} \fmfbottom{b1,b2}
			\fmf{quark,label=$\hat e_t$,label.side=left,tension=4}{t1,v1}
			\fmf{quark,label=$\hat e_s$,label.side=left,tension=4}{v2,t2}
			\fmf{plain,label=$\hat \nu_r$,label.side=right,tension=4}{b1,v3}
			\fmf{plain,label=$\hat \nu_p$,label.side=left,tension=4}{b2,v4}
			\fmf{quark,label=$e_u$,label.side=right}{v1,v3}
			\fmf{quark,label=$e_v$,label.side=right}{v4,v2}
			\fmf{wboson,label=$\W^+$,label.side=right}{v4,v3}
			\fmf{photon,label=$\Z$,label.side=left}{v1,v2}
		\end{fmfgraph*}
	\end{gathered} \quad \times 2
\qquad\quad
	\begin{gathered}
		\begin{fmfgraph*}(55,60)
			\fmfset{curly_len}{2mm}
			\fmftop{t1,t2} \fmfbottom{b1,b2}
			\fmf{quark,label=$\hat e_t$,label.side=left,tension=4}{t1,v1}
			\fmf{quark,label=$\hat e_s$,label.side=left,tension=4}{v2,t2}
			\fmf{plain,label=$\hat \nu_r$,label.side=right,tension=4}{b1,v3}
			\fmf{plain,label=$\hat \nu_p$,label.side=left,tension=4}{b2,v4}
			\fmf{plain,label=$\nu_u$,label.side=right}{v1,v2}
			\fmf{quark,label=$e_v$,label.side=right}{v3,v4}
			\fmf{wboson,label=$\W^+$,label.side=left}{v3,v1}
			\fmf{wboson,label=$\W^+$,label.side=left}{v2,v4}
		\end{fmfgraph*}
	\end{gathered} \quad\,\, \times 2
\qquad\quad
	\begin{gathered}
		\begin{fmfgraph*}(55,60)
			\fmfset{curly_len}{2mm}
			\fmftop{t1,t2} \fmfbottom{b1,b2}
			\fmf{quark,label=$\hat e_t$,label.side=left,tension=4}{t1,v1}
			\fmf{quark,label=$\hat e_s$,label.side=left,tension=4}{v2,t2}
			\fmf{plain,label=$\hat \nu_r$,label.side=right,tension=4}{b1,v3}
			\fmf{plain,label=$\hat \nu_p$,label.side=left,tension=4}{b2,v4}
			\fmf{quark,label=$e_u$,label.side=right}{v1,v2}
			\fmf{plain,label=$\nu_v$,label.side=right}{v3,v4}
			\fmf{photon,label=$\Z$,label.side=left}{v3,v1}
			\fmf{photon,label=$\Z$,label.side=left}{v2,v4}
		\end{fmfgraph*}
	\end{gathered} \quad \times 2
\nnw
	& \begin{gathered}
		\begin{fmfgraph*}(55,60)
			\fmfset{curly_len}{2mm}
			\fmftop{t1,t2} \fmfbottom{b1,b2}
			\fmf{quark,label=$\hat e_t$,label.side=left,tension=4}{t1,v1}
			\fmf{quark,label=$\hat e_s$,label.side=left,tension=4}{v2,t2}
			\fmf{plain,label=$\hat \nu_r$,label.side=right,tension=4}{b1,v3}
			\fmf{plain,label=$\hat \nu_p$,label.side=left,tension=4}{b2,v4}
			\fmf{plain,label=$\nu_u$,label.side=right}{v1,v3}
			\fmf{plain,label=$\nu_v$,label.side=right}{v4,v2}
			\fmf{photon,label=$\Z$,label.side=right}{v3,v4}
			\fmf{wboson,label=$\W^+$,label.side=left}{v2,v1}
		\end{fmfgraph*}
	\end{gathered} \quad \times 2
\qquad\quad
	\begin{gathered}
		\begin{fmfgraph*}(55,60)
			\fmfset{curly_len}{2mm}
			\fmftop{t1,t2} \fmfbottom{b1,b2}
			\fmf{quark,label=$\hat e_t$,label.side=left,tension=4}{t1,v1}
			\fmf{plain,label=$\hat \nu_r$,label.side=right,tension=4}{b1,v3}
			\fmf{phantom,label=$\hat e_s$,label.side=right,tension=4}{v2,t2}
			\fmf{phantom,label=$\hat \nu_p$,label.side=right,tension=4}{b2,v4}
			\fmf{plain,label=$\nu_u$,label.side=right}{v2,v1}
			\fmf{quark,label=$e_v$,label.side=right}{v3,v4}
			\fmf{photon,label=$\Z$,label.side=right}{v2,v4}
			\fmf{wboson,label=$\W^+$,label.side=left}{v3,v1}
			\fmffreeze
			\fmf{quark}{v4,t2}
			\fmf{plain}{b2,v2}
		\end{fmfgraph*}
	\end{gathered} \quad \times 2
\qquad\quad
	\begin{gathered}
		\begin{fmfgraph*}(55,60)
			\fmfset{curly_len}{2mm}
			\fmftop{t1,t2} \fmfbottom{b1,b2}
			\fmf{quark,label=$\hat e_t$,label.side=left,tension=4}{t1,v1}
			\fmf{plain,label=$\hat \nu_r$,label.side=right,tension=4}{b1,v3}
			\fmf{phantom,label=$\hat e_s$,label.side=right,tension=4}{v2,t2}
			\fmf{phantom,label=$\hat \nu_p$,label.side=right,tension=4}{b2,v4}
			\fmf{quark,label=$e_u$,label.side=left}{v1,v2}
			\fmf{plain,label=$\nu_v$,label.side=right}{v3,v4}
			\fmf{photon,label=$\Z$,label.side=right}{v1,v3}
			\fmf{wboson,label=$\W^+$,label.side=left}{v4,v2}
			\fmffreeze
			\fmf{quark}{v4,t2}
			\fmf{plain}{b2,v2}
		\end{fmfgraph*}
	\end{gathered} \quad \times 2
\qquad\quad
	\begin{gathered}
		\begin{fmfgraph*}(55,60)
			\fmfset{curly_len}{2mm}
			\fmftop{t1,t2} \fmfbottom{b1,b2}
			\fmf{quark,label=$\hat e_t$,label.side=left,label.dist=3,tension=3}{t1,v2}
			\fmf{quark,label=$\hat e_s$,label.side=right,label.dist=4,tension=3}{v1,t2}
			\fmf{plain,label=$\hat \nu_r$,label.side=right,label.dist=3,tension=3}{b1,v3}
			\fmf{plain,label=$\hat \nu_p$,label.side=right,label.dist=4,tension=3}{b2,v1}
			\fmf{photon,label=$\A$,label.side=right,label.dist=3,tension=1}{v1,v2}
			\fmf{wboson,label=$\W^+$,label.side=left,label.dist=3,tension=1}{v1,v3}
			\fmffreeze
			\fmf{quark,label=$e_u$,label.side=right}{v2,v3}
			\fmfv{decoration.shape=square,decoration.size=1.5mm}{v1}
		\end{fmfgraph*}
	\end{gathered} \quad \times 2
\nnw
	& \begin{gathered}
		\begin{fmfgraph*}(55,60)
			\fmfset{curly_len}{2mm}
			\fmftop{t1,t2} \fmfbottom{b1,b2}
			\fmf{quark,label=$\hat e_t$,label.side=left,label.dist=3,tension=3}{t1,v2}
			\fmf{quark,label=$\hat e_s$,label.side=right,label.dist=4,tension=3}{v1,t2}
			\fmf{plain,label=$\hat \nu_r$,label.side=right,label.dist=3,tension=3}{b1,v3}
			\fmf{plain,label=$\hat \nu_p$,label.side=right,label.dist=4,tension=3}{b2,v1}
			\fmf{photon,label=$\Z$,label.side=right,label.dist=3,tension=1}{v1,v2}
			\fmf{wboson,label=$\W^+$,label.side=left,label.dist=3,tension=1}{v1,v3}
			\fmffreeze
			\fmf{quark,label=$e_u$,label.side=right}{v2,v3}
			\fmfv{decoration.shape=square,decoration.size=1.5mm}{v1}
		\end{fmfgraph*}
	\end{gathered} \quad \times 2
\qquad\quad
	\begin{gathered}
		\begin{fmfgraph*}(55,60)
			\fmfset{curly_len}{2mm}
			\fmftop{t1,t2} \fmfbottom{b1,b2}
			\fmf{quark,label=$\hat e_t$,label.side=left,label.dist=3,tension=3}{t1,v2}
			\fmf{quark,label=$\hat e_s$,label.side=right,label.dist=4,tension=3}{v1,t2}
			\fmf{plain,label=$\hat \nu_r$,label.side=right,label.dist=3,tension=3}{b1,v3}
			\fmf{plain,label=$\hat \nu_p$,label.side=right,label.dist=4,tension=3}{b2,v1}
			\fmf{wboson,label=$\W^+$,label.side=right,label.dist=3,tension=1}{v1,v2}
			\fmf{photon,label=$\Z$,label.side=left,label.dist=3,tension=1}{v1,v3}
			\fmffreeze
			\fmf{plain,label=$\nu_u$,label.side=right}{v2,v3}
			\fmfv{decoration.shape=square,decoration.size=1.5mm}{v1}
		\end{fmfgraph*}
	\end{gathered} \quad \times 2
\qquad\quad
	\begin{gathered}
		\begin{fmfgraph*}(55,60)
			\fmfset{curly_len}{2mm}
			\fmftop{t1,t2} \fmfbottom{b1,b2}
			\fmf{quark,label=$\hat e_t$,label.side=left,label.dist=4,tension=3}{t1,v1}
			\fmf{quark,label=$\hat e_s$,label.side=left,label.dist=4,tension=3}{v1,t2}
			\fmf{plain,label=$\hat \nu_r$,label.side=left,label.dist=3,tension=3}{b1,v2}
			\fmf{plain,label=$\hat \nu_p$,label.side=right,label.dist=3,tension=3}{b2,v3}
			\fmf{wboson,label=$\W^+$,label.side=left,label.dist=3,tension=1}{v2,v1,v3}
			\fmffreeze
			\fmf{quark,label=$e_u$,label.side=right}{v2,v3}
			\fmfv{decoration.shape=square,decoration.size=1.5mm}{v1}
		\end{fmfgraph*}
	\end{gathered} \quad \times 2
\qquad\quad
	\begin{gathered}
		\begin{fmfgraph*}(55,60)
			\fmfset{curly_len}{2mm}
			\fmftop{t1,t2} \fmfbottom{b1,b2}
			\fmf{quark,label=$\hat e_t$,label.side=right,label.dist=4,tension=3}{t1,v1}
			\fmf{quark,label=$\hat e_s$,label.side=left,label.dist=3,tension=3}{v2,t2}
			\fmf{plain,label=$\hat \nu_r$,label.side=left,label.dist=4,tension=3}{b1,v1}
			\fmf{plain,label=$\hat \nu_p$,label.side=left,label.dist=3,tension=3}{b2,v3}
			\fmf{photon,label=$\A$,label.side=left,label.dist=3,tension=1}{v1,v2}
			\fmf{wboson,label=$\W^+$,label.side=left,label.dist=3,tension=1}{v3,v1}
			\fmffreeze
			\fmf{quark,label=$e_u$,label.side=right}{v3,v2}
			\fmfv{decoration.shape=square,decoration.size=1.5mm}{v1}
		\end{fmfgraph*}
	\end{gathered} \quad \times 2
\nnw
	& \begin{gathered}
		\begin{fmfgraph*}(55,60)
			\fmfset{curly_len}{2mm}
			\fmftop{t1,t2} \fmfbottom{b1,b2}
			\fmf{quark,label=$\hat e_t$,label.side=right,label.dist=4,tension=3}{t1,v1}
			\fmf{quark,label=$\hat e_s$,label.side=left,label.dist=3,tension=3}{v2,t2}
			\fmf{plain,label=$\hat \nu_r$,label.side=left,label.dist=4,tension=3}{b1,v1}
			\fmf{plain,label=$\hat \nu_p$,label.side=left,label.dist=3,tension=3}{b2,v3}
			\fmf{photon,label=$\Z$,label.side=left,label.dist=3,tension=1}{v1,v2}
			\fmf{wboson,label=$\W^+$,label.side=left,label.dist=3,tension=1}{v3,v1}
			\fmffreeze
			\fmf{quark,label=$e_u$,label.side=right}{v3,v2}
			\fmfv{decoration.shape=square,decoration.size=1.5mm}{v1}
		\end{fmfgraph*}
	\end{gathered} \quad \times 2
\qquad\quad
	\begin{gathered}
		\begin{fmfgraph*}(55,60)
			\fmfset{curly_len}{2mm}
			\fmftop{t1,t2} \fmfbottom{b1,b2}
			\fmf{quark,label=$\hat e_t$,label.side=right,label.dist=4,tension=3}{t1,v1}
			\fmf{quark,label=$\hat e_s$,label.side=left,label.dist=3,tension=3}{v2,t2}
			\fmf{plain,label=$\hat \nu_r$,label.side=left,label.dist=4,tension=3}{b1,v1}
			\fmf{plain,label=$\hat \nu_p$,label.side=left,label.dist=3,tension=3}{b2,v3}
			\fmf{wboson,label=$\W^+$,label.side=right,label.dist=3,tension=1}{v2,v1}
			\fmf{photon,label=$\Z$,label.side=left,label.dist=3,tension=1}{v3,v1}
			\fmffreeze
			\fmf{plain,label=$\nu_u$,label.side=right}{v3,v2}
			\fmfv{decoration.shape=square,decoration.size=1.5mm}{v1}
		\end{fmfgraph*}
	\end{gathered} \quad \times 2
\qquad\quad
	\begin{gathered}
		\begin{fmfgraph*}(55,60)
			\fmfset{curly_len}{2mm}
			\fmftop{t1,t2} \fmfbottom{b1,b2}
			\fmf{quark,label=$\hat e_t$,label.side=left,label.dist=3,tension=6}{t1,v2}
			\fmf{quark,label=$\hat e_s$,label.side=right,label.dist=3,tension=2}{v1,t2}
			\fmf{plain,label=$\hat \nu_r$,label.side=right,label.dist=3,tension=6}{b1,v3}
			\fmf{plain,label=$\hat \nu_p$,label.side=right,label.dist=3,tension=2}{b2,v1}
			\fmf{plain,label=$\nu_u$,label.side=right,label.dist=3,tension=3}{v1,v2}
			\fmf{quark,label=$e_v$,label.side=right,label.dist=3,tension=3}{v3,v1}
			\fmffreeze
			\fmf{wboson,label=$\W^+$,label.side=left,left=0.25}{v3,v2}
			\fmfv{decoration.shape=square,decoration.size=1.5mm}{v1}
		\end{fmfgraph*}
	\end{gathered} \quad \times 2
\qquad\quad
	\begin{gathered}
		\begin{fmfgraph*}(55,60)
			\fmfset{curly_len}{2mm}
			\fmftop{t1,t2} \fmfbottom{b1,b2}
			\fmf{quark,label=$\hat e_t$,label.side=left,label.dist=3,tension=6}{t1,v2}
			\fmf{quark,label=$\hat e_s$,label.side=right,label.dist=3,tension=2}{v1,t2}
			\fmf{plain,label=$\hat \nu_r$,label.side=right,label.dist=3,tension=6}{b1,v3}
			\fmf{plain,label=$\hat \nu_p$,label.side=right,label.dist=3,tension=2}{b2,v1}
			\fmf{quark,label=$e_u$,label.side=left,label.dist=3,tension=3}{v2,v1}
			\fmf{plain,label=$\nu_v$,label.side=right,label.dist=3,tension=3}{v3,v1}
			\fmffreeze
			\fmf{photon,label=$\Z$,label.side=left,left=0.25}{v3,v2}
			\fmfv{decoration.shape=square,decoration.size=1.5mm}{v1}
		\end{fmfgraph*}
	\end{gathered} \quad \times 2
\nnw
	& \begin{gathered}
		\begin{fmfgraph*}(55,60)
			\fmfset{curly_len}{2mm}
			\fmftop{t1,t2} \fmfbottom{b1,b2}
			\fmf{quark,label=$\hat e_t$,label.side=right,label.dist=3,tension=2}{t1,v1}
			\fmf{quark,label=$\hat e_s$,label.side=left,label.dist=3,tension=6}{v2,t2}
			\fmf{plain,label=$\hat \nu_r$,label.side=left,label.dist=3,tension=2}{b1,v1}
			\fmf{plain,label=$\hat \nu_p$,label.side=left,label.dist=3,tension=6}{b2,v3}
			\fmf{plain,label=$\nu_u$,label.side=right,label.dist=3,tension=3}{v2,v1}
			\fmf{quark,label=$e_v$,label.side=right,label.dist=3,tension=3}{v1,v3}
			\fmffreeze
			\fmf{wboson,label=$\W^+$,label.side=left,left=0.25}{v2,v3}
			\fmfv{decoration.shape=square,decoration.size=1.5mm}{v1}
		\end{fmfgraph*}
	\end{gathered} \qquad \times 2
\quad\quad\,\,
	\begin{gathered}
		\begin{fmfgraph*}(55,60)
			\fmfset{curly_len}{2mm}
			\fmftop{t1,t2} \fmfbottom{b1,b2}
			\fmf{quark,label=$\hat e_t$,label.side=right,label.dist=3,tension=2}{t1,v1}
			\fmf{quark,label=$\hat e_s$,label.side=left,label.dist=3,tension=6}{v2,t2}
			\fmf{plain,label=$\hat \nu_r$,label.side=left,label.dist=3,tension=2}{b1,v1}
			\fmf{plain,label=$\hat \nu_p$,label.side=left,label.dist=3,tension=6}{b2,v3}
			\fmf{quark,label=$e_u$,label.side=left,label.dist=3,tension=3}{v1,v2}
			\fmf{plain,label=$\nu_v$,label.side=right,label.dist=3,tension=3}{v1,v3}
			\fmffreeze
			\fmf{photon,label=$\Z$,label.side=left,left=0.25}{v2,v3}
			\fmfv{decoration.shape=square,decoration.size=1.5mm}{v1}
		\end{fmfgraph*}
	\end{gathered} \quad \times 2
\qquad\quad
	\begin{gathered}
		\begin{fmfgraph*}(55,60)
			\fmfset{curly_len}{2mm}
			\fmftop{t1,t2} \fmfbottom{b1,b2}
			\fmf{quark,label=$\hat e_t$,label.side=left,label.dist=3,tension=2}{t1,v1}
			\fmf{quark,label=$\hat e_s$,label.side=left,label.dist=3,tension=2}{v1,t2}
			\fmf{plain,label=$\hat \nu_r$,label.side=left,label.dist=3,tension=6}{b1,v2}
			\fmf{plain,label=$\hat \nu_p$,label.side=right,label.dist=3,tension=6}{b2,v3}
			\fmf{quark,label=$e_u$,label.side=left,label.dist=3,tension=3}{v2,v1}
			\fmf{quark,label=$e_v$,label.side=left,label.dist=3,tension=3}{v1,v3}
			\fmffreeze
			\fmf{wboson,label=$\W^+$,label.side=right,right=0.25}{v2,v3}
			\fmfv{decoration.shape=square,decoration.size=1.5mm}{v1}
		\end{fmfgraph*}
	\end{gathered} \quad \times 2
\qquad\quad
	\begin{gathered}
		\begin{fmfgraph*}(55,60)
			\fmfset{curly_len}{2mm}
			\fmftop{t1,t2} \fmfbottom{b1,b2}
			\fmf{quark,label=$\hat e_t$,label.side=left,label.dist=3,tension=2}{t1,v1}
			\fmf{quark,label=$\hat e_s$,label.side=left,label.dist=3,tension=2}{v1,t2}
			\fmf{plain,label=$\hat \nu_r$,label.side=left,label.dist=3,tension=6}{b1,v2}
			\fmf{plain,label=$\hat \nu_p$,label.side=right,label.dist=3,tension=6}{b2,v3}
			\fmf{plain,label=$\nu_u$,label.side=left,label.dist=3,tension=3}{v2,v1}
			\fmf{plain,label=$\nu_v$,label.side=left,label.dist=3,tension=3}{v1,v3}
			\fmffreeze
			\fmf{photon,label=$\Z$,label.side=right,right=0.25}{v2,v3}
			\fmfv{decoration.shape=square,decoration.size=1.5mm}{v1}
		\end{fmfgraph*}
	\end{gathered} \quad
\nnw
	& \begin{gathered}
		\begin{fmfgraph*}(55,60)
			\fmfset{curly_len}{2mm}
			\fmftop{t1,t2} \fmfbottom{b1,b2}
			\fmf{quark,label=$\hat e_t$,label.side=right,label.dist=3,tension=6}{t1,v2}
			\fmf{quark,label=$\hat e_s$,label.side=right,label.dist=3,tension=6}{v3,t2}
			\fmf{plain,label=$\hat \nu_r$,label.side=right,label.dist=3,tension=2}{b1,v1}
			\fmf{plain,label=$\hat \nu_p$,label.side=left,label.dist=3,tension=2}{b2,v1}
			\fmf{plain,label=$\nu_u$,label.side=right,label.dist=3,tension=3}{v2,v1}
			\fmf{plain,label=$\nu_v$,label.side=right,label.dist=3,tension=3}{v1,v3}
			\fmffreeze
			\fmf{wboson,label=$\W^+$,label.side=right,right=0.25}{v3,v2}
			\fmfv{decoration.shape=square,decoration.size=1.5mm}{v1}
		\end{fmfgraph*}
	\end{gathered} \quad
\qquad\quad
	\begin{gathered}
		\begin{fmfgraph*}(55,60)
			\fmfset{curly_len}{2mm}
			\fmftop{t1,t2} \fmfbottom{b1,b2}
			\fmf{quark,label=$\hat e_t$,label.side=right,label.dist=3,tension=6}{t1,v2}
			\fmf{quark,label=$\hat e_s$,label.side=right,label.dist=3,tension=6}{v3,t2}
			\fmf{plain,label=$\hat \nu_r$,label.side=right,label.dist=3,tension=2}{b1,v1}
			\fmf{plain,label=$\hat \nu_p$,label.side=left,label.dist=3,tension=2}{b2,v1}
			\fmf{quark,label=$e_u$,label.side=right,label.dist=3,tension=3}{v2,v1}
			\fmf{quark,label=$e_v$,label.side=right,label.dist=3,tension=3}{v1,v3}
			\fmffreeze
			\fmf{photon,label=$\Z$,label.side=right,right=0.25}{v3,v2}
			\fmfv{decoration.shape=square,decoration.size=1.5mm}{v1}
		\end{fmfgraph*}
	\end{gathered}
\end{align*}


\subsubsection[$\nu \nu \bar uu$]{\boldmath$\nu \nu \bar uu$}

\begin{align*}
	& \begin{gathered}
		\begin{fmfgraph*}(55,60)
			\fmfset{curly_len}{2mm}
			\fmftop{t1,t2} \fmfbottom{b1,b2}
			\fmf{quark,label=$\hat u_t$,label.side=left,tension=4}{t1,v1}
			\fmf{quark,label=$\hat u_s$,label.side=left,tension=4}{v2,t2}
			\fmf{plain,label=$\hat \nu_r$,label.side=right,tension=4}{b1,v3}
			\fmf{plain,label=$\hat \nu_p$,label.side=left,tension=4}{b2,v4}
			\fmf{quark,label=$d_u$,label.side=right}{v1,v2}
			\fmf{quark,label=$e_v$,label.side=left}{v4,v3}
			\fmf{wboson,label=$\W^+$,label.side=right}{v1,v3}
			\fmf{wboson,label=$\W^+$,label.side=right}{v4,v2}
		\end{fmfgraph*}
	\end{gathered} \qquad \times 2
\qquad\quad
	\begin{gathered}
		\begin{fmfgraph*}(55,60)
			\fmfset{curly_len}{2mm}
			\fmftop{t1,t2} \fmfbottom{b1,b2}
			\fmf{quark,label=$\hat u_t$,label.side=left,tension=4}{t1,v1}
			\fmf{quark,label=$\hat u_s$,label.side=left,tension=4}{v2,t2}
			\fmf{plain,label=$\hat \nu_r$,label.side=right,tension=4}{b1,v3}
			\fmf{plain,label=$\hat \nu_p$,label.side=left,tension=4}{b2,v4}
			\fmf{quark,label=$u_u$,label.side=right}{v1,v2}
			\fmf{plain,label=$\nu_v$,label.side=right}{v3,v4}
			\fmf{photon,label=$\Z$,label.side=left}{v3,v1}
			\fmf{photon,label=$\Z$,label.side=left}{v2,v4}
		\end{fmfgraph*}
	\end{gathered} \quad \times 2
\qquad\quad
	\begin{gathered}
		\begin{fmfgraph*}(55,60)
			\fmfset{curly_len}{2mm}
			\fmftop{t1,t2} \fmfbottom{b1,b2}
			\fmf{quark,label=$\hat u_t$,label.side=left,label.dist=4,tension=3}{t1,v1}
			\fmf{quark,label=$\hat u_s$,label.side=left,label.dist=4,tension=3}{v1,t2}
			\fmf{plain,label=$\hat \nu_r$,label.side=left,label.dist=3,tension=3}{b1,v2}
			\fmf{plain,label=$\hat \nu_p$,label.side=right,label.dist=3,tension=3}{b2,v3}
			\fmf{wboson,label=$\W^+$,label.side=left,label.dist=3,tension=1}{v2,v1,v3}
			\fmffreeze
			\fmf{quark,label=$e_u$,label.side=right}{v2,v3}
			\fmfv{decoration.shape=square,decoration.size=1.5mm}{v1}
		\end{fmfgraph*}
	\end{gathered} \quad \times 2
\nnw
	& \begin{gathered}
		\begin{fmfgraph*}(55,60)
			\fmfset{curly_len}{2mm}
			\fmftop{t1,t2} \fmfbottom{b1,b2}
			\fmf{quark,label=$\hat u_t$,label.side=left,label.dist=3,tension=6}{t1,v2}
			\fmf{quark,label=$\hat u_s$,label.side=right,label.dist=3,tension=2}{v1,t2}
			\fmf{plain,label=$\hat \nu_r$,label.side=right,label.dist=3,tension=6}{b1,v3}
			\fmf{plain,label=$\hat \nu_p$,label.side=right,label.dist=3,tension=2}{b2,v1}
			\fmf{quark,label=$d_u$,label.side=left,label.dist=3,tension=3}{v2,v1}
			\fmf{quark,label=$e_v$,label.side=left,label.dist=3,tension=3}{v1,v3}
			\fmffreeze
			\fmf{wboson,label=$\W^+$,label.side=right,right=0.25}{v2,v3}
			\fmfv{decoration.shape=square,decoration.size=1.5mm}{v1}
		\end{fmfgraph*}
	\end{gathered} \quad \times 2
\qquad\quad
	\begin{gathered}
		\begin{fmfgraph*}(55,60)
			\fmfset{curly_len}{2mm}
			\fmftop{t1,t2} \fmfbottom{b1,b2}
			\fmf{quark,label=$\hat u_t$,label.side=left,label.dist=3,tension=6}{t1,v2}
			\fmf{quark,label=$\hat u_s$,label.side=right,label.dist=3,tension=2}{v1,t2}
			\fmf{plain,label=$\hat \nu_r$,label.side=right,label.dist=3,tension=6}{b1,v3}
			\fmf{plain,label=$\hat \nu_p$,label.side=right,label.dist=3,tension=2}{b2,v1}
			\fmf{quark,label=$u_u$,label.side=left,label.dist=3,tension=3}{v2,v1}
			\fmf{plain,label=$\nu_v$,label.side=right,label.dist=3,tension=3}{v3,v1}
			\fmffreeze
			\fmf{photon,label=$\Z$,label.side=left,left=0.25}{v3,v2}
			\fmfv{decoration.shape=square,decoration.size=1.5mm}{v1}
		\end{fmfgraph*}
	\end{gathered} \quad \times 2
\qquad\quad
	\begin{gathered}
		\begin{fmfgraph*}(55,60)
			\fmfset{curly_len}{2mm}
			\fmftop{t1,t2} \fmfbottom{b1,b2}
			\fmf{quark,label=$\hat u_t$,label.side=right,label.dist=3,tension=2}{t1,v1}
			\fmf{quark,label=$\hat u_s$,label.side=left,label.dist=3,tension=6}{v2,t2}
			\fmf{plain,label=$\hat \nu_r$,label.side=left,label.dist=3,tension=2}{b1,v1}
			\fmf{plain,label=$\hat \nu_p$,label.side=left,label.dist=3,tension=6}{b2,v3}
			\fmf{quark,label=$d_u$,label.side=left,label.dist=3,tension=3}{v1,v2}
			\fmf{quark,label=$e_v$,label.side=left,label.dist=3,tension=3}{v3,v1}
			\fmffreeze
			\fmf{wboson,label=$\W^+$,label.side=right,right=0.25}{v3,v2}
			\fmfv{decoration.shape=square,decoration.size=1.5mm}{v1}
		\end{fmfgraph*}
	\end{gathered} \qquad \times 2
\qquad\quad
	\begin{gathered}
		\begin{fmfgraph*}(55,60)
			\fmfset{curly_len}{2mm}
			\fmftop{t1,t2} \fmfbottom{b1,b2}
			\fmf{quark,label=$\hat u_t$,label.side=right,label.dist=3,tension=2}{t1,v1}
			\fmf{quark,label=$\hat u_s$,label.side=left,label.dist=3,tension=6}{v2,t2}
			\fmf{plain,label=$\hat \nu_r$,label.side=left,label.dist=3,tension=2}{b1,v1}
			\fmf{plain,label=$\hat \nu_p$,label.side=left,label.dist=3,tension=6}{b2,v3}
			\fmf{quark,label=$u_u$,label.side=left,label.dist=3,tension=3}{v1,v2}
			\fmf{plain,label=$\nu_v$,label.side=right,label.dist=3,tension=3}{v1,v3}
			\fmffreeze
			\fmf{photon,label=$\Z$,label.side=left,left=0.25}{v2,v3}
			\fmfv{decoration.shape=square,decoration.size=1.5mm}{v1}
		\end{fmfgraph*}
	\end{gathered} \quad \times 2
\nnw
	& \begin{gathered}
		\begin{fmfgraph*}(55,60)
			\fmfset{curly_len}{2mm}
			\fmftop{t1,t2} \fmfbottom{b1,b2}
			\fmf{quark,label=$\hat u_t$,label.side=left,label.dist=3,tension=2}{t1,v1}
			\fmf{quark,label=$\hat u_s$,label.side=left,label.dist=3,tension=2}{v1,t2}
			\fmf{plain,label=$\hat \nu_r$,label.side=left,label.dist=3,tension=6}{b1,v2}
			\fmf{plain,label=$\hat \nu_p$,label.side=right,label.dist=3,tension=6}{b2,v3}
			\fmf{quark,label=$e_u$,label.side=left,label.dist=3,tension=3}{v2,v1}
			\fmf{quark,label=$e_v$,label.side=left,label.dist=3,tension=3}{v1,v3}
			\fmffreeze
			\fmf{wboson,label=$\W^+$,label.side=right,right=0.25}{v2,v3}
			\fmfv{decoration.shape=square,decoration.size=1.5mm}{v1}
		\end{fmfgraph*}
	\end{gathered} \quad \times 2
\qquad\quad
	\begin{gathered}
		\begin{fmfgraph*}(55,60)
			\fmfset{curly_len}{2mm}
			\fmftop{t1,t2} \fmfbottom{b1,b2}
			\fmf{quark,label=$\hat u_t$,label.side=left,label.dist=3,tension=2}{t1,v1}
			\fmf{quark,label=$\hat u_s$,label.side=left,label.dist=3,tension=2}{v1,t2}
			\fmf{plain,label=$\hat \nu_r$,label.side=left,label.dist=3,tension=6}{b1,v2}
			\fmf{plain,label=$\hat \nu_p$,label.side=right,label.dist=3,tension=6}{b2,v3}
			\fmf{plain,label=$\nu_u$,label.side=left,label.dist=3,tension=3}{v2,v1}
			\fmf{plain,label=$\nu_v$,label.side=left,label.dist=3,tension=3}{v1,v3}
			\fmffreeze
			\fmf{photon,label=$\Z$,label.side=right,right=0.25}{v2,v3}
			\fmfv{decoration.shape=square,decoration.size=1.5mm}{v1}
		\end{fmfgraph*}
	\end{gathered} \quad
\qquad\quad
	\begin{gathered}
		\begin{fmfgraph*}(55,60)
			\fmfset{curly_len}{2mm}
			\fmftop{t1,t2} \fmfbottom{b1,b2}
			\fmf{quark,label=$\hat u_t$,label.side=right,label.dist=3,tension=6}{t1,v2}
			\fmf{quark,label=$\hat u_s$,label.side=right,label.dist=3,tension=6}{v3,t2}
			\fmf{plain,label=$\hat \nu_r$,label.side=right,label.dist=3,tension=2}{b1,v1}
			\fmf{plain,label=$\hat \nu_p$,label.side=left,label.dist=3,tension=2}{b2,v1}
			\fmf{quark,label=$d_u$,label.side=right,label.dist=3,tension=3}{v2,v1}
			\fmf{quark,label=$d_v$,label.side=right,label.dist=3,tension=3}{v1,v3}
			\fmffreeze
			\fmf{wboson,label=$\W^+$,label.side=left,left=0.25}{v2,v3}
			\fmfv{decoration.shape=square,decoration.size=1.5mm}{v1}
		\end{fmfgraph*}
	\end{gathered} \quad
\qquad\quad
	\begin{gathered}
		\begin{fmfgraph*}(55,60)
			\fmfset{curly_len}{2mm}
			\fmftop{t1,t2} \fmfbottom{b1,b2}
			\fmf{quark,label=$\hat u_t$,label.side=right,label.dist=3,tension=6}{t1,v2}
			\fmf{quark,label=$\hat u_s$,label.side=right,label.dist=3,tension=6}{v3,t2}
			\fmf{plain,label=$\hat \nu_r$,label.side=right,label.dist=3,tension=2}{b1,v1}
			\fmf{plain,label=$\hat \nu_p$,label.side=left,label.dist=3,tension=2}{b2,v1}
			\fmf{quark,label=$u_u$,label.side=right,label.dist=3,tension=3}{v2,v1}
			\fmf{quark,label=$u_v$,label.side=right,label.dist=3,tension=3}{v1,v3}
			\fmffreeze
			\fmf{photon,label=$\Z$,label.side=right,right=0.25}{v3,v2}
			\fmfv{decoration.shape=square,decoration.size=1.5mm}{v1}
		\end{fmfgraph*}
	\end{gathered}
\end{align*}


\subsubsection[$\nu \nu \bar dd$]{\boldmath$\nu \nu \bar dd$}

\begin{align*}
	& \begin{gathered}
		\begin{fmfgraph*}(55,60)
			\fmfset{curly_len}{2mm}
			\fmftop{t1,t2} \fmfbottom{b1,b2}
			\fmf{quark,label=$\hat d_t$,label.side=left,tension=4}{t1,v1}
			\fmf{quark,label=$\hat d_s$,label.side=left,tension=4}{v2,t2}
			\fmf{plain,label=$\hat \nu_r$,label.side=right,tension=4}{b1,v3}
			\fmf{plain,label=$\hat \nu_p$,label.side=left,tension=4}{b2,v4}
			\fmf{quark,label=$u_u$,label.side=right}{v1,v2}
			\fmf{quark,label=$e_v$,label.side=right}{v3,v4}
			\fmf{wboson,label=$\W^+$,label.side=left}{v3,v1}
			\fmf{wboson,label=$\W^+$,label.side=left}{v2,v4}
		\end{fmfgraph*}
	\end{gathered} \quad\,\, \times 2
\qquad\quad
	\begin{gathered}
		\begin{fmfgraph*}(55,60)
			\fmfset{curly_len}{2mm}
			\fmftop{t1,t2} \fmfbottom{b1,b2}
			\fmf{quark,label=$\hat d_t$,label.side=left,tension=4}{t1,v1}
			\fmf{quark,label=$\hat d_s$,label.side=left,tension=4}{v2,t2}
			\fmf{plain,label=$\hat \nu_r$,label.side=right,tension=4}{b1,v3}
			\fmf{plain,label=$\hat \nu_p$,label.side=left,tension=4}{b2,v4}
			\fmf{quark,label=$t$,label.side=right}{v1,v2}
			\fmf{quark,label=$e_v$,label.side=right}{v3,v4}
			\fmf{scalar,label=$G^+$,label.side=left}{v3,v1}
			\fmf{wboson,label=$\W^+$,label.side=left}{v2,v4}
			\fmfv{decoration.shape=square,decoration.size=1.5mm}{v3}
		\end{fmfgraph*}
	\end{gathered} \quad\,\, \times 2
\qquad\quad
	\begin{gathered}
		\begin{fmfgraph*}(55,60)
			\fmfset{curly_len}{2mm}
			\fmftop{t1,t2} \fmfbottom{b1,b2}
			\fmf{quark,label=$\hat d_t$,label.side=left,tension=4}{t1,v1}
			\fmf{quark,label=$\hat d_s$,label.side=left,tension=4}{v2,t2}
			\fmf{plain,label=$\hat \nu_r$,label.side=right,tension=4}{b1,v3}
			\fmf{plain,label=$\hat \nu_p$,label.side=left,tension=4}{b2,v4}
			\fmf{quark,label=$t$,label.side=right}{v1,v2}
			\fmf{quark,label=$e_v$,label.side=right}{v3,v4}
			\fmf{wboson,label=$\W^+$,label.side=left}{v3,v1}
			\fmf{scalar,label=$G^+$,label.side=left}{v2,v4}
			\fmfv{decoration.shape=square,decoration.size=1.5mm}{v4}
		\end{fmfgraph*}
	\end{gathered} \quad\,\, \times 2
\qquad\quad
	\begin{gathered}
		\begin{fmfgraph*}(55,60)
			\fmfset{curly_len}{2mm}
			\fmftop{t1,t2} \fmfbottom{b1,b2}
			\fmf{quark,label=$\hat d_t$,label.side=left,tension=4}{t1,v1}
			\fmf{quark,label=$\hat d_s$,label.side=left,tension=4}{v2,t2}
			\fmf{plain,label=$\hat \nu_r$,label.side=right,tension=4}{b1,v3}
			\fmf{plain,label=$\hat \nu_p$,label.side=left,tension=4}{b2,v4}
			\fmf{quark,label=$d_u$,label.side=right}{v1,v2}
			\fmf{plain,label=$\nu_v$,label.side=right}{v3,v4}
			\fmf{photon,label=$\Z$,label.side=left}{v3,v1}
			\fmf{photon,label=$\Z$,label.side=left}{v2,v4}
		\end{fmfgraph*}
	\end{gathered} \quad \times 2
\nnw
	& \begin{gathered}
		\begin{fmfgraph*}(55,60)
			\fmfset{curly_len}{2mm}
			\fmftop{t1,t2} \fmfbottom{b1,b2}
			\fmf{quark,label=$\hat d_t$,label.side=left,label.dist=4,tension=3}{t1,v1}
			\fmf{quark,label=$\hat d_s$,label.side=left,label.dist=4,tension=3}{v1,t2}
			\fmf{plain,label=$\hat \nu_r$,label.side=left,label.dist=3,tension=3}{b1,v2}
			\fmf{plain,label=$\hat \nu_p$,label.side=right,label.dist=3,tension=3}{b2,v3}
			\fmf{wboson,label=$\W^+$,label.side=left,label.dist=3,tension=1}{v2,v1,v3}
			\fmffreeze
			\fmf{quark,label=$e_u$,label.side=right}{v2,v3}
			\fmfv{decoration.shape=square,decoration.size=1.5mm}{v1}
		\end{fmfgraph*}
	\end{gathered} \quad \times 2
\qquad\quad
	\begin{gathered}
		\begin{fmfgraph*}(55,60)
			\fmfset{curly_len}{2mm}
			\fmftop{t1,t2} \fmfbottom{b1,b2}
			\fmf{quark,label=$\hat d_t$,label.side=right,label.dist=3,tension=3}{t1,v2}
			\fmf{quark,label=$\hat d_s$,label.side=right,label.dist=3,tension=3}{v3,t2}
			\fmf{plain,label=$\hat \nu_r$,label.side=right,label.dist=3,tension=3}{b1,v1}
			\fmf{plain,label=$\hat \nu_p$,label.side=left,label.dist=3,tension=3}{b2,v1}
			\fmf{wboson,label=$\W^+$,label.side=left,label.dist=3,tension=1}{v1,v2}
			\fmf{scalar,label=$G^+$,label.side=left,label.dist=3,tension=1}{v3,v1}
			\fmffreeze
			\fmf{quark,label=$t$,label.side=left}{v2,v3}
			\fmfv{decoration.shape=square,decoration.size=1.5mm}{v1}
		\end{fmfgraph*}
	\end{gathered} \quad
\qquad\quad
	\begin{gathered}
		\begin{fmfgraph*}(55,60)
			\fmfset{curly_len}{2mm}
			\fmftop{t1,t2} \fmfbottom{b1,b2}
			\fmf{quark,label=$\hat d_t$,label.side=right,label.dist=3,tension=3}{t1,v2}
			\fmf{quark,label=$\hat d_s$,label.side=right,label.dist=3,tension=3}{v3,t2}
			\fmf{plain,label=$\hat \nu_r$,label.side=right,label.dist=3,tension=3}{b1,v1}
			\fmf{plain,label=$\hat \nu_p$,label.side=left,label.dist=3,tension=3}{b2,v1}
			\fmf{scalar,label=$G^+$,label.side=left,label.dist=3,tension=1}{v1,v2}
			\fmf{wboson,label=$\W^+$,label.side=left,label.dist=3,tension=1}{v3,v1}
			\fmffreeze
			\fmf{quark,label=$t$,label.side=left}{v2,v3}
			\fmfv{decoration.shape=square,decoration.size=1.5mm}{v1}
		\end{fmfgraph*}
	\end{gathered} \quad
\qquad\quad
	\begin{gathered}
		\begin{fmfgraph*}(55,60)
			\fmfset{curly_len}{2mm}
			\fmftop{t1,t2} \fmfbottom{b1,b2}
			\fmf{quark,label=$\hat d_t$,label.side=right,label.dist=3,tension=3}{t1,v2}
			\fmf{quark,label=$\hat d_s$,label.side=right,label.dist=3,tension=3}{v3,t2}
			\fmf{plain,label=$\hat \nu_r$,label.side=right,label.dist=3,tension=3}{b1,v1}
			\fmf{plain,label=$\hat \nu_p$,label.side=left,label.dist=3,tension=3}{b2,v1}
			\fmf{scalar,label=$G^+$,label.side=left,label.dist=3,tension=1}{v1,v2}
			\fmf{scalar,label=$G^+$,label.side=left,label.dist=3,tension=1}{v3,v1}
			\fmffreeze
			\fmf{quark,label=$t$,label.side=left}{v2,v3}
			\fmfv{decoration.shape=square,decoration.size=1.5mm}{v1}
		\end{fmfgraph*}
	\end{gathered} \qquad
\nnw
	& \begin{gathered}
		\begin{fmfgraph*}(55,60)
			\fmfset{curly_len}{2mm}
			\fmftop{t1,t2} \fmfbottom{b1,b2}
			\fmf{quark,label=$\hat d_t$,label.side=left,label.dist=3,tension=6}{t1,v2}
			\fmf{quark,label=$\hat d_s$,label.side=right,label.dist=3,tension=2}{v1,t2}
			\fmf{plain,label=$\hat \nu_r$,label.side=right,label.dist=3,tension=6}{b1,v3}
			\fmf{plain,label=$\hat \nu_p$,label.side=right,label.dist=3,tension=2}{b2,v1}
			\fmf{quark,label=$u_u$,label.side=left,label.dist=3,tension=3}{v2,v1}
			\fmf{quark,label=$e_v$,label.side=right,label.dist=3,tension=3}{v3,v1}
			\fmffreeze
			\fmf{wboson,label=$\W^+$,label.side=left,left=0.25}{v3,v2}
			\fmfv{decoration.shape=square,decoration.size=1.5mm}{v1}
		\end{fmfgraph*}
	\end{gathered} \quad \times 2
\qquad\quad
	\begin{gathered}
		\begin{fmfgraph*}(55,60)
			\fmfset{curly_len}{2mm}
			\fmftop{t1,t2} \fmfbottom{b1,b2}
			\fmf{quark,label=$\hat d_t$,label.side=left,label.dist=3,tension=6}{t1,v2}
			\fmf{quark,label=$\hat d_s$,label.side=right,label.dist=3,tension=2}{v1,t2}
			\fmf{plain,label=$\hat \nu_r$,label.side=right,label.dist=3,tension=6}{b1,v3}
			\fmf{plain,label=$\hat \nu_p$,label.side=right,label.dist=3,tension=2}{b2,v1}
			\fmf{quark,label=$d_u$,label.side=left,label.dist=3,tension=3}{v2,v1}
			\fmf{plain,label=$\nu_v$,label.side=right,label.dist=3,tension=3}{v3,v1}
			\fmffreeze
			\fmf{photon,label=$\Z$,label.side=left,left=0.25}{v3,v2}
			\fmfv{decoration.shape=square,decoration.size=1.5mm}{v1}
		\end{fmfgraph*}
	\end{gathered} \quad \times 2
\qquad\quad
	\begin{gathered}
		\begin{fmfgraph*}(55,60)
			\fmfset{curly_len}{2mm}
			\fmftop{t1,t2} \fmfbottom{b1,b2}
			\fmf{quark,label=$\hat d_t$,label.side=right,label.dist=3,tension=2}{t1,v1}
			\fmf{quark,label=$\hat d_s$,label.side=left,label.dist=3,tension=6}{v2,t2}
			\fmf{plain,label=$\hat \nu_r$,label.side=left,label.dist=3,tension=2}{b1,v1}
			\fmf{plain,label=$\hat \nu_p$,label.side=left,label.dist=3,tension=6}{b2,v3}
			\fmf{quark,label=$u_u$,label.side=left,label.dist=3,tension=3}{v1,v2}
			\fmf{quark,label=$e_v$,label.side=right,label.dist=3,tension=3}{v1,v3}
			\fmffreeze
			\fmf{wboson,label=$\W^+$,label.side=left,left=0.25}{v2,v3}
			\fmfv{decoration.shape=square,decoration.size=1.5mm}{v1}
		\end{fmfgraph*}
	\end{gathered} \qquad \times 2
\qquad\quad
	\begin{gathered}
		\begin{fmfgraph*}(55,60)
			\fmfset{curly_len}{2mm}
			\fmftop{t1,t2} \fmfbottom{b1,b2}
			\fmf{quark,label=$\hat d_t$,label.side=right,label.dist=3,tension=2}{t1,v1}
			\fmf{quark,label=$\hat d_s$,label.side=left,label.dist=3,tension=6}{v2,t2}
			\fmf{plain,label=$\hat \nu_r$,label.side=left,label.dist=3,tension=2}{b1,v1}
			\fmf{plain,label=$\hat \nu_p$,label.side=left,label.dist=3,tension=6}{b2,v3}
			\fmf{quark,label=$d_u$,label.side=left,label.dist=3,tension=3}{v1,v2}
			\fmf{plain,label=$\nu_v$,label.side=right,label.dist=3,tension=3}{v1,v3}
			\fmffreeze
			\fmf{photon,label=$\Z$,label.side=left,left=0.25}{v2,v3}
			\fmfv{decoration.shape=square,decoration.size=1.5mm}{v1}
		\end{fmfgraph*}
	\end{gathered} \quad \times 2
\nnw
	& \begin{gathered}
		\begin{fmfgraph*}(55,60)
			\fmfset{curly_len}{2mm}
			\fmftop{t1,t2} \fmfbottom{b1,b2}
			\fmf{quark,label=$\hat d_t$,label.side=left,label.dist=3,tension=2}{t1,v1}
			\fmf{quark,label=$\hat d_s$,label.side=left,label.dist=3,tension=2}{v1,t2}
			\fmf{plain,label=$\hat \nu_r$,label.side=left,label.dist=3,tension=6}{b1,v2}
			\fmf{plain,label=$\hat \nu_p$,label.side=right,label.dist=3,tension=6}{b2,v3}
			\fmf{quark,label=$e_u$,label.side=left,label.dist=3,tension=3}{v2,v1}
			\fmf{quark,label=$e_v$,label.side=left,label.dist=3,tension=3}{v1,v3}
			\fmffreeze
			\fmf{wboson,label=$\W^+$,label.side=right,right=0.25}{v2,v3}
			\fmfv{decoration.shape=square,decoration.size=1.5mm}{v1}
		\end{fmfgraph*}
	\end{gathered} \quad \times 2
\qquad\quad
	\begin{gathered}
		\begin{fmfgraph*}(55,60)
			\fmfset{curly_len}{2mm}
			\fmftop{t1,t2} \fmfbottom{b1,b2}
			\fmf{quark,label=$\hat d_t$,label.side=left,label.dist=3,tension=2}{t1,v1}
			\fmf{quark,label=$\hat d_s$,label.side=left,label.dist=3,tension=2}{v1,t2}
			\fmf{plain,label=$\hat \nu_r$,label.side=left,label.dist=3,tension=6}{b1,v2}
			\fmf{plain,label=$\hat \nu_p$,label.side=right,label.dist=3,tension=6}{b2,v3}
			\fmf{plain,label=$\nu_u$,label.side=left,label.dist=3,tension=3}{v2,v1}
			\fmf{plain,label=$\nu_v$,label.side=left,label.dist=3,tension=3}{v1,v3}
			\fmffreeze
			\fmf{photon,label=$\Z$,label.side=right,right=0.25}{v2,v3}
			\fmfv{decoration.shape=square,decoration.size=1.5mm}{v1}
		\end{fmfgraph*}
	\end{gathered}
\qquad\qquad
	\begin{gathered}
		\begin{fmfgraph*}(55,60)
			\fmfset{curly_len}{2mm}
			\fmftop{t1,t2} \fmfbottom{b1,b2}
			\fmf{quark,label=$\hat d_t$,label.side=right,label.dist=3,tension=6}{t1,v2}
			\fmf{quark,label=$\hat d_s$,label.side=right,label.dist=3,tension=6}{v3,t2}
			\fmf{plain,label=$\hat \nu_r$,label.side=right,label.dist=3,tension=2}{b1,v1}
			\fmf{plain,label=$\hat \nu_p$,label.side=left,label.dist=3,tension=2}{b2,v1}
			\fmf{quark,label=$u_u$,label.side=right,label.dist=3,tension=3}{v2,v1}
			\fmf{quark,label=$u_v$,label.side=right,label.dist=3,tension=3}{v1,v3}
			\fmffreeze
			\fmf{wboson,label=$\W^+$,label.side=right,right=0.25}{v3,v2}
			\fmfv{decoration.shape=square,decoration.size=1.5mm}{v1}
		\end{fmfgraph*}
	\end{gathered}
\qquad\qquad
	\begin{gathered}
		\begin{fmfgraph*}(55,60)
			\fmfset{curly_len}{2mm}
			\fmftop{t1,t2} \fmfbottom{b1,b2}
			\fmf{quark,label=$\hat d_t$,label.side=right,label.dist=3,tension=6}{t1,v2}
			\fmf{quark,label=$\hat d_s$,label.side=right,label.dist=3,tension=6}{v3,t2}
			\fmf{plain,label=$\hat \nu_r$,label.side=right,label.dist=3,tension=2}{b1,v1}
			\fmf{plain,label=$\hat \nu_p$,label.side=left,label.dist=3,tension=2}{b2,v1}
			\fmf{quark,label=$t$,label.side=right,label.dist=3,tension=3}{v2,v1}
			\fmf{quark,label=$t$,label.side=right,label.dist=3,tension=3}{v1,v3}
			\fmffreeze
			\fmf{scalar,label=$G^+$,label.side=right,right=0.25}{v3,v2}
			\fmfv{decoration.shape=square,decoration.size=1.5mm}{v1}
		\end{fmfgraph*}
	\end{gathered}
\nnw
	& \begin{gathered}
		\begin{fmfgraph*}(55,60)
			\fmfset{curly_len}{2mm}
			\fmftop{t1,t2} \fmfbottom{b1,b2}
			\fmf{quark,label=$\hat d_t$,label.side=right,label.dist=3,tension=6}{t1,v2}
			\fmf{quark,label=$\hat d_s$,label.side=right,label.dist=3,tension=6}{v3,t2}
			\fmf{plain,label=$\hat \nu_r$,label.side=right,label.dist=3,tension=2}{b1,v1}
			\fmf{plain,label=$\hat \nu_p$,label.side=left,label.dist=3,tension=2}{b2,v1}
			\fmf{quark,label=$d_u$,label.side=right,label.dist=3,tension=3}{v2,v1}
			\fmf{quark,label=$d_v$,label.side=right,label.dist=3,tension=3}{v1,v3}
			\fmffreeze
			\fmf{photon,label=$\Z$,label.side=right,right=0.25}{v3,v2}
			\fmfv{decoration.shape=square,decoration.size=1.5mm}{v1}
		\end{fmfgraph*}
	\end{gathered}
\end{align*}


\subsubsection[$\bar ee \bar ee$]{\boldmath$\bar ee \bar ee$}

We do not consider diagrams consisting of lepton-number-violating parts, because they are proportional to $C_5$ or $m_\nu$.
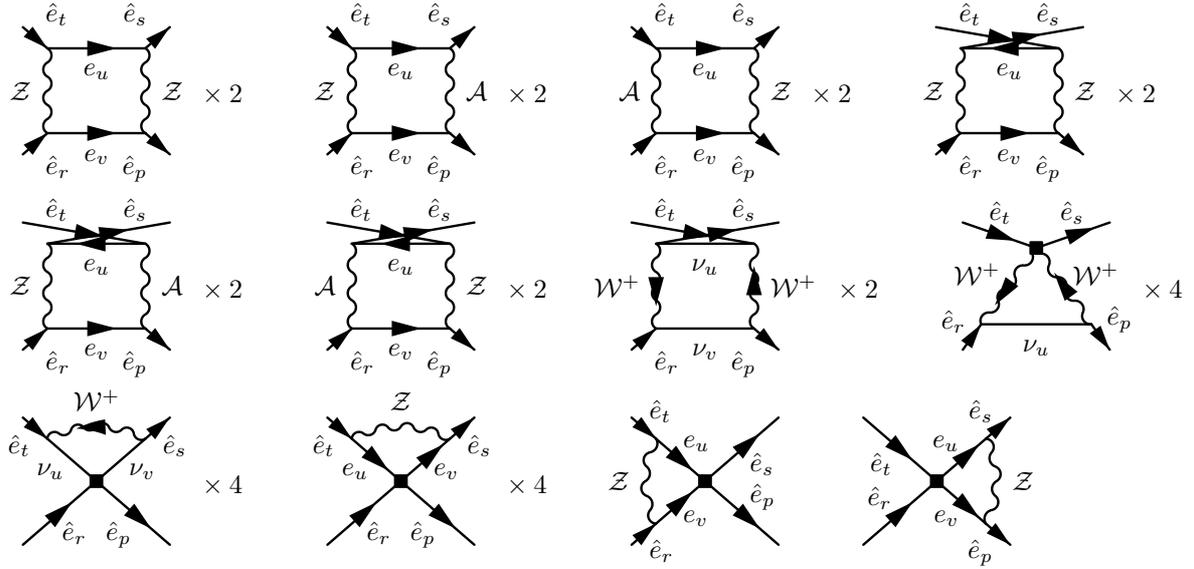
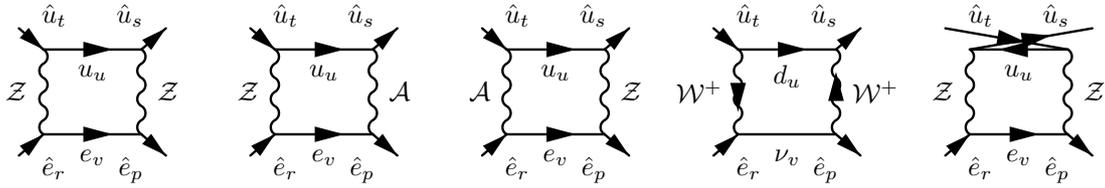
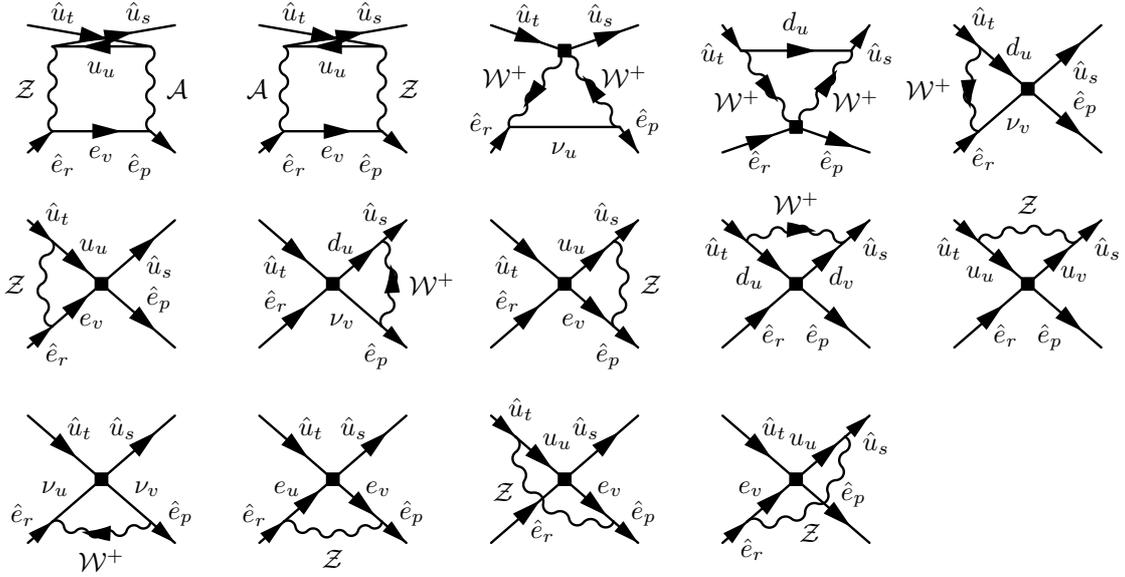
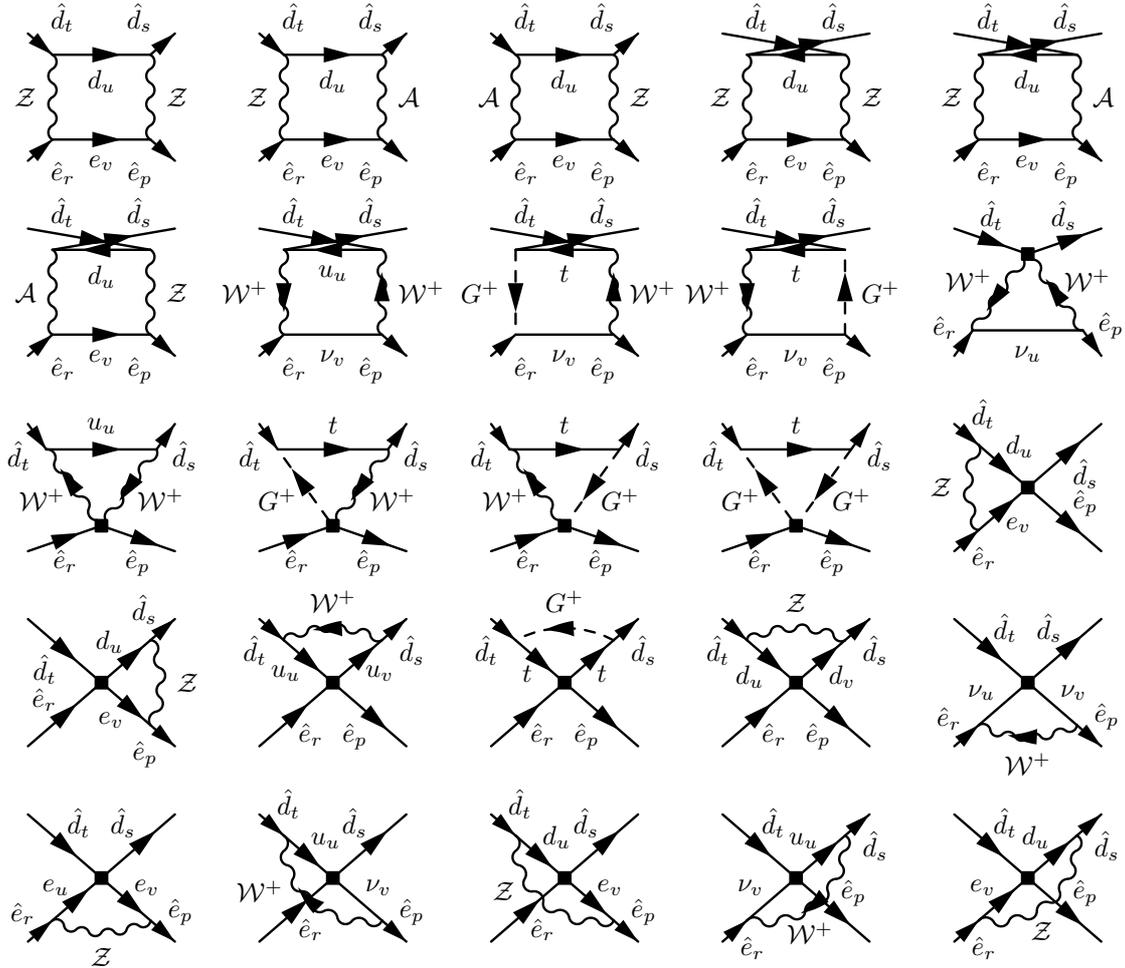
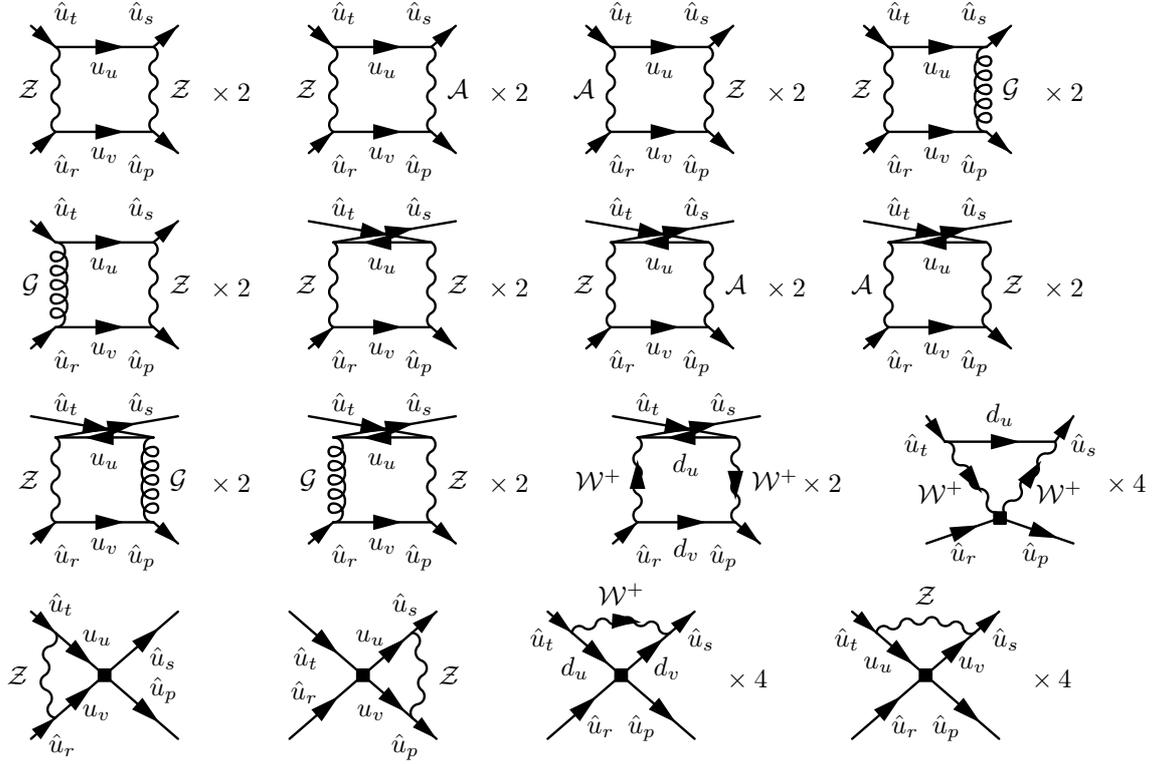
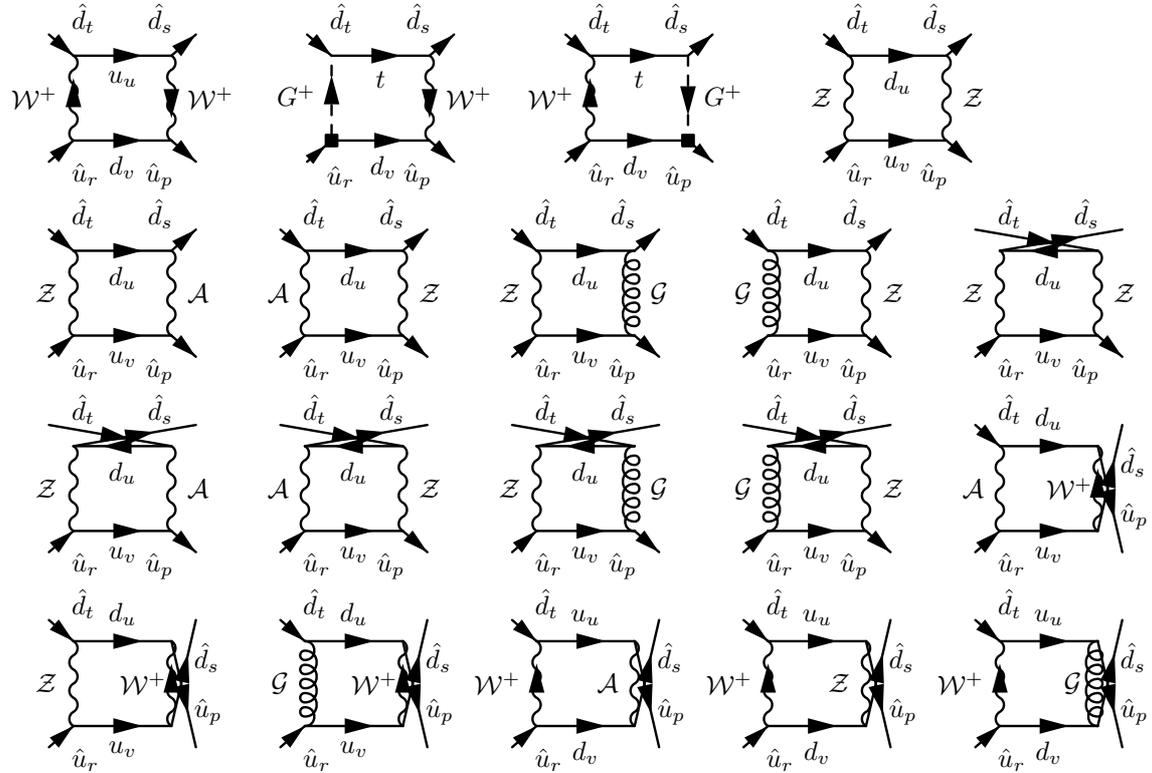
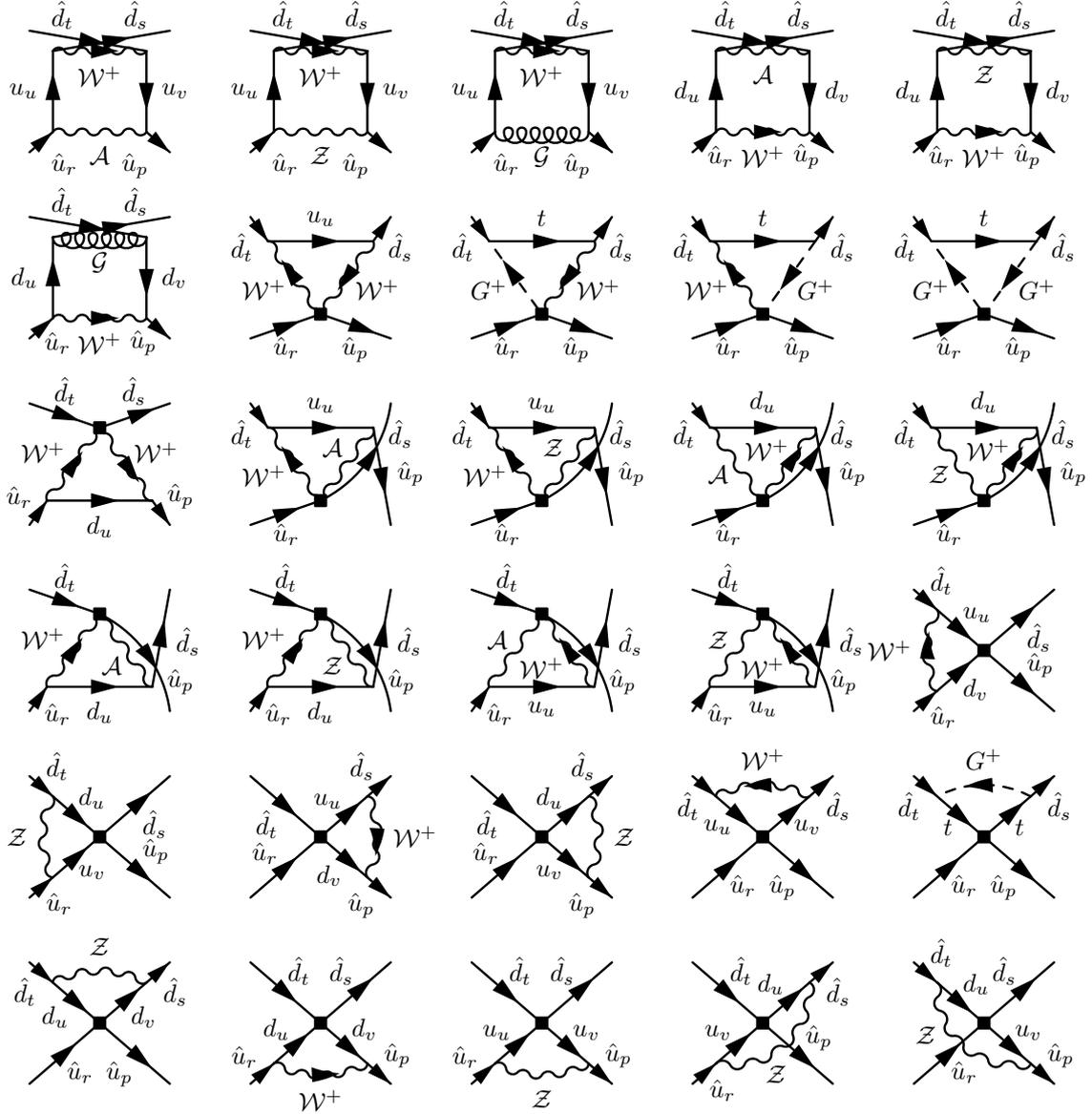
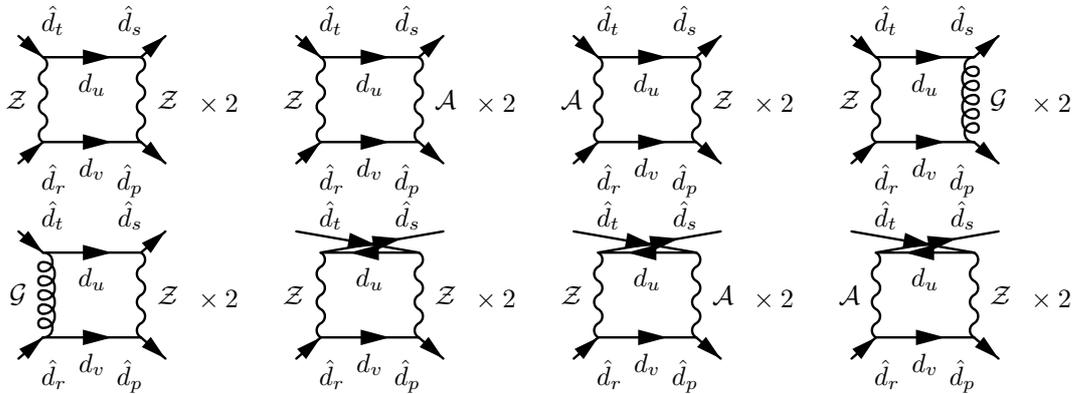
\begin{align*}
	& \begin{gathered}
		\begin{fmfgraph*}(55,60)
			\fmfset{curly_len}{2mm}
			\fmftop{t1,t2} \fmfbottom{b1,b2}
			\fmf{quark,label=$\hat e_t$,label.side=left,tension=4}{t1,v1}
			\fmf{quark,label=$\hat e_s$,label.side=left,tension=4}{v2,t2}
			\fmf{quark,label=$\hat e_r$,label.side=right,tension=4}{b1,v3}
			\fmf{quark,label=$\hat e_p$,label.side=right,tension=4}{v4,b2}
			\fmf{quark,label=$e_u$,label.side=right}{v1,v2}
			\fmf{quark,label=$e_v$,label.side=right}{v3,v4}
			\fmf{photon,label=$\Z$,label.side=left}{v3,v1}
			\fmf{photon,label=$\Z$,label.side=left}{v2,v4}
		\end{fmfgraph*}
	\end{gathered} \quad \times 2
\qquad\quad
	\begin{gathered}
		\begin{fmfgraph*}(55,60)
			\fmfset{curly_len}{2mm}
			\fmftop{t1,t2} \fmfbottom{b1,b2}
			\fmf{quark,label=$\hat e_t$,label.side=left,tension=4}{t1,v1}
			\fmf{quark,label=$\hat e_s$,label.side=left,tension=4}{v2,t2}
			\fmf{quark,label=$\hat e_r$,label.side=right,tension=4}{b1,v3}
			\fmf{quark,label=$\hat e_p$,label.side=right,tension=4}{v4,b2}
			\fmf{quark,label=$e_u$,label.side=right}{v1,v2}
			\fmf{quark,label=$e_v$,label.side=right}{v3,v4}
			\fmf{photon,label=$\Z$,label.side=left}{v3,v1}
			\fmf{photon,label=$\A$,label.side=left}{v2,v4}
		\end{fmfgraph*}
	\end{gathered} \quad \times 2
\qquad\quad
	\begin{gathered}
		\begin{fmfgraph*}(55,60)
			\fmfset{curly_len}{2mm}
			\fmftop{t1,t2} \fmfbottom{b1,b2}
			\fmf{quark,label=$\hat e_t$,label.side=left,tension=4}{t1,v1}
			\fmf{quark,label=$\hat e_s$,label.side=left,tension=4}{v2,t2}
			\fmf{quark,label=$\hat e_r$,label.side=right,tension=4}{b1,v3}
			\fmf{quark,label=$\hat e_p$,label.side=right,tension=4}{v4,b2}
			\fmf{quark,label=$e_u$,label.side=right}{v1,v2}
			\fmf{quark,label=$e_v$,label.side=right}{v3,v4}
			\fmf{photon,label=$\A$,label.side=left}{v3,v1}
			\fmf{photon,label=$\Z$,label.side=left}{v2,v4}
		\end{fmfgraph*}
	\end{gathered} \quad \times 2
\qquad\quad
	\begin{gathered}
		\begin{fmfgraph*}(55,60)
			\fmfset{curly_len}{2mm}
			\fmftop{t1,t2} \fmfbottom{b1,b2}
			\fmf{phantom,label=$\hat e_t$,label.side=left,tension=4}{t1,v1}
			\fmf{phantom,label=$\hat e_s$,label.side=left,tension=4}{v2,t2}
			\fmf{quark,label=$\hat e_r$,label.side=right,tension=4}{b1,v3}
			\fmf{quark,label=$\hat e_p$,label.side=right,tension=4}{v4,b2}
			\fmf{quark,label=$e_u$,label.side=left}{v2,v1}
			\fmf{quark,label=$e_v$,label.side=right}{v3,v4}
			\fmf{photon,label=$\Z$,label.side=left}{v3,v1}
			\fmf{photon,label=$\Z$,label.side=left}{v2,v4}
			\fmffreeze
			\fmf{quark}{t1,v2}
			\fmf{quark}{v1,t2}
		\end{fmfgraph*}
	\end{gathered} \quad \times 2
\nnw
	& \begin{gathered}
		\begin{fmfgraph*}(55,60)
			\fmfset{curly_len}{2mm}
			\fmftop{t1,t2} \fmfbottom{b1,b2}
			\fmf{phantom,label=$\hat e_t$,label.side=left,tension=4}{t1,v1}
			\fmf{phantom,label=$\hat e_s$,label.side=left,tension=4}{v2,t2}
			\fmf{quark,label=$\hat e_r$,label.side=right,tension=4}{b1,v3}
			\fmf{quark,label=$\hat e_p$,label.side=right,tension=4}{v4,b2}
			\fmf{quark,label=$e_u$,label.side=left}{v2,v1}
			\fmf{quark,label=$e_v$,label.side=right}{v3,v4}
			\fmf{photon,label=$\Z$,label.side=left}{v3,v1}
			\fmf{photon,label=$\A$,label.side=left}{v2,v4}
			\fmffreeze
			\fmf{quark}{t1,v2}
			\fmf{quark}{v1,t2}
		\end{fmfgraph*}
	\end{gathered} \quad \times 2
\qquad\quad
	\begin{gathered}
		\begin{fmfgraph*}(55,60)
			\fmfset{curly_len}{2mm}
			\fmftop{t1,t2} \fmfbottom{b1,b2}
			\fmf{phantom,label=$\hat e_t$,label.side=left,tension=4}{t1,v1}
			\fmf{phantom,label=$\hat e_s$,label.side=left,tension=4}{v2,t2}
			\fmf{quark,label=$\hat e_r$,label.side=right,tension=4}{b1,v3}
			\fmf{quark,label=$\hat e_p$,label.side=right,tension=4}{v4,b2}
			\fmf{quark,label=$e_u$,label.side=left}{v2,v1}
			\fmf{quark,label=$e_v$,label.side=right}{v3,v4}
			\fmf{photon,label=$\A$,label.side=left}{v3,v1}
			\fmf{photon,label=$\Z$,label.side=left}{v2,v4}
			\fmffreeze
			\fmf{quark}{t1,v2}
			\fmf{quark}{v1,t2}
		\end{fmfgraph*}
	\end{gathered} \quad \times 2
\qquad\quad
	\begin{gathered}
		\begin{fmfgraph*}(55,60)
			\fmfset{curly_len}{2mm}
			\fmftop{t1,t2} \fmfbottom{b1,b2}
			\fmf{phantom,label=$\hat e_t$,label.side=left,tension=4}{t1,v1}
			\fmf{phantom,label=$\hat e_s$,label.side=left,tension=4}{v2,t2}
			\fmf{quark,label=$\hat e_r$,label.side=right,tension=4}{b1,v3}
			\fmf{quark,label=$\hat e_p$,label.side=right,tension=4}{v4,b2}
			\fmf{plain,label=$\nu_u$,label.side=left}{v2,v1}
			\fmf{plain,label=$\nu_v$,label.side=right}{v3,v4}
			\fmf{wboson,label=$\W^+$,label.side=right}{v1,v3}
			\fmf{wboson,label=$\W^+$,label.side=right}{v4,v2}
			\fmffreeze
			\fmf{quark}{t1,v2}
			\fmf{quark}{v1,t2}
		\end{fmfgraph*}
	\end{gathered} \qquad \times 2
\qquad\quad
	\begin{gathered}
		\begin{fmfgraph*}(55,60)
			\fmfset{curly_len}{2mm}
			\fmftop{t1,t2} \fmfbottom{b1,b2}
			\fmf{quark,label=$\hat e_t$,label.side=left,label.dist=4,tension=3}{t1,v1}
			\fmf{quark,label=$\hat e_s$,label.side=left,label.dist=4,tension=3}{v1,t2}
			\fmf{quark,label=$\hat e_r$,label.side=left,label.dist=3,tension=3}{b1,v2}
			\fmf{quark,label=$\hat e_p$,label.side=left,label.dist=3,tension=3}{v3,b2}
			\fmf{wboson,label=$\W^+$,label.side=right,label.dist=3,tension=1}{v3,v1,v2}
			\fmffreeze
			\fmf{plain,label=$\nu_u$,label.side=right}{v2,v3}
			\fmfv{decoration.shape=square,decoration.size=1.5mm}{v1}
		\end{fmfgraph*}
	\end{gathered} \quad \times 4
\nnw
	& \begin{gathered}
		\begin{fmfgraph*}(55,60)
			\fmfset{curly_len}{2mm}
			\fmftop{t1,t2} \fmfbottom{b1,b2}
			\fmf{quark,label=$\hat e_t$,label.side=right,label.dist=3,tension=6}{t1,v2}
			\fmf{quark,label=$\hat e_s$,label.side=right,label.dist=3,tension=6}{v3,t2}
			\fmf{quark,label=$\hat e_r$,label.side=right,label.dist=3,tension=2}{b1,v1}
			\fmf{quark,label=$\hat e_p$,label.side=right,label.dist=3,tension=2}{v1,b2}
			\fmf{plain,label=$\nu_u$,label.side=right,label.dist=3,tension=3}{v2,v1}
			\fmf{plain,label=$\nu_v$,label.side=right,label.dist=3,tension=3}{v1,v3}
			\fmffreeze
			\fmf{wboson,label=$\W^+$,label.side=right,right=0.25}{v3,v2}
			\fmfv{decoration.shape=square,decoration.size=1.5mm}{v1}
		\end{fmfgraph*}
	\end{gathered} \quad \times 4
\qquad\quad
	\begin{gathered}
		\begin{fmfgraph*}(55,60)
			\fmfset{curly_len}{2mm}
			\fmftop{t1,t2} \fmfbottom{b1,b2}
			\fmf{quark,label=$\hat e_t$,label.side=right,label.dist=3,tension=6}{t1,v2}
			\fmf{quark,label=$\hat e_s$,label.side=right,label.dist=3,tension=6}{v3,t2}
			\fmf{quark,label=$\hat e_r$,label.side=right,label.dist=3,tension=2}{b1,v1}
			\fmf{quark,label=$\hat e_p$,label.side=right,label.dist=3,tension=2}{v1,b2}
			\fmf{quark,label=$e_u$,label.side=right,label.dist=3,tension=3}{v2,v1}
			\fmf{quark,label=$e_v$,label.side=right,label.dist=3,tension=3}{v1,v3}
			\fmffreeze
			\fmf{photon,label=$\Z$,label.side=right,right=0.25}{v3,v2}
			\fmfv{decoration.shape=square,decoration.size=1.5mm}{v1}
		\end{fmfgraph*}
	\end{gathered} \quad \times 4
\qquad\quad
	\begin{gathered}
		\begin{fmfgraph*}(55,60)
			\fmfset{curly_len}{2mm}
			\fmftop{t1,t2} \fmfbottom{b1,b2}
			\fmf{quark,label=$\hat e_t$,label.side=left,label.dist=3,tension=6}{t1,v2}
			\fmf{quark,label=$\hat e_s$,label.side=right,label.dist=3,tension=2}{v1,t2}
			\fmf{quark,label=$\hat e_r$,label.side=right,label.dist=3,tension=6}{b1,v3}
			\fmf{quark,label=$\hat e_p$,label.side=left,label.dist=3,tension=2}{v1,b2}
			\fmf{quark,label=$e_u$,label.side=left,label.dist=3,tension=3}{v2,v1}
			\fmf{quark,label=$e_v$,label.side=right,label.dist=3,tension=3}{v3,v1}
			\fmffreeze
			\fmf{photon,label=$\Z$,label.side=left,left=0.25}{v3,v2}
			\fmfv{decoration.shape=square,decoration.size=1.5mm}{v1}
		\end{fmfgraph*}
	\end{gathered}
\qquad\quad
	\begin{gathered}
		\begin{fmfgraph*}(55,60)
			\fmfset{curly_len}{2mm}
			\fmftop{t1,t2} \fmfbottom{b1,b2}
			\fmf{quark,label=$\hat e_t$,label.side=right,label.dist=3,tension=2}{t1,v1}
			\fmf{quark,label=$\hat e_s$,label.side=left,label.dist=3,tension=6}{v2,t2}
			\fmf{quark,label=$\hat e_r$,label.side=left,label.dist=3,tension=2}{b1,v1}
			\fmf{quark,label=$\hat e_p$,label.side=right,label.dist=3,tension=6}{v3,b2}
			\fmf{quark,label=$e_u$,label.side=left,label.dist=3,tension=3}{v1,v2}
			\fmf{quark,label=$e_v$,label.side=right,label.dist=3,tension=3}{v1,v3}
			\fmffreeze
			\fmf{photon,label=$\Z$,label.side=left,left=0.25}{v2,v3}
			\fmfv{decoration.shape=square,decoration.size=1.5mm}{v1}
		\end{fmfgraph*}
	\end{gathered}
\end{align*}


\subsubsection[$\bar ee \bar uu$]{\boldmath$\bar ee \bar uu$}

\begin{align*}
	& \begin{gathered}
		\begin{fmfgraph*}(55,60)
			\fmfset{curly_len}{2mm}
			\fmftop{t1,t2} \fmfbottom{b1,b2}
			\fmf{quark,label=$\hat u_t$,label.side=left,tension=4}{t1,v1}
			\fmf{quark,label=$\hat u_s$,label.side=left,tension=4}{v2,t2}
			\fmf{quark,label=$\hat e_r$,label.side=right,tension=4}{b1,v3}
			\fmf{quark,label=$\hat e_p$,label.side=right,tension=4}{v4,b2}
			\fmf{quark,label=$u_u$,label.side=right}{v1,v2}
			\fmf{quark,label=$e_v$,label.side=right}{v3,v4}
			\fmf{photon,label=$\Z$,label.side=left}{v3,v1}
			\fmf{photon,label=$\Z$,label.side=left}{v2,v4}
		\end{fmfgraph*}
	\end{gathered}
\qquad\quad
	\begin{gathered}
		\begin{fmfgraph*}(55,60)
			\fmfset{curly_len}{2mm}
			\fmftop{t1,t2} \fmfbottom{b1,b2}
			\fmf{quark,label=$\hat u_t$,label.side=left,tension=4}{t1,v1}
			\fmf{quark,label=$\hat u_s$,label.side=left,tension=4}{v2,t2}
			\fmf{quark,label=$\hat e_r$,label.side=right,tension=4}{b1,v3}
			\fmf{quark,label=$\hat e_p$,label.side=right,tension=4}{v4,b2}
			\fmf{quark,label=$u_u$,label.side=right}{v1,v2}
			\fmf{quark,label=$e_v$,label.side=right}{v3,v4}
			\fmf{photon,label=$\Z$,label.side=left}{v3,v1}
			\fmf{photon,label=$\A$,label.side=left}{v2,v4}
		\end{fmfgraph*}
	\end{gathered}
\qquad\quad
	\begin{gathered}
		\begin{fmfgraph*}(55,60)
			\fmfset{curly_len}{2mm}
			\fmftop{t1,t2} \fmfbottom{b1,b2}
			\fmf{quark,label=$\hat u_t$,label.side=left,tension=4}{t1,v1}
			\fmf{quark,label=$\hat u_s$,label.side=left,tension=4}{v2,t2}
			\fmf{quark,label=$\hat e_r$,label.side=right,tension=4}{b1,v3}
			\fmf{quark,label=$\hat e_p$,label.side=right,tension=4}{v4,b2}
			\fmf{quark,label=$u_u$,label.side=right}{v1,v2}
			\fmf{quark,label=$e_v$,label.side=right}{v3,v4}
			\fmf{photon,label=$\A$,label.side=left}{v3,v1}
			\fmf{photon,label=$\Z$,label.side=left}{v2,v4}
		\end{fmfgraph*}
	\end{gathered}
\qquad\quad
	\begin{gathered}
		\begin{fmfgraph*}(55,60)
			\fmfset{curly_len}{2mm}
			\fmftop{t1,t2} \fmfbottom{b1,b2}
			\fmf{quark,label=$\hat u_t$,label.side=left,tension=4}{t1,v1}
			\fmf{quark,label=$\hat u_s$,label.side=left,tension=4}{v2,t2}
			\fmf{quark,label=$\hat e_r$,label.side=right,tension=4}{b1,v3}
			\fmf{quark,label=$\hat e_p$,label.side=right,tension=4}{v4,b2}
			\fmf{quark,label=$d_u$,label.side=right}{v1,v2}
			\fmf{plain,label=$\nu_v$,label.side=right}{v3,v4}
			\fmf{wboson,label=$\W^+$,label.side=right}{v1,v3}
			\fmf{wboson,label=$\W^+$,label.side=right}{v4,v2}
		\end{fmfgraph*}
	\end{gathered}
\qquad\quad
	\begin{gathered}
		\begin{fmfgraph*}(55,60)
			\fmfset{curly_len}{2mm}
			\fmftop{t1,t2} \fmfbottom{b1,b2}
			\fmf{phantom,label=$\hat u_t$,label.side=left,tension=4}{t1,v1}
			\fmf{phantom,label=$\hat u_s$,label.side=left,tension=4}{v2,t2}
			\fmf{quark,label=$\hat e_r$,label.side=right,tension=4}{b1,v3}
			\fmf{quark,label=$\hat e_p$,label.side=right,tension=4}{v4,b2}
			\fmf{quark,label=$u_u$,label.side=left}{v2,v1}
			\fmf{quark,label=$e_v$,label.side=right}{v3,v4}
			\fmf{photon,label=$\Z$,label.side=left}{v3,v1}
			\fmf{photon,label=$\Z$,label.side=left}{v2,v4}
			\fmffreeze
			\fmf{quark}{t1,v2}
			\fmf{quark}{v1,t2}
		\end{fmfgraph*}
	\end{gathered}
\nnw
	& \begin{gathered}
		\begin{fmfgraph*}(55,60)
			\fmfset{curly_len}{2mm}
			\fmftop{t1,t2} \fmfbottom{b1,b2}
			\fmf{phantom,label=$\hat u_t$,label.side=left,tension=4}{t1,v1}
			\fmf{phantom,label=$\hat u_s$,label.side=left,tension=4}{v2,t2}
			\fmf{quark,label=$\hat e_r$,label.side=right,tension=4}{b1,v3}
			\fmf{quark,label=$\hat e_p$,label.side=right,tension=4}{v4,b2}
			\fmf{quark,label=$u_u$,label.side=left}{v2,v1}
			\fmf{quark,label=$e_v$,label.side=right}{v3,v4}
			\fmf{photon,label=$\Z$,label.side=left}{v3,v1}
			\fmf{photon,label=$\A$,label.side=left}{v2,v4}
			\fmffreeze
			\fmf{quark}{t1,v2}
			\fmf{quark}{v1,t2}
		\end{fmfgraph*}
	\end{gathered}
\qquad\quad
	\begin{gathered}
		\begin{fmfgraph*}(55,60)
			\fmfset{curly_len}{2mm}
			\fmftop{t1,t2} \fmfbottom{b1,b2}
			\fmf{phantom,label=$\hat u_t$,label.side=left,tension=4}{t1,v1}
			\fmf{phantom,label=$\hat u_s$,label.side=left,tension=4}{v2,t2}
			\fmf{quark,label=$\hat e_r$,label.side=right,tension=4}{b1,v3}
			\fmf{quark,label=$\hat e_p$,label.side=right,tension=4}{v4,b2}
			\fmf{quark,label=$u_u$,label.side=left}{v2,v1}
			\fmf{quark,label=$e_v$,label.side=right}{v3,v4}
			\fmf{photon,label=$\A$,label.side=left}{v3,v1}
			\fmf{photon,label=$\Z$,label.side=left}{v2,v4}
			\fmffreeze
			\fmf{quark}{t1,v2}
			\fmf{quark}{v1,t2}
		\end{fmfgraph*}
	\end{gathered}
\qquad\quad
	\begin{gathered}
		\begin{fmfgraph*}(55,60)
			\fmfset{curly_len}{2mm}
			\fmftop{t1,t2} \fmfbottom{b1,b2}
			\fmf{quark,label=$\hat u_t$,label.side=left,label.dist=4,tension=3}{t1,v1}
			\fmf{quark,label=$\hat u_s$,label.side=left,label.dist=4,tension=3}{v1,t2}
			\fmf{quark,label=$\hat e_r$,label.side=left,label.dist=3,tension=3}{b1,v2}
			\fmf{quark,label=$\hat e_p$,label.side=left,label.dist=3,tension=3}{v3,b2}
			\fmf{wboson,label=$\W^+$,label.side=right,label.dist=3,tension=1}{v3,v1,v2}
			\fmffreeze
			\fmf{plain,label=$\nu_u$,label.side=right}{v2,v3}
			\fmfv{decoration.shape=square,decoration.size=1.5mm}{v1}
		\end{fmfgraph*}
	\end{gathered}
\qquad\quad
	\begin{gathered}
		\begin{fmfgraph*}(55,60)
			\fmfset{curly_len}{2mm}
			\fmftop{t1,t2} \fmfbottom{b1,b2}
			\fmf{quark,label=$\hat u_t$,label.side=right,label.dist=3,tension=3}{t1,v2}
			\fmf{quark,label=$\hat u_s$,label.side=right,label.dist=3,tension=3}{v3,t2}
			\fmf{quark,label=$\hat e_r$,label.side=right,label.dist=4,tension=3}{b1,v1}
			\fmf{quark,label=$\hat e_p$,label.side=right,label.dist=4,tension=3}{v1,b2}
			\fmf{wboson,label=$\W^+$,label.side=right,label.dist=3,tension=1}{v2,v1}
			\fmf{wboson,label=$\W^+$,label.side=right,label.dist=3,tension=1}{v1,v3}
			\fmffreeze
			\fmf{quark,label=$d_u$,label.side=left}{v2,v3}
			\fmfv{decoration.shape=square,decoration.size=1.5mm}{v1}
		\end{fmfgraph*}
	\end{gathered}
\qquad\quad
	\begin{gathered}
		\begin{fmfgraph*}(55,60)
			\fmfset{curly_len}{2mm}
			\fmftop{t1,t2} \fmfbottom{b1,b2}
			\fmf{quark,label=$\hat u_t$,label.side=left,label.dist=3,tension=6}{t1,v2}
			\fmf{quark,label=$\hat u_s$,label.side=right,label.dist=3,tension=2}{v1,t2}
			\fmf{quark,label=$\hat e_r$,label.side=right,label.dist=3,tension=6}{b1,v3}
			\fmf{quark,label=$\hat e_p$,label.side=left,label.dist=3,tension=2}{v1,b2}
			\fmf{quark,label=$d_u$,label.side=left,label.dist=3,tension=3}{v2,v1}
			\fmf{plain,label=$\nu_v$,label.side=right,label.dist=3,tension=3}{v3,v1}
			\fmffreeze
			\fmf{wboson,label=$\W^+$,label.side=right,right=0.25}{v2,v3}
			\fmfv{decoration.shape=square,decoration.size=1.5mm}{v1}
		\end{fmfgraph*}
	\end{gathered}
\nnw
	& \begin{gathered}
		\begin{fmfgraph*}(55,60)
			\fmfset{curly_len}{2mm}
			\fmftop{t1,t2} \fmfbottom{b1,b2}
			\fmf{quark,label=$\hat u_t$,label.side=left,label.dist=3,tension=6}{t1,v2}
			\fmf{quark,label=$\hat u_s$,label.side=right,label.dist=3,tension=2}{v1,t2}
			\fmf{quark,label=$\hat e_r$,label.side=right,label.dist=3,tension=6}{b1,v3}
			\fmf{quark,label=$\hat e_p$,label.side=left,label.dist=3,tension=2}{v1,b2}
			\fmf{quark,label=$u_u$,label.side=left,label.dist=3,tension=3}{v2,v1}
			\fmf{quark,label=$e_v$,label.side=right,label.dist=3,tension=3}{v3,v1}
			\fmffreeze
			\fmf{photon,label=$\Z$,label.side=right,right=0.25}{v2,v3}
			\fmfv{decoration.shape=square,decoration.size=1.5mm}{v1}
		\end{fmfgraph*}
	\end{gathered}
\qquad\quad
	\begin{gathered}
		\begin{fmfgraph*}(55,60)
			\fmfset{curly_len}{2mm}
			\fmftop{t1,t2} \fmfbottom{b1,b2}
			\fmf{quark,label=$\hat u_t$,label.side=right,label.dist=3,tension=2}{t1,v1}
			\fmf{quark,label=$\hat u_s$,label.side=left,label.dist=3,tension=6}{v2,t2}
			\fmf{quark,label=$\hat e_r$,label.side=left,label.dist=3,tension=2}{b1,v1}
			\fmf{quark,label=$\hat e_p$,label.side=right,label.dist=3,tension=6}{v3,b2}
			\fmf{quark,label=$d_u$,label.side=left,label.dist=3,tension=3}{v1,v2}
			\fmf{plain,label=$\nu_v$,label.side=right,label.dist=3,tension=3}{v1,v3}
			\fmffreeze
			\fmf{wboson,label=$\W^+$,label.side=right,right=0.25}{v3,v2}
			\fmfv{decoration.shape=square,decoration.size=1.5mm}{v1}
		\end{fmfgraph*}
	\end{gathered}
\qquad\quad
	\begin{gathered}
		\begin{fmfgraph*}(55,60)
			\fmfset{curly_len}{2mm}
			\fmftop{t1,t2} \fmfbottom{b1,b2}
			\fmf{quark,label=$\hat u_t$,label.side=right,label.dist=3,tension=2}{t1,v1}
			\fmf{quark,label=$\hat u_s$,label.side=left,label.dist=3,tension=6}{v2,t2}
			\fmf{quark,label=$\hat e_r$,label.side=left,label.dist=3,tension=2}{b1,v1}
			\fmf{quark,label=$\hat e_p$,label.side=right,label.dist=3,tension=6}{v3,b2}
			\fmf{quark,label=$u_u$,label.side=left,label.dist=3,tension=3}{v1,v2}
			\fmf{quark,label=$e_v$,label.side=right,label.dist=3,tension=3}{v1,v3}
			\fmffreeze
			\fmf{photon,label=$\Z$,label.side=left,left=0.25}{v2,v3}
			\fmfv{decoration.shape=square,decoration.size=1.5mm}{v1}
		\end{fmfgraph*}
	\end{gathered}
\qquad\quad
	\begin{gathered}
		\begin{fmfgraph*}(55,60)
			\fmfset{curly_len}{2mm}
			\fmftop{t1,t2} \fmfbottom{b1,b2}
			\fmf{quark,label=$\hat u_t$,label.side=right,label.dist=3,tension=6}{t1,v2}
			\fmf{quark,label=$\hat u_s$,label.side=right,label.dist=3,tension=6}{v3,t2}
			\fmf{quark,label=$\hat e_r$,label.side=right,label.dist=3,tension=2}{b1,v1}
			\fmf{quark,label=$\hat e_p$,label.side=right,label.dist=3,tension=2}{v1,b2}
			\fmf{quark,label=$d_u$,label.side=right,label.dist=3,tension=3}{v2,v1}
			\fmf{quark,label=$d_v$,label.side=right,label.dist=3,tension=3}{v1,v3}
			\fmffreeze
			\fmf{wboson,label=$\W^+$,label.side=left,left=0.25}{v2,v3}
			\fmfv{decoration.shape=square,decoration.size=1.5mm}{v1}
		\end{fmfgraph*}
	\end{gathered}
\qquad\quad
	\begin{gathered}
		\begin{fmfgraph*}(55,60)
			\fmfset{curly_len}{2mm}
			\fmftop{t1,t2} \fmfbottom{b1,b2}
			\fmf{quark,label=$\hat u_t$,label.side=right,label.dist=3,tension=6}{t1,v2}
			\fmf{quark,label=$\hat u_s$,label.side=right,label.dist=3,tension=6}{v3,t2}
			\fmf{quark,label=$\hat e_r$,label.side=right,label.dist=3,tension=2}{b1,v1}
			\fmf{quark,label=$\hat e_p$,label.side=right,label.dist=3,tension=2}{v1,b2}
			\fmf{quark,label=$u_u$,label.side=right,label.dist=3,tension=3}{v2,v1}
			\fmf{quark,label=$u_v$,label.side=right,label.dist=3,tension=3}{v1,v3}
			\fmffreeze
			\fmf{photon,label=$\Z$,label.side=left,left=0.25}{v2,v3}
			\fmfv{decoration.shape=square,decoration.size=1.5mm}{v1}
		\end{fmfgraph*}
	\end{gathered}
\nnw
	& \begin{gathered}
		\begin{fmfgraph*}(55,60)
			\fmfset{curly_len}{2mm}
			\fmftop{t1,t2} \fmfbottom{b1,b2}
			\fmf{quark,label=$\hat u_t$,label.side=left,label.dist=3,tension=2}{t1,v1}
			\fmf{quark,label=$\hat u_s$,label.side=left,label.dist=3,tension=2}{v1,t2}
			\fmf{quark,label=$\hat e_r$,label.side=left,label.dist=3,tension=6}{b1,v2}
			\fmf{quark,label=$\hat e_p$,label.side=left,label.dist=3,tension=6}{v3,b2}
			\fmf{plain,label=$\nu_u$,label.side=left,label.dist=3,tension=3}{v2,v1}
			\fmf{plain,label=$\nu_v$,label.side=left,label.dist=3,tension=3}{v1,v3}
			\fmffreeze
			\fmf{wboson,label=$\W^+$,label.side=left,left=0.25}{v3,v2}
			\fmfv{decoration.shape=square,decoration.size=1.5mm}{v1}
		\end{fmfgraph*}
	\end{gathered}
\qquad\quad
	\begin{gathered}
		\begin{fmfgraph*}(55,60)
			\fmfset{curly_len}{2mm}
			\fmftop{t1,t2} \fmfbottom{b1,b2}
			\fmf{quark,label=$\hat u_t$,label.side=left,label.dist=3,tension=2}{t1,v1}
			\fmf{quark,label=$\hat u_s$,label.side=left,label.dist=3,tension=2}{v1,t2}
			\fmf{quark,label=$\hat e_r$,label.side=left,label.dist=3,tension=6}{b1,v2}
			\fmf{quark,label=$\hat e_p$,label.side=left,label.dist=3,tension=6}{v3,b2}
			\fmf{quark,label=$e_u$,label.side=left,label.dist=3,tension=3}{v2,v1}
			\fmf{quark,label=$e_v$,label.side=left,label.dist=3,tension=3}{v1,v3}
			\fmffreeze
			\fmf{photon,label=$\Z$,label.side=left,left=0.25}{v3,v2}
			\fmfv{decoration.shape=square,decoration.size=1.5mm}{v1}
		\end{fmfgraph*}
	\end{gathered}
\qquad\quad
	\begin{gathered}
		\begin{fmfgraph*}(55,60)
			\fmfset{curly_len}{2mm}
			\fmftop{t1,t2} \fmfbottom{b1,b2}
			\fmf{quark,label=$\hat u_t$,label.side=left,label.dist=3,tension=6}{t1,v2}
			\fmf{quark,label=$\hat u_s$,label.side=left,label.dist=3,tension=2}{v1,t2}
			\fmf{quark,label=$\hat e_r$,label.side=right,label.dist=3,tension=2}{b1,v1}
			\fmf{quark,label=$\hat e_p$,label.side=left,label.dist=3,tension=6}{v3,b2}
			\fmf{quark,label=$u_u$,label.side=left,label.dist=3,tension=3}{v2,v1}
			\fmf{quark,label=$e_v$,label.side=left,label.dist=3,tension=3}{v1,v3}
			\fmffreeze
			\fmf{photon,right=0.5}{v2,v3}
			\fmf{phantom,label=$\Z$,label.side=right,left=0.5}{v2,b1}
			\fmfv{decoration.shape=square,decoration.size=1.5mm}{v1}
		\end{fmfgraph*}
	\end{gathered}
\qquad\quad
	\begin{gathered}
		\begin{fmfgraph*}(55,60)
			\fmfset{curly_len}{2mm}
			\fmftop{t1,t2} \fmfbottom{b1,b2}
			\fmf{quark,label=$\hat u_t$,label.side=left,label.dist=3,tension=2}{t1,v1}
			\fmf{quark,label=$\hat u_s$,label.side=right,label.dist=3,tension=6}{v3,t2}
			\fmf{quark,label=$\hat e_r$,label.side=right,label.dist=3,tension=6}{b1,v2}
			\fmf{quark,label=$\hat e_p$,label.side=left,label.dist=3,tension=2}{v1,b2}
			\fmf{quark,label=$u_u$,label.side=left,label.dist=3,tension=3}{v1,v3}
			\fmf{quark,label=$e_v$,label.side=left,label.dist=3,tension=3}{v2,v1}
			\fmffreeze
			\fmf{photon,right=0.5}{v2,v3}
			\fmf{phantom,label=$\Z$,label.side=right,left=0.4}{v2,b2}
			\fmfv{decoration.shape=square,decoration.size=1.5mm}{v1}
		\end{fmfgraph*}
	\end{gathered}
\end{align*}


\subsubsection[$\bar ee \bar dd$]{\boldmath$\bar ee \bar dd$}

\begin{align*}
	& \begin{gathered}
		\begin{fmfgraph*}(55,60)
			\fmfset{curly_len}{2mm}
			\fmftop{t1,t2} \fmfbottom{b1,b2}
			\fmf{quark,label=$\hat d_t$,label.side=left,tension=4}{t1,v1}
			\fmf{quark,label=$\hat d_s$,label.side=left,tension=4}{v2,t2}
			\fmf{quark,label=$\hat e_r$,label.side=right,tension=4}{b1,v3}
			\fmf{quark,label=$\hat e_p$,label.side=right,tension=4}{v4,b2}
			\fmf{quark,label=$d_u$,label.side=right}{v1,v2}
			\fmf{quark,label=$e_v$,label.side=right}{v3,v4}
			\fmf{photon,label=$\Z$,label.side=left}{v3,v1}
			\fmf{photon,label=$\Z$,label.side=left}{v2,v4}
		\end{fmfgraph*}
	\end{gathered}
\qquad\quad
	\begin{gathered}
		\begin{fmfgraph*}(55,60)
			\fmfset{curly_len}{2mm}
			\fmftop{t1,t2} \fmfbottom{b1,b2}
			\fmf{quark,label=$\hat d_t$,label.side=left,tension=4}{t1,v1}
			\fmf{quark,label=$\hat d_s$,label.side=left,tension=4}{v2,t2}
			\fmf{quark,label=$\hat e_r$,label.side=right,tension=4}{b1,v3}
			\fmf{quark,label=$\hat e_p$,label.side=right,tension=4}{v4,b2}
			\fmf{quark,label=$d_u$,label.side=right}{v1,v2}
			\fmf{quark,label=$e_v$,label.side=right}{v3,v4}
			\fmf{photon,label=$\Z$,label.side=left}{v3,v1}
			\fmf{photon,label=$\A$,label.side=left}{v2,v4}
		\end{fmfgraph*}
	\end{gathered}
\qquad\quad
	\begin{gathered}
		\begin{fmfgraph*}(55,60)
			\fmfset{curly_len}{2mm}
			\fmftop{t1,t2} \fmfbottom{b1,b2}
			\fmf{quark,label=$\hat d_t$,label.side=left,tension=4}{t1,v1}
			\fmf{quark,label=$\hat d_s$,label.side=left,tension=4}{v2,t2}
			\fmf{quark,label=$\hat e_r$,label.side=right,tension=4}{b1,v3}
			\fmf{quark,label=$\hat e_p$,label.side=right,tension=4}{v4,b2}
			\fmf{quark,label=$d_u$,label.side=right}{v1,v2}
			\fmf{quark,label=$e_v$,label.side=right}{v3,v4}
			\fmf{photon,label=$\A$,label.side=left}{v3,v1}
			\fmf{photon,label=$\Z$,label.side=left}{v2,v4}
		\end{fmfgraph*}
	\end{gathered}
\qquad\quad
	\begin{gathered}
		\begin{fmfgraph*}(55,60)
			\fmfset{curly_len}{2mm}
			\fmftop{t1,t2} \fmfbottom{b1,b2}
			\fmf{phantom,label=$\hat d_t$,label.side=left,tension=4}{t1,v1}
			\fmf{phantom,label=$\hat d_s$,label.side=left,tension=4}{v2,t2}
			\fmf{quark,label=$\hat e_r$,label.side=right,tension=4}{b1,v3}
			\fmf{quark,label=$\hat e_p$,label.side=right,tension=4}{v4,b2}
			\fmf{quark,label=$d_u$,label.side=left}{v2,v1}
			\fmf{quark,label=$e_v$,label.side=right}{v3,v4}
			\fmf{photon,label=$\Z$,label.side=left}{v3,v1}
			\fmf{photon,label=$\Z$,label.side=left}{v2,v4}
			\fmffreeze
			\fmf{quark}{t1,v2}
			\fmf{quark}{v1,t2}
		\end{fmfgraph*}
	\end{gathered}
\qquad\quad
	\begin{gathered}
		\begin{fmfgraph*}(55,60)
			\fmfset{curly_len}{2mm}
			\fmftop{t1,t2} \fmfbottom{b1,b2}
			\fmf{phantom,label=$\hat d_t$,label.side=left,tension=4}{t1,v1}
			\fmf{phantom,label=$\hat d_s$,label.side=left,tension=4}{v2,t2}
			\fmf{quark,label=$\hat e_r$,label.side=right,tension=4}{b1,v3}
			\fmf{quark,label=$\hat e_p$,label.side=right,tension=4}{v4,b2}
			\fmf{quark,label=$d_u$,label.side=left}{v2,v1}
			\fmf{quark,label=$e_v$,label.side=right}{v3,v4}
			\fmf{photon,label=$\Z$,label.side=left}{v3,v1}
			\fmf{photon,label=$\A$,label.side=left}{v2,v4}
			\fmffreeze
			\fmf{quark}{t1,v2}
			\fmf{quark}{v1,t2}
		\end{fmfgraph*}
	\end{gathered}
\nnw
	& \begin{gathered}
		\begin{fmfgraph*}(55,60)
			\fmfset{curly_len}{2mm}
			\fmftop{t1,t2} \fmfbottom{b1,b2}
			\fmf{phantom,label=$\hat d_t$,label.side=left,tension=4}{t1,v1}
			\fmf{phantom,label=$\hat d_s$,label.side=left,tension=4}{v2,t2}
			\fmf{quark,label=$\hat e_r$,label.side=right,tension=4}{b1,v3}
			\fmf{quark,label=$\hat e_p$,label.side=right,tension=4}{v4,b2}
			\fmf{quark,label=$d_u$,label.side=left}{v2,v1}
			\fmf{quark,label=$e_v$,label.side=right}{v3,v4}
			\fmf{photon,label=$\A$,label.side=left}{v3,v1}
			\fmf{photon,label=$\Z$,label.side=left}{v2,v4}
			\fmffreeze
			\fmf{quark}{t1,v2}
			\fmf{quark}{v1,t2}
		\end{fmfgraph*}
	\end{gathered}
\qquad\quad
	\begin{gathered}
		\begin{fmfgraph*}(55,60)
			\fmfset{curly_len}{2mm}
			\fmftop{t1,t2} \fmfbottom{b1,b2}
			\fmf{phantom,label=$\hat d_t$,label.side=left,tension=4}{t1,v1}
			\fmf{phantom,label=$\hat d_s$,label.side=left,tension=4}{v2,t2}
			\fmf{quark,label=$\hat e_r$,label.side=right,tension=4}{b1,v3}
			\fmf{quark,label=$\hat e_p$,label.side=right,tension=4}{v4,b2}
			\fmf{quark,label=$u_u$,label.side=left}{v2,v1}
			\fmf{plain,label=$\nu_v$,label.side=right}{v3,v4}
			\fmf{wboson,label=$\W^+$,label.side=right}{v1,v3}
			\fmf{wboson,label=$\W^+$,label.side=right}{v4,v2}
			\fmffreeze
			\fmf{quark}{t1,v2}
			\fmf{quark}{v1,t2}
		\end{fmfgraph*}
	\end{gathered}
\qquad\quad
	\begin{gathered}
		\begin{fmfgraph*}(55,60)
			\fmfset{curly_len}{2mm}
			\fmftop{t1,t2} \fmfbottom{b1,b2}
			\fmf{phantom,label=$\hat d_t$,label.side=left,tension=4}{t1,v1}
			\fmf{phantom,label=$\hat d_s$,label.side=left,tension=4}{v2,t2}
			\fmf{quark,label=$\hat e_r$,label.side=right,tension=4}{b1,v3}
			\fmf{quark,label=$\hat e_p$,label.side=right,tension=4}{v4,b2}
			\fmf{quark,label=$t$,label.side=left}{v2,v1}
			\fmf{plain,label=$\nu_v$,label.side=right}{v3,v4}
			\fmf{scalar,label=$G^+$,label.side=right}{v1,v3}
			\fmf{wboson,label=$\W^+$,label.side=right}{v4,v2}
			\fmffreeze
			\fmf{quark}{t1,v2}
			\fmf{quark}{v1,t2}
		\end{fmfgraph*}
	\end{gathered}
\qquad\quad
	\begin{gathered}
		\begin{fmfgraph*}(55,60)
			\fmfset{curly_len}{2mm}
			\fmftop{t1,t2} \fmfbottom{b1,b2}
			\fmf{phantom,label=$\hat d_t$,label.side=left,tension=4}{t1,v1}
			\fmf{phantom,label=$\hat d_s$,label.side=left,tension=4}{v2,t2}
			\fmf{quark,label=$\hat e_r$,label.side=right,tension=4}{b1,v3}
			\fmf{quark,label=$\hat e_p$,label.side=right,tension=4}{v4,b2}
			\fmf{quark,label=$t$,label.side=left}{v2,v1}
			\fmf{plain,label=$\nu_v$,label.side=right}{v3,v4}
			\fmf{wboson,label=$\W^+$,label.side=right}{v1,v3}
			\fmf{scalar,label=$G^+$,label.side=right}{v4,v2}
			\fmffreeze
			\fmf{quark}{t1,v2}
			\fmf{quark}{v1,t2}
		\end{fmfgraph*}
	\end{gathered}
\qquad\quad
	\begin{gathered}
		\begin{fmfgraph*}(55,60)
			\fmfset{curly_len}{2mm}
			\fmftop{t1,t2} \fmfbottom{b1,b2}
			\fmf{quark,label=$\hat d_t$,label.side=left,label.dist=4,tension=3}{t1,v1}
			\fmf{quark,label=$\hat d_s$,label.side=left,label.dist=4,tension=3}{v1,t2}
			\fmf{quark,label=$\hat e_r$,label.side=left,label.dist=3,tension=3}{b1,v2}
			\fmf{quark,label=$\hat e_p$,label.side=left,label.dist=3,tension=3}{v3,b2}
			\fmf{wboson,label=$\W^+$,label.side=right,label.dist=3,tension=1}{v3,v1,v2}
			\fmffreeze
			\fmf{plain,label=$\nu_u$,label.side=right}{v2,v3}
			\fmfv{decoration.shape=square,decoration.size=1.5mm}{v1}
		\end{fmfgraph*}
	\end{gathered}
\nnw
	& \begin{gathered}
		\begin{fmfgraph*}(55,60)
			\fmfset{curly_len}{2mm}
			\fmftop{t1,t2} \fmfbottom{b1,b2}
			\fmf{quark,label=$\hat d_t$,label.side=right,label.dist=3,tension=3}{t1,v2}
			\fmf{quark,label=$\hat d_s$,label.side=right,label.dist=3,tension=3}{v3,t2}
			\fmf{quark,label=$\hat e_r$,label.side=right,label.dist=4,tension=3}{b1,v1}
			\fmf{quark,label=$\hat e_p$,label.side=right,label.dist=4,tension=3}{v1,b2}
			\fmf{wboson,label=$\W^+$,label.side=left,label.dist=3,tension=1}{v1,v2}
			\fmf{wboson,label=$\W^+$,label.side=left,label.dist=3,tension=1}{v3,v1}
			\fmffreeze
			\fmf{quark,label=$u_u$,label.side=left}{v2,v3}
			\fmfv{decoration.shape=square,decoration.size=1.5mm}{v1}
		\end{fmfgraph*}
	\end{gathered}
\qquad\quad
	\begin{gathered}
		\begin{fmfgraph*}(55,60)
			\fmfset{curly_len}{2mm}
			\fmftop{t1,t2} \fmfbottom{b1,b2}
			\fmf{quark,label=$\hat d_t$,label.side=right,label.dist=3,tension=3}{t1,v2}
			\fmf{quark,label=$\hat d_s$,label.side=right,label.dist=3,tension=3}{v3,t2}
			\fmf{quark,label=$\hat e_r$,label.side=right,label.dist=4,tension=3}{b1,v1}
			\fmf{quark,label=$\hat e_p$,label.side=right,label.dist=4,tension=3}{v1,b2}
			\fmf{scalar,label=$G^+$,label.side=left,label.dist=3,tension=1}{v1,v2}
			\fmf{wboson,label=$\W^+$,label.side=left,label.dist=3,tension=1}{v3,v1}
			\fmffreeze
			\fmf{quark,label=$t$,label.side=left}{v2,v3}
			\fmfv{decoration.shape=square,decoration.size=1.5mm}{v1}
		\end{fmfgraph*}
	\end{gathered}
\qquad\quad
	\begin{gathered}
		\begin{fmfgraph*}(55,60)
			\fmfset{curly_len}{2mm}
			\fmftop{t1,t2} \fmfbottom{b1,b2}
			\fmf{quark,label=$\hat d_t$,label.side=right,label.dist=3,tension=3}{t1,v2}
			\fmf{quark,label=$\hat d_s$,label.side=right,label.dist=3,tension=3}{v3,t2}
			\fmf{quark,label=$\hat e_r$,label.side=right,label.dist=4,tension=3}{b1,v1}
			\fmf{quark,label=$\hat e_p$,label.side=right,label.dist=4,tension=3}{v1,b2}
			\fmf{wboson,label=$\W^+$,label.side=left,label.dist=3,tension=1}{v1,v2}
			\fmf{scalar,label=$G^+$,label.side=left,label.dist=3,tension=1}{v3,v1}
			\fmffreeze
			\fmf{quark,label=$t$,label.side=left}{v2,v3}
			\fmfv{decoration.shape=square,decoration.size=1.5mm}{v1}
		\end{fmfgraph*}
	\end{gathered}
\qquad\quad
	\begin{gathered}
		\begin{fmfgraph*}(55,60)
			\fmfset{curly_len}{2mm}
			\fmftop{t1,t2} \fmfbottom{b1,b2}
			\fmf{quark,label=$\hat d_t$,label.side=right,label.dist=3,tension=3}{t1,v2}
			\fmf{quark,label=$\hat d_s$,label.side=right,label.dist=3,tension=3}{v3,t2}
			\fmf{quark,label=$\hat e_r$,label.side=right,label.dist=4,tension=3}{b1,v1}
			\fmf{quark,label=$\hat e_p$,label.side=right,label.dist=4,tension=3}{v1,b2}
			\fmf{scalar,label=$G^+$,label.side=left,label.dist=3,tension=1}{v1,v2}
			\fmf{scalar,label=$G^+$,label.side=left,label.dist=3,tension=1}{v3,v1}
			\fmffreeze
			\fmf{quark,label=$t$,label.side=left}{v2,v3}
			\fmfv{decoration.shape=square,decoration.size=1.5mm}{v1}
		\end{fmfgraph*}
	\end{gathered}
\qquad\quad
	\begin{gathered}
		\begin{fmfgraph*}(55,60)
			\fmfset{curly_len}{2mm}
			\fmftop{t1,t2} \fmfbottom{b1,b2}
			\fmf{quark,label=$\hat d_t$,label.side=left,label.dist=3,tension=6}{t1,v2}
			\fmf{quark,label=$\hat d_s$,label.side=right,label.dist=3,tension=2}{v1,t2}
			\fmf{quark,label=$\hat e_r$,label.side=right,label.dist=3,tension=6}{b1,v3}
			\fmf{quark,label=$\hat e_p$,label.side=left,label.dist=3,tension=2}{v1,b2}
			\fmf{quark,label=$d_u$,label.side=left,label.dist=3,tension=3}{v2,v1}
			\fmf{quark,label=$e_v$,label.side=right,label.dist=3,tension=3}{v3,v1}
			\fmffreeze
			\fmf{photon,label=$\Z$,label.side=right,right=0.25}{v2,v3}
			\fmfv{decoration.shape=square,decoration.size=1.5mm}{v1}
		\end{fmfgraph*}
	\end{gathered}
\nnw
	& \begin{gathered}
		\begin{fmfgraph*}(55,60)
			\fmfset{curly_len}{2mm}
			\fmftop{t1,t2} \fmfbottom{b1,b2}
			\fmf{quark,label=$\hat d_t$,label.side=right,label.dist=3,tension=2}{t1,v1}
			\fmf{quark,label=$\hat d_s$,label.side=left,label.dist=3,tension=6}{v2,t2}
			\fmf{quark,label=$\hat e_r$,label.side=left,label.dist=3,tension=2}{b1,v1}
			\fmf{quark,label=$\hat e_p$,label.side=right,label.dist=3,tension=6}{v3,b2}
			\fmf{quark,label=$d_u$,label.side=left,label.dist=3,tension=3}{v1,v2}
			\fmf{quark,label=$e_v$,label.side=right,label.dist=3,tension=3}{v1,v3}
			\fmffreeze
			\fmf{photon,label=$\Z$,label.side=left,left=0.25}{v2,v3}
			\fmfv{decoration.shape=square,decoration.size=1.5mm}{v1}
		\end{fmfgraph*}
	\end{gathered}
\qquad\quad
	\begin{gathered}
		\begin{fmfgraph*}(55,60)
			\fmfset{curly_len}{2mm}
			\fmftop{t1,t2} \fmfbottom{b1,b2}
			\fmf{quark,label=$\hat d_t$,label.side=right,label.dist=3,tension=6}{t1,v2}
			\fmf{quark,label=$\hat d_s$,label.side=right,label.dist=3,tension=6}{v3,t2}
			\fmf{quark,label=$\hat e_r$,label.side=right,label.dist=3,tension=2}{b1,v1}
			\fmf{quark,label=$\hat e_p$,label.side=right,label.dist=3,tension=2}{v1,b2}
			\fmf{quark,label=$u_u$,label.side=right,label.dist=3,tension=3}{v2,v1}
			\fmf{quark,label=$u_v$,label.side=right,label.dist=3,tension=3}{v1,v3}
			\fmffreeze
			\fmf{wboson,label=$\W^+$,label.side=right,right=0.25}{v3,v2}
			\fmfv{decoration.shape=square,decoration.size=1.5mm}{v1}
		\end{fmfgraph*}
	\end{gathered}
\qquad\quad
	\begin{gathered}
		\begin{fmfgraph*}(55,60)
			\fmfset{curly_len}{2mm}
			\fmftop{t1,t2} \fmfbottom{b1,b2}
			\fmf{quark,label=$\hat d_t$,label.side=right,label.dist=3,tension=6}{t1,v2}
			\fmf{quark,label=$\hat d_s$,label.side=right,label.dist=3,tension=6}{v3,t2}
			\fmf{quark,label=$\hat e_r$,label.side=right,label.dist=3,tension=2}{b1,v1}
			\fmf{quark,label=$\hat e_p$,label.side=right,label.dist=3,tension=2}{v1,b2}
			\fmf{quark,label=$t$,label.side=right,label.dist=3,tension=3}{v2,v1}
			\fmf{quark,label=$t$,label.side=right,label.dist=3,tension=3}{v1,v3}
			\fmffreeze
			\fmf{scalar,label=$G^+$,label.side=right,right=0.25}{v3,v2}
			\fmfv{decoration.shape=square,decoration.size=1.5mm}{v1}
		\end{fmfgraph*}
	\end{gathered}
\qquad\quad
	\begin{gathered}
		\begin{fmfgraph*}(55,60)
			\fmfset{curly_len}{2mm}
			\fmftop{t1,t2} \fmfbottom{b1,b2}
			\fmf{quark,label=$\hat d_t$,label.side=right,label.dist=3,tension=6}{t1,v2}
			\fmf{quark,label=$\hat d_s$,label.side=right,label.dist=3,tension=6}{v3,t2}
			\fmf{quark,label=$\hat e_r$,label.side=right,label.dist=3,tension=2}{b1,v1}
			\fmf{quark,label=$\hat e_p$,label.side=right,label.dist=3,tension=2}{v1,b2}
			\fmf{quark,label=$d_u$,label.side=right,label.dist=3,tension=3}{v2,v1}
			\fmf{quark,label=$d_v$,label.side=right,label.dist=3,tension=3}{v1,v3}
			\fmffreeze
			\fmf{photon,label=$\Z$,label.side=left,left=0.25}{v2,v3}
			\fmfv{decoration.shape=square,decoration.size=1.5mm}{v1}
		\end{fmfgraph*}
	\end{gathered}
\qquad\quad
	\begin{gathered}
		\begin{fmfgraph*}(55,60)
			\fmfset{curly_len}{2mm}
			\fmftop{t1,t2} \fmfbottom{b1,b2}
			\fmf{quark,label=$\hat d_t$,label.side=left,label.dist=3,tension=2}{t1,v1}
			\fmf{quark,label=$\hat d_s$,label.side=left,label.dist=3,tension=2}{v1,t2}
			\fmf{quark,label=$\hat e_r$,label.side=left,label.dist=3,tension=6}{b1,v2}
			\fmf{quark,label=$\hat e_p$,label.side=left,label.dist=3,tension=6}{v3,b2}
			\fmf{plain,label=$\nu_u$,label.side=left,label.dist=3,tension=3}{v2,v1}
			\fmf{plain,label=$\nu_v$,label.side=left,label.dist=3,tension=3}{v1,v3}
			\fmffreeze
			\fmf{wboson,label=$\W^+$,label.side=left,left=0.25}{v3,v2}
			\fmfv{decoration.shape=square,decoration.size=1.5mm}{v1}
		\end{fmfgraph*}
	\end{gathered}
\nnw
	& \begin{gathered}
		\begin{fmfgraph*}(55,60)
			\fmfset{curly_len}{2mm}
			\fmftop{t1,t2} \fmfbottom{b1,b2}
			\fmf{quark,label=$\hat d_t$,label.side=left,label.dist=3,tension=2}{t1,v1}
			\fmf{quark,label=$\hat d_s$,label.side=left,label.dist=3,tension=2}{v1,t2}
			\fmf{quark,label=$\hat e_r$,label.side=left,label.dist=3,tension=6}{b1,v2}
			\fmf{quark,label=$\hat e_p$,label.side=left,label.dist=3,tension=6}{v3,b2}
			\fmf{quark,label=$e_u$,label.side=left,label.dist=3,tension=3}{v2,v1}
			\fmf{quark,label=$e_v$,label.side=left,label.dist=3,tension=3}{v1,v3}
			\fmffreeze
			\fmf{photon,label=$\Z$,label.side=left,left=0.25}{v3,v2}
			\fmfv{decoration.shape=square,decoration.size=1.5mm}{v1}
		\end{fmfgraph*}
	\end{gathered}
\qquad\quad
	\begin{gathered}
		\begin{fmfgraph*}(55,60)
			\fmfset{curly_len}{2mm}
			\fmftop{t1,t2} \fmfbottom{b1,b2}
			\fmf{quark,label=$\hat d_t$,label.side=left,label.dist=3,tension=6}{t1,v2}
			\fmf{quark,label=$\hat d_s$,label.side=left,label.dist=3,tension=2}{v1,t2}
			\fmf{quark,label=$\hat e_r$,label.side=right,label.dist=3,tension=2}{b1,v1}
			\fmf{quark,label=$\hat e_p$,label.side=left,label.dist=3,tension=6}{v3,b2}
			\fmf{quark,label=$u_u$,label.side=left,label.dist=3,tension=3}{v2,v1}
			\fmf{plain,label=$\nu_v$,label.side=left,label.dist=3,tension=3}{v1,v3}
			\fmffreeze
			\fmf{wboson,left=0.5}{v3,v2}
			\fmf{phantom,label=$\W^+$,label.side=right,left=0.5}{v2,b1}
			\fmfv{decoration.shape=square,decoration.size=1.5mm}{v1}
		\end{fmfgraph*}
	\end{gathered}
\qquad\quad
	\begin{gathered}
		\begin{fmfgraph*}(55,60)
			\fmfset{curly_len}{2mm}
			\fmftop{t1,t2} \fmfbottom{b1,b2}
			\fmf{quark,label=$\hat d_t$,label.side=left,label.dist=3,tension=6}{t1,v2}
			\fmf{quark,label=$\hat d_s$,label.side=left,label.dist=3,tension=2}{v1,t2}
			\fmf{quark,label=$\hat e_r$,label.side=right,label.dist=3,tension=2}{b1,v1}
			\fmf{quark,label=$\hat e_p$,label.side=left,label.dist=3,tension=6}{v3,b2}
			\fmf{quark,label=$d_u$,label.side=left,label.dist=3,tension=3}{v2,v1}
			\fmf{quark,label=$e_v$,label.side=left,label.dist=3,tension=3}{v1,v3}
			\fmffreeze
			\fmf{photon,right=0.5}{v2,v3}
			\fmf{phantom,label=$\Z$,label.side=right,left=0.5}{v2,b1}
			\fmfv{decoration.shape=square,decoration.size=1.5mm}{v1}
		\end{fmfgraph*}
	\end{gathered}
\qquad\quad
	\begin{gathered}
		\begin{fmfgraph*}(55,60)
			\fmfset{curly_len}{2mm}
			\fmftop{t1,t2} \fmfbottom{b1,b2}
			\fmf{quark,label=$\hat d_t$,label.side=left,label.dist=3,tension=2}{t1,v1}
			\fmf{quark,label=$\hat d_s$,label.side=right,label.dist=3,tension=6}{v3,t2}
			\fmf{quark,label=$\hat e_r$,label.side=right,label.dist=3,tension=6}{b1,v2}
			\fmf{quark,label=$\hat e_p$,label.side=left,label.dist=3,tension=2}{v1,b2}
			\fmf{quark,label=$u_u$,label.side=left,label.dist=3,tension=3}{v1,v3}
			\fmf{plain,label=$\nu_v$,label.side=left,label.dist=3,tension=3}{v2,v1}
			\fmffreeze
			\fmf{wboson,left=0.5}{v3,v2}
			\fmf{phantom,label=$\W^+$,label.side=right,left=0.4}{v2,b2}
			\fmfv{decoration.shape=square,decoration.size=1.5mm}{v1}
		\end{fmfgraph*}
	\end{gathered}
\qquad\quad
	\begin{gathered}
		\begin{fmfgraph*}(55,60)
			\fmfset{curly_len}{2mm}
			\fmftop{t1,t2} \fmfbottom{b1,b2}
			\fmf{quark,label=$\hat d_t$,label.side=left,label.dist=3,tension=2}{t1,v1}
			\fmf{quark,label=$\hat d_s$,label.side=right,label.dist=3,tension=6}{v3,t2}
			\fmf{quark,label=$\hat e_r$,label.side=right,label.dist=3,tension=6}{b1,v2}
			\fmf{quark,label=$\hat e_p$,label.side=left,label.dist=3,tension=2}{v1,b2}
			\fmf{quark,label=$d_u$,label.side=left,label.dist=3,tension=3}{v1,v3}
			\fmf{quark,label=$e_v$,label.side=left,label.dist=3,tension=3}{v2,v1}
			\fmffreeze
			\fmf{photon,right=0.5}{v2,v3}
			\fmf{phantom,label=$\Z$,label.side=right,left=0.4}{v2,b2}
			\fmfv{decoration.shape=square,decoration.size=1.5mm}{v1}
		\end{fmfgraph*}
	\end{gathered}
\end{align*}


\subsubsection[$\bar uu \bar uu$]{\boldmath$\bar uu \bar uu$}

\begin{align*}
	& \begin{gathered}
		\begin{fmfgraph*}(55,60)
			\fmfset{curly_len}{2mm}
			\fmftop{t1,t2} \fmfbottom{b1,b2}
			\fmf{quark,label=$\hat u_t$,label.side=left,tension=4}{t1,v1}
			\fmf{quark,label=$\hat u_s$,label.side=left,tension=4}{v2,t2}
			\fmf{quark,label=$\hat u_r$,label.side=right,tension=4}{b1,v3}
			\fmf{quark,label=$\hat u_p$,label.side=right,tension=4}{v4,b2}
			\fmf{quark,label=$u_u$,label.side=right}{v1,v2}
			\fmf{quark,label=$u_v$,label.side=right}{v3,v4}
			\fmf{photon,label=$\Z$,label.side=left}{v3,v1}
			\fmf{photon,label=$\Z$,label.side=left}{v2,v4}
		\end{fmfgraph*}
	\end{gathered} \quad \times 2
\quad\quad
	\begin{gathered}
		\begin{fmfgraph*}(55,60)
			\fmfset{curly_len}{2mm}
			\fmftop{t1,t2} \fmfbottom{b1,b2}
			\fmf{quark,label=$\hat u_t$,label.side=left,tension=4}{t1,v1}
			\fmf{quark,label=$\hat u_s$,label.side=left,tension=4}{v2,t2}
			\fmf{quark,label=$\hat u_r$,label.side=right,tension=4}{b1,v3}
			\fmf{quark,label=$\hat u_p$,label.side=right,tension=4}{v4,b2}
			\fmf{quark,label=$u_u$,label.side=right}{v1,v2}
			\fmf{quark,label=$u_v$,label.side=right}{v3,v4}
			\fmf{photon,label=$\Z$,label.side=left}{v3,v1}
			\fmf{photon,label=$\A$,label.side=left}{v2,v4}
		\end{fmfgraph*}
	\end{gathered} \quad \times 2
\quad\quad
	\begin{gathered}
		\begin{fmfgraph*}(55,60)
			\fmfset{curly_len}{2mm}
			\fmftop{t1,t2} \fmfbottom{b1,b2}
			\fmf{quark,label=$\hat u_t$,label.side=left,tension=4}{t1,v1}
			\fmf{quark,label=$\hat u_s$,label.side=left,tension=4}{v2,t2}
			\fmf{quark,label=$\hat u_r$,label.side=right,tension=4}{b1,v3}
			\fmf{quark,label=$\hat u_p$,label.side=right,tension=4}{v4,b2}
			\fmf{quark,label=$u_u$,label.side=right}{v1,v2}
			\fmf{quark,label=$u_v$,label.side=right}{v3,v4}
			\fmf{photon,label=$\A$,label.side=left}{v3,v1}
			\fmf{photon,label=$\Z$,label.side=left}{v2,v4}
		\end{fmfgraph*}
	\end{gathered} \quad \times 2
\quad\quad
	\begin{gathered}
		\begin{fmfgraph*}(55,60)
			\fmfset{curly_len}{2mm}
			\fmftop{t1,t2} \fmfbottom{b1,b2}
			\fmf{quark,label=$\hat u_t$,label.side=left,tension=4}{t1,v1}
			\fmf{quark,label=$\hat u_s$,label.side=left,tension=4}{v2,t2}
			\fmf{quark,label=$\hat u_r$,label.side=right,tension=4}{b1,v3}
			\fmf{quark,label=$\hat u_p$,label.side=right,tension=4}{v4,b2}
			\fmf{quark,label=$u_u$,label.side=right}{v1,v2}
			\fmf{quark,label=$u_v$,label.side=right}{v3,v4}
			\fmf{photon,label=$\Z$,label.side=left}{v3,v1}
			\fmf{gluon,label=$\G$,label.side=left}{v2,v4}
		\end{fmfgraph*}
	\end{gathered} \quad \times 2
\nnw
	& \begin{gathered}
		\begin{fmfgraph*}(55,60)
			\fmfset{curly_len}{2mm}
			\fmftop{t1,t2} \fmfbottom{b1,b2}
			\fmf{quark,label=$\hat u_t$,label.side=left,tension=4}{t1,v1}
			\fmf{quark,label=$\hat u_s$,label.side=left,tension=4}{v2,t2}
			\fmf{quark,label=$\hat u_r$,label.side=right,tension=4}{b1,v3}
			\fmf{quark,label=$\hat u_p$,label.side=right,tension=4}{v4,b2}
			\fmf{quark,label=$u_u$,label.side=right}{v1,v2}
			\fmf{quark,label=$u_v$,label.side=right}{v3,v4}
			\fmf{gluon,label=$\G$,label.side=left}{v3,v1}
			\fmf{photon,label=$\Z$,label.side=left}{v2,v4}
		\end{fmfgraph*}
	\end{gathered} \quad \times 2
\quad\quad
	\begin{gathered}
		\begin{fmfgraph*}(55,60)
			\fmfset{curly_len}{2mm}
			\fmftop{t1,t2} \fmfbottom{b1,b2}
			\fmf{phantom,label=$\hat u_t$,label.side=left,tension=4}{t1,v1}
			\fmf{phantom,label=$\hat u_s$,label.side=left,tension=4}{v2,t2}
			\fmf{quark,label=$\hat u_r$,label.side=right,tension=4}{b1,v3}
			\fmf{quark,label=$\hat u_p$,label.side=right,tension=4}{v4,b2}
			\fmf{quark,label=$u_u$,label.side=left}{v2,v1}
			\fmf{quark,label=$u_v$,label.side=right}{v3,v4}
			\fmf{photon,label=$\Z$,label.side=left}{v3,v1}
			\fmf{photon,label=$\Z$,label.side=left}{v2,v4}
			\fmffreeze
			\fmf{quark}{t1,v2}
			\fmf{quark}{v1,t2}
		\end{fmfgraph*}
	\end{gathered} \quad \times 2
\quad\quad
	\begin{gathered}
		\begin{fmfgraph*}(55,60)
			\fmfset{curly_len}{2mm}
			\fmftop{t1,t2} \fmfbottom{b1,b2}
			\fmf{phantom,label=$\hat u_t$,label.side=left,tension=4}{t1,v1}
			\fmf{phantom,label=$\hat u_s$,label.side=left,tension=4}{v2,t2}
			\fmf{quark,label=$\hat u_r$,label.side=right,tension=4}{b1,v3}
			\fmf{quark,label=$\hat u_p$,label.side=right,tension=4}{v4,b2}
			\fmf{quark,label=$u_u$,label.side=left}{v2,v1}
			\fmf{quark,label=$u_v$,label.side=right}{v3,v4}
			\fmf{photon,label=$\Z$,label.side=left}{v3,v1}
			\fmf{photon,label=$\A$,label.side=left}{v2,v4}
			\fmffreeze
			\fmf{quark}{t1,v2}
			\fmf{quark}{v1,t2}
		\end{fmfgraph*}
	\end{gathered} \quad \times 2
\quad\quad
	\begin{gathered}
		\begin{fmfgraph*}(55,60)
			\fmfset{curly_len}{2mm}
			\fmftop{t1,t2} \fmfbottom{b1,b2}
			\fmf{phantom,label=$\hat u_t$,label.side=left,tension=4}{t1,v1}
			\fmf{phantom,label=$\hat u_s$,label.side=left,tension=4}{v2,t2}
			\fmf{quark,label=$\hat u_r$,label.side=right,tension=4}{b1,v3}
			\fmf{quark,label=$\hat u_p$,label.side=right,tension=4}{v4,b2}
			\fmf{quark,label=$u_u$,label.side=left}{v2,v1}
			\fmf{quark,label=$u_v$,label.side=right}{v3,v4}
			\fmf{photon,label=$\A$,label.side=left}{v3,v1}
			\fmf{photon,label=$\Z$,label.side=left}{v2,v4}
			\fmffreeze
			\fmf{quark}{t1,v2}
			\fmf{quark}{v1,t2}
		\end{fmfgraph*}
	\end{gathered} \quad \times 2
\nnw
	& \begin{gathered}
		\begin{fmfgraph*}(55,60)
			\fmfset{curly_len}{2mm}
			\fmftop{t1,t2} \fmfbottom{b1,b2}
			\fmf{phantom,label=$\hat u_t$,label.side=left,tension=4}{t1,v1}
			\fmf{phantom,label=$\hat u_s$,label.side=left,tension=4}{v2,t2}
			\fmf{quark,label=$\hat u_r$,label.side=right,tension=4}{b1,v3}
			\fmf{quark,label=$\hat u_p$,label.side=right,tension=4}{v4,b2}
			\fmf{quark,label=$u_u$,label.side=left}{v2,v1}
			\fmf{quark,label=$u_v$,label.side=right}{v3,v4}
			\fmf{photon,label=$\Z$,label.side=left}{v3,v1}
			\fmf{gluon,label=$\G$,label.side=left}{v2,v4}
			\fmffreeze
			\fmf{quark}{t1,v2}
			\fmf{quark}{v1,t2}
		\end{fmfgraph*}
	\end{gathered} \quad \times 2
\quad\quad
	\begin{gathered}
		\begin{fmfgraph*}(55,60)
			\fmfset{curly_len}{2mm}
			\fmftop{t1,t2} \fmfbottom{b1,b2}
			\fmf{phantom,label=$\hat u_t$,label.side=left,tension=4}{t1,v1}
			\fmf{phantom,label=$\hat u_s$,label.side=left,tension=4}{v2,t2}
			\fmf{quark,label=$\hat u_r$,label.side=right,tension=4}{b1,v3}
			\fmf{quark,label=$\hat u_p$,label.side=right,tension=4}{v4,b2}
			\fmf{quark,label=$u_u$,label.side=left}{v2,v1}
			\fmf{quark,label=$u_v$,label.side=right}{v3,v4}
			\fmf{gluon,label=$\G$,label.side=left}{v3,v1}
			\fmf{photon,label=$\Z$,label.side=left}{v2,v4}
			\fmffreeze
			\fmf{quark}{t1,v2}
			\fmf{quark}{v1,t2}
		\end{fmfgraph*}
	\end{gathered} \quad \times 2
\qquad\quad
	\begin{gathered}
		\begin{fmfgraph*}(55,60)
			\fmfset{curly_len}{2mm}
			\fmftop{t1,t2} \fmfbottom{b1,b2}
			\fmf{phantom,label=$\hat u_t$,label.side=left,tension=4}{t1,v1}
			\fmf{phantom,label=$\hat u_s$,label.side=left,tension=4}{v2,t2}
			\fmf{quark,label=$\hat u_r$,label.side=right,tension=4}{b1,v3}
			\fmf{quark,label=$\hat u_p$,label.side=right,tension=4}{v4,b2}
			\fmf{quark,label=$d_u$,label.side=left}{v2,v1}
			\fmf{quark,label=$d_v$,label.side=right}{v3,v4}
			\fmf{wboson,label=$\W^+$,label.side=left}{v3,v1}
			\fmf{wboson,label=$\W^+$,label.side=left}{v2,v4}
			\fmffreeze
			\fmf{quark}{t1,v2}
			\fmf{quark}{v1,t2}
		\end{fmfgraph*}
	\end{gathered} \quad\,\, \times 2
\qquad\quad
	\begin{gathered}
		\begin{fmfgraph*}(55,60)
			\fmfset{curly_len}{2mm}
			\fmftop{t1,t2} \fmfbottom{b1,b2}
			\fmf{quark,label=$\hat u_t$,label.side=right,label.dist=3,tension=3}{t1,v2}
			\fmf{quark,label=$\hat u_s$,label.side=right,label.dist=3,tension=3}{v3,t2}
			\fmf{quark,label=$\hat u_r$,label.side=right,label.dist=4,tension=3}{b1,v1}
			\fmf{quark,label=$\hat u_p$,label.side=right,label.dist=4,tension=3}{v1,b2}
			\fmf{wboson,label=$\W^+$,label.side=right,label.dist=3,tension=1}{v2,v1}
			\fmf{wboson,label=$\W^+$,label.side=right,label.dist=3,tension=1}{v1,v3}
			\fmffreeze
			\fmf{quark,label=$d_u$,label.side=left}{v2,v3}
			\fmfv{decoration.shape=square,decoration.size=1.5mm}{v1}
		\end{fmfgraph*}
	\end{gathered} \quad \times 4
\nnw
	& \begin{gathered}
		\begin{fmfgraph*}(55,60)
			\fmfset{curly_len}{2mm}
			\fmftop{t1,t2} \fmfbottom{b1,b2}
			\fmf{quark,label=$\hat u_t$,label.side=left,label.dist=3,tension=6}{t1,v2}
			\fmf{quark,label=$\hat u_s$,label.side=right,label.dist=3,tension=2}{v1,t2}
			\fmf{quark,label=$\hat u_r$,label.side=right,label.dist=3,tension=6}{b1,v3}
			\fmf{quark,label=$\hat u_p$,label.side=left,label.dist=3,tension=2}{v1,b2}
			\fmf{quark,label=$u_u$,label.side=left,label.dist=3,tension=3}{v2,v1}
			\fmf{quark,label=$u_v$,label.side=right,label.dist=3,tension=3}{v3,v1}
			\fmffreeze
			\fmf{photon,label=$\Z$,label.side=right,right=0.25}{v2,v3}
			\fmfv{decoration.shape=square,decoration.size=1.5mm}{v1}
		\end{fmfgraph*}
	\end{gathered}
\qquad\qquad
	\begin{gathered}
		\begin{fmfgraph*}(55,60)
			\fmfset{curly_len}{2mm}
			\fmftop{t1,t2} \fmfbottom{b1,b2}
			\fmf{quark,label=$\hat u_t$,label.side=right,label.dist=3,tension=2}{t1,v1}
			\fmf{quark,label=$\hat u_s$,label.side=left,label.dist=3,tension=6}{v2,t2}
			\fmf{quark,label=$\hat u_r$,label.side=left,label.dist=3,tension=2}{b1,v1}
			\fmf{quark,label=$\hat u_p$,label.side=right,label.dist=3,tension=6}{v3,b2}
			\fmf{quark,label=$u_u$,label.side=left,label.dist=3,tension=3}{v1,v2}
			\fmf{quark,label=$u_v$,label.side=right,label.dist=3,tension=3}{v1,v3}
			\fmffreeze
			\fmf{photon,label=$\Z$,label.side=left,left=0.25}{v2,v3}
			\fmfv{decoration.shape=square,decoration.size=1.5mm}{v1}
		\end{fmfgraph*}
	\end{gathered}
\qquad\qquad
	\begin{gathered}
		\begin{fmfgraph*}(55,60)
			\fmfset{curly_len}{2mm}
			\fmftop{t1,t2} \fmfbottom{b1,b2}
			\fmf{quark,label=$\hat u_t$,label.side=right,label.dist=3,tension=6}{t1,v2}
			\fmf{quark,label=$\hat u_s$,label.side=right,label.dist=3,tension=6}{v3,t2}
			\fmf{quark,label=$\hat u_r$,label.side=right,label.dist=3,tension=2}{b1,v1}
			\fmf{quark,label=$\hat u_p$,label.side=right,label.dist=3,tension=2}{v1,b2}
			\fmf{quark,label=$d_u$,label.side=right,label.dist=3,tension=3}{v2,v1}
			\fmf{quark,label=$d_v$,label.side=right,label.dist=3,tension=3}{v1,v3}
			\fmffreeze
			\fmf{wboson,label=$\W^+$,label.side=left,left=0.25}{v2,v3}
			\fmfv{decoration.shape=square,decoration.size=1.5mm}{v1}
		\end{fmfgraph*}
	\end{gathered} \quad \times 4
\qquad\quad
	\begin{gathered}
		\begin{fmfgraph*}(55,60)
			\fmfset{curly_len}{2mm}
			\fmftop{t1,t2} \fmfbottom{b1,b2}
			\fmf{quark,label=$\hat u_t$,label.side=right,label.dist=3,tension=6}{t1,v2}
			\fmf{quark,label=$\hat u_s$,label.side=right,label.dist=3,tension=6}{v3,t2}
			\fmf{quark,label=$\hat u_r$,label.side=right,label.dist=3,tension=2}{b1,v1}
			\fmf{quark,label=$\hat u_p$,label.side=right,label.dist=3,tension=2}{v1,b2}
			\fmf{quark,label=$u_u$,label.side=right,label.dist=3,tension=3}{v2,v1}
			\fmf{quark,label=$u_v$,label.side=right,label.dist=3,tension=3}{v1,v3}
			\fmffreeze
			\fmf{photon,label=$\Z$,label.side=left,left=0.25}{v2,v3}
			\fmfv{decoration.shape=square,decoration.size=1.5mm}{v1}
		\end{fmfgraph*}
	\end{gathered} \quad \times 4
\end{align*}


\subsubsection[$\bar uu \bar dd$]{\boldmath$\bar uu \bar dd$}

\begin{align*}
	& \begin{gathered}
		\begin{fmfgraph*}(55,60)
			\fmfset{curly_len}{2mm}
			\fmftop{t1,t2} \fmfbottom{b1,b2}
			\fmf{quark,label=$\hat d_t$,label.side=left,tension=4}{t1,v1}
			\fmf{quark,label=$\hat d_s$,label.side=left,tension=4}{v2,t2}
			\fmf{quark,label=$\hat u_r$,label.side=right,tension=4}{b1,v3}
			\fmf{quark,label=$\hat u_p$,label.side=right,tension=4}{v4,b2}
			\fmf{quark,label=$u_u$,label.side=right}{v1,v2}
			\fmf{quark,label=$d_v$,label.side=right}{v3,v4}
			\fmf{wboson,label=$\W^+$,label.side=left}{v3,v1}
			\fmf{wboson,label=$\W^+$,label.side=left}{v2,v4}
		\end{fmfgraph*}
	\end{gathered}
\qquad\qquad
	\begin{gathered}
		\begin{fmfgraph*}(55,60)
			\fmfset{curly_len}{2mm}
			\fmftop{t1,t2} \fmfbottom{b1,b2}
			\fmf{quark,label=$\hat d_t$,label.side=left,tension=4}{t1,v1}
			\fmf{quark,label=$\hat d_s$,label.side=left,tension=4}{v2,t2}
			\fmf{quark,label=$\hat u_r$,label.side=right,tension=4}{b1,v3}
			\fmf{quark,label=$\hat u_p$,label.side=right,tension=4}{v4,b2}
			\fmf{quark,label=$t$,label.side=right}{v1,v2}
			\fmf{quark,label=$d_v$,label.side=right}{v3,v4}
			\fmf{scalar,label=$G^+$,label.side=left}{v3,v1}
			\fmf{wboson,label=$\W^+$,label.side=left}{v2,v4}
			\fmfv{decoration.shape=square,decoration.size=1.5mm}{v3}
		\end{fmfgraph*}
	\end{gathered}
\qquad\qquad
	\begin{gathered}
		\begin{fmfgraph*}(55,60)
			\fmfset{curly_len}{2mm}
			\fmftop{t1,t2} \fmfbottom{b1,b2}
			\fmf{quark,label=$\hat d_t$,label.side=left,tension=4}{t1,v1}
			\fmf{quark,label=$\hat d_s$,label.side=left,tension=4}{v2,t2}
			\fmf{quark,label=$\hat u_r$,label.side=right,tension=4}{b1,v3}
			\fmf{quark,label=$\hat u_p$,label.side=right,tension=4}{v4,b2}
			\fmf{quark,label=$t$,label.side=right}{v1,v2}
			\fmf{quark,label=$d_v$,label.side=right}{v3,v4}
			\fmf{wboson,label=$\W^+$,label.side=left}{v3,v1}
			\fmf{scalar,label=$G^+$,label.side=left}{v2,v4}
			\fmfv{decoration.shape=square,decoration.size=1.5mm}{v4}
		\end{fmfgraph*}
	\end{gathered}
\qquad\qquad
	\begin{gathered}
		\begin{fmfgraph*}(55,60)
			\fmfset{curly_len}{2mm}
			\fmftop{t1,t2} \fmfbottom{b1,b2}
			\fmf{quark,label=$\hat d_t$,label.side=left,tension=4}{t1,v1}
			\fmf{quark,label=$\hat d_s$,label.side=left,tension=4}{v2,t2}
			\fmf{quark,label=$\hat u_r$,label.side=right,tension=4}{b1,v3}
			\fmf{quark,label=$\hat u_p$,label.side=right,tension=4}{v4,b2}
			\fmf{quark,label=$d_u$,label.side=right}{v1,v2}
			\fmf{quark,label=$u_v$,label.side=right}{v3,v4}
			\fmf{photon,label=$\Z$,label.side=left}{v3,v1}
			\fmf{photon,label=$\Z$,label.side=left}{v2,v4}
		\end{fmfgraph*}
	\end{gathered}
\nnw
	& \begin{gathered}
		\begin{fmfgraph*}(55,60)
			\fmfset{curly_len}{2mm}
			\fmftop{t1,t2} \fmfbottom{b1,b2}
			\fmf{quark,label=$\hat d_t$,label.side=left,tension=4}{t1,v1}
			\fmf{quark,label=$\hat d_s$,label.side=left,tension=4}{v2,t2}
			\fmf{quark,label=$\hat u_r$,label.side=right,tension=4}{b1,v3}
			\fmf{quark,label=$\hat u_p$,label.side=right,tension=4}{v4,b2}
			\fmf{quark,label=$d_u$,label.side=right}{v1,v2}
			\fmf{quark,label=$u_v$,label.side=right}{v3,v4}
			\fmf{photon,label=$\Z$,label.side=left}{v3,v1}
			\fmf{photon,label=$\A$,label.side=left}{v2,v4}
		\end{fmfgraph*}
	\end{gathered}
\qquad\quad
	\begin{gathered}
		\begin{fmfgraph*}(55,60)
			\fmfset{curly_len}{2mm}
			\fmftop{t1,t2} \fmfbottom{b1,b2}
			\fmf{quark,label=$\hat d_t$,label.side=left,tension=4}{t1,v1}
			\fmf{quark,label=$\hat d_s$,label.side=left,tension=4}{v2,t2}
			\fmf{quark,label=$\hat u_r$,label.side=right,tension=4}{b1,v3}
			\fmf{quark,label=$\hat u_p$,label.side=right,tension=4}{v4,b2}
			\fmf{quark,label=$d_u$,label.side=right}{v1,v2}
			\fmf{quark,label=$u_v$,label.side=right}{v3,v4}
			\fmf{photon,label=$\A$,label.side=left}{v3,v1}
			\fmf{photon,label=$\Z$,label.side=left}{v2,v4}
		\end{fmfgraph*}
	\end{gathered}
\qquad\quad
	\begin{gathered}
		\begin{fmfgraph*}(55,60)
			\fmfset{curly_len}{2mm}
			\fmftop{t1,t2} \fmfbottom{b1,b2}
			\fmf{quark,label=$\hat d_t$,label.side=left,tension=4}{t1,v1}
			\fmf{quark,label=$\hat d_s$,label.side=left,tension=4}{v2,t2}
			\fmf{quark,label=$\hat u_r$,label.side=right,tension=4}{b1,v3}
			\fmf{quark,label=$\hat u_p$,label.side=right,tension=4}{v4,b2}
			\fmf{quark,label=$d_u$,label.side=right}{v1,v2}
			\fmf{quark,label=$u_v$,label.side=right}{v3,v4}
			\fmf{photon,label=$\Z$,label.side=left}{v3,v1}
			\fmf{gluon,label=$\G$,label.side=left}{v2,v4}
		\end{fmfgraph*}
	\end{gathered}
\qquad\quad
	\begin{gathered}
		\begin{fmfgraph*}(55,60)
			\fmfset{curly_len}{2mm}
			\fmftop{t1,t2} \fmfbottom{b1,b2}
			\fmf{quark,label=$\hat d_t$,label.side=left,tension=4}{t1,v1}
			\fmf{quark,label=$\hat d_s$,label.side=left,tension=4}{v2,t2}
			\fmf{quark,label=$\hat u_r$,label.side=right,tension=4}{b1,v3}
			\fmf{quark,label=$\hat u_p$,label.side=right,tension=4}{v4,b2}
			\fmf{quark,label=$d_u$,label.side=right}{v1,v2}
			\fmf{quark,label=$u_v$,label.side=right}{v3,v4}
			\fmf{gluon,label=$\G$,label.side=left}{v3,v1}
			\fmf{photon,label=$\Z$,label.side=left}{v2,v4}
		\end{fmfgraph*}
	\end{gathered}
\qquad\quad
	\begin{gathered}
		\begin{fmfgraph*}(55,60)
			\fmfset{curly_len}{2mm}
			\fmftop{t1,t2} \fmfbottom{b1,b2}
			\fmf{phantom,label=$\hat d_t$,label.side=left,tension=4}{t1,v1}
			\fmf{phantom,label=$\hat d_s$,label.side=left,tension=4}{v2,t2}
			\fmf{quark,label=$\hat u_r$,label.side=right,tension=4}{b1,v3}
			\fmf{quark,label=$\hat u_p$,label.side=right,tension=4}{v4,b2}
			\fmf{quark,label=$d_u$,label.side=left}{v2,v1}
			\fmf{quark,label=$u_v$,label.side=right}{v3,v4}
			\fmf{photon,label=$\Z$,label.side=left}{v3,v1}
			\fmf{photon,label=$\Z$,label.side=left}{v2,v4}
			\fmffreeze
			\fmf{quark}{t1,v2}
			\fmf{quark}{v1,t2}
		\end{fmfgraph*}
	\end{gathered}
\nnw
	& \begin{gathered}
		\begin{fmfgraph*}(55,60)
			\fmfset{curly_len}{2mm}
			\fmftop{t1,t2} \fmfbottom{b1,b2}
			\fmf{phantom,label=$\hat d_t$,label.side=left,tension=4}{t1,v1}
			\fmf{phantom,label=$\hat d_s$,label.side=left,tension=4}{v2,t2}
			\fmf{quark,label=$\hat u_r$,label.side=right,tension=4}{b1,v3}
			\fmf{quark,label=$\hat u_p$,label.side=right,tension=4}{v4,b2}
			\fmf{quark,label=$d_u$,label.side=left}{v2,v1}
			\fmf{quark,label=$u_v$,label.side=right}{v3,v4}
			\fmf{photon,label=$\Z$,label.side=left}{v3,v1}
			\fmf{photon,label=$\A$,label.side=left}{v2,v4}
			\fmffreeze
			\fmf{quark}{t1,v2}
			\fmf{quark}{v1,t2}
		\end{fmfgraph*}
	\end{gathered}
\qquad\quad
	\begin{gathered}
		\begin{fmfgraph*}(55,60)
			\fmfset{curly_len}{2mm}
			\fmftop{t1,t2} \fmfbottom{b1,b2}
			\fmf{phantom,label=$\hat d_t$,label.side=left,tension=4}{t1,v1}
			\fmf{phantom,label=$\hat d_s$,label.side=left,tension=4}{v2,t2}
			\fmf{quark,label=$\hat u_r$,label.side=right,tension=4}{b1,v3}
			\fmf{quark,label=$\hat u_p$,label.side=right,tension=4}{v4,b2}
			\fmf{quark,label=$d_u$,label.side=left}{v2,v1}
			\fmf{quark,label=$u_v$,label.side=right}{v3,v4}
			\fmf{photon,label=$\A$,label.side=left}{v3,v1}
			\fmf{photon,label=$\Z$,label.side=left}{v2,v4}
			\fmffreeze
			\fmf{quark}{t1,v2}
			\fmf{quark}{v1,t2}
		\end{fmfgraph*}
	\end{gathered}
\qquad\quad
	\begin{gathered}
		\begin{fmfgraph*}(55,60)
			\fmfset{curly_len}{2mm}
			\fmftop{t1,t2} \fmfbottom{b1,b2}
			\fmf{phantom,label=$\hat d_t$,label.side=left,tension=4}{t1,v1}
			\fmf{phantom,label=$\hat d_s$,label.side=left,tension=4}{v2,t2}
			\fmf{quark,label=$\hat u_r$,label.side=right,tension=4}{b1,v3}
			\fmf{quark,label=$\hat u_p$,label.side=right,tension=4}{v4,b2}
			\fmf{quark,label=$d_u$,label.side=left}{v2,v1}
			\fmf{quark,label=$u_v$,label.side=right}{v3,v4}
			\fmf{photon,label=$\Z$,label.side=left}{v3,v1}
			\fmf{gluon,label=$\G$,label.side=left}{v2,v4}
			\fmffreeze
			\fmf{quark}{t1,v2}
			\fmf{quark}{v1,t2}
		\end{fmfgraph*}
	\end{gathered}
\qquad\quad
	\begin{gathered}
		\begin{fmfgraph*}(55,60)
			\fmfset{curly_len}{2mm}
			\fmftop{t1,t2} \fmfbottom{b1,b2}
			\fmf{phantom,label=$\hat d_t$,label.side=left,tension=4}{t1,v1}
			\fmf{phantom,label=$\hat d_s$,label.side=left,tension=4}{v2,t2}
			\fmf{quark,label=$\hat u_r$,label.side=right,tension=4}{b1,v3}
			\fmf{quark,label=$\hat u_p$,label.side=right,tension=4}{v4,b2}
			\fmf{quark,label=$d_u$,label.side=left}{v2,v1}
			\fmf{quark,label=$u_v$,label.side=right}{v3,v4}
			\fmf{gluon,label=$\G$,label.side=left}{v3,v1}
			\fmf{photon,label=$\Z$,label.side=left}{v2,v4}
			\fmffreeze
			\fmf{quark}{t1,v2}
			\fmf{quark}{v1,t2}
		\end{fmfgraph*}
	\end{gathered}
\qquad\quad
	\begin{gathered}
		\begin{fmfgraph*}(55,60)
			\fmfset{curly_len}{2mm}
			\fmftop{t1,t2} \fmfbottom{b1,b2}
			\fmf{quark,label=$\hat d_t$,label.side=left,tension=4}{t1,v1}
			\fmf{quark,label=$\hat u_r$,label.side=right,tension=4}{b1,v3}
			\fmf{phantom,label=$\hat d_s$,label.side=right,tension=4}{v2,t2}
			\fmf{phantom,label=$\hat u_p$,label.side=left,tension=4}{v4,b2}
			\fmf{quark,label=$d_u$,label.side=left}{v1,v2}
			\fmf{quark,label=$u_v$,label.side=right}{v3,v4}
			\fmf{photon,label=$\A$,label.side=left}{v3,v1}
			\fmf{wboson,label=$\W^+$,label.side=left,label.dist=2}{v4,v2}
			\fmffreeze
			\fmf{quark}{v4,t2}
			\fmf{quark}{v2,b2}
		\end{fmfgraph*}
	\end{gathered}
\nnw
	& \begin{gathered}
		\begin{fmfgraph*}(55,60)
			\fmfset{curly_len}{2mm}
			\fmftop{t1,t2} \fmfbottom{b1,b2}
			\fmf{quark,label=$\hat d_t$,label.side=left,tension=4}{t1,v1}
			\fmf{quark,label=$\hat u_r$,label.side=right,tension=4}{b1,v3}
			\fmf{phantom,label=$\hat d_s$,label.side=right,tension=4}{v2,t2}
			\fmf{phantom,label=$\hat u_p$,label.side=left,tension=4}{v4,b2}
			\fmf{quark,label=$d_u$,label.side=left}{v1,v2}
			\fmf{quark,label=$u_v$,label.side=right}{v3,v4}
			\fmf{photon,label=$\Z$,label.side=left}{v3,v1}
			\fmf{wboson,label=$\W^+$,label.side=left,label.dist=2}{v4,v2}
			\fmffreeze
			\fmf{quark}{v4,t2}
			\fmf{quark}{v2,b2}
		\end{fmfgraph*}
	\end{gathered}
\qquad\quad
	\begin{gathered}
		\begin{fmfgraph*}(55,60)
			\fmfset{curly_len}{2mm}
			\fmftop{t1,t2} \fmfbottom{b1,b2}
			\fmf{quark,label=$\hat d_t$,label.side=left,tension=4}{t1,v1}
			\fmf{quark,label=$\hat u_r$,label.side=right,tension=4}{b1,v3}
			\fmf{phantom,label=$\hat d_s$,label.side=right,tension=4}{v2,t2}
			\fmf{phantom,label=$\hat u_p$,label.side=left,tension=4}{v4,b2}
			\fmf{quark,label=$d_u$,label.side=left}{v1,v2}
			\fmf{quark,label=$u_v$,label.side=right}{v3,v4}
			\fmf{gluon,label=$\G$,label.side=left}{v3,v1}
			\fmf{wboson,label=$\W^+$,label.side=left,label.dist=2}{v4,v2}
			\fmffreeze
			\fmf{quark}{v4,t2}
			\fmf{quark}{v2,b2}
		\end{fmfgraph*}
	\end{gathered}
\qquad\quad
	\begin{gathered}
		\begin{fmfgraph*}(55,60)
			\fmfset{curly_len}{2mm}
			\fmftop{t1,t2} \fmfbottom{b1,b2}
			\fmf{quark,label=$\hat d_t$,label.side=left,tension=4}{t1,v1}
			\fmf{quark,label=$\hat u_r$,label.side=right,tension=4}{b1,v3}
			\fmf{phantom,label=$\hat d_s$,label.side=right,tension=4}{v2,t2}
			\fmf{phantom,label=$\hat u_p$,label.side=left,tension=4}{v4,b2}
			\fmf{quark,label=$u_u$,label.side=left}{v1,v2}
			\fmf{quark,label=$d_v$,label.side=right}{v3,v4}
			\fmf{wboson,label=$\W^+$,label.side=left}{v3,v1}
			\fmf{photon,label=$\A$,label.side=right}{v2,v4}
			\fmffreeze
			\fmf{quark}{v4,t2}
			\fmf{quark}{v2,b2}
		\end{fmfgraph*}
	\end{gathered}
\qquad\quad
	\begin{gathered}
		\begin{fmfgraph*}(55,60)
			\fmfset{curly_len}{2mm}
			\fmftop{t1,t2} \fmfbottom{b1,b2}
			\fmf{quark,label=$\hat d_t$,label.side=left,tension=4}{t1,v1}
			\fmf{quark,label=$\hat u_r$,label.side=right,tension=4}{b1,v3}
			\fmf{phantom,label=$\hat d_s$,label.side=right,tension=4}{v2,t2}
			\fmf{phantom,label=$\hat u_p$,label.side=left,tension=4}{v4,b2}
			\fmf{quark,label=$u_u$,label.side=left}{v1,v2}
			\fmf{quark,label=$d_v$,label.side=right}{v3,v4}
			\fmf{wboson,label=$\W^+$,label.side=left}{v3,v1}
			\fmf{photon,label=$\Z$,label.side=right}{v2,v4}
			\fmffreeze
			\fmf{quark}{v4,t2}
			\fmf{quark}{v2,b2}
		\end{fmfgraph*}
	\end{gathered}
\qquad\quad
	\begin{gathered}
		\begin{fmfgraph*}(55,60)
			\fmfset{curly_len}{2mm}
			\fmftop{t1,t2} \fmfbottom{b1,b2}
			\fmf{quark,label=$\hat d_t$,label.side=left,tension=4}{t1,v1}
			\fmf{quark,label=$\hat u_r$,label.side=right,tension=4}{b1,v3}
			\fmf{phantom,label=$\hat d_s$,label.side=right,tension=4}{v2,t2}
			\fmf{phantom,label=$\hat u_p$,label.side=left,tension=4}{v4,b2}
			\fmf{quark,label=$u_u$,label.side=left}{v1,v2}
			\fmf{quark,label=$d_v$,label.side=right}{v3,v4}
			\fmf{wboson,label=$\W^+$,label.side=left}{v3,v1}
			\fmf{gluon,label=$\G$,label.side=right}{v2,v4}
			\fmffreeze
			\fmf{quark}{v4,t2}
			\fmf{quark}{v2,b2}
		\end{fmfgraph*}
	\end{gathered}
\nnw
	& \begin{gathered}
		\begin{fmfgraph*}(55,60)
			\fmfset{curly_len}{2mm}
			\fmftop{t1,t2} \fmfbottom{b1,b2}
			\fmf{phantom,label=$\hat d_t$,label.side=left,tension=4}{t1,v1}
			\fmf{phantom,label=$\hat d_s$,label.side=left,tension=4}{v2,t2}
			\fmf{quark,label=$\hat u_r$,label.side=right,tension=4}{b1,v3}
			\fmf{quark,label=$\hat u_p$,label.side=right,tension=4}{v4,b2}
			\fmf{wboson,label=$\W^+$,label.side=right}{v1,v2}
			\fmf{photon,label=$\A$,label.side=right}{v3,v4}
			\fmf{quark,label=$u_u$,label.side=left}{v3,v1}
			\fmf{quark,label=$u_v$,label.side=left}{v2,v4}
			\fmffreeze
			\fmf{quark}{t1,v2}
			\fmf{quark}{v1,t2}
		\end{fmfgraph*}
	\end{gathered}
\qquad\quad
	\begin{gathered}
		\begin{fmfgraph*}(55,60)
			\fmfset{curly_len}{2mm}
			\fmftop{t1,t2} \fmfbottom{b1,b2}
			\fmf{phantom,label=$\hat d_t$,label.side=left,tension=4}{t1,v1}
			\fmf{phantom,label=$\hat d_s$,label.side=left,tension=4}{v2,t2}
			\fmf{quark,label=$\hat u_r$,label.side=right,tension=4}{b1,v3}
			\fmf{quark,label=$\hat u_p$,label.side=right,tension=4}{v4,b2}
			\fmf{wboson,label=$\W^+$,label.side=right}{v1,v2}
			\fmf{photon,label=$\Z$,label.side=right}{v3,v4}
			\fmf{quark,label=$u_u$,label.side=left}{v3,v1}
			\fmf{quark,label=$u_v$,label.side=left}{v2,v4}
			\fmffreeze
			\fmf{quark}{t1,v2}
			\fmf{quark}{v1,t2}
		\end{fmfgraph*}
	\end{gathered}
\qquad\quad
	\begin{gathered}
		\begin{fmfgraph*}(55,60)
			\fmfset{curly_len}{2mm}
			\fmftop{t1,t2} \fmfbottom{b1,b2}
			\fmf{phantom,label=$\hat d_t$,label.side=left,tension=4}{t1,v1}
			\fmf{phantom,label=$\hat d_s$,label.side=left,tension=4}{v2,t2}
			\fmf{quark,label=$\hat u_r$,label.side=right,tension=4}{b1,v3}
			\fmf{quark,label=$\hat u_p$,label.side=right,tension=4}{v4,b2}
			\fmf{wboson,label=$\W^+$,label.side=right}{v1,v2}
			\fmf{gluon,label=$\G$,label.side=right}{v3,v4}
			\fmf{quark,label=$u_u$,label.side=left}{v3,v1}
			\fmf{quark,label=$u_v$,label.side=left}{v2,v4}
			\fmffreeze
			\fmf{quark}{t1,v2}
			\fmf{quark}{v1,t2}
		\end{fmfgraph*}
	\end{gathered}
\qquad\quad
	\begin{gathered}
		\begin{fmfgraph*}(55,60)
			\fmfset{curly_len}{2mm}
			\fmftop{t1,t2} \fmfbottom{b1,b2}
			\fmf{phantom,label=$\hat d_t$,label.side=left,tension=4}{t1,v1}
			\fmf{phantom,label=$\hat d_s$,label.side=left,tension=4}{v2,t2}
			\fmf{quark,label=$\hat u_r$,label.side=right,label.dist=2,tension=4}{b1,v3}
			\fmf{quark,label=$\hat u_p$,label.side=right,label.dist=1,tension=4}{v4,b2}
			\fmf{photon,label=$\A$,label.side=right}{v1,v2}
			\fmf{wboson,label=$\W^+$,label.side=right}{v3,v4}
			\fmf{quark,label=$d_u$,label.side=left}{v3,v1}
			\fmf{quark,label=$d_v$,label.side=left}{v2,v4}
			\fmffreeze
			\fmf{quark}{t1,v2}
			\fmf{quark}{v1,t2}
		\end{fmfgraph*}
	\end{gathered}
\qquad\quad
	\begin{gathered}
		\begin{fmfgraph*}(55,60)
			\fmfset{curly_len}{2mm}
			\fmftop{t1,t2} \fmfbottom{b1,b2}
			\fmf{phantom,label=$\hat d_t$,label.side=left,tension=4}{t1,v1}
			\fmf{phantom,label=$\hat d_s$,label.side=left,tension=4}{v2,t2}
			\fmf{quark,label=$\hat u_r$,label.side=right,label.dist=2,tension=4}{b1,v3}
			\fmf{quark,label=$\hat u_p$,label.side=right,label.dist=1,tension=4}{v4,b2}
			\fmf{photon,label=$\Z$,label.side=right}{v1,v2}
			\fmf{wboson,label=$\W^+$,label.side=right}{v3,v4}
			\fmf{quark,label=$d_u$,label.side=left}{v3,v1}
			\fmf{quark,label=$d_v$,label.side=left}{v2,v4}
			\fmffreeze
			\fmf{quark}{t1,v2}
			\fmf{quark}{v1,t2}
		\end{fmfgraph*}
	\end{gathered}
\nnw
	& \begin{gathered}
		\begin{fmfgraph*}(55,60)
			\fmfset{curly_len}{2mm}
			\fmftop{t1,t2} \fmfbottom{b1,b2}
			\fmf{phantom,label=$\hat d_t$,label.side=left,tension=4}{t1,v1}
			\fmf{phantom,label=$\hat d_s$,label.side=left,tension=4}{v2,t2}
			\fmf{quark,label=$\hat u_r$,label.side=right,label.dist=2,tension=4}{b1,v3}
			\fmf{quark,label=$\hat u_p$,label.side=right,label.dist=1,tension=4}{v4,b2}
			\fmf{gluon,label=$\G$,label.side=right}{v1,v2}
			\fmf{wboson,label=$\W^+$,label.side=right}{v3,v4}
			\fmf{quark,label=$d_u$,label.side=left}{v3,v1}
			\fmf{quark,label=$d_v$,label.side=left}{v2,v4}
			\fmffreeze
			\fmf{quark}{t1,v2}
			\fmf{quark}{v1,t2}
		\end{fmfgraph*}
	\end{gathered}
\qquad\quad
	\begin{gathered}
		\begin{fmfgraph*}(55,60)
			\fmfset{curly_len}{2mm}
			\fmftop{t1,t2} \fmfbottom{b1,b2}
			\fmf{quark,label=$\hat d_t$,label.side=right,label.dist=3,tension=3}{t1,v2}
			\fmf{quark,label=$\hat d_s$,label.side=right,label.dist=3,tension=3}{v3,t2}
			\fmf{quark,label=$\hat u_r$,label.side=right,label.dist=4,tension=3}{b1,v1}
			\fmf{quark,label=$\hat u_p$,label.side=right,label.dist=4,tension=3}{v1,b2}
			\fmf{wboson,label=$\W^+$,label.side=left,label.dist=3,tension=1}{v1,v2}
			\fmf{wboson,label=$\W^+$,label.side=left,label.dist=3,tension=1}{v3,v1}
			\fmffreeze
			\fmf{quark,label=$u_u$,label.side=left}{v2,v3}
			\fmfv{decoration.shape=square,decoration.size=1.5mm}{v1}
		\end{fmfgraph*}
	\end{gathered}
\qquad\quad
	\begin{gathered}
		\begin{fmfgraph*}(55,60)
			\fmfset{curly_len}{2mm}
			\fmftop{t1,t2} \fmfbottom{b1,b2}
			\fmf{quark,label=$\hat d_t$,label.side=right,label.dist=3,tension=3}{t1,v2}
			\fmf{quark,label=$\hat d_s$,label.side=right,label.dist=3,tension=3}{v3,t2}
			\fmf{quark,label=$\hat u_r$,label.side=right,label.dist=4,tension=3}{b1,v1}
			\fmf{quark,label=$\hat u_p$,label.side=right,label.dist=4,tension=3}{v1,b2}
			\fmf{scalar,label=$G^+$,label.side=left,label.dist=3,tension=1}{v1,v2}
			\fmf{wboson,label=$\W^+$,label.side=left,label.dist=3,tension=1}{v3,v1}
			\fmffreeze
			\fmf{quark,label=$t$,label.side=left}{v2,v3}
			\fmfv{decoration.shape=square,decoration.size=1.5mm}{v1}
		\end{fmfgraph*}
	\end{gathered}
\qquad\quad
	\begin{gathered}
		\begin{fmfgraph*}(55,60)
			\fmfset{curly_len}{2mm}
			\fmftop{t1,t2} \fmfbottom{b1,b2}
			\fmf{quark,label=$\hat d_t$,label.side=right,label.dist=3,tension=3}{t1,v2}
			\fmf{quark,label=$\hat d_s$,label.side=right,label.dist=3,tension=3}{v3,t2}
			\fmf{quark,label=$\hat u_r$,label.side=right,label.dist=4,tension=3}{b1,v1}
			\fmf{quark,label=$\hat u_p$,label.side=right,label.dist=4,tension=3}{v1,b2}
			\fmf{wboson,label=$\W^+$,label.side=left,label.dist=3,tension=1}{v1,v2}
			\fmf{scalar,label=$G^+$,label.side=left,label.dist=3,tension=1}{v3,v1}
			\fmffreeze
			\fmf{quark,label=$t$,label.side=left}{v2,v3}
			\fmfv{decoration.shape=square,decoration.size=1.5mm}{v1}
		\end{fmfgraph*}
	\end{gathered}
\qquad\quad
	\begin{gathered}
		\begin{fmfgraph*}(55,60)
			\fmfset{curly_len}{2mm}
			\fmftop{t1,t2} \fmfbottom{b1,b2}
			\fmf{quark,label=$\hat d_t$,label.side=right,label.dist=3,tension=3}{t1,v2}
			\fmf{quark,label=$\hat d_s$,label.side=right,label.dist=3,tension=3}{v3,t2}
			\fmf{quark,label=$\hat u_r$,label.side=right,label.dist=4,tension=3}{b1,v1}
			\fmf{quark,label=$\hat u_p$,label.side=right,label.dist=4,tension=3}{v1,b2}
			\fmf{scalar,label=$G^+$,label.side=left,label.dist=3,tension=1}{v1,v2}
			\fmf{scalar,label=$G^+$,label.side=left,label.dist=3,tension=1}{v3,v1}
			\fmffreeze
			\fmf{quark,label=$t$,label.side=left}{v2,v3}
			\fmfv{decoration.shape=square,decoration.size=1.5mm}{v1}
		\end{fmfgraph*}
	\end{gathered}
\nnw
	& \begin{gathered}
		\begin{fmfgraph*}(55,60)
			\fmfset{curly_len}{2mm}
			\fmftop{t1,t2} \fmfbottom{b1,b2}
			\fmf{quark,label=$\hat d_t$,label.side=left,label.dist=4,tension=3}{t1,v1}
			\fmf{quark,label=$\hat d_s$,label.side=left,label.dist=4,tension=3}{v1,t2}
			\fmf{quark,label=$\hat u_r$,label.side=left,label.dist=3,tension=3}{b1,v2}
			\fmf{quark,label=$\hat u_p$,label.side=left,label.dist=3,tension=3}{v3,b2}
			\fmf{wboson,label=$\W^+$,label.side=left,label.dist=3,tension=1}{v2,v1,v3}
			\fmffreeze
			\fmf{quark,label=$d_u$,label.side=right}{v2,v3}
			\fmfv{decoration.shape=square,decoration.size=1.5mm}{v1}
		\end{fmfgraph*}
	\end{gathered}
\qquad\quad
	\begin{gathered}
		\begin{fmfgraph*}(55,60)
			\fmfset{curly_len}{2mm}
			\fmftop{t1,t2} \fmfbottom{b1,b2}
			\fmf{quark,label=$\hat d_t$,label.side=right,label.dist=3,tension=3}{t1,v2}
			\fmf{phantom,label=$\hat d_s$,label.side=right,label.dist=3,tension=3}{v3,t2}
			\fmf{quark,label=$\hat u_r$,label.side=right,label.dist=4,tension=3}{b1,v1}
			\fmf{phantom,tension=3}{v1,b2}
			\fmf{wboson,label=$\W^+$,label.side=left,label.dist=3,tension=1}{v1,v2}
			\fmf{photon,label=$\A$,label.side=right,label.dist=3,tension=1}{v3,v1}
			\fmffreeze
			\fmf{quark,label=$u_u$,label.side=left}{v2,v3}
			\fmf{quark,right=0.25}{v1,t2}
			\fmf{quark,label=$\hat u_p$,label.side=left}{v3,b2}
			\fmfv{decoration.shape=square,decoration.size=1.5mm}{v1}
		\end{fmfgraph*}
	\end{gathered}
\qquad\quad
	\begin{gathered}
		\begin{fmfgraph*}(55,60)
			\fmfset{curly_len}{2mm}
			\fmftop{t1,t2} \fmfbottom{b1,b2}
			\fmf{quark,label=$\hat d_t$,label.side=right,label.dist=3,tension=3}{t1,v2}
			\fmf{phantom,label=$\hat d_s$,label.side=right,label.dist=3,tension=3}{v3,t2}
			\fmf{quark,label=$\hat u_r$,label.side=right,label.dist=4,tension=3}{b1,v1}
			\fmf{phantom,tension=3}{v1,b2}
			\fmf{wboson,label=$\W^+$,label.side=left,label.dist=3,tension=1}{v1,v2}
			\fmf{photon,label=$\Z$,label.side=right,label.dist=3,tension=1}{v3,v1}
			\fmffreeze
			\fmf{quark,label=$u_u$,label.side=left}{v2,v3}
			\fmf{quark,right=0.25}{v1,t2}
			\fmf{quark,label=$\hat u_p$,label.side=left}{v3,b2}
			\fmfv{decoration.shape=square,decoration.size=1.5mm}{v1}
		\end{fmfgraph*}
	\end{gathered}
\qquad\quad
	\begin{gathered}
		\begin{fmfgraph*}(55,60)
			\fmfset{curly_len}{2mm}
			\fmftop{t1,t2} \fmfbottom{b1,b2}
			\fmf{quark,label=$\hat d_t$,label.side=right,label.dist=3,tension=3}{t1,v2}
			\fmf{phantom,label=$\hat d_s$,label.side=right,label.dist=3,tension=3}{v3,t2}
			\fmf{quark,label=$\hat u_r$,label.side=right,label.dist=4,tension=3}{b1,v1}
			\fmf{phantom,tension=3}{v1,b2}
			\fmf{photon,label=$\A$,label.side=left,label.dist=3,tension=1}{v1,v2}
			\fmf{wboson,label=$\W^+$,label.side=left,label.dist=1,tension=1}{v1,v3}
			\fmffreeze
			\fmf{quark,label=$d_u$,label.side=left}{v2,v3}
			\fmf{quark,right=0.25}{v1,t2}
			\fmf{quark,label=$\hat u_p$,label.side=left}{v3,b2}
			\fmfv{decoration.shape=square,decoration.size=1.5mm}{v1}
		\end{fmfgraph*}
	\end{gathered}
\qquad\quad
	\begin{gathered}
		\begin{fmfgraph*}(55,60)
			\fmfset{curly_len}{2mm}
			\fmftop{t1,t2} \fmfbottom{b1,b2}
			\fmf{quark,label=$\hat d_t$,label.side=right,label.dist=3,tension=3}{t1,v2}
			\fmf{phantom,label=$\hat d_s$,label.side=right,label.dist=3,tension=3}{v3,t2}
			\fmf{quark,label=$\hat u_r$,label.side=right,label.dist=4,tension=3}{b1,v1}
			\fmf{phantom,tension=3}{v1,b2}
			\fmf{photon,label=$\Z$,label.side=left,label.dist=3,tension=1}{v1,v2}
			\fmf{wboson,label=$\W^+$,label.side=left,label.dist=1,tension=1}{v1,v3}
			\fmffreeze
			\fmf{quark,label=$d_u$,label.side=left}{v2,v3}
			\fmf{quark,right=0.25}{v1,t2}
			\fmf{quark,label=$\hat u_p$,label.side=left}{v3,b2}
			\fmfv{decoration.shape=square,decoration.size=1.5mm}{v1}
		\end{fmfgraph*}
	\end{gathered}
\nnw
	& \begin{gathered}
		\begin{fmfgraph*}(55,60)
			\fmfset{curly_len}{2mm}
			\fmftop{t1,t2} \fmfbottom{b1,b2}
			\fmf{quark,label=$\hat d_t$,label.side=left,label.dist=4,tension=3}{t1,v1}
			\fmf{phantom,tension=3}{v1,t2}
			\fmf{quark,label=$\hat u_r$,label.side=right,label.dist=3,tension=3}{b1,v2}
			\fmf{phantom,label=$\hat u_p$,label.side=left,label.dist=3,tension=3}{v3,b2}
			\fmf{wboson,label=$\W^+$,label.side=left,label.dist=3,tension=1}{v2,v1}
			\fmf{photon,label=$\A$,label.side=left,label.dist=3,tension=1}{v3,v1}
			\fmffreeze
			\fmf{quark,label=$d_u$,label.side=right}{v2,v3}
			\fmf{quark,left=0.25}{v1,b2}
			\fmf{quark,label=$\hat d_s$,label.side=right}{v3,t2}
			\fmfv{decoration.shape=square,decoration.size=1.5mm}{v1}
		\end{fmfgraph*}
	\end{gathered}
\qquad\quad
	\begin{gathered}
		\begin{fmfgraph*}(55,60)
			\fmfset{curly_len}{2mm}
			\fmftop{t1,t2} \fmfbottom{b1,b2}
			\fmf{quark,label=$\hat d_t$,label.side=left,label.dist=4,tension=3}{t1,v1}
			\fmf{phantom,tension=3}{v1,t2}
			\fmf{quark,label=$\hat u_r$,label.side=right,label.dist=3,tension=3}{b1,v2}
			\fmf{phantom,label=$\hat u_p$,label.side=left,label.dist=3,tension=3}{v3,b2}
			\fmf{wboson,label=$\W^+$,label.side=left,label.dist=3,tension=1}{v2,v1}
			\fmf{photon,label=$\Z$,label.side=left,label.dist=3,tension=1}{v3,v1}
			\fmffreeze
			\fmf{quark,label=$d_u$,label.side=right}{v2,v3}
			\fmf{quark,left=0.25}{v1,b2}
			\fmf{quark,label=$\hat d_s$,label.side=right}{v3,t2}
			\fmfv{decoration.shape=square,decoration.size=1.5mm}{v1}
		\end{fmfgraph*}
	\end{gathered}
\qquad\quad
	\begin{gathered}
		\begin{fmfgraph*}(55,60)
			\fmfset{curly_len}{2mm}
			\fmftop{t1,t2} \fmfbottom{b1,b2}
			\fmf{quark,label=$\hat d_t$,label.side=left,label.dist=4,tension=3}{t1,v1}
			\fmf{phantom,tension=3}{v1,t2}
			\fmf{quark,label=$\hat u_r$,label.side=right,label.dist=3,tension=3}{b1,v2}
			\fmf{phantom,label=$\hat u_p$,label.side=left,label.dist=3,tension=3}{v3,b2}
			\fmf{photon,label=$\A$,label.side=left,label.dist=3,tension=1}{v2,v1}
			\fmf{wboson,label=$\W^+$,label.side=left,label.dist=3,tension=1}{v3,v1}
			\fmffreeze
			\fmf{quark,label=$u_u$,label.side=right}{v2,v3}
			\fmf{quark,left=0.25}{v1,b2}
			\fmf{quark,label=$\hat d_s$,label.side=right}{v3,t2}
			\fmfv{decoration.shape=square,decoration.size=1.5mm}{v1}
		\end{fmfgraph*}
	\end{gathered}
\qquad\quad
	\begin{gathered}
		\begin{fmfgraph*}(55,60)
			\fmfset{curly_len}{2mm}
			\fmftop{t1,t2} \fmfbottom{b1,b2}
			\fmf{quark,label=$\hat d_t$,label.side=left,label.dist=4,tension=3}{t1,v1}
			\fmf{phantom,tension=3}{v1,t2}
			\fmf{quark,label=$\hat u_r$,label.side=right,label.dist=3,tension=3}{b1,v2}
			\fmf{phantom,label=$\hat u_p$,label.side=left,label.dist=3,tension=3}{v3,b2}
			\fmf{photon,label=$\Z$,label.side=left,label.dist=3,tension=1}{v2,v1}
			\fmf{wboson,label=$\W^+$,label.side=left,label.dist=3,tension=1}{v3,v1}
			\fmffreeze
			\fmf{quark,label=$u_u$,label.side=right}{v2,v3}
			\fmf{quark,left=0.25}{v1,b2}
			\fmf{quark,label=$\hat d_s$,label.side=right}{v3,t2}
			\fmfv{decoration.shape=square,decoration.size=1.5mm}{v1}
		\end{fmfgraph*}
	\end{gathered}
\qquad\quad
	\begin{gathered}
		\begin{fmfgraph*}(55,60)
			\fmfset{curly_len}{2mm}
			\fmftop{t1,t2} \fmfbottom{b1,b2}
			\fmf{quark,label=$\hat d_t$,label.side=left,label.dist=3,tension=6}{t1,v2}
			\fmf{quark,label=$\hat d_s$,label.side=right,label.dist=3,tension=2}{v1,t2}
			\fmf{quark,label=$\hat u_r$,label.side=right,label.dist=3,tension=6}{b1,v3}
			\fmf{quark,label=$\hat u_p$,label.side=left,label.dist=3,tension=2}{v1,b2}
			\fmf{quark,label=$u_u$,label.side=left,label.dist=3,tension=3}{v2,v1}
			\fmf{quark,label=$d_v$,label.side=right,label.dist=3,tension=3}{v3,v1}
			\fmffreeze
			\fmf{wboson,label=$\W^+$,label.side=left,left=0.25}{v3,v2}
			\fmfv{decoration.shape=square,decoration.size=1.5mm}{v1}
		\end{fmfgraph*}
	\end{gathered}
\nnw
	& \begin{gathered}
		\begin{fmfgraph*}(55,60)
			\fmfset{curly_len}{2mm}
			\fmftop{t1,t2} \fmfbottom{b1,b2}
			\fmf{quark,label=$\hat d_t$,label.side=left,label.dist=3,tension=6}{t1,v2}
			\fmf{quark,label=$\hat d_s$,label.side=right,label.dist=3,tension=2}{v1,t2}
			\fmf{quark,label=$\hat u_r$,label.side=right,label.dist=3,tension=6}{b1,v3}
			\fmf{quark,label=$\hat u_p$,label.side=left,label.dist=3,tension=2}{v1,b2}
			\fmf{quark,label=$d_u$,label.side=left,label.dist=3,tension=3}{v2,v1}
			\fmf{quark,label=$u_v$,label.side=right,label.dist=3,tension=3}{v3,v1}
			\fmffreeze
			\fmf{photon,label=$\Z$,label.side=right,right=0.25}{v2,v3}
			\fmfv{decoration.shape=square,decoration.size=1.5mm}{v1}
		\end{fmfgraph*}
	\end{gathered}
\qquad\quad
	\begin{gathered}
		\begin{fmfgraph*}(55,60)
			\fmfset{curly_len}{2mm}
			\fmftop{t1,t2} \fmfbottom{b1,b2}
			\fmf{quark,label=$\hat d_t$,label.side=right,label.dist=3,tension=2}{t1,v1}
			\fmf{quark,label=$\hat d_s$,label.side=left,label.dist=3,tension=6}{v2,t2}
			\fmf{quark,label=$\hat u_r$,label.side=left,label.dist=3,tension=2}{b1,v1}
			\fmf{quark,label=$\hat u_p$,label.side=right,label.dist=3,tension=6}{v3,b2}
			\fmf{quark,label=$u_u$,label.side=left,label.dist=3,tension=3}{v1,v2}
			\fmf{quark,label=$d_v$,label.side=right,label.dist=3,tension=3}{v1,v3}
			\fmffreeze
			\fmf{wboson,label=$\W^+$,label.side=left,left=0.25}{v2,v3}
			\fmfv{decoration.shape=square,decoration.size=1.5mm}{v1}
		\end{fmfgraph*}
	\end{gathered}
\qquad\quad
	\begin{gathered}
		\begin{fmfgraph*}(55,60)
			\fmfset{curly_len}{2mm}
			\fmftop{t1,t2} \fmfbottom{b1,b2}
			\fmf{quark,label=$\hat d_t$,label.side=right,label.dist=3,tension=2}{t1,v1}
			\fmf{quark,label=$\hat d_s$,label.side=left,label.dist=3,tension=6}{v2,t2}
			\fmf{quark,label=$\hat u_r$,label.side=left,label.dist=3,tension=2}{b1,v1}
			\fmf{quark,label=$\hat u_p$,label.side=right,label.dist=3,tension=6}{v3,b2}
			\fmf{quark,label=$d_u$,label.side=left,label.dist=3,tension=3}{v1,v2}
			\fmf{quark,label=$u_v$,label.side=right,label.dist=3,tension=3}{v1,v3}
			\fmffreeze
			\fmf{photon,label=$\Z$,label.side=left,left=0.25}{v2,v3}
			\fmfv{decoration.shape=square,decoration.size=1.5mm}{v1}
		\end{fmfgraph*}
	\end{gathered}
\qquad\quad
	\begin{gathered}
		\begin{fmfgraph*}(55,60)
			\fmfset{curly_len}{2mm}
			\fmftop{t1,t2} \fmfbottom{b1,b2}
			\fmf{quark,label=$\hat d_t$,label.side=right,label.dist=3,tension=6}{t1,v2}
			\fmf{quark,label=$\hat d_s$,label.side=right,label.dist=3,tension=6}{v3,t2}
			\fmf{quark,label=$\hat u_r$,label.side=right,label.dist=3,tension=2}{b1,v1}
			\fmf{quark,label=$\hat u_p$,label.side=right,label.dist=3,tension=2}{v1,b2}
			\fmf{quark,label=$u_u$,label.side=right,label.dist=3,tension=3}{v2,v1}
			\fmf{quark,label=$u_v$,label.side=right,label.dist=3,tension=3}{v1,v3}
			\fmffreeze
			\fmf{wboson,label=$\W^+$,label.side=right,right=0.25}{v3,v2}
			\fmfv{decoration.shape=square,decoration.size=1.5mm}{v1}
		\end{fmfgraph*}
	\end{gathered}
\qquad\quad
	\begin{gathered}
		\begin{fmfgraph*}(55,60)
			\fmfset{curly_len}{2mm}
			\fmftop{t1,t2} \fmfbottom{b1,b2}
			\fmf{quark,label=$\hat d_t$,label.side=right,label.dist=3,tension=6}{t1,v2}
			\fmf{quark,label=$\hat d_s$,label.side=right,label.dist=3,tension=6}{v3,t2}
			\fmf{quark,label=$\hat u_r$,label.side=right,label.dist=3,tension=2}{b1,v1}
			\fmf{quark,label=$\hat u_p$,label.side=right,label.dist=3,tension=2}{v1,b2}
			\fmf{quark,label=$t$,label.side=right,label.dist=3,tension=3}{v2,v1}
			\fmf{quark,label=$t$,label.side=right,label.dist=3,tension=3}{v1,v3}
			\fmffreeze
			\fmf{scalar,label=$G^+$,label.side=right,right=0.25}{v3,v2}
			\fmfv{decoration.shape=square,decoration.size=1.5mm}{v1}
		\end{fmfgraph*}
	\end{gathered}
\nnw
	& \begin{gathered}
		\begin{fmfgraph*}(55,60)
			\fmfset{curly_len}{2mm}
			\fmftop{t1,t2} \fmfbottom{b1,b2}
			\fmf{quark,label=$\hat d_t$,label.side=right,label.dist=3,tension=6}{t1,v2}
			\fmf{quark,label=$\hat d_s$,label.side=right,label.dist=3,tension=6}{v3,t2}
			\fmf{quark,label=$\hat u_r$,label.side=right,label.dist=3,tension=2}{b1,v1}
			\fmf{quark,label=$\hat u_p$,label.side=right,label.dist=3,tension=2}{v1,b2}
			\fmf{quark,label=$d_u$,label.side=right,label.dist=3,tension=3}{v2,v1}
			\fmf{quark,label=$d_v$,label.side=right,label.dist=3,tension=3}{v1,v3}
			\fmffreeze
			\fmf{photon,label=$\Z$,label.side=left,left=0.25}{v2,v3}
			\fmfv{decoration.shape=square,decoration.size=1.5mm}{v1}
		\end{fmfgraph*}
	\end{gathered}
\qquad\quad
	\begin{gathered}
		\begin{fmfgraph*}(55,60)
			\fmfset{curly_len}{2mm}
			\fmftop{t1,t2} \fmfbottom{b1,b2}
			\fmf{quark,label=$\hat d_t$,label.side=left,label.dist=3,tension=2}{t1,v1}
			\fmf{quark,label=$\hat d_s$,label.side=left,label.dist=3,tension=2}{v1,t2}
			\fmf{quark,label=$\hat u_r$,label.side=left,label.dist=3,tension=6}{b1,v2}
			\fmf{quark,label=$\hat u_p$,label.side=left,label.dist=3,tension=6}{v3,b2}
			\fmf{quark,label=$d_u$,label.side=left,label.dist=3,tension=3}{v2,v1}
			\fmf{quark,label=$d_v$,label.side=left,label.dist=3,tension=3}{v1,v3}
			\fmffreeze
			\fmf{wboson,label=$\W^+$,label.side=right,right=0.25}{v2,v3}
			\fmfv{decoration.shape=square,decoration.size=1.5mm}{v1}
		\end{fmfgraph*}
	\end{gathered}
\qquad\quad
	\begin{gathered}
		\begin{fmfgraph*}(55,60)
			\fmfset{curly_len}{2mm}
			\fmftop{t1,t2} \fmfbottom{b1,b2}
			\fmf{quark,label=$\hat d_t$,label.side=left,label.dist=3,tension=2}{t1,v1}
			\fmf{quark,label=$\hat d_s$,label.side=left,label.dist=3,tension=2}{v1,t2}
			\fmf{quark,label=$\hat u_r$,label.side=left,label.dist=3,tension=6}{b1,v2}
			\fmf{quark,label=$\hat u_p$,label.side=left,label.dist=3,tension=6}{v3,b2}
			\fmf{quark,label=$u_u$,label.side=left,label.dist=3,tension=3}{v2,v1}
			\fmf{quark,label=$u_v$,label.side=left,label.dist=3,tension=3}{v1,v3}
			\fmffreeze
			\fmf{photon,label=$\Z$,label.side=right,right=0.25}{v2,v3}
			\fmfv{decoration.shape=square,decoration.size=1.5mm}{v1}
		\end{fmfgraph*}
	\end{gathered}
\qquad\quad
	\begin{gathered}
		\begin{fmfgraph*}(55,60)
			\fmfset{curly_len}{2mm}
			\fmftop{t1,t2} \fmfbottom{b1,b2}
			\fmf{quark,label=$\hat d_t$,label.side=left,label.dist=3,tension=2}{t1,v1}
			\fmf{quark,label=$\hat d_s$,label.side=right,label.dist=3,tension=6}{v3,t2}
			\fmf{quark,label=$\hat u_r$,label.side=right,label.dist=3,tension=6}{b1,v2}
			\fmf{quark,label=$\hat u_p$,label.side=left,label.dist=3,tension=2}{v1,b2}
			\fmf{quark,label=$d_u$,label.side=left,label.dist=3,tension=3}{v1,v3}
			\fmf{quark,label=$u_v$,label.side=left,label.dist=3,tension=3}{v2,v1}
			\fmffreeze
			\fmf{photon,right=0.5}{v2,v3}
			\fmf{phantom,label=$\Z$,label.side=right,left=0.4}{v2,b2}
			\fmfv{decoration.shape=square,decoration.size=1.5mm}{v1}
		\end{fmfgraph*}
	\end{gathered}
\qquad\quad
	\begin{gathered}
		\begin{fmfgraph*}(55,60)
			\fmfset{curly_len}{2mm}
			\fmftop{t1,t2} \fmfbottom{b1,b2}
			\fmf{quark,label=$\hat d_t$,label.side=left,label.dist=3,tension=6}{t1,v2}
			\fmf{quark,label=$\hat d_s$,label.side=left,label.dist=3,tension=2}{v1,t2}
			\fmf{quark,label=$\hat u_r$,label.side=right,label.dist=3,tension=2}{b1,v1}
			\fmf{quark,label=$\hat u_p$,label.side=left,label.dist=3,tension=6}{v3,b2}
			\fmf{quark,label=$d_u$,label.side=left,label.dist=3,tension=3}{v2,v1}
			\fmf{quark,label=$u_v$,label.side=left,label.dist=3,tension=3}{v1,v3}
			\fmffreeze
			\fmf{photon,right=0.5}{v2,v3}
			\fmf{phantom,label=$\Z$,label.side=right,left=0.5}{v2,b1}
			\fmfv{decoration.shape=square,decoration.size=1.5mm}{v1}
		\end{fmfgraph*}
	\end{gathered}
\end{align*}


\subsubsection[$\bar dd \bar dd$]{\boldmath$\bar dd \bar dd$}

\begin{align*}
	& \begin{gathered}
		\begin{fmfgraph*}(55,60)
			\fmfset{curly_len}{2mm}
			\fmftop{t1,t2} \fmfbottom{b1,b2}
			\fmf{quark,label=$\hat d_t$,label.side=left,tension=4}{t1,v1}
			\fmf{quark,label=$\hat d_s$,label.side=left,tension=4}{v2,t2}
			\fmf{quark,label=$\hat d_r$,label.side=right,tension=4}{b1,v3}
			\fmf{quark,label=$\hat d_p$,label.side=right,tension=4}{v4,b2}
			\fmf{quark,label=$d_u$,label.side=right}{v1,v2}
			\fmf{quark,label=$d_v$,label.side=right}{v3,v4}
			\fmf{photon,label=$\Z$,label.side=left}{v3,v1}
			\fmf{photon,label=$\Z$,label.side=left}{v2,v4}
		\end{fmfgraph*}
	\end{gathered} \quad \times 2
\quad\quad
	\begin{gathered}
		\begin{fmfgraph*}(55,60)
			\fmfset{curly_len}{2mm}
			\fmftop{t1,t2} \fmfbottom{b1,b2}
			\fmf{quark,label=$\hat d_t$,label.side=left,tension=4}{t1,v1}
			\fmf{quark,label=$\hat d_s$,label.side=left,tension=4}{v2,t2}
			\fmf{quark,label=$\hat d_r$,label.side=right,tension=4}{b1,v3}
			\fmf{quark,label=$\hat d_p$,label.side=right,tension=4}{v4,b2}
			\fmf{quark,label=$d_u$,label.side=right}{v1,v2}
			\fmf{quark,label=$d_v$,label.side=right}{v3,v4}
			\fmf{photon,label=$\Z$,label.side=left}{v3,v1}
			\fmf{photon,label=$\A$,label.side=left}{v2,v4}
		\end{fmfgraph*}
	\end{gathered} \quad \times 2
\quad\quad
	\begin{gathered}
		\begin{fmfgraph*}(55,60)
			\fmfset{curly_len}{2mm}
			\fmftop{t1,t2} \fmfbottom{b1,b2}
			\fmf{quark,label=$\hat d_t$,label.side=left,tension=4}{t1,v1}
			\fmf{quark,label=$\hat d_s$,label.side=left,tension=4}{v2,t2}
			\fmf{quark,label=$\hat d_r$,label.side=right,tension=4}{b1,v3}
			\fmf{quark,label=$\hat d_p$,label.side=right,tension=4}{v4,b2}
			\fmf{quark,label=$d_u$,label.side=right}{v1,v2}
			\fmf{quark,label=$d_v$,label.side=right}{v3,v4}
			\fmf{photon,label=$\A$,label.side=left}{v3,v1}
			\fmf{photon,label=$\Z$,label.side=left}{v2,v4}
		\end{fmfgraph*}
	\end{gathered} \quad \times 2
\quad\quad
	\begin{gathered}
		\begin{fmfgraph*}(55,60)
			\fmfset{curly_len}{2mm}
			\fmftop{t1,t2} \fmfbottom{b1,b2}
			\fmf{quark,label=$\hat d_t$,label.side=left,tension=4}{t1,v1}
			\fmf{quark,label=$\hat d_s$,label.side=left,tension=4}{v2,t2}
			\fmf{quark,label=$\hat d_r$,label.side=right,tension=4}{b1,v3}
			\fmf{quark,label=$\hat d_p$,label.side=right,tension=4}{v4,b2}
			\fmf{quark,label=$d_u$,label.side=right}{v1,v2}
			\fmf{quark,label=$d_v$,label.side=right}{v3,v4}
			\fmf{photon,label=$\Z$,label.side=left}{v3,v1}
			\fmf{gluon,label=$\G$,label.side=left}{v2,v4}
		\end{fmfgraph*}
	\end{gathered} \quad \times 2
\nnw
	& \begin{gathered}
		\begin{fmfgraph*}(55,60)
			\fmfset{curly_len}{2mm}
			\fmftop{t1,t2} \fmfbottom{b1,b2}
			\fmf{quark,label=$\hat d_t$,label.side=left,tension=4}{t1,v1}
			\fmf{quark,label=$\hat d_s$,label.side=left,tension=4}{v2,t2}
			\fmf{quark,label=$\hat d_r$,label.side=right,tension=4}{b1,v3}
			\fmf{quark,label=$\hat d_p$,label.side=right,tension=4}{v4,b2}
			\fmf{quark,label=$d_u$,label.side=right}{v1,v2}
			\fmf{quark,label=$d_v$,label.side=right}{v3,v4}
			\fmf{gluon,label=$\G$,label.side=left}{v3,v1}
			\fmf{photon,label=$\Z$,label.side=left}{v2,v4}
		\end{fmfgraph*}
	\end{gathered} \quad \times 2
\quad\quad
	\begin{gathered}
		\begin{fmfgraph*}(55,60)
			\fmfset{curly_len}{2mm}
			\fmftop{t1,t2} \fmfbottom{b1,b2}
			\fmf{phantom,label=$\hat d_t$,label.side=left,tension=4}{t1,v1}
			\fmf{phantom,label=$\hat d_s$,label.side=left,tension=4}{v2,t2}
			\fmf{quark,label=$\hat d_r$,label.side=right,tension=4}{b1,v3}
			\fmf{quark,label=$\hat d_p$,label.side=right,tension=4}{v4,b2}
			\fmf{quark,label=$d_u$,label.side=left}{v2,v1}
			\fmf{quark,label=$d_v$,label.side=right}{v3,v4}
			\fmf{photon,label=$\Z$,label.side=left}{v3,v1}
			\fmf{photon,label=$\Z$,label.side=left}{v2,v4}
			\fmffreeze
			\fmf{quark}{t1,v2}
			\fmf{quark}{v1,t2}
		\end{fmfgraph*}
	\end{gathered} \quad \times 2
\quad\quad
	\begin{gathered}
		\begin{fmfgraph*}(55,60)
			\fmfset{curly_len}{2mm}
			\fmftop{t1,t2} \fmfbottom{b1,b2}
			\fmf{phantom,label=$\hat d_t$,label.side=left,tension=4}{t1,v1}
			\fmf{phantom,label=$\hat d_s$,label.side=left,tension=4}{v2,t2}
			\fmf{quark,label=$\hat d_r$,label.side=right,tension=4}{b1,v3}
			\fmf{quark,label=$\hat d_p$,label.side=right,tension=4}{v4,b2}
			\fmf{quark,label=$d_u$,label.side=left}{v2,v1}
			\fmf{quark,label=$d_v$,label.side=right}{v3,v4}
			\fmf{photon,label=$\Z$,label.side=left}{v3,v1}
			\fmf{photon,label=$\A$,label.side=left}{v2,v4}
			\fmffreeze
			\fmf{quark}{t1,v2}
			\fmf{quark}{v1,t2}
		\end{fmfgraph*}
	\end{gathered} \quad \times 2
\quad\quad
	\begin{gathered}
		\begin{fmfgraph*}(55,60)
			\fmfset{curly_len}{2mm}
			\fmftop{t1,t2} \fmfbottom{b1,b2}
			\fmf{phantom,label=$\hat d_t$,label.side=left,tension=4}{t1,v1}
			\fmf{phantom,label=$\hat d_s$,label.side=left,tension=4}{v2,t2}
			\fmf{quark,label=$\hat d_r$,label.side=right,tension=4}{b1,v3}
			\fmf{quark,label=$\hat d_p$,label.side=right,tension=4}{v4,b2}
			\fmf{quark,label=$d_u$,label.side=left}{v2,v1}
			\fmf{quark,label=$d_v$,label.side=right}{v3,v4}
			\fmf{photon,label=$\A$,label.side=left}{v3,v1}
			\fmf{photon,label=$\Z$,label.side=left}{v2,v4}
			\fmffreeze
			\fmf{quark}{t1,v2}
			\fmf{quark}{v1,t2}
		\end{fmfgraph*}
	\end{gathered} \quad \times 2
\nnw
	& \begin{gathered}
		\begin{fmfgraph*}(55,60)
			\fmfset{curly_len}{2mm}
			\fmftop{t1,t2} \fmfbottom{b1,b2}
			\fmf{phantom,label=$\hat d_t$,label.side=left,tension=4}{t1,v1}
			\fmf{phantom,label=$\hat d_s$,label.side=left,tension=4}{v2,t2}
			\fmf{quark,label=$\hat d_r$,label.side=right,tension=4}{b1,v3}
			\fmf{quark,label=$\hat d_p$,label.side=right,tension=4}{v4,b2}
			\fmf{quark,label=$d_u$,label.side=left}{v2,v1}
			\fmf{quark,label=$d_v$,label.side=right}{v3,v4}
			\fmf{photon,label=$\Z$,label.side=left}{v3,v1}
			\fmf{gluon,label=$\G$,label.side=left}{v2,v4}
			\fmffreeze
			\fmf{quark}{t1,v2}
			\fmf{quark}{v1,t2}
		\end{fmfgraph*}
	\end{gathered} \quad \times 2
\quad\quad
	\begin{gathered}
		\begin{fmfgraph*}(55,60)
			\fmfset{curly_len}{2mm}
			\fmftop{t1,t2} \fmfbottom{b1,b2}
			\fmf{phantom,label=$\hat d_t$,label.side=left,tension=4}{t1,v1}
			\fmf{phantom,label=$\hat d_s$,label.side=left,tension=4}{v2,t2}
			\fmf{quark,label=$\hat d_r$,label.side=right,tension=4}{b1,v3}
			\fmf{quark,label=$\hat d_p$,label.side=right,tension=4}{v4,b2}
			\fmf{quark,label=$d_u$,label.side=left}{v2,v1}
			\fmf{quark,label=$d_v$,label.side=right}{v3,v4}
			\fmf{gluon,label=$\G$,label.side=left}{v3,v1}
			\fmf{photon,label=$\Z$,label.side=left}{v2,v4}
			\fmffreeze
			\fmf{quark}{t1,v2}
			\fmf{quark}{v1,t2}
		\end{fmfgraph*}
	\end{gathered} \quad \times 2
\qquad\quad
	\begin{gathered}
		\begin{fmfgraph*}(55,60)
			\fmfset{curly_len}{2mm}
			\fmftop{t1,t2} \fmfbottom{b1,b2}
			\fmf{phantom,label=$\hat d_t$,label.side=left,tension=4}{t1,v1}
			\fmf{phantom,label=$\hat d_s$,label.side=left,tension=4}{v2,t2}
			\fmf{quark,label=$\hat d_r$,label.side=right,tension=4}{b1,v3}
			\fmf{quark,label=$\hat d_p$,label.side=right,tension=4}{v4,b2}
			\fmf{quark,label=$u_u$,label.side=left}{v2,v1}
			\fmf{quark,label=$u_v$,label.side=right}{v3,v4}
			\fmf{wboson,label=$\W^+$,label.side=right}{v1,v3}
			\fmf{wboson,label=$\W^+$,label.side=right}{v4,v2}
			\fmffreeze
			\fmf{quark}{t1,v2}
			\fmf{quark}{v1,t2}
		\end{fmfgraph*}
	\end{gathered} \quad\,\, \times 2
\qquad\quad
	\begin{gathered}
		\begin{fmfgraph*}(55,60)
			\fmfset{curly_len}{2mm}
			\fmftop{t1,t2} \fmfbottom{b1,b2}
			\fmf{phantom,label=$\hat d_t$,label.side=left,tension=4}{t1,v1}
			\fmf{phantom,label=$\hat d_s$,label.side=left,tension=4}{v2,t2}
			\fmf{quark,label=$\hat d_r$,label.side=right,tension=4}{b1,v3}
			\fmf{quark,label=$\hat d_p$,label.side=right,tension=4}{v4,b2}
			\fmf{quark,label=$u_u$,label.side=left}{v2,v1}
			\fmf{quark,label=$u_v$,label.side=right}{v3,v4}
			\fmf{scalar,label=$G^+$,label.side=right}{v1,v3}
			\fmf{wboson,label=$\W^+$,label.side=right}{v4,v2}
			\fmffreeze
			\fmf{quark}{t1,v2}
			\fmf{quark}{v1,t2}
		\end{fmfgraph*}
	\end{gathered} \quad\,\, \times 2
\nnw
	& \begin{gathered}
		\begin{fmfgraph*}(55,60)
			\fmfset{curly_len}{2mm}
			\fmftop{t1,t2} \fmfbottom{b1,b2}
			\fmf{phantom,label=$\hat d_t$,label.side=left,tension=4}{t1,v1}
			\fmf{phantom,label=$\hat d_s$,label.side=left,tension=4}{v2,t2}
			\fmf{quark,label=$\hat d_r$,label.side=right,tension=4}{b1,v3}
			\fmf{quark,label=$\hat d_p$,label.side=right,tension=4}{v4,b2}
			\fmf{quark,label=$u_u$,label.side=left}{v2,v1}
			\fmf{quark,label=$u_v$,label.side=right}{v3,v4}
			\fmf{wboson,label=$\W^+$,label.side=right}{v1,v3}
			\fmf{scalar,label=$G^+$,label.side=right}{v4,v2}
			\fmffreeze
			\fmf{quark}{t1,v2}
			\fmf{quark}{v1,t2}
		\end{fmfgraph*}
	\end{gathered} \quad \times 2
\quad\quad
	\begin{gathered}
		\begin{fmfgraph*}(55,60)
			\fmfset{curly_len}{2mm}
			\fmftop{t1,t2} \fmfbottom{b1,b2}
			\fmf{phantom,label=$\hat d_t$,label.side=left,tension=4}{t1,v1}
			\fmf{phantom,label=$\hat d_s$,label.side=left,tension=4}{v2,t2}
			\fmf{quark,label=$\hat d_r$,label.side=right,tension=4}{b1,v3}
			\fmf{quark,label=$\hat d_p$,label.side=right,tension=4}{v4,b2}
			\fmf{quark,label=$t$,label.side=left}{v2,v1}
			\fmf{quark,label=$t$,label.side=right}{v3,v4}
			\fmf{scalar,label=$G^+$,label.side=right}{v1,v3}
			\fmf{scalar,label=$G^+$,label.side=right}{v4,v2}
			\fmffreeze
			\fmf{quark}{t1,v2}
			\fmf{quark}{v1,t2}
		\end{fmfgraph*}
	\end{gathered} \quad \times 2
\quad\quad
	\begin{gathered}
		\begin{fmfgraph*}(55,60)
			\fmfset{curly_len}{2mm}
			\fmftop{t1,t2} \fmfbottom{b1,b2}
			\fmf{quark,label=$\hat d_t$,label.side=right,label.dist=3,tension=3}{t1,v2}
			\fmf{quark,label=$\hat d_s$,label.side=right,label.dist=3,tension=3}{v3,t2}
			\fmf{quark,label=$\hat d_r$,label.side=right,label.dist=4,tension=3}{b1,v1}
			\fmf{quark,label=$\hat d_p$,label.side=right,label.dist=4,tension=3}{v1,b2}
			\fmf{wboson,label=$\W^+$,label.side=left,label.dist=3,tension=1}{v3,v1,v2}
			\fmffreeze
			\fmf{quark,label=$u_u$,label.side=left}{v2,v3}
			\fmfv{decoration.shape=square,decoration.size=1.5mm}{v1}
		\end{fmfgraph*}
	\end{gathered} \quad \times 4
\quad\quad
	\begin{gathered}
		\begin{fmfgraph*}(55,60)
			\fmfset{curly_len}{2mm}
			\fmftop{t1,t2} \fmfbottom{b1,b2}
			\fmf{quark,label=$\hat d_t$,label.side=right,label.dist=3,tension=3}{t1,v2}
			\fmf{quark,label=$\hat d_s$,label.side=right,label.dist=3,tension=3}{v3,t2}
			\fmf{quark,label=$\hat d_r$,label.side=right,label.dist=4,tension=3}{b1,v1}
			\fmf{quark,label=$\hat d_p$,label.side=right,label.dist=4,tension=3}{v1,b2}
			\fmf{wboson,label=$\W^+$,label.side=left,label.dist=3,tension=1}{v3,v1}
			\fmf{scalar,label=$G^+$,label.side=left,label.dist=3,tension=1}{v1,v2}
			\fmffreeze
			\fmf{quark,label=$t$,label.side=left}{v2,v3}
			\fmfv{decoration.shape=square,decoration.size=1.5mm}{v1}
		\end{fmfgraph*}
	\end{gathered} \quad \times 4
\nnw
	& \begin{gathered}
		\begin{fmfgraph*}(55,60)
			\fmfset{curly_len}{2mm}
			\fmftop{t1,t2} \fmfbottom{b1,b2}
			\fmf{quark,label=$\hat d_t$,label.side=right,label.dist=3,tension=3}{t1,v2}
			\fmf{quark,label=$\hat d_s$,label.side=right,label.dist=3,tension=3}{v3,t2}
			\fmf{quark,label=$\hat d_r$,label.side=right,label.dist=4,tension=3}{b1,v1}
			\fmf{quark,label=$\hat d_p$,label.side=right,label.dist=4,tension=3}{v1,b2}
			\fmf{scalar,label=$G^+$,label.side=left,label.dist=3,tension=1}{v3,v1}
			\fmf{wboson,label=$\W^+$,label.side=left,label.dist=3,tension=1}{v1,v2}
			\fmffreeze
			\fmf{quark,label=$t$,label.side=left}{v2,v3}
			\fmfv{decoration.shape=square,decoration.size=1.5mm}{v1}
		\end{fmfgraph*}
	\end{gathered} \quad \times 4
\quad\quad
	\begin{gathered}
		\begin{fmfgraph*}(55,60)
			\fmfset{curly_len}{2mm}
			\fmftop{t1,t2} \fmfbottom{b1,b2}
			\fmf{quark,label=$\hat d_t$,label.side=right,label.dist=3,tension=3}{t1,v2}
			\fmf{quark,label=$\hat d_s$,label.side=right,label.dist=3,tension=3}{v3,t2}
			\fmf{quark,label=$\hat d_r$,label.side=right,label.dist=4,tension=3}{b1,v1}
			\fmf{quark,label=$\hat d_p$,label.side=right,label.dist=4,tension=3}{v1,b2}
			\fmf{scalar,label=$G^+$,label.side=left,label.dist=3,tension=1}{v3,v1}
			\fmf{scalar,label=$G^+$,label.side=left,label.dist=3,tension=1}{v1,v2}
			\fmffreeze
			\fmf{quark,label=$t$,label.side=left}{v2,v3}
			\fmfv{decoration.shape=square,decoration.size=1.5mm}{v1}
		\end{fmfgraph*}
	\end{gathered} \quad \times 4
\quad\quad
	\begin{gathered}
		\begin{fmfgraph*}(55,60)
			\fmfset{curly_len}{2mm}
			\fmftop{t1,t2} \fmfbottom{b1,b2}
			\fmf{quark,label=$\hat d_t$,label.side=left,label.dist=3,tension=6}{t1,v2}
			\fmf{quark,label=$\hat d_s$,label.side=right,label.dist=3,tension=2}{v1,t2}
			\fmf{quark,label=$\hat d_r$,label.side=right,label.dist=3,tension=6}{b1,v3}
			\fmf{quark,label=$\hat d_p$,label.side=left,label.dist=3,tension=2}{v1,b2}
			\fmf{quark,label=$d_u$,label.side=left,label.dist=3,tension=3}{v2,v1}
			\fmf{quark,label=$d_v$,label.side=right,label.dist=3,tension=3}{v3,v1}
			\fmffreeze
			\fmf{photon,label=$\Z$,label.side=right,right=0.25}{v2,v3}
			\fmfv{decoration.shape=square,decoration.size=1.5mm}{v1}
		\end{fmfgraph*}
	\end{gathered}
\qquad\quad
	\begin{gathered}
		\begin{fmfgraph*}(55,60)
			\fmfset{curly_len}{2mm}
			\fmftop{t1,t2} \fmfbottom{b1,b2}
			\fmf{quark,label=$\hat d_t$,label.side=right,label.dist=3,tension=2}{t1,v1}
			\fmf{quark,label=$\hat d_s$,label.side=left,label.dist=3,tension=6}{v2,t2}
			\fmf{quark,label=$\hat d_r$,label.side=left,label.dist=3,tension=2}{b1,v1}
			\fmf{quark,label=$\hat d_p$,label.side=right,label.dist=3,tension=6}{v3,b2}
			\fmf{quark,label=$d_u$,label.side=left,label.dist=3,tension=3}{v1,v2}
			\fmf{quark,label=$d_v$,label.side=right,label.dist=3,tension=3}{v1,v3}
			\fmffreeze
			\fmf{photon,label=$\Z$,label.side=left,left=0.25}{v2,v3}
			\fmfv{decoration.shape=square,decoration.size=1.5mm}{v1}
		\end{fmfgraph*}
	\end{gathered}
\nnw
	& \begin{gathered}
		\begin{fmfgraph*}(55,60)
			\fmfset{curly_len}{2mm}
			\fmftop{t1,t2} \fmfbottom{b1,b2}
			\fmf{quark,label=$\hat d_t$,label.side=right,label.dist=3,tension=6}{t1,v2}
			\fmf{quark,label=$\hat d_s$,label.side=right,label.dist=3,tension=6}{v3,t2}
			\fmf{quark,label=$\hat d_r$,label.side=right,label.dist=3,tension=2}{b1,v1}
			\fmf{quark,label=$\hat d_p$,label.side=right,label.dist=3,tension=2}{v1,b2}
			\fmf{quark,label=$u_u$,label.side=right,label.dist=3,tension=3}{v2,v1}
			\fmf{quark,label=$u_v$,label.side=right,label.dist=3,tension=3}{v1,v3}
			\fmffreeze
			\fmf{wboson,label=$\W^+$,label.side=right,right=0.25}{v3,v2}
			\fmfv{decoration.shape=square,decoration.size=1.5mm}{v1}
		\end{fmfgraph*}
	\end{gathered} \quad \times 4
\qquad\quad
	\begin{gathered}
		\begin{fmfgraph*}(55,60)
			\fmfset{curly_len}{2mm}
			\fmftop{t1,t2} \fmfbottom{b1,b2}
			\fmf{quark,label=$\hat d_t$,label.side=right,label.dist=3,tension=6}{t1,v2}
			\fmf{quark,label=$\hat d_s$,label.side=right,label.dist=3,tension=6}{v3,t2}
			\fmf{quark,label=$\hat d_r$,label.side=right,label.dist=3,tension=2}{b1,v1}
			\fmf{quark,label=$\hat d_p$,label.side=right,label.dist=3,tension=2}{v1,b2}
			\fmf{quark,label=$t$,label.side=right,label.dist=3,tension=3}{v2,v1}
			\fmf{quark,label=$t$,label.side=right,label.dist=3,tension=3}{v1,v3}
			\fmffreeze
			\fmf{scalar,label=$G^+$,label.side=right,right=0.25}{v3,v2}
			\fmfv{decoration.shape=square,decoration.size=1.5mm}{v1}
		\end{fmfgraph*}
	\end{gathered} \quad \times 4
\qquad\quad
	\begin{gathered}
		\begin{fmfgraph*}(55,60)
			\fmfset{curly_len}{2mm}
			\fmftop{t1,t2} \fmfbottom{b1,b2}
			\fmf{quark,label=$\hat d_t$,label.side=right,label.dist=3,tension=6}{t1,v2}
			\fmf{quark,label=$\hat d_s$,label.side=right,label.dist=3,tension=6}{v3,t2}
			\fmf{quark,label=$\hat d_r$,label.side=right,label.dist=3,tension=2}{b1,v1}
			\fmf{quark,label=$\hat d_p$,label.side=right,label.dist=3,tension=2}{v1,b2}
			\fmf{quark,label=$d_u$,label.side=right,label.dist=3,tension=3}{v2,v1}
			\fmf{quark,label=$d_v$,label.side=right,label.dist=3,tension=3}{v1,v3}
			\fmffreeze
			\fmf{photon,label=$\Z$,label.side=left,left=0.25}{v2,v3}
			\fmfv{decoration.shape=square,decoration.size=1.5mm}{v1}
		\end{fmfgraph*}
	\end{gathered} \quad \times 4
\end{align*}


\subsubsection[$\nu e \bar du$]{\boldmath$\nu e \bar du$}

\begin{align*}
	& \begin{gathered}
		\begin{fmfgraph*}(55,60)
			\fmfset{curly_len}{2mm}
			\fmftop{t1,t2} \fmfbottom{b1,b2}
			\fmf{quark,label=$\hat u_t$,label.side=left,tension=4}{t1,v1}
			\fmf{quark,label=$\hat d_s$,label.side=left,tension=4}{v2,t2}
			\fmf{quark,label=$\hat e_r$,label.side=right,tension=4}{b1,v3}
			\fmf{plain,label=$\hat \nu_p$,label.side=right,tension=4}{v4,b2}
			\fmf{quark,label=$u_u$,label.side=right}{v1,v2}
			\fmf{quark,label=$e_v$,label.side=right}{v3,v4}
			\fmf{photon,label=$\A$,label.side=left}{v3,v1}
			\fmf{wboson,label=$\W^+$,label.side=left}{v2,v4}
		\end{fmfgraph*}
	\end{gathered}
\qquad\quad
	\begin{gathered}
		\begin{fmfgraph*}(55,60)
			\fmfset{curly_len}{2mm}
			\fmftop{t1,t2} \fmfbottom{b1,b2}
			\fmf{quark,label=$\hat u_t$,label.side=left,tension=4}{t1,v1}
			\fmf{quark,label=$\hat d_s$,label.side=left,tension=4}{v2,t2}
			\fmf{quark,label=$\hat e_r$,label.side=right,tension=4}{b1,v3}
			\fmf{plain,label=$\hat \nu_p$,label.side=right,tension=4}{v4,b2}
			\fmf{quark,label=$u_u$,label.side=right}{v1,v2}
			\fmf{quark,label=$e_v$,label.side=right}{v3,v4}
			\fmf{photon,label=$\Z$,label.side=left}{v3,v1}
			\fmf{wboson,label=$\W^+$,label.side=left}{v2,v4}
		\end{fmfgraph*}
	\end{gathered}
\qquad\quad\,\,
	\begin{gathered}
		\begin{fmfgraph*}(55,60)
			\fmfset{curly_len}{2mm}
			\fmftop{t1,t2} \fmfbottom{b1,b2}
			\fmf{quark,label=$\hat u_t$,label.side=left,tension=4}{t1,v1}
			\fmf{quark,label=$\hat d_s$,label.side=left,tension=4}{v2,t2}
			\fmf{quark,label=$\hat e_r$,label.side=right,tension=4}{b1,v3}
			\fmf{plain,label=$\hat \nu_p$,label.side=right,tension=4}{v4,b2}
			\fmf{quark,label=$d_u$,label.side=right}{v1,v2}
			\fmf{plain,label=$\nu_v$,label.side=right}{v3,v4}
			\fmf{wboson,label=$\W^+$,label.side=right}{v1,v3}
			\fmf{photon,label=$\Z$,label.side=left}{v2,v4}
		\end{fmfgraph*}
	\end{gathered}
\qquad\quad
	\begin{gathered}
		\begin{fmfgraph*}(55,60)
			\fmfset{curly_len}{2mm}
			\fmftop{t1,t2} \fmfbottom{b1,b2}
			\fmf{phantom,label=$\hat u_t$,label.side=left,tension=4}{t1,v1}
			\fmf{phantom,label=$\hat d_s$,label.side=left,tension=4}{v2,t2}
			\fmf{quark,label=$\hat e_r$,label.side=right,tension=4}{b1,v3}
			\fmf{plain,label=$\hat \nu_p$,label.side=right,tension=4}{v4,b2}
			\fmf{quark,label=$d_u$,label.side=left}{v2,v1}
			\fmf{quark,label=$e_v$,label.side=right}{v3,v4}
			\fmf{photon,label=$\A$,label.side=left}{v3,v1}
			\fmf{wboson,label=$\W^+$,label.side=left}{v2,v4}
			\fmffreeze
			\fmf{quark}{t1,v2}
			\fmf{quark}{v1,t2}
		\end{fmfgraph*}
	\end{gathered}
\qquad\quad
	\begin{gathered}
		\begin{fmfgraph*}(55,60)
			\fmfset{curly_len}{2mm}
			\fmftop{t1,t2} \fmfbottom{b1,b2}
			\fmf{phantom,label=$\hat u_t$,label.side=left,tension=4}{t1,v1}
			\fmf{phantom,label=$\hat d_s$,label.side=left,tension=4}{v2,t2}
			\fmf{quark,label=$\hat e_r$,label.side=right,tension=4}{b1,v3}
			\fmf{plain,label=$\hat \nu_p$,label.side=right,tension=4}{v4,b2}
			\fmf{quark,label=$d_u$,label.side=left}{v2,v1}
			\fmf{quark,label=$e_v$,label.side=right}{v3,v4}
			\fmf{photon,label=$\Z$,label.side=left}{v3,v1}
			\fmf{wboson,label=$\W^+$,label.side=left}{v2,v4}
			\fmffreeze
			\fmf{quark}{t1,v2}
			\fmf{quark}{v1,t2}
		\end{fmfgraph*}
	\end{gathered}
\nnw
	& \begin{gathered}
		\begin{fmfgraph*}(55,60)
			\fmfset{curly_len}{2mm}
			\fmftop{t1,t2} \fmfbottom{b1,b2}
			\fmf{phantom,label=$\hat u_t$,label.side=left,tension=4}{t1,v1}
			\fmf{phantom,label=$\hat d_s$,label.side=left,tension=4}{v2,t2}
			\fmf{quark,label=$\hat e_r$,label.side=right,tension=4}{b1,v3}
			\fmf{plain,label=$\hat \nu_p$,label.side=right,tension=4}{v4,b2}
			\fmf{quark,label=$u_u$,label.side=left}{v2,v1}
			\fmf{plain,label=$\nu_v$,label.side=right}{v3,v4}
			\fmf{wboson,label=$\W^+$,label.side=right}{v1,v3}
			\fmf{photon,label=$\Z$,label.side=left}{v2,v4}
			\fmffreeze
			\fmf{quark}{t1,v2}
			\fmf{quark}{v1,t2}
		\end{fmfgraph*}
	\end{gathered}
\qquad\quad
	\begin{gathered}
		\begin{fmfgraph*}(55,60)
			\fmfset{curly_len}{2mm}
			\fmftop{t1,t2} \fmfbottom{b1,b2}
			\fmf{quark,label=$\hat u_t$,label.side=right,label.dist=3,tension=3}{t1,v2}
			\fmf{quark,label=$\hat d_s$,label.side=right,label.dist=3,tension=3}{v3,t2}
			\fmf{quark,label=$\hat e_r$,label.side=right,label.dist=4,tension=3}{b1,v1}
			\fmf{plain,label=$\hat \nu_p$,label.side=right,label.dist=4,tension=3}{v1,b2}
			\fmf{photon,label=$\A$,label.side=left,label.dist=3,tension=1}{v1,v2}
			\fmf{wboson,label=$\W^+$,label.side=left,label.dist=3,tension=1}{v3,v1}
			\fmffreeze
			\fmf{quark,label=$u_u$,label.side=left}{v2,v3}
			\fmfv{decoration.shape=square,decoration.size=1.5mm}{v1}
		\end{fmfgraph*}
	\end{gathered}
\qquad\quad
	\begin{gathered}
		\begin{fmfgraph*}(55,60)
			\fmfset{curly_len}{2mm}
			\fmftop{t1,t2} \fmfbottom{b1,b2}
			\fmf{quark,label=$\hat u_t$,label.side=right,label.dist=3,tension=3}{t1,v2}
			\fmf{quark,label=$\hat d_s$,label.side=right,label.dist=3,tension=3}{v3,t2}
			\fmf{quark,label=$\hat e_r$,label.side=right,label.dist=4,tension=3}{b1,v1}
			\fmf{plain,label=$\hat \nu_p$,label.side=right,label.dist=4,tension=3}{v1,b2}
			\fmf{photon,label=$\Z$,label.side=left,label.dist=3,tension=1}{v1,v2}
			\fmf{wboson,label=$\W^+$,label.side=left,label.dist=3,tension=1}{v3,v1}
			\fmffreeze
			\fmf{quark,label=$u_u$,label.side=left}{v2,v3}
			\fmfv{decoration.shape=square,decoration.size=1.5mm}{v1}
		\end{fmfgraph*}
	\end{gathered}
\qquad\quad
	\begin{gathered}
		\begin{fmfgraph*}(55,60)
			\fmfset{curly_len}{2mm}
			\fmftop{t1,t2} \fmfbottom{b1,b2}
			\fmf{quark,label=$\hat u_t$,label.side=right,label.dist=3,tension=3}{t1,v2}
			\fmf{quark,label=$\hat d_s$,label.side=right,label.dist=3,tension=3}{v3,t2}
			\fmf{quark,label=$\hat e_r$,label.side=right,label.dist=4,tension=3}{b1,v1}
			\fmf{plain,label=$\hat \nu_p$,label.side=right,label.dist=4,tension=3}{v1,b2}
			\fmf{wboson,label=$\W^+$,label.side=right,label.dist=3,tension=1}{v2,v1}
			\fmf{photon,label=$\A$,label.side=left,label.dist=3,tension=1}{v3,v1}
			\fmffreeze
			\fmf{quark,label=$d_u$,label.side=left}{v2,v3}
			\fmfv{decoration.shape=square,decoration.size=1.5mm}{v1}
		\end{fmfgraph*}
	\end{gathered}
\qquad\quad
	\begin{gathered}
		\begin{fmfgraph*}(55,60)
			\fmfset{curly_len}{2mm}
			\fmftop{t1,t2} \fmfbottom{b1,b2}
			\fmf{quark,label=$\hat u_t$,label.side=right,label.dist=3,tension=3}{t1,v2}
			\fmf{quark,label=$\hat d_s$,label.side=right,label.dist=3,tension=3}{v3,t2}
			\fmf{quark,label=$\hat e_r$,label.side=right,label.dist=4,tension=3}{b1,v1}
			\fmf{plain,label=$\hat \nu_p$,label.side=right,label.dist=4,tension=3}{v1,b2}
			\fmf{wboson,label=$\W^+$,label.side=right,label.dist=3,tension=1}{v2,v1}
			\fmf{photon,label=$\Z$,label.side=left,label.dist=3,tension=1}{v3,v1}
			\fmffreeze
			\fmf{quark,label=$d_u$,label.side=left}{v2,v3}
			\fmfv{decoration.shape=square,decoration.size=1.5mm}{v1}
		\end{fmfgraph*}
	\end{gathered}
\nnw
	& \begin{gathered}
		\begin{fmfgraph*}(55,60)
			\fmfset{curly_len}{2mm}
			\fmftop{t1,t2} \fmfbottom{b1,b2}
			\fmf{quark,label=$\hat u_t$,label.side=left,label.dist=4,tension=3}{t1,v1}
			\fmf{quark,label=$\hat d_s$,label.side=left,label.dist=4,tension=3}{v1,t2}
			\fmf{quark,label=$\hat e_r$,label.side=left,label.dist=3,tension=3}{b1,v2}
			\fmf{plain,label=$\hat \nu_p$,label.side=left,label.dist=3,tension=3}{v3,b2}
			\fmf{photon,label=$\A$,label.side=left,label.dist=3,tension=1}{v2,v1}
			\fmf{wboson,label=$\W^+$,label.side=left,label.dist=3,tension=1}{v1,v3}
			\fmffreeze
			\fmf{quark,label=$e_u$,label.side=right}{v2,v3}
			\fmfv{decoration.shape=square,decoration.size=1.5mm}{v1}
		\end{fmfgraph*}
	\end{gathered}
\qquad\quad
	\begin{gathered}
		\begin{fmfgraph*}(55,60)
			\fmfset{curly_len}{2mm}
			\fmftop{t1,t2} \fmfbottom{b1,b2}
			\fmf{quark,label=$\hat u_t$,label.side=left,label.dist=4,tension=3}{t1,v1}
			\fmf{quark,label=$\hat d_s$,label.side=left,label.dist=4,tension=3}{v1,t2}
			\fmf{quark,label=$\hat e_r$,label.side=left,label.dist=3,tension=3}{b1,v2}
			\fmf{plain,label=$\hat \nu_p$,label.side=left,label.dist=3,tension=3}{v3,b2}
			\fmf{photon,label=$\Z$,label.side=left,label.dist=3,tension=1}{v2,v1}
			\fmf{wboson,label=$\W^+$,label.side=left,label.dist=3,tension=1}{v1,v3}
			\fmffreeze
			\fmf{quark,label=$e_u$,label.side=right}{v2,v3}
			\fmfv{decoration.shape=square,decoration.size=1.5mm}{v1}
		\end{fmfgraph*}
	\end{gathered}
\qquad\quad
	\begin{gathered}
		\begin{fmfgraph*}(55,60)
			\fmfset{curly_len}{2mm}
			\fmftop{t1,t2} \fmfbottom{b1,b2}
			\fmf{quark,label=$\hat u_t$,label.side=left,label.dist=4,tension=3}{t1,v1}
			\fmf{quark,label=$\hat d_s$,label.side=left,label.dist=4,tension=3}{v1,t2}
			\fmf{quark,label=$\hat e_r$,label.side=left,label.dist=3,tension=3}{b1,v2}
			\fmf{plain,label=$\hat \nu_p$,label.side=left,label.dist=3,tension=3}{v3,b2}
			\fmf{wboson,label=$\W^+$,label.side=right,label.dist=3,tension=1}{v1,v2}
			\fmf{photon,label=$\Z$,label.side=left,label.dist=3,tension=1}{v1,v3}
			\fmffreeze
			\fmf{plain,label=$\nu_u$,label.side=right}{v2,v3}
			\fmfv{decoration.shape=square,decoration.size=1.5mm}{v1}
		\end{fmfgraph*}
	\end{gathered}
\qquad\quad
	\begin{gathered}
		\begin{fmfgraph*}(55,60)
			\fmfset{curly_len}{2mm}
			\fmftop{t1,t2} \fmfbottom{b1,b2}
			\fmf{quark,label=$\hat u_t$,label.side=left,label.dist=3,tension=6}{t1,v2}
			\fmf{quark,label=$\hat d_s$,label.side=right,label.dist=3,tension=2}{v1,t2}
			\fmf{quark,label=$\hat e_r$,label.side=right,label.dist=3,tension=6}{b1,v3}
			\fmf{plain,label=$\hat \nu_p$,label.side=left,label.dist=3,tension=2}{v1,b2}
			\fmf{quark,label=$d_u$,label.side=left,label.dist=3,tension=3}{v2,v1}
			\fmf{plain,label=$\nu_v$,label.side=right,label.dist=3,tension=3}{v3,v1}
			\fmffreeze
			\fmf{wboson,label=$\W^+$,label.side=right,right=0.25}{v2,v3}
			\fmfv{decoration.shape=square,decoration.size=1.5mm}{v1}
		\end{fmfgraph*}
	\end{gathered}
\qquad\quad
	\begin{gathered}
		\begin{fmfgraph*}(55,60)
			\fmfset{curly_len}{2mm}
			\fmftop{t1,t2} \fmfbottom{b1,b2}
			\fmf{quark,label=$\hat u_t$,label.side=left,label.dist=3,tension=6}{t1,v2}
			\fmf{quark,label=$\hat d_s$,label.side=right,label.dist=3,tension=2}{v1,t2}
			\fmf{quark,label=$\hat e_r$,label.side=right,label.dist=3,tension=6}{b1,v3}
			\fmf{plain,label=$\hat \nu_p$,label.side=left,label.dist=3,tension=2}{v1,b2}
			\fmf{quark,label=$u_u$,label.side=left,label.dist=3,tension=3}{v2,v1}
			\fmf{quark,label=$e_v$,label.side=right,label.dist=3,tension=3}{v3,v1}
			\fmffreeze
			\fmf{photon,label=$\Z$,label.side=right,right=0.25}{v2,v3}
			\fmfv{decoration.shape=square,decoration.size=1.5mm}{v1}
		\end{fmfgraph*}
	\end{gathered}
\nnw
	& \begin{gathered}
		\begin{fmfgraph*}(55,60)
			\fmfset{curly_len}{2mm}
			\fmftop{t1,t2} \fmfbottom{b1,b2}
			\fmf{quark,label=$\hat u_t$,label.side=right,label.dist=3,tension=6}{t1,v2}
			\fmf{quark,label=$\hat d_s$,label.side=right,label.dist=3,tension=6}{v3,t2}
			\fmf{quark,label=$\hat e_r$,label.side=right,label.dist=3,tension=2}{b1,v1}
			\fmf{plain,label=$\hat \nu_p$,label.side=right,label.dist=3,tension=2}{v1,b2}
			\fmf{quark,label=$u_u$,label.side=right,label.dist=3,tension=3}{v2,v1}
			\fmf{quark,label=$d_v$,label.side=right,label.dist=3,tension=3}{v1,v3}
			\fmffreeze
			\fmf{photon,label=$\Z$,label.side=left,left=0.25}{v2,v3}
			\fmfv{decoration.shape=square,decoration.size=1.5mm}{v1}
		\end{fmfgraph*}
	\end{gathered}
\qquad\quad
	\begin{gathered}
		\begin{fmfgraph*}(55,60)
			\fmfset{curly_len}{2mm}
			\fmftop{t1,t2} \fmfbottom{b1,b2}
			\fmf{quark,label=$\hat u_t$,label.side=left,label.dist=3,tension=2}{t1,v1}
			\fmf{quark,label=$\hat d_s$,label.side=left,label.dist=3,tension=2}{v1,t2}
			\fmf{quark,label=$\hat e_r$,label.side=left,label.dist=3,tension=6}{b1,v2}
			\fmf{plain,label=$\hat \nu_p$,label.side=left,label.dist=3,tension=6}{v3,b2}
			\fmf{quark,label=$e_u$,label.side=left,label.dist=3,tension=3}{v2,v1}
			\fmf{plain,label=$\nu_v$,label.side=left,label.dist=3,tension=3}{v1,v3}
			\fmffreeze
			\fmf{photon,label=$\Z$,label.side=right,right=0.25}{v2,v3}
			\fmfv{decoration.shape=square,decoration.size=1.5mm}{v1}
		\end{fmfgraph*}
	\end{gathered}
\qquad\quad
	\begin{gathered}
		\begin{fmfgraph*}(55,60)
			\fmfset{curly_len}{2mm}
			\fmftop{t1,t2} \fmfbottom{b1,b2}
			\fmf{quark,label=$\hat u_t$,label.side=right,label.dist=3,tension=2}{t1,v1}
			\fmf{quark,label=$\hat d_s$,label.side=left,label.dist=3,tension=6}{v2,t2}
			\fmf{quark,label=$\hat e_r$,label.side=left,label.dist=3,tension=2}{b1,v1}
			\fmf{plain,label=$\hat \nu_p$,label.side=right,label.dist=3,tension=6}{v3,b2}
			\fmf{quark,label=$u_u$,label.side=left,label.dist=3,tension=3}{v1,v2}
			\fmf{quark,label=$e_v$,label.side=right,label.dist=3,tension=3}{v1,v3}
			\fmffreeze
			\fmf{wboson,label=$\W^+$,label.side=left,left=0.25}{v2,v3}
			\fmfv{decoration.shape=square,decoration.size=1.5mm}{v1}
		\end{fmfgraph*}
	\end{gathered}
\qquad\quad
	\begin{gathered}
		\begin{fmfgraph*}(55,60)
			\fmfset{curly_len}{2mm}
			\fmftop{t1,t2} \fmfbottom{b1,b2}
			\fmf{quark,label=$\hat u_t$,label.side=right,label.dist=3,tension=2}{t1,v1}
			\fmf{quark,label=$\hat d_s$,label.side=left,label.dist=3,tension=6}{v2,t2}
			\fmf{quark,label=$\hat e_r$,label.side=left,label.dist=3,tension=2}{b1,v1}
			\fmf{plain,label=$\hat \nu_p$,label.side=right,label.dist=3,tension=6}{v3,b2}
			\fmf{quark,label=$d_u$,label.side=left,label.dist=3,tension=3}{v1,v2}
			\fmf{plain,label=$\nu_v$,label.side=right,label.dist=3,tension=3}{v1,v3}
			\fmffreeze
			\fmf{photon,label=$\Z$,label.side=left,left=0.25}{v2,v3}
			\fmfv{decoration.shape=square,decoration.size=1.5mm}{v1}
		\end{fmfgraph*}
	\end{gathered}
\qquad\quad
	\begin{gathered}
		\begin{fmfgraph*}(55,60)
			\fmfset{curly_len}{2mm}
			\fmftop{t1,t2} \fmfbottom{b1,b2}
			\fmf{quark,label=$\hat u_t$,label.side=left,label.dist=3,tension=6}{t1,v2}
			\fmf{quark,label=$\hat d_s$,label.side=left,label.dist=3,tension=2}{v1,t2}
			\fmf{quark,label=$\hat e_r$,label.side=right,label.dist=3,tension=2}{b1,v1}
			\fmf{plain,label=$\hat \nu_p$,label.side=left,label.dist=3,tension=6}{v3,b2}
			\fmf{quark,label=$d_u$,label.side=left,label.dist=3,tension=3}{v2,v1}
			\fmf{quark,label=$e_v$,label.side=left,label.dist=3,tension=3}{v1,v3}
			\fmffreeze
			\fmf{wboson,right=0.5}{v2,v3}
			\fmf{phantom,label=$\W^+$,label.side=right,left=0.5}{v2,b1}
			\fmfv{decoration.shape=square,decoration.size=1.5mm}{v1}
		\end{fmfgraph*}
	\end{gathered}
\nn
	& \begin{gathered}
		\begin{fmfgraph*}(55,60)
			\fmfset{curly_len}{2mm}
			\fmftop{t1,t2} \fmfbottom{b1,b2}
			\fmf{quark,label=$\hat u_t$,label.side=left,label.dist=3,tension=6}{t1,v2}
			\fmf{quark,label=$\hat d_s$,label.side=left,label.dist=3,tension=2}{v1,t2}
			\fmf{quark,label=$\hat e_r$,label.side=right,label.dist=3,tension=2}{b1,v1}
			\fmf{plain,label=$\hat \nu_p$,label.side=left,label.dist=3,tension=6}{v3,b2}
			\fmf{quark,label=$u_u$,label.side=left,label.dist=3,tension=3}{v2,v1}
			\fmf{plain,label=$\nu_v$,label.side=left,label.dist=3,tension=3}{v1,v3}
			\fmffreeze
			\fmf{photon,right=0.5}{v2,v3}
			\fmf{phantom,label=$\Z$,label.side=right,left=0.5}{v2,b1}
			\fmfv{decoration.shape=square,decoration.size=1.5mm}{v1}
		\end{fmfgraph*}
	\end{gathered}
\qquad\quad
	\begin{gathered}
		\begin{fmfgraph*}(55,60)
			\fmfset{curly_len}{2mm}
			\fmftop{t1,t2} \fmfbottom{b1,b2}
			\fmf{quark,label=$\hat u_t$,label.side=left,label.dist=3,tension=2}{t1,v1}
			\fmf{quark,label=$\hat d_s$,label.side=right,label.dist=3,tension=6}{v3,t2}
			\fmf{quark,label=$\hat e_r$,label.side=right,label.dist=3,tension=6}{b1,v2}
			\fmf{plain,label=$\hat \nu_p$,label.side=left,label.dist=3,tension=2}{v1,b2}
			\fmf{quark,label=$u_u$,label.side=left,label.dist=3,tension=3}{v1,v3}
			\fmf{plain,label=$\nu_v$,label.side=left,label.dist=3,tension=3}{v2,v1}
			\fmffreeze
			\fmf{wboson,left=0.5}{v3,v2}
			\fmf{phantom,label=$\W^+$,label.side=right,left=0.4}{v2,b2}
			\fmfv{decoration.shape=square,decoration.size=1.5mm}{v1}
		\end{fmfgraph*}
	\end{gathered}
\qquad\quad
	\begin{gathered}
		\begin{fmfgraph*}(55,60)
			\fmfset{curly_len}{2mm}
			\fmftop{t1,t2} \fmfbottom{b1,b2}
			\fmf{quark,label=$\hat u_t$,label.side=left,label.dist=3,tension=2}{t1,v1}
			\fmf{quark,label=$\hat d_s$,label.side=right,label.dist=3,tension=6}{v3,t2}
			\fmf{quark,label=$\hat e_r$,label.side=right,label.dist=3,tension=6}{b1,v2}
			\fmf{plain,label=$\hat \nu_p$,label.side=left,label.dist=3,tension=2}{v1,b2}
			\fmf{quark,label=$d_u$,label.side=left,label.dist=3,tension=3}{v1,v3}
			\fmf{quark,label=$e_v$,label.side=left,label.dist=3,tension=3}{v2,v1}
			\fmffreeze
			\fmf{photon,left=0.5}{v3,v2}
			\fmf{phantom,label=$\Z$,label.side=right,left=0.4}{v2,b2}
			\fmfv{decoration.shape=square,decoration.size=1.5mm}{v1}
		\end{fmfgraph*}
	\end{gathered}
\end{align*}


\subsubsection[$\nu udd$ ($\Delta B = \pm \Delta L = 1$)]{\boldmath$\nu udd$ ($\Delta B = \pm \Delta L = 1$)}

\begin{align*}
	& \begin{gathered}
		\begin{fmfgraph*}(55,60)
			\fmfset{curly_len}{2mm}
			\fmftop{t1,t2} \fmfbottom{b1,b2}
			\fmf{quark,label=$\hat d_p$,label.side=right,label.dist=3,tension=2}{b2,v1}
			\fmf{quark,label=$\hat d_r$,label.side=right,label.dist=3,tension=6}{b1,v2}
			\fmf{quark,label=$\hat u_s$,label.side=left,label.dist=3,tension=2}{t2,v1}
			\fmf{plain,label=$\hat \nu_t$,label.side=right,label.dist=3,tension=6}{v3,t1}
			\fmf{quark,label=$u_v$,label.side=right,label.dist=3,tension=3}{v2,v1}
			\fmf{quark,label=$e_u$,label.side=left,label.dist=3,tension=3}{v3,v1}
			\fmffreeze
			\fmf{wboson,label=$\W^+$,label.side=right,right=0.25}{v3,v2}
			\fmfv{decoration.shape=square,decoration.size=1.5mm}{v1}
		\end{fmfgraph*}
	\end{gathered}
\qquad\quad
	\begin{gathered}
		\begin{fmfgraph*}(55,60)
			\fmfset{curly_len}{2mm}
			\fmftop{t1,t2} \fmfbottom{b1,b2}
			\fmf{quark,label=$\hat d_p$,label.side=right,label.dist=3,tension=2}{b2,v1}
			\fmf{quark,label=$\hat d_r$,label.side=right,label.dist=3,tension=6}{b1,v2}
			\fmf{quark,label=$\hat u_s$,label.side=left,label.dist=3,tension=2}{t2,v1}
			\fmf{plain,label=$\hat \nu_t$,label.side=right,label.dist=3,tension=6}{v3,t1}
			\fmf{quark,label=$d_v$,label.side=right,label.dist=3,tension=3}{v2,v1}
			\fmf{plain,label=$\nu_u$,label.side=right,label.dist=3,tension=3}{v1,v3}
			\fmffreeze
			\fmf{photon,label=$\Z$,label.side=left,left=0.25}{v2,v3}
			\fmfv{decoration.shape=square,decoration.size=1.5mm}{v1}
		\end{fmfgraph*}
	\end{gathered}
\qquad\quad
	\begin{gathered}
		\begin{fmfgraph*}(55,60)
			\fmfset{curly_len}{2mm}
			\fmftop{t1,t2} \fmfbottom{b1,b2}
			\fmf{quark,label=$\hat d_p$,label.side=left,label.dist=3,tension=2}{b2,v1}
			\fmf{quark,label=$\hat d_r$,label.side=right,label.dist=3,tension=2}{b1,v1}
			\fmf{quark,label=$\hat u_s$,label.side=left,label.dist=3,tension=6}{t2,v2}
			\fmf{plain,label=$\hat \nu_t$,label.side=left,label.dist=3,tension=6}{v3,t1}
			\fmf{quark,label=$u_v$,label.side=left,label.dist=3,tension=3}{v2,v1}
			\fmf{plain,label=$\nu_u$,label.side=left,label.dist=3,tension=3}{v1,v3}
			\fmffreeze
			\fmf{photon,label=$\Z$,label.side=right,right=0.25}{v2,v3}
			\fmfv{decoration.shape=square,decoration.size=1.5mm}{v1}
		\end{fmfgraph*}
	\end{gathered}
\qquad\quad
	\begin{gathered}
		\begin{fmfgraph*}(55,60)
			\fmfset{curly_len}{2mm}
			\fmftop{t1,t2} \fmfbottom{b1,b2}
			\fmf{quark,label=$\hat d_p$,label.side=right,label.dist=3,tension=6}{b2,v2}
			\fmf{quark,label=$\hat d_r$,label.side=left,label.dist=3,tension=6}{b1,v3}
			\fmf{quark,label=$\hat u_s$,label.side=right,label.dist=3,tension=2}{t2,v1}
			\fmf{plain,label=$\hat \nu_t$,label.side=right,label.dist=2,tension=2}{v1,t1}
			\fmf{quark,label=$d_v$,label.side=right,label.dist=3,tension=3}{v2,v1}
			\fmf{quark,label=$d_u$,label.side=left,label.dist=3,tension=3}{v3,v1}
			\fmffreeze
			\fmf{photon,label=$\Z$,label.side=left,left=0.25}{v2,v3}
			\fmfv{decoration.shape=square,decoration.size=1.5mm}{v1}
		\end{fmfgraph*}
	\end{gathered}
\qquad\quad
	\begin{gathered}
		\begin{fmfgraph*}(55,60)
			\fmfset{curly_len}{2mm}
			\fmftop{t1,t2} \fmfbottom{b1,b2}
			\fmf{quark,label=$\hat d_p$,label.side=left,label.dist=3,tension=6}{b2,v2}
			\fmf{quark,label=$\hat d_r$,label.side=left,label.dist=3,tension=2}{b1,v1}
			\fmf{quark,label=$\hat u_s$,label.side=right,label.dist=3,tension=6}{t2,v3}
			\fmf{plain,label=$\hat \nu_t$,label.side=left,label.dist=1,tension=2}{v1,t1}
			\fmf{quark,label=$u_v$,label.side=left,label.dist=3,tension=3}{v2,v1}
			\fmf{quark,label=$d_u$,label.side=right,label.dist=3,tension=3}{v3,v1}
			\fmffreeze
			\fmf{wboson,label=$\W^+$,label.side=left,left=0.25}{v3,v2}
			\fmfv{decoration.shape=square,decoration.size=1.5mm}{v1}
		\end{fmfgraph*}
	\end{gathered}
\nnw
	& \begin{gathered}
		\begin{fmfgraph*}(55,60)
			\fmfset{curly_len}{2mm}
			\fmftop{t1,t2} \fmfbottom{b1,b2}
			\fmf{quark,label=$\hat d_p$,label.side=left,label.dist=3,tension=6}{b2,v2}
			\fmf{quark,label=$\hat d_r$,label.side=left,label.dist=3,tension=2}{b1,v1}
			\fmf{quark,label=$\hat u_s$,label.side=right,label.dist=3,tension=6}{t2,v3}
			\fmf{plain,label=$\hat \nu_t$,label.side=left,label.dist=1,tension=2}{v1,t1}
			\fmf{quark,label=$d_v$,label.side=left,label.dist=3,tension=3}{v2,v1}
			\fmf{quark,label=$u_u$,label.side=right,label.dist=3,tension=3}{v3,v1}
			\fmffreeze
			\fmf{photon,label=$\Z$,label.side=right,right=0.25}{v2,v3}
			\fmfv{decoration.shape=square,decoration.size=1.5mm}{v1}
		\end{fmfgraph*}
	\end{gathered}
\qquad\quad
	\begin{gathered}
		\begin{fmfgraph*}(55,60)
			\fmfset{curly_len}{2mm}
			\fmftop{t1,t2} \fmfbottom{b1,b2}
			\fmf{quark,label=$\hat d_p$,label.side=left,label.dist=3,tension=6}{b2,v2}
			\fmf{quark,label=$\hat d_r$,label.side=right,label.dist=3,tension=2}{b1,v1}
			\fmf{quark,label=$\hat u_s$,label.side=right,label.dist=3,tension=2}{t2,v1}
			\fmf{plain,label=$\hat \nu_t$,label.side=left,label.dist=3,tension=6}{v3,t1}
			\fmf{quark,label=$u_v$,label.side=left,label.dist=3,tension=3}{v2,v1}
			\fmf{quark,label=$e_u$,label.side=right,label.dist=3,tension=3}{v3,v1}
			\fmffreeze
			\fmf{wboson,left=0.5}{v3,v2}
			\fmf{phantom,label=$\W^+$,label.side=right,left=0.5}{v2,t2}
			\fmfv{decoration.shape=square,decoration.size=1.5mm}{v1}
		\end{fmfgraph*}
	\end{gathered}
\qquad\quad
	\begin{gathered}
		\begin{fmfgraph*}(55,60)
			\fmfset{curly_len}{2mm}
			\fmftop{t1,t2} \fmfbottom{b1,b2}
			\fmf{quark,label=$\hat d_p$,label.side=left,label.dist=3,tension=6}{b2,v2}
			\fmf{quark,label=$\hat d_r$,label.side=right,label.dist=3,tension=2}{b1,v1}
			\fmf{quark,label=$\hat u_s$,label.side=right,label.dist=3,tension=2}{t2,v1}
			\fmf{plain,label=$\hat \nu_t$,label.side=left,label.dist=3,tension=6}{v3,t1}
			\fmf{quark,label=$d_v$,label.side=left,label.dist=3,tension=3}{v2,v1}
			\fmf{plain,label=$\nu_u$,label.side=left,label.dist=3,tension=3}{v1,v3}
			\fmffreeze
			\fmf{photon,right=0.5}{v2,v3}
			\fmf{phantom,label=$\Z$,label.side=right,left=0.5}{v2,t2}
			\fmfv{decoration.shape=square,decoration.size=1.5mm}{v1}
		\end{fmfgraph*}
	\end{gathered}
\qquad\quad
	\begin{gathered}
		\begin{fmfgraph*}(55,60)
			\fmfset{curly_len}{2mm}
			\fmftop{t1,t2} \fmfbottom{b1,b2}
			\fmf{quark,label=$\hat d_p$,label.side=left,label.dist=3,tension=2}{b2,v1}
			\fmf{quark,label=$\hat d_r$,label.side=right,label.dist=3,tension=6}{b1,v3}
			\fmf{quark,label=$\hat u_s$,label.side=right,label.dist=3,tension=6}{t2,v2}
			\fmf{plain,label=$\hat \nu_t$,label.side=left,label.dist=3,tension=2}{v1,t1}
			\fmf{quark,label=$u_v$,label.side=right,label.dist=3,tension=3}{v3,v1}
			\fmf{quark,label=$d_u$,label.side=left,label.dist=3,tension=3}{v2,v1}
			\fmffreeze
			\fmf{wboson,right=0.5}{v2,v3}
			\fmf{phantom,label=$\W^+$,label.side=right,left=0.4}{v2,t1}
			\fmfv{decoration.shape=square,decoration.size=1.5mm}{v1}
		\end{fmfgraph*}
	\end{gathered}
\qquad\quad
	\begin{gathered}
		\begin{fmfgraph*}(55,60)
			\fmfset{curly_len}{2mm}
			\fmftop{t1,t2} \fmfbottom{b1,b2}
			\fmf{quark,label=$\hat d_p$,label.side=left,label.dist=3,tension=2}{b2,v1}
			\fmf{quark,label=$\hat d_r$,label.side=right,label.dist=3,tension=6}{b1,v3}
			\fmf{quark,label=$\hat u_s$,label.side=right,label.dist=3,tension=6}{t2,v2}
			\fmf{plain,label=$\hat \nu_t$,label.side=left,label.dist=3,tension=2}{v1,t1}
			\fmf{quark,label=$d_v$,label.side=right,label.dist=3,tension=3}{v3,v1}
			\fmf{quark,label=$u_u$,label.side=left,label.dist=3,tension=3}{v2,v1}
			\fmffreeze
			\fmf{photon,left=0.5}{v3,v2}
			\fmf{phantom,label=$\Z$,label.side=right,left=0.4}{v2,t1}
			\fmfv{decoration.shape=square,decoration.size=1.5mm}{v1}
		\end{fmfgraph*}
	\end{gathered}
\end{align*}


\subsubsection[$e duu$ ($\Delta B = \Delta L = 1$)]{\boldmath$e duu$ ($\Delta B = \Delta L = 1$)}

\begin{align*}
	& \begin{gathered}
		\begin{fmfgraph*}(55,60)
			\fmfset{curly_len}{2mm}
			\fmftop{t1,t2} \fmfbottom{b1,b2}
			\fmf{quark,label=$\hat u_p$,label.side=right,label.dist=3,tension=2}{b2,v1}
			\fmf{quark,label=$\hat u_r$,label.side=right,label.dist=3,tension=6}{b1,v2}
			\fmf{quark,label=$\hat d_s$,label.side=left,label.dist=3,tension=2}{t2,v1}
			\fmf{quark,label=$\hat e_t$,label.side=left,label.dist=3,tension=6}{t1,v3}
			\fmf{quark,label=$d_v$,label.side=right,label.dist=3,tension=3}{v2,v1}
			\fmf{plain,label=$\nu_u$,label.side=left,label.dist=3,tension=3}{v3,v1}
			\fmffreeze
			\fmf{wboson,label=$\W^+$,label.side=left,left=0.25}{v2,v3}
			\fmfv{decoration.shape=square,decoration.size=1.5mm}{v1}
		\end{fmfgraph*}
	\end{gathered}
\qquad\quad
	\begin{gathered}
		\begin{fmfgraph*}(55,60)
			\fmfset{curly_len}{2mm}
			\fmftop{t1,t2} \fmfbottom{b1,b2}
			\fmf{quark,label=$\hat u_p$,label.side=right,label.dist=3,tension=2}{b2,v1}
			\fmf{quark,label=$\hat u_r$,label.side=right,label.dist=3,tension=6}{b1,v2}
			\fmf{quark,label=$\hat d_s$,label.side=left,label.dist=3,tension=2}{t2,v1}
			\fmf{quark,label=$\hat e_t$,label.side=left,label.dist=3,tension=6}{t1,v3}
			\fmf{quark,label=$u_v$,label.side=right,label.dist=3,tension=3}{v2,v1}
			\fmf{quark,label=$e_u$,label.side=left,label.dist=3,tension=3}{v3,v1}
			\fmffreeze
			\fmf{photon,label=$\Z$,label.side=left,left=0.25}{v2,v3}
			\fmfv{decoration.shape=square,decoration.size=1.5mm}{v1}
		\end{fmfgraph*}
	\end{gathered}
\qquad\quad
	\begin{gathered}
		\begin{fmfgraph*}(55,60)
			\fmfset{curly_len}{2mm}
			\fmftop{t1,t2} \fmfbottom{b1,b2}
			\fmf{quark,label=$\hat u_p$,label.side=left,label.dist=3,tension=2}{b2,v1}
			\fmf{quark,label=$\hat u_r$,label.side=right,label.dist=3,tension=2}{b1,v1}
			\fmf{quark,label=$\hat d_s$,label.side=left,label.dist=3,tension=6}{t2,v2}
			\fmf{quark,label=$\hat e_t$,label.side=right,label.dist=3,tension=6}{t1,v3}
			\fmf{quark,label=$d_v$,label.side=left,label.dist=3,tension=3}{v2,v1}
			\fmf{quark,label=$e_u$,label.side=right,label.dist=3,tension=3}{v3,v1}
			\fmffreeze
			\fmf{photon,label=$\Z$,label.side=right,right=0.25}{v2,v3}
			\fmfv{decoration.shape=square,decoration.size=1.5mm}{v1}
		\end{fmfgraph*}
	\end{gathered}
\qquad\quad
	\begin{gathered}
		\begin{fmfgraph*}(55,60)
			\fmfset{curly_len}{2mm}
			\fmftop{t1,t2} \fmfbottom{b1,b2}
			\fmf{quark,label=$\hat u_p$,label.side=right,label.dist=3,tension=6}{b2,v2}
			\fmf{quark,label=$\hat u_r$,label.side=left,label.dist=3,tension=6}{b1,v3}
			\fmf{quark,label=$\hat d_s$,label.side=right,label.dist=3,tension=2}{t2,v1}
			\fmf{quark,label=$\hat e_t$,label.side=left,label.dist=3,tension=2}{t1,v1}
			\fmf{quark,label=$u_v$,label.side=right,label.dist=3,tension=3}{v2,v1}
			\fmf{quark,label=$u_u$,label.side=left,label.dist=3,tension=3}{v3,v1}
			\fmffreeze
			\fmf{photon,label=$\Z$,label.side=left,left=0.25}{v2,v3}
			\fmfv{decoration.shape=square,decoration.size=1.5mm}{v1}
		\end{fmfgraph*}
	\end{gathered}
\qquad\quad
	\begin{gathered}
		\begin{fmfgraph*}(55,60)
			\fmfset{curly_len}{2mm}
			\fmftop{t1,t2} \fmfbottom{b1,b2}
			\fmf{quark,label=$\hat u_p$,label.side=left,label.dist=3,tension=6}{b2,v2}
			\fmf{quark,label=$\hat u_r$,label.side=left,label.dist=3,tension=2}{b1,v1}
			\fmf{quark,label=$\hat d_s$,label.side=right,label.dist=3,tension=6}{t2,v3}
			\fmf{quark,label=$\hat e_t$,label.side=right,label.dist=3,tension=2}{t1,v1}
			\fmf{quark,label=$d_v$,label.side=left,label.dist=3,tension=3}{v2,v1}
			\fmf{quark,label=$u_u$,label.side=right,label.dist=3,tension=3}{v3,v1}
			\fmffreeze
			\fmf{wboson,label=$\W^+$,label.side=right,right=0.25}{v2,v3}
			\fmfv{decoration.shape=square,decoration.size=1.5mm}{v1}
		\end{fmfgraph*}
	\end{gathered}
\nnw
	& \begin{gathered}
		\begin{fmfgraph*}(55,60)
			\fmfset{curly_len}{2mm}
			\fmftop{t1,t2} \fmfbottom{b1,b2}
			\fmf{quark,label=$\hat u_p$,label.side=left,label.dist=3,tension=6}{b2,v2}
			\fmf{quark,label=$\hat u_r$,label.side=left,label.dist=3,tension=2}{b1,v1}
			\fmf{quark,label=$\hat d_s$,label.side=right,label.dist=3,tension=6}{t2,v3}
			\fmf{quark,label=$\hat e_t$,label.side=right,label.dist=3,tension=2}{t1,v1}
			\fmf{quark,label=$u_v$,label.side=left,label.dist=3,tension=3}{v2,v1}
			\fmf{quark,label=$d_u$,label.side=right,label.dist=3,tension=3}{v3,v1}
			\fmffreeze
			\fmf{photon,label=$\Z$,label.side=right,right=0.25}{v2,v3}
			\fmfv{decoration.shape=square,decoration.size=1.5mm}{v1}
		\end{fmfgraph*}
	\end{gathered}
\qquad\quad
	\begin{gathered}
		\begin{fmfgraph*}(55,60)
			\fmfset{curly_len}{2mm}
			\fmftop{t1,t2} \fmfbottom{b1,b2}
			\fmf{quark,label=$\hat u_p$,label.side=left,label.dist=3,tension=6}{b2,v2}
			\fmf{quark,label=$\hat u_r$,label.side=right,label.dist=3,tension=2}{b1,v1}
			\fmf{quark,label=$\hat d_s$,label.side=right,label.dist=3,tension=2}{t2,v1}
			\fmf{quark,label=$\hat e_t$,label.side=right,label.dist=3,tension=6}{t1,v3}
			\fmf{quark,label=$d_v$,label.side=left,label.dist=3,tension=3}{v2,v1}
			\fmf{plain,label=$\nu_u$,label.side=right,label.dist=3,tension=3}{v3,v1}
			\fmffreeze
			\fmf{wboson,right=0.5}{v2,v3}
			\fmf{phantom,label=$\W^+$,label.side=right,left=0.5}{v2,t2}
			\fmfv{decoration.shape=square,decoration.size=1.5mm}{v1}
		\end{fmfgraph*}
	\end{gathered}
\qquad\quad
	\begin{gathered}
		\begin{fmfgraph*}(55,60)
			\fmfset{curly_len}{2mm}
			\fmftop{t1,t2} \fmfbottom{b1,b2}
			\fmf{quark,label=$\hat u_p$,label.side=left,label.dist=3,tension=6}{b2,v2}
			\fmf{quark,label=$\hat u_r$,label.side=right,label.dist=3,tension=2}{b1,v1}
			\fmf{quark,label=$\hat d_s$,label.side=right,label.dist=3,tension=2}{t2,v1}
			\fmf{quark,label=$\hat e_t$,label.side=right,label.dist=3,tension=6}{t1,v3}
			\fmf{quark,label=$u_v$,label.side=left,label.dist=3,tension=3}{v2,v1}
			\fmf{quark,label=$e_u$,label.side=right,label.dist=3,tension=3}{v3,v1}
			\fmffreeze
			\fmf{photon,right=0.5}{v2,v3}
			\fmf{phantom,label=$\Z$,label.side=right,left=0.5}{v2,t2}
			\fmfv{decoration.shape=square,decoration.size=1.5mm}{v1}
		\end{fmfgraph*}
	\end{gathered}
\qquad\quad
	\begin{gathered}
		\begin{fmfgraph*}(55,60)
			\fmfset{curly_len}{2mm}
			\fmftop{t1,t2} \fmfbottom{b1,b2}
			\fmf{quark,label=$\hat u_p$,label.side=left,label.dist=3,tension=2}{b2,v1}
			\fmf{quark,label=$\hat u_r$,label.side=right,label.dist=3,tension=6}{b1,v3}
			\fmf{quark,label=$\hat d_s$,label.side=right,label.dist=3,tension=6}{t2,v2}
			\fmf{quark,label=$\hat e_t$,label.side=right,label.dist=3,tension=2}{t1,v1}
			\fmf{quark,label=$d_v$,label.side=right,label.dist=3,tension=3}{v3,v1}
			\fmf{quark,label=$u_u$,label.side=left,label.dist=3,tension=3}{v2,v1}
			\fmffreeze
			\fmf{wboson,left=0.5}{v3,v2}
			\fmf{phantom,label=$\W^+$,label.side=right,left=0.4}{v2,t1}
			\fmfv{decoration.shape=square,decoration.size=1.5mm}{v1}
		\end{fmfgraph*}
	\end{gathered}
\qquad\quad
	\begin{gathered}
		\begin{fmfgraph*}(55,60)
			\fmfset{curly_len}{2mm}
			\fmftop{t1,t2} \fmfbottom{b1,b2}
			\fmf{quark,label=$\hat u_p$,label.side=left,label.dist=3,tension=2}{b2,v1}
			\fmf{quark,label=$\hat u_r$,label.side=right,label.dist=3,tension=6}{b1,v3}
			\fmf{quark,label=$\hat d_s$,label.side=right,label.dist=3,tension=6}{t2,v2}
			\fmf{quark,label=$\hat e_t$,label.side=right,label.dist=3,tension=2}{t1,v1}
			\fmf{quark,label=$u_v$,label.side=right,label.dist=3,tension=3}{v3,v1}
			\fmf{quark,label=$d_u$,label.side=left,label.dist=3,tension=3}{v2,v1}
			\fmffreeze
			\fmf{photon,left=0.5}{v3,v2}
			\fmf{phantom,label=$\Z$,label.side=right,left=0.4}{v2,t1}
			\fmfv{decoration.shape=square,decoration.size=1.5mm}{v1}
		\end{fmfgraph*}
	\end{gathered}
\end{align*}


\subsubsection[$\bar e ddd$ ($\Delta B = -\Delta L = 1$)]{\boldmath$\bar e ddd$ ($\Delta B = -\Delta L = 1$)}

There is no SMEFT contribution in this class up to dimension six.

\end{myfmf}


\section{SMEFT operator basis}
\label{sec:SMEFTBasis}

In this appendix, we reproduce the SMEFT operator basis from~\cite{Grzadkowski:2010es}.



\begin{table}[H]
\begin{center}
\small
\begin{minipage}[t]{5.2cm}
\vspace{-0.2cm}
\renewcommand{\arraystretch}{1.5}
\begin{tabular}[t]{c|c}
\multicolumn{2}{c}{\boldmath{$\Delta L = 2 \qquad (LL)HH+\hc$}} \\
\hline
$Q_{5}$      & $\epsilon^{ij} \epsilon^{k\ell} (l_{ip}^T C l_{kr} ) H_j H_\ell  $  \\
\end{tabular}
\vspace{-0.4cm}
\end{minipage}
\end{center}
\caption{Dimension-five $\Delta L=2$ operator $Q_5$ in SMEFT.  There is also the Hermitian conjugate $\Delta L = -2$ operator $Q_5^\dagger$, as indicated by ${}+\hc$ in the table heading.
Subscripts $p$ and $r$ are weak-eigenstate indices.}
\label{tab:smeft5ops}
\end{table}


\begin{table}[H]
\hspace{-0.5cm}
\begin{center}
\begin{adjustbox}{width=0.9\textwidth,left}
\small
\begin{minipage}[t]{4.45cm}
\renewcommand{\arraystretch}{1.5}
\begin{tabular}[t]{c|c}
\multicolumn{2}{c}{\boldmath$1:X^3$} \\
\hline
$Q_G$                & $f^{ABC} G_\mu^{A\nu} G_\nu^{B\rho} G_\rho^{C\mu} $ \\
$Q_{\widetilde G}$          & $f^{ABC} \widetilde G_\mu^{A\nu} G_\nu^{B\rho} G_\rho^{C\mu} $ \\
$Q_W$                & $\epsilon^{IJK} W_\mu^{I\nu} W_\nu^{J\rho} W_\rho^{K\mu}$ \\ 
$Q_{\widetilde W}$          & $\epsilon^{IJK} \widetilde W_\mu^{I\nu} W_\nu^{J\rho} W_\rho^{K\mu}$ \\
\end{tabular}
\end{minipage}
\begin{minipage}[t]{2.7cm}
\renewcommand{\arraystretch}{1.5}
\begin{tabular}[t]{c|c}
\multicolumn{2}{c}{\boldmath$2:H^6$} \\
\hline
$Q_H$       & $(H^\dag H)^3$ 
\end{tabular}
\end{minipage}
\begin{minipage}[t]{5.1cm}
\renewcommand{\arraystretch}{1.5}
\begin{tabular}[t]{c|c}
\multicolumn{2}{c}{\boldmath$3:H^4 D^2$} \\
\hline
$Q_{H\Box}$ & $(H^\dag H)\Box(H^\dag H)$ \\
$Q_{H D}$   & $\ \left(H^\dag D^\mu H\right)^* \left(H^\dag D_\mu H\right)$ 
\end{tabular}
\end{minipage}
\begin{minipage}[t]{2.7cm}
\renewcommand{\arraystretch}{1.5}
\begin{tabular}[t]{c|c}
\multicolumn{2}{c}{\boldmath$5: \psi^2H^3 + \hc$} \\
\hline
$Q_{eH}$           & $(H^\dag H)(\bar l_p e_r H)$ \\
$Q_{uH}$          & $(H^\dag H)(\bar q_p u_r \widetilde H )$ \\
$Q_{dH}$           & $(H^\dag H)(\bar q_p d_r H)$\\
\end{tabular}
\end{minipage}

\end{adjustbox}

\vspace{0.35cm}

\begin{adjustbox}{width=0.9\textwidth,left}

\begin{minipage}[t]{4.7cm}
\renewcommand{\arraystretch}{1.5}
\begin{tabular}[t]{c|c}
\multicolumn{2}{c}{\boldmath$4:X^2H^2$} \\
\hline
$Q_{H G}$     & $H^\dag H\, G^A_{\mu\nu} G^{A\mu\nu}$ \\
$Q_{H\widetilde G}$         & $H^\dag H\, \widetilde G^A_{\mu\nu} G^{A\mu\nu}$ \\
$Q_{H W}$     & $H^\dag H\, W^I_{\mu\nu} W^{I\mu\nu}$ \\
$Q_{H\widetilde W}$         & $H^\dag H\, \widetilde W^I_{\mu\nu} W^{I\mu\nu}$ \\
$Q_{H B}$     & $ H^\dag H\, B_{\mu\nu} B^{\mu\nu}$ \\
$Q_{H\widetilde B}$         & $H^\dag H\, \widetilde B_{\mu\nu} B^{\mu\nu}$ \\
$Q_{H WB}$     & $ H^\dag \tau^I H\, W^I_{\mu\nu} B^{\mu\nu}$ \\
$Q_{H\widetilde W B}$         & $H^\dag \tau^I H\, \widetilde W^I_{\mu\nu} B^{\mu\nu}$ 
\end{tabular}
\end{minipage}
\begin{minipage}[t]{5.2cm}
\renewcommand{\arraystretch}{1.5}
\begin{tabular}[t]{c|c}
\multicolumn{2}{c}{\boldmath$6:\psi^2 XH+\hc$} \\
\hline
$Q_{eW}$      & $(\bar l_p \sigma^{\mu\nu} \tau^I e_r) H W_{\mu\nu}^I$ \\
$Q_{eB}$        & $(\bar l_p \sigma^{\mu\nu} e_r) H B_{\mu\nu}$ \\
$Q_{uG}$        & $(\bar q_p \sigma^{\mu\nu} T^A u_r) \widetilde H \, G_{\mu\nu}^A$ \\
$Q_{uW}$        & $(\bar q_p \sigma^{\mu\nu} \tau^I u_r) \widetilde H \, W_{\mu\nu}^I$ \\
$Q_{uB}$        & $(\bar q_p \sigma^{\mu\nu} u_r) \widetilde H \, B_{\mu\nu}$ \\
$Q_{dG}$        & $(\bar q_p \sigma^{\mu\nu} T^A d_r) H\, G_{\mu\nu}^A$ \\
$Q_{dW}$         & $(\bar q_p \sigma^{\mu\nu} \tau^I d_r) H\, W_{\mu\nu}^I$ \\
$Q_{dB}$        & $(\bar q_p \sigma^{\mu\nu} d_r) H\, B_{\mu\nu}$ 
\end{tabular}
\end{minipage}
\begin{minipage}[t]{5.4cm}
\renewcommand{\arraystretch}{1.5}
\begin{tabular}[t]{c|c}
\multicolumn{2}{c}{\boldmath$7:\psi^2H^2 D$} \\
\hline
$Q_{H l}^{(1)}$      & $(H^\dag i\overleftrightarrow{D}_\mu H)(\bar l_p \gamma^\mu l_r)$\\
$Q_{H l}^{(3)}$      & $(H^\dag i\overleftrightarrow{D}^I_\mu H)(\bar l_p \tau^I \gamma^\mu l_r)$\\
$Q_{H e}$            & $(H^\dag i\overleftrightarrow{D}_\mu H)(\bar e_p \gamma^\mu e_r)$\\
$Q_{H q}^{(1)}$      & $(H^\dag i\overleftrightarrow{D}_\mu H)(\bar q_p \gamma^\mu q_r)$\\
$Q_{H q}^{(3)}$      & $(H^\dag i\overleftrightarrow{D}^I_\mu H)(\bar q_p \tau^I \gamma^\mu q_r)$\\
$Q_{H u}$            & $(H^\dag i\overleftrightarrow{D}_\mu H)(\bar u_p \gamma^\mu u_r)$\\
$Q_{H d}$            & $(H^\dag i\overleftrightarrow{D}_\mu H)(\bar d_p \gamma^\mu d_r)$\\
$Q_{H u d}$ + h.c.   & $i(\widetilde H ^\dag D_\mu H)(\bar u_p \gamma^\mu d_r)$\\
\end{tabular}
\end{minipage}
\end{adjustbox}
\end{center}

\begin{center}
\begin{adjustbox}{width=0.9\textwidth,left}
\begin{minipage}[t]{4.75cm}
\renewcommand{\arraystretch}{1.5}
\begin{tabular}[t]{c|c}
\multicolumn{2}{c}{\boldmath$8:(\bar L L)(\bar L L)$} \\
\hline
$Q_{ll}$        & $(\bar l_p \gamma^\mu l_r)(\bar l_s \gamma_\mu l_t)$ \\
$Q_{qq}^{(1)}$  & $(\bar q_p \gamma^\mu q_r)(\bar q_s \gamma_\mu q_t)$ \\
$Q_{qq}^{(3)}$  & $(\bar q_p \gamma^\mu \tau^I q_r)(\bar q_s \gamma_\mu \tau^I q_t)$ \\
$Q_{lq}^{(1)}$                & $(\bar l_p \gamma^\mu l_r)(\bar q_s \gamma_\mu q_t)$ \\
$Q_{lq}^{(3)}$                & $(\bar l_p \gamma^\mu \tau^I l_r)(\bar q_s \gamma_\mu \tau^I q_t)$ 
\end{tabular}
\end{minipage}
\hspace{0.2cm}
\begin{minipage}[t]{5.25cm}
\renewcommand{\arraystretch}{1.5}
\begin{tabular}[t]{c|c}
\multicolumn{2}{c}{\boldmath$8:(\bar R R)(\bar R R)$} \\
\hline
$Q_{ee}$               & $(\bar e_p \gamma^\mu e_r)(\bar e_s \gamma_\mu e_t)$ \\
$Q_{uu}$        & $(\bar u_p \gamma^\mu u_r)(\bar u_s \gamma_\mu u_t)$ \\
$Q_{dd}$        & $(\bar d_p \gamma^\mu d_r)(\bar d_s \gamma_\mu d_t)$ \\
$Q_{eu}$                      & $(\bar e_p \gamma^\mu e_r)(\bar u_s \gamma_\mu u_t)$ \\
$Q_{ed}$                      & $(\bar e_p \gamma^\mu e_r)(\bar d_s\gamma_\mu d_t)$ \\
$Q_{ud}^{(1)}$                & $(\bar u_p \gamma^\mu u_r)(\bar d_s \gamma_\mu d_t)$ \\
$Q_{ud}^{(8)}$                & $(\bar u_p \gamma^\mu T^A u_r)(\bar d_s \gamma_\mu T^A d_t)$ \\
\end{tabular}
\end{minipage}
\hspace{0.2cm}
\begin{minipage}[t]{4.75cm}
\renewcommand{\arraystretch}{1.5}
\begin{tabular}[t]{c|c}
\multicolumn{2}{c}{\boldmath$8:(\bar L L)(\bar R R)$} \\
\hline
$Q_{le}$               & $(\bar l_p \gamma^\mu l_r)(\bar e_s \gamma_\mu e_t)$ \\
$Q_{lu}$               & $(\bar l_p \gamma^\mu l_r)(\bar u_s \gamma_\mu u_t)$ \\
$Q_{ld}$               & $(\bar l_p \gamma^\mu l_r)(\bar d_s \gamma_\mu d_t)$ \\
$Q_{qe}$               & $(\bar q_p \gamma^\mu q_r)(\bar e_s \gamma_\mu e_t)$ \\
$Q_{qu}^{(1)}$         & $(\bar q_p \gamma^\mu q_r)(\bar u_s \gamma_\mu u_t)$ \\ 
$Q_{qu}^{(8)}$         & $(\bar q_p \gamma^\mu T^A q_r)(\bar u_s \gamma_\mu T^A u_t)$ \\ 
$Q_{qd}^{(1)}$ & $(\bar q_p \gamma^\mu q_r)(\bar d_s \gamma_\mu d_t)$ \\
$Q_{qd}^{(8)}$ & $(\bar q_p \gamma^\mu T^A q_r)(\bar d_s \gamma_\mu T^A d_t)$\\
\end{tabular}
\end{minipage}

\end{adjustbox}

\vspace{0.25cm}

\begin{adjustbox}{width=0.9\textwidth,left}

\begin{minipage}[t]{3.75cm}
\renewcommand{\arraystretch}{1.5}
\begin{tabular}[t]{c|c}
\multicolumn{2}{c}{\boldmath$8:(\bar LR)(\bar RL)+\hc$} \\
\hline
$Q_{ledq}$ & $(\bar l_p^j e_r)(\bar d_s q_{tj})$ 
\end{tabular}
\end{minipage}
\hspace{0.4cm}
\begin{minipage}[t]{5.5cm}
\renewcommand{\arraystretch}{1.5}
\begin{tabular}[t]{c|c}
\multicolumn{2}{c}{\boldmath$8:(\bar LR)(\bar L R)+\hc$} \\
\hline
$Q_{quqd}^{(1)}$ & $(\bar q_p^j u_r) \epsilon_{jk} (\bar q_s^k d_t)$ \\
$Q_{quqd}^{(8)}$ & $(\bar q_p^j T^A u_r) \epsilon_{jk} (\bar q_s^k T^A d_t)$ \\
$Q_{lequ}^{(1)}$ & $(\bar l_p^j e_r) \epsilon_{jk} (\bar q_s^k u_t)$ \\
$Q_{lequ}^{(3)}$ & $(\bar l_p^j \sigma_{\mu\nu} e_r) \epsilon_{jk} (\bar q_s^k \sigma^{\mu\nu} u_t)$
\end{tabular}
\end{minipage}
\hspace{0.4cm}
\begin{minipage}[t]{5.2cm}
\renewcommand{\arraystretch}{1.5}
\begin{tabular}[t]{c|c}
\multicolumn{2}{c}{\boldmath{$\Delta B = \Delta L = 1 +\hc$}} \\
\hline
$Q_{duql}$      & $\epsilon^{\alpha \beta \gamma} \epsilon^{ij} (d^T_{\alpha p} C u_{\beta r} ) (q^T_{\gamma i s} C l_{jt})  $  \\
$Q_{qque}$      & $\epsilon^{\alpha \beta \gamma} \epsilon^{ij} (q^T_{\alpha i p} C q_{\beta j r} ) (u^T_{\gamma s} C e_{t})  $  \\
$Q_{qqql}$      & $\epsilon^{\alpha \beta \gamma} \epsilon^{i\ell} \epsilon^{jk} (q^T_{\alpha i p} C q_{\beta j r} ) (q^T_{\gamma k s} C l_{\ell t})  $  \\
$Q_{duue}$      & $\epsilon^{\alpha \beta \gamma} (d^T_{\alpha p} C u_{\beta r} ) (u^T_{\gamma s} C e_{t})  $  \\
\end{tabular}
\end{minipage}
\end{adjustbox}

\end{center}
\caption{The dimension-six operators in SMEFT. The operators that conserve baryon and lepton number are divided into eight classes according to their field content. The class-8 $\psi^4$ four-fermion operators are further divided into subclasses according to their chiral properties. Operators with ${}+\hc$ have Hermitian conjugates. The subscripts $p, r, s, t$ are weak-eigenstate indices.}
\label{tab:smeft6ops}
\end{table}


\clearpage

\section{LEFT operator basis}
\label{sec:LEFTBasis}

We reproduce the list of LEFT operators up to dimension six from~\cite{Jenkins:2017jig}. 

\begin{table}[H]
\capstart
\begin{adjustbox}{width=0.85\textwidth,center}
\begin{minipage}[t]{3cm}
\renewcommand{\arraystretch}{1.51}
\small
\begin{align*}
\begin{array}[t]{c|c}
\multicolumn{2}{c}{\boldsymbol{\nu \nu+\hc}} \\
\hline
\O_{\nu} & (\nu_{Lp}^T C \nu_{Lr})  \\
\end{array}
\end{align*}
\end{minipage}
%
\begin{minipage}[t]{3cm}
\renewcommand{\arraystretch}{1.51}
\small
\begin{align*}
\begin{array}[t]{c|c}
\multicolumn{2}{c}{\boldsymbol{(\nu \nu) X+\hc}} \\
\hline
\O_{\nu \gamma} & (\nu_{Lp}^T C   \sigma^{\mu \nu}  \nu_{Lr})  F_{\mu \nu}  \\
\end{array}
\end{align*}
\end{minipage}
\begin{minipage}[t]{3cm}
\renewcommand{\arraystretch}{1.51}
\small
\begin{align*}
\begin{array}[t]{c|c}
\multicolumn{2}{c}{\boldsymbol{(\overline L R ) X+\hc}} \\
\hline
\O_{e \gamma} & \bar e_{Lp}   \sigma^{\mu \nu} e_{Rr}\, F_{\mu \nu}  \\
\O_{u \gamma} & \bar u_{Lp}   \sigma^{\mu \nu}  u_{Rr}\, F_{\mu \nu}   \\
\O_{d \gamma} & \bar d_{Lp}  \sigma^{\mu \nu} d_{Rr}\, F_{\mu \nu}  \\
\O_{u G} & \bar u_{Lp}   \sigma^{\mu \nu}  T^A u_{Rr}\,  G_{\mu \nu}^A  \\
\O_{d G} & \bar d_{Lp}   \sigma^{\mu \nu} T^A d_{Rr}\,  G_{\mu \nu}^A \\
\end{array}
\end{align*}
\end{minipage}
\begin{minipage}[t]{3cm}
\renewcommand{\arraystretch}{1.51}
\small
\begin{align*}
\begin{array}[t]{c|c}
\multicolumn{2}{c}{\boldsymbol{X^3}} \\
\hline
\O_G     & f^{ABC} G_\mu^{A\nu} G_\nu^{B\rho} G_\rho^{C\mu}  \\
\O_{\widetilde G} & f^{ABC} \widetilde G_\mu^{A\nu} G_\nu^{B\rho} G_\rho^{C\mu}   \\
\end{array}
\end{align*}
\end{minipage}
\end{adjustbox}
%

%
\mbox{}\\[-0.5cm]

\begin{adjustbox}{width=1.05\textwidth,center}
\begin{minipage}[t]{3cm}
\renewcommand{\arraystretch}{1.51}
\small
\begin{align*}
\begin{array}[t]{c|c}
\multicolumn{2}{c}{\boldsymbol{(\overline L L)(\overline L  L)}} \\
\hline
\op{\nu\nu}{V}{LL} & (\bar \nu_{Lp}  \gamma^\mu \nu_{Lr} )(\bar \nu_{Ls} \gamma_\mu \nu_{Lt})   \\
\op{ee}{V}{LL}       & (\bar e_{Lp}  \gamma^\mu e_{Lr})(\bar e_{Ls} \gamma_\mu e_{Lt})   \\
\op{\nu e}{V}{LL}       & (\bar \nu_{Lp} \gamma^\mu \nu_{Lr})(\bar e_{Ls}  \gamma_\mu e_{Lt})  \\
\op{\nu u}{V}{LL}       & (\bar \nu_{Lp} \gamma^\mu \nu_{Lr}) (\bar u_{Ls}  \gamma_\mu u_{Lt})  \\
\op{\nu d}{V}{LL}       & (\bar \nu_{Lp} \gamma^\mu \nu_{Lr})(\bar d_{Ls} \gamma_\mu d_{Lt})     \\
\op{eu}{V}{LL}      & (\bar e_{Lp}  \gamma^\mu e_{Lr})(\bar u_{Ls} \gamma_\mu u_{Lt})   \\
\op{ed}{V}{LL}       & (\bar e_{Lp}  \gamma^\mu e_{Lr})(\bar d_{Ls} \gamma_\mu d_{Lt})  \\
\op{\nu edu}{V}{LL}      & (\bar \nu_{Lp} \gamma^\mu e_{Lr}) (\bar d_{Ls} \gamma_\mu u_{Lt})  + \hc   \\
\op{uu}{V}{LL}        & (\bar u_{Lp} \gamma^\mu u_{Lr})(\bar u_{Ls} \gamma_\mu u_{Lt})    \\
\op{dd}{V}{LL}   & (\bar d_{Lp} \gamma^\mu d_{Lr})(\bar d_{Ls} \gamma_\mu d_{Lt})    \\
\op{ud}{V1}{LL}     & (\bar u_{Lp} \gamma^\mu u_{Lr}) (\bar d_{Ls} \gamma_\mu d_{Lt})  \\
\op{ud}{V8}{LL}     & (\bar u_{Lp} \gamma^\mu T^A u_{Lr}) (\bar d_{Ls} \gamma_\mu T^A d_{Lt})   \\[-0.5cm]
\end{array}
\end{align*}
\renewcommand{\arraystretch}{1.51}
\small
\begin{align*}
\begin{array}[t]{c|c}
\multicolumn{2}{c}{\boldsymbol{(\overline R  R)(\overline R R)}} \\
\hline
\op{ee}{V}{RR}     & (\bar e_{Rp} \gamma^\mu e_{Rr})(\bar e_{Rs} \gamma_\mu e_{Rt})  \\
\op{eu}{V}{RR}       & (\bar e_{Rp}  \gamma^\mu e_{Rr})(\bar u_{Rs} \gamma_\mu u_{Rt})   \\
\op{ed}{V}{RR}     & (\bar e_{Rp} \gamma^\mu e_{Rr})  (\bar d_{Rs} \gamma_\mu d_{Rt})   \\
\op{uu}{V}{RR}      & (\bar u_{Rp} \gamma^\mu u_{Rr})(\bar u_{Rs} \gamma_\mu u_{Rt})  \\
\op{dd}{V}{RR}      & (\bar d_{Rp} \gamma^\mu d_{Rr})(\bar d_{Rs} \gamma_\mu d_{Rt})    \\
\op{ud}{V1}{RR}       & (\bar u_{Rp} \gamma^\mu u_{Rr}) (\bar d_{Rs} \gamma_\mu d_{Rt})  \\
\op{ud}{V8}{RR}    & (\bar u_{Rp} \gamma^\mu T^A u_{Rr}) (\bar d_{Rs} \gamma_\mu T^A d_{Rt})  \\
\end{array}
\end{align*}
\end{minipage}
%
%
\begin{minipage}[t]{3cm}
\renewcommand{\arraystretch}{1.51}
\small
\begin{align*}
\begin{array}[t]{c|c}
\multicolumn{2}{c}{\boldsymbol{(\overline L  L)(\overline R  R)}} \\
\hline
\op{\nu e}{V}{LR}     & (\bar \nu_{Lp} \gamma^\mu \nu_{Lr})(\bar e_{Rs}  \gamma_\mu e_{Rt})  \\
\op{ee}{V}{LR}       & (\bar e_{Lp}  \gamma^\mu e_{Lr})(\bar e_{Rs} \gamma_\mu e_{Rt}) \\
\op{\nu u}{V}{LR}         & (\bar \nu_{Lp} \gamma^\mu \nu_{Lr})(\bar u_{Rs}  \gamma_\mu u_{Rt})    \\
\op{\nu d}{V}{LR}         & (\bar \nu_{Lp} \gamma^\mu \nu_{Lr})(\bar d_{Rs} \gamma_\mu d_{Rt})   \\
\op{eu}{V}{LR}        & (\bar e_{Lp}  \gamma^\mu e_{Lr})(\bar u_{Rs} \gamma_\mu u_{Rt})   \\
\op{ed}{V}{LR}        & (\bar e_{Lp}  \gamma^\mu e_{Lr})(\bar d_{Rs} \gamma_\mu d_{Rt})   \\
\op{ue}{V}{LR}        & (\bar u_{Lp} \gamma^\mu u_{Lr})(\bar e_{Rs}  \gamma_\mu e_{Rt})   \\
\op{de}{V}{LR}         & (\bar d_{Lp} \gamma^\mu d_{Lr}) (\bar e_{Rs} \gamma_\mu e_{Rt})   \\
\op{\nu edu}{V}{LR}        & (\bar \nu_{Lp} \gamma^\mu e_{Lr})(\bar d_{Rs} \gamma_\mu u_{Rt})  +\hc \\
\op{uu}{V1}{LR}        & (\bar u_{Lp} \gamma^\mu u_{Lr})(\bar u_{Rs} \gamma_\mu u_{Rt})   \\
\op{uu}{V8}{LR}       & (\bar u_{Lp} \gamma^\mu T^A u_{Lr})(\bar u_{Rs} \gamma_\mu T^A u_{Rt})    \\ 
\op{ud}{V1}{LR}       & (\bar u_{Lp} \gamma^\mu u_{Lr}) (\bar d_{Rs} \gamma_\mu d_{Rt})  \\
\op{ud}{V8}{LR}       & (\bar u_{Lp} \gamma^\mu T^A u_{Lr})  (\bar d_{Rs} \gamma_\mu T^A d_{Rt})  \\
\op{du}{V1}{LR}       & (\bar d_{Lp} \gamma^\mu d_{Lr})(\bar u_{Rs} \gamma_\mu u_{Rt})   \\
\op{du}{V8}{LR}       & (\bar d_{Lp} \gamma^\mu T^A d_{Lr})(\bar u_{Rs} \gamma_\mu T^A u_{Rt}) \\
\op{dd}{V1}{LR}      & (\bar d_{Lp} \gamma^\mu d_{Lr})(\bar d_{Rs} \gamma_\mu d_{Rt})  \\
\op{dd}{V8}{LR}   & (\bar d_{Lp} \gamma^\mu T^A d_{Lr})(\bar d_{Rs} \gamma_\mu T^A d_{Rt}) \\
\op{uddu}{V1}{LR}   & (\bar u_{Lp} \gamma^\mu d_{Lr})(\bar d_{Rs} \gamma_\mu u_{Rt})  + \hc  \\
\op{uddu}{V8}{LR}      & (\bar u_{Lp} \gamma^\mu T^A d_{Lr})(\bar d_{Rs} \gamma_\mu T^A  u_{Rt})  + \hc \\
\end{array}
\end{align*}
\end{minipage}

\begin{minipage}[t]{3cm}
\renewcommand{\arraystretch}{1.51}
\small
\begin{align*}
\begin{array}[t]{c|c}
\multicolumn{2}{c}{\boldsymbol{(\overline L R)(\overline L R)+\hc}} \\
\hline
\op{ee}{S}{RR} 		& (\bar e_{Lp}   e_{Rr}) (\bar e_{Ls} e_{Rt})   \\
\op{eu}{S}{RR}  & (\bar e_{Lp}   e_{Rr}) (\bar u_{Ls} u_{Rt})   \\
\op{eu}{T}{RR} & (\bar e_{Lp}   \sigma^{\mu \nu}   e_{Rr}) (\bar u_{Ls}  \sigma_{\mu \nu}  u_{Rt})  \\
\op{ed}{S}{RR}  & (\bar e_{Lp} e_{Rr})(\bar d_{Ls} d_{Rt})  \\
\op{ed}{T}{RR} & (\bar e_{Lp} \sigma^{\mu \nu} e_{Rr}) (\bar d_{Ls} \sigma_{\mu \nu} d_{Rt})   \\
\op{\nu edu}{S}{RR} & (\bar   \nu_{Lp} e_{Rr})  (\bar d_{Ls} u_{Rt} ) \\
\op{\nu edu}{T}{RR} &  (\bar  \nu_{Lp}  \sigma^{\mu \nu} e_{Rr} )  (\bar  d_{Ls}  \sigma_{\mu \nu} u_{Rt} )   \\
\op{uu}{S1}{RR}  & (\bar u_{Lp}   u_{Rr}) (\bar u_{Ls} u_{Rt})  \\
\op{uu}{S8}{RR}   & (\bar u_{Lp}   T^A u_{Rr}) (\bar u_{Ls} T^A u_{Rt})  \\
\op{ud}{S1}{RR}   & (\bar u_{Lp} u_{Rr})  (\bar d_{Ls} d_{Rt})   \\
\op{ud}{S8}{RR}  & (\bar u_{Lp} T^A u_{Rr})  (\bar d_{Ls} T^A d_{Rt})  \\
\op{dd}{S1}{RR}   & (\bar d_{Lp} d_{Rr}) (\bar d_{Ls} d_{Rt}) \\
\op{dd}{S8}{RR}  & (\bar d_{Lp} T^A d_{Rr}) (\bar d_{Ls} T^A d_{Rt})  \\
\op{uddu}{S1}{RR} &  (\bar u_{Lp} d_{Rr}) (\bar d_{Ls}  u_{Rt})   \\
\op{uddu}{S8}{RR}  &  (\bar u_{Lp} T^A d_{Rr}) (\bar d_{Ls}  T^A u_{Rt})  \\[-0.5cm]
\end{array}
\end{align*}
\renewcommand{\arraystretch}{1.51}
\small
\begin{align*}
\begin{array}[t]{c|c}
\multicolumn{2}{c}{\boldsymbol{(\overline L R)(\overline R L) +\hc}} \\
\hline
\op{eu}{S}{RL}  & (\bar e_{Lp} e_{Rr}) (\bar u_{Rs}  u_{Lt})  \\
\op{ed}{S}{RL} & (\bar e_{Lp} e_{Rr}) (\bar d_{Rs} d_{Lt}) \\
\op{\nu edu}{S}{RL}  & (\bar \nu_{Lp} e_{Rr}) (\bar d_{Rs}  u_{Lt})  \\
\end{array}
\end{align*}
\end{minipage}
\end{adjustbox}
\setlength{\belowcaptionskip}{-3cm}
\caption{LEFT operators of dimension three and five, as well as LEFT operators of dimension six that conserve baryon and lepton number.}
\label{tab:oplist1}
\end{table}

\begin{table}[H]
\capstart
%
\centering
\begin{minipage}[t]{3cm}
\renewcommand{\arraystretch}{1.5}
\small
\begin{align*}
\begin{array}[t]{c|c}
\multicolumn{2}{c}{\boldsymbol{\Delta L = 4 + \hc}}  \\
\hline
\op{\nu\nu}{S}{LL} &  (\nu_{Lp}^T C \nu_{Lr}^{}) (\nu_{Ls}^T C \nu_{Lt}^{} )  \\
\end{array}
\end{align*}
\end{minipage}

\begin{adjustbox}{width=\textwidth,center}
\begin{minipage}[t]{3cm}
\renewcommand{\arraystretch}{1.5}
\small
\begin{align*}
\begin{array}[t]{c|c}
\multicolumn{2}{c}{\boldsymbol{\Delta L =2 + \hc}}  \\
\hline
\op{\nu e}{S}{LL}  &  (\nu_{Lp}^T C \nu_{Lr}) (\bar e_{Rs} e_{Lt})   \\
\op{\nu e}{T}{LL} &  (\nu_{Lp}^T C \sigma^{\mu \nu} \nu_{Lr}) (\bar e_{Rs}\sigma_{\mu \nu} e_{Lt} )  \\
\op{\nu e}{S}{LR} &  (\nu_{Lp}^T C \nu_{Lr}) (\bar e_{Ls} e_{Rt} )  \\
\op{\nu u}{S}{LL}  &  (\nu_{Lp}^T C \nu_{Lr}) (\bar u_{Rs} u_{Lt} )  \\
\op{\nu u}{T}{LL}  &  (\nu_{Lp}^T C \sigma^{\mu \nu} \nu_{Lr}) (\bar u_{Rs} \sigma_{\mu \nu} u_{Lt} ) \\
\op{\nu u}{S}{LR}  &  (\nu_{Lp}^T C \nu_{Lr}) (\bar u_{Ls} u_{Rt} )  \\
\op{\nu d}{S}{LL}   &  (\nu_{Lp}^T C \nu_{Lr}) (\bar d_{Rs} d_{Lt} ) \\
\op{\nu d}{T}{LL}   &  (\nu_{Lp}^T C \sigma^{\mu \nu}  \nu_{Lr}) (\bar d_{Rs} \sigma_{\mu \nu} d_{Lt} ) \\
\op{\nu d}{S}{LR}  &  (\nu_{Lp}^T C \nu_{Lr}) (\bar d_{Ls} d_{Rt} ) \\
\op{\nu edu}{S}{LL} &  (\nu_{Lp}^T C e_{Lr}) (\bar d_{Rs} u_{Lt} )  \\
\op{\nu edu}{T}{LL}  & (\nu_{Lp}^T C  \sigma^{\mu \nu} e_{Lr}) (\bar d_{Rs}  \sigma_{\mu \nu} u_{Lt} ) \\
\op{\nu edu}{S}{LR}   & (\nu_{Lp}^T C e_{Lr}) (\bar d_{Ls} u_{Rt} ) \\
\op{\nu edu}{V}{RL}   & (\nu_{Lp}^T C \gamma^\mu e_{Rr}) (\bar d_{Ls} \gamma_\mu u_{Lt} )  \\
\op{\nu edu}{V}{RR}   & (\nu_{Lp}^T C \gamma^\mu e_{Rr}) (\bar d_{Rs} \gamma_\mu u_{Rt} )  \\
\end{array}
\end{align*}
\end{minipage}
%
\begin{minipage}[t]{3cm}
\renewcommand{\arraystretch}{1.5}
\small
\begin{align*}
\begin{array}[t]{c|c}
\multicolumn{2}{c}{\boldsymbol{\Delta B = \Delta L = 1 + \hc}} \\
\hline
\op{udd}{S}{LL} &  \epsilon_{\alpha\beta\gamma}  (u_{Lp}^{\alpha T} C d_{Lr}^{\beta}) (d_{Ls}^{\gamma T} C \nu_{Lt}^{})   \\
\op{duu}{S}{LL} & \epsilon_{\alpha\beta\gamma}  (d_{Lp}^{\alpha T} C u_{Lr}^{\beta}) (u_{Ls}^{\gamma T} C e_{Lt}^{})  \\
\op{uud}{S}{LR} & \epsilon_{\alpha\beta\gamma}  (u_{Lp}^{\alpha T} C u_{Lr}^{\beta}) (d_{Rs}^{\gamma T} C e_{Rt}^{})  \\
\op{duu}{S}{LR} & \epsilon_{\alpha\beta\gamma}  (d_{Lp}^{\alpha T} C u_{Lr}^{\beta}) (u_{Rs}^{\gamma T} C e_{Rt}^{})   \\
\op{uud}{S}{RL} & \epsilon_{\alpha\beta\gamma}  (u_{Rp}^{\alpha T} C u_{Rr}^{\beta}) (d_{Ls}^{\gamma T} C e_{Lt}^{})   \\
\op{duu}{S}{RL} & \epsilon_{\alpha\beta\gamma}  (d_{Rp}^{\alpha T} C u_{Rr}^{\beta}) (u_{Ls}^{\gamma T} C e_{Lt}^{})   \\
\op{dud}{S}{RL} & \epsilon_{\alpha\beta\gamma}  (d_{Rp}^{\alpha T} C u_{Rr}^{\beta}) (d_{Ls}^{\gamma T} C \nu_{Lt}^{})   \\
\op{ddu}{S}{RL} & \epsilon_{\alpha\beta\gamma}  (d_{Rp}^{\alpha T} C d_{Rr}^{\beta}) (u_{Ls}^{\gamma T} C \nu_{Lt}^{})   \\
\op{duu}{S}{RR}  & \epsilon_{\alpha\beta\gamma}  (d_{Rp}^{\alpha T} C u_{Rr}^{\beta}) (u_{Rs}^{\gamma T} C e_{Rt}^{})  \\
\end{array}
\end{align*}
\end{minipage}
%
\begin{minipage}[t]{3cm}
\renewcommand{\arraystretch}{1.5}
\small
\begin{align*}
\begin{array}[t]{c|c}
\multicolumn{2}{c}{\boldsymbol{\Delta B = - \Delta L = 1 + \hc}}  \\
\hline
\op{ddd}{S}{LL} & \epsilon_{\alpha\beta\gamma}  (d_{Lp}^{\alpha T} C d_{Lr}^{\beta}) (\bar e_{Rs}^{} d_{Lt}^\gamma )  \\
\op{udd}{S}{LR}  & \epsilon_{\alpha\beta\gamma}  (u_{Lp}^{\alpha T} C d_{Lr}^{\beta}) (\bar \nu_{Ls}^{} d_{Rt}^\gamma )  \\
\op{ddu}{S}{LR} & \epsilon_{\alpha\beta\gamma}  (d_{Lp}^{\alpha T} C d_{Lr}^{\beta})  (\bar \nu_{Ls}^{} u_{Rt}^\gamma )  \\
\op{ddd}{S}{LR} & \epsilon_{\alpha\beta\gamma}  (d_{Lp}^{\alpha T} C d_{Lr}^{\beta}) (\bar e_{Ls}^{} d_{Rt}^\gamma ) \\
\op{ddd}{S}{RL}  & \epsilon_{\alpha\beta\gamma}  (d_{Rp}^{\alpha T} C d_{Rr}^{\beta}) (\bar e_{Rs}^{} d_{Lt}^\gamma )  \\
\op{udd}{S}{RR}  & \epsilon_{\alpha\beta\gamma}  (u_{Rp}^{\alpha T} C d_{Rr}^{\beta}) (\bar \nu_{Ls}^{} d_{Rt}^\gamma )  \\
\op{ddd}{S}{RR}  & \epsilon_{\alpha\beta\gamma}  (d_{Rp}^{\alpha T} C d_{Rr}^{\beta}) (\bar e_{Ls}^{} d_{Rt}^\gamma )  \\
\end{array}
\end{align*}
\end{minipage}
\end{adjustbox}
%
\caption{LEFT operators of dimension six that violate baryon and/or lepton number.}
\label{tab:oplist2}
\end{table}


\section{Conventions for the supplemental material}

\label{app:code}

The complete results for the one-loop matching are provided in digital form as supplemental material, containing the coefficients of all LEFT operators listed in App.~\ref{sec:LEFTBasis} as well as the mass matrices, gauge couplings, and theta parameters. This appendix is a short documentation of the conventions and variable names that we use. The supplemental material consists of a \textsc{Mathematica} file and a subdirectory containing all the results in text form. The \textsc{Mathematica} file can be used to directly read in and further process the results.

The results for the LEFT mass matrices and dipole operators carry external flavor indices $p$ and $r$ (denoted by \verb$pi$ and \verb$ri$), the four-fermion operators carry flavor indices $p$, $r$, $s$, and $t$ (denoted by \verb$pi$, \verb$ri$, \verb$si$, \verb$ti$), see Table~\ref{tab:CodeIndices}.
Most standard variables and functions follow \textsc{Mathematica} conventions. Further code conventions for SMEFT variables are listed in Table~\ref{tab:CodeVariables}, conventions for flavor indices are given in Table~\ref{tab:CodeIndices}, and examples for the labelling of Wilson coefficients are shown in Table~\ref{tab:CodeWilsonCoefficients}, which also specifies the conventions for the ordering of the flavor indices of the coefficients of Hermitian conjugate operators.

In the results we include the contribution from the tree-level matching. The parameters in the tree-level expressions are the renormalized running \msbar{} parameters at the scale $\mu_W$. An implicit dependence of all the parameters on the matching scale is understood. The results are expressed in terms of masses, gauge couplings, and the vev, even if~\eqref{eq:GaugeBosonMasses} could be used to eliminate certain parameters. A mixed form is chosen with the intention to simplify the expressions.

In order to enable a direct comparison of the code conventions, here we print the text form of~\eqref{eq:DipoleResult} as it is given in the supplemental material:
\begin{lstlisting}[frame=trBL]
(vT*CuG[pi, ri])/Sqrt[2] + ((g3bar*mt*Cqu1[pi, t, t, ri])/16 - 
(g3bar*mt*Cqu8[pi, t, t, ri])/(32*Nc) + 
CuW[pi, ri]*(-(g3bar*mW*(8*mW^2 + mZ^2))/(48*Sqrt[2]*mZ^2) - 
(g3bar*mW*Log[\[Mu]W^2/mW^2])/(4*Sqrt[2]) + 
(g3bar*mW*(-8*mW^2 + 5*mZ^2)*Log[\[Mu]W^2/mZ^2])/(24*Sqrt[2]*mZ^2)) + 
CuG[pi, ri]*((54*mH^4*mZ^2 + mH^2*(32*mW^4 + 14*mW^2*mZ^2 + 35*mZ^4) + 
36*mZ^2*(2*mW^4 + mZ^4 - 4*mt^4*Nc))/(576*Sqrt[2]*mH^2*mZ^2*vT) + 
(3*mH^2*Log[\[Mu]W^2/mH^2])/(32*Sqrt[2]*vT) + 
(-(mt^4*Nc)/(4*Sqrt[2]*mH^2*vT) - (g3bar^2*vT)/(48*Sqrt[2]))*
Log[\[Mu]W^2/mt^2] + (3*mW^4*Log[\[Mu]W^2/mW^2])/(8*Sqrt[2]*mH^2*vT) + 
((27*mZ^6 - 4*mH^2*(4*mW^4 - 5*mW^2*mZ^2 + mZ^4))*Log[\[Mu]W^2/mZ^2])/
(144*Sqrt[2]*mH^2*mZ^2*vT)) + CuB[pi, ri]*
((g1bar*g3bar*(8*mW^2 - 5*mZ^2)*vT)/(96*Sqrt[2]*mZ^2) + 
(g1bar*g3bar*(8*mW^2 - 5*mZ^2)*vT*Log[\[Mu]W^2/mZ^2])/
(48*Sqrt[2]*mZ^2)) + (g3bar*(-mW^2 + mZ^2)*CHu[pi, ri]*mu[pi])/
(18*mZ^2) + kd[pi, ri]*((CHD*g3bar*(-17 + (32*mW^4)/mZ^4)*mu[pi])/864 + 
(g3bar*(32*mW^4 - 40*mW^2*mZ^2 + 35*mZ^4)*mu[pi])/(432*mZ^4*vT^2) + 
(CHWB*g1bar*g3bar*(8*mW^3 - 5*mW*mZ^2)*vT*mu[pi])/(108*mZ^4) + 
CHGt*(((3*I)/16)*g3bar*mu[pi] + (I/8)*g3bar*Log[\[Mu]W^2/mH^2]*mu[pi]) + 
CHG*((3*g3bar*mu[pi])/16 + (g3bar*Log[\[Mu]W^2/mH^2]*mu[pi])/8)) + 
(g3bar*(-4*mW^2 + mZ^2)*CHq1[pi, ri]*mu[ri])/(72*mZ^2) + 
(g3bar*(4*mW^2 + 5*mZ^2)*CHq3[pi, ri]*mu[ri])/(72*mZ^2))/Pi^2
\end{lstlisting}

\begin{table}[H]
	\centering
	\begin{tabular}{lllc}
		\toprule
		variable							& code name				& explanation						& equation ref.	 \\
		\midrule
		$\mu_W$ 							& \verb$\[Mu]W$			& matching scale \\
		\midrule
		$\bar g_1$, \ldots					& \verb$g1bar$, \ldots 		& rescaled gauge couplings			& \eqref{eq:GaugeFieldRescaling} \\
		$\bar \theta_1$, \ldots				& \verb$\[Theta]1bar$, \ldots	& rescaled theta parameters			& \eqref{eq:ThetaParameterRescaling} \\
		\midrule
		$M_\W, M_\Z$						& \verb$mW$, \verb$mZ$		& $\W$- and $\Z$-boson mass			& \eqref{eq:GaugeBosonMasses} \\
		$M_H$							& \verb$mH$				& $H$-boson mass					& \eqref{eq:HiggsMass} \\
		$m_t$							& \verb$mt$				& $t$-quark mass					& \eqref{eq:TreeLevelSMEFTMasses} \\
		$\{m_u,m_c,m_t\}_i$					& \verb$mu[i]$				& up-type-quark mass				& \eqref{eq:TreeLevelSMEFTMasses} \\
		$\{m_d,m_s,m_b\}_i$				& \verb$md[i]$				& down-type-quark mass				& \eqref{eq:TreeLevelSMEFTMasses} \\
		$\{m_e,m_\mu,m_\tau\}_i$			& \verb$me[i]$				& lepton mass						& \eqref{eq:TreeLevelSMEFTMasses} \\
		$\{m_{\nu_1},m_{\nu_2},m_{\nu_3}\}_i$	& \verb$m\[Nu][i]$			& neutrino mass					& \eqref{eq:TreeLevelSMEFTMasses} \\
		$v_T$							& \verb$vT$				& Higgs vev						& \eqref{eq:Vev} \\
		$V, V^\dagger$						& \verb$V$, \verb$Vdag$		& unitary quark-mixing matrix			& \eqref{eq:BasisChange} \\
		$U, U^\dagger$						& \verb$U$, \verb$Udag$		& unitary lepton-mixing matrix			& \eqref{eq:BasisChange} \\
		\midrule
		$N_c = 3$							& \verb$Nc$				& number of colors \\
		$C_F$							& \verb$cf$				& $SU(3)_c$ color factor				& \eqref{eq:ColorFactors} \\
		\midrule
		$a_\mathrm{ev}$, \ldots				& \verb$aEvan$, \ldots		& parameters defining evanescent scheme	& \eqref{eq:EvanescentOperators} \\
		\bottomrule
	\end{tabular}
	\caption{SMEFT variables appearing in the code with the one-loop matching results, which is provided as supplemental material.}
	\label{tab:CodeVariables}
\end{table}

\begin{table}[H]
	\centering
	\begin{tabular}{lll}
		\toprule
		formula							& code name				& explanation						\\
		\midrule
		$\{\}_{prst}$						& \verb$[pi,ri,si,ti]$			& fixed external flavor indices, no implicit or explicit sum \\
		$\{\}_{u}$							& \verb$[ui]$				& internal flavor index, appearing multiple times, always \\
																&& together with an explicit sum over three generations \\
																&& (including the top quark) \\
		$\sum_u$							& \verb$sum[ui]$			& explicit sum over three generations \\
		$\{\}_{uvwx}$						& \verb$[uu,vv,ww,xx]$		& internal flavor indices, always appearing in pairs, \\
																&& implicit sum over three generations \\
		$\{\}_{t}$							& \verb$[t]$				& fixed top-quark flavor index \\
		$\delta_{pr}$						& \verb$kd[pi,ri]$			& Kronecker delta \\
		\bottomrule
	\end{tabular}
	\caption{Code conventions for flavor indices.}
	\label{tab:CodeIndices}
\end{table}

\begin{table}[H]
	\centering
	\begin{tabular}{ll}
		\toprule
		coefficient													& code name			 \\
		\midrule
		$\cwc{H\widetilde WB}{}$										& \verb$CHWtB$		 \\[0.2cm]
		$\cwc{H\Box}{}$											& \verb$CHBox$		 \\[0.2cm]
		$\cwc{uB}{\dagger}[][pr]:=\Big( \cwc{uB}{}[][rp]\Big)^*$				& \verb$CuBDag[pi,ri]$		 \\[0.3cm]
		$\cwc{quqd}{(8)}[][prst]$										& \verb$Cquqd8[pi,ri,si,ti]$			\\[0.3cm]
		$\cwc{quqd}{(1)\dagger}[][prst] := \Big( \cwc{quqd}{(1)}[][rpts]\Big)^*$		& \verb$Cquqd1Dag[pi,ri,si,ti]$ \\[0.3cm]
		$\sum_u \cwc{5}{\dagger}[][pu] \cwc{5}{}[][ur]$						& \verb$C5DagC5[pi,ri]$ \\
		\ldots \\
		\bottomrule
	\end{tabular}
	\caption{Examples of SMEFT Wilson coefficients appearing in the code with the one-loop matching results, provided as supplemental material.}
	\label{tab:CodeWilsonCoefficients}
\end{table}

	
	\addcontentsline{toc}{section}{\numberline{}References}
	\bibliographystyle{utphysmod}
	\bibliography{Literature}
	
\end{document}